\pdfoutput=1
\documentclass[a4paper,11pt,twoside,openright]{book} 
\usepackage[utf8]{inputenc}
\usepackage{etex}

\usepackage{xspace} 
\usepackage[
nospace,
sort,
compress
]{cite}
\usepackage{ifpdf}

\usepackage{feynmp}
\usepackage{xkeyval} 

\usepackage{anyfontsize} 	
\usepackage{stackrel}		

\usepackage{moresize}		
\usepackage{emptypage}		

\usepackage[ddmmyyyy]{datetime}

\ifpdf
\else
\fi

\usepackage[usenames,dvipsnames]{xcolor}	

\usepackage{mathrsfs}
\usepackage{textcomp}

\usepackage{amsmath,amsfonts,latexsym, amstext, amssymb, amsthm}
\usepackage[english]{babel}
\usepackage{braket}
\usepackage{dsfont}
\usepackage{tabularx}
\usepackage{ltablex}
\usepackage{float}
\usepackage{longtable}
\usepackage{comment}
\usepackage[
tableposition=top,font=small,labelfont=sl,textfont=sl
]{caption}
\usepackage{pstricks}
\usepackage{pst-node}
\usepackage{pst-plot}
\usepackage{pst-node}
\usepackage{pst-coil}

\usepackage{youngtab}
\usepackage{mathtools}

\usepackage{multirow}

\usepackage{etoolbox}
\usepackage{subfig}
\usepackage[subfigure]{tocloft}

 \usepackage{fancyhdr}
 
\usepackage{calc}

\fancypagestyle{main}{
\fancyheadoffset[RO,LE]{0cm}
\fancyhf{}
\fancyhead[LE]{\bfseries\rightmark}								
\fancyhead[RO]{\bfseries\leftmark}
\fancyfoot[RO, LE]{\thepage}

}
\fancypagestyle{back}{
	\fancyheadoffset[RO,LE]{0cm}
	\fancyhf{}
	\fancyhead[RO]{\nouppercase{\bfseries\rightmark}}
	\fancyfoot[RO, LE]{\thepage}
	
}
\fancypagestyle{plain}{%
	\fancyhead{} 
	\fancyhf{}
	\fancyheadoffset[RO,LE]{0cm}
	\fancyfoot[RO, LE]{\thepage}
}

\usepackage[a4paper,top=3cm,bottom=3cm,
inner=3.5cm,outer=2.1cm,
marginparsep=0.25cm, 
marginparwidth=2.25cm 
]{geometry}

\preto\section{%
	\ifnum\value{section}=0\addtocontents{toc}{\vskip 10pt}
	\else\addtocontents{toc}{\vskip 4pt}
	\fi
}

\usepackage{longtable}
\usepackage{multirow}
\usepackage{pdflscape}
\usepackage[textsize=footnotesize]{todonotes}
\usepackage{mathtools}

\usepackage{latexsym,amsmath,amssymb,graphics,stmaryrd}

\usepackage{feynmp}
\usepackage{xkeyval}

\makeatletter
\define@key{Blob}{blobscale}{\def\Blobscale{#1}}%
\define@key{Blob}{blobtype}{\def\Blobtype{#1}}%
\presetkeys{Blob}{%
blobscale=0.2,%
blobtype="white"%
}{}
\makeatother

\usepackage{ifpdf}
\ifpdf
\fi

\newlength{\eqoff}
\newlength{\unitlengthhalf}
\newcommand{\marrow}[5]{%
	\fmfcmd{style_def marrow#1
		expr p = drawarrow subpath (1/4, 3/4) of p shifted 6 #2 withpen pencircle scaled 0.4;
		label.#3(btex #4 etex, point 0.5 of p shifted 6 #2);
		enddef;}
	\fmf{marrow#1,tension=0}{#5}
}

\newcommand{\Marrow}[6]{%
	\fmfcmd{style_def marrow#1
		expr p = drawarrow subpath (1/4, 3/4) of p shifted #6 #2 withpen pencircle scaled 0.4;
		label.#3(#4, point 0.5 of p shifted #6 #2);
		enddef;}
	\fmf{marrow#1,tension=0}{#5}}

\newcommand{\MarrowShift}[5]{%
	\fmfcmd{style_def marrow#1
		expr p = drawarrow subpath (1/4, 3/4) of p shifted #2 withpen pencircle scaled 0.4;
		label.#3(#4, point 0.5 of p shifted #2);
		enddef;}
	\fmf{marrow#1,tension=0}{#5}}

\newcommand{\Marrowi}[6]{%
	\fmfcmd{style_def marrow#1
		expr p = drawarrow subpath (1/4, 3/4) of p shifted #6 #2 withpen pencircle scaled 0.4;
		label.#3(#4, point 0.5 of p shifted #6 #2);
		enddef;}
	\fmfi{marrow#1,tension=0}{#5}}

\newcommand{\MarrowShifti}[5]{%
	\fmfcmd{style_def marrow#1
		expr p = drawarrow subpath (1/4, 3/4) of p shifted #2 withpen pencircle scaled 0.4;
		label.#3(#4, point 0.5 of p shifted #2);
		enddef;}
	\fmfi{marrow#1,tension=0}{#5}}

\newcommand{\propagator}[3]{
	\settoheight{\eqoff}{$\times$}%
	\setlength{\eqoff}{0.5\eqoff}%
	\addtolength{\eqoff}{-5.75\unitlength}%
	\raisebox{\eqoff}{%
		\fmfframe(5,2)(5,2){%
			\begin{fmfchar*}(15,7.5)
				\fmfforce{0 w,0.5 h}{v1}
				\fmfforce{1 w,0.5 h}{v2}
				\fmf{#1}{v1,v2}
				\fmffreeze
				\fmfposition
				\marrow{a}{down}{bot}{$p$}{v1,v2}
				\fmfiv{label=$\scriptstyle#2$,l.dist=2}{vloc(__v1)}
				\fmfiv{label=$\scriptstyle#3$,l.dist=2}{vloc(__v2)}
			\end{fmfchar*}
		}
	}
}
\newcommand{\propagatorR}[3]{
	\settoheight{\eqoff}{$\times$}%
	\setlength{\eqoff}{0.5\eqoff}%
	\addtolength{\eqoff}{-5.75\unitlength}%
	\raisebox{\eqoff}{%
		\fmfframe(5,2)(5,2){%
			\begin{fmfchar*}(15,7.5)
				\fmfforce{0 w,0.5 h}{v1}
				\fmfforce{1 w,0.5 h}{v2}
				\fmf{#1}{v1,v2}
				\fmffreeze
				\fmfposition
				\marrow{a}{down}{bot}{$p$}{v2,v1}
				\fmfiv{label=$\scriptstyle#2$,l.dist=2}{vloc(__v1)}
				\fmfiv{label=$\scriptstyle#3$,l.dist=2}{vloc(__v2)}
			\end{fmfchar*}
		}
	}
}

\newcommand{\twovertex}[4]{
	\settoheight{\eqoff}{$\times$}%
	\setlength{\eqoff}{0.5\eqoff}%
	\addtolength{\eqoff}{-4.0\unitlength}%
	\raisebox{\eqoff}{%
		\fmfframe(0,0)(-1,0){%
			\begin{fmfchar*}(20,7.5)
				\fmfforce{0 w,0.5 h}{v1}
				\fmfforce{1 w,0.5 h}{v2}
				\fmfforce{0.5 w,0.5 h}{vc}
				\fmfv{decor.shape=hexacross,decor.size=10 thin}{vc}
				\fmf{#1}{v1,vc}
				\fmf{#3}{vc,v2}
				\fmffreeze
				\fmfposition
				\fmfiv{label=$\scriptstyle#2$,label.angle=-150,label.dist=9}{vloc(__vc)}
				\fmfiv{label=$\scriptstyle#4$,label.angle=-30,label.dist=9}{vloc(__vc)}
			\end{fmfchar*}
		}
	}
}

\newcommand{\threevertexJ}[6]{
	\settoheight{\eqoff}{$\times$}%
	\setlength{\eqoff}{0.5\eqoff}%
	\addtolength{\eqoff}{-9\unitlength}%
	\raisebox{\eqoff}{%
		\fmfframe(5,1)(0,1){%
			\begin{fmfchar*}(14,14)
				\fmfforce{0 w,1 h}{v1}
				\fmfforce{1 w,0.5 h}{v2}
				\fmfforce{0 w,0 h}{v3}
				\fmf{#1,label=#2,l.side=left,l.dist=2,l.dist=2}{v1,vc}
				\fmf{#3,label=#4,l.side=right,l.dist=2,l.dist=2}{vc,v2}
				\fmf{#5,label=#6,l.side=left,l.dist=2,l.dist=2}{v3,vc}
			\end{fmfchar*}
		}
	}
}

\newcommand{\fourvertextwo}[8]{
	\settoheight{\eqoff}{$\times$}%
	\setlength{\eqoff}{0.5\eqoff}%
	\addtolength{\eqoff}{-9\unitlength}%
	\raisebox{\eqoff}{%
		\fmfframe(5,1)(0,1){%
			\begin{fmfchar*}(14,14)
				\fmfforce{0 w,1 h}{v1}
				\fmfforce{1 w,1 h}{v2}
				\fmfforce{1 w,0 h}{v3}
				\fmfforce{0 w,0 h}{v4}
				\fmf{#1}{v1,vc}
				\fmf{#3}{vc,v2}
				\fmf{#5}{vc,v3}
				\fmf{#7}{v4,vc}
				\fmffreeze
				\fmfposition
				\fmfiv{label=#2,label.angle=-20,label.dist=9}{vloc(__v1)}
				\fmfiv{label=#4,label.angle=-100,label.dist=9}{vloc(__v2)}
				\fmfiv{label=#6,label.angle=160,label.dist=9}{vloc(__v3)}
				\fmfiv{label=#8,label.angle=80,label.dist=9}{vloc(__v4)}
			\end{fmfchar*}
		}
	}
}

\newcommand{\swfone}[4][]{%
	\settoheight{\eqoff}{$\times$}%
	\setlength{\eqoff}{0.5\eqoff}%
	\addtolength{\eqoff}{-5\unitlength}%
	\raisebox{\eqoff}{%
		\fmfframe(1,1)(-1,1){%
			\begin{fmfchar*}(15,7.5)
				\fmfleft{v1}
				\fmfright{v2}
				\fmffixed{(0.5w,0)}{vc1,vc2}
				\fmf{#2}{v1,vc1}
				\fmf{#2}{vc2,v2}
				\fmf{#3}{vc1,vc2}
				\fmf{#4}{vc2,vc1}
				\fmffreeze
				\fmfposition
				\fmfipath{p[]}
				\fmfipair{vm[]}
				\fmfiset{p1}{vpath(__v1,__vc1)}
				\fmfiset{p2}{vpath(__vc1,__vc2)}
				\fmfiset{p3}{reverse vpath(__vc2,__vc1) }
				\fmfiset{p4}{vpath(__vc2,__v2)}
				\svertex{vm1}{p2}
				\svertex{vm2}{p3}
				{#1}
			\end{fmfchar*}
		}
	}
}

\newgray{ogray}{0.85}
\newgray{hatchgray}{1}
\newgray{sgray}{0.8}
\newgray{hiddengray}{0.9}

\newcommand{\olcolor}{ogray}
\newcommand{\olfillstyle}{crosshatch*}
\newcommand{\olhatchcolor}{ogray}
\newcommand{\orcolor}{ogray}
\newcommand{\orfillstyle}{crosshatch*}
\newcommand{\orhatchcolor}{ogray}
\newcommand{\sfillstyle}{solid}

\newlength{\unit}
\newlength{\rad}
\newlength{\roff}
\newlength{\ri}
\psset{xunit=\unit,yunit=\unit,runit=\unit}
\newlength{\linew}
\newlength{\blacklinew}
\newlength{\dlinewidth}
\newlength{\doublesep}
\psset{doublesep=\doublesep}
\psset{linewidth=\linew}
\psset{dotscale=0.8}
\newlength{\auxlen}
\newlength{\linearc}
\newlength{\flinearc}
\newlength{\xa}
\newlength{\ya}
\newlength{\xb}
\newlength{\yb}
\newlength{\xc}
\newlength{\yc}
\newlength{\xd}
\newlength{\yd}
\newlength{\xe}
\newlength{\ye}
\newlength{\xf}
\newlength{\yf}

\newlength{\yg}

\newlength{\yh}

\newcommand{\uoneprop}[4][white]{%
	\setlength{\xa}{0\unit}
	\addtolength{\xa}{-1\unit}
	\setlength{\xb}{-.6\unit}
	\addtolength{\xb}{-0.5\dlinewidth}
	\setlength{\xc}{0.6\unit}
	\addtolength{\xc}{0.5\dlinewidth}
	\setlength{\xd}{0\unit}
	\addtolength{\xd}{1\unit}
	\setlength{\ya}{0\unit}
	\addtolength{\ya}{-1\unit}
	\setlength{\yb}{0\unit}
	\addtolength{\yb}{-0.5\dlinewidth}
	\setlength{\yc}{0\unit}
	\addtolength{\yc}{0.5\dlinewidth}
	\setlength{\yd}{0\unit}
	\addtolength{\yd}{1\unit}
	\psset{doubleline=false}
	\rput{0}(#2\unit,#3\unit){%
		\pscustom[fillstyle=\sfillstyle,fillcolor=#1,linecolor=#1,linewidth=0pt]{%
			\rotate{#4}
			\psbezier[liftpen=1,linearc=\linearc](\xb,\yc)(0,\yc)(0,\yb)(\xb,\yb)
		}
		\pscustom[fillstyle=\sfillstyle,fillcolor=#1,linecolor=#1,linewidth=0pt]{%
			\rotate{#4}
			\psbezier[liftpen=2,linearc=\linearc](\xc,\yb)(0,\yb)(0,\yc)(\xc,\yc)
		}
		\pscustom{%
			\rotate{#4}
			\psbezier[linearc=\linearc](\xb,\yc)(-0.1,\yc)(-0.1,\yb)(\xb,\yb)
			\psbezier[liftpen=2,linearc=\linearc](\xc,\yb)(0.1,\yb)(0.1,\yc)(\xc,\yc)
		}
	}
}
	
\newcommand{\ulinsert}[3][white]{%
	\setlength{\xa}{#2\unit}
	\addtolength{\xa}{0\unit}
	\setlength{\xb}{#2\unit}
	\addtolength{\xb}{1.5\unit}
	\setlength{\xc}{#2\unit}
	\addtolength{\xc}{3\unit}
	\setlength{\ya}{#3\unit}
	\addtolength{\ya}{0.5\dlinewidth}
	\setlength{\yb}{#3\unit}
	\addtolength{\yb}{-0.5\dlinewidth}
	\setlength{\yc}{#3\unit}
	\addtolength{\yc}{-1.5\unit}
	\psset{doubleline=false}
	\pscustom[fillstyle=\sfillstyle,fillcolor=#1,linecolor=#1,linewidth=0pt]{%
		\psline[liftpen=1,linearc=\linearc](\xc,\yb)(\xb,\yb)(\xb,\yc)
		\psline[liftpen=1](\xb,\ya)(\xc,\ya)}
	\pscustom[fillstyle=\olfillstyle,fillcolor=\olcolor,hatchcolor=\olhatchcolor,
	linecolor=\olcolor,linewidth=\linew]{%
		\psline[linearc=\linearc](\xb,\yc)(\xb,\ya)
		\psline[liftpen=1,linearc=2\linearc](\xb,\ya)(\xa,\ya)(\xa,\yc)}
	\psline[linearc=\linearc,linewidth=\blacklinew](\xb,\yc)(\xb,\yb)(\xc,\yb)
	\psline[linearc=2\linearc,linewidth=\blacklinew](\xa,\yc)(\xa,\ya)(\xb,\ya)
	\psline[linewidth=\blacklinew](\xb,\ya)(\xc,\ya)
}

\newcommand{\dlinsert}[3][white]{%
	\setlength{\xa}{#2\unit}
	\addtolength{\xa}{0\unit}
	\setlength{\xb}{#2\unit}
	\addtolength{\xb}{1.5\unit}
	\setlength{\xc}{#2\unit}
	\addtolength{\xc}{3\unit}
	\setlength{\ya}{#3\unit}
	\addtolength{\ya}{-0.5\dlinewidth}
	\setlength{\yb}{#3\unit}
	\addtolength{\yb}{0.5\dlinewidth}
	\setlength{\yc}{#3\unit}
	\addtolength{\yc}{1.5\unit}
	\psset{doubleline=false}
	\pscustom[fillstyle=\sfillstyle,fillcolor=#1,linecolor=#1,linewidth=0pt]{%
		\psline[linearc=\linearc](\xc,\yb)(\xb,\yb)(\xb,\yc)
		\psline(\xb,\ya)(\xc,\ya)}
	\pscustom[fillstyle=\olfillstyle,fillcolor=\olcolor,hatchcolor=\olhatchcolor,
	linecolor=\olcolor,linewidth=\linew]{%
		\psline[linearc=\linearc](\xb,\yc)(\xb,\ya)
		\psline[liftpen=1,linearc=2\linearc](\xb,\ya)(\xa,\ya)(\xa,\yc)}
	\psline[linearc=\linearc,linewidth=\blacklinew](\xb,\yc)(\xb,\yb)(\xc,\yb)
	\psline[linearc=2\linearc,linewidth=\blacklinew](\xa,\yc)(\xa,\ya)(\xb,\ya)
	\psline[linewidth=\blacklinew](\xb,\ya)(\xc,\ya)
}
\newcommand{\drinsert}[3][white]{%
	\setlength{\xa}{#2\unit}
	\addtolength{\xa}{0\unit}
	\setlength{\xb}{#2\unit}
	\addtolength{\xb}{-1.5\unit}
	\setlength{\xc}{#2\unit}
	\addtolength{\xc}{-3\unit}
	\setlength{\ya}{#3\unit}
	\addtolength{\ya}{-0.5\dlinewidth}
	\setlength{\yb}{#3\unit}
	\addtolength{\yb}{0.5\dlinewidth}
	\setlength{\yc}{#3\unit}
	\addtolength{\yc}{1.5\unit}
	\psset{doubleline=false}
	\pscustom[fillstyle=\sfillstyle,fillcolor=#1,linecolor=#1,linewidth=0pt]{%
		\psline[linearc=\linearc](\xc,\yb)(\xb,\yb)(\xb,\yc)
		\psline(\xb,\ya)(\xc,\ya)}
	\pscustom[fillstyle=\orfillstyle,fillcolor=\orcolor,hatchcolor=\orhatchcolor,
	linecolor=\orcolor,linewidth=\linew]{%
		\psline[linearc=\linearc](\xb,\yc)(\xb,\ya)
		\psline[liftpen=1,linearc=2\linearc](\xb,\ya)(\xa,\ya)(\xa,\yc)}
	\psline[linearc=\linearc,linewidth=\blacklinew](\xb,\yc)(\xb,\yb)(\xc,\yb)
	\psline[linearc=2\linearc,linewidth=\blacklinew](\xa,\yc)(\xa,\ya)(\xb,\ya)
	\psline[linewidth=\blacklinew](\xb,\ya)(\xc,\ya)
}
\newcommand{\urinsert}[3][white]{%
	\setlength{\xa}{#2\unit}
	\addtolength{\xa}{0\unit}
	\setlength{\xb}{#2\unit}
	\addtolength{\xb}{-1.5\unit}
	\setlength{\xc}{#2\unit}
	\addtolength{\xc}{-3\unit}
	\setlength{\ya}{#3\unit}
	\addtolength{\ya}{0.5\dlinewidth}
	\setlength{\yb}{#3\unit}
	\addtolength{\yb}{-0.5\dlinewidth}
	\setlength{\yc}{#3\unit}
	\addtolength{\yc}{-1.5\unit}
	\psset{doubleline=false}
	\pscustom[fillstyle=\sfillstyle,fillcolor=#1,linecolor=#1,linewidth=0pt]{%
		\psline[linearc=\linearc](\xc,\yb)(\xb,\yb)(\xb,\yc)
		\psline[liftpen=1](\xb,\ya)(\xc,\ya)}
	\pscustom[fillstyle=\orfillstyle,fillcolor=\orcolor,hatchcolor=\orhatchcolor,
	linecolor=\orcolor,linewidth=\linew]{%
		\psline[linearc=\linearc](\xb,\yc)(\xb,\ya)
		\psline[liftpen=1,linearc=2\linearc](\xb,\ya)(\xa,\ya)(\xa,\yc)}
	\psline[linearc=\linearc,linewidth=\blacklinew](\xb,\yc)(\xb,\yb)(\xc,\yb)
	\psline[linearc=2\linearc,linewidth=\blacklinew](\xa,\yc)(\xa,\ya)(\xb,\ya)
	\psline[linewidth=\blacklinew](\xb,\ya)(\xc,\ya)
}
\newcommand{\olvertex}[3][white]{%
	\setlength{\xa}{#2\unit}
	\addtolength{\xa}{0\unit}
	\setlength{\xb}{#2\unit}
	\addtolength{\xb}{1.5\unit}
	\setlength{\xc}{#2\unit}
	\addtolength{\xc}{3\unit}
	\setlength{\ya}{#3\unit}
	\addtolength{\ya}{1.5\unit}
	\setlength{\yb}{#3\unit}
	\addtolength{\yb}{0.5\dlinewidth}
	\setlength{\yc}{#3\unit}
	\addtolength{\yc}{-0.5\dlinewidth}
	\setlength{\yd}{#3\unit}
	\addtolength{\yd}{-1.5\unit}
	\psset{doubleline=false}
	\pscustom[fillstyle=\sfillstyle,fillcolor=#1,linecolor=#1,linewidth=0pt]{%
		\psline[linearc=\linearc](\xc,\yb)(\xb,\yb)(\xb,\ya)
		\psline[liftpen=1,linearc=\linearc](\xb,\yd)(\xb,\yc)(\xc,\yc)}
	\pscustom[fillstyle=\olfillstyle,fillcolor=\olcolor,hatchcolor=\olhatchcolor,
	linecolor=\olcolor,linewidth=0pt]{%
		\psline[liftpen=0](\xa,\yd)(\xb,\yd)
		\psline[liftpen=0](\xb,\ya)(\xa,\ya)
	}
	\psline[linecolor=\olcolor,linewidth=\linew](\xb,\ya)(\xb,\yd)
	\psline[linearc=\linearc,linewidth=\blacklinew]{-C}(\xc,\yb)(\xb,\yb)(\xb,\ya)
	\psline[liftpen=1,linearc=\linearc,linewidth=\blacklinew](\xb,\yd)(\xb,\yc)(\xc,\yc)
	
	\psline[linewidth=\blacklinew]{C-}(\xa,\ya)(\xa,\yd)
}
\newcommand{\orvertex}[3][white]{%
	\setlength{\xa}{#2\unit}
	\addtolength{\xa}{0\unit}
	\setlength{\xb}{#2\unit}
	\addtolength{\xb}{-1.5\unit}
	\setlength{\xc}{#2\unit}
	\addtolength{\xc}{-3\unit}
	\setlength{\ya}{#3\unit}
	\addtolength{\ya}{1.5\unit}
	\setlength{\yb}{#3\unit}
	\addtolength{\yb}{0.5\dlinewidth}
	\setlength{\yc}{#3\unit}
	\addtolength{\yc}{-0.5\dlinewidth}
	\setlength{\yd}{#3\unit}
	\addtolength{\yd}{-1.5\unit}
	\psset{doubleline=false}
	\pscustom[fillstyle=\sfillstyle,fillcolor=#1,linecolor=#1,linewidth=0pt]{%
		\psline[linearc=\linearc](\xc,\yb)(\xb,\yb)(\xb,\ya)
		\psline[liftpen=1,linearc=\linearc](\xb,\yd)(\xb,\yc)(\xc,\yc)}
	\pscustom[fillstyle=\orfillstyle,fillcolor=\orcolor,hatchcolor=\orhatchcolor,
	linecolor=\orcolor,linewidth=0pt]{%
		\psline[liftpen=0](\xb,\ya)(\xa,\ya)
		\psline[liftpen=0](\xa,\yd)(\xb,\yd)
		\psline[liftpen=0](\xb,\ya)(\xa,\ya)
	}
	\psline[linecolor=\orcolor,linewidth=\linew](\xb,\ya)(\xb,\yd)
	\psline[linearc=\linearc,linewidth=\blacklinew]{-C}(\xc,\yb)(\xb,\yb)(\xb,\ya)
	\psline[liftpen=1,linearc=\linearc,linewidth=\blacklinew](\xb,\yd)(\xb,\yc)(\xc,\yc)
	\psline[linewidth=\blacklinew]{C-}(\xa,\ya)(\xa,\yd)
}

\newcommand{\threevertex}[4][white]{%
	\setlength{\xa}{0\unit}
	\addtolength{\xa}{-0.5\dlinewidth}
	\setlength{\xb}{0\unit}
	\addtolength{\xb}{0.5\dlinewidth}
	\setlength{\xc}{0\unit}
	\addtolength{\xc}{1\unit}
	\setlength{\ya}{0\unit}
	\addtolength{\ya}{1\unit}
	\setlength{\yb}{0\unit}
	\addtolength{\yb}{0.5\dlinewidth}
	\setlength{\yc}{0\unit}
	\addtolength{\yc}{-0.5\dlinewidth}
	\setlength{\yd}{0\unit}
	\addtolength{\yd}{-1\unit}
	\psset{doubleline=false}
	\rput{0}(#2\unit,#3\unit){%
		\pscustom[fillstyle=\sfillstyle,fillcolor=#1,linecolor=#1,linewidth=0pt]{%
			\rotate{#4}
			\psline[liftpen=1,linearc=\linearc](\xb,\ya)(\xb,\yb)(\xc,\yb)
			\psline[liftpen=1,linearc=\linearc](\xc,\yc)(\xb,\yc)(\xb,\yd)
			\psline[liftpen=1](\xa,\yd)(\xa,\ya)}
		\pscustom{%
			\rotate{#4}
			\psline[liftpen=2,linearc=\linearc](\xb,\ya)(\xb,\yb)(\xc,\yb)
			\psline[liftpen=2,linearc=\linearc](\xc,\yc)(\xb,\yc)(\xb,\yd)
			\psline[liftpen=2](\xa,\yd)(\xa,\ya)
		}
	}
}

\newcommand{\recthreevertex}[4][white]{%
	\setlength{\xa}{-0.7071\auxlen}
	\addtolength{\xa}{-0.7071\dlinewidth}
	\setlength{\xb}{0\unit}
	\addtolength{\xb}{-0.7071\auxlen}
	\setlength{\xc}{-0.7071\auxlen}
	\addtolength{\xc}{0.7071\unit}
	\setlength{\xd}{-0.7071\auxlen}
	\addtolength{\xd}{1.2071\unit}
	
	\setlength{\ya}{-0.7071\auxlen}
	\addtolength{\ya}{-0.7071\dlinewidth}
	\setlength{\yb}{0\unit}
	\addtolength{\yb}{-0.7071\auxlen}
	\setlength{\yc}{-0.7071\auxlen}
	\addtolength{\yc}{-0.7071\dlinewidth}
	\addtolength{\yc}{0.7071\unit}
	\setlength{\yd}{0.7071\auxlen}
	\addtolength{\yd}{0.7071\dlinewidth}
	\addtolength{\yd}{-0.7071\unit}
	\setlength{\ye}{0\unit}
	\addtolength{\ye}{0.7071\auxlen}
	\setlength{\yf}{0.7071\auxlen}
	\addtolength{\yf}{0.7071\dlinewidth}
	\psset{doubleline=false}
	\rput{0}(#2\unit,#3\unit){%
		\pscustom[fillstyle=\sfillstyle,fillcolor=#1,linecolor=#1,linewidth=0pt]{%
			\rotate{#4}
			\psline[liftpen=1,linearc=\linearc](\xa,\yb)(-0.7071\dlinewidth,0)(\xa,\ye)
			\psline[liftpen=1,linearc=\linearc](\xb,\yf)(\xc,\yd)(\xd,\yd)
			\psline[liftpen=1,linearc=\linearc](\xd,\yc)(\xc,\yc)(\xb,\ya)}
		\pscustom{%
			\rotate{#4}
			\psline[liftpen=1,linearc=\linearc](\xa,\yb)(-0.7071\dlinewidth,0)(\xa,\ye)
			\psline[liftpen=2,linearc=\linearc](\xb,\yf)(\xc,\yd)(\xd,\yd)
			\psline[liftpen=2,linearc=\linearc](\xd,\yc)(\xc,\yc)(\xb,\ya)
		}
	}
}

\newcommand{\recthreevertexoneside}[4][white]{%
	\setlength{\xa}{-0.7071\auxlen}
	\addtolength{\xa}{-0.7071\dlinewidth}
	\setlength{\xb}{0\unit}
	\addtolength{\xb}{-0.7071\auxlen}
	\setlength{\xc}{-0.7071\auxlen}
	\addtolength{\xc}{-0.353505\unit}
	\setlength{\xd}{-\auxlen}
	\addtolength{\xd}{-0.5\dlinewidth}
	\setlength{\ya}{-0.7071\auxlen}
	\addtolength{\ya}{-0.7071\dlinewidth}
	\setlength{\yb}{0\unit}
	\addtolength{\yb}{-0.7071\auxlen}
	\setlength{\yc}{-0.7071\auxlen}
	\addtolength{\yc}{-0.7071\dlinewidth}
	\addtolength{\yc}{0.7071\unit}
	\setlength{\yd}{0.7071\auxlen}
	\addtolength{\yd}{0.7071\dlinewidth}
	\addtolength{\yd}{-0.7071\unit}
	\setlength{\ye}{0\unit}
	\addtolength{\ye}{0.7071\auxlen}
	\setlength{\yf}{0.7071\auxlen}
	\addtolength{\yf}{0.7071\dlinewidth}
	\psset{doubleline=false}
	\rput{0}(#2\unit,#3\unit){%
		\pscustom[fillstyle=\sfillstyle,fillcolor=#1,linecolor=#1,linewidth=0pt]{%
			\rotate{#4}
			\psline[liftpen=1,linearc=\linearc](\xb,\yf)(0.7071\dlinewidth,0)(\xb,\ya)
			\psline[liftpen=1,linearc=0.2\linearc](\xa,\yb)(\xc,\yc)(\xd,\yc)
			\psline[liftpen=1,linearc=0.2\linearc](\xd,\yd)(\xc,\yd)(\xa,\ye)}
		\pscustom{%
			\rotate{#4}
			\psline[liftpen=1,linearc=\linearc](\xb,\yf)(0.7071\dlinewidth,0)(\xb,\ya)
			\psline[liftpen=2,linearc=0.2\linearc](\xa,\yb)(\xc,\yc)(\xd,\yc)
			\psline[liftpen=2,linearc=0.2\linearc](\xd,\yd)(\xc,\yd)(\xa,\ye)
		}
	}
}
\newcommand{\fourvertex}[4][white]{%
	\setlength{\xa}{0\unit}
	\addtolength{\xa}{-1\unit}
	\setlength{\xb}{0\unit}
	\addtolength{\xb}{-0.5\dlinewidth}
	\setlength{\xc}{0\unit}
	\addtolength{\xc}{0.5\dlinewidth}
	\setlength{\xd}{0\unit}
	\addtolength{\xd}{1\unit}
	\setlength{\ya}{0\unit}
	\addtolength{\ya}{-1\unit}
	\setlength{\yb}{0\unit}
	\addtolength{\yb}{-0.5\dlinewidth}
	\setlength{\yc}{0\unit}
	\addtolength{\yc}{0.5\dlinewidth}
	\setlength{\yd}{0\unit}
	\addtolength{\yd}{1\unit}
	\psset{doubleline=false}
	\rput{0}(#2\unit,#3\unit){%
		\pscustom[fillstyle=\sfillstyle,fillcolor=#1,linecolor=#1,linewidth=0pt]{%
			\rotate{#4}
			\psline[liftpen=1,linearc=\linearc](\xc,\ya)(\xc,\yb)(\xd,\yb)
			\psline[liftpen=1,linearc=\linearc](\xd,\yc)(\xc,\yc)(\xc,\yd)
			\psline[liftpen=1,linearc=\linearc](\xb,\yd)(\xb,\yc)(\xa,\yc)
			\psline[liftpen=1,linearc=\linearc](\xa,\yb)(\xb,\yb)(\xb,\ya)}
		\pscustom{%
			\rotate{#4}
			\psline[liftpen=1,linearc=\linearc](\xc,\ya)(\xc,\yb)(\xd,\yb)
			\psline[liftpen=2,linearc=\linearc](\xd,\yc)(\xc,\yc)(\xc,\yd)
			\psline[liftpen=2,linearc=\linearc](\xb,\yd)(\xb,\yc)(\xa,\yc)
			\psline[liftpen=2,linearc=\linearc](\xa,\yb)(\xb,\yb)(\xb,\ya)
		}
	}
}

\newcommand{\fourvertexdbltr}[4][white]{%
	\setlength{\xa}{0\unit}
	\addtolength{\xa}{-1\unit}
	\setlength{\xb}{0\unit}
	\addtolength{\xb}{-0.5\dlinewidth}
	\setlength{\xc}{0\unit}
	\addtolength{\xc}{0.5\dlinewidth}
	\setlength{\xd}{0\unit}
	\addtolength{\xd}{1\unit}
	\setlength{\ya}{0\unit}
	\addtolength{\ya}{-1\unit}
	\setlength{\yb}{0\unit}
	\addtolength{\yb}{-0.5\dlinewidth}
	\setlength{\yc}{0\unit}
	\addtolength{\yc}{0.5\dlinewidth}
	\setlength{\yd}{0\unit}
	\addtolength{\yd}{1\unit}
	\psset{doubleline=false}
	\rput{0}(#2\unit,#3\unit){%
		\pscustom[fillstyle=\sfillstyle,fillcolor=#1,linecolor=#1,linewidth=0pt]{%
			\rotate{#4}
			\psline[liftpen=1,linearc=1.5\linearc](\xc,\ya)(0,0)(\xa,\yc)
			\psline[liftpen=1,linearc=\linearc](\xd,\yc)(\xc,\yc)(\xc,\yd)
			\psline[liftpen=1,linearc=1.5\linearc](\xb,\yd)(0,0)(\xa,\yc)
			\psline[liftpen=1,linearc=\linearc](\xa,\yb)(\xb,\yb)(\xb,\ya)}
		\pscustom{%
			\rotate{#4}
			\psline[liftpen=1,linearc=1.5\linearc](\xc,\ya)(0,0)(\xa,\yc)
			\psline[liftpen=2,linearc=\linearc](\xd,\yc)(\xc,\yc)(\xc,\yd)
			\psline[liftpen=2,linearc=1.5\linearc](\xb,\yd)(0,0)(\xd,\yb)
			\psline[liftpen=2,linearc=\linearc](\xa,\yb)(\xb,\yb)(\xb,\ya)
		}
	}
}

\newcommand{\uoutex}[3][white]{%
	\setlength{\xb}{#2\unit}
	\addtolength{\xb}{-1.5\unit}
	\setlength{\xc}{#2\unit}
	\setlength{\ya}{#3\unit}
	\addtolength{\ya}{0.5\dlinewidth}
	\setlength{\yb}{#3\unit}
	\addtolength{\yb}{-0.5\dlinewidth}
	\setlength{\yc}{#3\unit}
	\addtolength{\yc}{-1.5\unit}
	\psset{doubleline=false}
	\pscustom[fillstyle=\sfillstyle,fillcolor=#1,linecolor=#1]{%
		\psline[liftpen=1,linearc=\linearc](\xc,\yb)(\xb,\yb)(\xb,\yc)
		\psline[liftpen=1](\xb,\ya)(\xc,\ya)}
	\psline[linearc=\linearc,linewidth=\blacklinew](\xb,\yc)(\xb,\yb)(\xc,\yb)
	\psline[linewidth=\blacklinew](\xb,\ya)(\xc,\ya)
	
}
\newcommand{\doutex}[3][white]{%
	\setlength{\xb}{#2\unit}
	\addtolength{\xb}{-1.5\unit}
	\setlength{\xc}{#2\unit}
	\setlength{\ya}{#3\unit}
	\addtolength{\ya}{-0.5\dlinewidth}
	\setlength{\yb}{#3\unit}
	\addtolength{\yb}{0.5\dlinewidth}
	\setlength{\yc}{#3\unit}
	\addtolength{\yc}{1.5\unit}
	\psset{doubleline=false}
	\pscustom[fillstyle=\sfillstyle,fillcolor=#1,linecolor=#1]{%
		\psline[linearc=\linearc](\xc,\yb)(\xb,\yb)(\xb,\yc)
		\psline(\xb,\ya)(\xc,\ya)}
	\psline[linearc=\linearc,linewidth=\blacklinew](\xb,\yc)(\xb,\yb)(\xc,\yb)
	\psline[linewidth=\blacklinew](\xb,\ya)(\xc,\ya)
}
\newcommand{\dinex}[3][white]{%
	\setlength{\xb}{#2\unit}
	\addtolength{\xb}{1.5\unit}
	\setlength{\xc}{#2\unit}
	\setlength{\ya}{#3\unit}
	\addtolength{\ya}{-0.5\dlinewidth}
	\setlength{\yb}{#3\unit}
	\addtolength{\yb}{0.5\dlinewidth}
	\setlength{\yc}{#3\unit}
	\addtolength{\yc}{1.5\unit}
	\psset{doubleline=false}
	\pscustom[fillstyle=\sfillstyle,fillcolor=#1,linecolor=#1]{%
		\psline[linearc=\linearc](\xc,\yb)(\xb,\yb)(\xb,\yc)
		\psline(\xb,\ya)(\xc,\ya)}
	\psline[linearc=\linearc,linewidth=\blacklinew](\xb,\yc)(\xb,\yb)(\xc,\yb)
	\psline[linewidth=\blacklinew](\xb,\ya)(\xc,\ya)
}
\newcommand{\uinex}[3][white]{%
	\setlength{\xb}{#2\unit}
	\addtolength{\xb}{1.5\unit}
	\setlength{\xc}{#2\unit}
	\setlength{\ya}{#3\unit}
	\addtolength{\ya}{0.5\dlinewidth}
	\setlength{\yb}{#3\unit}
	\addtolength{\yb}{-0.5\dlinewidth}
	\setlength{\yc}{#3\unit}
	\addtolength{\yc}{-1.5\unit}
	\psset{doubleline=false}
	\pscustom[fillstyle=\sfillstyle,fillcolor=#1,linecolor=#1]{%
		\psline[linearc=\linearc](\xc,\yb)(\xb,\yb)(\xb,\yc)
		\psline[liftpen=1](\xb,\ya)(\xc,\ya)}
	\psline[linearc=\linearc,linewidth=\blacklinew](\xb,\yc)(\xb,\yb)(\xc,\yb)
	\psline[linewidth=\blacklinew](\xb,\ya)(\xc,\ya)
}
\newcommand{\ioutex}[3][white]{%
	\setlength{\xb}{#2\unit}
	\addtolength{\xb}{-1.5\unit}
	\setlength{\xc}{#2\unit}
	\setlength{\ya}{#3\unit}
	\addtolength{\ya}{1.5\unit}
	\setlength{\yb}{#3\unit}
	\addtolength{\yb}{0.5\dlinewidth}
	\setlength{\yc}{#3\unit}
	\addtolength{\yc}{-0.5\dlinewidth}
	\setlength{\yd}{#3\unit}
	\addtolength{\yd}{-1.5\unit}
	\psset{doubleline=false}
	\pscustom[fillstyle=\sfillstyle,fillcolor=#1,linecolor=#1]{%
		\psline[linearc=\linearc](\xc,\yb)(\xb,\yb)(\xb,\ya)
		\psline[liftpen=1,linearc=\linearc](\xb,\yd)(\xb,\yc)(\xc,\yc)}
	\psline[linecolor=#1,linewidth=\blacklinew](\xb,\ya)(\xb,\yd)
	\psline[linearc=\linearc,linewidth=\blacklinew]{-C}(\xc,\yb)(\xb,\yb)(\xb,\ya)
	\psline[liftpen=1,linearc=\linearc,linewidth=\blacklinew](\xb,\yd)(\xb,\yc)(\xc,\yc)
}
\newcommand{\iinex}[3][white]{%
	\setlength{\xb}{#2\unit}
	\addtolength{\xb}{1.5\unit}
	\setlength{\xc}{#2\unit}
	\setlength{\ya}{#3\unit}
	\addtolength{\ya}{1.5\unit}
	\setlength{\yb}{#3\unit}
	\addtolength{\yb}{0.5\dlinewidth}
	\setlength{\yc}{#3\unit}
	\addtolength{\yc}{-0.5\dlinewidth}
	\setlength{\yd}{#3\unit}
	\addtolength{\yd}{-1.5\unit}
	\psset{doubleline=false}
	\pscustom[fillstyle=\sfillstyle,fillcolor=#1,linecolor=#1,linewidth=0pt]{%
		\psline[linearc=\linearc](\xc,\yb)(\xb,\yb)(\xb,\ya)
		\psline[liftpen=1,linearc=\linearc](\xb,\yd)(\xb,\yc)(\xc,\yc)}
	\psline[linecolor=#1,linewidth=\blacklinew](\xb,\ya)(\xb,\yd)
	\psline[linearc=\linearc,linewidth=\blacklinew]{-C}(\xc,\yb)(\xb,\yb)(\xb,\ya)
	\psline[liftpen=1,linearc=\linearc,linewidth=\blacklinew](\xb,\yd)(\xb,\yc)(\xc,\yc)
}

\usepackage{rotating}
\newlength{\arlength}
\newlength{\arheight}

\DeclareMathOperator{\diladensity}{\mathfrak{D}}

\newcommand{\EulerPhi}{\varphi_{\text{E}}}
\newcommand{\expectationvalue}{expectation value\xspace}

\newcommand{\parderiv}[2][]{\frac{\partial #1}{\partial #2}}

\DeclareMathOperator{\one}{\mathds{1}}
\DeclareMathOperator{\Uop}{U}
\newcommand{\onee}[1]{\one_{(#1)}}
\newcommand{\order}[1]{\mathcal{O}\left(#1\right)}

\newcommand{\WR}{\text{WR}}

\DeclarePairedDelimiter\floor{\lfloor}{\rfloor}

\newcommand{\tM}{{\text{M}}}

\newcommand{\RR}{\ensuremath{\mathbb{R}}}
\newcommand{\CC}{\ensuremath{\mathbb{C}}}
\newcommand{\ZZ}{\ensuremath{\mathbb{Z}}}
\newcommand{\NN}{\ensuremath{\mathbb{N}}}

\newcommand{\G}{\ensuremath{G}\xspace}
\newcommand{\UN}{\ensuremath{\text{U}(N)}\xspace}
\newcommand{\SUN}{\ensuremath{\text{SU}(N)}\xspace}
\newcommand{\U}[1]{\ensuremath{\text{U}(#1)}\xspace}
\newcommand{\SU}[1]{\ensuremath{\text{SU}(#1)}\xspace}
\newcommand{\SO}[1]{\ensuremath{\text{SO}(#1)}\xspace}

\newcommand{\PSLs}[2]{\ensuremath{\text{PSL}(#1|#2)}\xspace}

\newcommand{\su}[1]{\ensuremath{\mathfrak{su}(#1)}\xspace}
\newcommand{\so}[1]{\ensuremath{\mathfrak{so}(#1)}\xspace}
\newcommand{\spl}[1]{\ensuremath{\mathfrak{sl}(#1)}\xspace}
\newcommand{\splbar}[1]{\ensuremath{\ol{\mathfrak{sl}}(#1)}\xspace}

\newcommand{\clifford}{\ensuremath{\text{C}\ell}\xspace}
\newcommand{\spin}[1]{\ensuremath{\mathfrak{spin}(#1)}\xspace}

\newcommand{\complexi}{\mathnormal{i}}

\newcommand{\diag}{\text{diag}}
\newcommand{\phan}[1]{\phantom{#1}}

\newcommand{\ol}[1]{\overline{#1}}

\DeclareMathOperator{\tr}{tr}

\newcommand{\cstar}{\ensuremath{\ast}}
\newcommand{\gammaE}{\gamma_{\text{E}}}

\newcommand{\ev}[1]{\langle #1 \rangle}
\newcommand{\abs}[1]{|#1|}

\newcommand{\vac}{|0\rangle}
\newcommand{\vacl}{\langle 0|}

\newcommand{\eqndot}{\, .}
\newcommand{\eqncom}{\, ,}

\newcommand{\de}{\operatorname{d}\!}

\newcommand{\appref}[1]{appendix~\ref{#1}}

\newcommand{\chapref}[1]{chapter~\ref{#1}}

\newcommand{\secref}[1]{section~\ref{#1}}

\newcommand{\subsecref}[1]{subsection~\ref{#1}}
\newcommand{\tabref}[1]{table~\ref{#1}}
\newcommand{\figref}[1]{figure~\ref{#1}}

\newcommand{\cA}{\mathcal{A}}

\newcommand{\cF}{\mathcal{F}}

\newcommand{\cN}{\mathcal{N}}
\newcommand{\cO}{\mathcal{O}}
\newcommand{\cP}{\mathcal{P}}

\newcommand{\cT}{\mathcal{T}}

\newcommand{\cZ}{\mathcal{Z}}

\newcommand{\ba}{\mathbf{a}}
\newcommand{\bb}{\mathbf{b}}
\newcommand{\bc}{\mathbf{c}}
\newcommand{\bq}{\mathbf{q}}

\newcommand{\alphadot}{\dot{\alpha}}
\newcommand{\betadot}{\dot{\beta}}

\newcommand{\rhs}{r.h.s.\xspace}
\newcommand{\lhs}{l.h.s.\xspace}
\newcommand{\cf}{c.f.\xspace}
\newcommand{\eom}{e.o.m.\xspace}
\newcommand{\RGE}{RGE\xspace}
\newcommand{\QFT}{QFT\xspace}
\newcommand{\QFTs}{QFTs\xspace}
\newcommand{\dof}{d.o.f.\xspace}
\newcommand{\tHooft}{'t~Hooft\xspace} 
\newcommand{\CFT}{CFT\xspace} 
\newcommand{\CFTs}{CFTs\xspace} 
\newcommand{\THag}{\ensuremath{T_{\text{H}}}}
\renewcommand{\H}{{\text{H}}}
\newcommand{\QCD}{{\text{QCD}}}

\newcommand{\AdSCFT}{AdS/CFT\xspace}
\newcommand{\AdSCFTc}{AdS/CFT correspondence\xspace}
\newcommand{\Polya}{P\'{o}lya\xspace}
\newcommand{\Poincare}{Poincar\'{e}\xspace}
\newcommand{\RxSt}{\ensuremath{\text{S}^3\times \mathbb{R}}\xspace}
\newcommand{\Nfour}{$\mathcal{N}=4$\xspace}
\newcommand{\NfSYM}{\Nfour SYM\xspace}
\newcommand{\NfSYMt}{\Nfour SYM theory\xspace}
\newcommand{\AdS}[1]{\ensuremath{\text{AdS}_{#1}}\xspace}

\newcommand{\MSbar}{\ensuremath{\overline{\text{MS}}}\xspace}

\newcommand{\aosc}{\ba}
\newcommand{\aoscdag}{\aosc^\dagger}
\newcommand{\bosc}{\bb}
\newcommand{\boscdag}{\bosc^\dagger}
\newcommand{\cosc}{\bc}

\newcommand{\akind}[1][ ]{\ifthenelse{\equal{#1}{}}{a}{a^{#1}}}
\newcommand{\bkind}[1][ ]{\ifthenelse{\equal{#1}{}}{b}{b^{#1}}}
\newcommand{\ckind}[1][ ]{\ifthenelse{\equal{#1}{}}{c}{c^{#1}}}
\newcommand{\dkind}[1][ ]{\ifthenelse{\equal{#1}{}}{d}{d^{#1}}}
\newcommand{\atkind}[1][ ]{{\tilde{a}^{#1}}}
\newcommand{\btkind}[1][ ]{{\tilde{b}^{#1}}}
\newcommand{\ctkind}[1][ ]{{\tilde{c}^{#1}}}
\newcommand{\akindsite}[2][ ]{a^{#1}_{(#2)}}
\newcommand{\bkindsite}[2][ ]{b^{#1}_{(#2)}}
\newcommand{\ckindsite}[2][ ]{c^{#1}_{(#2)}}

\DeclareMathOperator{\Kop}{K}
\DeclareMathOperator{\Rop}{R}

\DeclareMathOperator{\T}{T}

\newcommand{\chap}[1]{chapter #1}
\newcommand{\app}[1]{appendix #1}
\newcommand{\indups}[1]{_{\mathrm{\scriptscriptstyle #1}}}
\newcommand{\gym}{g\indups{YM}}
\newcommand{\YM}{{\mathrm{\scriptscriptstyle YM}}}
\newcommand{\maxset}[1]{\max{(#1)}}
\newcommand{\minset}[1]{\min{(#1)}}

\DeclareMathOperator{\phaneq}{\phantom{{}=}}
\newcommand{\phaneqtimes}{\qquad \times}

\newcommand{\colors}{s}

\newcommand{\e}{\operatorname{e}}
\newcommand{\numberdot}[1]{#1}

\newcommand{\mumu}{\mu\!\mu}

\DeclareMathOperator{\dop}{d}
\newcommand{\measure}[1]{\dop\! #1}
\newcommand{\Diff}[2]{\frac{\dop^{#2}}{\dop\!{#1}^{#2}}}
\newcommand{\IntOp}[3]{\int^{#3}_{#2} \!\!\! \dop\! #1 \,}

\newcommand{\ff}{\text{{\scriptsize ff}}\ensuremath{{\scriptstyle[\_]}}}
\newcommand{\fs}{\text{{\scriptsize fs}}\ensuremath{{\scriptstyle[\_]}}}
\newcommand{\expr}{\text{{\scriptsize exp}}\ensuremath{\_}}
\newcommand{\num}[1]{\text{{\scriptsize n#1}}\ensuremath{\_}}
\newcommand{\numn}{\text{{\scriptsize n}}\ensuremath{\_}}
\newcommand{\muindex}{\ensuremath{{\scriptstyle\mu[\_]}}}
\newcommand{\muind}[1]{\ensuremath{{\scriptstyle\mu[#1]}}}
\newcommand{\cindex}{\text{{\scriptsize c}}\ensuremath{{\scriptstyle[\_]}}}
\newcommand{\alphaindex}{\ensuremath{{\scriptstyle\alpha[\_]}}}
\newcommand{\betadcindex}{\ensuremath{{\scriptstyle\beta}}\text{{\scriptsize d}}\ensuremath{{\scriptstyle[\_]}}}

\newcommand{\ttt}[1]{\text{{\tt #1}}}

\DeclareMathOperator{\D}{D}

\newcommand{\svertex}[2]{%
\fmfiequ{#1}{point length(#2)/2 of (#2)}
}

\makeatletter

\define@boolkey{FDiagram}{schannel}[true]{}
\define@boolkey{FDiagram}{tchannel}[true]{}
\define@boolkey{FDiagram}{xchannel}[true]{}
\define@boolkey{FDiagram}{leftSE}[true]{}
\define@boolkey{FDiagram}{rightSE}[true]{}
\define@boolkey{FDiagram}{long}[true]{}
\define@boolkey{FDiagram}{longup}[true]{}
\define@boolkey{FDiagram}{oldlabels}[true]{}
\define@key{FDiagram}{styleleftbottom}%
{\def\FDiagramstyleleftbottom{#1}}
\define@key{FDiagram}{stylerightbottom}%
{\def\FDiagramstylerightbottom{#1}}
\define@key{FDiagram}{stylelefttop}%
{\def\FDiagramstylelefttop{#1}}
\define@key{FDiagram}{stylerighttop}%
{\def\FDiagramstylerighttop{#1}}
\define@key{FDiagram}{stylemid}%
{\def\FDiagramstylemid{#1}}
\define@key{FDiagram}{labelleftbottom}%
{\def\FDiagramlabelleftbottom{#1}}
\define@key{FDiagram}{labelrightbottom}%
{\def\FDiagramlabelrightbottom{#1}}
\define@key{FDiagram}{labellefttop}%
{\def\FDiagramlabellefttop{#1}}
\define@key{FDiagram}{labelrighttop}%
{\def\FDiagramlabelrighttop{#1}}
\define@key{FDiagram}{labelmid}%
{\def\FDiagramlabelmid{#1}}
\presetkeys{FDiagram}{%
	schannel=false,%
	tchannel=false,%
	xchannel=false,%
	leftSE=false,%
	rightSE=false,%
	long=false,%
	longup=false,%
	oldlabels=false,%
	labelleftbottom=$\phantom{0}$,%
	labelrightbottom=$\phantom{0}$,%
	labellefttop=$\phantom{0}$,%
	labelrighttop=$\phantom{0}$,%
	labelmid=$\phantom{0}$,%
	styleleftbottom=plain,%
	styleleftbottom=plain,%
	stylerightbottom=plain,%
	stylelefttop=plain,%
	stylerighttop=plain,%
	stylemid=plain}{}
\newcommand*\FDiagram[6][]{%
	\setkeys{FDiagram}{#1}%
	\settoheight{\eqoff}{$\times$}%
	\setlength{\eqoff}{0.5\eqoff}%
	\addtolength{\eqoff}{-12.0\unitlength}%
	\raisebox{\eqoff}{%
		\fmfframe(2,2)(2,2){%
			\begin{fmfchar*}(12,20)
				\fmfbottom{vb1,vbb1,v1,vb2,vbb2,vb3,vb4,v2,vbb5,vb5}
				\fmftop{vt1,vtt1,v3,vt2,vtt2,vt3,vt4,v4,vtt5,vt5}
				\ifKV@FDiagram@schannel
				\fmf{#2,left=0.3,tension=1}{v1,vc1}
				\fmf{#3,right=0.3,tension=1}{v2,vc1}
				\fmf{#4,tension=2}{vc1,vc2}
				\fmf{#5,left=0.3,foreground=(0.65,,0.65,,0.65)}{vc2,v3}
				\fmf{#6,right=0.3,foreground=(0.65,,0.65,,0.65)}{vc2,v4}
				\else\fi
				\ifKV@FDiagram@tchannel
				\fmf{#2,left=0,tension=1}{v1,vc1}
				\fmf{#3,right=0,tension=1}{v2,vc2}
				\fmf{#4,tension=0}{vc1,vc2}
				\fmf{#5,foreground=(0.65,,0.65,,0.65)}{vc1,v3}
				\fmf{#6,foreground=(0.65,,0.65,,0.65)}{vc2,v4}
				\else\fi
				\ifKV@FDiagram@xchannel
				\fmf{#2,left=0.25,tension=1}{v1,vc1}
				\fmf{#3,right=0.25,tension=1}{v2,vc1}
				\fmf{#5,left=0.25,foreground=(0.65,,0.65,,0.65)}{vc1,v3}
				\fmf{#6,right=0.25,foreground=(0.65,,0.65,,0.65)}{vc1,v4}
				\else\fi
				\ifKV@FDiagram@leftSE
				\fmf{#2,left=0,tension=1}{v1,vc1}
				\fmf{#2,right=0,tension=1,foreground=(0.65,,0.65,,0.65)}{vc1,v3}
				\fmf{#3,foreground=(0.65,,0.65,,0.65)}{v2,v4}
				\fmfv{decor.shape=circle,decor.filled=shaded,decor.size=10thin}{vc1}
				\else\fi
				\ifKV@FDiagram@rightSE
				\fmf{#3,left=0,tension=1}{v2,vc1}
				\fmf{#3,right=0,tension=1,foreground=(0.65,,0.65,,0.65)}{vc1,v4}
				\fmf{#2,foreground=(0.65,,0.65,,0.65)}{v1,v3}
				\fmfv{decor.shape=circle,decor.filled=shaded,decor.size=10thin}{vc1}
				\else\fi
				\fmffreeze
				\fmfposition
				\fmfipath{p[]}
				\fmfipair{vm[]}
				\fmfcmd{pair verta, vertb, vertc, vertd, vertca, vertcb; verta = vloc(__v1); vertb = vloc(__v2); vertc = vloc(__v3); vertd = vloc(__v4); vertca = vloc(__vc1); vertcb = vloc(__vc2);}
				\ifKV@FDiagram@oldlabels
				\fmfiv{label=\FDiagramlabelleftbottom,l.a=+120,l.dist=0.07w}{verta}
				\fmfiv{label=\FDiagramlabelrightbottom,l.a=+60,l.dist=0.07w}{vertb}
				\ifKV@FDiagram@tchannel
				\fmfiv{label=\FDiagramlabelmid,l.a=30,l.dist=0.18w}{vertca}
				\else
				\fmfiv{label=\FDiagramlabelmid,l.a=60,l.dist=0.20w}{vertca}
				\fi
				\fmfiv{label=\FDiagramlabellefttop,l.a=-120,l.dist=0.07w}{vertc}
				\fmfiv{label=\FDiagramlabelrighttop,l.a=-60,l.dist=0.07w}{vertd}
				\else
				\fmfiv{label=\FDiagramlabelleftbottom,l.a=+120,l.dist=0.07w}{verta}
				\fmfiv{label=\FDiagramlabelrightbottom,l.a=+60,l.dist=0.07w}{vertb}
				\ifKV@FDiagram@tchannel
				\fmfiv{label=\FDiagramlabellefttop,l.a=+120,l.dist=0.07w}{vertca}
				\fmfiv{label=\FDiagramlabelrighttop,l.a=+60,l.dist=0.07w}{vertcb}
				\else
				\fi
				\ifKV@FDiagram@schannel
				\fmfiv{label=\FDiagramlabellefttop,l.a=+165,l.dist=0.22w}{vertcb}
				\fmfiv{label=\FDiagramlabelrighttop,l.a=+15,l.dist=0.22w}{vertcb}
				\else
				\fi
				\ifKV@FDiagram@xchannel
				\fmfiv{label=\FDiagramlabellefttop,l.a=+158,l.dist=0.20w}{vertca+(0,0.025h)}
				\fmfiv{label=\FDiagramlabelrighttop,l.a=+22,l.dist=0.20w}{vertca+(0,0.025h)}
				\else
				\fi
				\ifKV@FDiagram@leftSE
				\fmfiv{label=\FDiagramlabellefttop,l.a=+120,l.dist=0.07w}{verta+(0,0.6125h)}
				\else
				\fi
				\ifKV@FDiagram@rightSE
				\fmfiv{label=\FDiagramlabelrighttop,l.a=+60,l.dist=0.07w}{vertb+(0,0.6125h)}
				\else
				\fi
				\fi
				\fmfdraw
				\ifKV@FDiagram@long
				\fmf{plain,width=1mm}{vb1,vb5}
				\else 
				\fmf{plain,width=1mm}{v1,v2}
				\fi
				\ifKV@FDiagram@longup
				\fmf{plain,width=1mm,foreground=(0.65,,0.65,,0.65)}{vt1,vt5}
				\fi
			\end{fmfchar*}%
		}
	}%
}
\makeatother

\makeatletter
\def\thickhrulefill{\leavevmode \leaders \hrule height 1ex \hfill \kern \z@}
\def\@makechapterhead#1{%
	\reset@font
	\vspace*{10\p@}%
	{\parindent \z@ 
		\begin{flushleft} 
			\reset@font \slshape   {\color{lightblue}\fontsize{100}{120}\selectfont \thechapter} \par 
		\end{flushleft}
		\vskip -5\p@
		\noindent\rule{8cm}{0.7pt}		
		\begin{flushleft} 
			\reset@font \huge \strut {\color{black} \bfseries
				#1} \strut \par
		\end{flushleft}
		\vskip 50\p@
	}}

\def\@makeschapterhead#1{%
	\reset@font
	\vspace*{10\p@}%
	{\parindent \z@ 
		\noindent\rule{8cm}{0.7pt}		
		\begin{flushleft} 
			\reset@font \huge \strut {\color{black} \bfseries
				#1} \strut \par
		\end{flushleft}
		\vskip 50\p@
	}}
\makeatother

\makeatletter
\let\old@makeschapterhead\@makeschapterhead
\def\fake@makeschapterhead#1{%
	\reset@font
	\vspace*{10\p@}%
	{\parindent \z@ 
		\begin{flushleft} 
			\reset@font \huge \strut{\color{black} \bfseries
				#1} \strut \par
		\end{flushleft}
		\vskip 50\p@
	}
}
\makeatother
\makeatletter
\newcommand{\newchapterhead}{\let\@makeschapterhead\fake@makeschapterhead}
\newcommand{\restorechapterhead}{\let\@makeschapterhead\old@makeschapterhead}
\makeatother

\numberwithin{equation}{section}

\usepackage{etex}
\makeatletter
\DeclareRobustCommand*{\bfseries}{%
  \not@math@alphabet\bfseries\mathbf
  \fontseries\bfdefault\selectfont
  \boldmath
}
\makeatother

\makeatletter
\def\footnoterule{\kern-5\p@
	\hrule \@width 2in \kern 7.6\p@} 
\makeatother

\newcommand{\myprenote}{The final list of numbers in each bibliography entry indicates the pages which refer to the particular reference.}



\definecolor{lightblue}{rgb}{0,0.4,0.9} 	
\usepackage{ocg-p}

\usepackage[ocgcolorlinks,		
backref=page				
]{hyperref}
\hypersetup{
	unicode=false,          
	pdftoolbar=true,        
	pdfmenubar=true,        
	pdffitwindow=false,     
	pdfstartview={FitH},    
	pdftitle={A hitchhiker's guide to quantum field theoretic aspects of \texorpdfstring{\NfSYMt}{N=4 SYM theory} and its deformations},    
	pdfauthor={Jan Fokken},     
	pdfsubject={Quantum field theoretic aspects of \texorpdfstring{\NfSYMt}{N=4 SYM theory} and its deformations},   
	pdfproducer={Jan Fokken}, 	
	pdfkeywords={Partition Functions} {N=4 Super Yang-Mills theory} {Deformation} {Feynman rules} {correlation functions} {psu(2,2|4) algebra}, 
	pdfnewwindow=true,      	
	colorlinks=true,       	
	citecolor=ForestGreen,
	urlcolor=ForestGreen,
	linkcolor=lightblue,
	linktocpage=true,
	breaklinks=true,					
	anchorcolor=Red
}

\makeatletter
\Hy@colorlinkstrue
\Hy@ocgcolorlinksfalse
\newcommand*{\reenable@ocglinks@pdftex}{%
	\Hy@AtBeginDocument{%
		\def\Hy@colorlink##1{%
			\begingroup
			\def\Hy@ocgcolor{##1}%
			\setbox0=\hbox\bgroup\color@begingroup
		}%
		\def\Hy@endcolorlink{%
			\color@endgroup\egroup
			\mbox{%
				\pdfliteral page{/OC/OCPrint BDC}%
				\rlap{\copy0}%
				\pdfliteral page{EMC/OC/OCView BDC}%
				\begingroup
				\expandafter\HyColor@UseColor\Hy@ocgcolor
				\box0 %
				\endgroup
				\pdfliteral page{EMC}%
			}%
			\endgroup
		}%
	}%
}
\newcommand*{\reenable@ocglinks@dvipdfm}{%
	\Hy@AtBeginDocument{%
		\def\Hy@colorlink##1{%
			\begingroup
			\def\Hy@ocgcolor{##1}%
			\setbox0=\hbox\bgroup\color@begingroup
		}%
		\def\Hy@endcolorlink{%
			\color@endgroup\egroup
			\mbox{%
				\@pdfm@mark{content /OC/OCPrint BDC}%
				\rlap{\copy0}%
				\@pdfm@mark{content EMC/OC/OCView BDC}%
				\begingroup
				\expandafter\HyColor@UseColor\Hy@ocgcolor
				\box0 %
				\endgroup
				\@pdfm@mark{content EMC}%
			}%
			\endgroup
		}%
	}%
}
\def\Hy@temp{hpdftex}
\ifx\Hy@driver\Hy@temp
\reenable@ocglinks@pdftex
\else
\def\Hy@temp{hdvipdfm}
\ifx\Hy@driver\Hy@temp
\reenable@ocglinks@dvipdfm
\else
\def\Hy@temp{hxetex}
\ifx\Hy@driver\Hy@temp
\reenable@ocglinks@dvipdfm
\fi
\fi
\fi
\@ocgp@newocg{View}{View}{1}{printocg=never,listintoolbar=never}
\@ocgp@newocg{Print}{Print}{0}{printocg=always,listintoolbar=never}
\makeatother

\title{A hitchhiker's guide to quantum field theoretic aspects of \texorpdfstring{\NfSYMt}{N=4 SYM theory} and its deformations}
\author{Jan Fokken}

\begin{document}
\pagestyle{plain} 
\frontmatter

\begingroup\parindent0pt
\vspace*{4em}
\centering
\begingroup\LARGE
\bf
A hitchhiker's guide to \\
quantum field theoretic aspects 
of \NfSYMt 
and its deformations
\par\endgroup
\vspace{4em}
based on my
\vspace{\baselineskip}
\\
D i s s e r t a t i o n

\vspace{\baselineskip}

\vspace{\baselineskip}

\vspace{\baselineskip}
%
\vspace{\baselineskip}

eingereicht an der 
\vspace{\baselineskip}

Mathematisch-Naturwissenschaftlichen Fakultät

der Humboldt-Universität zu Berlin 
\vspace{\baselineskip}

von 
\vspace{\baselineskip}

\textbf{Jan Fokken}
\vspace{\baselineskip}
\\
fokken@physik.hu-berlin.de
\vspace{\baselineskip}
\\
{\it
	Institut für Mathematik und Institut für Physik, Humboldt-Universität zu Berlin,
	IRIS-Adlershof, Zum Großen Windkanal 6, 12489 Berlin, Germany
	}
\vspace{4\baselineskip}


\vspace{\baselineskip}


\vspace{\baselineskip}

\endgroup


\thispagestyle{empty}

%
%

\newpage
\thispagestyle{empty}
\cleardoublepage
\setcounter{page}{3}

\thispagestyle{empty}
\vspace*{16cm}
{\slshape\Large
\noindent\hspace*{8cm}{Für sie, die mich fing,}\\
\noindent\hspace*{8cm}{den Sirenen entriss,}\\
\noindent\hspace*{8cm}{auf mich wartete und ging,}\\
\noindent\hspace*{8cm}{und doch niemals verließ.}\\
}

\cleardoublepage

\section*{Zusammenfassung}
\addcontentsline{toc}{section}{\protect\numberline{}Zusammenfassung}%
In den vergangenen Jahrzehnten gab es enormen Fortschritt im Verständnis der Struktur der \NfSYM Theorie in vier Raumzeitdimensionen, welcher viele Werkzeuge für die effiziente Berechnung von Observablen hervorgebracht hat. 
Mit Hilfe von Integrabilitätsmethoden wurde es prinzipiell möglich die anomalen Dimensionen zusammengesetzter Operatoren im \tHooft Limes exakt zu berechnen.
Inspiriert durch diese Fortschritte gehen wir der Frage nach, welche Voraussetzungen erfüllt sein müssen, damit Observablen einer Theorie mit Hilfe dieser neuen Werkzeuge berechnet werden können. 
Insbesondere untersuchen wir am Beispiel der einparametrischen $\beta$- und dreiparametrischen $\gamma_i$-deformierten Abkömmlinge der \NfSYM Theorie, ob die anomalen Dimensionen zusammengesetzter Operatoren auch in diesen weniger symmetrischen Theorien durch Integrabilitätsmethoden erhalten werden können. 

Für die deformierten Theorien stellt sich heraus, dass nicht alle ihre Wechselwirkungen als Abkömmlinge der undeformierten Wechselwirkungen verstanden werden können. Um persistente Divergenzen in perturbativen Entwicklungen zu vermeiden, führen wir zusätzliche sogenannte Mehrspurwechselwirkungen ein. 
Für die $\gamma_i$-Deformation zeigen wir durch feynmandiagrammatische Berechnung der relevanten Einschleifenkorrekturen im \tHooft Limes, dass diese nichtvererbten Wechselwirkungen laufende Kopplungskonstanten besitzen, welche die konforme Invarianz der quantisierten Theorie brechen.
Darüber hinaus untersuchen wir den Einfluss der nichtvererbten Wechselwirkungen auf die anomalen Dimensionen zusammengesetzter Operatoren am Beispiel der Operatoren $\tr\bigl(\phi_i^L)$, indem wir ihre anomalen Dimensionen bis zur führenden Wickel (\glqq Wrapping\grqq) Schleifenordnung $K=L$ berechnen. Für $L\geq 3$ lassen sich so die Ergebnisse von integrabilitätsbasierten Methoden reproduzieren. Für $L=2$ finden wir jedoch die endliche und renormierungsschemenabhängige anomale Dimension im Kontrast zum divergenten integrabilitätsbasierten Ergebnis.
Basierend auf den feldtheoretischen Daten aus der $\beta$- und der $\gamma_i$-Deformation schlagen wir einen Test vor, welcher klären soll ob Supersymmetrie und/oder exakte konforme Invarianz notwendige Bedingungen für die in der \NfSYM Theorie gefundene Quantenintegrabilität sind.

Auch für die $\beta$-Deformation analysieren wir das Auftreten von nichtvererbten Mehr\-spur\-beiträgen. Aus der vollständigen Wechselwirkungsstruktur leiten wir einen Algorithmus ab, der erlaubt den Einfluss von Mehrspurkopplungen auf die anomalen Dimensionen zusammengesetzter Operatoren auf Einschleifenebene in der \tHooft Kopplung konsistent im Spinkettenbild abzubilden. Hiermit konstruieren wir den vollständigen Dilatationsoperator der konformen $\beta$-Deformation im \tHooft Limes auf Einschleifenebene.

Abschließend nutzen wir unsere Ergebnisse, um den p\'{o}lyatheoretischen Ansatz zur Berechnung der thermalen Einschleifen-Zustandssumme auf dem kompakten Raum \RxSt in den deformierten Theorien nutzbar zu machen. Unsere Ergebnisse zeigen, dass die \glqq Deconfinement\grqq\,-Phasenübergangstemperatur der deformierten Theorien auf Einschleifenniveau mit jener der undeformierten \NfSYM Theorie übereinstimmt und wir vermuten, dass dieser Befund sogar im nichtperturbativen Bereich Bestand hat.

Zusätzlich zu den Forschungsergebnissen enthält diese Arbeit die vollständige Wirkung inklusive der Symmetriegeneratoren der \NfSYM Theorie, der $\beta$- und der $\gamma_i$-Deformation.
Wir wiederholen allgemeine Techniken zur Renormierung zusammengesetzter Operatoren und elementarer Felder, gehen auf das weit verbreitete dimensionale Reduktionsschema ein und wie relevante UV Divergenzen in Niederschleifenintegralen effizient bestimmt werden können.
In diesem Zusammenhang leiten wir die Feynman Regeln aller untersuchten Theorien her und stellen das Werkzeug \ttt{FokkenFeynPackage} vor, welches diese Regeln in {\tt Mathematica} implementiert. Alle Rechnungen in dieser Dissertation wurden mit \ttt{FokkenFeynPackage} durchgeführt, so dass diese Arbeit einem unabhängigen Test aller Feynman-diagrammatischen Rechnungen in den Publikationen \cite{Fokken:2013aea,Fokken:2013mza,Fokken:2014soa,Fokken:2014moa} darstellt. 
\enlargethispage{10\baselineskip}

\newpage

\section*{Abstract}
\addcontentsline{toc}{section}{\protect\numberline{}Abstract}%
Over the last decades tremendous progress was made in understanding the structure of \NfSYMt in four-dimensional spacetime and many tools for the efficient calculation of observables in this theory were developed. The anomalous dimensions of composite operators in the \tHooft limit became in principle accessible by means of integrability-based methods. Inspired by these findings, we investigate which prerequisites must be fulfilled for observables of a theory to be calculable by the means of these new tools. In particular, we focus on the one-parameter $\beta$- and the three-parameter $\gamma_i$-deformed descendents of \NfSYMt to analyse whether the anomalous dimensions of composite operators in these less symmetric theories can also be obtained by the means of integrability. 

In the deformed theories it turns out that not all interactions originate from the interactions in the undeformed theory. Additionally, we have to include so-called multi-trace interactions to prevent persistent divergences in perturbative expansions. For the $\gamma_i$-deformation, we show by an explicit feynman-diagrammatic one-loop calculation that these non-inherited interactions have running coupling constants which spoil the conformal invariance of the quantised theory, even in the \tHooft limit. Furthermore, we investigate the impact of these non-inherited interactions on the anomalous dimensions of composite operators, by perturbatively calculating the $K=L$ loop leading order wrapping corrections to the operators $\tr\bigl(\phi_i^L)$. We reproduce the findings from integrability for $L\geq 3$ and find the finite renormalisation-scheme-dependent anomalous dimension of the $L=2$ states in contrast to the integrability-based methods which yield a divergent result. Based on the field-theoretic data from the $\beta$- and $\gamma_i$-deformation, we propose a test to determine whether supersymmetry and/or exact conformal invariance are necessary prerequisites of the quantum integrability found for \NfSYMt. 

For the $\beta$-deformation, we also analyse the occurrence of non-inherited multi-trace contributions. From the full interaction structure, we derive an algorithm which allows to consistently include multi-trace couplings that affect anomalous dimensions of composite operators at one-loop order in the \tHooft coupling in the spin-chain picture. This leads to the complete one-loop dilatation operator of the conformal $\beta$-deformation in the \tHooft limit.

Finally, we employ our findings to generalise the \Polya-theoretic approach to the thermal one-loop partition function of \NfSYMt on \RxSt to be also applicable in the deformed theories.  We find that the deconfinement phase-transition temperature in the deformed theories is the same as in the undeformed \NfSYMt at one-loop level and we conjecture that it remains the same even non-perturbatively.

In the context of this thesis, we employ various field-theoretic aspects of \NfSYMt and its deformations. Therefore, we provide the action and symmetry generators of \NfSYMt, the $\beta$-, and the $\gamma_i$-deformation. Furthermore, we review the general techniques for the renormalisation of elementary fields and composite operators in a unified setting and discuss the relation to the dilatation operator. We include a detailed description of the widely used dimensional reduction scheme and discuss how the UV divergence of logarithmically divergent integrals may be extracted with relatively little effort. In this context, we derive the Feynman rules for \NfSYMt, the $\beta$- and the $\gamma_i$-deformation and present the tool \ttt{FokkenFeynPackage} which implements these rules into {\tt Mathematica}. All calculations in this thesis are carried out using this tool and hence it provides an independent test of all Feynman-diagrammatic calculations in \cite{Fokken:2013aea,Fokken:2013mza,Fokken:2014soa,Fokken:2014moa}.

\newpage
\section*{List of own publications}
\addcontentsline{toc}{section}{\protect\numberline{}List of own publications}%
This thesis is based on the following publications:
\begin{itemize}
	\item[\cite{Fokken:2013aea}]
	J.~Fokken, C.~Sieg and M.~Wilhelm, \emph{{Non-conformality of ${{\gamma}_{i}}$-deformed N = 4 SYM theory}},
	\href{http://dx.doi.org/10.1088/1751-8113/47/45/455401}{\emph{J.Phys.} {\bf A47} (2014) 455401},  [\href{http://arxiv.org/abs/1308.4420}{{arXiv:1308.4420 [hep-th]}}].
	\item[\cite{Fokken:2013mza}]
	J.~Fokken, C.~Sieg and M.~Wilhelm, \emph{{The complete one-loop dilatation operator of planar real $\beta$-deformed $ \mathcal{N} = 4$ SYM theory}},
	\href{http://dx.doi.org/10.1007/JHEP07(2014)150}{\emph{JHEP} {\bf 1407}	(2014) 150}, 
	[\href{http://arxiv.org/abs/1312.2959}{{ arXiv:1312.2959 [hep-th]}}].
	\item[\cite{Fokken:2014soa}]
	J.~Fokken, C.~Sieg and M.~Wilhelm, \emph{{A piece of cake: the ground-state	energies in $\gamma_{i}$-deformed $ \mathcal{N} = 4$ SYM theory at leading wrapping order}},
	\href{http://dx.doi.org/10.1007/JHEP09(2014)078}{\emph{JHEP} {\bf 1409}	(2014) 78},  
	[\href{http://arxiv.org/abs/1405.6712}{{ arXiv:1405.6712 [hep-th]}}].
	\item[\cite{Fokken:2014moa}]
	J.~Fokken and M.~Wilhelm, \emph{{One-Loop Partition Functions in Deformed $\mathcal{N}=4$ SYM Theory}},
	\href{http://dx.doi.org/10.1007/JHEP03(2015)018}{\emph{JHEP} {\bf 03} (2015) 018},  
	[\href{http://arxiv.org/abs/1411.7695}{{ arXiv:1411.7695 [hep-th]}}].
\end{itemize}
\begin{itemize}
	\item In addition, this thesis contains the presentation and manual of the {\tt Mathematica} tool {\tt FokkenFeynPackage}. I developed the {\tt FokkenFeynPackage} to provide an efficient {\tt Mathematica} implementation of the Feynman rules of \NfSYMt and its deformations that were presented in this thesis. This tool will be made publicly available together with this thesis.
\end{itemize}

%
\setcounter{tocdepth}{2}
\tableofcontents
\clearpage
\thispagestyle{empty}
\null\newpage


\begin{fmffile}{./graphs/graphs}
\fmfcmd{input hatching;}
\fmfcmd{%
	style_def plain_sarrow expr p =
	cdraw p;
	shrink (0.4); 
	cfill (arrow p);
	endshrink;
	enddef;
	style_def dashes_sarrow expr p =
	draw_dashes p;
	shrink (0.4);
	cfill (arrow p);
	endshrink;
	enddef;
	style_def plain_srarrow expr p =
	cdraw p;
	shrink (0.4);
	cfill (arrow (reverse p));
	endshrink;
	enddef;
	style_def dashes_srarrow expr p =
	draw_dashes p;
	shrink (0.4);
	cfill (arrow (reverse p));
	endshrink;
	enddef;
	style_def dots_sarrow expr p =
	draw_dots p;
	shrink (0.55);
	cfill (arrow p);
	endshrink;
	enddef;
	style_def dots_srarrow expr p =
	draw_dots p;
	shrink (0.55);
	cfill (arrow (reverse p));
	endshrink;
	enddef;
}

\fmfcmd{%
	style_def plain_n expr p =
	linecap:=butt;
	cdraw p;
	linecap:=rounded;
	enddef;
	style_def plain_h expr p =
	linecap:=butt;
	cdraw subpath (0, 1length p ) of p;
	linecap:=rounded;
	cdraw subpath (0.95length p ,length p ) of p;
	enddef;
	style_def plain_t expr p =
	linecap:=rounded;
	cdraw subpath (0, 0.05length p ) of p;
	linecap:=butt;
	cdraw subpath (0,length p ) of p;
	linecap:=rounded;
	enddef;
	style_def plain_ht expr p =
	linecap:=rounded;
	cdraw p;
	enddef;
	style_def iplain_n expr p =
	linecap:=squared;
	cdraw p;
	linecap:=rounded;
	enddef;
	style_def iplain_h expr p =
	linecap:=squared;
	cdraw subpath (0, 0.5length p ) of p;
	linecap:=rounded;
	cdraw subpath (0.5length p ,length p ) of p;
	enddef;
	style_def iplain_t expr p =
	linecap:=rounded;
	cdraw subpath (0, 0.5length p ) of p;
	linecap:=squared;
	cdraw subpath (0.5length p,length p ) of p;
	linecap:=rounded;
	enddef;
	style_def iplain_ht expr p =
	linecap:=rounded;
	cdraw p;
	enddef;
	style_def interrupted_plain_n expr p =
	draw_plain_h subpath (0, 0.25length p ) of p;
	draw_plain_t subpath (0.75length p ,length p ) of p;
	enddef;
	style_def interrupted_plain_h expr p =
	draw_plain_h subpath (0, 0.25length p ) of p;
	draw_plain_ht subpath (0.75length p ,length p ) of p;
	enddef;
	style_def interrupted_plain_t expr p =
	draw_plain_ht subpath (0, 0.25length p ) of p;
	draw_plain_t subpath (0.75length p ,length p ) of p;
	enddef;
	style_def interrupted_plain_ht expr p =
	draw_plain_ht subpath (0, 0.25length p ) of p;
	draw_plain_ht subpath (0.75length p ,length p ) of p;
	enddef;
	style_def iinterrupted_plain_n expr p =
	draw_iplain_h subpath (0, 0.25length p ) of p;
	draw_iplain_t subpath (0.75length p ,length p ) of p;
	enddef;
	style_def iinterrupted_plain_h expr p =
	draw_iplain_h subpath (0, 0.25length p ) of p;
	draw_iplain_ht subpath (0.75length p ,length p ) of p;
	enddef;
	style_def iinterrupted_plain_t expr p =
	draw_iplain_ht subpath (0, 0.25length p ) of p;
	draw_iplain_t subpath (0.75length p ,length p ) of p;
	enddef;
	style_def iinterrupted_plain_ht expr p =
	draw_iplain_ht subpath (0, 0.25length p ) of p;
	draw_iplain_ht subpath (0.75length p ,length p ) of p;
	enddef;
	vardef shift_p expr p =  
	p  shifted 4.5 (unitvector direction (length (p)/2) of p rotated +90)
	enddef;
	style_def leftrightarrows expr p =
	shrink 0.5;
	cfill arrow (shift_p (p));
	cfill arrow (shift_p (reverse p));
	endshrink;
	enddef;%
	style_def leftrightarrows_interrupted expr p =
	shrink 0.5;
	cfill arrow (shift_p (subpath (0, 0.4length p ) of p));
	cfill arrow (shift_p (reverse (subpath (0, 0.4length p ) of p)));
	cfill arrow (shift_p (subpath (0.6length p ,length p ) of p));
	cfill arrow (shift_p (reverse (subpath (0.6length p ,length p ) of p)));
	endshrink;
	enddef;%
	style_def leftrightarrows_interruptedold expr p =
	shrink 0.5;
	cfill arrow (shift_p (subpath (0, 0.25length p ) of p));
	cfill arrow (shift_p (reverse (subpath (0, 0.25length p ) of p)));
	cfill arrow (shift_p (subpath (0.75length p ,length p ) of p));
	cfill arrow (shift_p (reverse (subpath (0.75length p ,length p ) of p)));
	endshrink;
	enddef;%
	style_def leftrightarrows_start expr p =
	shrink 0.5;
	cfill arrow (shift_p (subpath (0, 0.1length p ) of p));
	cfill arrow (shift_p (reverse (subpath (0, 0.1length p ) of p)));
	endshrink;
	enddef;
}

\fmfcmd{%
style_def plain_ar_end expr p =
cdraw (wiggly p);
shrink (1);
cfill (arrow p);
endshrink;
enddef;}

\fmfcmd{%
vardef cross_bar (expr p, len, ang) =
((-len/2,0)--(len/2,0))
rotated (ang + angle direction length(p)/2 of p)
shifted point length(p)/2 of p
enddef;
style_def crossed expr p =
cdraw p;
ccutdraw cross_bar (p, 5mm, 45);
ccutdraw cross_bar (p, 5mm, -45)
enddef;}

\fmfcmd{%
thin := 1pt; 
thick := 2thin;
arrow_len := 4mm;
arrow_ang := 15;
curly_len := 3mm;
dash_len := 1.5mm; 
dot_len := 1mm; 
wiggly_len := 2mm; 
wiggly_slope := 60;
zigzag_len := 2mm;
zigzag_width := 2thick;
decor_size := 5mm;
dot_size := 2thick;
}


\fmfcmd{%
marksize=2mm;
def draw_mark(expr p,a) =
  begingroup
    save t,tip,dma,dmb; pair tip,dma,dmb;
    t=arctime a of p;
    tip =marksize*unitvector direction t of p;
    dma =marksize*unitvector direction t of p rotated -45;
    dmb =marksize*unitvector direction t of p rotated 45;
    linejoin:=beveled;
    draw (-.5dma.. .5tip-- -.5dmb) shifted point t of p;
  endgroup
enddef;
style_def derplain expr p =
    save amid;
    amid=.5*arclength p;
    draw_mark(p, amid);
    draw p;
enddef;
def draw_marks(expr p,a) =
  begingroup
    save t,tip,dma,dmb,dmo; pair tip,dma,dmb,dmo;
    t=arctime a of p;
    tip =marksize*unitvector direction t of p;
    dma =marksize*unitvector direction t of p rotated -45;
    dmb =marksize*unitvector direction t of p rotated 45;
    dmo =marksize*unitvector direction t of p rotated 90;
    linejoin:=beveled;
    draw (-.5dma.. .5tip-- -.5dmb) shifted point t of p withcolor 0white;
    draw (-.5dmo.. .5dmo) shifted point t of p;
  endgroup
enddef;
style_def derplains expr p =
    save amid;
    amid=.5*arclength p;
    draw_marks(p, amid);
    draw p;
enddef;
def draw_markss(expr p,a) =
  begingroup
    save t,tip,dma,dmb,dmo; pair tip,dma,dmb,dmo;
    t=arctime a of p;
    tip =marksize*unitvector direction t of p;
    dma =marksize*unitvector direction t of p rotated -45;
    dmb =marksize*unitvector direction t of p rotated 45;
    dmo =marksize*unitvector direction t of p rotated 90;
    linejoin:=beveled;
    draw (-.5dma.. .5tip-- -.5dmb) shifted point t of p withcolor 0white;
    draw (-.5dmo.. .5dmo) shifted point arctime a+0.25 mm of p of p;
    draw (-.5dmo.. .5dmo) shifted point arctime a-0.25 mm of p of p;
  endgroup
enddef;
style_def derplainss expr p =
    save amid;
    amid=.5*arclength p;
    draw_markss(p, amid);
    draw p;
enddef;
style_def dblderplain expr p =
    save amidm;
    save amidp;
    amidm=.5*arclength p-0.75mm;
    amidp=.5*arclength p+0.75mm;
    draw_mark(p, amidm);
    draw_mark(p, amidp);
    draw p;
enddef;
style_def dblderplains expr p =
    save amidm;
    save amidp;
    amidm=.5*arclength p-0.75mm;
    amidp=.5*arclength p+0.75mm;
    draw_mark(p, amidm);
    draw_marks(p, amidp);
    draw p;
enddef;
style_def dblderplainss expr p =
    save amidm;
    save amidp;
    amidm=.5*arclength p-0.75mm;
    amidp=.5*arclength p+0.75mm;
    draw_mark(p, amidm);
    draw_markss(p, amidp);
    draw p;
enddef;
style_def dblderplainsss expr p =
    save amidm;
    save amidp;
    amidm=.5*arclength p-0.75mm;
    amidp=.5*arclength p+0.75mm;
    draw_marks(p, amidm);
    draw_markss(p, amidp);
    draw p;
enddef;
}

\fmfcmd{%
style_def plain_ar expr p =
  cdraw p;
  shrink (0.6);
  cfill (arrow p);
  endshrink;
enddef;
style_def plain_rar expr p =
  cdraw p; 
  shrink (0.6);
  cfill (arrow reverse(p));
  endshrink;
enddef;
style_def dashes_ar expr p =
  draw_dashes p;
  shrink (0.6);
  cfill (arrow p);
  endshrink;
enddef;
style_def dashes_rar expr p =
  draw_dashes p;
  shrink (0.6);
  cfill (arrow reverse(p));
  endshrink;
enddef;
style_def dots_ar expr p =
  draw_dots p;
  shrink (0.6);
  cfill (arrow p);
  endshrink;
enddef;
style_def dots_rar expr p =
  draw_dots p;
  shrink (0.6);
  cfill (arrow reverse(p));
  endshrink;
enddef;
}

\fmfcmd{%
style_def phantom_cross expr p =
    save amid,ang;
    amid=.5*length p;
    ang= angle direction amid of p;
    draw ((polycross 4) scaled 8 rotated ang) shifted point amid of p;
enddef;
}

\mainmatter
\pagestyle{main} 

\setcounter{page}{13}

\chapter{Introduction and Overview}\label{chap:Introduction}

\section{Introduction}
Over the last century, tremendous progress was made in understanding the structure underlying our physical world. The developments have largely been sparked by the advent of quantum mechanics in 1901 and the formulation of general relativity in 1915. The former has continuously been advanced to the current quantum field theoretic understanding of microscopic systems. In this process the weak and strong forces have been discovered and their descriptions have been unified with the description of electromagnetism to form the standard model of particle physics. Both, the standard model at microscopic scales and general relativity at macroscopic scales, give astonishingly accurate predictions of physical phenomena.\footnote{For examples concerning general tests of the standard model \cite{Renton2002}, the calculation of the anomalous electric moment of muons \cite{Prades:2001}, and the experimental confirmation of the Higgs field \cite{Aad:2012tfa,Chatrchyan:2012xdj}. For tests of general relativity see e.g.\ \cite{Turyshev:2008dr} and for the recently discovered gravitational waves \cite{Abbott:2016blz}.} Despite the success of both theories, there are still open questions in the realm of fundamental physical theories, see e.g.\ \cite{Troitsky:2011xj,1063-7869-42-4-A03}. On the one hand, these concern practical questions, e.g.\ how quantitative predictions can be obtained from the fundamental theories in non-perturbative regimes. On the other hand, these also concern conceptual issues like the hierarchy problem, colour confinement, and the lacking description of quantum gravity, which prevents a unified description of all fundamental forces.

While a long-term goal in theoretical physics certainly is to resolve open questions beyond the standard model and general relativity, which both deliver descriptions of actual physical phenomena, this goal seems too ambitious to be tackled immediately. Instead, it proved very fruitful to investigate open questions in a simpler, i.e.\ more symmetric, setting and generalise finding to less symmetric and more realistic settings. In this light, the maximally supersymmetric non-abelian gauge theory\footnote{It is the maximally supersymmetric non-abelian gauge theory in flat four-dimensional Minkowski space. For gauge theories, we also assume that particles have a maximal spin of one.} (\NfSYMt) found in \cite{Brink197777} became very prominent in theoretical particle physics. This theory can be thought of as a highly symmetric relative of quantum chromo dynamics (QCD) which, in addition to \Poincare symmetry, exhibits an internal \SU{4} flavour or $R$-symmetry, $\cN=4$ supersymmetries\footnote{In fact, there are four spin-$\frac12$ supersymmetry generators and conjugates, summing to a total of sixteen supersymmetries.} (SUSY) and a conformal invariance that remains unbroken in the quantised theory \cite{Caswell:1980yi,SOHNIUS1981245,Mandelstam:1982cb,Brink:1982wv,HOWE1984125,Seiberg:1988ur}. In addition, \NfSYMt is conjectured to be also invariant under $\text{SL}(2,\ZZ)$ duality transformation \cite{MONTONEN1977117,Girardello1995127}. In contrast to QCD, all its elementary fields live in the adjoint representation of the gauge group \UN or \SUN. All symmetries combine to the $\text{PSU}(2,2|4)$ supersymmetry group of the theory \cite{D'Hoker:2002aw,Beisert:2010jr} and observables are characterised in terms of the conserved charges, the number of colours $N$ and a single complex coupling constant $\tau=\frac{4\pi}{\gym^2}+\frac{\complexi\theta}{2\pi}$ which combines the gauge-theory coupling with the imaginary one that accounts for topological contributions. The exact conformal invariance implies that this conformal field theory (CFT) has no inherent scale and the coupling constant is not renormalised. For local gauge-invariant composite operators $\mathcal{O}(x)$, this non-renormalisation of the coupling guarantees that the operators' scaling dimensions $\Delta_{\mathcal{O}}$ are independent of the renormalisation scheme and in particular observable. In the interacting theory, the classical scaling dimension $\Delta_{\mathcal{O}}^0$, which is obtained from naive dimensional analysis is supplemented by an anomalous piece $\gamma_{\mathcal{O}}$, which originates from the renormalisation of the external composite operator. The scaling dimensions enter the two- and three-point correlation functions of composite operators $\mathcal{O}_i$ at positions $x_i$ in a \CFT as\footnote{Note that we suppressed the infinitesimal imaginary factors that are required to make the Minkowski space correlation functions unambiguous. In position space, they are obtained by replacing every distance in \eqref{eq:2_3_pt_function} by $x_{ij}^2\Rightarrow x_{ij}^2+\complexi \epsilon$, see \appref{sec:Conventions} for further details concerning our Minkowski space conventions.}
\begin{equation}\label{eq:2_3_pt_function}
\vacl\T\mathcal{O}_1(x_1)\mathcal{O}_2(x_2)\vac=\frac{\delta_{\Delta_{1}\Delta_{2}}}{x_{12}^{2\Delta_{1}}}\eqncom
\quad
\vacl\T\mathcal{O}_1(x_1)\mathcal{O}_2(x_2)\mathcal{O}_3(x_3)\vac=\frac{C_{123}}{
	x_{12}^{\Delta-2\Delta_{3}}x_{23}^{\Delta-2\Delta_{1}}x_{31}^{\Delta-2\Delta_{2}}
	}\eqncom 
\end{equation}
where the distance between two operator insertions is $x_{ij}^2=\abs{x_i-x_j}^2$ and $\Delta=\Delta_{1}+\Delta_2+\Delta_3$, see \cite{CFT,D'Hoker:2002aw} for details. Like the scaling dimensions, the structure constants $C_{123}$ that are characteristic to each three-point function also receive perturbative corrections \cite{D'Hoker:1999}. Higher-point functions can be related to the two- and three-point functions by using the operator product expansion (OPE) 
\begin{equation}\label{eq:OPE_intro}
\mathcal{O}_1(x_1)\mathcal{O}_2(x_2)=\sum_J \frac{C_{12J}}{
	x_{12}^{\Delta_1+\Delta_2-\Delta_{J}}}\mathcal{O}_J(x_2)\eqncom
\end{equation}
to expand products of two operators within a correlation function in terms of the basis operators $\mathcal{O}_J$, see \cite{PhysRev.179.1499,Dolan:2000ut,Pappadopulo:2012jk}.

Apart from the high degree of symmetry, a very interesting feature of \NfSYMt is its conjectured duality to type $\text{II}\,\text{B}$ superstring theory on the background $\AdS{5}\times \text{S}^5$, which is a prime example of the \AdSCFTc proposed in \cite{Maldacena:1997re,Witten:1998qj,Witten:1998zw,Gubser:1998bc}, see also \cite{Aharony:1999ti,D'Hoker:2002aw} for detailed introductions. This correspondence can be motivated by analysing $\text{II}\,\text{B}$ superstring theory in flat ten-dimensional Minkowski space with $N$ coincident D3-branes and a string coupling constant $g_s$ at energies much smaller than the string energy scale $\tfrac{1}{\ell_s}$. As discussed in \cite{Zwiebach:2004tj,D'Hoker:1999}, in this limit \NfSYMt arises in the $g_s N\ll 1$ regime, whereas $\text{II}\,\text{B}$ superstring theory on the background geometry $\AdS{5}\times \text{S}^5$ appears in the $g_s N\gg 1$ regime. According to the \AdSCFTc, the parameters of this string and gauge theory are related as
\begin{equation}\label{eq:AdSCFT_parameters}
4\pi g_{\text{s}}=\gym^2\eqncom\qquad \text{and}\qquad
\frac{R^4}{\ell_{\text{s}}^4}=\gym^2 N\eqncom
\end{equation}
where $\frac{R}{\ell_{\text{s}}}$ is the radius of the $\AdS{5}$ and the $\text{S}^5$ factors in units of the string-length $\ell_{\text{s}}$. The symmetries match upon noting that the isometries of the string background  $\SO{2,4}\times \SO{6}$ combine with the 32 supersymmetries to form the $\text{PSU}(2,2|4)$ symmetry group of the string theory. Apart from the parameters, also the generating functionals of the gauge and the string theory are connected as
\begin{equation}
Z_{\text{\CFT}}[\{J\}]=Z_{\text{string}}[\{\phi| \phi_{\partial \text{AdS}}=J\}]\eqncom
\end{equation}
where $J$ are the sources of composite operator insertions in the \CFT and $\phi$ are string sources whose values on the boundary $\partial\text{AdS}_5$ are equal to $J$, see e.g.\ \cite{Zaffaroni:2000vh,Ramallo:2013bua} for a detailed discussion. 
This relation implies that the scaling dimension of a composite operator in \NfSYMt can also be obtained by calculating the energy of the corresponding state in the string theory. Proving the \AdSCFTc is notoriously difficult, since the perturbatively accessible regimes of both theories have no overlap: the gauge theory is weakly coupled for $\gym^2 \ll 1$, while the string theory reduces to the tractable supergravity system in the limits\footnote{Sending the string coupling to zero allows to use string perturbation theory and sending the ratio $\frac{\alpha^\prime}{R^2}$ to zero allows to use small curvature approximations of the $\AdS{5}\times \text{S}^5$ space.} $g_{\text{s}}\rightarrow 0$ and $\frac{\ell_{\text{s}}}{R}\rightarrow 0$, see \cite{Zwiebach:2004tj,D'Hoker:1999} for details.

Despite the above obstacle, the \AdSCFTc can be tested by analysing observables in the \tHooft limit, which is given by $\gym^2\rightarrow 0$ and $N\rightarrow \infty$ while keeping the product $\lambda=\gym^2 N$ fixed. In this limit, the string theory becomes free\footnote{While it becomes free, its solution is still non-trivial, since classical string solutions on the $\AdS{5}\times\text{S}^5$ with radius $\lambda=\bigl(\frac{R}{\ell_s}\bigr)^4$ have to be found.} and on the gauge-theory side only planar diagrams\footnote{A diagram in which all colour lines are closed is planar when it can be drawn on a plane without intersecting lines.} contribute in gauge-invariant correlation functions. In addition, the fission and fusion of colour traces within composite operators in correlation functions is suppressed by powers of $N$ in this limit. As a consequence, the properties of composite operators with multiple colour traces (multi-trace operators) can be deduced from the properties of composite operators with a single colour trace (single-trace operators). In particular, the scaling dimensions of composite operators with multiple colour traces are given by the sum of scaling dimensions of its single-trace constituents. The simplifications of the \tHooft limit allow for highly non-trivial tests of the \AdSCFTc that go beyond observables protected by symmetries, see \cite{Berenstein:2002jq,Constable:2002vq,Beisert:2002bb,Roiban:2002xr,Pearson:2002zs} for examples in the BMN limit\footnote{In this limit, chiral primary operators $\mathcal{O}$ with one \su{4} Cartan charge $J\rightarrow \infty$ are investigated, while the \tHooft coupling is simultaneously dialled up $\lambda\rightarrow\infty$, so that $\frac{\lambda}{J^2}$ and $\Delta_{\mathcal{O}}^0-J$ remain constant.}.

An even more striking discovery was made for \NfSYMt, when it was noted that the problem to calculate one-loop anomalous dimensions of composite operators in the \tHooft limit can be mapped to an integrable system\footnote{Integrability in a quantum field theoretic context is not trivially defined due to the infinitely many degrees of freedom of the system. For a discussion of quantum integrability from a mathematical perspective see \cite{Clemente-Gallardo:2011}.} \cite{Minahan:2002ve,Beisert:2003yb}. In this interpretation, single-trace operators with $L$ elementary constituent fields are identified with a cyclic spin-chain state of length $L$, where an individual $\text{PSU}(2,2|4)$ spin at each site characterises the $L$ elementary fields of the operator. In a planar one-loop process, which has a maximal interaction range of two, only two neighbouring spins of such a spin-chain state can be affected. Combining all such interactions into the (spin-chain) Hamiltonian allows to write a two-point correlation function as a Hamiltonian sandwiched between an initial and a final spin-chain state, see \cite{Beisert:2010jr} and references therein for a comprehensive overview. When acting on an eigenvector of all possible spin-chain states, the Hamiltonian takes the form of a diagonal matrix whose entries form the one-loop spectrum of anomalous dimensions in the \tHooft limit. Due to the found integrability, the entries of this matrix do not have to be calculated using perturbative Feynman diagram techniques, but they can be obtained by solving a set of functional relations. The latter approach is known as the Bethe ansatz\footnote{In the original Bethe ansatz the spin-chain states are built from $\su{2}$ spins \cite{Bethe}.} and the functional relations are obtained by treating the spin-chain state with maximal spins on each site as a spin-chain groundstate. On this groundstate it is then analysed how other spins as possible excitations at individual sites may propagate around the spin chain, see \cite{Staudacher:2010jz,Frassek:2014bya} and references therein for a modern introduction. For \NfSYMt, the appropriate Bethe ansatz was given in \cite{Beisert:2003yb} and the complete one-loop Hamiltonian was constructed in \cite{Beisert:2003jj}. Beyond the one-loop case, asymptotic Bethe-ansatz techniques can be employed, see e.g.\ \cite{Ahn:2010ka,deLeeuw:2010nd} for a detailed discussion. Asymptotic in this context means that the length of the spin-chain state exceeds the 
maximal interaction range of the Hamiltonian. When the interaction range of the Hamiltonian meets the length of the spin-chain state, finite-size effects limit the applicability of the asymptotic Bethe-ansatz. These effects encapsulate seemingly non-planar contributions to the Hamiltonian that nevertheless contribute in the \tHooft limit. Wrapping contributions, which are contributions that only become planar when the spin-chain states are connected to the Hamiltonian, are an example of such finite-size contributions. For length-preserving processes, they start when the loop order $K$ meets the length $L$ of the spin-chain state and they were studied in \cite{Sieg:2005kd} and \cite{Ambjorn:2005wa} from a field- and string-theoretic perspective, respectively. In the integrability approach, wrapping corrections can be incorporated into the asymptotic Bethe ansatz by means of  L\"uscher corrections, Y-system and the thermodynamic Bethe ansatz (TBA), see \cite{Bombardelli:2009ns,Arutyunov:2009ur,Gromov:2009tv} for reviews. In light of the \AdSCFTc, the TBA can also be applied in the strong coupling regime of \NfSYMt. With its help, the energies of string states in the non-linear $\sigma$-model describing the string theory can be found. These energies correspond to the anomalous dimensions of composite operators in the gauge theory, see \cite{vanTongeren:2013gva} and the references therein for a detailed introduction. Recently, the TBA and Y-system approach to finding anomalous dimensions in the context of \NfSYMt have been formalised further to the present quantum spectral curve (QSC) approach \cite{Gromov:2013pga,Gromov:2014caa,Gromov:2015dfa,Gromov:2015wca}, which in principle allows to determine the anomalous dimensions exactly. In practise, the anomalous dimension of the Konishi operator was determined up to an impressive $10^{\text{th}}$ loop order in the \tHooft coupling in \cite{Marboe:2014gma} with relatively little computational effort. There is, however, one caveat in the calculation of the entire spectrum of composite operators in \NfSYMt via integrability-based approaches: the anomalous dimension of the $L=2$ groundstate cannot directly be extracted from the TBA \cite{Frolov:2009in} directly and an additional regularisation needs to be included for the TBA to render the correct vanishing anomalous dimension \cite{deLeeuw:2012hp}. The origin of this divergence and how its ad hoc regularisation can applied in less symmetric examples of the \AdSCFTc is not yet clear.

To approach the question whether the entire spectrum of composite operators is calculable by the means of integrability, we focus on two less symmetric realisations of the \AdSCFTc. In both realisations similar divergences for the anomalous dimensions of $L=2$ states occur in the integrability-based approach but unlike in the \NfSYM case, they cannot be regularised and they affect the anomalous dimensions of all states they can be connected to via the OPE given in \eqref{eq:OPE_intro}.
The gauge theories of the two less symmetric realisations are called the $\beta$- and the $\gamma_i$-deformation and they can be obtained by adding marginal interactions to the action of the parent \NfSYMt, see \cite{Zoubos:2010kh} for a review. The $\beta$-deformation is a real one-parameter deformation which breaks the original \SU{4} flavour symmetry to its $\U{1}^{\times 3}$ Cartan subgroup, while leaving a single  $\cN=1$ supersymmetry of the parent theory intact. Its string-theory dual is obtained by deforming the $\text{S}^5$ part of the $\AdS{5}\times \text{S}^5$ background via a T-duality, a shift and another T-duality (TsT) transformation \cite{Lunin:2005jy}, see \cite{vanTongeren:2013gva} for a review of integrability-based approaches in this context. The $\gamma_i$-deformation can be seen as a generalisation of the $\beta$-deformation, which introduces three instead of one real deformation parameter into the parent \NfSYMt. This theory also breaks the original \SU{4} flavour symmetry to its $\U{1}^{\times 3}$ Cartan subgroup but in addition supersymmetry is completely broken. Its string-theory dual is obtained by introducing three consecutive TsT transformations on the three grand circles of the $\text{S}^5$ part of the string background \cite{Frolov:2005dj}. The $\beta$- and the $\gamma_i$-deformation exhibit the new finite-size effect of prewrapping \cite{Fokken:2013aea,Fokken:2013mza} which, on the gauge-theory side, originates from new elementary interactions that contain multiple colour traces. It starts to affect two-point correlation functions of length-$L$ operators at loop order $K=L-1$ which is one order prior to wrapping corrections. Unlike in \NfSYMt, such interactions have to be included in the $\beta$- and the $\gamma_i$-deformation to render the renormalised theories finite. While these multi-trace interactions appear to be naively suppressed by powers of $N$ in the action, they do contribute in the \tHooft limit since they receive a proliferation in $N$ in certain correlation functions, as shown in \cite{Fokken:2013aea,Fokken:2013mza,Fokken:2014soa}. It is not yet clear how prewrapping can be incorporated into integrability-based descriptions and how it appears on the string-theory side of the \AdSCFTc.

In the $\beta$-deformation, the integrability-based approach and alongside the \AdSCFTc in this settings have been tested for so-called length-$L$ single-impurity operators $\tr\bigl(\phi_i^{L-1}\phi_j\bigr)$ which are built from two different complex scalars. For these operators, the field-theoretic results of \cite{Fiamberti:2008sn} for $L\leq 11$ were reproduced in \cite{Gromov:2010dy,Arutyunov:2010gu} for states with length $L\geq 3$. For the $L=2$ states, however, the integrability-based approach yields a divergent anomalous dimension. In the $\gamma_i$-deformation, where the groundstate receives quantum corrections, the integrability-based approach was also used to determine the anomalous dimensions of the length-$L$ groundstate up to next-to-leading wrapping order in \cite{Ahn:2011xq}. This approach also leads to a divergent anomalous dimension for the $L=2$ state. So, the integrability-based approach yields similar divergences for the anomalous dimensions of $L=2$ states in the undeformed theory, the $\beta$-deformation and the $\gamma_i$-deformation. All three theories exhibit different amounts of supersymmetry and only some are exactly conformally invariant. Hence, this is the perfect testing ground for the prerequisites necessary for integrability.

In this thesis, we will investigate the governing principles of quantum integrability in \NfSYMt, the $\beta$- and the $\gamma_i$-deformation from a field-theoretic perspective. In particular, we will try to identify the conditions that are necessary for the entire spectrum of composite operators to be calculable in the integrability-based approaches. We will analyse prewrapping contributions for the $\beta$- and $\gamma_i$-deformation in the \tHooft limit. For the former, which remains exactly conformal for gauge group \SUN, we construct the complete one-loop dilatation operator from this analysis. For the latter, we show that conformal invariance is broken by running multi-trace couplings. We can nevertheless calculate the anomalous dimensions of the length-$L$ groundstate up to leading wrapping order $K=L$ in the \tHooft limit. For $L\geq 3$, we reproduce the integrability-based result of \cite{Ahn:2011xq} and for $L=2$, we find a renormalisation-scheme dependent finite result in contrast to the divergence found in \cite{Ahn:2011xq}. Knowing the anomalous dimensions of the $L=2$ single-impurity state in the $\beta$-deformation and of the groundstate in the undeformed theory and the $\gamma_i$-deformation, we devise a test to clarify whether integrability-based methods rely on the \tHooft limit, conformal invariance, supersymmetry, the absence of prewrapping contributions or a combination of these aspects.

In the last part of this thesis, we turn to the problem of evaluating the phase diagram of \NfSYMt and its deformations perturbatively in the \tHooft limit. Following the original approach of \cite{Witten:1998qj,Witten98Mar}, we investigate the thermal properties of these theories on \RxSt, where they exhibit a non-trivial phase transition.\footnote{The space \RxSt is the universal cover of the space $\text{S}^3\times\text{S}^1$, which is the topological boundary of the $\AdS{5}$ factor that appears in the \AdSCFTc, see \cite{Witten:1998qj}. Due to conformal invariance, on $\text{S}^3\times\text{S}^1$ a phase transition can depend only on the ratio of radii of the two spheres. While this is a compact space, it still exhibits infinitely many states in the \tHooft limit where $N\rightarrow \infty$ and hence a sharp phase transition in this setting occurs. See \cite{Witten:1998qj,Witten98Mar} for details. In light of the \AdSCFTc, note that phase transitions were also investigated on the string-theory side in \cite{Caldarelli:1999ar,Landsteiner:1999gb,Gao:1998ww}.} In \cite{Sundborg:1999ue}, the thermal partition function of \NfSYMt at zero coupling was computed by the means of \Polya theory instead of using a Feynman diagrammatic approach. From this partition function the zero-order phase-transition temperature, at which the low-energy description of the system in terms of colour-neutral composite operators breaks down, was determined. This approach was even pushed to the first loop order in the \tHooft coupling in \cite{Spradlin:2004pp} and the respective first-order correction to the phase-transition temperature was determined. Since the \Polya-theoretic approach employs the conformality of \NfSYMt and the knowledge of its dilatation operator to efficiently calculate perturbative contributions to the thermal partition function, the question is whether this approach can also be applied to other theories. We rederive the \Polya-theoretic approach of \cite{Spradlin:2004pp} in an algorithmic way and make it applicable also for the $\beta$- and the $\gamma_i$-deformation, including a separate treatment of finite-size corrections from potentially running multi-trace couplings. From this, we calculate the thermal one-loop partition function as well as the phase-transition temperature of both deformed theories. Intriguingly, we can derive a closed expression for the temperature dependence of the thermal one-loop partition function without employing any concepts from integrability. The derivation only relies on \Polya-theoretic methods and a well chosen summation procedure over all low-energy states.

To make the perturbative field theoretic approach to \NfSYMt and its deformations more accessible, we also derive a large set of tools for the Feynman-diagrammatic calculation of observables in a unified presentation in the first part of this thesis. All calculations are carried out by the means of these tools and thereby this thesis provides an independent test of all Feynman-diagrammatic calculations in the publications \cite{Fokken:2013aea,Fokken:2013mza,Fokken:2014soa,Fokken:2014moa}. We give a detailed derivation of the classical actions and symmetry transformations of \NfSYMt and its deformations and, in the quantised theories, we discuss the general renormalisation program including the renormalisation of composite operators. From this presentation, we build the {\tt Mathematica} tool {\tt FokkenFeynPackage} which allows for an efficient calculation of low-loop perturbative contributions to correlation functions in \NfSYMt and its deformations.

\section{Overview}
This thesis is grouped into two parts. All used conventions and the defining table of abbreviations and symbols \ref{tab:abbreviations} are given in \appref{sec:Conventions}. 

The first part, which contains \chapref{chap:The_models} and \ref{chap:Renormalisation} and appendices \ref{sec:Clifford_algebras_in_various_dimensions} through \ref{sec:the-fourier-transformation-of-the-free-two-point-function} is dedicated to introducing a firm field-theoretic framework for \NfSYMt and its deformations. This framework contains a complete presentation of the respective actions including the symmetry generators and a detailed discussion of the \tHooft. We explicitly derive the Feynman rules and discuss techniques to evaluate the UV divergences of Feynman integrals. We also review the general renormalisation program, including the renormalisation of composite operators and the construction of the dilatation operator of the theory.

The second part contains \chapref{chap:applications} and the appendices \ref{app:harmonic_action} through \ref{app: summation identities} and it is based on my publications \cite{Fokken:2013aea,Fokken:2014soa,Fokken:2013mza,Fokken:2014moa}. In this part, we will employ the presented field-theoretic framework to investigate prewrapping contributions and aspects of conformality in the $\beta$- and $\gamma_i$-deformation. In our analysis we will construct the complete one-loop dilatation operator of the deformed theories up to non-conformal contributions and we will use it to calculate the thermal one-loop partition function of the deformed theories. 

\begin{itemize}
	\item In \chapref{chap:The_models}, we will construct the building blocks of the field theories that we are interested in. We will derive the Minkowski space action of \NfSYMt, the $\beta$- as well as the $\gamma_i$-deformation in the component field formulation. From this, we will construct the action of all symmetry generators on elementary fields and the symmetry algebra of the corresponding theory. In addition, we will give the definition of gauge invariant composite operators.
	\item In \chapref{chap:Renormalisation}, we will review the general concepts of renormalisation, including the renormalisation of composite operators using explicit one-loop examples. We will also discuss the perturbative corrections to the dilatation generator and give a detailed discussion of the \tHooft limit.
	\item In \chapref{chap:applications}, we will first show that the $\gamma_i$-deformation is not conformally invariant in the \tHooft limit. Second, we will calculate the leading wrapping corrections to the integrability-vacuum state $\tr\bigl(\phi_i^L\big)$, thereby providing a test of integrability for the $\gamma_i$-deformation. Third, we will characterise prewrapping contributions in the $\beta$-deformation and derive the complete one-loop dilatation operator in the \tHooft limit. Fourth, we will use the dilatation operator of the $\beta$- and $\gamma_i$-deformation to calculate the thermal one-loop partition function of both deformations on \RxSt.
	\item In \chapref{chap:Conclusion_outlook}, we will present our conclusion and summary.
	\item In \appref{sec:Conventions}, we gather our conventions, $\sigma$-matrix identities, and the table of abbreviations and symbols used in this thesis.
	\item In \appref{sec:Clifford_algebras_in_various_dimensions}, we present our representation of Clifford algebra generators in $n\in \NN$ dimensions. We explicitly construct the $\gamma$-matrices of $\RR^{(3,1)}$, $\RR^{6}$, and $\RR^{(9,1)}$.
	\item In \appref{sec:spinor_in_various_dimensions}, we discuss properties of spinors in four- and ten-dimensional Minkowski space, as well as six-dimensional Euclidean space.
	\item In \appref{app:Kaluza_Klein_compactification}, we present details of the Kaluza-Klein reduction.
	\item In \appref{sec:conformal_algebra}, we give the action of the conformal generators on coordinates and primary fields of a given theory.
	\item In \appref{sec:comparison-of-field-and-oscillator-representation}, we give explicit examples how the action of symmetry generators on elementary fields is mapped to the oscillator representation.
	\item In \appref{app:Feynman_rules}, we derive the Feynman rules of non-abelian gauge theories in four-dimensional Minkowski space with fields in the adjoint representation of the gauge group \UN or \SUN. In addition to the gauge fields, the theory contains complex or real scalars and Weyl fermions.
	\item In \appref{app:EOM_Bianchi}, we present the Bianchi identity and classical \eom of elementary fields compatible with our conventions of \chapref{chap:The_models}.
	\item In \appref{sec:Feynman_rules_Mathematica}, we present the manual of the \ttt{Mathematica} package \ttt{FokkenFeynPackage}, which can be used to generate generic integrands of Feynman diagrams in \NfSYMt and its deformations. At one- and two-loop order, scalar one-scale integrals can also be solved explicitly with this tool.
	\item In \appref{sec:Renormalisation_schemes}, we give a precise definition of the renormalisation schemes that alter the spacetime dimension. These include the MS, $\ol{\text{MS}}$, DR, and $\ol{\text{DR}}$ schemes.	
	\item In \appref{app:Evaluating_Feynman_integrals}, we discuss general techniques to evaluate Feynman integrals. In particular, we discuss how the UV divergence of a given graph may be extracted by choosing special kinematics. In this discussion we also briefly touch the question how spurious IR divergences in these kinematics may be removed. Finally, we present the general result of one-loop propagator-type tensor integrals.
	\item In \appref{sec:the-fourier-transformation-of-the-free-two-point-function}, we calculate the Fourier transformation of the two-point correlation function of composite operators in a free theory.
	\item In \appref{app:harmonic_action}, we present an explicit form of the harmonic action of \NfSYMt in the oscillator representation.
	\item In \appref{app:oneloopse}, we present the one-loop self-energy contribution to the scalar propagator in our conventions.
	\item In \appref{sec:Coupling_tensor_identities}, we give the coupling-tensor identities that are need for the evaluation of Feynman integrals in \secref{sec:non-conformal_double_trace_coupling} and \ref{sec:cake}.
	\item Finally, appendices \ref{app: PD2 calculation}, \ref{app: corrections}, and \ref{app: summation identities} contain the computational details needed for the calculation of the one-loop thermal partition function of the $\beta$- and $\gamma_i$-deformation in \secref{sec:the-thermal-one-loop-partition-functions-of-the-deformed-theories}.
\end{itemize}

\chapter{The classical theories}\label{chap:The_models}
In this chapter, we introduce classical aspects of the gauge theories and fundamental objects that we are most concerned with in this thesis. 

We derive the action of \NfSYMt as it arises when an $\cN=1$ supersymmetric Yang-Mills theory is dimensionally reduced from ten- to four-dimensional Minkowski space. In addition, we discuss the symmetries of the \NfSYM action and derive the symmetry algebra from the action of all symmetry generators on elementary fields. We also give an explicit mapping to the spinor representation from which the dilatation operator in the spin-chain picture is constructed. Using the fully fixed action and symmetry algebra of the undeformed theory, we introduce the action of the $\beta$- and $\gamma_i$-deformation and discuss which symmetries prevail after the deformation procedure. In addition to the single-trace action, which is entirely fixed from the undeformed theory, we also include all renormalisable multi-trace interactions compatible with the symmetries of the theories in the definition of the deformations.

After having constructed the actions and symmetry generators of interest, we discuss local gauge invariant composite operators, which are objects built from the elementary fields of the theory that all reside at a single point in spacetime in a normal-ordered fashion. As discussed in the introduction, the scaling dimensions of such objects are observables in a \CFT -- even in the quantised theory that we will focus on in \chapref{chap:Renormalisation}. Unlike the elementary interactions, composite operators are added into correlation functions by hand and their properties are not restricted by the symmetries of the underlying field theory. We fix the alphabet from which composite operators can be built up to perturbative quantum corrections to the equations of motion (\eom) of elementary fields. We also discuss the mapping of such operators to the spin-chain picture and finally we give the action of symmetry generators on composite operators.

\newpage

\section{\texorpdfstring{$\mathcal{N}=1$ SYM theory in ten dimensions}{N=1 SYM theory in ten dimensions}}\label{sec:N1SYM_10D_to4D}
In this section, we construct the action of \NfSYMt in flat four-dimensional Minkowski space by dimensionally reducing a classical Yang-Mills theory with a simple $\mathcal{N}=1$ supersymmetry in flat ten-dimensional Minkowski space. While \NfSYMt in four-dimensional Minkowski space is entirely fixed by symmetries, its construction via the dimensional reduction from a ten-dimensional theory has the advantage to fix all occurring tensor structures in a straightforward way, once the ten-dimensional Clifford algebra is fixed. 

We start with a supersymmetric ten-dimensional theory on ten-dimensional Minkowski space with mostly plus metric ($\RR^{(9,1)}$) with fermions of only one chirality. Note that this theory is therefore only classically consistent \cite{Seiberg:1997ax}. We then compactify six Euclidean dimensions on a torus in a Kaluza-Klein reduction to obtain the four-dimensional theory on $\RR^{(3,1)}$. The compactification procedure not only lifts the problem of fermions with a single chirality, but also 
enhances the symmetry, so that the dimensionally reduced theory is superconformally invariant at the quantum level \cite{Caswell:1980yi,SOHNIUS1981245,Mandelstam:1982cb,Brink:1982wv,HOWE1984125,Seiberg:1988ur} with $\cN=4$ supersymmetries. This derivation was first done in \cite{Brink197777} and the Kaluza-Klein reduction in this context goes back to \cite{Cremmer1976409,Schwarz1981321}.

\subsection{The action}
A supersymmetric theory must have the same number of bosonic and fermionic real on-shell degrees of freedom (\dof) if translations are an invertible operation \cite{VanProeyen:1999ni}. This greatly reduces the possible dimensions in which supersymmetric theories may exist, as bosonic and fermionic \dof scale differently with the space-time dimension $d$. Real scalars $\phi$ give 1 \dof, gauge fields\footnote{One \dof is eliminated by the gauge symmetry and one by the mass-shell condition $k^m A_m$.} $A^m$ give $(d-2)$ \dof and real Dirac spinors\footnote{A $d$-dimensional Dirac spinor can be regarded as a $2\times 2^{d/2}$-dimensional real vector. The mass-shell condition $\Gamma^m p_m\psi=0$ cuts the \dof in half.} $\Psi$ yield $2^{d/2}$ \dof There are two additional constraints that may reduce the fermionic \dof First, for massless spinors in even dimensions the spinor representation can be reduced to the Weyl representation, which cuts the fermionic \dof in half. In the Weyl representation\footnote{Weyl spinors in $d$ dimensions are eigenvectors of the $\gamma$-matrix $\Gamma^{d+1}=\complexi^{d/2}\prod_{i=1}^d\Gamma^i$.}, spinors are grouped according to their chirality. Using the projectors $P^+$ and $P^-$, a $2^{d/2}$-dimensional Dirac spinor splits into a plus- and minus-chirality part as
\begin{equation}
\Psi=\begin{pmatrix}
\psi\\
\chi^{\dagger}
\end{pmatrix}\eqncom\qquad
\Psi^-=P^- \Psi=\begin{pmatrix}
\psi\\
0
\end{pmatrix}\eqncom\qquad
\Psi^+=P^+ \Psi=\begin{pmatrix}
0\\
\chi^{\dagger}
\end{pmatrix}\eqncom
\end{equation}
where $\psi$ and $\chi$ are $2^{d/2-1}$-dimensional Weyl spinors and we used the dagger to distinguish a plus- from a minus-chirality spinor. See \appref{sec:Clifford_algebra_construction} for a derivation of the $d$-dimensional Weyl representation. 
Second, in certain spacetime dimensions the spinors may be chosen to be in a Majorana representation, which also reduces the fermionic \dof by a factor of $\frac 12$. Majorana spinors $\Psi^\tM$ are invariant under the Majorana conjugation
\begin{equation}\label{eq:Majorana_condition}
\Psi^\tM=(\Psi^\tM)^C=B(\Psi^{\tM})^*\eqncom
\end{equation}
where $B$ is determined from the $\Gamma$-matrices in $d$ dimensions. If the latter are purely real, the Majorana spinors are purely real as well and $B=\one$. See \appref{sec:spinor_in_various_dimensions} for details on spinors in various dimensions and their irreducible representations. 

In ten-dimensional Minkowski space, the Majorana- and Weyl-conditions are mutually compatible and we can form Majorana-Weyl spinors with $8$ \dof This is exactly the number of \dof that the ten-dimensional gauge field has and therefore we can build a supersymmetric Yang-Mills theory in ten dimensions with a gauge field $A_{10}^m$ coupled to a Majorana-Weyl-fermion field $\Psi_{10}^{\tM -}$. We take the corresponding action to have the form\footnote{The factor of $\tfrac{1}{2}$ in front of the fermion term appears as we present the action in terms of Majorana spinors instead of the Dirac spinors.}
\begin{equation}\label{eq:10d_action}
S_{10}=\int\de^{10}x \tr\left[-\frac 14F^{mn}F_{mn}
+\frac 12\ol{\Psi_{10}^{\tM-}}\Gamma^m\complexi\D_m\Psi_{10}^{\tM-}
\right]\eqncom
\end{equation}
where $\tr[\,\cdot\,]$ indicates a trace over the fundamental indices of the gauge group, the $\Gamma^m$ are the ten-dimensional $\Gamma$-matrices, $m,n\in\{0,1,\dots,9\}$ are  spacetime indices and the bar indicates the usual Dirac conjugate\footnote{Numerically $\beta_{10}$ is the same as $\Gamma^0$ and we only differentiate between them to account for the particular index positions within $\beta$ that are needed for the Dirac conjugate in our conventions.} $\ol{X} =X^\dagger \beta_{10}$. All fields transform in the adjoint representation of the gauge group \UN or \SUN, i.e.\ a field $X$ has the form
\begin{equation}
X=\sum_{a=s}^{N^2-1} X^a\T^a\eqncom \qquad \text{with} \qquad
a=\begin{cases}
s=1 & \text{for }\SUN\\
s=0 & \text{for }\UN
\end{cases}\eqndot
\end{equation} 
The covariant derivative and field strength are respectively given by
\begin{equation}
\D_m\Psi=[D_m,\Psi]=(\partial_m\Psi)-\complexi g_{10} [A_m,\Psi]\eqncom\quad F_{mn}=(\partial_m A_n)-(\partial_n A_m) -\complexi g_{10} [A_m,A_n]
\end{equation}
and the mass dimensions of the fields and parameters are
\begin{equation}
[A_m]=4\eqncom\qquad [\Psi^{\tM-}]=\frac 92\eqncom\qquad [g_{10}]=-3\eqndot
\end{equation}

The action \eqref{eq:10d_action} is invariant under the a SUSY transformation which is generated by the supercharges $\mathfrak{Q}_{10}^{\tM +}$ and $\ol{\mathfrak{Q}^{\tM +}_{10}}$, see e.g.\ \cite{deWit:1997sz} for a classification of supersymmetries in dimensions $d\leq 11$. This transformation can be realised via the unitary operators
\begin{equation}\label{eq:unitary_operator_SUSY_10}
U_{(\xi \cdot Q)}=\e^{-\complexi\delta_{(\xi \cdot Q)}}\eqncom\qquad
\delta_{(\xi \cdot Q)}=\frac 12(\ol{\xi^{\tM -}}\mathfrak{Q}^{\tM +}_{10}+\ol{\mathfrak{Q}^{\tM +}_{10}}\xi^{\tM -})\eqncom
\end{equation}
where the constant fermionic parameters $\xi^{\tM -}$ have a mass dimension $[\xi^{\tM -}]=\tfrac 12$. The fields in the action \eqref{eq:10d_action} transform according to
\begin{equation}
f^\prime=U_{(\xi \mathfrak{Q})}^{-1}f\,U_{(\xi \mathfrak{Q})}=f+\complexi [\delta_{(\xi \cdot Q)},f]+\order{\xi^2}\eqncom
\end{equation}
with the non-vanishing field variations $\delta f=f^\prime-f$ given by
\begin{equation}\label{eq:susy_trafo_10D}
\begin{aligned}
\delta\Psi^{\tM-}_{10}&=\complexi[\delta_{(\xi \cdot Q)},\Psi^{\tM-}_{10}]
=-\frac{\complexi}{2} \{\ol{\mathfrak{Q}_{10}^{\tM +}},\Psi^{\tM-}_{10}\}\xi^{\tM -}=
-\frac{\complexi}{2} F^{mn}M_{mn}\xi^{\tM-}\eqncom\\
\delta\ol{\Psi^{\tM-}_{10}}&=
\complexi [\delta_{(\xi \cdot Q)},\ol{\Psi^{\tM-}_{10}}]
=\frac{\complexi}{2} \ol{\xi^{\tM -}}\{\mathfrak{Q}_{10}^{\tM +},\ol{\Psi^{\tM-}_{10}}\}
=\frac{\complexi}{2}\ol{\xi^{\tM-}}F^{mn}M_{mn}\eqncom\\
\delta A_n&=\complexi[\delta_{(\xi \cdot Q)},A_n]=
\frac{\complexi}{2}\bigl(\ol{\Psi^{\tM -}_{10}}\,\Gamma_n\xi^{\tM -}-\ol{\xi^{\tM -}}\,\Gamma_n\Psi^{\tM -}_{10}\bigr)\eqndot
\end{aligned}
\end{equation}
The generators $M_{mn}=\frac{\complexi}{4}\bigl[\Gamma_m,\Gamma_n\bigr]$ are the generators of the spinor representation, \cf \appref{sec:spinor_in_various_dimensions} for details.
Under this variation the action turns into a total divergence (t.d.)\footnote{For the second equality we used the $\Gamma$-matrix identities 
	$-\frac{\complexi}{2}\eta_{mp} \Gamma_n+\frac{\complexi}{2}\eta_{np} \Gamma_m+M_{mn}\Gamma_p=\frac{\complexi}{3!}\Gamma_{[m}\Gamma_n\Gamma_{p]}$ and 
	$-\frac{\complexi}{2} \Gamma_m\eta_{pn}+\frac{\complexi}{2} \Gamma_n\eta_{mp}+\Gamma_pM_{mn}=\frac{\complexi}{3!}\Gamma_{[p}\Gamma_m\Gamma_{n]}$.}
\begin{equation}
\begin{aligned}\label{eq:variation_S10}
\delta S_{10} 
&=-\frac 14\int\de^{10}x \tr\Bigl[
\ol{\xi^{\tM -}}(\D^pF^{mn})\bigl(\complexi\eta_{mp} \Gamma_n - M_{mn}\Gamma_p\bigr)\Psi^{\tM -}_{10}\\
&\phan{=-\frac 14}-\ol{\Psi^{\tM -}_{10}}\bigl(\complexi\eta_{mp}\Gamma_n+\Gamma_pM_{mn}\bigr)(\D^pF^{mn})\xi^{\tM -} 
+\frac{g_{10}}{2}\ol{\Psi^{\tM -}_{10}}\Gamma^m\bigl[\delta A^m,\Psi^{\tM -}_{10}\bigr]
+\text{t.d.}\Bigr]\\
&=\frac{1}{24}\int\de^{10}x \tr\Bigl[
\ol{\xi^{\tM -}}(\D^pF^{mn})\Gamma_{[m}\Gamma_n\Gamma_{p]}\Psi^{\tM -}_{10}
+\ol{\Psi^{\tM -}_{10}}\Gamma_{[m}\Gamma_n\Gamma_{p]}(\D^pF^{mn})\xi^{\tM -}\\
&\phan{=-\frac{1}{24}\int\de^{10}x \tr\Bigl[}
-3!g_{10}\ol{\Psi^{\tM -}_{10}}\Gamma^m\bigl[(\ol{\xi^{\tM -}}\,\Gamma_m\Psi^{\tM-}_{10})-(\ol{\Psi^{\tM-}_{10}}\,\Gamma_m\xi^{\tM -}),\Psi^{\tM -}_{10}\bigr]
+\text{t.d.}\Bigr]\\
&\sim\int\de^{10}x\,\bigl(\text{t.d.}\bigr)
\eqncom
\end{aligned}
\end{equation}
where the total divergence terms appear in various partial integrations. To arrive at the last equality, note that the totally antisymmetric product of $\Gamma$-matrices allows us to also antisymmetrise the terms $\D^pF^{mn}$, which turns the latter ones into the Bianchi identity\footnote{The Bianchi identity in non-abelian gauge theories is conceptually equivalent to the second Bianchi identity for Riemannian curvature tensors \cite{Hubsch:2015vpa}.}. Therefore, the first and second term in the second equality vanish identically. The vanishing of three-fermion terms is guaranteed by the Fierz identities and hence implicitly depends on the spacetime dimension as well as Majorana and/or Weyl constraints that the fermions fulfil. For Majorana-Weyl fermions in ten dimensions both terms vanish as was shown in \cite{Brink197777}. 

\subsection{Dimensional reduction to four dimensions}\label{subsec:Dimensional_reduction_to_four_dimensions}
To get from the ten-dimensional action to a four-dimensional one, we need to eliminate the dependence of all fields on six extra-dimensions. A consistent way to do this is to perform a Kaluza-Klein reduction from ten to four dimensions, c.f.\ \cite{Cheng:2010pt, Witten_KK, Duff19861} for detailed overviews. We will only discuss the key ideas for a free scalar field in ten dimensions here and refer the reader to \appref{app:Kaluza_Klein_compactification} and the references therein for the analogous discussion in the case of gauge fields and fermions.

Let us assume that six dimensions of our ten-dimensional spacetime are the periodic directions of a six-torus $\T^6=(S^1)^{\times 6}$, where each $S^1$ has radius $R$. We can write the dependence of a real scalar field $X$ in ten dimensions on the six compact coordinates $y^j\in \T^6$ in terms of the following Fourier expansion
\begin{equation}\label{eq:scalar_inKK_modes}
X(x^\mu,y^j)=\sum_{\bf{n}\in \ZZ^6}X_{(\bf{n})}(x^\mu)\frac{\e^{\frac{\complexi\bf{n}\bf{y}}{R}}}{(2\pi R)^3}\eqncom
\end{equation}
where $x^\mu\in\RR^{(3,1)}$ is a coordinate of four-dimensional Minkowski-space and the Fourier mode has dimension $[X_{(\bf{n})}]=1$. Note that a derivative of $X$ with respect to one of the $\T^6$ coordinates $\frac{\partial}{\partial y_j}$ now simply generates a factor of $\frac{\complexi n^j}{R}$ which is expected since the Dirac operator has discrete eigenvalues in this space that scale as $\frac{\bf{n}}{R}$. With this choice, the action of a free real scalar field in ten dimensions becomes
\begin{equation}
\begin{aligned}
S&=-\frac 12\int\de^{4}x \de^6y\,\tr\left(\partial^m X\partial_m X\right)\\
&=-\frac 12\sum_{\bf{l},\bf{n}\in\ZZ^6}\int\de^{4}x \de^6y\,\frac{\e^{\frac{\complexi(\bf{l}+\bf{n})\bf{y}}{R}}}{(2\pi R)^6}
\tr\left(\partial^\mu X_{(\bf{l})}\partial_\mu X_{\bf{(n})}-\frac{l^jn_j}{R^2}X_{(\bf{l})}X_{(\bf{n})}\right)\\
&=\int\de^{4}x\,
\tr\left(-\frac 12 \partial^\mu X_{(\bf{0})}\partial_\mu X_{\bf{(0})}-\sum_{\bf{n}> \bf{0}}\Bigl(\partial^\mu X_{(\bf{-n})}\partial_\mu X_{\bf{(n})}+\frac{\bf{n}^2}{R^2}X_{(\bf{-n})}X_{(\bf{n})}\Bigr)\right)\eqncom
\end{aligned}
\end{equation}
where the reality of $X$ implies that $(X_{(\bf{n})})^\dagger=X_{(-\bf{n})}$. From a four-dimensional perspective, the ten-dimensional model acquires an infinite tower of free complex scalar fields $X_{(\bf{n})}$ with masses $m_{X_{(\bf{n})}}=\frac{\bf{n}^2}{R^2}$, called Kaluza-Klein tower. If we now take the torus to have vanishing radius $R\rightarrow 0$, all scalar fields $X_{(\bf{n}\neq \bf{0})}$ become infinitely heavy. From the four-dimensional perspective, they become invisible as long as we only probe distances that are much larger than $R$. Hence, in the low energy limit, the free scalar field in ten dimensions is seen as a single real scalar field in four dimensions. 

In the action \eqref{eq:10d_action}, which involves gauge fields and fermions, massive Kaluza-Klein modes appear in a very similar way when we expand the fields in analogy to \eqref{eq:scalar_inKK_modes}. Since $R$ is the only scale of the torus, the mass of the fermionic and scalar Kaluza-Klein modes must be proportional to $R^{-1}$ and $R^{-2}$, respectively. Therefore, when we take the strict\footnote{We want to construct a conformal four-dimensional theory, in which the notion of probing certain distances is not well defined. Therefore we must take $R=0$, strictly.} limit $R\rightarrow 0$, these modes are not accessible in the four-dimensional theory. The dimensionally reduced action \eqref{eq:10d_action} becomes
\begin{equation}\label{eq:10d_action_in_4D_KK}
\begin{aligned}
S&=\int\de^{4}x \tr\Biggl(-\frac 14F^{\mu\nu}F_{\mu\nu}
-\frac 12\bigl(\D^{\mu}\varphi^j\bigr)\bigl(\D_\mu \varphi_j\bigr)
+\frac{\gym^2}{4}\bigl[\varphi^i,\varphi^j\bigr]\bigl[\varphi_i,\varphi_j\bigr]\Biggr)\\
&\phan{=}
+\int\de^{4}x\tr\Biggl(
\frac 12\ol{\psi_{10}^{\tM-}}\,\Gamma^\mu\complexi\D_\mu\psi_{10}^{\tM-}
+\frac{\gym}{2} \ol{\psi_{10}^{\tM-}}\,\Gamma^{j+3}\bigl[\varphi_{j},\psi_{10}^{\tM-}\bigr]
\Biggr)\eqncom
\end{aligned}
\end{equation}
where the indices run over $\mu,\nu\in\{0,1,2,3\}$ and $i,j\in\{1,2,\dots6\}$. The six gauge-field components along the vanishing $\T^6$ directions\footnote{The metric on this space is Euclidean and therefore the distinction of upper and lower indices for the $\varphi^j$ is purely conventional.} become the six real scalars $\varphi^j$ and the coupling constant $\gym$ is related to the ten-dimensional one as $\gym=\frac{g_{10}}{(2\pi R)^3}$. The dimensionally reduced Fourier modes have the following mass dimensions
\begin{equation}\label{eq:mass_dim}
[A_\mu]=1\eqncom\qquad[\varphi_j]=1\eqncom\qquad[\psi_{10}^{\tM-}]=\frac 32\eqncom\qquad
[\gym]=0\eqncom
\end{equation}
and the covariant derivative and field strength in four dimensions are respectively given by
\begin{equation}
\D_\mu X=[\D_\mu,X]=(\partial_\mu X)-\complexi \gym [A_\mu,X]\eqncom\quad F_{\mu\nu}=\partial_\mu A_\nu-\partial_\mu A_\nu -\complexi \gym [A_\mu,A_\nu]\eqndot
\end{equation}

To arrive at an entirely four-dimensional action, we still need to express the ten-di\-men\-sion\-al $\Gamma$-matrices and Majorana-Weyl fermions in \eqref{eq:10d_action_in_4D_KK} in terms of four-di\-men\-sion\-al quantities. Several aspects concerning the Clifford-algebra in four, six, and ten dimensions can be found in \appref{sec:Clifford_algebras_in_various_dimensions} and the references therein. Likewise, the corresponding spinor representations for the fermions are discussed in \appref{sec:spinor_in_various_dimensions} and the references therein. We only give a brief discussion here and refer the reader to these appendices for details. We start with the Minkowski-space Clifford algebra in ten dimensions with mostly plus metric\footnote{The first and second argument of \clifford indicate how many positive and respectively negative entries the metric has.} $\clifford(9,1)$. It is generated by the $\Gamma$-matrices that fulfil
\begin{equation}\label{eq:gamma_10}
\{\Gamma^m,\Gamma^n\}=-2\eta^{mn}\eqncom\qquad
\eta^{mn}=\diag(-1,+1,\dots,+1)\eqncom\qquad
 0\leq m,n\leq 9\eqndot
\end{equation}
From the periodicity properties of Clifford algebras over real vector spaces the isomorphism $\clifford(9,1)\simeq\clifford(0,6)\otimes \clifford(3,1)$ can be deduced, as shown e.g.\ in \cite{OFarrill}. So the ten-dimensional algebra can be constructed from a six-dimensional Euclidean-space algebra with generators $\rho^a$ and 
four-dimensional Minkowski-space algebra with generators $\gamma^\mu$, that respectively fulfil\footnote{The additional sign in the six-dimensional case occurs, as the $\rho^a$ are generators of $\clifford(0,6)$ which has a negative Euclidean metric $-\delta^{ab}$.}
\begin{equation}\label{eq:4_6_clifford_relation}
\{\rho^a,\rho^b\}=2\delta^{ab}\eqncom\qquad 
\{\gamma^\mu,\gamma^\nu\}=-2\eta^{\mu\nu}\eqndot
\end{equation}
We take $\rho^a$ and $\gamma^\mu$ to be in the Weyl representation of the explicit form 
\begin{equation}\label{eq:4_6_gamma_matrices}
\rho^a=\begin{pmatrix}
0& \Sigma^a\\
\bar\Sigma^a&0
\end{pmatrix}\eqncom\quad
\rho^7=
\begin{pmatrix}
-\onee{4}&0\\
0&\onee{4}
\end{pmatrix}\eqncom\quad
\gamma^\mu=
\begin{pmatrix}
0&\sigma^\mu\\
\bar{\sigma}^\mu&0
\end{pmatrix}\eqncom\quad
\gamma^5=\begin{pmatrix}
-\onee{2}&0\\
0&\onee{2}
\end{pmatrix}\eqncom
\end{equation}
where $\sigma^\mu$, $\bar{\sigma}^\mu$ and $\Sigma^a$, $\bar{\Sigma}^a$ are the reduced $\Gamma$-matrices defined in \eqref{eq:sigma_matrices} and \eqref{eq:Sigma_6}, respectively. The new generators $\rho^7$ and $\gamma^5$ are the chirality operators in six and four dimensions, respectively. Using \eqref{eq:4_6_clifford_relation} and \eqref{eq:4_6_gamma_matrices}, we can construct the generators of $\clifford(9,1)$ that fulfil \eqref{eq:gamma_10} as
\begin{equation}\label{eq:Gamma_10}
\Gamma^m=\begin{cases}
\onee{8}\otimes \gamma^m	& 0\leq m\leq 3\\
\rho^{m-3}\otimes\complexi \gamma^5& 4\leq m\leq 9
\end{cases}\eqncom\qquad
\Gamma^{11}=\complexi \prod_{m=0}^9\Gamma^{m}=-\rho^7\otimes\gamma^5\eqndot
\end{equation}
With this realisation of $\gamma$-matrices we can construct the matrix $B$ for the Majorana condition \eqref{eq:Majorana_condition}. It realises the complex conjugation of $\gamma$-matrices as $(\gamma^m)^*= \pm B^{-1}\gamma^m B$ and is proportional to the product of $\gamma$-matrices with complex entries
\begin{equation}\label{eq:B_in_all_dimensions}
B_4=-\complexi \gamma^2
=\begin{pmatrix}
0&-\complexi \sigma^2\\
\complexi\sigma^2&0
\end{pmatrix}\eqncom\qquad
B_6=\complexi \rho^1\rho^4\rho^6
=\begin{pmatrix}
0&\onee{4}\\
\onee{4}&0
\end{pmatrix}\eqncom\qquad
B_{10}=B_6\otimes B_4\eqncom
\end{equation}
where $\sigma^2$ is the second Pauli matrix. 
The projectors to chirality eigenstates are built from $\gamma^5$, $\rho^7$, and $\Gamma^{11}$ in the usual way
\begin{equation}\label{eq:P_in_all_dimensions}
P_4^\pm=\frac 12(\onee{4}\pm \gamma^5)\eqncom\qquad
P_6^\pm=\frac 12(\onee{8}\pm \rho^7)\eqncom\qquad
P_{10}^\pm=\frac 12(\onee{8}\pm \Gamma^{11})
=P_6^-\otimes P_4^\pm+P_6^\mp\otimes P_4^+\eqncom
\end{equation}
where we reexpressed $P_{10}^\pm$ in terms of the lower-dimensional projectors in the last equality for later convenience. Finally, we construct generators of the spinor representation in four, six, and ten dimensions from the $\gamma$-matrices as
\begin{equation}\label{eq:Lorentz_generator_D_dim}
M^{mn}=\frac{\complexi}{4}[g^m,g^n] \eqncom \quad \text{with}\quad g=\{\rho,\gamma,\Gamma\}\eqndot
\end{equation}
Spinors transform as vectors under this representation and therefore a ten-dimensional Dirac-spinor decomposes into a six- and a four-dimensional one, analogously to the decomposition of the ten-dimensional $\Gamma$-matrices:
\begin{equation}\label{eq:10D_spinors}
\psi_{10}=\psi_6\otimes\psi_4\eqncom \qquad
\ol{\psi_{10}}=\psi_{10}^\dagger\beta_{10}=(\psi_6^\dagger\beta_6\otimes\psi_4^\dagger\beta_4)\eqncom
\end{equation}
where $\beta$ realises the hermitian conjugation of $\gamma$-matrices as $(\gamma^i)^\dagger=\beta\gamma^i\beta^{-1}$. We can now construct a Majorana-Weyl spinor $\psi^{\tM \pm}_{10}$ in ten dimensions that fulfils the Majorana and chirality constraints
\begin{equation}
\psi^{\tM \pm}_{10}=B_{10}(\psi^{\tM \pm}_{10})^*\eqncom\qquad
P^{\pm}_{10}\psi^{\tM \pm}_{10}=\pm\psi^{\tM \pm}_{10}\eqncom
\end{equation}
with $B_{10}$ and $P^{\pm}_{10}$ from \eqref{eq:B_in_all_dimensions} and \eqref{eq:P_in_all_dimensions}, respectively. When we take negative-chirality fields in four and six dimensions to have lower indices, the negative Majorana-Weyl spinors and their conjugates in ten dimensions have the explicit form\footnote{Note that the $*$ indicates the conjugation of the field in the six-dimensional Euclidean space, analogously to the $\dagger$ in four-dimensional Minkowski space. See \appref{sec:spinors-in-six-dimensional-euclidean-space} for details.}
\begin{equation}\label{eq:10D_MW_spinors}
\begin{aligned}
\psi_{10}^{\tM -}&=
\begin{pmatrix}
\psi_A\\
0
\end{pmatrix}
\otimes
\begin{pmatrix}
\chi_\alpha\\
0
\end{pmatrix}+
\begin{pmatrix}
0\\
\psi^{*A}
\end{pmatrix}
\otimes
\begin{pmatrix}
0\\
\chi^{\dagger\dot\alpha}
\end{pmatrix}\eqncom\\
\ol{\psi_{10}^{\tM -}}&=
(\psi^{*A},0)\otimes(0,\chi^{\dagger}_{\dot{\alpha}})+
(0,\psi_A)\otimes(\chi^\alpha,0)\eqndot
\end{aligned}
\end{equation}
Using the crucial isomorphism $\spin{6}\simeq\su{4}$, the spinors $\psi_A$ and $\psi^{*A}$ transform in the fundamental and anti-fundamental representation of \su{4}, respectively. The spinors $\chi_\alpha$ and $\chi^{\dagger\dot{\alpha}}$ transform in the fundamental and anti-fundamental representations of \spl{2,\CC}, which we label by $\spl{2}$ and $\splbar{2}$, respectively.\footnote{The group $\text{SL}(2,\CC)$ is the universal cover of the Lorentz group \SO{3,1}. More concretely, the proper orthochronous Lorentz group is $\SO{3,1}\simeq \text{SL}(2,\CC)/\ZZ_2$ and the precise connections are nicely presented in \cite[\chap{7A}]{tanedo2013flight}.}
We can get rid of the redundancies of the ten-dimensional Dirac representation, by introducing the eight component spinors
\begin{equation}\label{eq:10D_MW_spinors_2}
\begin{aligned}
\lambda_{A\alpha}=
\psi_A\otimes\chi_\alpha\eqncom\quad
\ol{\lambda}^{ A\dot\alpha}=
\psi^{*A}\otimes\chi^{\dagger\dot\alpha}\eqncom\quad
\lambda^A_{\alpha}=
\psi^{* A}
\otimes
\chi_\alpha\eqncom\quad
\ol{\lambda}_A^{\dot\alpha}=
\psi_{A}\otimes\chi^{\dagger\dot\alpha}\eqncom
\end{aligned}
\end{equation}
where the spinors with both upper or lower indices have negative-chirality and the remaining ones have positive-chirality in ten dimensions. Note that the mass dimension of these fermions follows from \eqref{eq:mass_dim} to be $[\lambda]=\frac 32$. We can now insert the definitions of $\Gamma$-matrices \eqref{eq:Gamma_10} and spinors \eqref{eq:10D_MW_spinors} into the action \eqref{eq:10d_action_in_4D_KK} and by resolving the redundant matrix structure we find the action of \NfSYMt with real scalars and spinors from \eqref{eq:10D_MW_spinors_2} to be
\begin{equation}
\begin{aligned}\label{eq:N4_action_real_scalars}
S&=\int\de^{4}x \tr\Bigl(-\frac 14F^{\mu\nu}F_{\mu\nu}
-\frac 12\bigl(\D^{\mu}\varphi^j\bigr)\bigl(\D_\mu \varphi_j\bigr)
+\frac{\complexi}{2}\bigl(\lambda_A^\alpha(\sigma^\mu)_{\alpha\dot{\beta}}\D_\mu\ol{\lambda}^{A\dot{\beta}}
+\ol{\lambda}^A_{\dot{\alpha}}(\bar\sigma^\mu)^{\dot{\alpha}\beta}\D_\mu\lambda_{A\beta}\bigr)
\\
&\phan{=\int\de^{4}x\tr\Biggl(}
-\frac{\complexi\gym}{2}\bigl(
\bar\Sigma^{jAB}\lambda^{\alpha}_A\bigl[\varphi_j,\lambda_{B\alpha}\bigr]
-\Sigma^j_{AB}\ol{\lambda}^A_{\dot{\alpha}}\bigl[\varphi_j,\ol{\lambda}^{B\dot{\alpha}}\bigr]
\bigr)
+\frac{\gym^2}{4}\bigl[\varphi^i,\varphi^j\bigr]\bigl[\varphi_i,\varphi_j\bigr]
\Bigr)\eqndot
\end{aligned}
\end{equation}
Note that in the \su{4} representation, the plus- and minus-chirality fermions couple as a singlet to the gauge fields and as a real vector to the scalars.\footnote{In the perspective of representation theory, the fermions couple to the gauge field via the singlet in the decomposition $4\otimes \ol{4}=15\oplus 1$, whereas they couple to the scalars via the $6$ in the two decompositions $4\otimes 4=10\oplus 6$ and $\ol{4}\otimes \ol{4}=\ol{10}\oplus \ol{6}=\ol{10}\oplus 6$.}

\section{\texorpdfstring{\NfSYMt in four dimensions}{N=4 SYM theory in four dimensions}}\label{sec:N4SYM_4D}
In the last section we have derived the action of a supersymmetric four-dimensional model from an $\mathcal{N}=1$ SYM theory in ten dimensions. With the definitions 
\begin{equation}
\varphi_{AB}\equiv\frac{\complexi}{2}\varphi_j\Sigma^j_{AB}\eqncom\qquad
\varphi^{AB}\equiv\frac{\complexi}{2}\varphi_j\bar\Sigma^{jAB}
\end{equation} 
for antisymmetric scalar fields, this model in flat Minkowski space $\RR^{(3,1)}$ takes the form\footnote{For the transformation of the scalars we used $-4\delta^i_j=\Sigma^i_{AB}\bar{\Sigma}_j^{AB}$. The two kinetic fermion terms from \eqref{eq:N4_action_real_scalars} are combined via a partial integration and the identities \eqref{eq:sigma_matrix_identities} and \eqref{eq:Weyl_spinor_raise_lower}. For a presentation of this action in the $\cN=1$ superspace formulation and a general introduction to the superspace approach see \cite[\chap{4}]{Gates:1983nr} and \cite{Siegnotes}.}
\begin{equation}
\begin{aligned}\label{eq:N4_action_antisym_scalars2}
S&=\int\de^{4}x \tr\Bigl(-\frac 14F^{\mu\nu}F_{\mu\nu}
-\frac 12\bigl(\D^{\mu}\varphi^{AB}\bigr)\bigl(\D_\mu \varphi_{AB}\bigr)
+\complexi\ol{\lambda}^A_{\dot{\alpha}}(\bar\sigma^\mu)^{\dot{\alpha}\beta}\D_\mu\lambda_{A\beta}
\\
&\phan{=\int\de^{4}x\tr\Bigl(}
-\gym\bigl(
\lambda^{\alpha}_A\bigl[\varphi^{AB},\lambda_{B\alpha}\bigr]
-\ol{\lambda}^A_{\dot{\alpha}}\bigl[\varphi_{AB},\ol{\lambda}^{B\dot{\alpha}}\bigr]
\bigr)
+\frac{\gym^2}{4}\bigl[\varphi^{AB},\varphi^{CD}\bigr]\bigl[\varphi_{AB},\varphi_{CD}\bigr]
\Bigr)\eqncom
\end{aligned}
\end{equation}
with one non-abelian gauge field $A_\mu$, four negative-chirality Weyl fermions $\lambda_A$ and their conjugates, six antisymmetric scalars $\varphi_{AB}$, and the dimensionless coupling constant $\gym$. Under hermitian conjugation the fields transform according to\footnote{The conjugate transformations can be obtained by requiring $(X^\dagger)^\dagger=X$ for consistency.}
\begin{equation}\label{eq:field_conjugations}
\begin{aligned}
(\lambda_{A\alpha})^\dagger&=\ol{\lambda}^A_{\dot\alpha}\eqncom&
(\ol{\lambda}^{ A\dot\alpha})^\dagger&=\lambda^\alpha_A\eqncom&
(\varphi^{AB})^\dagger &=\varphi_{AB}\eqncom&
(\D^\mu)^\dagger&=\D^\mu\eqncom&
(F^{\mu\nu})^\dagger=F^{\mu\nu}
\end{aligned}
\end{equation}
and the canonical index contraction of \spl{2} and $\splbar{2}$ spinor indices is
\begin{equation}
\psi \chi\equiv\psi^\alpha\chi_\alpha=\varepsilon^{\alpha\beta}\psi_\beta\chi_\alpha\eqncom\qquad
\chi^\dagger\psi^\dagger\equiv\chi^\dagger_{\dot{\alpha}}\psi^{\dagger\dot{\alpha}}=\varepsilon^{\dot\alpha\dot\beta}\chi^{\dagger}_{\dot\alpha}\psi^{\dagger}_{\dot\beta}\eqndot
\end{equation}
The $(\sigma^\mu)_{\alpha\dot{\alpha}}$ and $(\bar\sigma^\mu)^{\dot{\alpha}\alpha}$ matrices are their own hermitian conjugates. They are defined according to \cite{Srednicki:2007}, c.f.\ \subsecref{sec:4D_Clifford_algebra} for details concerning the raising and lowering of \spl{2} spinor indices. The coupling tensors $\Sigma^i$ and $\bar\Sigma^i$ are explicitly constructed in \subsecref{sec:6D_Clifford_algebra}. These tensors are hermitian conjugates of each other and fulfil the following relations\footnote{Bear in mind that capital latin indices are \su{4} indices and not four-dimensional Minkowski space indices.}
\begin{equation}
\begin{aligned}\label{eq:identities_Sigma}
\epsilon_{ABCD}\bar\Sigma^{jAB}&=2\Sigma^j_{AB}\eqncom&
\epsilon^{ABCD}\Sigma^j_{AB}&=2\bar{\Sigma}^{jAB}\eqncom&\\
\Sigma^j_{AB}\Sigma_{jCD}&=-2\epsilon_{ABCD}\eqncom&
\bar{\Sigma}^{jAB}\bar{\Sigma}_j^{CD}&=-2\epsilon^{ABCD}\eqncom&\\
\Sigma^j_{AB}\bar\Sigma^{CD}_{j}&=-2(\delta_A^C\delta_B^D-\delta_A^D\delta_B^C)\eqncom&
2\delta_{ij}\delta^A_C&=\Sigma_i^{AB}\bar{\Sigma}_{jBC}+\Sigma_j^{AB}\bar{\Sigma}_{iBC}&
\end{aligned}
\end{equation}
with the normalisation $\epsilon_{1234}=\epsilon^{1234}=1$ of the Euclidean Levi-Civita tensor.

In addition, to make contact with the spinorial representation, we define a vector field in terms of $\spl{2}\times\splbar{2}$ spinor indices to be
\begin{equation}\label{eq:vectorfield_spinorial}
\begin{aligned}
X_{\alpha\dot{\beta}}&=-\complexi(\sigma_\mu)_{\alpha\dot{\beta}}X^\mu\eqncom\qquad&
X^{\dot\alpha\beta}&=\varepsilon^{\dot{\alpha}\dot{\gamma}}\varepsilon^{\beta\omega}X_{\omega\dot{\gamma}}=-\complexi(\bar\sigma_\mu)^{\dot\alpha\beta}X^\mu\eqndot&
\end{aligned}
\end{equation}
Since $(X^\mu)^\dagger=X^\mu$, in our spinorial representation vector fields become anti-hermitian as $(X_{\alpha\dot{\beta}})^\dagger=-X_{\beta\dot{\alpha}}$. This applies in particular to the covariant derivative $\D^\mu$ and $\D_{\alpha\dot{\alpha}}$. We also introduce the (anti-) selfdual field strength as\footnote{The projectors $\sigma_{\mu\nu}$ and $\bar\sigma_{\mu\nu}$ are explicitly given in \eqref{eq:sigma_mu_nu} and the (anti-) self-dual projectors in spacetime indices are given in \eqref{eq:selfdual_projectors}.}
\begin{equation}
\begin{aligned}\label{eq:selfdual_fieldstrength}
\cF_{\alpha\beta}&=\frac{1}{2}F^{\mu\nu}\varepsilon_{\beta\gamma}(\sigma_{\mu\nu})_\alpha^{\phan{\alpha}\gamma}
=\frac{\complexi}{4}\bigl(
\D_{\alpha\dot\alpha}A^{\dot{\alpha}}_{\phan{\alpha}\beta}+\D_{\beta\dot\alpha}A^{\dot{\alpha}}_{\phan{\alpha}\alpha}\bigr)\eqncom&\\
\bar\cF_{\dot\alpha\dot\beta}&=\frac{1}{2}F^{\mu\nu}\varepsilon_{\dot\beta\dot\gamma}(\bar\sigma_{\mu\nu})^{\dot\gamma}_{\phan{\alpha}\dot\alpha}
=\frac{\complexi}{4}\bigl(\D^{\phan{\alpha}\alpha}_{\dot\alpha}A_{\alpha\dot\beta}+\D^{\phan{\alpha}\alpha}_{\dot\beta}A_{\alpha\dot\alpha}\bigr)\eqncom&
\end{aligned}
\end{equation}
which are connected through hermitian conjugation as $(\cF_{\alpha\beta})^\dagger=\bar{\cF}_{\dot\beta\dot\alpha}$. Upon rewriting the action \eqref{eq:N4_action_antisym_scalars2} in terms of spinor indices we obtain the hermitian spinorial representation
\begin{equation}
\begin{aligned}\label{eq:N4_action_osci}
S&=\int\de^{4}x \tr\Bigl(\frac 12(\cF_{\alpha\beta}\cF^{\alpha\beta}+\bar\cF_{\dot\alpha\dot\beta}\bar\cF^{\dot\alpha\dot\beta})
-\frac 14\bigl(\D^{\dot\alpha \alpha}\varphi^{AB}\bigr)\bigl(\D_{\alpha \dot\alpha} \varphi_{AB}\bigr)
-\ol{\lambda}^A_{\dot{\alpha}}\D^{\dot{\alpha}\beta}\lambda_{A\beta}
\\
&\phan{=\int\de^{4}x\tr\Bigl(}
+\gym\bigl(\ol{\lambda}^A_{\dot{\alpha}}\bigl[\varphi_{AB},\ol{\lambda}^{B\dot{\alpha}}\bigr]
-\lambda^{\alpha}_A\bigl[\varphi^{AB},\lambda_{B\alpha}\bigr]\bigr)
+\frac{\gym^2}{4}\bigl[\varphi^{AB},\varphi^{CD}\bigr]\bigl[\varphi_{AB},\varphi_{CD}\bigr]
\Bigr)\eqncom
\end{aligned}
\end{equation}
where we used the identities for contracting $\sigma$-matrices, c.f.\ \appref{sec:Conventions}.

\section{Symmetries of \texorpdfstring{\NfSYMt}{N=4 SYM theory}}\label{sec:symmetries}
\NfSYMt is the maximally supersymmetric gauge theory in four dimensions. In the dimensional reduction from the ten-dimensional $\mathcal{N}=1$ SYM theory, six spacetime directions are transformed to internal (flavour or $R$-symmetry) \dof which enlarges the simple to a fourfold supersymmetry. In addition, the four-dimensional theory exhibits a conformal symmetry, that combines with the supersymmetry to render the theory superconformally invariant.\footnote{This combination of an internal symmetry with the \Poincare symmetry is possible for theories with fermionic supercharges. This was shown in \cite{Haag1975257} as a generalisation of the Coleman-Mandula no-go theorem of \cite{Coleman:1967ad}, which restricts the possibility to combine an internal with the \Poincare symmetry to the trivial combination.} For an introduction see e.g.\ \cite{thesis:Genovese, CFT} for conformal symmetry and e.g.\ \cite{Sohnius:1985qm,FigueroaO'Farrill:2001tr,Dolan:2002zh} for a discussion in the context of \NfSYMt. In contrast to the ten-dimensional theory, the quantisation of \NfSYMt does not introduce anomalies \cite{Caswell:1980yi,SOHNIUS1981245,Mandelstam:1982cb,Brink:1982wv,HOWE1984125,Seiberg:1988ur} and the superconformal invariance is unbroken for observables of the quantum theory. Apart from the manifest\footnote{This does not mean manifest symmetries in the mathematical sense. Mathematically, only the Lorentz and $R$-symmetry are manifest symmetries of the action.} symmetries, \NfSYMt is expected to exhibit further hidden symmetries which we will not explicitly discuss in this section. These hidden symmetries include the $\text{SL}(2,\ZZ)$ duality which relates the strong to the weak coupling regime \cite{MONTONEN1977117,Girardello1995127} and the exact quantum integrability \cite{Minahan:2002ve,Beisert:2003yb,Beisert:2005fw} of the theory.

Following \cite{Beisert:2010kp}, the generators of the symmetry algebra of \NfSYMt can be neatly packaged into a $\mathfrak{psu}(2,2|4)$ supermatrix of the form
\begin{equation}
\begin{pmatrix}
\multicolumn{2}{c}{\multirow{ 2}{*}{\su{2,2}}}&\mathfrak{Q}\\
	&	&\ol{\mathfrak{S}}\\
\ol{\mathfrak{Q}}&\mathfrak{S}& \mathfrak{R}
\end{pmatrix}
\eqncom
\end{equation}
with the conformal symmetry algebra\footnote{In four dimensional Minkowski space, the conformal group is locally isomorphic to \SU{2,2}, see \cite{mack1977}.} $\su{2,2}\simeq\so{4,2}$, the internal symmetry algebra $\mathfrak{R}$, and the fermionic SUSY and special conformal SUSY generators $\mathfrak{Q}$ and $\mathfrak{S}$. We will first discuss the action of the symmetry generators of $\mathfrak{so}(4,2)$, $\mathfrak{R}$, $\mathfrak{Q}$, and $\mathfrak{S}$ on the elementary fields of the theory, i.e.\ fermions, scalars, and field strengths. Afterwards, we give the resulting commutation relations between all generators. We express a symmetry transformation in terms of some infinitesimal parameter $\alpha_m$ contracted with a generator $g^m$. Macroscopic symmetry transformations, i.e.\ elements of the symmetry group, can then be reached via the exponential map and we write them in terms of the unitary operators $U_{\alpha\cdot g}=\e^{-\complexi \alpha \cdot g}$, see e.g.\ \cite[chapter 2]{Georgi:514148}, \cite[chapter 1]{goodman2009symmetry} for further details.
The change of an elementary field $f\rightarrow \hat{f}$ under the symmetry transformation yields
\begin{equation}\label{eq:general_symmetry_trafo}
\hat{f} = U^{-1}_{\alpha\cdot g}f\,U_{\alpha\cdot g}
=\sum_{n=0}^{\infty}\frac{\complexi^n}{n!}[\alpha\cdot g,f]_{(n)}\eqncom\qquad
\hat{f}= f+\complexi \alpha_m[g^m,f]+\order{\alpha^2}
\end{equation}
where we abbreviate an $n$-fold commutator as $[a,b]_{(n)}=[a,[a,\dots,[a,b]\dots]]$ and we denote the contraction of open indices as in the linearised case with a central dot.

\subsection{Conformal symmetry algebra \texorpdfstring{$\so{4,2}$}{so(4,2)}}\label{sec:conf_symmetry_N4}
Let us first discuss the conformal symmetry, which combines the Poincar\'e symmetry with the dilatation or scaling symmetry and the special conformal symmetry. We keep its discussion to a minimum here and refer the reader to \appref{sec:conformal_algebra} and the references therein for a more detailed presentation. 
On coordinates in flat space, the conformal generators act as
\begin{equation}\label{eq:conformal_generators_2}
\begin{aligned}
&\text{translations:}& \mathcal{P}_\mu&=-\complexi\partial_\mu\eqncom&\\
&\text{Lorentz transformations:}& \mathcal{M}_{\mu\nu}&=\complexi (x_\mu\partial_\nu-x_\nu\partial_\mu)\eqncom&\\
&\text{Dilatations:}& \mathcal{D}&=\complexi x_\mu\partial^\mu\eqncom&\\
&\text{special conformal transformations:} &\mathcal{K}_\mu&=\frac{\complexi}{2}\Bigl(x_\mu x_\nu\partial^\nu-\frac{x^2}{2}\partial_\mu\Bigr)\eqndot&
\end{aligned}
\end{equation}
Together with their mutual commutation relations \eqref{eq:Poincare_algebra} and \eqref{eq:conformal_algebra} they form the conformal algebra. If our theory also contains a gauge connection, the generator of translations realises the parallel transport of a field and hence it acts as a covariant derivative $\mathcal{P}_\mu=-\complexi \D_\mu$, instead of the partial derivative above, c.f.\ \cite[chapter 2.4]{VanProeyen:1999ni} for a detailed discussion. For a field $f_A$, transforming in some representation of the Lorentz group labelled by $A$, we require\footnote{For scalar fields this requirement is plausible. Let us take for example the scalar temperature field $T(x)$, created by some source at the origin of a coordinate system. If we shift the coordinate system by some parameter $a$, then $T(x+a)$ will look different. However, if we shift the source by the same parameter $a$, then the shifted field with shifted origin will look the same as the original one.} that a conformal transformation of the coordinates $x\rightarrow \hat{x}= C\cdot x$ is compensated by a conformal transformation of the field $f_A\rightarrow \hat{f}_A$  itself up to an overall conformal transformation acting on the representation label $A$. Furthermore, let us split the conformal coordinate transformation in $d\in \NN$ dimensions into an angle $\tilde C_\mu^{\phan{\mu}\nu}$ and a scale contribution $\e^{-\alpha}=(\det(C))^{1/d}$, so that we have $C_\mu^{\phan{\mu}\nu}=\e^{-\alpha}  \tilde C_\mu^{\phan{\mu}\nu}$. In this representation, coordinates and fields transform according to
\begin{equation}
\begin{aligned}\label{eq:symmetry_trafo_elementary_field}
\widehat{x^\mu}=\e^{-\alpha} \tilde C^{\mu}_{\phan{\mu}\nu}x^\nu\eqncom\qquad
\hat{f}_A(\hat{x})=\e^{\alpha\Delta_{f_A}}L_A^{\phan{A}B}f_B(x)\eqncom
\end{aligned}
\end{equation}
where $L=L(\tilde C)$ realises the angle transformation\footnote{Effectively this pure angle transformation is a Lorentz transformation in the appropriate representation of the Lorentz group, compare \eqref{eq:unitary_trafo_fields_2} with the coordinate transformations $\mathcal{P}$, $\mathcal{M}$, $\mathcal{K}$, $\mathcal{D}$, and the scale transformation $\Delta$ set to zero.} of the Lorentz-representation index $A$ and $\Delta_{f_A}$ is the scaling dimension of $f_A$, which classically equals its mass dimension. Note that the scale is related to the Jacobian that arises in a coordinate change from $x^\prime$ to $x$ as $\e^{-\alpha d}=\left|\partial \hat{x}/\partial x\right|$. Therefore, the action of a set of fields $f_i$ that interact in a Lorentz-invariant Lagrangian of the form $\mathcal{L}(\{f_i(x),\D_\mu f_i(x)\})$ transforms under a conformal transformation as
\begin{equation}
\begin{aligned}
\hat{S}
=\int\de^d x\, \e^{-\alpha d}\mathcal{L}\Bigl(
\bigl\{\e^{\alpha \Delta_{f_i}}(Lf_i(x)),\,\e^{\alpha\Delta_{f_i}-1}(\tilde{C}^{-1}\D) (Lf_i(x))\bigr\}\Bigr)\eqncom
\end{aligned}
\end{equation}
where we have suppressed the spacetime and representation indices. This implies that a Lorentz-invariant action in four dimensions without dimensionful parameters and classical scaling dimensions of fields $\Delta^0_\varphi=\Delta^0_{\D}=\Delta^0_A=1$ and $\Delta^0_\lambda=\Delta^0_{\ol{\lambda}}=\frac 32$ is also conformally invariant\footnote{Strictly speaking, this implies only the scaling invariance of the theory. Only if the virial field is expressible as a total derivative of some local quantity without the use of the \eom, also conformal invariance is guaranteed \cite{Coleman1971552,Fortin:2011sz}. For a textbook treatment see \cite[\chap{4}]{CFT}.} at the classical level\footnote{In perturbative calculations the classical scaling dimensions receive quantum corrections -- the anomalous dimensions. Their occurrence will be discussed in detail in \chapref{chap:Renormalisation}.}. In particular the classical invariance of the action \eqref{eq:N4_action_antisym_scalars2} can easily be verified for each term. As we will make use of the symmetry algebra of \NfSYMt, let us also introduce the action of the conformal generators $\mathfrak{g}=\{\mathfrak{P}^\mu,\mathfrak{M}^{\mu\nu},\mathfrak{K}^\mu,\mathfrak{D}\}$ on fields. For the transformation \eqref{eq:general_symmetry_trafo}, we need the commutators
\begin{equation}
\begin{aligned}\label{eq:unitary_trafo_fields_2}
\bigl[\mathfrak{M}^{\mu\nu},f_a(x)\bigr]
&=(S^{\mu\nu})_a^{\phan{a}b}f_b(x)-M^{\mu\nu}f_a(x)\eqncom\\
\bigl[\mathfrak{P}^{\mu},f_a(x)\bigr]
&=P^{\mu}f_a(x)-\gym [A^\mu,f_a(x)]=-\complexi \D^\mu f_a(x)\eqncom\\
\bigl[\mathfrak{K}^{\mu},f_a(x)\bigr]
&=-\frac 12x_\nu (S^{\mu\nu})_a^{\phan{a}b}f_b(x)+\frac{\complexi}{2} x^\mu \Delta_{f_a} f_a(x)+K^{\mu}f_a(x)\eqncom\\
\bigl[\mathfrak{D},f_a(x)\bigr]
&=-\complexi\Delta_{f_a} f_a(x)- df_a(x)\eqncom
\end{aligned}
\end{equation}
where the conformal generators on coordinates are given in \eqref{eq:conformal_generators_2}. The Lorentz transformation $S^{\mu\nu}$ acts on Weyl-spinors $\lambda$ and $\ol{\lambda}$, vectors $A$, and scalars $\phi$ as\footnote{The transformations for fermions imply that switching spinor index position gives an additional sign as $\lambda^\alpha(S^{\mu\nu})_\alpha^{\phan{\alpha}\beta}=-(\sigma^{\mu\nu})_\alpha^{\phan{\alpha}\beta}\lambda^\alpha$ and analogously for $\ol{\lambda}$.}
\begin{equation}
\begin{aligned}
(S^{\mu\nu})_\alpha^{\phan{\alpha}\beta}\lambda_\beta
&=(\sigma^{\mu\nu})_\alpha^{\phan{\alpha}\beta}\lambda_\beta
\eqncom&
S^{\mu\nu}A^\rho&=-\complexi (\eta^{\mu\rho}A^\nu-\eta^{\nu\rho}A^\mu)\eqncom&\\
(S^{\mu\nu})^{\dot\alpha}_{\phan{\alpha}\dot\beta}\ol{\lambda}^{\dot{\beta}}
&=(\bar\sigma^{\mu\nu})^{\dot\alpha}_{\phan{\alpha}\dot\beta}\ol{\lambda}^{\dot{\beta}}
\eqncom&
S^{\mu\nu}\phi&=0\eqncom&
\end{aligned}
\end{equation}
with $\sigma^{\mu\nu}=\frac{\complexi}{4}\sigma^{[\mu}\bar{\sigma}^{\nu]}$ and $\bar\sigma^{\mu\nu}=\frac{\complexi}{4}\bar\sigma^{[\mu}\sigma^{\nu]}$, see \appref{sec:Conventions}. Further details concerning the Weyl representation can be found in \appref{sec:spinor_in_various_dimensions}. 

\subsection{Internal symmetry algebra \texorpdfstring{$\mathfrak{R}$}{R}}\label{sec:R_symmetry}
As a second closed subalgebra, we have the $R$-symmetry of \NfSYMt with generators $\mathfrak{R}$. This symmetry arises in the compactification procedure from the ten to the four dimensional theory as\footnote{When the Clifford algebra $\clifford(p,q)$ is restricted to its invertible elements, the spin algebra $\spin{p,q}$ is obtained. A detailed derivation of the algebra decomposition can be found in \cite{OFarrill}.}
\begin{equation}
\spin{9,1}\rightarrow\spin{3,1}\times \spin{0,6}\simeq\spin{3,1}\times\su{4}\eqncom
\end{equation}
where we used the isomorphism $\spin{6}\simeq\su{4}$. Therefore the $R$-symmetry algebra of \NfSYMt is $\su{4}$ and it acts on the fields in the appropriate representations: trivial (gauge fields), fundamental (fermions), real antisymmetric (scalars), and anti-fundamental (conjugate fermions). We choose to label the $R$-symmetry generators by two indices $A,B\in\{1,2,3,4\}$, so that the generators can be represented in a canonical matrix representation
\begin{equation}\label{eq:R_symmetry_generators}
\begin{aligned}
\bigl(\mathfrak{R}_B^{\phan{A}A}\bigr)^\dagger&=\mathfrak{R}_A^{\phan{A}B}=\hat{e}_A^{\phan{A}B}\eqncom\qquad&
&\forall A\neq B\eqncom&\\
\mathfrak{R}_j^{\phan{A}j}&=\frac 14(4\hat{e}_j^{\phan{A}j}-\one)\eqncom\qquad&
& \forall j=\{1,2,3,4\}\eqncom&
\end{aligned}
\end{equation}
where $\hat{e}_A^{\phan{A}B}$ is a matrix with one in the $A^{\text{th}}$ row and $B^{\text{th}}$ column and zeros else. Note that only three of our four diagonal generators are linearly independent, which is necessary since we have the trace constraint for \su{4} generators, reducing the number of generators to fifteen. The three independent diagonal generators form the Cartan subalgebra of \su{4} and their eigenvalues in two different bases\footnote{The matrix representation of the Cartan generators in these two bases can be read off from the eigenvalues of the fundamental fermions, e.g.\ $\mathfrak{R}_1|_q=\frac 12 \diag (+1,-1,-1,+1)$ or $\mathfrak{R}_1|_Q=\diag (1,-1,0,0)$. The four diagonal generators of \eqref{eq:R_symmetry_generators} can be written as linear combinations of the Cartan generators in the $q$ basis as $\mathfrak{R}_4^{\phan{A}4}=\frac 12(\mathfrak{R}_1+\mathfrak{R}_2+\mathfrak{R}_3)|_{q}$ and $\mathfrak{R}_j^{\phan{A}j}=(\mathfrak{R}_j-\mathfrak{R}_4^{\phan{A}4})|_{q}$ with $j\in\{1,2,3\}$.} $(q^1,q^2,q^3)$ and $(Q^1,Q^2,r)$ will be given in \tabref{tab: su(4) charges}. Fields $f_C$ and their duals $f^C$ with fundamental $R$-symmetry index $C$ transform under the generators as $[\mathfrak{R},f]=\mathfrak{R}\cdot f$, where the \rhs follows from \eqref{eq:R_symmetry_generators} and explicitly reads
\begin{equation}
\begin{aligned}\label{eq:R_symmetry_on_fields}
\mathfrak{R}_A^{\phan{A}B}f_C&=\bigl(\delta_C^B\delta_A^D-\frac 14\delta_A^B\delta_C^D\bigr)f_D\eqncom\quad&
\mathfrak{R}_A^{\phan{A}B}f^C&=-\bigl(\delta_A^C\delta_D^B-\frac 14\delta_A^B\delta_D^C\bigr)f^D\eqndot&
\end{aligned}
\end{equation}
This concludes the discussion of the bosonic symmetry subalgebras of \NfSYMt.

\subsection{SUSY transformations involving \texorpdfstring{$\mathfrak{Q}$}{Q}}\label{sec:SUSY_trafos}
Next, we investigate the SUSY transformations generated by the fermionic symmetry generators $\mathfrak{Q}$. They follow from the ten-dimensional transformations \eqref{eq:susy_trafo_10D}. Analogously to \eqref{eq:10D_MW_spinors_2}, we write the SUSY generators and parameters in terms of $\su{4}\otimes \spl{2,\CC}$ quantities. In four dimensions the unitary operator \eqref{eq:unitary_operator_SUSY_10} which realises the $\mathcal{N}=4$ SUSY transformations becomes
$U_{(\epsilon\cdot \mathfrak{Q})}=\e^{-\complexi\delta_{\epsilon\cdot\mathfrak{Q}}}$. Upon writing the four-dimensional SUSY generators and their corresponding free parameters in terms of \eqref{eq:10D_MW_spinors_2}, the exponent is given by
\begin{equation}
\delta_{\epsilon\cdot\mathfrak{Q}}=\frac 1 2\bigl(
\mathfrak{Q}^{A\alpha}\epsilon_{A\alpha}+\ol{\mathfrak{Q}}_{A\dot{\alpha}}\ol{\epsilon}^{A\dot{\alpha}}
+\ol{\epsilon}^A_{\dot{\alpha}}\ol{\mathfrak{Q}}_A^{\dot\alpha}+\epsilon_A^{\alpha}\mathfrak{Q}^A_{\alpha}
\bigr)
= \bigl(\epsilon_A^{\alpha}\mathfrak{Q}^A_{\alpha}+\ol{\mathfrak{Q}}_{A\dot{\alpha}}\ol{\epsilon}^{A\dot{\alpha}}\bigr)\eqndot
\end{equation}
The ten-dimensional SUSY variations from \eqref{eq:susy_trafo_10D} can be written as $\delta_{\epsilon,\ol{\epsilon}} X=\complexi[\delta_{\epsilon\cdot\mathfrak{Q}},X]$ and for the gauge fields and real scalars in four dimensions we find
\begin{equation}
\begin{aligned}\label{eq:SUSY_trafo_scalar}
\delta_{\epsilon,\ol{\epsilon}} A_\mu& 
= 
\frac{1}{2}\Bigl(-\complexi\ol{\xi^{\tM -}}\,(\onee{8}\otimes\gamma_\mu)\psi^{\tM -}_{10}+\,\text{h.c.}\Bigr)
=\complexi(-\ol{\epsilon}^A_{\dot{\alpha}}(\bar{\sigma}_\mu)^{\dot{\alpha}\alpha}\lambda_{A\alpha}
+\ol{\lambda}_{\dot{\alpha}}^A(\bar{\sigma}_\mu)^{\dot{\alpha}\alpha}\epsilon_{A\alpha})\eqncom\\
\delta_{\epsilon,\ol{\epsilon}}\varphi_{j}&=\frac{1}{2}
\Bigl(-\complexi \ol{\xi^{\tM -}}\,(\rho_j\otimes\complexi\gamma^5)\psi^{\tM -}_{10}+\,\text{h.c.}\Bigr)
=\Bigl(\ol{\epsilon}^A_{\dot{\alpha}}\Sigma_{jAB}\ol{\lambda}^{B\dot{\alpha}}-\epsilon^\alpha_A\bar\Sigma_j^{AB}\lambda_{B\alpha}\Bigr)\eqncom
\end{aligned}
\end{equation}
where we used \eqref{eq:susy_trafo_10D}, \eqref{eq:4_6_gamma_matrices}, and \eqref{eq:Gamma_10}. In spinorial indices, these variations become
\begin{equation}
\begin{aligned}\label{eq:scalar_variations_spinorial}
\delta_{\epsilon,\ol{\epsilon}} A_{\alpha\dot{\alpha}}& =-\complexi(\sigma^\mu)_{\alpha\dot{\alpha}}
\delta_{\epsilon,\ol{\epsilon}} A_\mu
=2\Bigl(\epsilon^\beta_{A}\varepsilon_{\alpha\beta}\ol{\lambda}_{\dot{\alpha}}^A
-\varepsilon_{\dot{\alpha}\dot{\beta}}\lambda_{A\alpha}\ol{\epsilon}^{A\dot{\beta}}\Bigr)\eqncom\\
\delta_{\epsilon,\ol{\epsilon}} \varphi_{AB}&=
\frac{\complexi}{2}\Sigma^j_{AB}\delta_{\epsilon,\ol{\epsilon}}\varphi_{j}
=\complexi\Bigl(\epsilon_A^\alpha\lambda_{B\alpha}-\epsilon_B^\alpha\lambda_{A\alpha}
-\epsilon_{ABCD}\ol{\lambda}^{D}_{\dot{\alpha}}\ol{\epsilon}^{C\dot{\alpha}}\Bigr)\eqncom\\
\delta_{\epsilon,\ol{\epsilon}} \varphi^{AB}&=(\delta_{\epsilon,\ol{\epsilon}} \varphi_{AB})^\dagger=
\complexi\Bigl(-\ol{\lambda}^B_{\dot\alpha}\ol{\epsilon}^{A\dot\alpha}+\ol{\lambda}^A_{\dot\alpha}\ol{\epsilon}^{B\dot\alpha}
+\epsilon^{ABCD}\epsilon^\alpha_{C}\lambda_{D\alpha}\Bigr)\eqndot
\end{aligned}
\end{equation}
For the transformation of fermions, we split the ten-dimensional transformation \eqref{eq:susy_trafo_10D} into three contributions
\begin{equation}
\begin{aligned}\label{eq:fermion_trafo}
\delta_{\epsilon,\ol{\epsilon}}\psi^{\tM -}_{10}=\complexi[\delta_{\epsilon\cdot \mathfrak{Q}},\psi^{\tM -}_{10}]&= -\frac{\complexi}{2}(F_{10}^{\mu\nu}M_{\mu\nu}+ F_{10}^{jk}M_{jk}+2 F_{10}^{\mu j}M_{\mu j}) \xi^{\tM -}\\
&=-\frac{\complexi}{2}\bigl( F^{\mu\nu}M_{10\mu\nu}-\complexi\gym[\varphi^j,\varphi^k]M_{10jk}+2 (D^\mu\varphi^j)M_{10\mu j}\bigr)\xi^{\tM -}
\eqncom
\end{aligned}
\end{equation}
where $\mu,\nu$ are four-dimensional Minkowski-space indices and $j,k$ are six-dimensional Euclidean-space indices. Using \eqref{eq:4_6_gamma_matrices} and \eqref{eq:Lorentz_generator_D_dim}, the ten-dimensional Lorentz generators decompose into
\begin{equation}
\begin{aligned}
M_{10\mu\nu}=\onee{8}\otimes \frac{\complexi}{4} [\gamma_\mu,\gamma_\nu]\eqncom\qquad
M_{10jk}=\frac{\complexi}{4} [\rho_j,\rho_k]\otimes\onee{4}\eqncom\qquad
M_{10\mu j}=\frac 14 \rho_j\otimes [\gamma^5,\gamma_\mu]\eqndot
\end{aligned}
\end{equation}
This allows us to write \eqref{eq:fermion_trafo} in terms of the two contributions\footnote{For the multiplication, note that the negative-chirality parameter $\xi^{\tM -}$ has the same structure as the negative-chirality fermion given in \eqref{eq:10D_MW_spinors}.}
\begin{equation}
\begin{aligned}\label{eq:SUSY_trafo_psi}
\delta_{\epsilon,\ol{\epsilon}}\lambda_{A\alpha} 
&=
-\frac{\complexi}{2}F^{\mu\nu}(\sigma_{\mu\nu})_{\alpha}^{\phan{\alpha}\beta}\epsilon_{A\beta}
+\complexi\gym [\varphi_{AB},\varphi^{BC}]\epsilon_{C\alpha}
+(\sigma_\mu)_{\alpha\dot\beta}\D^\mu\varphi_{AB}\ol{\epsilon}^{B\dot\beta}\\
&=\complexi\Bigl(
\epsilon_{A}^\beta\cF_{\beta\alpha}
-\epsilon^\gamma_{C}\,\gym\varepsilon_{\alpha\gamma} [\varphi^{CB},\varphi_{BA}]
-\D_{\alpha\dot{\beta}}\varphi_{BA}\ol{\epsilon}^{B\dot\beta}\Bigr)\eqncom\\
\delta_{\epsilon,\ol{\epsilon}}\ol{\lambda}^{A\dot\alpha} 
&=
-\frac{\complexi}{2}F^{\mu\nu}(\bar\sigma_{\mu\nu})^{\dot\alpha}_{\phan{\alpha}\dot\beta}\ol{\epsilon}^{A\dot\beta}
+\complexi\gym [\varphi^{AB},\varphi_{BC}]\ol{\epsilon}^{C\dot\alpha}
- (\bar\sigma_\mu)^{\dot\alpha\beta}\D^\mu\varphi^{AB}\epsilon_{B\beta}\\
&=
\complexi\Bigl(-\bar{\cF}^{\dot\alpha}_{\phan{\alpha}\dot\beta}\ol{\epsilon}^{A\dot\beta}
+\gym [\varphi^{AB},\varphi_{BC}]\ol{\epsilon}^{C\dot\alpha}
+\epsilon_{B}^{\gamma}\varepsilon_{\beta\gamma}\D^{\dot{\alpha}\beta}\varphi^{BA}\Bigr)\eqncom
\end{aligned}
\end{equation}
where we used \eqref{eq:selfdual_fieldstrength} for the field-strength representation. Note that the variation of the conjugate fermion can also be obtained as $\delta_{\epsilon,\ol{\epsilon}}\ol{\lambda}^{A\dot\gamma} =\varepsilon^{\dot{\gamma}\dot{\alpha}}(\delta_{\epsilon,\ol{\epsilon}}\lambda_{A\alpha} )^\dagger$, which is required for consistency. Analogously the transformations of the positive-chirality fields $\lambda_A^\alpha$ and $\ol{\lambda}^A_{\dot{\alpha}}$ are obtained from \eqref{eq:SUSY_trafo_psi} by raising or lowering the respective spinor indices with $\varepsilon$-tensors. To complete the picture we also include the variation of the (anti-)selfdual field strength. Using the variations of $\cF$ and the spinorial gauge field respectively given in \eqref{eq:selfdual_field_variation} and \eqref{eq:scalar_variations_spinorial}, we find
\begin{equation}
\begin{aligned}\label{eq:fieldstrength_variation}
\delta_{\epsilon,\ol{\epsilon}}\cF_{\alpha\beta}&=
\varepsilon_{\beta\gamma}\delta_{\epsilon,\ol{\epsilon}}\cF_{\alpha}^{\phan{\alpha}\gamma}
=-\frac{\complexi}{4}
\bigl(\delta_{\alpha}^{\rho}\delta_{\beta}^\omega
+\delta^{\rho}_\beta\delta_{\alpha}^\omega\bigr)\varepsilon^{\dot{\alpha}\dot{\beta}}
\D_{\rho\dot{\alpha}}\delta_{\epsilon,\ol{\epsilon}} A_{\omega\dot{\beta}}\\
&=\complexi\Bigl(\frac{1}{2}\bigl(
\D_{\alpha\dot{\alpha}}\lambda_{A\beta}+\D_{\beta\dot{\alpha}}\lambda_{A\alpha}\bigr)\ol{\epsilon}^{A\dot{\alpha}}
-\frac{1}{2}\epsilon^{\omega}_A
\bigl(
\varepsilon_{\alpha\omega}\D_{\beta\dot\alpha}
+\varepsilon_{\beta\omega}\D_{\alpha\dot\alpha}
\bigr)\ol{\lambda}^{A\dot{\alpha}}\Bigr)\eqncom\\
\delta_{\epsilon,\ol{\epsilon}}\bar{\cF}_{\dot{\beta}\dot\alpha}&=
\bigl(\delta_{\epsilon,\ol{\epsilon}}\cF_{\alpha\beta}\bigr)^\dagger
=\frac{\complexi}{4}
\bigl(\delta_{\dot\alpha}^{\dot\rho}\delta_{\dot\beta}^{\dot\omega}
+\delta^{\dot\rho}_{\dot\beta}\delta_{\dot\alpha}^{\dot\omega}\bigr)\varepsilon^{\alpha\beta}
\D_{\alpha\dot{\rho}}\delta_{\epsilon,\ol{\epsilon}} A_{\beta\dot{\omega}}\\
&=\complexi\Bigl(\frac{1}{2}\epsilon^\alpha_{A}\bigl(
\D_{\dot{\alpha}\alpha}\ol{\lambda}_{\dot{\beta}}^A
+\D_{\dot{\beta}\alpha}\ol{\lambda}_{\dot{\alpha}}^A
\bigr)
-\frac{1}{2}\bigl(
\varepsilon_{\dot\alpha\dot\omega}\D_{\alpha\dot\beta}
+\varepsilon_{\dot\beta\dot\omega}\D_{\alpha\dot\alpha}
\bigr)
\lambda^\alpha_{A}\ol{\epsilon}^{A\dot{\omega}}\Bigr)\eqndot
\end{aligned}
\end{equation}
The variations \eqref{eq:scalar_variations_spinorial}, \eqref{eq:SUSY_trafo_psi}, and \eqref{eq:fieldstrength_variation} are consistent with the field variations in the oscillator representation discussed in \subsecref{subsec:the-oscillator-representation} and \appref{sec:comparison-of-field-and-oscillator-representation}.

\subsection{Special SUSY transformations involving \texorpdfstring{$\mathfrak{S}$}{S}}
Finally, the only generators missing are those of the special conformal SUSY transformations $\mathfrak{S}^{A\alpha}$ and $\ol{\mathfrak{S}}_A^{\dot\alpha}$. We define them as a commutator of a special conformal generator with a SUSY generator as
\begin{equation}\label{eq:special_conformal_SUSY_generators}
\mathfrak{S}_A^{\alpha}=-\frac 12 (\bar\sigma_\mu)^{\dot\alpha\alpha}[\mathfrak{K}^\mu,\ol{\mathfrak{Q}}_{A\dot{\alpha}}]
\eqncom\qquad
\ol{\mathfrak{S}}^{A\dot{\alpha}}=
-\frac 12 (\bar\sigma_\mu)^{\dot\alpha\alpha}
[\mathfrak{K}^\mu,\mathfrak{Q}^A_{\alpha}]
\eqndot
\end{equation}
The commutation relations with fields are obtained by replacing the special conformal generators by the respective \rhs of \eqref{eq:special_conformal_SUSY_generators}. The resulting double (anti-)commutator can be solved by the means of the graded Jacobi identity
\begin{equation}\label{eq:graded_Jacobi_identity}
[g_A,[g_B,g_C\}\}+
\omega^{2\mathfrak{D}_0(g_A)\mathfrak{D}_0(g_B+g_C)}
[g_B,[g_C,g_A\}\}+
\omega^{2\mathfrak{D}_0(g_C)\mathfrak{D}_0(g_A+g_B)}
[g_C,[g_A,g_B\}\}=0\eqncom
\end{equation}
where the graded commutator is 
\begin{equation}
[X,Y\}=XY-\omega^{2\mathfrak{D}_0(X)\mathfrak{D}_0(Y)}YX\eqndot
\end{equation}
The prefactors account for minus signs that arise from commuting two fermionic objects. The generator $\mathfrak{D}_0$ gives the scaling dimension\footnote{The scaling dimensions of the generators are given in the following paragraph.} to $0^{\text{th}}$ order in perturbation theory, which for bosonic and fermionic objects is integer- and half-integer-valued, respectively. This combines with the quantity $\omega$, which must be evaluated after its complete exponent has been determined\footnote{We choose this realisation of fermionic signs for compatibility with the treatment in \secref{sec:the-thermal-one-loop-partition-functions-of-the-deformed-theories}.} and has the formal property $\sqrt\omega=-1$. The commutator of $\mathfrak{S}_A^{\alpha}$ with a given field $X$ then becomes
\begin{equation}\label{eq:action_of_special_conf_generators}
[\mathfrak{S}_A^{\alpha},X\}
= -\frac{1}{2}
(\bar\sigma_\mu)^{\dot\alpha\alpha}\Bigl(
[\mathfrak{K}^{\mu},[\ol{\mathfrak{Q}}_{A\dot{\alpha}},X\}]
- 
[\ol{\mathfrak{Q}}_{A\dot{\alpha}},[\mathfrak{K}^{\mu},X]\}
\Bigr)\eqndot
\end{equation}
The factors of $\omega$ vanish when we take into account the scaling dimensions of generators which will be defined after \eqref{eq:action_dilatation_operator}. The analogous relation is obtained for the conjugate generator by replacing $\mathfrak{S}_A^{\alpha}\rightarrow\ol{\mathfrak{S}}^{A\dot{\alpha}}$ and $\ol{\mathfrak{Q}}_{A\dot{\alpha}}\rightarrow \mathfrak{Q}^A_{\alpha}$.

\subsection{Commutation relations of the symmetry algebra of \texorpdfstring{$\mathfrak{psu}(2,2|4)$}{psu(2,2|4)}}\label{sec:SUSY_algebra_N4}
Finally, we can use the Jacobi identity \eqref{eq:graded_Jacobi_identity} and the action of all symmetry generators on elementary fields \eqref{eq:unitary_trafo_fields_2}, \eqref{eq:R_symmetry_on_fields}, \eqref{eq:SUSY_trafo_scalar} and \eqref{eq:SUSY_trafo_psi} to determine the mutual commutation relations of all symmetry generators. For the conformal and $R$-symmetry subalgebras, we find the following commutation relations
\begin{equation}
\begin{aligned}\label{eq:symmetry_algebra_1}
[\mathfrak{R}_A^{\phan{A}B},\mathfrak{R}_C^{\phan{A}D}]&=
\delta_C^B\mathfrak{R}_A^{\phan{A}D}-\delta_A^D\mathfrak{R}_C^{\phan{A}B}\eqncom&\\
[\mathfrak{M}_{\nu\rho},\mathfrak{V}_\mu]&=-\complexi(\eta_{\mu\nu}\mathfrak{V}_\rho-\eta_{\mu\rho}\mathfrak{V}_\nu)\eqncom&\\
[\mathfrak{K}_\mu,\mathfrak{P}_\nu]&=\frac{\complexi}{2}(\eta_{\mu\nu}\mathfrak{D}+ \mathfrak{M}_{\mu\nu})\eqncom&\\
[\mathfrak{M}_{\mu\nu},\mathfrak{M}_{\rho\sigma}]&=
\complexi(\eta_{\mu\sigma}\mathfrak{M}_{\nu\rho}
+\eta_{\nu\rho}\mathfrak{M}_{\mu\sigma}
-\eta_{\mu\rho}\mathfrak{M}_{\nu\sigma}
-\eta_{\nu\sigma}\mathfrak{M}_{\mu\rho}
)\eqncom&
\end{aligned}
\end{equation}
where $\mathfrak{V}_\mu$ is a generator transforming in the vector representation, e.g.\ $\mathfrak{P}_\mu$ or $\mathfrak{K}_\mu$. The commutation relations with the dilatation generator are
\begin{equation}\label{eq:action_dilatation_operator}
[\mathfrak{D},\mathfrak{J}]=-\complexi\Delta_\mathfrak{J}\mathfrak{J}\eqncom
\end{equation}
with the non-vanishing scaling dimensions $\Delta_{\mathfrak{R}}=\Delta_{\mathfrak{P}}=-\Delta_{\mathfrak{K}}=1$ and $\Delta_{\mathfrak{Q}}=\Delta_{\mathfrak{\ol{Q}}}=-\Delta_{\mathfrak{S}}=-\Delta_{\mathfrak{\ol{S}}}=\frac 12$. On the supercharges the $R$-symmetry and Lorentz generators act as rotations of the form
\begin{equation}
\begin{aligned}
\bigl[\mathfrak{M}^{\mu\nu},\mathfrak{Q}^A_{\alpha}\bigr]
&= (\sigma^{\mu\nu})_\alpha^{\phan{\alpha}\beta}\mathfrak{Q}^A_{\beta}\eqncom&
\bigl[\mathfrak{M}^{\mu\nu},\ol{\mathfrak{Q}}_{A\dot{\alpha}}\bigr]
&=-\ol{\mathfrak{Q}}_{A\dot{\beta}} (\bar\sigma^{\mu\nu})^{\dot{\beta}}_{\phan{\alpha}\dot\alpha}\eqncom&\\
\bigl[\mathfrak{M}^{\mu\nu},\mathfrak{S}_A^{\alpha}\bigr]
&=-\mathfrak{S}_A^{\beta}(\sigma^{\mu\nu})_\beta^{\phan{\alpha}\alpha}\eqncom&
\bigl[\mathfrak{M}^{\mu\nu},\ol{\mathfrak{S}}^{A\dot{\alpha}}\bigr]
&=(\bar\sigma^{\mu\nu})^{\dot{\alpha}}_{\phan{\alpha}\dot\beta}\ol{\mathfrak{S}}^{A\dot{\beta}}\eqncom&\\
\bigl[\mathfrak{R}_A^{\phan{A}B},\mathfrak{Q}^C_{\alpha}\bigr]
&= -\delta^C_A\mathfrak{Q}^B_{\beta}+\frac 14\delta_A^B\mathfrak{Q}^C_{\beta}\eqncom&
\bigl[\mathfrak{R}_A^{\phan{A}B},\ol{\mathfrak{Q}}_{C\dot{\alpha}}\bigr]
&=\delta^B_C\ol{\mathfrak{Q}}_{A\dot{\beta}}-\frac 14\delta^B_A \ol{\mathfrak{Q}}_{C\dot{\beta}}\eqncom&\\
\bigl[\mathfrak{R}_A^{\phan{A}B},\mathfrak{S}_C^{\alpha}\bigr]
&=\delta^B_C\mathfrak{S}_A^{\beta}-\frac 14\delta_A^B\mathfrak{S}_C^{\beta}\eqncom&
\bigl[\mathfrak{R}_A^{\phan{A}B},\ol{\mathfrak{S}}^{C\dot{\alpha}}\bigr]
&=  -\delta^C_A\ol{\mathfrak{S}}^{B\dot{\beta}}+\frac 14 \delta_A^B\ol{\mathfrak{S}}^{C\dot{\beta}}\eqndot&
\end{aligned}	
\end{equation}
The special conformal and ordinary translations yield the following commutation relations with the supercharges
\begin{equation}
\begin{aligned}\label{eq:symmetry_algebra_middle}
[\mathfrak{P}^\mu,\mathfrak{S}_A^{\alpha}]&=\frac 12 
(\bar\sigma^\mu)^{\dot\alpha\alpha}\ol{\mathfrak{Q}}_{A\dot{\alpha}}\eqncom&
[\mathfrak{K}^\mu,\ol{\mathfrak{Q}}_{A\dot{\alpha}}]&=\frac 12 
(\sigma^\mu)_{\alpha\dot\alpha}\mathfrak{S}_A^{\alpha}\eqncom&\\
[\mathfrak{P}^\mu,\ol{\mathfrak{S}}^{A\dot\alpha}]&= \frac 12 
(\bar\sigma^\mu)^{\dot\alpha\alpha}\mathfrak{Q}^A_{\alpha}\eqncom&
[\mathfrak{K}^\mu,\mathfrak{Q}^A_{\alpha}]&=\frac 12 
(\sigma^\mu)_{\alpha\dot{\alpha}}\ol{\mathfrak{S}}^{A\dot{\alpha}}\eqncom&\\
\bigl\{\mathfrak{Q}^A_{\alpha},\ol{\mathfrak{Q}}_{B\dot{\alpha}}\bigr\}
&=\delta^A_B(\sigma^\mu)_{\alpha\dot\alpha}\mathfrak{P}_{\mu}\eqncom&
\bigl\{\mathfrak{S}_A^{\alpha},\ol{\mathfrak{S}}^{B\dot\alpha}\bigr\}
&=-\delta^B_A (\bar\sigma^\mu)^{\dot\alpha\alpha}\mathfrak{K}_\mu\eqncom& 
\end{aligned}	
\end{equation}
and finally the non-vanishing anti-commutation relations between the supercharges are\footnote{To obtain this relation we enforce the graded Jacobi identity for e.g.\ $[Q,[S,Q\}\}+\dots=0$ to fix the coefficients on the \rhs}
\begin{equation}
\begin{aligned}\label{eq:symmetry_algebra_last}
\bigl\{\mathfrak{Q}^A_{\alpha},\mathfrak{S}_B^{\beta}\bigr\}
&= 
-\frac{1}{2}\mathfrak{M}^{\mu\nu}(\sigma_{\mu\nu})_\alpha^{\phan{\alpha}\beta}\delta^A_B-\delta^\beta_\alpha \mathfrak{R}^{\phan{A}A}_{B}+\frac{\complexi}{2} \delta_B^A\delta_\alpha^\beta \mathfrak{D}
\eqncom&\\
\bigl\{\ol{\mathfrak{S}}^{B\dot\beta},\ol{\mathfrak{Q}}_{A\dot\alpha}\bigr\}
&= 
\frac{-1}{2}\mathfrak{M}^{\mu\nu}(\bar{\sigma}_{\mu\nu})^{\dot\beta}_{\phan{\alpha}\dot\alpha}\delta_A^B-\delta^{\dot\beta}_{\dot\alpha} \mathfrak{R}^{\phan{A}B}_{A}+\frac{\complexi}{2} \delta_A^B\delta^{\dot\beta}_{\dot\alpha} \mathfrak{D}
\eqndot&
\end{aligned}	
\end{equation}
All remaining (anti-)commutators of the symmetry algebra vanish. Note, however, that the action of symmetry generators on elementary fields will be supplemented with coupling-dependent corrections. These corrections originate from the dilatation generator which, in the interacting theory, is only well defined on renormalised fields and becomes coupling-dependent, \cf \secref{sec:the-quantum-dilatation-operator-on-composite-operators}. Since the commutation relations of the algebra \eqref{eq:symmetry_algebra_1} -- \eqref{eq:symmetry_algebra_last} must remain valid, the action of the special conformal and special conformal SUSY generators on fields must becomes coupling-dependent.

\subsection{The spinor or oscillator representation}\label{subsec:the-oscillator-representation}
With the explicit field representation of symmetry generators presented so far in this section, we can also present the spinor or oscillator representation \cite{Gunaydin82,Gunaydin:1998sw,Beisert:2003jj} which is compatible with our conventions. We first give the mapping of the generators to the spinor representation and then we follow the presentation of \cite{Beisert:2004ry} to write all elementary fields and spinor-representation symmetry generators in terms of creation and annihilation operators of the symmetry algebra $\mathfrak{psu}(2,2|4)$. For an explicit calculatory verification of the mapping we refer to \appref{sec:comparison-of-field-and-oscillator-representation}.

All generators given so far in this section can be written explicitly in terms of \spl{2} and \splbar{2} spinor indices. While the SUSY, special conformal SUSY and $R$-symmetry generators are already given in this representation, the conformal symmetry generators take the spinor representation form
\begin{equation}
\begin{aligned}\label{eq:generators_in_osci_rep}
\mathfrak{L}_{\alpha}^{\phan{\alpha}\beta}&=\frac 12 \mathfrak{M}^{\mu\nu}(\sigma_{\mu\nu})_{\alpha}^{\phan{\alpha}\beta}\eqncom&
\ol{\mathfrak{L}}^{\dot\alpha}_{\phan{\alpha}\dot\beta}&=-\frac 12 \mathfrak{M}^{\mu\nu}(\bar\sigma_{\mu\nu})^{\dot\alpha}_{\phan{\alpha}\dot\beta}\eqncom&
\mathfrak{D}_{\text{osci}}&=\complexi\mathfrak{D}\eqncom&\\
\mathfrak{P}_{\alpha\dot{\alpha}}&=(\sigma_\mu)_{\alpha\dot{\alpha}}\mathfrak{P}^\mu\eqncom&
\mathfrak{K}^{\dot\alpha\alpha}&=-(\bar\sigma_\mu)^{\dot\alpha\alpha}\mathfrak{K}^\mu\eqndot&
\end{aligned}
\end{equation}

In the oscillator representation, the elementary fields of \NfSYMt and its deformations can be represented via two sets of bosonic oscillators $\aosc^{\dagger}_\alpha$ ($\alpha=1,2$) and $\bosc^{\dagger}_{\dot{\alpha}}$ ($\alphadot=1,2$) and one set of fermionic oscillators $\cosc^{\dagger}_A$ ($A=1,2,3,4$) acting on the oscillator vacuum $\vac$. These oscillators obey the usual (anti-)commutation relations:
\begin{equation}\label{eq:osci_commutation_relations}
[\aosc^{\alpha},\aosc^{\dagger}_\beta]=\delta^\alpha_\beta \eqncom \qquad
[\bosc^{\dot\alpha},\bosc^{\dagger}_{\dot\beta}]=\delta^{\dot\alpha}_{\dot\beta} \eqncom \qquad
\{\cosc^A,\cosc^{\dagger}_B\}=\delta^A_B\eqncom  
\end{equation}
with all other (anti-)commutators vanishing. Under hermitian conjugation, indicated by $f^{\dagger_{\text{hc}}}$, the oscillators transform as
\begin{equation}
\begin{aligned}\label{eq:oscillator_hermitian_conjugation}
(\aosc^\alpha)^{\dagger_{\text{hc}}}&=\bosc^{\dot{\alpha}}\eqncom\qquad& (\aosc^\dagger_\alpha)^{\dagger_{\text{hc}}}&=\bosc_{\dot{\alpha}}^\dagger\eqncom\qquad&
(\cosc^A)^{\dagger_{\text{hc}}}&=\cosc_A^\dagger\eqncom& 
\end{aligned}
\end{equation}
where the remaining relations are obtained from $(f^{\dagger_{\text{hc}}})^{\dagger_{\text{hc}}}=f$. The unusual transformations for the $\aosc$ and $\bosc$ oscillators under hermitian conjugation arise since these oscillators really characterise \spl{2} and \splbar{2} representations, which are connected by hermitian conjugation. In terms of the oscillators with suppressed spinor indices, the (anti-) self-dual field strength, (anti-) Weyl fermions and antisymmetric scalars are
\begin{equation}\label{eq:fields_osci}
\begin{aligned}
\D^k \cF_{\phantom{ABC}} &\mathrel{\widehat{=}} 
(\aoscdag)^{k+2} 
(\boscdag)^{k\phantom{+0}}
\vac \eqncom \\
\D^k \lambda_{A\phantom{BC}} &\mathrel{\widehat{=}}     
(\aoscdag)^{k+1} 
(\boscdag)^{k\phantom{+0}}
\cosc^{\dagger}_A 
\vac \eqncom \\
\D^k \varphi_{AB\phantom{C}} &\mathrel{\widehat{=}}     
(\aoscdag)^{k\phantom{+0}} 
(\boscdag)^{k\phantom{+0}} 
\cosc^{\dagger}_A \cosc^{\dagger}_B 
\vac \eqncom \\
\D^k \ol{\lambda}_{ABC} &\mathrel{\widehat{=}} 
(\aoscdag)^{k\phantom{+0}} 
(\boscdag)^{k+1} 
\cosc^{\dagger}_A \cosc^{\dagger}_B \cosc^{\dagger}_C
\vac \eqncom \\
\D^k \bar{\cF}_{\phantom{ABC}} &\mathrel{\widehat{=}}   
(\aoscdag)^{k\phantom{+0}}
(\boscdag)^{k+2} 
\cosc^{\dagger}_1 \cosc^{\dagger}_2 \cosc^{\dagger}_3 \cosc^{\dagger}_4 
\vac \eqncom
\end{aligned}
\end{equation}
with $\varphi^{AB\phantom{C}} =\frac{1}{2!}\varepsilon^{ABCD}\varphi_{CD}$ and $\ol{\lambda}^{A}=\frac{1}{3!}\varepsilon^{ABCD}\ol{\lambda}_{BCD}$ and the Levi-Civita tensor\footnote{For calculations note the useful relations $\cosc^{\dagger}_1 \cosc^{\dagger}_2 \cosc^{\dagger}_3 \cosc^{\dagger}_4=\frac{1}{4!}\varepsilon^{ABCD}\cosc^{\dagger}_A \cosc^{\dagger}_B \cosc^{\dagger}_C \cosc^{\dagger}_D$ and $\ol{\lambda}_{BCD}=\varepsilon_{ABCD}\ol{\lambda}^A$.} in Euclidean space is taken to be normalised as $\varepsilon^{1234}=\varepsilon_{1234}=1$. Note that all $\spl{2}$ and $\splbar{2}$ spinor indices that may occur on the \lhs of \eqref{eq:fields_osci} are totally symmetrised, see \subsecref{sec:Building_blocs} for details.

The symmetry generators can also be represented in terms of creation and annihilation operators. The dilatation, $R$-symmetry, diagonal Lorentz and central charge generators have the form
\begin{equation}
\begin{aligned}\label{eq:symmetry_generators_osci_LR}
\mathfrak{D}_0&\mathrel{\widehat{=}}1+\frac 12 \aosc^{\dagger}_{\gamma} \aosc^\gamma+\frac 12 \bosc^{\dagger}_{\dot\gamma} \bosc^{\dot\gamma}&
\mathfrak{R}^{A}_{B}&\mathrel{\widehat{=}}\cosc_B^{\dagger}\cosc^{A}-\frac 14\delta_B^A \cosc_D^{\dagger}\cosc^{D}\eqncom&\\
\mathfrak{L}^{\alpha}_{\beta}&\mathrel{\widehat{=}}\aosc^{\dagger}_{\beta}\aosc^{\alpha}-\frac 12\delta_\beta^\alpha \aosc^{\dagger}_{\gamma}\aosc^{\gamma}\eqncom&
\ol{\mathfrak{L}}^{\dot\alpha}_{\dot\beta}&\mathrel{\widehat{=}}\bosc^{\dagger}_{\dot\beta}\bosc^{\dot\alpha}-\frac 12\delta^{\dot\beta}_{\dot\alpha} \bosc^{\dagger}_{\dot\gamma}\bosc^{\dot\gamma}\eqncom&\\
\mathfrak{C}&\mathrel{\widehat{=}}1-\frac 12 \aosc^{\dagger}_{\gamma} \aosc^\gamma+\frac 12 \bosc^{\dagger}_{\dot\gamma} \bosc^{\dot\gamma}
-\frac 12 \cosc^\dagger_A \cosc^A&&&
\end{aligned}
\end{equation}
and the remaining off-diagonal translation, special conformal translation, SUSY, and special conformal SUSY generators are
\begin{equation}
\begin{aligned}\label{eq:symmetry_generators_osci_QSP}
\mathfrak{P}_{\alpha\dot{\alpha}}&\mathrel{\widehat{=}}  \aosc^{\dagger}_{\alpha}\bosc^{\dagger}_{\dot\alpha}\eqncom
&\mathfrak{Q}^{A}_\alpha&\mathrel{\widehat{=}}  \cosc^{A}\aosc^{\dagger}_{\alpha}\eqncom
&\mathfrak{S}^\alpha_{A}&\mathrel{\widehat{=}}  \cosc^{\dagger}_A\aosc^{\alpha}\eqncom
\\
\mathfrak{K}^{\dot{\alpha}\alpha}&\mathrel{\widehat{=}} \aosc^{\alpha}\bosc^{\dot\alpha}\eqncom
&\ol{\mathfrak{Q}}_{A\dot{\alpha}}&\mathrel{\widehat{=}}  \cosc^{\dagger}_A\bosc^{\dagger}_{\dot\alpha}\eqncom
&\ol{\mathfrak{S}}^{A\dot{\alpha}}&\mathrel{\widehat{=}}  \cosc^{A}\bosc^{\dot\alpha}\eqndot
\end{aligned}
\end{equation}
The action of symmetry generators on fields in the oscillator representation is determined by writing them left to the fields and commuting the annihilation operators to the right, using \eqref{eq:osci_commutation_relations} and for an explicit verification of the present oscillator representation of symmetry generators see \appref{sec:comparison-of-field-and-oscillator-representation}.
The symmetry generators in the oscillator representation have the same transformation properties\footnote{In the oscillator representation, we use \eqref{eq:oscillator_hermitian_conjugation} for the hermitian conjugation and for the generators in the field representation we use \eqref{eq:sigma_hermitian_conjugation}, the hermiticity of $\mathfrak{P}^\mu$, $\mathfrak{K}^\mu$, $\mathfrak{M}^{\mu\nu}$, and the anti-hermiticity of $\mathfrak{D}$. Note also that $\ol{\mathfrak{M}}^{\dot{\alpha}}_{\phan{\alpha}\dot{\beta}}$ transforms an upper index, while $\ol{\mathcal{L}}^{\dot{\alpha}}_{\dot{\beta}}$ transforms a lower one, which explains the additional sign.}
under hermitian conjugation as the generators given in \eqref{eq:generators_in_osci_rep}.

\section{Deformations of \texorpdfstring{\NfSYMt}{N=4 SYM theory}}\label{sec:The_deformations}
In this section, we introduce classical actions and symmetries of the two less symmetric relatives of \NfSYMt mentioned in \chapref{chap:Introduction}: the $\beta$- and the $\gamma_i$-deformation. We will investigate the properties of these theories in detail in \chapref{chap:applications}. The presentation in this section is based on the analysis in my publications \cite{Fokken:2013aea,Fokken:2013mza,Fokken:2014soa}. 

We define a classical deformation of some parent gauge theory to have the same field content as the parent theory. In addition, some or all of the parent theory's symmetries may be broken in the deformations by altering the existing interactions of the parent theory in a given deformation procedure. Finally, we include all possible renormalisable interactions\footnote{If some of these newly introduced couplings turn out to be running, exact conformal invariance is broken for the corresponding deformation. Therefore, the exact conformal invariance can only be checked for the quantised theories.} that are compatible with the remaining symmetries of the deformed theory. This last step guarantees that quantum corrections do not add further couplings in the renormalisation of the deformed theories which we will discuss in \chapref{chap:Renormalisation} and \ref{chap:applications}. In practical examples any of the additional couplings that do not receive quantum corrections may also be set to zero.

In the remainder of this section, we will introduce the deformation procedure that we will impose on \NfSYMt. Then we will present the single-trace part of the deformed actions and the new renormalisable multi-trace trace interactions that are compatible with the surviving symmetries. Finally, we discuss how the symmetry generators of \secref{sec:symmetries} must be modified for the deformed theories.

\subsection{The deformations}\label{sec:deformation}
On the string theory side, the deformed theories are obtained by altering the $\text{S}^5$ part in the $\AdS{5}\times \text{S}^5$ background, as mentioned above. On the gauge theory side, this procedure can be mimicked by altering the internal space\footnote{In \subsecref{subsec:Dimensional_reduction_to_four_dimensions}, this is the six-dimensional space that is compactified to a torus with vanishing radius.} of the parent \NfSYMt to a non-commutative space with a Moyal-like $\ast$-product which also breaks the $\su{4}_R$-symmetry algebra to its $\mathfrak{u}(1)^{\times 3}$ Cartan subalgebra, see \cite{Szabo:2001kg} for a review on the connections to noncommutative field theory.
The $\ast$-product for two fields $A$ and $B$ is realised as
\begin{equation}\label{eq:starproduct}
A\cstar B= A B \e^{\frac{\complexi}{2}\mathbf{q}_{A}\wedge\mathbf{q}_{B}} \eqncom
\end{equation}
where $\mathbf{q}_{X}=(q_X^1,q_X^2,q_X^3)$ is the Cartan charge vector of the field $X$. For fields with definite Cartan charges, i.e.\ complex scalars $\phi_i$ or antisymmetric scalars $\varphi_{ij}$, Weyl fermions $\lambda_{A\alpha}$, and gauge fields $A_\mu$, the charges are given in table \ref{tab: su(4) charges}. The antisymmetric product of the charge vectors in the $\gamma_i$-deformation is given by 
\begin{equation}\label{eq: antisymmetric product}
\mathbf{q}_A\wedge \mathbf{q}_B=(\mathbf{q}_A)^{T}\mathbf{C}\,\mathbf{q}_B
\eqncom \qquad
\mathbf{C}=\begin{pmatrix}
0 & -\gamma_3 & \gamma_2 \\
\gamma_3 & 0 & -\gamma_1 \\
-\gamma_2 & \gamma_1 & 0 
\end{pmatrix}
\eqndot
\end{equation}
While this definition is sufficient, we choose to work with the following linear combinations of the deformation parameters: 
\begin{equation}\label{eq:gamma_pm}
\gamma^\pm_i=\pm\frac 12 (\gamma_{i+1}\pm\gamma_{i+2})\eqncom
\end{equation}
where cyclic identification $i+3\sim i$ is understood. For the $\beta$-deformation, the antisymmetric product is obtained by setting $\gamma_i^+=\gamma_i=\beta$ and $\gamma_i^-=0$. This renders the simplification
\begin{equation}\label{eq: antisymmetric product_beta}
\mathbf{q}_{A}\wedge \mathbf{q}_{B}\Bigr|_{\beta\text{-def.}}= -\beta \sum_{a,b,c=1}^3 \varepsilon_{abc} q_A^a q_B^b 
=-\beta\sum_{a,b=1}^3 \varepsilon_{ab3}Q^a_AQ^b_B\eqncom
\end{equation}
where the last equality holds for the second basis choice given in \tabref{tab: su(4) charges}. From this second representation, which was originally used in \cite{Lunin:2005jy}, we also see that the $\beta$-deformation only depends on the two charges $Q^1=q^1-q^2$ and $Q^2=q^2-q^3$ and is insensitive to the third $r=\frac{2}{3}(q^1+q^2+q^3)$. 
\begin{table}[ht] 
	\centering
	\caption{Cartan charges of the fields in two different bases, which are related as $Q^1=q^1-q^2$, $Q^2=q^2-q^3$, and $r=\frac{2}{3}(q^1+q^2+q^3)$. The charges of antisymmetric scalars can be obtained from the complex scalars using the mapping that will be defined in \eqref{eq:anti_sym_to_complex_scalar}. The respective anti-fields carry the opposite charges.
	}\label{tab: su(4) charges}
	$\begin{array}{|c|c|ccc|cccc| 
	}
	\hline
	B&A_{\mu}&\phi_1&\phi_2&\phi_3 & \lambda_{1\alpha}&\lambda_{2\alpha}&\lambda_{3\alpha}&\lambda_{4\alpha} 
	\\
	& & & & & & & & 
	\\[-0.4cm]
	\hline
	& & & & & & & & 
	\\[-0.4cm]
	q^1_B & 0 & 1 & 0 & 0 &  +\frac12 & -\frac12 & -\frac12 & +\frac12 
	\\
	& & & & & & & & 
	\\[-0.4cm]
	q^2_B &  0 & 0 & 1 & 0 &-\frac12 & +\frac12 & -\frac12 & +\frac12 
	\\
	& & & & & & & & 
	\\[-0.4cm]
	q^3_B &  0 & 0 & 0 & 1 &-\frac12 & -\frac12 & +\frac12 & +\frac12 
	\\
	& & & & & & & & 
	\\[-0.4cm] \hline
	& & & & & & & & 
	\\[-0.4cm]
	Q^1_B &  0 & 1 & -1 & 0 &1 & -1 & 0 & 0 
	\\
	& & & & & & & & 
	\\[-0.4cm]
	Q^2_B &  0 & 0 & 1 & -1 & 0 & 1 & -1 & 0 
	\\
	& & & & & & & & 
	\\[-0.4cm]
	r_B &  0 & \frac{2}{3} & \frac{2}{3} & \frac{2}{3} &-\frac{1}{3} & -\frac{1}{3} & -\frac{1}{3} & 1 
	\\
	\hline
	\end{array}$
\end{table}

While we use a Moyal-like $\ast$-product, we must mention one important difference to the Moyal-product used in non-commutative field theory \cite{Filk96}. For the latter, the deformation phase as well as the ordering principle in interactions depends on one quantity: the spacetime coordinate. In our case, however, the deformation phase of the Moyal-like $\ast$-product 
depends on the Cartan charges of $\su{4}_R$ (the flavours of elementary fields), while it alters ordinary products within colour traces. Hence, the phase depends on the flavour of fields, while their ordering is dictated by the colour trace, in contrast to the non-commutative field theory case. An immediate consequence of the separation of ordering and deformation phase is that $\ast$-deformed traces of fields are in general not cyclically invariant any more, e.g.
\begin{equation}
\begin{aligned}\label{eq:star_product_trace_LN}
\tr\bigl(f_1\ast f_2\ast\dots\ast f_N\bigr)&=
\e^{\frac{\complexi}{2}\sum_{j\leq k=1}^N\mathbf{q}_{f_j}\wedge \mathbf{q}_{f_k}}\tr\bigl(f_1 f_2\dots f_N\bigr)\eqncom\\
\tr\bigl(f_N\ast f_1\ast f_2\dots\ast f_{N-1}\bigr)&=
\e^{\frac{\complexi}{2}\bigl(\sum_{k=1}^{N-1}\mathbf{q}_{f_N}\wedge\mathbf{q}_{f_k}+\sum_{j\leq k=1}^{N-1}\mathbf{q}_{f_j}\wedge \mathbf{q}_{f_k}\bigr)}\tr\bigl(f_1 f_2\dots f_N\bigr)\eqncom
\end{aligned}
\end{equation}
where the two phase factors are not equal for generic fields $f_i$. However, if the sum of Cartan charges in a trace vanishes, we can express the charge vector of the $N^{\text{th}}$ field as $\mathbf{q}_{f_N}=-\sum_{j=1}^{N-1}\mathbf{q}_{f_j}$. Together with the antisymmetry of the $\wedge$-product, we then find that both traces in \eqref{eq:star_product_trace_LN} are equal which restores the cyclic invariance of the trace in this case.

\subsection{The deformed single-trace action}\label{sec:the-deformed-single-trace-action}
The single-trace parts of both gauge theory deformations inherit the properties of the parent \NfSYMt \cite{Mauri:2005pa,Ananth:2006ac,AKS07}. Note, however, that the arguments in these publications do not hold for potential multi-trace parts of the actions, since such terms are absent in the parent theory and we will determine the additional multi-trace interactions separately in the following subsection. To construct the single-trace part from the parent theory, we have to write all elementary fields in representations with definite $\mathfrak{u}(1)^{\times 3}$ charge vectors. For comparability with the $\mathcal{N}=1$ superspace formulation \cite{Gates:1983nr}, we choose to work with complex scalars, which are obtained from the antisymmetric ones in \eqref{eq:N4_action_antisym_scalars2} as\footnote{%
	In terms of real scalars this mapping corresponds to the choice
	\begin{equation}
	\begin{aligned}
	\phi_1=-\frac{1}{\sqrt{2}}(\varphi_3+\complexi \varphi_4)\eqncom\qquad
	\phi_2=-\frac{1}{\sqrt{2}}(\varphi_2-\complexi \varphi_1)\eqncom\qquad
	\phi_3=\frac{1}{\sqrt{2}}(\varphi_5-\complexi \varphi_6)\eqndot
	\end{aligned}
	\end{equation}
} %
\begin{equation}
\begin{aligned}\label{eq:anti_sym_to_complex_scalar}
\varphi_{ij}&=\frac{\complexi}{\sqrt{2}}\varepsilon_{ijk4}\ol{\phi}^k\eqncom&
\varphi_{4i}&=\frac{\complexi}{\sqrt{2}}\phi_i\eqncom&
\varphi^{ij}&=-\frac{\complexi}{\sqrt{2}}\varepsilon^{ijk4}\phi_k\eqncom&
\varphi^{4i}&=-\frac{\complexi}{\sqrt{2}}\ol{\phi}^i\eqncom&
\end{aligned}
\end{equation}
where the complex conjugate scalar fields are $\ol{\phi}^i=(\phi_i)^\dagger$. As before, we have $\varepsilon_{1234}=1$ and under hermitian conjugation it turns into its dual according to $(\varepsilon_{ijk4})^\dagger=\varepsilon^{ijk4}$. We can now build the single-trace part of the deformed theories from the action \eqref{eq:N4_action_antisym_scalars2}. We employ \eqref{eq:anti_sym_to_complex_scalar} to obtain the Minkowski-space action with complex scalars and promote all ordinary products between fields to the $\ast$-products defined in \eqref{eq:starproduct}. This yields
\begin{equation}
\begin{aligned}\label{eq:deformed_action_complex_scalars2}
S_{\text{s.t.}}&=\int\de^{4}x \tr\Bigl(-\frac 14F^{\mu\nu}F_{\mu\nu}
-\bigl(\D^{\mu}\ol{\phi}^{j}\bigr)\bigl(\D_\mu \phi_{j}\bigr)
+\complexi\ol{\lambda}^A_{\dot{\alpha}}(\bar\sigma^\mu)^{\dot{\alpha}\beta}\D_\mu\lambda_{A\beta}
\\
&\phan{=\int\de^{4}x\tr\Bigl(}
-\frac{\gym}{\sqrt{2}}\Bigl(
\complexi\varepsilon^{ijk4}
\lambda^{\alpha}_i\bigl[\phi_{j},\lambda_{k\alpha}\bigr]_{\ast}
+2\complexi\lambda^{\alpha}_i\bigl[\ol{\phi}^{i},\lambda_{4\alpha}\bigr]_{\ast}
+\text{h.c.}
\Bigr)\\
&\phan{=\int\de^{4}x\tr\Bigl(}
+\gym^2\Bigl(
\bigl[\phi_{j},\phi_{k}\bigr]_{\ast}\bigl[\ol{\phi}^{j},\ol{\phi}^{k}\bigr]_{\ast}
-\frac 12\bigl[\phi_{j},\ol{\phi}^{j}\bigr]\bigl[\phi_{k},\ol{\phi}^{k}\bigr]
\Bigr)
\Bigr)\eqncom
\end{aligned}
\end{equation}
where $\ast$-deformed commutators are defined by replacing ordinary products within them by by $\ast$-deformed ones. To obtain this action, we also used that all terms in the single-trace action have a vanishing total $\su{4}$ Cartan charge and that the $\ast$-product of two fields with opposite charge vectors reduces to the ordinary product. In case of the $\beta$-deformation, also commutators involving the gluino $\lambda_{4\alpha}$ reduce to ordinary ones. Since the $\ast$-products within commutators in \eqref{eq:deformed_action_complex_scalars2} depend on the order of fields, the deformation explicitly breaks the commutator-type structure of interaction it appears in. While all interactions in \NfSYMt are of commutator-type, some of these are altered in the deformations. As a consequence, both deformations distinguish between the gauge groups \SUN and \UN, since \U1 modes do not decouple any more in non-commutator-type interactions, see \cite{Frolov:2005iq}.

Up to ghost and gauge-fixing terms the action \eqref{eq:deformed_action_complex_scalars2} can be cast into the single-trace part of \eqref{eq:action_Feynman_rules} with coupling tensors given in \eqref{eq:coupling_tensors_N4} and \eqref{eq:coupling_tensors_gammai}. This is done by expanding all $\ast$-commutators, using the (anti-)cyclicity of the trace and absorbing the deformation parameters into appropriate coupling tensors analogously to the following example
\begin{equation}
\begin{aligned}
\frac{\varepsilon^{ijk4}}{\complexi\sqrt{2}}\tr(\lambda^{\alpha}_i\bigl[\phi_{j},\lambda_{k\alpha}\bigr]_{\ast})&=
\frac{\varepsilon^{ijk4}}{\complexi\sqrt{2}}\Bigl(
\e^{\frac{\complexi}{2}\mathbf{q}_{\phi_j}\wedge\mathbf{q}_{\lambda_k}}\tr(\lambda^{\alpha}_i\phi_{j}\lambda_{k\alpha})
-\e^{\frac{\complexi}{2}\mathbf{q}_{\lambda_k}\wedge\mathbf{q}_{\phi_j}}\tr(\lambda^{\alpha}_i\lambda_{k\alpha}\phi_{j})
\Bigr)\\
&=
\frac{\varepsilon^{ijk4}}{\complexi\sqrt{2}}\Bigl(
\e^{-\frac{\complexi}{2}\mathbf{q}_{\lambda_i}\wedge\mathbf{q}_{\lambda_k}}\tr(\lambda^{\alpha}_i\phi_{j}\lambda_{k\alpha})
-\e^{-\frac{\complexi}{2}\mathbf{q}_{\lambda_k}\wedge\mathbf{q}_{\lambda_i}}\tr(\lambda^\alpha_{k}\phi_{j}\lambda_{i\alpha})
\Bigr)\\
&=
(\rho^j)^{ki}\tr(\lambda^{\alpha}_i\phi_{j}\lambda_{k\alpha})\eqncom
\end{aligned} 
\end{equation}
where we used $\mathbf{q}_{\lambda_k}+\mathbf{q}_{\lambda_j}+\mathbf{q}_{\phi_i}=0$ in the second and implicitly defined via the last line as $(\rho^j)^{ki}=-\complexi\sqrt{2}\varepsilon^{jki4}\e^{\frac{\complexi}{2}\mathbf{q}_{\lambda_k}\wedge\mathbf{q}_{\lambda_i}}$. The mapping between the current conventions and the one used in \cite{Fokken:2013aea} is given in \appref{app:Feynman_rules}.

\subsection{Multi-trace parts of the action}\label{sec:multi-trace-parts-of-the-action}
In this subsection, we discuss which multi-trace interactions can be added to the single-trace action \eqref{eq:deformed_action_complex_scalars2} of the deformed theories. In principle, all multi-trace couplings whose individual traces factors have non-vanishing $\mathbf{q}$-charge are not captured in the non-planar inheritance principle of \cite{Jin:2012np} and hence do not appear in the deformed action \eqref{eq:deformed_action_complex_scalars2}.\footnote{In orbifolded theories such interactions are constructed from the twisted sectors \cite{Dymarsky:2005uh}.} Such interactions can, however, appear in loop corrections in the quantised theory and, as discussed at the beginning of this section, we include all additional couplings that are compatible with the remaining symmetries of the deformed theories into the respective classical actions of the deformations. In case of the conformally invariant $\beta$-deformation, it is in fact mandatory to add a multi-trace structure, as the theory does not preserve conformal invariance otherwise, \cf \secref{sec:beta_paper}.

In principle, we are free to choose any type of renormalisable multi-trace structure for the deformed models. However, since we are interested in deformations that have the same field content as the parent theory, a well defined \tHooft limit, natural limits for identical or vanishing deformation parameters and at best also preserve conformal invariance, we restrict each multi-trace coupling to fulfil the following requirements:
\begin{enumerate}
	\item \label{U1preserve}
	the three global $\mathfrak{u}(1)$ charges are preserved (it has vanishing total $\mathfrak{u}(1)^3$ charge),
	\item \label{existplanar}
	it does not spoil the \tHooft limit (no proliferation of $N$-power beyond the planar order occurs in perturbative calculations),
	\item \label{betalimit} for gauge group \SUN, in the limit $\gamma_i^+=\beta$ and $\gamma_i^-=0$ the action of the $\mathcal{N}=1$ supersymmetric $\beta$-deformation is obtained,
	\item \label{N4limit} in the limit $\gamma_i^\pm=0$ the action of the undeformed \NfSYMt is obtained,
	\item \label{renormalisability}
	it is marginal (its classical scaling dimension is four).
\end{enumerate}
In addition, we restrict to theories with gauge group \UN or \SUN. Note that requirement \ref{renormalisability} is a necessary but not sufficient requirement for the deformed theory to be conformally invariant beyond the classical level, see \chapref{chap:Renormalisation} for details and \secref{sec:non-conformal_double_trace_coupling} for an explicit example of broken conformality in the $\gamma_i$-deformation. For calculational purposes, we also want to avoid a perturbative mixing of the expansion in deformation angles $\gamma_i^\pm$ on the one hand with the expansion in the \tHooft or effective planar coupling constant
\begin{equation}\label{eq:coupldef}
\lambda=\gym^2N\eqncom\qquad 
g=\frac{\sqrt{\lambda}}{4\pi}
\end{equation}
on the other hand. For the $\beta$- and $\gamma_i$-deformation we hence choose $\gamma_i^-=0$ with $\gamma_i^+$ not of order $\lambda$ and $\gamma_i^\pm$ not of order $\lambda$, respectively.

\subsubsection{Gauge group \texorpdfstring{\SUN}{SU(N)}}
For gauge group \SUN, the tracelessness of the colour generators together with requirement \ref{renormalisability} ensures that the only possible multi-trace structure is a product of two traces which each contain two scalar fields. The only double-trace action that fulfils the requirements \ref{U1preserve} -- \ref{renormalisability} hence is\footnote{The Yang-Mills coupling is factored out of the coupling tensors to simplify a perturbative expansion in one of the couplings in \eqref{eq:coupldef}.}
\begin{equation}\label{eq:dtc}
S_{\text{d.t.}}=\int \de^4x\Bigl(
-\frac{\gym^2}{2N}\big[Q^{ij}_{\text{F}\,kl}\tr(\phi_i\phi_j)\tr(\bar\phi^k\bar\phi^l)
+Q^{ij}_{\text{D}\,kl}\tr(\bar\phi^k\phi_i)\tr(\bar\phi^l\phi_j)\big]\Bigr)\eqncom
\end{equation}
where the coupling tensors $Q_{\text{F}}$ and $Q_{\text{D}}$ contain the dependence on $\gamma_i^\pm$ and are compatible with the Feynman rules derivation in \appref{app:Feynman_rules}. Since we want the action \eqref{eq:dtc} to be real the coupling tensors must fulfil
\begin{equation}\label{eq:conjQFD}
\begin{aligned}
(Q^{ij}_{\text{F}\,kl})^\ast&=Q^{lk}_{\text{F}\,ji}\eqncom\qquad
(Q^{ij}_{\text{D}\,kl})^\ast=Q^{lk}_{\text{D}\,ji}\eqndot
\end{aligned}
\end{equation}
Note that requirement \ref{existplanar} in general restricts the $N$-power that can occur in multi-trace couplings: compared to the single-trace coupling an $n$-trace coupling must be suppressed by a factor of at least $N^{1-n}$, see \subsecref{sec:finite-size-effects} for details.

In case of the $\beta$-deformation with gauge group \SUN, there is an alternative derivation that fixes the structure of the double-trace terms entirely. Starting from the $\mathcal{N}=4$ SYM action in its $\mathcal{N}=1$ superspace formulation, the superpotential can be deformed by altering all ordinary products of chiral superfields to superspace $\star$-deformed ones \cite{Fokken:2013aea}. This deformation must be carried out before auxiliary fields are integrated out. Only after the deformation is completed, the auxiliary fields may be integrated out. This procedure gives the $\beta$-deformation in elementary fields as a special case of the conformally invariant Leigh-Strassler deformations \cite{Leigh:1995ep}. The double-trace term obtained in this procedure has a vanishing $\beta$-function \cite{Mauri:2005pa} and takes the form
\begin{equation}\label{eq:2_trace_action}
S^\beta_{\text{d.t.}}=-\frac{\gym^2}{N}\int \de^4x
\tr\bigl(\bigl[\phi_{j},\phi_{k}\bigr]_{\ast}\bigr)\tr\bigl(\bigl[\ol{\phi}^{j},\ol{\phi}^{k}\bigr]_{\ast}\bigr)
=4\frac{\gym^2}{N}\sin^2\frac{\beta}{2}\int \de^4x\sum_{j\neq k}
\tr\bigl(\phi_{j}\phi_{k}\bigr)\tr\bigl(\ol{\phi}^{j}\ol{\phi}^{k}\bigr)
\eqncom
\end{equation}
where we used the explicit structure of the $\ast$-product in the second equality.

\subsubsection{Gauge group \texorpdfstring{\UN}{U(N)}}
For gauge group \UN, there are many more possible couplings, since we can now have a single colour generator of the \U{1} mode inside a single trace. In couplings we separate the \U{1} modes off explicitly: each \U{1} component is written as a trace over the respective \UN field, whereas traces of more than one field are understood to contain only the \SUN components. Note that we only make this distinction in the action in order to efficiently present all couplings that are unique to the \UN theory. In calculations in \chapref{chap:Renormalisation} and \ref{chap:applications} all gauge group components can occur in generic colour traces. All cubic terms that fulfil the requirements \ref{U1preserve} -- \ref{renormalisability} in this case are combined in 
\begin{equation}
\begin{aligned}\label{eq:3_trace_action}
S^3_{\U{1}}&=\int\de^4 x\Bigl(
\frac{\gym}{N}\big[
(\rho^i_{\lambda})^{BA}\tr(\lambda^{\alpha}_A)\tr(\phi_i\lambda_{B\alpha})
+(\rho^i_{\phi})^{BA}\tr(\phi_i)\tr(\lambda^{\alpha}_B\lambda_{A\alpha})
\\
&\hphantom{=\int\de^4 x\Bigl(\frac{\gym}{N}\big[}
+(\rho_{\bar\lambda\,i})_{BA}\tr(\bar\lambda^{A}_{\dot\alpha})\tr(\bar\phi^i\bar\lambda^{B\dot\alpha})
+(\rho_{\bar\phi\,i})_{BA}\tr(\bar\phi^i)\tr(\bar\lambda^{B}_{\dot\alpha}\bar\lambda^{A\dot\alpha})
\\
&\hphantom{=\int\de^4 x\Bigl(\frac{\gym}{N}\big[}
+(\tilde\rho_{\bar\lambda}^i)_{BA}\tr(\bar\lambda^{A}_{\dot\alpha})\tr(\phi_i\bar\lambda^{B\dot\alpha})
+(\tilde\rho_{\bar\phi}^i)_{BA}\tr(\phi_i)\tr(\bar\lambda^{B}_{\dot\alpha}\bar\lambda^{A\dot\alpha})
\\
&\hphantom{=\int\de^4 x\Bigl(\frac{\gym}{N}\big[}
+(\tilde\rho_{\lambda\,i})^{BA}\tr(\lambda_A^{\alpha})\tr(\bar\phi^i\lambda_{B\alpha})
+(\tilde\rho_{\bar\phi\,i})^{BA}\tr(\bar\phi^i)\tr(\lambda_B^{\alpha}\lambda_{A\alpha})
\big]
\\
&\hphantom{=\int\de^4 x\Bigl(}
+\frac{\gym}{N^2}\big[
(\rho_{3}^i)^{BA}\tr(\lambda_A^{\alpha})\tr(\phi_i)\tr(\lambda_{B\alpha})
+(\rho_{3\,i})_{BA}\tr(\bar\lambda^{A}_{\dot\alpha})\tr(\bar\phi^i)\tr(\bar\lambda_{B}^{\dot\alpha})\\
&\hphantom{=\int\de^4 x\Bigl({}+{}\frac{\gym}{N^2}\big[}
+(\tilde\rho_{3}^i)_{BA}\tr(\bar\lambda^{A}_{\dot\alpha})\tr(\phi_i)\tr(\bar\lambda^{B\dot\alpha})
+(\tilde\rho_{3\,i})^{BA}\tr(\lambda_A^{\alpha})\tr(\bar\phi^i)\tr(\lambda_{B\alpha})
\big]\Bigr)
\eqndot
\end{aligned}
\end{equation}
In addition, the quartic scalar interaction may be supplemented with the following terms
\begin{equation}
\begin{aligned}\label{eq:4_trace_action}
S^4_{\U{1}}&=\int\de^4 x\Bigl(
-\frac{\gym^2}{N}\big[
Q^{ij}_{\bar\phi\,kl}\tr(\phi_i\phi_j\bar\phi^l)\tr(\bar\phi^k)
+Q^{ij}_{\phi\,kl}\tr(\phi_i)\tr(\phi_j\bar\phi^k\bar\phi^l)\big]\\
&\hphantom{=\int\de^4 x\Bigl(}
-\frac{\gym^2}{N^2}\big[
Q^{ij}_{\bar\phi\bar\phi\,kl}\tr(\phi_i\phi_j)\tr(\bar\phi^k)\tr(\bar\phi^l)
+Q^{ij}_{\phi\phi\,kl}\tr(\phi_i)\tr(\phi_j)\tr(\bar\phi^k\bar\phi^l)\\
&\hphantom{=\int\de^4 x\Bigl({}-{}\frac{\gym^2}{N^2}\big[}+Q^{ij}_{\bar\phi\phi\,kl}\tr(\phi_i)\tr(\phi_j\bar\phi^k)\tr(\bar\phi^l)\big]\\
&\hphantom{=\int\de^4 x\Bigl(}
-\frac{\gym^2}{N^3}
Q^{ij}_{4\,kl}\tr(\phi_i)\tr(\phi_j)\tr(\bar\phi^k)\tr(\bar\phi^l)\Bigr)
\eqndot
\end{aligned}
\end{equation}
Like in the \SUN case, by requiring that the action is hermitian the transformation rules of all \UN coupling tensors under complex conjugation can be obtained. In addition, the requirements \ref{U1preserve} and \ref{N4limit} further restrict the \UN coupling tensors. 

For the $\beta$-deformation with gauge group \UN, we could follow the same logic as earlier in the \SUN case. This leads to $S^\beta_{\text{d.t.}}=S_{\U{1}}^3=S_{\U{1}}^4=0$. However, in \cite{Hollowood:2004ek} it was found that the \UN $\beta$-deformation is not conformally invariant and flows to the \SUN $\beta$-deformation in the infrared (IR). Therefore, couplings in the \UN $\beta$-deformation receive non-vanishing UV-divergent quantum corrections and a priori we could add any coupling within \eqref{eq:2_trace_action}, \eqref{eq:3_trace_action} or \eqref{eq:4_trace_action} that vanishes perturbatively at the IR fixed point. Whether the last requirement is fulfilled is, however, subject to explicit perturbative calculations.

\subsection{Symmetries of the deformed models}\label{sec:classical-symmetries-of-the-deformed-models}
The deformations that we discussed so far in this section are obtained by breaking the $R$-symmetry of the parent \NfSYMt partially or completely. Hence, many symmetry properties of the deformed models can immediately be adopted from the discussion in \secref{sec:symmetries}. For the classical discussion in this section, we assume that any added multi-trace coupling in the deformed theories does not break additional symmetries. Be warned, however, that this assumption is not true in general and in \secref{sec:non-conformal_double_trace_coupling} we will explicitly show that certain double-trace couplings receive UV-divergent quantum corrections that break the dilatation symmetry and hence conformal invariance. 

At the classical level, the conformal invariance of the parent \NfSYMt described in \subsecref{sec:conf_symmetry_N4} is inherited to the deformed models. This is clear, since the four-dimensional Minkowski-space structure and the classical scaling properties of all fields are untouched. In addition, only marginal interactions are added to the actions of the deformed models.

The $R$-symmetry is explicitly broken for the deformed models. The deformations described in \subsecref{sec:deformation} break the original $\su{4}_R$ symmetry algebra down to its $\mathfrak{u}(1)^{\times 3}$ Cartan subalgebra. Hence, from the generators $\mathfrak{R}_A^{\phan{A}B}$ in \subsecref{sec:R_symmetry}, only the three diagonal Cartan elements\footnote{Their matrix representation in the $(q^1,q^2,q^3)$ or the $(Q^1,Q^2,r)$ can be read off from the eigenvalues of fundamental fermions in \tabref{tab: su(4) charges}, e.g.\ $\mathfrak{R}_1|_q=\frac 12 \diag (+1,-1,-1,+1)$ or $\mathfrak{R}_1|_Q=\diag (1,-1,0,0)$.} $\mathfrak{R}_1$, $\mathfrak{R}_2$, and $\mathfrak{R}_3$ survive.

Finally, coming to the supersymmetry, the discussion is rather short for the $\gamma_i$-de\-for\-ma\-tion: this symmetry is entirely broken and hence all generators $\mathfrak{Q}^A_\alpha$, $\ol{\mathfrak{Q}}_{A\dot\alpha}$, $\mathfrak{S}_A^\alpha$, and $\ol{\mathfrak{S}}^{A\dot\alpha}$ from \secref{sec:symmetries} are absent. For the $\beta$-deformation, the $\ast$-product only depends on the two Cartan charges $Q^1$ and $Q^2$ but not on $r$, see \eqref{eq: antisymmetric product_beta}.\footnote{The remaining simple supersymmetry of the $\beta$-deformation becomes manifest, when the model is expressed in an $\mathcal{N}=1$ superspace, where fundamental fermions and scalars with the same $(Q^1,Q^2)$ charge are combined in one chiral superfield.} Since $A_\mu$ and $\lambda_{4\alpha}$ are not affected by the deformation \eqref{eq: antisymmetric product_beta}, the respective SUSY generators $\mathfrak{Q}^4_\alpha$ and $\ol{\mathfrak{Q}}_{4\dot{\alpha}}$ and the associated special conformal generators $\mathfrak{S}^\alpha_4$ and $\ol{\mathfrak{S}}^{4\dot{\alpha}}$ survive in the $\beta$-deformation. The superconformal algebra of the $\beta$-deformation is obtained by taking the one of \NfSYMt given in \subsecref{sec:SUSY_algebra_N4}, restricting the $R$-symmetry generators to the three Cartan generators and restricting the SUSY and special conformal SUSY generators to the remaining ones with spinor index $A=4$.

\section{Composite operators}\label{sec:composite-operators}
Apart from the fundamental properties of theories, we are most interested in the properties of composite operators that we may insert into correlation functions of external states, like in the case of two- and three-point functions in \eqref{eq:2_3_pt_function}. Following \cite[lecture 3]{Witten98book}, we define composite operators to be gauge invariant\footnote{We restrict the definition to gauge-invariant objects because these objects do not mix with gauge-dependent ones -- not even under renormalisation, see the discussion in \secref{sec:composite-operator-insertionsN4}.} products of elementary fields, possibly with covariant derivatives acting on them, which all reside at the same point in spacetime. An immediate example of composite operators are the fundamental interactions in the action, e.g.\ \eqref{eq:deformed_action_complex_scalars2}. They are, however, special since they must obey all symmetries of the \QFT. 

In this section, we discuss how finite composite operators can be defined in a free \QFT, then we introduce the elementary building blocks from which they can be constructed for \NfSYMt and its deformations, and finally we present the action of symmetry generators on composite operators in the free theory.

\subsection{Normal ordering}\label{sec:Normal_ordering}
In the quantised theory, naive products of multiple fields at a coincident point are not well defined, since they develop divergences in correlation functions. In order to define such products and hence composite operators in the quantised (free) theory, it is therefore necessary to subtract the occurring divergences. For correlation functions in which a composite operator of $L$ elementary fields is connected to $r$ external fields, this amounts to subtracting all $\floor{\frac{L-r}{2}}$ loops that appear in all possible Wick contractions. In the canonical quantisation approach, this leads to the normal ordering of creation and annihilation operators introduced in \cite{PhysRev.80.268}. In \cite{Zimmermann1973570}, this normal-ordering procedure it was used to define finite composite operators with arbitrarily many constituent fields in the free quantised theory.

As a simple example of normal ordering, let us analyse the correlation function of two complex scalars $\Delta(x,y)=\vacl \T \phi(x)\ol{\phi}(y)\vac$, which is the Green's function of the d'Alembert operator $-\partial^2_x$. In the conventions of \cite[\chap{8}]{Srednicki:2007}, it can be represented in four-dimensional Minkowski space by 
\begin{equation}
\Delta(x,y)=\int\frac{\de^4p}{(2\pi)^4}\frac{\e^{\complexi p(x-y)}}{p^2-\complexi\epsilon}\eqncom
\end{equation}
which is divergent for $y\rightarrow x$. Therefore, the naive definition $\mathcal{O}^{(\phi,\ol{\phi})}(x)=\phi(x)\ol{\phi}(x)$ is ill-defined in the quantum theory and we use the alternative definition, which explicitly subtracts the divergent part as
\begin{equation}\label{eq:O2_normal_ordering}
\mathcal{O}^{(\phi,\ol{\phi})}(x)=\lim_{y\rightarrow x}\left(\phi(x)\ol{\phi}(y)-\vacl \T \phi(x)\ol{\phi}(y)\vac\text{id}\right)\eqndot
\end{equation}
Note that for a correlation function with no external fields ($r=0$), this definition implies that the vacuum expectation value of the operator vanishes $\vacl\mathcal{O}^{(\phi,\ol{\phi})}(x)\vac=0$. 

In general, we define a composite operator with $L$ constituent fields in a free theory\footnote{In an interacting \QFT, the definition of a composite operator given here still contains divergences that need renormalisation, \cf \subsecref{sec:composite-operator-insertions}.} as 
\begin{equation}
\mathcal{O}^{(\vec{f}\,)}(x)=:f_1(x_1)f_2(x_2)\dots f_L(x_L):\Bigr|_{x_i=x}\eqncom
\end{equation}
where $:\cdot:$ indicates a normal-ordered product. Following \cite{Kehrein:2006ti}, the latter is defined via the three relations:
\begin{itemize}
	\item[1.] The normal-ordering of the identity operator $\text{id}$ is just itself, $:\text{id}:=\text{id}\eqncom$
	\item[2.] For operators with complex coefficients $c$, normal-ordering is linear:\\ $:c_f\mathcal{O}^{(\vec{f}\,)}+c_g\mathcal{O}^{(\vec{g}\,)}:= c_f:\mathcal{O}^{(\vec{f}\,)}:+c_g:\mathcal{O}^{(\vec{g}\,)}:\eqncom$
	\item[3.] A general normal-ordered product is defined recursively via\footnote{Note that the variation with respect to a fermionic field is Grassmann valued and hence commuting it to a certain position generates the signs for fermions.} 
	\begin{equation}\label{eq:normal_ordering}
	f_i(y):\mathcal{F}_L: =:f_i\mathcal{F}_L:+\int\de^d z\sum_j \vacl\T f_i(y) f_j(z)\vac:\frac{\delta \mathcal{F}_L}{\delta f_j(z)}:\eqncom
	\end{equation}
	where $f_i$ is an elementary field, $\mathcal{F}_L$ a collection of $L$ elementary fields at positions $x_1,\dots x_L$ and the functional variation fulfils $\frac{\delta f_i(x)}{\delta f_j(y)}=\delta_i^j\delta^{(4)}(x-y)$.
\end{itemize}
In \eqref{eq:normal_ordering} we explicitly see the separation of an ordinary product of operators into a normal-ordered part and a second part that contains all possible Wick contractions.

\subsection{Building blocks}\label{sec:Building_blocs}
Let us now introduce the basis elements from which all composite operators in \NfSYMt and its deformations can be constructed. 

In principle, these gauge-invariant\footnote{Since all these quantities transform in the adjoint representation of the gauge group, any product of them in a trace is gauge invariant, compare \eqref{eq:general_symmetry_trafo}.} operators are built from colour traces of the elementary fields: scalars, fermions, and (anti-)self-dual field strengths with an arbitrary number of covariant derivatives acting on them. However, not all operators that we construct in this way are independent. They may be connected to other composite via the Bianchi identity or the \eom, see \appref{app:EOM_Bianchi} for the realisations of these equations in our conventions. Since the Bianchi identity as well as the \eom always contain a term in which a covariant derivative acts on an elementary field, we can lift redundancies in our description, by imposing that no spinor indices may be contracted via the antisymmetric symbols $\varepsilon_{\alpha\beta}$ or $\varepsilon_{\dot{\alpha}\dot{\beta}}$, see \appref{app:EOM_Bianchi} for details. For spinor indices, this requirement translates to symmetrising all $\spl{2}$ and $\splbar{2}$ spinor indices of covariant derivatives and the fields they act on. This prescription yields the alphabet of composite operators
\begin{equation}\label{eq: alphabet}
\cA =\{ \D^k\phi_i, \D^k\ol\phi^i, \D^k\lambda_{A\alpha}, \D^k\ol{\lambda}^A_{\dot\alpha},\D^k\cF_{\alpha\beta},\D^k\bar\cF_{\dot\alpha\dot\beta} \}
\eqncom
\end{equation}
where the abbreviation $\D^k\lambda_{A\alpha}$ stands for an expression with $k\in \NN_0$ covariant derivatives $\D_{\gamma\dot{\gamma}}$ acting on $\lambda_{A\alpha}$ in which the $\spl{2}$ and $\splbar{2}$ indices are totally symmetrised. Single-trace composite operators can now be represented as a graded cyclic chain\footnote{The chain is graded cyclic to account for the anti-commutation property of fermion. If a fermion is shifted from last to first place in a composite operator which is bosonic overall, the operator acquires an additional sign, see e.g.\ the terms involving fermions in the action \eqref{eq:N4_action_osci}.} where each chain site is occupied by an element of the alphabet $\mathcal{A}_i$. Instead of a chain with sites, we can also think of the composite operator as a tensor product of fields taken from the alphabet $\cA$ with an appropriate equivalence relation $\sim$ realising the graded cyclic invariance. A length-$L$ operator then takes the form
\begin{equation}\label{eq:operator_as_tensorproduct}
\mathcal{O}^{(\vec{\cA}\,)}(x)=\frac{1}{\mathcal{N}}\tr\bigl(\mathcal{A}_1\mathcal{A}_2\dots \mathcal{A}_L\bigr)=\frac{1}{\mathcal{N}}\bigl(\mathcal{A}_1\otimes \mathcal{A}_2\otimes\dots \otimes \mathcal{A}_L\bigr)_\sim\eqncom
\end{equation}
where each field $\mathcal{A}_i$ is evaluated at position $x$ and the normalisation factor $\cN=N^{\frac L2} \cN_{\text{fl}}$ includes a colour normalisation and a flavour normalisation $\cN_{\text{fl}}$ to ensure that the two-point function in \eqref{eq:2_3_pt_function} is normalised.

With the oscillator representations given in \subsecref{subsec:the-oscillator-representation} for the alphabet \eqref{eq: alphabet}, we can represent a length-$L$ composite operator as a graded cyclic spin-chain state with the appropriate oscillator vacuum $\vac_L=\vac\otimes\dots\otimes \vac$ and $L$ families of collective creation oscillators $\mathbf{A}^\dagger_i=(\aosc^{\dagger}_{(i)\alpha},\bosc^{\dagger}_{(i)\dot\alpha},\cosc^{\dagger }_{(i)A})$ acting on the individual sites $i$. Instead of the collective creation oscillators we can also use the collective oscillator occupation numbers 
\begin{equation}\label{eq:occupation_numbers}
A_{i}=(\akindsite[1]{i},\akindsite[2]{i},\bkindsite[1]{i},\bkindsite[2]{i},
\ckindsite[1]{i},\ckindsite[2]{i},\ckindsite[3]{i},\ckindsite[4]{i})
\end{equation}
to characterise a composite operator in terms of the vector $\ket{A}=\ket{A_1,\dots,A_L}$. For the graded cyclic invariance, we define the graded shift operator $T$ which transforms a length-$L$ state as
\begin{equation}\label{eq:graded_shift_op}
T\ket{A_1,\dots, A_L}=\omega^{2\mathfrak{D}_0(A_L)\sum_{i=1}^{L-1}\mathfrak{D}_0(A_i
	)}\ket{A_L,A_1,\dots,A_{L-1}}\eqncom
\end{equation}
with the grading prefactor which was defined below \eqref{eq:graded_Jacobi_identity}. The classical dilatation operator in the oscillator representation is defined in \eqref{eq:symmetry_generators_osci_LR} and in terms of the occupation numbers in \eqref{eq: def classical dilatation op in osc language}. On a length-$L$ spin-chain state $\ket{A}$, graded cyclic invariance is then realised by including the length-$L$ projector
\begin{equation}\label{eq:graded_projetor}
\mathcal{P}_L=\frac 1L\sum_{i=0}^{L-1}T^i
\end{equation}
and hence the composite operator \eqref{eq:operator_as_tensorproduct} is realised in the oscillator representation by
\begin{equation}\label{eq:operator_as_spinchain}
\mathcal{O}^{(\vec{\cA}\,)}(x)=\frac{1}{\mathcal{N}}\mathcal{P}_L\ket{A_1,\dots,A_L}\eqndot
\end{equation}

\subsection{Symmetry transformations of composite operators}\label{subsec:symmetry-generators-and-composite-operators}
With composite operators at hand, let us discuss how symmetry generators act on them in the free theory. Note that this also covers the action of symmetry generators on the elementary interaction vertices, which can be understood as special composite operators.

In the previous subsection, we discussed that composite operators can be thought of as cyclic tensor products of elementary fields, possibly with traceless symmetric covariant derivatives acting on them. Symmetry generators act linearly on such products, i.e.\ on a length-$L$ operator $\mathcal{O}(x)$ the action of the symmetry generator $g_m$ is realised via the single-cite generator $\mathfrak{g}_m$ as
\begin{equation}\label{eq:symmetry_generator_on_comp_operators}
[g_m,\mathcal{O}(x)]= \sum_{i=1}^L\frac{\omega^{2\mathfrak{D}_0(\mathfrak{g}_m)\sum_{k=1}^{i-1}\mathfrak{D}_0(A_k)}}{\cN}\tr\bigl(
	\cA_1\dots\cA_{i-1}[\mathfrak{g}_m,\cA_i]\cA_{i+1}\dots\cA_L\bigr)\eqncom
\end{equation}
where the $\omega$ factor realises the grading for fermionic generators. In the context of composite operators we will call $g_m$ the symmetry generator and $\mathfrak{g}_m$ its density. In the case of \NfSYMt, the action of the symmetry generators on elementary fields was given in \subsecref{sec:symmetries} and the necessary adaptations for the $\beta$- and $\gamma_i$-deformation were discussed in \subsecref{sec:classical-symmetries-of-the-deformed-models}. We already see that the length of composite operators is only fixed in the free theory, since e.g.\ the SUSY transformations of fermions \eqref{eq:SUSY_trafo_psi} introduces length-changing $\gym$-dependent contributions in the field representation. More severe length-changing examples involve the dilatation generator $D$ which realises scaling transformations and which is directly affected by the introduction of a renormalisation scale in the quantised theory. Its coupling-dependent quantum corrections induce corrections to the realisation \eqref{eq:symmetry_generator_on_comp_operators}, see \subsecref{sec:the-quantum-dilatation-operator-on-composite-operators}. Length-changing contributions are also induced for generators that are constructed\footnote{The special conformal SUSY generators are explicitly given in terms of the dilatation generator in \eqref{eq:special_conformal_SUSY_generators}. The special conformal generator itself can be written as the linear combination $K_\mu=\frac{x^\nu}{4}(\eta_{\mu\nu}D+M_{\mu\nu})$, which can be seen from \eqref{eq:conformal_generators} and \eqref{eq:unitary_trafo_fields} with \eqref{eq:commutator_fields_conf_generators}.} from $D$, i.e.\ $K$, $\mathfrak{S}$ and $\ol{\mathfrak{S}}$.

Restricting to the dilatation generator of the free theory (be it \NfSYMt or its deformations) for the moment, its action on the composite operator $\mathcal{O}(x)$ is determined via \eqref{eq:symmetry_generator_on_comp_operators} and the commutation relations \eqref{eq:unitary_trafo_fields_2}. We find
\begin{equation}\label{eq:D_cl_on_O}
D\mathcal{O}(x)=-\complexi\bigl(\Delta_{\mathcal{O}}^0+x_\mu\partial^\mu)\mathcal{O}(x)\eqncom\qquad
\Delta^0_{\mathcal{O}}=\sum_{i=1}^L\Delta_{\mathcal{A}_i}\eqncom
\end{equation}
where $\Delta^0_{\mathcal{O}}$ is the classical scaling dimension of the operator $\mathcal{O}$ and the term $x_\mu\partial^\mu$ arises from the induced coordinate transformation, see \ref{sec:conf_symmetry_N4} and \ref{sec:conformal-transformations-of-fields} for details. The induced coordinate transformation can be stripped off by evaluating the action of the dilatation operator on composite operators at the origin, where we have
\begin{equation}
D\mathcal{O}(0)=-\complexi \Delta_{\mathcal{O}}\mathcal{O}(0)\eqndot
\end{equation}
Whenever we discuss properties of the dilatation operator, we mean its properties when acting on fields or composite operators at the origin. In the oscillator picture, the free dilatation operator density is given in \eqref{eq: def classical dilatation op in osc language}. 

The structure of the two- and three-point correlation functions of composite operators which was already mentioned in \eqref{eq:2_3_pt_function} can be fixed entirely from the properties of composite operators under symmetry transformations. In analogy to the macroscopic symmetry transformations of elementary fields in \eqref{eq:symmetry_trafo_elementary_field}, we find the macroscopic scaling transformation of composite operators by exponentiating \eqref{eq:D_cl_on_O}, which yields
\begin{equation}
\widehat{\mathcal{O}}(x)=\e^{\alpha \Delta_{\mathcal{O}}^0}\mathcal{O}(\e^\alpha x)\eqndot
\end{equation}
In a \CFT, the two-point correlation function must be invariant under rescaling and hence we have the condition
\begin{equation}
\e^{\alpha(\Delta^0_{\mathcal{O}_1}+\Delta^0_{\mathcal{O}_2})}
\vacl\T\mathcal{O}_1(\e^\alpha x)\mathcal{O}_2(\e^\alpha y)\vac =
\vacl\T\mathcal{O}_1(x_1)\mathcal{O}_2(x_2)\vac\eqndot
\end{equation}
Following \cite[\chap{4}]{CFT}, this equation can be solved by using the rotational and translational invariance as well as the Lorentz covariance of correlation functions to give the two-point function
\begin{equation}\label{eq:2pt_func_classical}
\vacl\T\mathcal{O}_1(x_1)\mathcal{O}_2(x_2)\vac=\frac{\delta_{\Delta^0_{\mathcal{O}_1}\Delta^0_{\mathcal{O}_2}}}{(|x_1-x_2|^2+\complexi \epsilon)^{\Delta^0_{\mathcal{O}_1}}}\eqndot
\end{equation}
The structure of the three-point correlation function in \eqref{eq:2_3_pt_function} can be determined analogously, \cf \cite[\chap{4}]{CFT} for details.

\chapter{Renormalisation and the quantised theories}\label{chap:Renormalisation}

In this chapter, we discuss the effects that quantisation has on the classical theories introduced in \chapref{chap:The_models}. The presentation in this chapter is inspired by \cite{Srednicki:2007, CFT, kleinert2001critical, Collins:1984xc, BROWN1980135} and we refer the reader there for further details.

We start with a brief review of the path integral approach and give the definitions of position and momentum space correlation functions in this context. Subsequently, we introduce the renormalisation program for $\varphi^3$-theory in six-dimensional Minkowski space. In this simplified setting, we fix our conventions and discuss general aspects of renormalisation for elementary fields, couplings, and composite operators that are also important for \NfSYMt and its deformations. In particular, we review the mixing of composite operators under renormalisation and how the associated anomalous dimensions arise.

With the general setup of the renormalisation program at hand, we then focus on the particularities that occur in the highly symmetric settings of \NfSYMt and its deformations. We review how the classical symmetries of these theories restrict the structure of correlation functions in the quantised theories and for \CFTs we discuss how the exact scale invariance influences the renormalisation. We also investigate how the anomalous dimensions, which are independent of the renormalisation scheme in a \CFT, affect the classical scaling symmetry of composite operators. This leads us to an extended definition of the dilatation generator in quantised \CFTs. We conclude our general discussion of the renormalisation program with an example of elementary field and composite operator renormalisation in \NfSYMt and we also use the examples to introduce the concrete notation that will be employed in perturbative calculations in \chapref{chap:applications}.

In addition to general aspects of renormalisation, we give a precise definition of the \tHooft limit in which we will perform all calculations in \chapref{chap:applications}. We discuss which types of diagrams may contribute in correlation functions of composite operators, including the finite-size wrapping and prewrapping contributions. In the final part of this chapter, we use this limit, to formally construct the (asymptotic) planar one-loop dilatation generator of \NfSYMt and its deformations in terms of a one-loop dilatation generator density.

\newpage

\section{The path integral approach}\label{sec:path_integral_approach}

In this section, we briefly discuss the concepts of the path integral approach that we explicitly need in the following sections and we refer to the literature for a general introduction, e.g.\ \cite{Peskin:1995ev, Srednicki:2007,Ticciati:1999}. For perturbative calculations of non-abelian gauge theories including Weyl fermions and scalars, a detailed derivation of Feynman rules from the path integral is presented in \appref{app:Feynman_rules} and its {\tt Mathematica} implementation in form of the package \ttt{FokkenFeynPackage} is presented in \appref{sec:Feynman_rules_Mathematica}.  

The partition function of a system is given by the sum over all possible states, weighted by a phase factor. If there are no external forces, the system will stay in the vacuum state and the sum of states in the quantised theory is given by a transition from the vacuum state to itself
\begin{equation}
Z=\langle 0 | 0 \rangle=\int\mathcal{D}\{\varphi\}\e^{\complexi S[\{\varphi\}]}\eqndot
\end{equation}
In the second equality we used the path integral representation which depends on the local $D$-dimensional action of the theory $S[X]=\int\de^Dx\mathcal{L}[X,\D^\mu X]$. The integration is taken over all possible field configurations with a suitable path integral measure indicated by $\mathcal{D}\{\varphi\}$. For the evaluation of more interesting transitions that include explicit field insertions, we must also incorporate a source action $S_{\text{source}}[\{\varphi,j\}]=\int\de^Dx j_i(x) \varphi^i(x)$ into the definition of the partition function. In doing so we arrive at the generating functional
\begin{equation}\label{eq:generating_functional}
Z[\{j\}]=\frac{1}{N_Z}\vacl \e^{\complexi S[\{\varphi\}]+\complexi S_{\text{source}}[\{\varphi,j\}]}\vac
=\frac{\int\mathcal{D}\{\varphi\}\e^{\complexi S[\{\varphi\}]+\complexi S_{\text{source}}[\{\varphi,j\}]}}{
	\int\mathcal{D}\{\varphi\}\e^{\complexi S[\{\varphi\}]}
	}\eqncom
\end{equation}
where the normalisation sets $Z[\{0\}]=1$ in the absence of external sources. We can now generate time-ordered correlation functions of $n$ elementary fields (also called $n$-point or Green's functions) from the normalised generating functional $Z[\{j\}]$ by taking functional derivatives with respect to the sources $j_m(x)$, using $\frac{\delta j_l(x)}{\delta j_m(y)}=\delta_l^m\delta^{(D)}(x-y)$ and setting all sources to zero in the end. We arrive at\footnote{If the theory contains fermions, additional signs appear in the functional differentiation, \cf \appref{app:Feynman_rules} and in particular \eqref{eq:field_as_source_der} therein for details.}
\begin{equation}
\begin{aligned}\label{eq:n_point_correlation_function}
G^{(\vec{i}\,)}(\vec{x})&\equiv\vacl \T \varphi^{i_1}(x_1)\varphi^{i_2}(x_2)\dots \varphi^{i_n}(x_n)\vac
=\left[\frac{\delta}{\complexi \delta j_{i_n}(x_n)}\dots \frac{\delta}{\complexi \delta j_{i_1}(x_1)} Z[\{j\}]\right]_{\{j\}=0}\\
&=\frac{1}{N_Z}
\int \mathcal{D}\{\varphi\}\, \varphi^{i_1}(x_1)\dots \varphi^{i_n}(x_n) \e^{\complexi S[\{\varphi\}]}\eqncom
\end{aligned}
\end{equation}
where the time ordering symbol $\T$ ensures that fields evaluated at later times occur to the left of fields evaluated at earlier times, \cf \eqref{eq:time_ordered_product}. Since we work with the momentum space Feynman rules of \appref{app:Feynman_rules}, we also need the Fourier transformed $n$-point functions\footnote{The time ordering symbol means that the Fourier transformation of the momentum space expression is a time-ordered position space expression, compare the definition in \appref{sec:Conventions}.}
\begin{equation}\label{eq:n_point_correlation_function_fourier}
\tilde{G}^{(\vec{i}\,)}(\vec{p})\equiv\vacl \T \tilde\varphi^{i_1}(p_1)\tilde\varphi^{i_2}(p_2)\dots \tilde\varphi^{i_n}(p_n)\vac
=\left[\frac{\delta}{\complexi \delta \tilde j_{i_n}(-p_n)}\dots \frac{\delta}{\complexi \delta \tilde j_{i_1}(-p_1)} \tilde Z[\{\tilde j\}]\right]_{\{\tilde j\}=0}\eqncom
\end{equation} 
which are obtained from \eqref{eq:n_point_correlation_function} using \eqref{eq:Fourier_transformation}. All calculations in this thesis are performed in momentum space and we will drop the tilde of Fourier transformed quantities from now on. For connected $n$-point functions, we can use the locality of interactions to factor out a momentum conservation factor, resulting in the reduced $n$-point function
\begin{equation}\label{eq:reduced_n_point_correlation_function}
G_{\text{c}}^{(\vec{i}\,)}(\vec{p})
=(2\pi)^D\delta^{(D)}\Bigl({\textstyle\sum\limits_{j=1}^n} p_j\Bigr)
\vacl \T \varphi^{i_1}(p_1)
\dots \varphi^{i_{n-1}}(p_{n-1})\varphi^{i_{n}}\Bigl(-{\textstyle\sum\limits_{j=1}^{n-1}}p_{j}\Bigr)\vac_{\complexi \cT}\eqndot
\end{equation}

Path integrals which enter our definition of correlation functions in the quantised theory in \eqref{eq:n_point_correlation_function} usually cannot be solved exactly but a prominent exception to this statement are path integrals of non-interacting theories. The free action $S_0$ of such theories takes a quadratic form in the field variables and the path integral can be solved via a generalisation of the Gaussian integral. In the presence of a source term, the free theory is still solvable and the generating functional can be written as
\begin{equation}\label{eq:free_generating_functional}
Z_0[\{j\}]=\frac{1}{N_Z}\int\mathcal{D}\{\varphi\}\e^{\complexi \bigl(S_0[\{\varphi\}]+ S_{\text{source}}[\{\varphi,j\}]\bigr)}
\sim \e^{\complexi \int\de^Dx\de^Dy j(x)\Delta(x,y)j(y)}\eqncom
\end{equation}
where $\Delta(x,y)$ is the propagator of the free theory. For interacting theories with a free action $S_0[\{\phi\}]$ and an interaction part $S_{\text{int}}[\{\phi\}]$, we can use this solution by expressing the interacting theory as a perturbation of a non-interacting one. For this to work, we assume that the interaction part depends on a set of coupling constants $g_i$, that allow for such a perturbative expansion. In the path integral approach, this idea amounts to splitting the generating functional into a free part whose solution takes the form of \eqref{eq:free_generating_functional} and an interaction part that acts on it. Following \cite[\chap{9}]{Srednicki:2007}, this procedure yields
\begin{equation}\label{eq:Zj_perturbative}
\begin{aligned}
Z[\{j\}]&=
\frac{1}{N_Z}\int\mathcal{D}\{\varphi\}\e^{\complexi S_{\text{int}}[\{\varphi\}]
	+\complexi \bigl(S_0[\{\varphi\}]+ S_{\text{source}}[\{\varphi,j\}]\bigr)}
=
\exp\left[\complexi S_{\text{int}}[\{\tfrac{\delta}{\complexi \delta j}\}]\right]Z_0[\{j\}]\eqndot
\end{aligned}
\end{equation}
For any correlation function, we can expand the interaction exponential up to a given order in the coupling constants $g_i$ and evaluate all occurring contributions.

\section{Massless \texorpdfstring{$\varphi^3$}{phi-cubed}-theory}\label{sec:phi3_theory}
In this section, we discuss how perturbative corrections to correlation functions can be obtained in the path integral approach. For the moment, we restrict to massless $\varphi^3$-theory (a non-conformal theory of real scalars with a single three-point interaction) to discuss general aspects of renormalisation and we will come back to \NfSYMt and its deformations in the next section. The $\varphi^3$-theory is intensively discussed in \cite[\chap{9}]{Srednicki:2007} and we refer the reader there for results and the Feynman rules used in this section. 
\subsection{The bare theory}\label{subsec:bare_phi_3}
Let us start with the bare or unrenormalised theory. In $d$-dimensional Minkowski space with mostly plus metric, we can write the action of $\varphi^3$-theory in terms of a free and an interaction part as
\begin{equation}\label{eq:phi3_action}
S_0[\varphi_{\text{B}},j_{\text{B}}]=\int\de^d x\Bigl(-\frac 12
\partial_\mu\varphi_{\text{B}}(x)\partial^\mu\varphi_{\text{B}}(x)
+j_{\text{B}}(x)\varphi_{\text{B}}(x)
\Bigr)
\eqncom\quad
S_{\text{int}}[\varphi_{\text{B}},g_{\text{B}}]=
\int\de^d x \frac{g_{\text{B}}}{3!}\varphi_{\text{B}}^3(x)\eqncom
\end{equation}
where the label B indicates bare (non-renormalised) quantities. From the discussion in \subsecref{sec:conf_symmetry_N4}, we know that this action is classically scale invariant if the classical scaling dimensions fulfil
\begin{equation}\label{eq:phi3_dimensions}
\Delta^0_\varphi=\frac 12(d-2)\eqncom\qquad \Delta^0_j=\frac 12(d+2)\eqncom\qquad \Delta^0_g=\frac 12(6-d)
\end{equation} 
and hence our coupling $g_{\text{B}}$ is dimensionless in $d=6$ dimensions. In this case, we also know the classical correlation function of two scalars which is given by taking \eqref{eq:2pt_func_classical} and replacing the two composite operator insertions by two elementary scalar fields.

Focussing on the two-point correlation function in $d=6$ dimensions, let us investigate its perturbative quantum corrections at lowest order in the coupling $g_{\text{B}}$. The reduced momentum space two-point function of the free theory can be obtained from the Fourier transformation of the free position space two-point function \eqref{eq:2pt_func_classical} and it is given by
\begin{equation}
\vacl\T\varphi_{\text{B}}(p)\varphi_{\text{B}}(-p)\vac_{\complexi\cT,\,\text{free}}=\frac{1}{\complexi}\Delta(p)=\frac{1}{\complexi}\frac{1}{p^2-\complexi \epsilon}\eqncom
\end{equation}
where the subscript $\complexi \cT$ means that we have dropped the momentum conservation factor $(2\pi)^d\delta^{(d)}(p+p_{\text{in}})$. The term $\complexi \epsilon$ in the last equality appears since we work in Minkowski space\footnote{Our conventions of the Wick rotation and Fourier transformations are given in \appref{sec:Conventions} and they are compatible with \cite{Srednicki:2007}.}. In the interacting theory, this free propagator is corrected by coupling dependent contributions and in terms of connected momentum space Feynman diagrams the lowest order contribution is
\begin{equation}
\begin{aligned}\label{eq:first_loop_2pt}
\vacl\T\varphi_{\text{B}}(p)\varphi_{\text{B}}(-p)\vac_{\complexi\cT}&=
\phan{\frac 1\complexi}
\settoheight{\eqoff}{$\times$}%
\setlength{\eqoff}{0.5\eqoff}%
\addtolength{\eqoff}{-3.5\unitlength}%
\raisebox{\eqoff}{%
	\fmfframe(1,1)(-1,1){%
		\begin{fmfchar*}(5,5)
		\fmfforce{0 w,0.5 h}{v1}
		\fmfforce{1 w,0.5 h}{v2}
		\fmf{plain}{v1,v2}
		\end{fmfchar*}
	}
}	
&&+\phan{\frac{1}{\complexi}\Delta(p)}
\settoheight{\eqoff}{$\times$}%
\setlength{\eqoff}{0.5\eqoff}%
\addtolength{\eqoff}{-3.5\unitlength}%
\raisebox{\eqoff}{%
	\fmfframe(1,1)(-1,1){%
		\begin{fmfchar*}(10,5)
		\fmfleft{v1}
		\fmfright{v2}
		\fmffixed{(0.5w,0)}{vc1,vc2}
		\fmf{plain}{v1,vc1}
		\fmf{plain}{vc2,v2}
		\fmf{plain,left=1}{vc1,vc2}
		\fmf{plain,left=1}{vc2,vc1}
		\end{fmfchar*}
	}
}
&&+\order{g_{\text{B}}^4}&\\
&=
\frac{1}{\complexi}\Delta(p)
&&+\frac{1}{\complexi}\Delta(p) \Bigl[\complexi \Pi(p)\Bigr]\frac{1}{\complexi}\Delta(p)
&&+\order{g_{\text{B}}^4}
\eqncom&
\end{aligned}
\end{equation}
where we depicted the free propagator by a straight line and each interaction with a power of $g_{\text{B}}$ by a three-point vertex. In the second line, we have divided the loop diagram into one-particle irreducible (1PI) parts, which separates the two external free propagators and the (amputated) self-energy contribution $\Pi(p)$. Using the Feynman rules of $\varphi^3$-theory \cite[\chap{9,14}]{Srednicki:2007}, the latter can be expressed in terms of the Minkowski space Feynman integral 
\begin{equation}
\begin{aligned}\label{eq:Pi_1}
\complexi\Pi^{(1)}_{\text{B}}(p)=
\Bigl(
\settoheight{\eqoff}{$\times$}%
\setlength{\eqoff}{0.5\eqoff}%
\addtolength{\eqoff}{-3.5\unitlength}%
\raisebox{\eqoff}{%
	\fmfframe(1,1)(-1,1){%
		\begin{fmfchar*}(10,5)
		\fmfleft{v1}
		\fmfright{v2}
		\fmffixed{(0.5w,0)}{vc1,vc2}
		\fmf{plain}{v1,vc1}
		\fmf{plain}{vc2,v2}
		\fmf{plain,left=1}{vc1,vc2}
		\fmf{plain,left=1}{vc2,vc1}
		\end{fmfchar*}
	}
}
\Bigr)_{\text{1PI}} 
&=\frac 12 g_{\text{B}}^2
\int\frac{\de^d l}{(2\pi)^d}\frac{1}{(l^2-\complexi \epsilon)((p-l)^2-\complexi \epsilon)} 
\eqncom
\end{aligned}
\end{equation}
where the superscript $(1)$ labels the one-loop contribution in $\Pi_{\text{B}}$ and the subscript 1PI indicates that all external propagators have been amputated. Note that we will drop the $\complexi \epsilon$ terms later in this thesis for purely notational reasons. The integral depends on the single external scale $p^2$ and, following \cite{Chetyrkin:1980pr,Vladimirov:1979zm}, it can be solved exactly in terms of the Euclidean space integral
\begin{equation}\label{eq:G_function}
\int\frac{\de^D\bar l}{(2\pi)^D}\frac{1}{\bar l^{2\alpha}(\bar p-\bar l)^{2\beta}}
=\frac{G(\alpha,\beta)}{(4\pi)^{\frac{D}{2}}\bar p^{2(\alpha+\beta-\frac{D}{2})}}\eqncom\quad
G(\alpha,\beta)=
\frac{
	\Gamma(\frac{D}{2}-\alpha)\Gamma(\frac{D}{2}-\beta)\Gamma(\alpha+\beta-\frac{D}{2})
}{\Gamma(\alpha)\Gamma(\beta)\Gamma(D-\alpha-\beta)}\eqncom
\end{equation}
with generalised parameters $D,\alpha,\beta\in \RR$ and Euclidean momenta indicated by a bar. Depending on the parameters, this integral develops divergences, which appear in the solution as poles in the $\Gamma$ functions. Whether an occurring divergence appears in the ultraviolet (UV) or in the infrared (IR) regime, can be determined from the Euclidean integrand in \eqref{eq:G_function}: UV divergences appear in the $\abs{\bar l}\rightarrow\infty$ regime and IR divergences in the $\abs{\bar l}\rightarrow 0$ or $\abs{\bar l}\rightarrow \abs{\bar p-\bar l}$ regime, see \appref{sec:UV_and_IR_div} and the references therein for a treatment of UV and IR divergences in Feynman integrals. Using the methods discussed in \appref{app:Evaluating_Feynman_integrals} to Wick rotate the integral in \eqref{eq:Pi_1} to Euclidean space and back to Minkowski space via the inverse operator $\WR^{-1}$, we find 
\begin{equation}\label{eq:example_div}
\complexi\Pi_{\text{B}}^{(1)}(p)=
\frac{\complexi}{2}g_{\text{B}}^2\WR^{-1}\int\frac{\de^d\bar l}{(2\pi)^d}\frac{1}{\bar l^{2}(\bar p-\bar l)^{2}}
=
\frac{\complexi}{2} \frac{g_{\text{B}}^2}{(4\pi)^{\frac d2}}
\frac{\Gamma^2(\frac d2-1)\Gamma(2-\frac d2)}{\Gamma(d-2)} 
\frac{1}{p^{2(2-\frac d2)}}\eqncom
\end{equation}
where the factor of $\complexi$ occurs in the Wick rotation. We see that the integral is UV divergent for even integer dimensions $d\geq 4$ and we will address the renormalisation of such divergences in the following subsection.

Next, we discuss the first order correction to the three-point function. For external momenta $p_1$ and $p_2$, we have 
\begin{equation}
\begin{aligned}\label{eq:3pt_corr_phi}
\vacl\T\varphi_{\text{B}}(p_1)\varphi_{\text{B}}(p_2)\varphi_{\text{B}}(-p_1-p_2)\vac_{\complexi\cT}&=
\settoheight{\eqoff}{$\times$}%
\setlength{\eqoff}{0.5\eqoff}%
\addtolength{\eqoff}{-4.5\unitlength}%
\raisebox{\eqoff}{%
	\fmfframe(1,1)(-1,1){%
		\begin{fmfchar*}(7,7)
		\fmfforce{0 w,1 h}{v1}
		\fmfforce{1 w,0.5 h}{v2}
		\fmfforce{0 w,0 h}{v3}
		\fmf{plain}{v1,vc}
		\fmf{plain}{vc,v2}
		\fmf{plain}{v3,vc}
		\end{fmfchar*}
	}
}
+
\settoheight{\eqoff}{$\times$}%
\setlength{\eqoff}{0.5\eqoff}%
\addtolength{\eqoff}{-4.5\unitlength}%
\raisebox{\eqoff}{%
	\fmfframe(1,1)(-1,1){%
		\begin{fmfchar*}(7,7)
		\fmfforce{0 w,1 h}{v1}
		\fmfforce{1 w,0.5 h}{v2}
		\fmfforce{0 w,0 h}{v3}
		\fmfforce{0.2 w,0.8 h}{vc1}
		\fmfforce{0.72 w,0.5 h}{vc2}
		\fmfforce{0.2 w,0.2 h}{vc3}
		\fmf{plain,left=0.6}{vc1,vc2}
		\fmf{plain,left=0.6}{vc2,vc3}
		\fmf{plain,left=0.6}{vc3,vc1}
		\fmf{plain}{v1,vc1}
		\fmf{plain}{vc2,v2}
		\fmf{plain}{vc3,v3}
		\end{fmfchar*}
	}
}
+
3\settoheight{\eqoff}{$\times$}%
\setlength{\eqoff}{0.5\eqoff}%
\addtolength{\eqoff}{-4.5\unitlength}%
\raisebox{\eqoff}{%
	\fmfframe(1,1)(-1,1){%
		\begin{fmfchar*}(10,7)
		\fmfforce{0 w,1 h}{v1}
		\fmfforce{1 w,0.5 h}{v2}
		\fmfforce{0 w,0 h}{v3}
		\fmfforce{0.3w, 0.5h}{vcl}
		\fmfforce{0.65w, 0.5h}{vcr}
		\fmf{plain}{v1,vcl,v2}
		\fmf{plain}{vcl,v3}
		\fmfv{d.sh=circle,d.f=empty,d.si=.4w}{vcr}
		\end{fmfchar*}
	}
}
+\order{g_{\text{B}}^5}\eqncom
\end{aligned}
\end{equation}
where the last diagram appears three times since we can have a one-loop self-energy insertion on each of the external legs. We can construct the third diagram by gluing \eqref{eq:Pi_1} together with two external propagators via the three-vertex of the theory and hence the second diagram renders the only new contribution. Following \cite[\chap{16}]{Srednicki:2007}, its 1PI part is given by
\begin{equation}\label{eq:V3_phi_bare}
\complexi V^{(1)}_{\varphi^3\text{B}}(p_1,p_2)=
\Bigl(
\settoheight{\eqoff}{$\times$}%
\setlength{\eqoff}{0.5\eqoff}%
\addtolength{\eqoff}{-4.5\unitlength}%
\raisebox{\eqoff}{%
	\fmfframe(1,1)(-1,1){%
		\begin{fmfchar*}(7,7)
		\fmfforce{0 w,1 h}{v1}
		\fmfforce{1 w,0.5 h}{v2}
		\fmfforce{0 w,0 h}{v3}
		\fmfforce{0.2 w,0.8 h}{vc1}
		\fmfforce{0.72 w,0.5 h}{vc2}
		\fmfforce{0.2 w,0.2 h}{vc3}
		\fmf{plain,left=0.6}{vc1,vc2}
		\fmf{plain,left=0.6}{vc2,vc3}
		\fmf{plain,left=0.6}{vc3,vc1}
		\fmf{plain}{v1,vc1}
		\fmf{plain}{vc2,v2}
		\fmf{plain}{vc3,v3}
		\end{fmfchar*}
	}
}
\Bigr)_{\text{1PI}}
=
\complexi g_{\text{B}}^3\WR^{-1}
\int\frac{\de^d \bar l}{(2\pi)^d}\frac{1}{\bar l^2(\bar{p}_1-\bar l)^2(\bar{p}_2+\bar l)^2}\eqncom
\end{equation}
where the Minkowski space integral is given in terms of the Euclidean space integral, analogously to the self-energy case in \eqref{eq:example_div}. While we cannot solve this integral exactly, we can still determine its UV and IR behaviour. In $d>2$ dimensions it is IR convergent, since it scales as $\int \de \abs{\bar l} \abs{\bar l}^{d-3}$ for non-vanishing external momenta $\bar{p}_1\neq \bar{p}_2$. In the UV regime, where it scales as $\int \de \abs{\bar l} \abs{\bar l}^{d-7}$, we find a divergence for $d\geq 6$. In particular, in the six-dimensional case the divergence is independent of the external kinematic since the integral is logarithmically divergent for $d=6$. Hence, we can choose the special kinematical point $p_2=0$, where the integral \eqref{eq:V3_phi_bare} becomes
\begin{equation}\label{eq:V3_phi_UV_div}
\complexi V^{(1)}_{\varphi^3\text{B}}(p_1,0)=
\complexi g_{\text{B}}^3\WR^{-1}
\int\frac{\de^d \bar l}{(2\pi)^{\frac d2}}\frac{1}{\bar l^4(\bar{p}_1-\bar l)^2}
=
\complexi \frac{g_{\text{B}}^3}{(4\pi)^{\frac d2}}
\frac{\Gamma(3-\frac d2)\Gamma(\frac d2-2)\Gamma(\frac d2-1)}{\Gamma(3-d)}
\frac{1}{p_1^{2(3-\frac d2)}}
\eqncom
\end{equation}
to extract the UV divergence\footnote{In $d=6$ dimensions, this integral has no IR divergence which follows from power counting in the Euclidean integrand.} analytically.

We stop the discussion of bare correlation functions in $\varphi^3$-theory here, since we are only concerned with questions of renormalisation and all higher-point functions are UV convergent for the $(d=6)$ dimensional theory.

\subsection{The renormalised theory}\label{sec:the-renormalised-theory}
In the previous subsection, we found that the two- and three-point correlation functions in bare $\varphi^3$-theory in $d=6$ dimensional Minkowski space yield UV-divergent contributions. In this subsection, we introduce a renormalisation scheme consisting of a regularisation and a subtraction procedure to rewrite the action \eqref{eq:phi3_action} which depends on bare quantities in terms of local\footnote{The locality of counterterms and hence the renormalisation constants was proven in \cite{PhysRev.118.838}. See also \cite{hahn1968} for a refined version of the proof.} 1PI renormalisation constants that absorb the UV divergences and renormalised quantities. The renormalised theory, in contrast to the bare theory, is finite when the regulated theory is transformed back to the original one. Here, we discuss aspects of renormalisation schemes explicitly needed in this thesis and we refer to \appref{sec:Renormalisation_schemes} and the references therein for a detailed definition.

For the regularisation procedure, we choose to analytically continue the spacetime dimension, so that we work with a regulated theory that lives in $D=6-2\epsilon$ dimensions instead of $d=6$ dimensions. All former divergences then appear as poles in $\epsilon$ in the regularised expressions. This change of dimension also affects the coupling constant which picks up a mass dimension of $\Delta^0_g=\epsilon$ as can be seen from \eqref{eq:phi3_dimensions}. However, to keep the expansion parameter dimensionless, we introduce a renormalisation scale $\mu$ that absorbs the dimensional shift of the coupling in the regularised theory as
\begin{equation}
g_{\text{B}}\underset{d\rightarrow D}{\longrightarrow}\mu^{\frac 12(d-D)}g_{\text{B}}\eqndot
\end{equation}

In the subtraction procedure, for each quantity $q$ in the action \eqref{eq:phi3_action} we define a 1PI counterterm $\delta_q$ which enters the 1PI renormalisation constant $Z_q$ and absorbs the poles in $\epsilon$ that are associated to this quantity in the regularised theory. We express the bare scalar field, source, and coupling in \eqref{eq:phi3_action} in terms of renormalised quantities as
\begin{equation}\label{eq:phi_3_renorm}
\varphi_{\text{B}}(x)=Z^{\frac 12}_\varphi\varphi(x)\eqncom\qquad
j_{\text{B}}(x)=Z^{-\frac 12}_\varphi j(x)\eqncom\qquad
g_{\text{B}}=Z_gZ^{-\frac 32}_\varphi\mu^\epsilon g\eqncom
\end{equation}
where the 1PI renormalisation constants are defined in terms of the respective 1PI counterterms as\footnote{The signs in this definition are chosen such that all minimal subtraction 1PI counterterms $\delta_X$ are given by $(-1)$ times the sum of UV-divergent 1PI contributions that involve $X$, regardless if $X$ is a field or a coupling.}
\begin{equation}
Z_\varphi=1-\delta_\phi\eqncom\qquad
Z_g=1+\delta_g\eqncom\qquad \text{with}\qquad \delta=\sum_{j=1}^\infty\delta^{(j)}\eqncom
\end{equation}
with $\delta^{(j)}$ depending on $g^j$. The counterterms and the renormalisation constants are expressed in terms of the renormalised coupling $g$ and hence they implicitly depend on the renormalisation scale $\mu$. The coefficients in each counterterm are determined perturbatively to a given order in $g$ so that observables in the regularised theory are finite when the limit to the renormalised theory with $\epsilon= 0$ is taken.

Let us now investigate the renormalised theory. Upon inserting  \eqref{eq:phi_3_renorm} into the action \eqref{eq:phi3_action}, we find the regularised action
\begin{equation}\label{eq:phi3_action_ren}
S_0[\varphi,j]=\int\de^D x\Bigl(-\frac 12 Z_\varphi
\partial_\mu\varphi(x)\partial^\mu\varphi(x)
+j(x)\varphi(x)
\Bigr)
\eqncom\quad
S_{\text{int}}[\varphi,g]=
\int\de^D x \frac{\mu^\epsilon g}{3!}Z_g\varphi^3(x)\eqncom
\end{equation}
in which the free field and the interaction vertex are renormalised by 1PI renormalisation constants. Note that the fields absorb the dimensional shift in the free action such that the partial derivatives still have dimension one. In the source term, 
the renormalisation constants cancel, so that we can use \eqref{eq:Zj_perturbative} as a generating functional for renormalised $\varphi^3$-theory in which all UV divergences are absent. Indeed, when we calculate the first correction to the renormalised two-point function, we find an additional counterterm contribution that must absorb the occurring divergence in \eqref{eq:first_loop_2pt}. Graphically, we depict the corresponding counterterm vertex as
\begin{equation}
\complexi p^2\delta_\varphi=
\complexi \Delta^{-1}(p)\delta_\varphi=
\Bigl(
\settoheight{\eqoff}{$\times$}%
\setlength{\eqoff}{0.5\eqoff}%
\addtolength{\eqoff}{-3.5\unitlength}%
\raisebox{\eqoff}{%
	\fmfframe(1,1)(-1,1){%
		\begin{fmfchar*}(5,5)
		\fmfforce{0 w,0.5 h}{v1}
		\fmfforce{1 w,0.5 h}{v2}
		\fmfforce{0.5w,0.5 h}{v3}
		\fmf{plain}{v1,v2}
		\fmfv{decor.shape=hexacross,decor.size=7thin}{v3}
		\end{fmfchar*}
	}
}
\Bigr)_{\text{1PI}}
\end{equation}
and with it the renormalised version of \eqref{eq:first_loop_2pt} becomes
\begin{equation}
\begin{aligned}\label{eq:first_loop_2pt_ren}
\vacl\T\varphi(p)\varphi(-p)\vac_{\complexi\cT}&=
Z_{\varphi}^{-1}\vacl\T\varphi_{\text{B}}(p)\varphi_{\text{B}}(-p)\vac_{\complexi\cT}
=
\settoheight{\eqoff}{$\times$}%
\setlength{\eqoff}{0.5\eqoff}%
\addtolength{\eqoff}{-3.5\unitlength}%
\raisebox{\eqoff}{%
	\fmfframe(1,1)(-1,1){%
		\begin{fmfchar*}(5,5)
		\fmfforce{0 w,0.5 h}{v1}
		\fmfforce{1 w,0.5 h}{v2}
		\fmf{plain}{v1,v2}
		\end{fmfchar*}
	}
}	
+
\settoheight{\eqoff}{$\times$}%
\setlength{\eqoff}{0.5\eqoff}%
\addtolength{\eqoff}{-3.5\unitlength}%
\raisebox{\eqoff}{%
	\fmfframe(1,1)(-1,1){%
		\begin{fmfchar*}(10,5)
		\fmfleft{v1}
		\fmfright{v2}
		\fmffixed{(0.5w,0)}{vc1,vc2}
		\fmf{plain}{v1,vc1}
		\fmf{plain}{vc2,v2}
		\fmf{plain,left=1}{vc1,vc2}
		\fmf{plain,left=1}{vc2,vc1}
		\end{fmfchar*}
	}
}
+
\settoheight{\eqoff}{$\times$}%
\setlength{\eqoff}{0.5\eqoff}%
\addtolength{\eqoff}{-3.5\unitlength}%
\raisebox{\eqoff}{%
	\fmfframe(1,1)(-1,1){%
		\begin{fmfchar*}(5,5)
		\fmfforce{0 w,0.5 h}{v1}
		\fmfforce{1 w,0.5 h}{v2}
		\fmfforce{0.5w,0.5 h}{v3}
		\fmf{plain}{v1,v2}
		\fmfv{decor.shape=hexacross,decor.size=7thin}{v3}
		\end{fmfchar*}
	}
}
+\order{g^4}
\end{aligned}
\end{equation}
where we expressed all bare quantities on the \rhs in terms of renormalised ones using \eqref{eq:phi_3_renorm}. Extracting the one-loop 1PI contributions, we find the renormalised one-loop self-energy in the regulated theory to be
\begin{equation}
\begin{aligned}
\complexi\Pi^{(1)}(p)&=
\Bigl(\settoheight{\eqoff}{$\times$}%
\setlength{\eqoff}{0.5\eqoff}%
\addtolength{\eqoff}{-3.5\unitlength}%
\raisebox{\eqoff}{%
	\fmfframe(1,1)(-1,1){%
		\begin{fmfchar*}(10,5)
		\fmfleft{v1}
		\fmfright{v2}
		\fmffixed{(0.5w,0)}{vc1,vc2}
		\fmf{plain}{v1,vc1}
		\fmf{plain}{vc2,v2}
		\fmf{plain,left=1}{vc1,vc2}
		\fmf{plain,left=1}{vc2,vc1}
		\end{fmfchar*}
	}
}
+
\settoheight{\eqoff}{$\times$}%
\setlength{\eqoff}{0.5\eqoff}%
\addtolength{\eqoff}{-3.5\unitlength}%
\raisebox{\eqoff}{%
	\fmfframe(1,1)(-1,1){%
		\begin{fmfchar*}(5,5)
		\fmfforce{0 w,0.5 h}{v1}
		\fmfforce{1 w,0.5 h}{v2}
		\fmfforce{0.5w,0.5 h}{v3}
		\fmf{plain}{v1,v2}
		\fmfv{decor.shape=hexacross,decor.size=7thin}{v3}
		\end{fmfchar*}
	}
}
\Bigr)_{\text{1PI}}
&=\complexi p^2\left(
\frac{1}{12}\frac{g^2}{(4\pi)^3}\left(-\frac{1}{\epsilon} 
-\frac 83+\log \frac{\e^{\gammaE} p^2}{4\pi\mu^2}
\right)
+\delta_\varphi^{(2)}\right)
+\order{\epsilon}
\eqndot
\end{aligned}
\end{equation}
This expression must be finite in the limit of vanishing regulator $\epsilon\rightarrow0$ which is guaranteed if the counterterm takes the form
\begin{equation}\label{eq:delta_varphi_1}
\delta_\varphi^{(2)}=\frac{-1}{\complexi p^2}\Kop\Bigl[
\bigl(
\settoheight{\eqoff}{$\times$}%
\setlength{\eqoff}{0.5\eqoff}%
\addtolength{\eqoff}{-3.5\unitlength}%
\raisebox{\eqoff}{%
	\fmfframe(1,1)(-1,1){%
		\begin{fmfchar*}(10,5)
		\fmfleft{v1}
		\fmfright{v2}
		\fmffixed{(0.5w,0)}{vc1,vc2}
		\fmf{plain}{v1,vc1}
		\fmf{plain}{vc2,v2}
		\fmf{plain,left=1}{vc1,vc2}
		\fmf{plain,left=1}{vc2,vc1}
		\end{fmfchar*}
	}
}
\bigr)_{\text{1PI}}\Bigr]
+\frac{g^2}{(4\pi)^3}c_{\text{s}}
=
\frac{g^2}{(4\pi)^3}\left(\frac{1}{12\epsilon}+c_{\text{s}}\right)+\order{\epsilon}\eqncom
\end{equation}
where the operator $\Kop$ extracts the divergence in $\epsilon$ of its regularised argument and the scheme constant $c_{\text{s}}$ is finite in $\epsilon$. The renormalisation scheme with a dimensional regularisation and $c_{\text{s}}=0$ is called the minimal subtraction (MS). We are, however, free to also absorb finite parts into the counterterms as long as these additional terms do not spoil any symmetries of the original theory. In the modified minimal subtraction ($\ol{\text{MS}}$) scheme also the factor $c_{\text{s}}=\frac{-c_{\ol{\text{MS}}}}{12}=\frac{-1}{12}\log\frac{\e^{\gammaE}}{4\pi}$ is absorbed and in kinematical subtraction\footnote{This renormalisation scheme is also known as momentum subtraction (MOM) scheme.} (KS) the counterterm is fixed by requiring that observables do not receive quantum corrections at a special kinematical point: here $c_{\text{s}}=\frac{1}{36}(8-3c_{\ol{\text{MS}}})$, which yields $\Pi^{(1)}(\mu)=0$.

Turning to the three-point function, we can fix the 1PI counterterm $\delta_g$ from the UV-divergent 1PI contribution \eqref{eq:V3_phi_bare}. In the renormalised theory, we have an additional 1PI contribution to \eqref{eq:V3_phi_bare} at order $g^3$ from the counterterm, which we depict as
\begin{equation}
\complexi g\mu^\epsilon\delta_g =
\Bigl(
\settoheight{\eqoff}{$\times$}%
\setlength{\eqoff}{0.5\eqoff}%
\addtolength{\eqoff}{-3.5\unitlength}%
\raisebox{\eqoff}{%
	\fmfframe(1,1)(-1,1){%
		\begin{fmfchar*}(5,5)
		\fmfforce{0 w,1 h}{v1}
		\fmfforce{1 w,0.5 h}{v2}
		\fmfforce{0 w,0 h}{v3}
		\fmf{plain}{v1,vc}
		\fmf{plain}{vc,v2}
		\fmf{plain}{v3,vc}
		\fmfv{decor.shape=hexacross,decor.size=7thin}{vc}
		\end{fmfchar*}
	}
}
\Bigr)_{\text{1PI}}\eqndot
\end{equation}
With this, the renormalised one-loop 1PI vertex function is given by
\begin{equation}
\begin{aligned}\label{eq:renormalised_phi3_vertex}
\complexi V^{(1)}_{\varphi^3}(p_1,p_2)=
\Bigl(
\settoheight{\eqoff}{$\times$}%
\setlength{\eqoff}{0.5\eqoff}%
\addtolength{\eqoff}{-4.5\unitlength}%
\raisebox{\eqoff}{%
	\fmfframe(1,1)(-1,1){%
		\begin{fmfchar*}(7,7)
		\fmfforce{0 w,1 h}{v1}
		\fmfforce{1 w,0.5 h}{v2}
		\fmfforce{0 w,0 h}{v3}
		\fmfforce{0.2 w,0.8 h}{vc1}
		\fmfforce{0.72 w,0.5 h}{vc2}
		\fmfforce{0.2 w,0.2 h}{vc3}
		\fmf{plain,left=0.6}{vc1,vc2}
		\fmf{plain,left=0.6}{vc2,vc3}
		\fmf{plain,left=0.6}{vc3,vc1}
		\fmf{plain}{v1,vc1}
		\fmf{plain}{vc2,v2}
		\fmf{plain}{vc3,v3}
		\end{fmfchar*}
	}
}
+
\settoheight{\eqoff}{$\times$}%
\setlength{\eqoff}{0.5\eqoff}%
\addtolength{\eqoff}{-3.5\unitlength}%
\raisebox{\eqoff}{%
	\fmfframe(1,1)(-1,1){%
		\begin{fmfchar*}(5,5)
		\fmfforce{0 w,1 h}{v1}
		\fmfforce{1 w,0.5 h}{v2}
		\fmfforce{0 w,0 h}{v3}
		\fmf{plain}{v1,vc}
		\fmf{plain}{vc,v2}
		\fmf{plain}{v3,vc}
		\fmfv{decor.shape=hexacross,decor.size=7thin}{vc}
		\end{fmfchar*}
	}
}
\Bigr)_{\text{1PI}}
=
\complexi g\mu^\epsilon\Bigl(g^2\WR^{-1}
\int\frac{\de^d \bar l}{(2\pi)^d}\frac{\mu^{2\epsilon}}{\bar l^2(\bar{p}_1-\bar l)^2(\bar{p}_2+\bar l)^2}
+\delta_g\Bigr)\eqndot
\end{aligned}
\end{equation}
We can fix the one-loop value $\delta_g^{(2)}$ in the MS and $\ol{\text{MS}}$ schemes\footnote{In the KS scheme, the vertex is renormalised at a non-exeptional momentum configuration, e.g.\ the symmetric renormalisation point where $V^{(1)}_{\varphi^3}(\mu,\mu)=0$ is enforced.} by requiring that the renormalised one-loop 1PI vertex function evaluated at the special kinematical point $p_2=0$ is free of UV divergences. Using \eqref{eq:V3_phi_UV_div}, this yields
\begin{equation}
\begin{aligned}\label{eq:delta_g_1}
\delta_g^{(2)}=\frac{-1}{\complexi g\mu^\epsilon}\Kop\Bigl[
\Bigl(
\settoheight{\eqoff}{$\times$}%
\setlength{\eqoff}{0.5\eqoff}%
\addtolength{\eqoff}{-4.5\unitlength}%
\raisebox{\eqoff}{%
	\fmfframe(1,1)(-1,1){%
		\begin{fmfchar*}(7,7)
		\fmfforce{0 w,1 h}{v1}
		\fmfforce{1 w,0.5 h}{v2}
		\fmfforce{0 w,0 h}{v3}
		\fmfforce{0.2 w,0.8 h}{vc1}
		\fmfforce{0.72 w,0.5 h}{vc2}
		\fmfforce{0.2 w,0.2 h}{vc3}
		\fmf{plain,left=0.6}{vc1,vc2}
		\fmf{plain,left=0.6}{vc2,vc3}
		\fmf{plain,left=0.6}{vc3,vc1}
		\fmf{plain}{v1,vc1}
		\fmf{plain}{vc2,v2}
		\fmf{plain}{vc3,v3}
		\end{fmfchar*}
	}
}
\Bigr)_{\text{1PI}}
\Bigr]
+
\frac{g^2}{(4\pi)^3}c_{\text{s}}
=\frac{g^2}{(4\pi)^3}\frac12\Bigl(-\frac{1}{\epsilon}+c_{\ol{\text{MS}}}\Bigr)\eqncom
\end{aligned}
\end{equation}
where the constant $c_{\ol{\text{MS}}}$ is only present in the $\ol{\text{MS}}$ scheme. The renormalised version of \eqref{eq:3pt_corr_phi} can now be expressed diagrammatically in analogy to renormalised two-point function \eqref{eq:first_loop_2pt_ren} in terms of unrenormalised graphs and counterterm insertions as
\begin{equation}
\begin{aligned}\label{eq:first_loop_3pt_ren}
\vacl\T\varphi(p_1)\varphi(p_2)\varphi(-p_1-p_2)\vac_{\complexi\cT}&=Z_\varphi^{-\frac 32}
\vacl\T\varphi_{\text{B}}(p_1)\varphi_{\text{B}}(p_2)\varphi_{\text{B}}(-p_1-p_2)\vac_{\complexi\cT}\\
&=
\settoheight{\eqoff}{$\times$}%
\setlength{\eqoff}{0.5\eqoff}%
\addtolength{\eqoff}{-4.5\unitlength}%
\raisebox{\eqoff}{%
	\fmfframe(1,1)(-1,1){%
		\begin{fmfchar*}(7,7)
		\fmfforce{0 w,1 h}{v1}
		\fmfforce{1 w,0.5 h}{v2}
		\fmfforce{0 w,0 h}{v3}
		\fmf{plain}{v1,vc}
		\fmf{plain}{vc,v2}
		\fmf{plain}{v3,vc}
		\end{fmfchar*}
	}
}
+
\settoheight{\eqoff}{$\times$}%
\setlength{\eqoff}{0.5\eqoff}%
\addtolength{\eqoff}{-4.5\unitlength}%
\raisebox{\eqoff}{%
	\fmfframe(1,1)(-1,1){%
		\begin{fmfchar*}(7,7)
		\fmfforce{0 w,1 h}{v1}
		\fmfforce{1 w,0.5 h}{v2}
		\fmfforce{0 w,0 h}{v3}
		\fmfforce{0.2 w,0.8 h}{vc1}
		\fmfforce{0.72 w,0.5 h}{vc2}
		\fmfforce{0.2 w,0.2 h}{vc3}
		\fmf{plain,left=0.6}{vc1,vc2}
		\fmf{plain,left=0.6}{vc2,vc3}
		\fmf{plain,left=0.6}{vc3,vc1}
		\fmf{plain}{v1,vc1}
		\fmf{plain}{vc2,v2}
		\fmf{plain}{vc3,v3}
		\end{fmfchar*}
	}
}
+
\settoheight{\eqoff}{$\times$}%
\setlength{\eqoff}{0.5\eqoff}%
\addtolength{\eqoff}{-4.5\unitlength}%
\raisebox{\eqoff}{%
	\fmfframe(1,1)(-1,1){%
		\begin{fmfchar*}(7,7)
		\fmfforce{0 w,1 h}{v1}
		\fmfforce{1 w,0.5 h}{v2}
		\fmfforce{0 w,0 h}{v3}
		\fmf{plain}{v1,vc}
		\fmf{plain}{vc,v2}
		\fmf{plain}{v3,vc}
		\fmfv{decor.shape=hexacross,decor.size=7thin}{vc}
		\end{fmfchar*}
	}
}
+
3\settoheight{\eqoff}{$\times$}%
\setlength{\eqoff}{0.5\eqoff}%
\addtolength{\eqoff}{-4.5\unitlength}%
\raisebox{\eqoff}{%
	\fmfframe(1,1)(-1,1){%
		\begin{fmfchar*}(10,7)
		\fmfforce{0 w,1 h}{v1}
		\fmfforce{1 w,0.5 h}{v2}
		\fmfforce{0 w,0 h}{v3}
		\fmfforce{0.3w, 0.5h}{vcl}
		\fmfforce{0.65w, 0.5h}{vcr}
		\fmf{plain}{v1,vcl,v2}
		\fmf{plain}{vcl,v3}
		\fmfv{d.sh=circle,d.f=empty,d.si=.4w}{vcr}
		\end{fmfchar*}
	}
}
+3
\raisebox{\eqoff}{%
	\fmfframe(1,1)(-1,1){%
		\begin{fmfchar*}(10,7)
		\fmfforce{0 w,1 h}{v1}
		\fmfforce{1 w,0.5 h}{v2}
		\fmfforce{0 w,0 h}{v3}
		\fmfforce{0.3w, 0.5h}{vcl}
		\fmfforce{0.65w, 0.5h}{vcr}
		\fmf{plain}{v1,vcl,v2}
		\fmf{plain}{vcl,v3}
		\fmfv{decor.shape=hexacross,decor.size=7thin}{vcr}
		\end{fmfchar*}
	}
}
+\order{g^5}\eqndot
\end{aligned}
\end{equation}

\subsection{Renormalisation group equation}\label{sec:renormalisation-group-equation}
In the renormalisation program via dimensional regularisation, we have introduced an arbitrary renormalisation scale $\mu$ in the previous subsection. However, quantities in the bare theory do not depend on this scale and we can use this fact to determine the $\mu$-dependence of renormalised quantities. This analysis leads to the renormalisation group equation (\RGE), to the anomalous dimensions $\gamma$ of fields and the $\beta$-functions of couplings.

Starting with the scalar field, its \RGE can be obtained by varying the first equation in \eqref{eq:phi_3_renorm} with respect to $\mu$. We find
\begin{equation}\label{eq:renormalised_field_mu_dep}
0=\mu\frac{\de}{\de\mu}\varphi_{\text{B}}(x)=Z_\varphi\left[\gamma_\varphi+\mu\frac{\de}{\de\mu}\right]\varphi(x)\eqncom
\qquad
\text{with}
\qquad 
\mu\frac{\de }{\de \mu}Z_\varphi=2\gamma_\varphi Z_\varphi\eqncom
\end{equation}
where the anomalous dimension $\gamma_\varphi$ measures the degree of homogeneity\footnote{In \cite{BROWN1980135}, it was shown for a renormalisable scalar field theory which is dimensionally regularised that the renormalisation constants are homogeneous functions in $\mu$.}  of the renormalisation constant $Z_\varphi$. The anomalous dimension appear since the scaling behaviour of $Z_\varphi$ affects the scaling behaviour of the renormalised field. Under a scaling transformation $\mu\rightarrow \e^{-\alpha}\mu$ we have $\widehat{Z}_\varphi=\e^{-2\gamma_\varphi\alpha}Z_\varphi$ and using the first equation in \eqref{eq:phi_3_renorm}, this implies 
\begin{equation}\label{eq:scalar_scaling}
\widehat{\varphi}(x)=
\widehat{Z}^{-\frac 12}_\varphi(\mu)\widehat{\varphi_{\text{B}}}(x)=\e^{\alpha\gamma_\varphi}Z^{\frac 12}_\varphi(\mu)\e^{\alpha\Delta_\varphi^0}\varphi_{\text{B}}(\e^\alpha x)=\e^{\alpha(\Delta_\varphi^0+\gamma_\varphi)}\varphi(\e^\alpha x)\eqncom
\end{equation}
where we used \eqref{eq:symmetry_trafo_elementary_field} for the rescaling of the bare scalar field. Hence, the classical scaling dimension of the scalar field $\Delta_\varphi^0$ is shifted by the anomalous piece $\gamma_\varphi$ in the renormalised theory.\footnote{For a detailed discussion how $\gamma_\varphi$ appears in the position space two-point function see e.g.\ \cite[\chap{10}]{kleinert2001critical}.} We find the one-loop coefficient of $\gamma_\varphi$ from the one-loop counterterm \eqref{eq:delta_varphi_1} and it is given by
\begin{equation}
\gamma_\varphi=\frac 12\frac{1}{Z_\varphi}\mu\frac{\de }{\de \mu}Z_\varphi=-\frac 12\mu\frac{\de }{\de \mu}(\delta_\varphi^{(2)})+\order{g^4}=\frac{1}{12}\frac{g^2}{(4\pi)^3}(1+\epsilon c_{\text{s}})+\order{g^4}\eqncom
\end{equation}
where we used \eqref{eq:phi_3_renorm} to take the derivative with respect to $\mu$. We see that the renormalisation scheme constant $c_{\text{s}}$ which was discussed after \eqref{eq:delta_varphi_1} vanishes at first loop order in the $\epsilon\rightarrow 0$ limit. This is, however, an artefact from the low loop order and from two-loop order onwards we find that the anomalous dimensions become renormalisation-scheme dependent in non-conformal theories.

Similarly to the renormalised field \RGE, we find the \RGE for the renormalised coupling from \eqref{eq:phi_3_renorm}. For this, we introduce the connected or complete coupling renormalisation constant which absorbs all renormalisation constants connected to the coupling renormalisation in \eqref{eq:phi_3_renorm} into one constant as
\begin{equation}
\cZ_g=1+\mathfrak{d}_g=Z_gZ_\varphi^{-\frac 32}=(1+\delta_g)(1-\delta_\varphi)^{-\frac 32}\eqncom 
\end{equation}
where we also introduced the connected counterterm-like object $\mathfrak{d}_g$. The \RGE for the coupling $g$ is now given by
\begin{equation}\label{eq:renormalised_coupling_mu_dep}
0=\mu\frac{\de}{\de\mu}g_{\text{B}}=\mathcal{Z}_g\mu^\epsilon\left[-\gamma_g+\epsilon+\mu\frac{\de}{\de\mu}\right]g\eqncom
\qquad
\text{with}
\qquad 
\mu\frac{\de }{\de \mu}\mathcal{Z}_g=-\gamma_g\cZ_g\eqncom
\end{equation}
where the complete renormalisation constant of $g$ is also a homogeneous function in $\mu$ with degree $\gamma_g$. Typically, this equation is written in terms of the coupling's $\beta$-function which determines the rate at which the renormalised coupling varies when the renormalisation scale is varied. The relation between the coupling's anomalous dimension and its $\beta$-function is
\begin{equation}\label{eq:beta_function}
\beta_g=\mu\frac{\de}{\de \mu}g= (\gamma_g-\epsilon)g\eqndot
\end{equation}
Hence, the one-loop $\beta$-function can be determined from $\gamma_g^{(2)}$ and using \eqref{eq:delta_varphi_1} and \eqref{eq:delta_g_1} we find\footnote{Our one-loop coefficient $\beta_g^{(2)}$ differs by a factor of $\tfrac 12$ from \cite[\chap{28}]{Srednicki:2007} since we calculate the $\beta$-function for the coupling $g$ and not its square $\frac{g^2}{(4\pi)^3}=\alpha$.}
\begin{equation}
\gamma_g= -\frac{1}{\cZ_g}\mu \frac{\de}{\de \mu}\cZ_g=-\frac{\de}{\de \mu}(\delta_g^{(2)}+\frac 32 \delta_\varphi^{(2)})+\order{g^4}
=-\frac 34\frac{g^2}{(4\pi)^3}(1-\epsilon c_{\ol{\text{MS}}})+\order{g^4}\eqncom
\end{equation}
where we have only given the MS scheme result with $c_{\ol{\text{MS}}}=0$ and the $\ol{\text{MS}}$ scheme result with $c_{\ol{\text{MS}}}=\log\frac{\e^{\gammaE}}{4\pi}$. Like in the case of $\gamma_\varphi$, this anomalous dimension and hence the coupling's $\beta$-function become renormalisation-scheme-dependent at higher loop orders. We will see an explicit example of the renormalisation scheme dependence in the $\gamma_i$-deformation in \secref{sec:cake}.

Finally, we can calculate the \RGE of any $n$-point correlation function \eqref{eq:n_point_correlation_function_fourier} involving only elementary fields. As we already implicitly used in \eqref{eq:first_loop_2pt_ren} and \eqref{eq:first_loop_3pt_ren}, the bare $n$-point function is related to the renormalised one via 
\begin{equation}
\begin{aligned}\label{eq:G_bare_to_G_ren}
G^{(n)}_{\text{B}}(\vec{p})
&=
\left(\frac{\delta j(k_n)}{ \delta j_{\text{B}}(p_n)}\dots \frac{\delta j(k_1)}{ \delta j_{\text{B}}(p_1)}\right)
\left[\frac{\delta}{\complexi\delta j(k_n)}\dots \frac{\delta}{\complexi \delta  j(k_1)} Z[j]\right]_{j=0}
=
Z_\varphi^{\frac n2} G^{(n)}(\vec{p})\eqndot
\end{aligned}
\end{equation}
Taking the derivative of this expression with respect to the renormalisation scale $\mu$, we find the \RGE for $n$-point functions in $\varphi^3$-theory
\begin{equation}\label{eq:renormalisation_group_equation_interacting}
0=\left[n\gamma_\varphi+\mu\frac{\de}{\de \mu}\right]G^{(n)}(\vec{p})
=\left[n\gamma_\varphi+\beta_{g}\frac{\partial}{\partial g}+\mu\frac{\partial}{\partial \mu}\right]G^{(n)}(\vec{p})\eqncom
\end{equation}
where the $\beta$-function appears since the $n$-point function depends on all couplings in the interacting theory, compare \eqref{eq:n_point_correlation_function_fourier}. Note that $n$-point functions in non-conformal theories, like the $\varphi^3$-theory discussed here, inherit the renormalisation-scheme dependence from the $\beta$-functions of the couplings and the anomalous dimensions of the elementary fields.

\subsection{Composite operator insertions}\label{sec:composite-operator-insertions}
With the structure of correlation functions of elementary fields fixed, let us discuss how composite operators can be inserted in correlation functions and how they behave under renormalisation. For further details, see also \cite[\chap{6,7}]{Collins:1984xc} where the massive version of the examples discussed here are treated.

In fact, for the simplest operator we already discussed this problem. In the beginning of this section, we included a source $j_{\text{B}}(x)$ in the partition function and we used this source to generate single-field operators $\varphi_{\text{B}}(x)$ in correlation functions. Analogously, we can generate bare composite operators by including the following source action in the partition function \eqref{eq:Zj_perturbative}:
\begin{equation}\label{eq:general_sources}
S_{\text{source}}[\{\cO_{\text{B}}\},\{j_{\text{B}}\}]=(2\pi)^{d}\int\de^d x\, 
\sum_{A}\sum_nj^{(A)n}_{\text{B}}(x)\mathcal{O}_{\text{B}\,n}^{(A)}(x)\eqncom
\end{equation}
where each composite operator is identified by its quantum numbers $A$ and an index $n$ that labels all distinct operators with the same quantum numbers. The classical scaling dimension of the composite operator $\Delta^0_{\cO^{(A)}}$ is given in \eqref{eq:D_cl_on_O} by the sum of its constituent-field scaling dimensions and to render the source action dimensionless, the classical scaling dimension of the corresponding source must take the form
\begin{equation}\label{eq:dimension_source}
\Delta^0_{j^{(A)}}=d-\Delta^0_{\cO^{(A)}}\eqndot
\end{equation}

With the source action \eqref{eq:general_sources}, the generalisation of \eqref{eq:n_point_correlation_function} to bare correlation functions that also contain bare composite operators is straightforwardly obtained by including the sources $j_{\text{B}\,n}^{(A)}(x)$ into the generating functional. For the momentum space realisation, note that the source action of a single length-$L$ composite operator $\cO_{\text{B}}(x)=\prod_{m=1}^{L}o_{\text{B}\,m}(x)$, built from elementary fields $o_{\text{B}\,m}$, in momentum space takes the form 
\begin{equation}
(2\pi)^{d}\int\de^d x\, j_{\text{B}}(x)\cO_{\text{B}}(x)=\int \left(\prod_{i=1}^L\frac{\de^d k_i}{(2\pi)^d}\right)\de^d p(2\pi)^d\delta^{(d)}(p+{\textstyle\sum\limits_{j=1}^L}k_j)j_{\text{B}}(p)\prod_{m=1}^Lo_{\text{B}\,m}(k_m)\eqndot
\end{equation}
To obtain the momentum space correlation functions \eqref{eq:n_point_correlation_function_fourier} with composite operators, we replace the elementary fields $o_{\text{B}\,m}(k_m)$ by variations with respect to their sources and follow the path integral approach of \secref{sec:path_integral_approach} for elementary fields and interactions, see also \appref{app:Feynman_rules} and the references therein for technical details. There is a slight complication for the momentum space representation: position space correlation functions of composite operators in the free theory already contain divergences\footnote{These divergences are absent in the free theory when we replace the multiple operators $\cO_i(x_i)$ at a coincident point $x_i\rightarrow x$ by a new composite operator which is built from all these operators as $\cO(x)=[:\prod_i\cO_i(x_i):]_{x_i= x}$, compare \subsecref{sec:Normal_ordering}.}, when multiple operators approach a coincident point. When we regulate these divergences by altering the spacetime dimension $d\rightarrow D=d-2\epsilon$, the Fourier transformation maps the divergences to poles in $\epsilon$ in the free momentum space correlation function. For example, let us calculate the free momentum space two-point function of the operator $\cO_{\text{B}}(p)=\frac 12\varphi_{\text{B}}^2(p)$ whose scaling dimension\footnote{The scaling dimension of the momentum space operator follows from the Fourier transformation of the respective position space operator which has scaling dimension $\Delta_{\cO(x)}=D-2$.} is $\Delta_{\cO(p)}=-2$. Separating off a momentum conserving $\delta$-distribution from the two-point function, we find the reduced two-point function in $D=6-2\epsilon$ dimensional Minkowski space
\begin{equation}\label{eq:free_2pt_comp_op}
\mu^{2\epsilon}\vacl\T \cO_{\text{B}}(p) \cO_{\text{B}}(-p)\vac_{\complexi \cT}\bigr|_{g_{\text{B}}=0} =
\mu^{2\epsilon}
\settoheight{\eqoff}{$\times$}%
\setlength{\eqoff}{0.5\eqoff}%
\addtolength{\eqoff}{-2.75\unitlength}%
\raisebox{\eqoff}{
	\fmfframe(0,0)(-1,0){
		\begin{fmfchar*}(10,5.0)
		\fmfforce{0 w, 0.5h}{v2}
		\fmfforce{1 w, 0.5h}{v1}
		\fmfv{decor.shape=circle,decor.filled=full,	decor.size=2thick}{v1,v2}
		\fmf{plain,left=0.6}{v1,v2,v1}
		\end{fmfchar*}
	}
}
=\frac{\complexi p^2}{2\complexi^2}\frac{G(1,1)}{(4\pi)^3}\Bigl(\frac{4\pi\mu^{2}}{p^{2}}\Bigr)^\epsilon
=\frac{p^2}{\complexi(4\pi)^3}\frac{-1}{12\epsilon}+\order{\epsilon^0}
\eqncom
\end{equation} 
where $\mu$ absorbs the non-integer dimensional shift from the separation of the $\delta$-distribution, the symmetry factor is $\frac 12$, the factor of $\frac{1}{\complexi^2}$ originates from the Feynman rules of the two propagators and the integral is evaluated using the same techniques as in \eqref{eq:example_div}. In this example, the divergence in $G(1,1)$ originates from the Fourier transformed position space correlation function $\vacl\T\cO_{\text{B}}(x)\cO_{\text{B}}(0)\vac\sim x^{-2(D-2)}$, given in \eqref{eq:2pt_func_classical}. In \appref{sec:the-fourier-transformation-of-the-free-two-point-function}, we calculate this Fourier transformation explicitly and find that its divergence at the origin matches that of \eqref{eq:free_2pt_comp_op} up to an overall normalisation. Following \cite[\chap{6}]{Collins:1984xc}, we can absorb the occurring divergence into a new local MS counterterm $c^{(0)}
=-\Kop\bigl[\mu^{2\epsilon}
\settoheight{\eqoff}{$\times$}%
\setlength{\eqoff}{0.5\eqoff}%
\addtolength{\eqoff}{-1.2\unitlength}%
\raisebox{\eqoff}{
	\fmfframe(0,0)(-1,0){
		\begin{fmfchar*}(5,2.5)
		\fmfforce{0 w, 0.5h}{v2}
		\fmfforce{1 w, 0.5h}{v1}
		\fmfv{decor.shape=circle,decor.filled=full,	decor.size=2thick}{v1,v2}
		\fmf{plain,left=0.6}{v1,v2,v1}
		\end{fmfchar*}
	}
}
\bigr]$ and define a renormalised two-point correlation function of composite operators as
\begin{equation}\label{eq:free_2pt_function_ren}
\bigl[\mu^{2\epsilon}\vacl\T \cO_{\text{B}}(p) \cO_{\text{B}}(-p)\vac_{\complexi \cT}|_{g_{\text{B}}=0}\bigr]_{\text{R}}=
\mu^{2\epsilon}\vacl\T \cO_{\text{B}}(p) \cO_{\text{B}}(-p)\vac_{\complexi \cT}|_{g_{\text{B}}=0}
+p^2c^{(0)}\vacl\one\vac\eqndot
\end{equation}
In the interacting theory, the local counterterm must be determined perturbatively for each loop order and we have $C(g)=\sum_{j=0}^\infty c^{(j)}(g)$.

\subsubsection{The renormalisation of composite operators}
When we turn to the interacting theory, we find yet more UV divergences in the correlation functions of composite operators which are similar to the ones that we encountered in the context of elementary fields and interactions in \subsecref{sec:the-renormalised-theory}. This is unsurprising since we can think of composite operators as additional interactions with couplings $j_{\text{B}\,n}^{(A)}(x)$ that are set to zero when the correlation functions are evaluated. Hence, we can introduce complete renormalisation constants for the sources of composite operators and mimic the coupling renormalisation of \subsecref{sec:the-renormalised-theory}. There is, however, an important difference in the renormalisation of composite operators: different operators with the same quantum numbers can be mixed into each other under renormalisation.\footnote{In \cite[\chap{6}]{Collins:1984xc} it was proven that operators with dimension $\Delta$ can be renormalised by adding operators with a maximal dimension of $\Delta$.} Therefore, for each $j_{\text{B}}^{(A)}(x)$ in \eqref{eq:general_sources} we must introduce a renormalisation matrix that absorbs all UV divergences associated to the composite operators with quantum numbers $A$. We explicitly choose the renormalisation matrix realisation
\begin{equation}\label{eq:renormalised_operators}
\mathcal{O}_{\text{B}\,m}^{(A)}(x)=\sum_n\mu^{s_\cO(m,n)}(\cZ^{-1}_{A})_m^{\phan{a}n}\mathcal{O}_n^{(A)}(x)\eqncom\qquad
j_{\text{B}}^{(A)m}(x)=\sum_n\mu^{s_j(m,n)}j^{(A)n}(x)(\cZ_{A})_n^{\phan{a}m}\eqncom
\end{equation}
where the function $s_\cO(m,n)=\Delta_{\cO^{(A)m}}-\Delta_{\cO^{(A)n}}$ accounts for the scaling dimension difference\footnote{In the original theory in $d$ dimensions the scaling dimension difference is $s_\cO(m,n)=0$, by construction.} of the mixing objects in the analytically continued $D$-dimensional theory. The complete renormalisation constant is given in terms of 1PI renormalisation constants of the operator $\mathcal{O}^{(A)}_n(x)=\prod_{i=1}^{L(n)}o_i(x)$ and its elementary constituent fields $o_i$ as
\begin{equation}\label{eq:comp_op_renormalisation}
(\cZ_{A})_n^{\phan{a}m}=(\one+\mathfrak{d}_{A})_n^{\phan{a}m}=\frac{(Z_{A})_n^{\phan{a}m}}{\prod_{i=1}^{L(n)}Z_{o_i}^{\frac 12}}
=\frac{(\one+\delta_{A})_n^{\phan{a}m}}{\prod_{i=1}^{L(n)}(1-\delta_{o_i})^{\frac 12}}
\eqncom\qquad
(\delta_{A})_n^{\phan{a}m}=\sum_{j=1}^\infty (\delta_{A}^{(j)})_n^{\phan{a}m}\eqncom
\end{equation}
where $(\delta_{A}^{(j)})_n^{\phan{a}m}$ depends on $g^j$. Note that the complete renormalisation matrix is of the form $\cZ_{A}=\one+\order{g}$ and we can use this to explicitly construct its inverse in \eqref{eq:renormalised_operators} as $\cZ^{-1}_A=\sum_{j=0}^\infty (-\mathfrak{d}_A)^j$.
In analogy to the situation in \subsecref{sec:the-renormalised-theory}, the bare quantities on each \lhs in \eqref{eq:renormalised_operators} are independent of the renormalisation scale $\mu$, while the renormalisation constants as well as the renormalised quantities on each \rhs depend on it.
When we insert \eqref{eq:renormalised_operators} into \eqref{eq:general_sources}, the complete renormalisation matrices cancel and we can use this source action in \eqref{eq:Zj_perturbative} to generate renormalised correlation functions involving renormalised composite operators. Like in the case of the two- and three-point function of elementary fields, we fix the new 1PI counterterm matrix $(\delta_{A})_n^{\phan{a}m}$ to absorb the divergences that occur in Feynman diagrams of bare quantities. 

To illustrate the renormalisation of composite operators, we follow \cite[\chap{6}]{Collins:1984xc} and calculate the one-loop 1PI counterterms that renormalise the operator $\cO^{(4)}_{\text{B}\,1}(x)=\frac 12\varphi_{\text{B}}^2(x)$ with classical scaling dimension $\Delta^0=4$ in the six-dimensional $\varphi^3$-theory. In this spacetime dimension, there is one further operator with the same quantum numbers $\cO^{(4)}_{\text{B}\,2}(x)=\partial^2 \varphi_{\text{B}2}(x)$ which can mix into the renormalisation of $\cO^{(4)}_{\text{B}\,1}$. To identify the UV divergent contributions, we analytically continue the spacetime dimension to $D=6-2\epsilon$ which alters the classical scaling dimensions of both operators to
\begin{equation}
\Delta_{\cO^{(4)}_1}=D-2\eqncom\qquad 
\Delta_{\cO^{(4)}_2}=\frac 12(D+2)\eqndot
\end{equation}
In the $D$-dimensional theory, we can now calculate the one-loop 1PI renormalisation constants from the following reduced correlation function
\begin{equation}
\begin{aligned}\label{eq:phi2_renormalisation}
\vacl\T\varphi(-p)&\varphi(0)\cO^{(4)}_{1}(p)\vac_{\complexi \cT}=
\sum_{n=1}^2
\frac{\mu^{-2-\Delta_{\cO^{(4)}_n}}(\cZ_4)^{\phan{a}n}_1}{Z_\varphi}\vacl\T\varphi_{\text{B}}(-p)\varphi_{\text{B}}(0)\cO^{(4)}_{\text{B}\,n}(p)\vac_{\complexi \cT}
\\
&=
\frac{(Z_4)^{\phan{a}1}_1}{Z^2_\varphi}
\settoheight{\eqoff}{$\times$}%
\setlength{\eqoff}{0.5\eqoff}%
\addtolength{\eqoff}{-3.75\unitlength}%
\raisebox{\eqoff}{
	\fmfframe(0,0)(-1,0){
		\begin{fmfchar*}(7.5,7.0)
		\fmfright{vop}
		\fmfforce{0 w, .9h}{v1}
		\fmfforce{0 w, .1h}{v3}
		\fmf{plain,right=0.2,tension=0.6}{vop,v1}
		\fmf{plain,left=0.2,tension=0.6}{vop,v3} 
		\fmffreeze
		\fmfposition
		\fmfipath{p[]}
		\fmfiset{p1}{vpath(__vop,__v1)}
		\fmfiset{p11}{subpath 2(0,length(p1)/5) of p1}
		\fmfiset{p12}{subpath 3(length(p1)/5,length(p1)) of p1}
		\fmfdraw
		\fmfiv{decor.shape=circle,decor.filled=full,decor.size=5}{vloc(__vop)}
		\end{fmfchar*}
	}
}
+
\settoheight{\eqoff}{$\times$}%
\setlength{\eqoff}{0.5\eqoff}%
\addtolength{\eqoff}{-3.75\unitlength}%
\raisebox{\eqoff}{
	\fmfframe(0,0)(-1,0){
		\begin{fmfchar*}(7.5,7.0)
		\fmfright{vop}
		\fmfforce{0 w, .9h}{v1}
		\fmfforce{0 w, .1h}{v3}
		\fmf{plain,right=0.2,tension=0.6}{vop,v1}
		\fmf{plain,left=0.2,tension=0.6}{vop,v3} 
		\fmffreeze
		\fmfposition
		\fmfipath{p[]}
		\fmfiset{p1}{vpath(__vop,__v1)}
		\fmfiset{p11}{subpath 2(0,length(p1)/5) of p1}
		\fmfiset{p12}{subpath 3(length(p1)/5,length(p1)) of p1}
		\fmfiset{p2}{vpath(__vop,__v3)}
		\fmfiset{p21}{subpath 2(0,length(p2)/5) of p2}
		\fmfiset{p22}{subpath 3(length(p2)/5,length(p2)) of p2}
		\fmfi{plain}{point 3length(p1)/5 of p1 -- point 3length(p2)/5 of p2}
		\fmfdraw
		\fmfiv{decor.shape=circle,decor.filled=full,decor.size=5}{vloc(__vop)}
		\end{fmfchar*}
	}
}
+
2
\settoheight{\eqoff}{$\times$}%
\setlength{\eqoff}{0.5\eqoff}%
\addtolength{\eqoff}{-3.75\unitlength}%
\raisebox{\eqoff}{
	\fmfframe(0,0)(-1,0){
		\begin{fmfchar*}(7.5,7.0)
		\fmfright{vop}
		\fmfforce{0 w, .9h}{v1}
		\fmfforce{0 w, .1h}{v3}
		\fmf{plain,right=0.2,tension=0.6}{vop,v1}
		\fmf{plain,left=0.2,tension=0.6}{vop,v3} 
		\fmffreeze
		\fmfposition
		\fmfipath{p[]}
		\fmfiset{p1}{vpath(__vop,__v1)}
		\fmfiset{p11}{subpath 2(0,length(p1)/5) of p1}
		\fmfiset{p12}{subpath 3(length(p1)/5,length(p1)) of p1}
		\fmfiset{p2}{vpath(__vop,__v3)}
		\fmfiset{p21}{subpath 2(0,length(p2)/5) of p2}
		\fmfiset{p22}{subpath 3(length(p2)/5,length(p2)) of p2}
		\fmfdraw
		\fmfiv{decor.shape=circle,decor.size=7 thin,decor.filled=empty}{point length(p1)/2 of p1}
		\fmfiv{decor.shape=circle,decor.filled=full,decor.size=5}{vloc(__vop)}
		\end{fmfchar*}
	}
}
+
\mu^{\frac{D-6}{2}}\frac{(Z_4)^{\phan{a}2}_1}{Z^{\frac 32}_\varphi}
\settoheight{\eqoff}{$\times$}%
\setlength{\eqoff}{0.5\eqoff}%
\addtolength{\eqoff}{-3.75\unitlength}%
\raisebox{\eqoff}{
	\fmfframe(0,0)(-1,0){
		\begin{fmfchar*}(7.5,7.0)
		\fmfright{vop}
		\fmfforce{0 w, .9h}{v1}
		\fmfforce{0 w, .1h}{v3}
		\fmfforce{0.2w, 0.5h}{v4}
		\fmf{plain}{vop,v4}
		\fmf{plain}{v1,v4,v3}
		\fmffreeze
		\fmfposition
		\fmfdraw
		\fmfiv{decor.shape=circle,decor.filled=full,decor.size=5}{vloc(__vop)}
		\end{fmfchar*}
	}
}
+
\settoheight{\eqoff}{$\times$}%
\setlength{\eqoff}{0.5\eqoff}%
\addtolength{\eqoff}{-3.75\unitlength}%
\raisebox{\eqoff}{
	\fmfframe(0,0)(-1,0){
		\begin{fmfchar*}(7.5,7.0)
		\fmfright{vop}
		\fmfforce{0 w, .9h}{v1}
		\fmfforce{0 w, .1h}{v3}
		\fmfforce{0.2w, 0.5h}{v4}
		\fmfforce{0.5w, 0.5h}{v5}
		\fmf{plain,right}{vop,v5}
		\fmf{plain,left}{vop,v5} 
		\fmf{plain}{v4,v5}
		\fmf{plain}{v1,v4,v3}
		\fmffreeze
		\fmfposition
		\fmfdraw
		\fmfiv{decor.shape=circle,decor.filled=full,decor.size=5}{vloc(__vop)}
		\end{fmfchar*}
	}
}
+\order{g^3}
\\
&=
\settoheight{\eqoff}{$\times$}%
\setlength{\eqoff}{0.5\eqoff}%
\addtolength{\eqoff}{-3.75\unitlength}%
\raisebox{\eqoff}{
	\fmfframe(0,0)(-1,0){
		\begin{fmfchar*}(7.5,7.0)
		\fmfright{vop}
		\fmfforce{0 w, .9h}{v1}
		\fmfforce{0 w, .1h}{v3}
		\fmf{plain,right=0.2,tension=0.6}{vop,v1}
		\fmf{plain,left=0.2,tension=0.6}{vop,v3} 
		\fmffreeze
		\fmfposition
		\fmfipath{p[]}
		\fmfiset{p1}{vpath(__vop,__v1)}
		\fmfiset{p11}{subpath 2(0,length(p1)/5) of p1}
		\fmfiset{p12}{subpath 3(length(p1)/5,length(p1)) of p1}
		\fmfdraw
		\fmfiv{decor.shape=circle,decor.filled=full,decor.size=5}{vloc(__vop)}
		\end{fmfchar*}
	}
}
+
\mu^{\frac{D-6}{2}}
\settoheight{\eqoff}{$\times$}%
\setlength{\eqoff}{0.5\eqoff}%
\addtolength{\eqoff}{-3.75\unitlength}%
\raisebox{\eqoff}{
	\fmfframe(0,0)(-1,0){
		\begin{fmfchar*}(7.5,7.0)
		\fmfright{vop}
		\fmfforce{0 w, .9h}{v1}
		\fmfforce{0 w, .1h}{v3}
		\fmfforce{0.2w, 0.5h}{v4}
		\fmf{plain}{vop,v4}
		\fmf{plain}{v1,v4,v3}
		\fmffreeze
		\fmfposition
		\fmfdraw
		\fmfiv{decor.shape=circle,decor.filled=full,decor.size=5}{vloc(__vop)}
		\end{fmfchar*}
	}
}
+
\Bigl(
\settoheight{\eqoff}{$\times$}%
\setlength{\eqoff}{0.5\eqoff}%
\addtolength{\eqoff}{-3.75\unitlength}%
\raisebox{\eqoff}{
	\fmfframe(0,0)(-1,0){
		\begin{fmfchar*}(7.5,7.0)
		\fmfright{vop}
		\fmfforce{0 w, .9h}{v1}
		\fmfforce{0 w, .1h}{v3}
		\fmf{plain,right=0.2,tension=0.6}{vop,v1}
		\fmf{plain,left=0.2,tension=0.6}{vop,v3} 
		\fmffreeze
		\fmfposition
		\fmfipath{p[]}
		\fmfiset{p1}{vpath(__vop,__v1)}
		\fmfiset{p11}{subpath 2(0,length(p1)/5) of p1}
		\fmfiset{p12}{subpath 3(length(p1)/5,length(p1)) of p1}
		\fmfiset{p2}{vpath(__vop,__v3)}
		\fmfiset{p21}{subpath 2(0,length(p2)/5) of p2}
		\fmfiset{p22}{subpath 3(length(p2)/5,length(p2)) of p2}
		\fmfi{plain}{point 3length(p1)/5 of p1 -- point 3length(p2)/5 of p2}
		\fmfdraw
		\fmfiv{decor.shape=circle,decor.filled=full,decor.size=5}{vloc(__vop)}
		\end{fmfchar*}
	}
}
+
\settoheight{\eqoff}{$\times$}%
\setlength{\eqoff}{0.5\eqoff}%
\addtolength{\eqoff}{-3.75\unitlength}%
\raisebox{\eqoff}{
	\fmfframe(0,0)(-1,0){
		\begin{fmfchar*}(7.5,7.0)
		\fmfright{vop}
		\fmfforce{0 w, .9h}{v1}
		\fmfforce{0 w, .1h}{v3}
		\fmf{plain,right=0.2,tension=0.6}{vop,v1}
		\fmf{plain,left=0.2,tension=0.6}{vop,v3} 
		\fmffreeze
		\fmfposition
		\fmfipath{p[]}
		\fmfiset{p1}{vpath(__vop,__v1)}
		\fmfiset{p11}{subpath 2(0,length(p1)/5) of p1}
		\fmfiset{p12}{subpath 3(length(p1)/5,length(p1)) of p1}
		\fmfdraw
		\fmfiv{decor.shape=circle,decor.filled=full,decor.size=5}{vloc(__vop)}
		\fmfdraw
		\fmfiv{decor.shape=hexacross,decor.size=9thin}{vloc(__vop)}
		\end{fmfchar*}
	}
}
\Bigr)
\\&\phan{=}
+
2
\Bigl(
\settoheight{\eqoff}{$\times$}%
\setlength{\eqoff}{0.5\eqoff}%
\addtolength{\eqoff}{-3.75\unitlength}%
\raisebox{\eqoff}{
	\fmfframe(0,0)(-1,0){
		\begin{fmfchar*}(7.5,7.0)
		\fmfright{vop}
		\fmfforce{0 w, .9h}{v1}
		\fmfforce{0 w, .1h}{v3}
		\fmf{plain,right=0.2,tension=0.6}{vop,v1}
		\fmf{plain,left=0.2,tension=0.6}{vop,v3} 
		\fmffreeze
		\fmfposition
		\fmfipath{p[]}
		\fmfiset{p1}{vpath(__vop,__v1)}
		\fmfiset{p11}{subpath 2(0,length(p1)/5) of p1}
		\fmfiset{p12}{subpath 3(length(p1)/5,length(p1)) of p1}
		\fmfiset{p2}{vpath(__vop,__v3)}
		\fmfiset{p21}{subpath 2(0,length(p2)/5) of p2}
		\fmfiset{p22}{subpath 3(length(p2)/5,length(p2)) of p2}
		\fmfdraw
		\fmfiv{decor.shape=circle,decor.size=7 thin,decor.filled=empty}{point length(p1)/2 of p1}
		\fmfiv{decor.shape=circle,decor.filled=full,decor.size=5}{vloc(__vop)}
		\end{fmfchar*}
	}
}
+
\settoheight{\eqoff}{$\times$}%
\setlength{\eqoff}{0.5\eqoff}%
\addtolength{\eqoff}{-3.75\unitlength}%
\raisebox{\eqoff}{
	\fmfframe(0,0)(-1,0){
		\begin{fmfchar*}(7.5,7.0)
		\fmfright{vop}
		\fmfforce{0 w, .9h}{v1}
		\fmfforce{0 w, .1h}{v3}
		\fmf{plain,right=0.2,tension=0.6}{vop,v1}
		\fmf{plain,left=0.2,tension=0.6}{vop,v3} 
		\fmffreeze
		\fmfposition
		\fmfipath{p[]}
		\fmfiset{p1}{vpath(__vop,__v1)}
		\fmfiset{p11}{subpath 2(0,length(p1)/5) of p1}
		\fmfiset{p12}{subpath 3(length(p1)/5,length(p1)) of p1}
		\fmfdraw
		\fmfiv{decor.shape=circle,decor.filled=full,decor.size=5}{vloc(__vop)}
		\fmfdraw
		\fmfiv{decor.shape=hexacross,decor.size=7thin}{point 1length(p1)/2 of p1}
		\end{fmfchar*}
	}
}
\Bigr)
+
\Bigl(
\settoheight{\eqoff}{$\times$}%
\setlength{\eqoff}{0.5\eqoff}%
\addtolength{\eqoff}{-3.75\unitlength}%
\raisebox{\eqoff}{
	\fmfframe(0,0)(-1,0){
		\begin{fmfchar*}(7.5,7.0)
		\fmfright{vop}
		\fmfforce{0 w, .9h}{v1}
		\fmfforce{0 w, .1h}{v3}
		\fmfforce{0.2w, 0.5h}{v4}
		\fmfforce{0.5w, 0.5h}{v5}
		\fmf{plain,right}{vop,v5}
		\fmf{plain,left}{vop,v5} 
		\fmf{plain}{v4,v5}
		\fmf{plain}{v1,v4,v3}
		\fmffreeze
		\fmfposition
		\fmfdraw
		\fmfiv{decor.shape=circle,decor.filled=full,decor.size=5}{vloc(__vop)}
		\end{fmfchar*}
	}
}
+
\mu^{\frac{D-6}{2}}
\settoheight{\eqoff}{$\times$}%
\setlength{\eqoff}{0.5\eqoff}%
\addtolength{\eqoff}{-3.75\unitlength}%
\raisebox{\eqoff}{
	\fmfframe(0,0)(-1,0){
		\begin{fmfchar*}(7.5,7.0)
		\fmfright{vop}
		\fmfforce{0 w, .9h}{v1}
		\fmfforce{0 w, .1h}{v3}
		\fmfforce{0.2w, 0.5h}{v4}
		\fmf{plain}{vop,v4}
		\fmf{plain}{v1,v4,v3}
		\fmffreeze
		\fmfposition
		\fmfdraw
		\fmfiv{decor.shape=circle,decor.filled=full,decor.size=5}{vloc(__vop)}
		\fmfdraw
		\fmfiv{decor.shape=hexacross,decor.size=9thin}{vloc(__vop)}
		\end{fmfchar*}
	}
}
\Bigr)
+\order{g^3}
\eqncom
\end{aligned}
\end{equation}
where the factor of two accounts for the self-energy insertion on each external leg and the external momentum is fused into the composite operator and extracted at the upper scalar leg. In the final line, we have expanded the renormalisation constants to order $g^2$ and indicated 1PI counterterm vertices for composite operators $(\delta_4)_m^{\phan{a}n} \cO_{\text{B}\,n}^{(4)}$ by a hexa-cross on the composite operator insertion. We see that the correlation function can only be made finite by the means of 1PI counterterms if we include the length-$1$ operator $\cO^{(4)}_{\text{B}\,2}(x)$ and hence we find that this operator mixes into the renormalisation of the operator $\cO^{(4)}_{\text{B}\,1}(x)$. The one-loop counterterms that render \eqref{eq:phi2_renormalisation} finite in the MS scheme\footnote{The corresponding counterterms in the $\ol{\text{MS}}$ scheme can directly be obtained as in \eqref{eq:delta_varphi_1} and \eqref{eq:delta_g_1}.} are concretely given by
\begin{equation}
\begin{aligned}\label{eq:1PI_counterterms_example}
g^2(\delta_{4}^{(2)})_1^{\phan{a}1}
=-\Kop\Bigl[\Bigl(
\settoheight{\eqoff}{$\times$}%
\setlength{\eqoff}{0.5\eqoff}%
\addtolength{\eqoff}{-3.75\unitlength}%
\raisebox{\eqoff}{
	\fmfframe(0,2)(1,0){
		\begin{fmfchar*}(7.5,7.0)
		\fmfright{vop}
		\fmfforce{0 w, .9h}{v1}
		\fmfforce{0 w, .1h}{v3}
		\fmf{plain,right=0.2,tension=0.6}{vop,v1}
		\fmf{plain,left=0.2,tension=0.6}{vop,v3} 
		\fmffreeze
		\fmfposition
		\fmfipath{p[]}
		\fmfiset{p1}{vpath(__vop,__v1)}
		\fmfiset{p11}{subpath 2(0,length(p1)/5) of p1}
		\fmfiset{p12}{subpath 3(length(p1)/5,length(p1)) of p1}
		\fmfiset{p2}{vpath(__vop,__v3)}
		\fmfiset{p21}{subpath 2(0,length(p2)/5) of p2}
		\fmfiset{p22}{subpath 3(length(p2)/5,length(p2)) of p2}
		\fmfi{plain}{point 3length(p1)/5 of p1 -- point 3length(p2)/5 of p2}
		\fmfdraw
		\fmfiv{decor.shape=circle,decor.filled=full,decor.size=5}{vloc(__vop)}
		\end{fmfchar*}
	}
}
\Bigr)_{\text{1PI}}
\Bigr]
=\frac{-1}{2\epsilon}\frac{g^2}{(4\pi)^3}\eqncom\quad
g(\delta_{4}^{(1)})_1^{\phan{a}2}
=-\Kop\Bigl[\Bigl(
\settoheight{\eqoff}{$\times$}%
\setlength{\eqoff}{0.5\eqoff}%
\addtolength{\eqoff}{-3.75\unitlength}%
\raisebox{\eqoff}{
	\fmfframe(0,2)(1,0){
		\begin{fmfchar*}(7.5,7.0)
		\fmfright{vop}
		\fmfforce{0.0w, 0.5h}{v4}
		\fmfforce{0.4w, 0.5h}{v5}
		\fmf{plain,right=.8}{vop,v5}
		\fmf{plain,left=.8}{vop,v5} 
		\fmf{plain}{v4,v5}
		\fmffreeze
		\fmfposition
		\fmfdraw
		\fmfiv{decor.shape=circle,decor.filled=full,decor.size=5}{vloc(__vop)}
		\end{fmfchar*}
	}
}
\Bigr)_{\text{1PI}}
\Bigr]
=
\frac{1}{12\epsilon}\frac{g}{(4\pi)^3}\eqncom
\end{aligned}
\end{equation}
together with the self-energy couterterm given in \eqref{eq:delta_varphi_1}. With the renormalised operator we can straightforwardly calculate the one-loop contribution to \eqref{eq:free_2pt_comp_op}. It is given by the following graphs
\begin{equation}\label{eq:comp_op_phi3_g2}
\begin{aligned}
\mu^{2\epsilon}\vacl\T\cO^{(4)}_{1}(p)\cO^{(4)}_{1}(-p)\vac_{\complexi \cT}&=
\sum_{n,m=1}^2
\mu^{\epsilon(2-\delta_{m,2}+\delta_{n,2})}(\cZ_4)^{\,m}_1(\cZ_4)^{\,n}_1\vacl\T
\cO^{(4)}_{\text{B}\,m}(p)\cO^{(4)}_{\text{B}\,n}(-p)\vac_{\complexi \cT}
\\
&=
\mu^{2\epsilon}\Bigl(
\bigl((\cZ_4)^{\phan{a}1}_1\bigr)^2
\settoheight{\eqoff}{$\times$}%
\setlength{\eqoff}{0.5\eqoff}%
\addtolength{\eqoff}{-2.75\unitlength}%
\raisebox{\eqoff}{
	\fmfframe(0,0)(-1,0){
		\begin{fmfchar*}(10,5.0)
		\fmfforce{0 w, 0.5h}{v2}
		\fmfforce{1 w, 0.5h}{v1}
		\fmfv{decor.shape=circle,decor.filled=full,	decor.size=2thick}{v1,v2}
		\fmf{plain,left=0.6}{v1,v2,v1}
		\end{fmfchar*}
	}
}
+
\settoheight{\eqoff}{$\times$}%
\setlength{\eqoff}{0.5\eqoff}%
\addtolength{\eqoff}{-2.75\unitlength}%
\raisebox{\eqoff}{
	\fmfframe(0,0)(-1,0){
		\begin{fmfchar*}(10,5.0)
		\fmfforce{0 w, 0.5h}{v2}
		\fmfforce{1 w, 0.5h}{v1}
		\fmfv{decor.shape=circle,decor.filled=full,
			decor.size=2thick}{v1,v2}
		\fmf{phantom,left=0.6}{v1,v2,v1}
		\fmffreeze
		\fmfposition
		\fmfipath{p[]}
		\fmfipair{v[]}
		\fmfiset{p1}{vpath(__v1,__v2)}
		\fmfiset{p2}{vpath(__v2,__v1)}
		\fmfi{plain}{p1}
		\fmfi{plain}{p2}
		\fmfi{plain}{point length(p1)/2 of p1 -- point length(p2)/2 of p2}
		\end{fmfchar*}
	}
}
+
\settoheight{\eqoff}{$\times$}%
\setlength{\eqoff}{0.5\eqoff}%
\addtolength{\eqoff}{-2.75\unitlength}%
\raisebox{\eqoff}{
	\fmfframe(0,0)(-1,0){
		\begin{fmfchar*}(10,5.0)
		\fmfforce{0 w, 0.5h}{v2}
		\fmfforce{1 w, 0.5h}{v1}
		\fmfv{decor.shape=circle,decor.filled=full,
			decor.size=2thick}{v1,v2}
		\fmf{phantom,left=0.6}{v1,v2,v1}
		\fmffreeze
		\fmfposition
		\fmfipath{p[]}
		\fmfipair{v[]}
		\fmfiset{p1}{vpath(__v1,__v2)}
		\fmfiset{p2}{vpath(__v2,__v1)}
		\fmfi{plain}{p1}
		\fmfi{plain}{p2}
		\fmfiv{d.sh=circle,d.f=empty,d.si=.3w}{point length(p2)/2 of p2}
		\end{fmfchar*}
	}
}
\\
&\phan{{}={}\mu^{2\epsilon}\Bigl(}
+
\settoheight{\eqoff}{$\times$}%
\setlength{\eqoff}{0.5\eqoff}%
\addtolength{\eqoff}{-3.75\unitlength}%
\raisebox{\eqoff}{
	\fmfframe(0,0)(-1,0){
		\begin{fmfchar*}(7.5,7.0)
		\fmfright{vop}
		\fmfleft{v3}
		\fmfforce{0.4w, 0.5h}{v1}
		\fmfforce{0.6w, 0.5h}{v2}
		\fmf{plain,right}{vop,v2}
		\fmf{plain,left}{vop,v2} 
		\fmf{plain}{v1,v2}
		\fmf{plain,right}{v3,v1}
		\fmf{plain,left}{v3,v1}
		\fmffreeze
		\fmfposition
		\fmfdraw
		\fmfiv{decor.shape=circle,decor.filled=full,decor.size=5}{vloc(__vop)}
		\fmfdraw
		\fmfiv{decor.shape=circle,decor.filled=full,decor.size=5}{vloc(__v3)}
		\end{fmfchar*}
	}
}
+
\mu^{-2\epsilon}
\settoheight{\eqoff}{$\times$}%
\setlength{\eqoff}{0.5\eqoff}%
\addtolength{\eqoff}{-3.75\unitlength}%
\raisebox{\eqoff}{
	\fmfframe(0,0)(-1,0){
		\begin{fmfchar*}(7.5,7.0)
		\fmfright{vop}
		\fmfforce{0 w, .5h}{v1}
		\fmf{plain}{v1,vop}
		\fmffreeze
		\fmfposition
		\fmfdraw
		\fmfiv{decor.shape=circle,decor.filled=full,decor.size=5}{vloc(__vop)}
		\fmfiv{decor.shape=circle,decor.filled=full,decor.size=5}{vloc(__v1)}
		\fmfdraw
		\fmfiv{decor.shape=hexacross,decor.size=9thin}{vloc(__vop)}
		\fmfiv{decor.shape=hexacross,decor.size=9thin}{vloc(__v1)}
		\end{fmfchar*}
	}
}
+
2\mu^{-\epsilon}
\settoheight{\eqoff}{$\times$}%
\setlength{\eqoff}{0.5\eqoff}%
\addtolength{\eqoff}{-3.75\unitlength}%
\raisebox{\eqoff}{
	\fmfframe(0,0)(-1,0){
		\begin{fmfchar*}(7.5,7.0)
		\fmfright{vop}
		\fmfforce{0 w, .5h}{v1}
		\fmfforce{0.5w, 0.5h}{v5}
		\fmf{plain,right}{vop,v5}
		\fmf{plain,left}{vop,v5} 
		\fmf{plain}{v1,v5}
		\fmffreeze
		\fmfposition
		\fmfdraw
		\fmfiv{decor.shape=circle,decor.filled=full,decor.size=5}{vloc(__vop)}
		\fmfiv{decor.shape=circle,decor.filled=full,decor.size=5}{vloc(__v1)}
		\fmfdraw
		\fmfiv{decor.shape=hexacross,decor.size=9thin}{vloc(__v1)}
		\end{fmfchar*}
	}
}
\Bigr)
+\order{g^4}\\
&=\frac{p^2}{\complexi (4\pi)^3}\left[\frac{-1}{12\epsilon}+\frac{g^2}{144(4\pi)^3}\left(\frac{5}{\epsilon^2}-\frac{1}{3\epsilon}\right)\right]+\order{g^4}+\order{\epsilon^0}\eqncom
\end{aligned}
\end{equation}
which is free of UV subdivergences\footnote{The two-loop diagrams can be evaluated using the techniques of \cite{Chetyrkin:1981qh} and we find
\begin{equation}
\begin{aligned} 
\Kop\bigl[
\settoheight{\eqoff}{$\times$}%
\setlength{\eqoff}{0.5\eqoff}%
\addtolength{\eqoff}{-1.35\unitlength}%
\raisebox{\eqoff}{
	\fmfframe(0,0)(-1,0){
		\begin{fmfchar*}(5,2.5)
		\fmfforce{0 w, 0.5h}{v2}
		\fmfforce{1 w, 0.5h}{v1}
		\fmfv{decor.shape=circle,decor.filled=full,
			decor.size=2thick}{v1,v2}
		\fmf{phantom,left=0.6}{v1,v2,v1}
		\fmffreeze
		\fmfposition
		\fmfipath{p[]}
		\fmfipair{v[]}
		\fmfiset{p1}{vpath(__v1,__v2)}
		\fmfiset{p2}{vpath(__v2,__v1)}
		\fmfi{plain}{p1}
		\fmfi{plain}{p2}
		\fmfi{plain}{point length(p1)/2 of p1 -- point length(p2)/2 of p2}
		\end{fmfchar*}
	}
}
+
2
\settoheight{\eqoff}{$\times$}%
\setlength{\eqoff}{0.5\eqoff}%
\addtolength{\eqoff}{-1.35\unitlength}%
\raisebox{\eqoff}{
	\fmfframe(0,0)(-1,0){
		\begin{fmfchar*}(5,2.5)
		\fmfforce{0 w, 0.5h}{v2}
		\fmfforce{1 w, 0.5h}{v1}
		\fmfv{decor.shape=circle,decor.filled=full,	decor.size=2thick}{v1,v2}
		\fmfdraw
		\fmfv{decor.shape=hexacross,decor.size=7thin}{v1}
		\fmf{plain,left=0.6}{v1,v2,v1}
		\end{fmfchar*}
	}
}
\bigr]
=
\frac{p^2g^2}{\complexi(4\pi)^6} \frac{1}{12}\Bigl(
\frac{1}{2\epsilon^2}-\frac{1}{3\epsilon}
\Bigr)\eqncom\qquad
\Kop\bigl[
\settoheight{\eqoff}{$\times$}%
\setlength{\eqoff}{0.5\eqoff}%
\addtolength{\eqoff}{-1.35\unitlength}%
\raisebox{\eqoff}{
	\fmfframe(0,0)(-1,0){
		\begin{fmfchar*}(5,2.5)
		\fmfforce{0 w, 0.5h}{v2}
		\fmfforce{1 w, 0.5h}{v1}
		\fmfv{decor.shape=circle,decor.filled=full,
			decor.size=2thick}{v1,v2}
		\fmf{phantom,left=0.6}{v1,v2,v1}
		\fmffreeze
		\fmfposition
		\fmfipath{p[]}
		\fmfipair{v[]}
		\fmfiset{p1}{vpath(__v1,__v2)}
		\fmfiset{p2}{vpath(__v2,__v1)}
		\fmfi{plain}{p1}
		\fmfi{plain}{p2}
		\fmfiv{d.sh=circle,d.f=empty,d.si=.3w}{point length(p1)/2 of p1}
		\end{fmfchar*}
	}
}
+
2\settoheight{\eqoff}{$\times$}%
\setlength{\eqoff}{0.5\eqoff}%
\addtolength{\eqoff}{-1.35\unitlength}%
\raisebox{\eqoff}{
	\fmfframe(0,0)(-1,0){
		\begin{fmfchar*}(5,2.5)
		\fmfforce{0 w, 0.5h}{v2}
		\fmfforce{1 w, 0.5h}{v1}
		\fmfv{decor.shape=circle,decor.filled=full,
			decor.size=2thick}{v1,v2}
		\fmf{phantom,left=0.6}{v1,v2,v1}
		\fmffreeze
		\fmfposition
		\fmfipath{p[]}
		\fmfipair{v[]}
		\fmfiset{p1}{vpath(__v1,__v2)}
		\fmfiset{p2}{vpath(__v2,__v1)}
		\fmfi{plain}{p1}
		\fmfi{plain}{p2}
		\fmfiv{d.sh=hexacross,decor.size=7thin}{point length(p1)/2 of p1}
		\end{fmfchar*}
	}
}
\bigr]
=
\frac{p^2g^2}{\complexi(4\pi)^6} \frac{1}{12}\Bigl(
-\frac{1}{12\epsilon^2}+\frac{11}{72\epsilon}
\Bigr)\eqndot
\end{aligned}
\end{equation}
} %
due to the one-loop counterterm contributions. The remaining local divergences can be absorbed into the two-point function counterterms $p^2(c^{(0)}+c^{(2)})$, analogously to the zero-loop counterterm $p^2c^{(0)}$ in \eqref{eq:free_2pt_function_ren}. This concludes our minimal one-loop example in $\varphi^3$-theory and we refer to the literature\footnote{See \cite{BROWN1980135}, for a calculation in massive $\varphi^4$-theroy or e.g.\ the detailed textbook discussions in \cite{Collins:1984xc,Muta,Zuber}.}  for further introductory calculations.

\subsubsection{The renormalisation group equation for composite operators}
So far, we introduced bare composite operators in correlation functions and absorbed all newly occurring UV divergences into complete renormalisation constant matrices that mix renormalised operators with the same quantum numbers. While the bare operators are independent of the renormalisation scale $\mu$, the renormalised composite operators and renormalisation constant matrices both depend on it. Hence, in analogy to the coupling and field renormalisation discussed in \subsecref{sec:renormalisation-group-equation}, we can exploit this fact to determine the $\mu$-dependence of renormalised quantities via the \RGE for composite operators.

For all composite operators labelled by the quantum numbers $A$, the \RGE matrix is obtained by varying the first equation in \eqref{eq:renormalised_operators} with respect to $\mu$. We find
\begin{equation}\label{eq:RGE_comp_operators}
0=\mu\frac{\de}{\de \mu}\cO^{(A)}_{\text{B}\,i} 
=(\cZ_A^{-1})_i^{\phan{a}j}\left[(\gamma_A)_j^{\phan{a}k}+\delta_j^{\phan{a}k}\,\mu\frac{\de}{\de \mu}\right]\cO_k^{(A)}\eqncom\qquad
\mu\frac{\de}{\de \mu}(\cZ_A)_i^{\phan{a}j}=-(\gamma_A)_i^{\phan{a}k}(\cZ_A)_k^{\phan{a}j}\eqncom
\end{equation}
where we absorbed the scale factor $\mu^{s_\cO}$ into the renormalisation constant. The matrix of anomalous dimensions $(\gamma_A)_i^{\phan{a}k}$ measures the degree of homogeneity in $\mu$ of the renormalisation matrix $\cZ_A$ and the sign in the second equality occurs since $\cZ_A$ corresponds to the source renormalisation matrix, compare \eqref{eq:renormalised_coupling_mu_dep} and \eqref{eq:renormalised_operators}. Like the anomalous dimensions of elementary fields and couplings, this anomalous dimension matrix is in general renormalisation-scheme dependent at higher loop orders and we will see an explicit example of this in \secref{sec:cake}. Note that the renormalised operator depends on the renormalised coupling of the theory and therefore the total derivative with respect to $\mu$ in \eqref{eq:RGE_comp_operators} induces contributions of the $\beta$-function, as in \eqref{eq:renormalisation_group_equation_interacting}. 

Generically, the matrices $\cZ_A$ and $\gamma_A$ are not diagonal which leads to the mixing of renormalised operators. Perturbatively, we may choose to diagonalise the renormalisation matrix $\tilde{\cZ}_A=U_z^{-1}\cZ_AU_{z}$ which leads to a different operator basis $\tilde{\cO}_{i}^{(A)}=(U_z^{-1})_i^{\phan{A}j}\cO_j^{(A)}$ for bare and renormalised operators. Note that the matrix $U_{z}$ depends on the renormalisation scale $\mu$ and hence we have to be careful when adjusting \eqref{eq:RGE_comp_operators}. We can, however, also diagonalise the first equation in \eqref{eq:RGE_comp_operators} directly. When the new operator basis is obtained by the transformation $\hat{\cO}_{i}^{(A)}=(U_\gamma^{-1})_i^{\phan{A}j}\cO_j^{(A)}$, this is possible if the diagonalised anomalous dimension matrix fulfils $\hat{\gamma}_A=U_\gamma^{-1}\gamma_AU_\gamma+U_\gamma^{-1}\mu\frac{\de}{\de \mu}U_\gamma$. Calculating $\hat{\gamma}_A$ from the second equation in \eqref{eq:RGE_comp_operators}, we find the relation between $U_z$ and $U_\gamma$ to be
\begin{equation}\label{eq:U_relation}
\hat{\gamma}_A=
U_\gamma^{-1}\gamma_AU_\gamma+\gamma_{U_\gamma}
=
U_z^{-1}\gamma_AU_z+[\gamma_{U_z},\tilde{\cZ}_{A}]\tilde{\cZ}^{-1}_{A}
\eqncom
\end{equation}
where $\gamma_{U}=U^{-1}\mu\frac{\de}{\de \mu}U$ measures the change of $U$ under a scale transformation.

To make this more explicit, let us return to the example in \eqref{eq:phi2_renormalisation}. Using the 1PI counterterms \eqref{eq:1PI_counterterms_example}, we can construct the one-loop renormalisation matrix to express the renormalised operators in \eqref{eq:renormalised_operators} in terms of bare ones to find
\begin{equation}
\begin{pmatrix}\label{eq:mixed_phi3}
\frac 12\varphi^2\\
p^2\varphi
\end{pmatrix}
=
\left[
\one+
\frac{1}{24\epsilon}\frac{g^2}{(4\pi)^3}
\begin{pmatrix}
-10 & \frac{2}{g\mu^\epsilon}\\
0 & 1
\end{pmatrix}
\right]
\begin{pmatrix}
\frac 12\varphi_{\text{B}}^2\\
p^2\varphi_{\text{B}}
\end{pmatrix}
\eqncom
\qquad
\Rightarrow
\gamma_4=
\frac{1}{12}\frac{g^2}{(4\pi)^3}
\begin{pmatrix}
-10 &\frac{2}{g \mu^{\epsilon}} \\
0 & 1
\end{pmatrix}\eqndot
\end{equation}
Note that the zero in the lower left corner of the renormalisation matrix indicates that the operator $p^2 \varphi$ does not receive divergent 1PI contributions from diagrams that involve the operator $\frac 12 \varphi^2$, which can directly be verified in $\varphi^3$-theory. The anomalous dimension matrix can be diagonalised via the transformation\footnote{In this example we have $\gamma_{U_\gamma}=0$ and thus \eqref{eq:U_relation} is trivially fulfilled.}
\begin{equation}
U_\gamma=
\begin{pmatrix}
1 & \frac{2}{11g\mu^\epsilon}\\
0 & 1
\end{pmatrix}\eqncom
\qquad
\hat{\cO}^{(4)}
=
U_\gamma^{-1}\cO^{(4)}
=
\begin{pmatrix}
\frac 12\varphi^2-\frac{2}{11 g\mu^\epsilon}p^2\varphi\\
p^2\varphi
\end{pmatrix}\eqncom
\end{equation}
where we also showed the one-loop diagonalised basis. In the latter basis, we can also calculate the one-loop scaling behaviour of the renormalised composite operators in analogy to \eqref{eq:scalar_scaling}. Under rescaling labelled by a wide hat, we find that the classical scaling dimensions of the composite operators are altered according to 
\begin{equation}\label{eq:phi3_scaling_comp_op}
\widehat{\hat{\cO}^{(4)}}(x)=\widehat{\cZ_4}(\mu)\widehat{\hat{\cO}^{(4)}_{\text{B}}}(x)=\e^{\alpha(4+\gamma_4)}\hat{\cO}^{(4)}(\e^\alpha x)\eqncom
\end{equation}
where we used \eqref{eq:symmetry_trafo_elementary_field} for the rescaling of the bare operator with classical dimension $\Delta_{\cO^{(4)}}^0=4$ in $d=6$ dimensions and \eqref{eq:RGE_comp_operators} for the rescaling of $\cZ_4$ under $\mu\rightarrow \e^{-\alpha}\mu$.

With the scaling properties of composite operators, we can also generalise the \RGE of $n$-point functions \eqref{eq:renormalisation_group_equation_interacting} to also include composite operators in a diagonal basis. When we label a correlation function by a vector of quantum numbers $\vec{A}$ and momenta $\vec{p}$ characterising $n$ operator insertions with mutually distinct momenta $p_i\neq p_j$ and which all renormalise diagonally, the \RGE is given by
\begin{equation}\label{eq:renormalisation_group_equation_operators}
0=\left[-\sum_{i=1}^n\hat{\gamma}_{A_i}+\mu\frac{\de}{\de \mu}\right]G^{(\vec{A})}(\vec{p})
=\left[-\sum_{i=1}^n\hat{\gamma}_{A_i}+\beta_{g}\frac{\partial}{\partial g}+\mu\frac{\partial}{\partial \mu}\right]G^{(\vec{A})}(\vec{p})\eqndot
\end{equation}

\section{\texorpdfstring{\NfSYM}{N=4 SYM} theory and its deformations}\label{sec:N4SYM_renormalisation}
In the previous section, we briefly introduced the program of renormalisation in the setting of $\varphi^3$-theory in six-dimensional Minkowski space. In this section, we discuss how this program is implemented for \NfSYMt and its deformations in four-dimensional Minkowski space. We focus on the differences and simplifications that occur in the settings of these highly symmetric theories and refer to the literature, e.g.\ \cite[\chap{71}]{Srednicki:2007} for the general treatment of non-abelian gauge theories in the path integral approach. As starting point, we use the classical actions introduced in \secref{sec:The_deformations} and replace all classical fields by bare quantum fields.

\subsection{The renormalised theories}\label{sec:N4_renormalised}
For \NfSYMt and its deformations, like for $\varphi^3$-theory, we can employ the path integral formalism of \secref{sec:path_integral_approach} to evaluate perturbative contributions to correlation functions of elementary fields and composite operators. In principle, the entire approach introduced in \secref{sec:phi3_theory} remains valid when we trade the action of $\varphi^3$-theory in \eqref{eq:phi3_action} for the action of deformed \NfSYMt introduced from \secref{sec:The_deformations}. The classical scaling dimensions of scalars, gauge fields, ghosts, and fermions in the regularised theory in $D=4-2\epsilon$ dimensional Minkowski space respectively are
\begin{equation}
\Delta_\phi^0=\Delta_A^0=\Delta_c^0=\frac 12(D-2)\eqncom\qquad
\Delta_\lambda^0=\frac 12(D-1)\eqndot
\end{equation}
The renormalised action of deformed \NfSYMt is obtained by replacing all bare fields $f_{\text{B}}$ and couplings $g_{\text{B}}$ in \secref{sec:The_deformations} with renormalised ones according to\footnote{The 1PI renormalisation constants are chosen to be compatible with our conventions from \appref{app:Feynman_rules}. All 1PI counterterms $\delta_X$ are given by $(-1)$ times the sum of divergent 1PI contributions that involve $X$, regardless whether $X$ is a field, operator, or coupling.}
\begin{equation}
f_\text{B}=Z_f^{\frac 12} f\eqncom\qquad
g_{\text{B}}=\mathcal{Z}_gg=\frac{\mu^{\Delta^0_{g}(4)-\Delta^0_{g}(D)}Z_{g}}{(Z_{f_1}Z_{f_2}\dots Z_{f_n})^{\frac 12}}g\eqncom\quad
Z_f=1-\delta_f\eqncom\quad
\cZ_g=1+\mathfrak{d}_g\eqncom\quad
Z_g=1+\delta_g\eqncom
\end{equation}
where the coupling occurs in an interaction term of the form $g_{\text{B}}\tr(f_{{\text{B}}\,1}\dots f_{{\text{B}}\,n})$ and it has the classical scaling dimension $\Delta^0_{g}(D)$ in $D$ dimensions. The connected and 1PI renormalisation constants are defined in terms of the connected and respectively 1PI counterterms as in \subsecref{sec:the-renormalised-theory}. The new action defines a new set of Feynman rules and consequently a new set of Feynman graphs that can contribute to a given correlation function and we refer to \appref{app:Feynman_rules} for a derivation of these rules. For \NfSYMt, we also derive the corresponding Feynman rules for the action \eqref{eq:N4_action_real_scalars} with real scalar fields in \appref{subsec:Feynman_real_scalars}.

As discussed in \chapref{chap:Introduction}, \NfSYMt is conformally invariant at the quantum level and in the renormalisation procedure this poses a significant difference compared to $\varphi^3$-theory. While the coupling in the latter theory is renormalised as we have explicitly seen in \subsecref{sec:the-renormalised-theory}, the couplings in \NfSYMt and its conformality preserving deformations are not. In the MS scheme, where only divergent contributions are absorbed into the renormalisation constants, this leads to the relations
\begin{equation}\label{eq:Z_relation_CFT}
1\stackrel{\text{MS}}{=}\cZ_{\gym}=\frac{Z_{\gym}}{(Z_{f_1}Z_{f_2}Z_{f_3})^{\frac{1}{2}}}
	=\frac{Z^2_{\gym}}{(Z_{f_1}Z_{f_2}Z_{f_3}Z_{f_4})^{\frac{1}{2}}}\eqncom
\end{equation}
for combinations of field renormalisation constants $Z_{f_i}$ that can be matched to one of the interaction terms in the \NfSYM action. Note that these relations imply that the connected counterterms $\mathfrak{d}$ vanish while they do not imply that the 1PI counterterms $\delta$ vanish. The latter counterterms do, however, fulfil exact relations such that \eqref{eq:Z_relation_CFT} remains true. In \subsecref{sec:calc_Greens_functions} we will see explicitly that the one-loop 1PI counterterm $\delta_\lambda^{(1)}$ is non-vanishing in \NfSYMt. We can combine \eqref{eq:Z_relation_CFT} with \eqref{eq:renormalised_coupling_mu_dep} and \eqref{eq:beta_function} to obtain an exact relation between the bare and renormalised coupling and the corresponding $\beta$-function in $D=4-2\epsilon$ dimensions. We find
\begin{equation}\label{eq:gYM}
g_{\scriptscriptstyle{\text{YM}}\,\text{B}}=\mu^\epsilon\gym\eqncom\qquad
\beta_{\gym}=-\epsilon \gym\eqncom
\end{equation}
which yields $\beta_{\gym}=0$ in strictly $d=4$ dimensions as is required for a \CFT. The $\epsilon$-dependence of the coupling and the $\beta$-function in the $D$-dimensional theory is related to the fact that dimensional regularisation does not preserve the conformal symmetry of the theory.

The exact conformal invariance also ensures that the anomalous dimensions become renormalisation scheme independent \footnote{To see this, note that two different renormalisation schemes can be related to one another via a finite function $\xi$ in the regulator $\epsilon$ and the couplings of the theory. The difference between anomalous dimensions calculated in two such schemes is proportional to the $\beta$-functions of the theory times a finite function in $\xi$. Hence this difference vanishes if all $\beta$-functions vanish, see \cite[\chap{7}]{Collins:1984xc} for details.} and in the MS scheme they can be written as
\begin{equation}\label{eq:gamma_fields_N4_MS}
\gamma_f=\lim_{\epsilon\rightarrow 0}\beta_{\gym}\frac{1}{2Z_f}\frac{\partial}{\partial \gym} Z_f=-\lim_{\epsilon\rightarrow 0}\frac{\epsilon\gym}{2} \frac{\partial}{\partial \gym}\log Z_f\eqndot
\end{equation}
Using \eqref{eq:gYM} and \eqref{eq:gamma_fields_N4_MS}, $n$-point correlation functions of elementary fields in \NfSYMt are straightforwardly obtained from \eqref{eq:renormalisation_group_equation_interacting}. It is important to keep in mind that \eqref{eq:Z_relation_CFT} -- \eqref{eq:gamma_fields_N4_MS} are only correct if the conformal invariance remains unbroken at the quantum level. For the deformations introduced in \secref{sec:The_deformations}, it is not guaranteed that the complete renormalisation constant of all newly introduced deformation-dependent couplings also vanishes as $\cZ_{\gym}$. Whether this is the case has to be checked explicitly, see \secref{sec:non-conformal_double_trace_coupling} for an example.

Correlation functions in \NfSYMt and its deformations must be compatible with the unbroken symmetries of the underlying theory. Therefore, $n$-point functions as given in \eqref{eq:n_point_correlation_function} must be invariant under symmetry variations generated by $g$ and a variation parameter $\alpha$, i.e.\ $\delta_{(\alpha\cdot g)}G^{(\vec{k}\,)}(\vec{x})=0$. For each symmetry generator of the theory, this yields a relation
\begin{equation}\label{eq:Ward_identity}
0= \complexi\vacl\T \bigl(\delta_{(\alpha\cdot g)}S\bigr)f_{k_1}(x_1)\dots f_{k_n}(x_n)\vac
+ \sum_{j=1}^n\vacl\T f_{k_1}(x_1)\dots \bigl(\delta_{(\alpha\cdot g)}f_{k_j}(x_j)\bigr) \dots f_{k_n}(x_n)\vac
\eqncom
\end{equation}
where $S$ is the action of the theory and the symmetry variations of elementary fields $\delta_{(\alpha\cdot g)}f=\complexi[\alpha\cdot g,f]$
are discussed in \secref{sec:symmetries} for \NfSYMt and in \subsecref{sec:classical-symmetries-of-the-deformed-models} for the $\beta$- and $\gamma_i$-deformation. When we write the variation of the action $\delta_{(\alpha\cdot g)}S$ 
in terms of a conserved Noether current and a surface term, these equations are known as the Ward-Takahashi identities \cite{Ward1950,Takahashi2008} for abelian gauge theories. For non-abelian gauge theories, like \NfSYMt and its deformations, there is the additional Becchi-Rouet-Stora-Tyutin (BRST) symmetry \cite{Becchi:1975nq,Tyutin:1975qk} under which $n$-point functions must be invariant. In the path integral approach, this is the rigid symmetry that remains when the classical non-abelian action $S$ is replaced by a gauge-fixed one $S+S_{\text{gf}}+S_{\text{gh}}$ with the gauge-fixing and ghost contributions $S_{\text{gf}}$ and $S_{\text{gh}}$, respectively. The explicit BRST variations compatible with our conventions in \chapref{chap:The_models} and \appref{app:Feynman_rules} can be found in \cite[\chap{74}]{Srednicki:2007}. When we write \eqref{eq:Ward_identity} including the identities from BRST transformations, we arrive at the Slavnov-Taylor identities \cite{Slavnov:1972fg,TAYLOR1971436}, which are the analoga of the Ward-Takahashi identities for non-abelian gauge theories\footnote{For abelian gauge theories, the ghost part in the action decouples from every other term, so that ghosts can be integrated out in this case which renders the BRST transformations trivial.}.

\subsection{Composite operator insertions}\label{sec:composite-operator-insertionsN4}
Composite operator insertions in correlation functions can also be discussed analogously to the $\varphi^3$-theory case. Specifically, we can follow \subsecref{sec:composite-operator-insertions} and adjust the alphabet from which composite operators are constructed to the one in \eqref{eq: alphabet} for \NfSYMt and its deformations. 

We renormalise composite operators in \NfSYMt and its deformations exactly as in \eqref{eq:renormalised_operators}. Hence, using \eqref{eq:RGE_comp_operators} and \eqref{eq:gYM} for \CFTs, we find the anomalous dimensions matrix of composite operators characterised by quantum numbers $A$ in the MS scheme to be
\begin{equation}\label{eq:anomalous_dim_matrix}
(\gamma_A)_i^{\phan{a}j}=-\lim_{\epsilon\rightarrow 0}\beta_{\gym}\left(\frac{\partial(\cZ_A)_i^{\phan{a}k}}{\partial \gym} \right)(\cZ_A^{-1})_k^{\phan{a}j}
=\lim_{\epsilon\rightarrow 0}\sum_{u=0}^\infty \epsilon \gym\left(\frac{\partial \mathfrak{d}_A}{\partial \gym}(-\mathfrak{d}_A)^u\right)_i^{\phan{a}j}\eqncom
\end{equation}
where we used $\cZ_A=\one+\mathfrak{d}_A$ and the series representation of $\cZ_A^{-1}$ in the second equality.\footnote{If $\cZ_A$ is diagonal, the anomalous dimension matrix can be written as $\gamma_A=\epsilon \gym \frac{\partial}{\partial \gym }\log \cZ_A$ which is compatible with the definition in \cite{Sieg:2010jt}.} Like in the case of elementary fields in \eqref{eq:gamma_fields_N4_MS}, this anomalous dimension matrix is independent of the renormalisation scheme due to the exact conformal invariance of the theory. Note that for some composite operators the complete counterterm vanishes in the interacting theory and for these operators the anomalous dimension matrix vanishes as well. A prime example of such operators are the so-called BPS operators\footnote{For a classification of BPS operators for \NfSYMt see e.g.\ \cite{Bianchi:2006ti}.} which preserve part of the supersymmetry and consequently do not receive perturbative quantum corrections \cite{WITTEN197897}. From \eqref{eq:variation_S10}, we know that the Lagrange density of \NfSYMt itself also belongs to these operators up to surface terms.

The complete renormalisation matrix $\cZ_A$ mixes composite operators with quantum numbers $A$ under renormalisation and consequently also the anomalous dimension matrix \eqref{eq:anomalous_dim_matrix} mixes the renormalised composite operators. Generically, in the construction of $\cZ_A$ gauge-invariant and gauge-noninvariant operators will contribute \cite{Joglekar:1975nu}. It was, however, shown\footnote{For a simplified version of this proof see \cite{Henneaux:1993jn}.} in \cite{Joglekar:1975nu} that there exists a basis in which the gauge-noninvariant operators decouple from the gauge-invariant ones such that renormalised gauge-invariant operators can be computed without the help of gauge-noninvariant operators. Therefore, the composite operators that we defined in \secref{sec:composite-operators} are only mixed among each other in a suitable basis and the corresponding matrix of anomalous dimensions is gauge-invariant.

When we turn to a diagonalised basis in which the composite operators do not mix under renormalisation, we can determine the scaling behaviour of renormalised composite operators in \NfSYMt and its deformations analogously to the situation in \eqref{eq:phi3_scaling_comp_op}. In this basis, the vector of composite operators $\hat{\cO}^{(A)}$ labelled by quantum numbers $A$ and the diagonalised renormalisation matrix $\hat{\cZ}_A$ transform under a rescaling $\mu\rightarrow \e^{-\alpha}\mu$ and $x\rightarrow \e^{\alpha}x$ as
\begin{equation}\label{eq:scaling_comp_op}
\widehat{\hat{\cO}^{(A)}}(x)=\widehat{\hat{\cZ}_A}(\mu)\widehat{\hat{\cO}^{(A)}_{\text{B}}}(x)=\e^{\alpha(\Delta_{A}^0\one +\hat{\gamma}_A)}\hat{\cO}^{(A)}(\e^\alpha x)\eqncom
\end{equation}
where $\Delta_{A}^0$ is the classical scaling dimension of the operators $\hat{\cO}^{(A)}$ and $\hat{\gamma}_A$ is the diagonal matrix with the anomalous dimensions $(\hat{\gamma}_A)_i^{\phan{a}i}$ as entries. We can use this scaling behaviour to determine the structure of the two-point function of renormalised composite operators in a \CFT. Following the analogous reasoning as we did in the classical theory at the end of \subsecref{subsec:symmetry-generators-and-composite-operators}, we find 
\begin{equation}\label{eq:2pt_func_quantum}
\vacl\T\hat{\mathcal{O}}^{(A)}(x_1)\hat{\mathcal{O}}^{(B)}(x_2)\vac=\frac{\delta_{AB}}{(|x_1-x_2|^2+\complexi \epsilon)^{\Delta_{\mathcal{O}^{(A)}}}}\eqncom
\end{equation}
where the complete scaling dimension $\Delta_{\mathcal{O}}=\Delta_{\mathcal{O}}^0+\gamma_{\mathcal{O}}$ is a combination of the classical piece $\Delta_{\mathcal{O}}^0$ and the anomalous piece $\gamma_{\mathcal{O}}$.

The anomalous scaling behaviour of renormalised composite operators as shown in \eqref{eq:scaling_comp_op} implies that the classical action of the dilatation generator \eqref{eq:symmetry_generator_on_comp_operators} is supplemented by an additional coupling-dependent part in the quantised theory. When we sort these quantum corrections according to their power in the coupling, the full dilatation generator can be written analogously to \eqref{eq:D_cl_on_O} as
\begin{equation}\label{eq:quantum_D}
	D=\sum_{j=0}^\infty \gym^j D_j\eqncom \qquad 
	\text{with}\qquad
	D_a^{\phan{a}b}\cO_b(x)=-\complexi\bigl(\delta_a^{\phan{a}b}(\Delta_\cO^0+x_\mu\partial^\mu)+(\gamma_\cO^{(j)})_a^{\phan{a}b}\bigr)\cO_b(x)\eqncom
\end{equation}
where $D_0$ is the classical dilatation generator and the remaining $D_j$ give the anomalous contributions\footnote{In \NfSYMt the $g^1D_1$ term is absent \cite{Beisert:2003jj}.} when acting on a composite operator. In general, the elements of the anomalous dimension matrix $(\gamma_\cO^{(j)})_a^{\phan{a}b}$ can be determined perturbatively as in \eqref{eq:RGE_comp_operators} or \eqref{eq:anomalous_dim_matrix} and in the basis of \eqref{eq:scaling_comp_op} this matrix is diagonal and immediately gives the anomalous dimension of the composite operator $\hat{\cO}$.

\subsection{Calculating Green's functions in \texorpdfstring{\NfSYMt}{N=4 SYM theory}}\label{sec:calc_Greens_functions}

With the techniques discussed so far in this chapter, we can in principle calculate any Green's function in \NfSYMt and its deformations in a perturbative approach. To exemplify the calculation of Green's functions, we present the evaluation of the fermionic self-energy and a composite operator insertion at low orders in the coupling constant $\gym$. We employ the momentum-space Feynman rules derived in \appref{app:Feynman_rules} for the action \eqref{eq:N4_action_real_scalars} of \NfSYMt with real scalars. For the \ttt{Mathematica} implementation of these rules, we use the \ttt{FokkenFeynPackage} described in \appref{sec:Feynman_rules_Mathematica}. The examples may easily be generalised to other theories by altering the Feynman rules appropriately. The occurring integrals are regularised in $D=4-2\epsilon$ dimensions in the dimensional reduction procedure introduced in \appref{sec:Renormalisation_schemes}. For the evaluation of occurring integrals we use the techniques presented in \appref{app:Evaluating_Feynman_integrals} and we express the final results in terms of the effective planar coupling constant $g^2=\frac{\gym^2 N}{(4\pi)^2}=\frac{\lambda}{(4\pi)^2}$.

\subsubsection{The fermionic two-point function}
Let us calculate the one-loop perturbative correction to the fermion propagator. We can construct it by combining the free fermionic propagator $S_{\alpha\dot{\beta}}(p)$ with the perturbatively evaluated 1PI contributions as 
\begin{equation}
\vacl\T \lambda_\alpha(p)\ol{\lambda}_{\dot{\alpha}}(-p)\vac_{\complexi\mathcal{T}}=
\frac{1}{\complexi}S_{\alpha\dot{\alpha}}(p)+\frac{1}{\complexi}S_{\alpha\dot{\beta}}(p)\bigl[\complexi\Sigma^{\dot{\beta}\beta}(p)\bigr]\frac{1}{\complexi} S_{\beta\dot{\alpha}}(p)+\order{\gym^4}\eqncom
\end{equation}
where we suppressed flavour and colour indices. From dimensional analysis we know that the reduced momentum space two-point function of two fermions on the \lhs has dimension $-1$. Hence, the free fermion propagator also has dimension $[S]=-1$ and the fermionic self-energy has dimension $[\Sigma]=1$. The 1PI one-loop self-energy contributions to this equation in \NfSYMt are given by
\begin{equation}
\begin{aligned}\label{eq:fermion_SE}
\complexi\Sigma 
(p)&=\Bigl(\vacl\T \ol{\lambda}^{\dot{\beta}b}_B(-p)\lambda^{A\alpha a}(p)\vac_{\complexi\mathcal{T}}\Bigr)_{\text{1PI}}\\
&=
\Bigl(
\settoheight{\eqoff}{$\times$}%
\setlength{\eqoff}{0.5\eqoff}%
\addtolength{\eqoff}{-4.\unitlength}%
\raisebox{\eqoff}{
	\fmfframe(4,0)(2,0){
		\begin{fmfchar*}(20,7.5)
		\fmfforce{0 w, 0.5h}{vout}
		\fmfforce{1 w, 0.5h}{vin}
		\fmfforce{0.33 w, 0.5h}{v2}
		\fmfforce{0.67 w, 0.5h}{v1}
		\fmf{dashes_ar}{vin,v1}
		\fmf{dashes_rar}{v1,v2}
		\fmf{dashes_ar}{v2,vout}
		\fmf{plain,right,tension=0}{v1,v2}
		\fmffreeze
		\fmfposition
		\fmfiv{label=$\scriptstyle pA \alpha a$,l.angle=-100,l.dist=0.07w}{vloc(__vin)}
		\fmfiv{label=$\scriptstyle -pB \dot\beta b$,l.angle=-80,l.dist=0.07w}{vloc(__vout)}
		\end{fmfchar*}
	}
}
+
\settoheight{\eqoff}{$\times$}%
\setlength{\eqoff}{0.5\eqoff}%
\addtolength{\eqoff}{-4.\unitlength}%
\raisebox{\eqoff}{
	\fmfframe(4,0)(2,0){
		\begin{fmfchar*}(20,7.5)
		\fmfforce{0 w, 0.5h}{vout}
		\fmfforce{1 w, 0.5h}{vin}
		\fmfforce{0.33 w, 0.5h}{v2}
		\fmfforce{0.67 w, 0.5h}{v1}
		\fmf{dashes_ar}{vin,v1}
		\fmf{dashes_ar}{v1,v2}
		\fmf{dashes_ar}{v2,vout}
		\fmf{photon,right,tension=0}{v1,v2}
		\fmffreeze
		\fmfposition
		\fmfiv{label=$\scriptstyle pA \alpha a$,l.angle=-100,l.dist=0.07w}{vloc(__vin)}
		\fmfiv{label=$\scriptstyle -pB \dot\beta b$,l.angle=-80,l.dist=0.07w}{vloc(__vout)}
		\end{fmfchar*}
	}
}
+ 
\settoheight{\eqoff}{$\times$}%
\setlength{\eqoff}{0.5\eqoff}%
\addtolength{\eqoff}{-4.0\unitlength}%
\raisebox{\eqoff}{%
	\fmfframe(4,0)(0,0){%
		\begin{fmfchar*}(20,7.5)
		\fmfforce{0 w,0.5 h}{v1}
		\fmfforce{1 w,0.5 h}{v2}
		\fmfforce{0.5 w,0.5 h}{vc}
		\fmfv{decor.shape=hexacross,decor.size=10 thin}{vc}
		\fmf{dashes_rar}{v1,vc}
		\fmf{dashes_rar}{vc,v2}
		\fmffreeze
		\fmfposition
		\fmfiv{label=$\scriptstyle -pB\dot\beta b$,label.angle=-150,label.dist=9}{vloc(__vc)}
		\fmfiv{label=$\scriptstyle pA\alpha a$,label.angle=-30,label.dist=9}{vloc(__vc)}
		\end{fmfchar*}
	}
}
\Bigr)_{\text{1PI}}
+\order{\gym^4}\eqncom
\end{aligned}
\end{equation}
where we suppressed all indices on $\Sigma$. Amputated legs are labelled like vertices, i.e.\ the shown indices do not label the line under which they stand but characterise which indices can be connected at this point. The first diagram is given by
\begin{equation}
\begin{aligned}
\Bigl(
\settoheight{\eqoff}{$\times$}%
\setlength{\eqoff}{0.5\eqoff}%
\addtolength{\eqoff}{-4.\unitlength}%
\raisebox{\eqoff}{
	\fmfframe(4,0)(2,0){
		\begin{fmfchar*}(20,7.5)
		\fmfforce{0 w, 0.5h}{vout}
		\fmfforce{1 w, 0.5h}{vin}
		\fmfforce{0.33 w, 0.5h}{v2}
		\fmfforce{0.67 w, 0.5h}{v1}
		\fmf{dashes_ar}{vin,v1}
		\fmf{dashes_rar}{v1,v2}
		\fmf{dashes_ar}{v2,vout}
		\fmf{plain,right,tension=0}{v1,v2}
		\fmffreeze
		\fmfposition
		\fmfiv{label=$\scriptstyle pA \alpha a$,l.angle=-100,l.dist=0.07w}{vloc(__vin)}
		\fmfiv{label=$\scriptstyle -pB \dot\beta b$,l.angle=-80,l.dist=0.07w}{vloc(__vout)}
		\end{fmfchar*}
	}
}
\Bigr)_{\text{1PI}}
&=
\text{nscf}(v_A,v_B)(T_\nu)^{\dot{\beta}\alpha}\hat{I}_{(1,1)}^\nu(p)\eqncom
\end{aligned}
\end{equation}
where we separated the diagram into momentum space integral $\hat{I}$, the tensor $T$ which combines spinor and spacetime indices from $\gamma$-matrices and a prefactor nscf that combines the remaining numerical, symmetry, colour, and flavour factors. 
For the prefactor we have
\begin{equation}
\begin{aligned}\label{eq:flavour_SElambda}
\text{nscf}(v_A,v_B)&=-\frac{\gym^2}{\complexi^2}\bar\Sigma^{jAC}\Sigma^j_{CB}(c_j[a,c_C])(c_j[c_C,b])
&=
12\gym^2N\delta^A_B\Bigl((ab)-\frac{(a)(b)}{N}\Bigr)\eqncom
\end{aligned}
\end{equation}
where we used the abbreviation $(a_1a_2\dots a_n)\equiv \tr\bigl[\T^{a_1}\T^{a_2}\dots \T^{a_n}\bigr]$ for the colour-trace factors. From the Feynman rules of \NfSYMt in \appref{sec:the-actual-feynman-rules}, we see that the tensor $T$ is given in general by an alternating product of $\sigma$- and $\bar{\sigma}$-matrices as
\begin{equation}\label{eq:sigma_matrix_product}
(T_{\nu_1\dots\nu_m})^{\dot{\alpha}\beta}=
(\bar{\sigma}_{\nu_1})^{\dot{\alpha}\gamma_1}(\sigma_{\nu_2})_{\gamma_1\dot{\gamma}_2}(\bar{\sigma}_{\nu_3})^{\dot{\gamma}_2\gamma_3}
\dots (\bar{\sigma}_{\nu_m})^{\dot{\gamma}_{m-1}\beta}\eqncom
\end{equation}
where the position of the spinor indices $\dot{\alpha}$ and $\beta$ defines which type of $\sigma$-matrix is at the beginning and at the end of the product. We work in the $\ol{\text{DR}}$ renormalisation scheme, where the objects within $T$ live in the quasi-four-dimensional space $\text{Q}_4\text{S}$ and the momentum space integral is evaluated in the quasi-$D$-dimensional space $\text{Q}_D\text{S}$, with $D=4-2\epsilon$. The regularised integral in $D$-dimensional Minkowski space is evaluated in Euclidean space, where it is given as a special case of \eqref{eq:G_function_rank_n}. Afterwards, it is Wick-rotated back to Minkowski space by the means of the operator $\WR^{-1}$ to yield
\begin{equation} 
\begin{aligned}
\hat{I}_{(1,1)}^{\nu}(p)&=\int\frac{\de^D l}{(2\pi)^D}
\frac{\mu^{2\epsilon}(-l^\nu)}{(l^2-\complexi \epsilon)((p-l)^2-\complexi\epsilon)}
=
\WR^{-1}\Bigl(\frac{\mu^{2\epsilon}}{\complexi}
\settoheight{\eqoff}{$\times$}%
\setlength{\eqoff}{0.5\eqoff}%
\addtolength{\eqoff}{-2.75\unitlength}%
\raisebox{\eqoff}{
	\fmfframe(0,0)(-1,0){
		\begin{fmfchar*}(15,5.0)
		\fmfleft{vout}
		\fmfright{vin}
		\fmfforce{0.15 w, 0.5h}{v2}
		\fmfforce{0.85 w, 0.5h}{v1}
		\fmfv{decor.shape=circle,decor.filled=full,
			decor.size=2thick}{v1,v2}
		\fmf{plain}{v2,vout}
		\fmf{plain}{vin,v1}
		\fmf{phantom,left=0.6}{v1,v2,v1}
		\fmffreeze
		\fmfposition
		\fmfipath{p[]}
		\fmfiset{p1}{vpath(__v2,__v1)}
		\fmfiset{p2}{vpath(__v1,__v2)}
		\fmfi{plain}{p2}
		\fmfi{derplain,left=0.6,label.dist=4,label.angle=-115,label=$\scriptscriptstyle \nu$}{p1}
		\end{fmfchar*}
	}
}
\Bigr)
&=\frac{G_{(1)}(1,1)}{\complexi(4\pi)^{2}} \frac{p^\nu}{L^{\epsilon}}\eqncom
\end{aligned}
\end{equation}
with $L=\frac{p^{2}}{4\pi\mu^{2}}$. From now on, we will suppress the $-i\epsilon$ terms in the denominators for purely notational reasons. Combining all three contributions and using the effective planar coupling \eqref{eq:coupldef}, we find
\begin{equation}\label{eq:fermion_SE_scalar_sol}
\Bigl(
\settoheight{\eqoff}{$\times$}%
\setlength{\eqoff}{0.5\eqoff}%
\addtolength{\eqoff}{-4.\unitlength}%
\raisebox{\eqoff}{
	\fmfframe(4,0)(2,0){
		\begin{fmfchar*}(20,7.5)
		\fmfforce{0 w, 0.5h}{vout}
		\fmfforce{1 w, 0.5h}{vin}
		\fmfforce{0.33 w, 0.5h}{v2}
		\fmfforce{0.67 w, 0.5h}{v1}
		\fmf{dashes_ar}{vin,v1}
		\fmf{dashes_rar}{v1,v2}
		\fmf{dashes_ar}{v2,vout}
		\fmf{plain,right,tension=0}{v1,v2}
		\fmffreeze
		\fmfposition
		\fmfiv{label=$\scriptstyle pA \alpha a$,l.angle=-100,l.dist=0.07w}{vloc(__vin)}
		\fmfiv{label=$\scriptstyle -pB \dot\beta b$,l.angle=-80,l.dist=0.07w}{vloc(__vout)}
		\end{fmfchar*}
	}
}
\Bigr)_{\text{1PI}}
=-12\complexi g^2 \delta^A_B\Bigl((ab)-\frac{(a)(b)}{N}\Bigr) G_{(1)}(1,1)\frac{p^{\nu}(\bar{\sigma}_\nu)^{\dot{\beta}\alpha}}{L^{\epsilon}}\eqncom
\end{equation}
where $p$ and $\bar\sigma$ both live in $\text{Q}_4\text{S}$.
For the second Feynman diagram, we define the linear combination 
\begin{equation}
\hat{I}_{(\alpha,\beta)}^{\mu_1\dots\mu_m,\nu}(p)\equiv p^\nu \hat{I}_{(\alpha,\beta)}^{\mu_1\dots\mu_m}(p)-\hat{I}_{(\alpha,\beta)}^{\nu\mu_1\dots\mu_m}(p)\eqncom
\end{equation}
where the Euclidean space versions of the integrals on the \rhs are defined in \eqref{eq:G_function_rank_n} and \eqref{eq:I_alpha_beta} for the case of $m=0$. The diagram is then given by
\begin{equation} 
\begin{aligned}\label{eq:fermion_SE_gluon}
\Bigl(
\settoheight{\eqoff}{$\times$}%
\setlength{\eqoff}{0.5\eqoff}%
\addtolength{\eqoff}{-4.\unitlength}%
\raisebox{\eqoff}{
	\fmfframe(4,0)(2,0){
		\begin{fmfchar*}(20,7.5)
		\fmfforce{0 w, 0.5h}{vout}
		\fmfforce{1 w, 0.5h}{vin}
		\fmfforce{0.33 w, 0.5h}{v2}
		\fmfforce{0.67 w, 0.5h}{v1}
		\fmf{dashes_ar}{vin,v1}
		\fmf{dashes_ar}{v1,v2}
		\fmf{dashes_ar}{v2,vout}
		\fmf{photon,right,tension=0}{v1,v2}
		\fmffreeze
		\fmfposition
		\fmfiv{label=$\scriptstyle pA \alpha a$,l.angle=-100,l.dist=0.07w}{vloc(__vin)}
		\fmfiv{label=$\scriptstyle -pB \dot\beta b$,l.angle=-80,l.dist=0.07w}{vloc(__vout)}
		\end{fmfchar*}
	}
}
\Bigr)_{\text{1PI}}
&=\text{nscf}(v_A,v_B)(T_{\nu_1\nu_2\nu_3})^{\dot{\beta}\alpha}\Bigl(
\hat{\eta}^{\nu_1\nu_3}\hat{I}^{,\nu_2}_{(1,1)}(p)
-(1-\xi)\hat{I}^{\nu_1\nu_3,\nu_2}_{(2,1)}(p)
\Bigr)
\end{aligned}
\end{equation}
and the first two terms take the form 
\begin{equation}
\begin{aligned}
\text{nscf}(v_A,v_B)&=-2\gym^2 \delta^A_B(c_j[a,c_C])(c_j[c_C,b])=-\gym^2N\delta^A_B\Bigl((ab)-\frac{1}{N}(a)(b)\Bigr)\eqncom\\
(T_{\nu_1\nu_2\nu_3})^{\dot{\beta}\alpha}&=
-\bigl(\eta_{\nu_1\nu_2}\eta_{\nu_3\rho}-\eta_{\nu_1\nu_3}\eta_{\nu_2\rho}+\eta_{\nu_2\nu_3}\eta_{\nu_1\rho}+\complexi\varepsilon_{\nu_1\nu_2\nu_3\rho}\bigr)(\bar{\sigma}^\rho)^{\dot{\beta}\alpha}\eqndot
\end{aligned}
\end{equation}
The two integrals evaluate to
\begin{equation}
\begin{aligned}
\hat{I}^{,\nu_2}_{(1,1)}(p)&=\frac{\complexi p^{\nu_2}}{(4\pi)^{2}L^{\epsilon}} \bigl(G(1,1)-G_{(1)}(1,1)\bigr)\eqncom\\
\hat{I}^{\nu_1\nu_3,\nu_2}_{(2,1)}(p)
&=\frac{\complexi}{(4\pi)^{2}L^\epsilon}\Bigl[
\bigl(G_{(2)}(2,1)-G_{(3)}(2,1)\bigr)\frac{p^{\nu_1}p^{\nu_2}p^{\nu_3}}{p^2}
+\frac{G(1,1)-G_{(2)}(2,1)}{D}p^{\nu_2}\hat{\eta}^{\nu_1\nu_3}\\
&\phan{=\frac{1}{\complexi(4\pi)^{2}\hat{L}^\epsilon}\Bigl[}
-\frac{G_{(1)}(1,1)-G_{(3)}(2,1)}{D+2}S(\hat{\eta}^{\nu_1\nu_2}p^{\nu_3})\Bigr]\eqncom
\end{aligned}
\end{equation}
where the operator\footnote{See also below \eqref{eq:p_traceless} for further defining comments.} $S$ symmetrises all spacetime indices of its argument, the metric tensor $\hat{\eta}$ lives in $\text{Q}_D\text{S}$ and the second equation is obtained using \eqref{eq:p_traceless}. Combining all contributions we find
\begin{equation}
\begin{aligned}\label{eq:fermion_SE_gluon_sol}
\Bigl(
\settoheight{\eqoff}{$\times$}%
\setlength{\eqoff}{0.5\eqoff}%
\addtolength{\eqoff}{-4.\unitlength}%
\raisebox{\eqoff}{
	\fmfframe(4,0)(2,0){
		\begin{fmfchar*}(20,7.5)
		\fmfforce{0 w, 0.5h}{vout}
		\fmfforce{1 w, 0.5h}{vin}
		\fmfforce{0.33 w, 0.5h}{v2}
		\fmfforce{0.67 w, 0.5h}{v1}
		\fmf{dashes_ar}{vin,v1}
		\fmf{dashes_ar}{v1,v2}
		\fmf{dashes_ar}{v2,vout}
		\fmf{photon,right,tension=0}{v1,v2}
		\fmffreeze
		\fmfposition
		\fmfiv{label=$\scriptstyle pA \alpha a$,l.angle=-100,l.dist=0.07w}{vloc(__vin)}
		\fmfiv{label=$\scriptstyle -pB \dot\beta b$,l.angle=-80,l.dist=0.07w}{vloc(__vout)}
		\end{fmfchar*}
	}
}
\Bigr)_{\text{1PI}}
&=
-2\complexi g^2 \xi
\delta^A_B\Bigl((ab)-\frac{1}{N}(a)(b)\Bigr)\frac{2^{-1+2\epsilon}(\epsilon-1)\pi^{\frac 32}}{\Gamma(\frac 32-\epsilon)\sin(\pi\epsilon)}\frac{p^{\nu}(\bar{\sigma}_{\nu})^{\dot{\beta}\alpha}}{L^\epsilon}
\end{aligned}
\end{equation}
where we used the definitions of $G$-functions \eqref{eq:G_function} and \eqref{eq:Gn_definition}. The regularised one-loop self-energy contributions to the fermion propagator displayed in \eqref{eq:fermion_SE} are obtained by combining \eqref{eq:fermion_SE_scalar_sol} and \eqref{eq:fermion_SE_gluon_sol}. Up to order $\epsilon^0$ we find
\begin{equation}
\begin{aligned}\label{eq:fermion_SE_res}
\Sigma_\text{B} 
(p)
&=-2g^2\Bigl[2+(3+\xi)
\Bigl(1+\frac{1}{\epsilon} -c_{\ol{\text{MS}}}-\log\bigl(\tfrac{p^2}{\mu^2}\bigr)\Bigr)\Bigr]
\delta^A_B\Bigl((ab)-\frac{(a)(b)}{N}\Bigr)\complexi p^{\dot{\beta}\alpha}\eqncom
\end{aligned}
\end{equation}
where $c_{\ol{\text{MS}}}=\log 4\pi-\gammaE$ and the spinorial momenta are defined as in \eqref{eq:vectorfield_spinorial}.

The counterterm can now easily be defined, since both diagrams are IR-finite and the divergence stems from the UV regime. We demand that the sum of $\complexi \Sigma_\text{B}$ and the counterterm given in \eqref{eq:counterterms} is finite to order $\gym^2$. Using the colour generator identities \eqref{eq:colour_generators}, we see from the colour part that only the \SUN modes of $\complexi \Sigma_\text{B}$ must be renormalised, as the \U{1} part vanishes in \NfSYMt. In the DR, $\ol{\text{DR}}$, and kinematical subtraction scheme this yields the following one-loop counterterms
\begin{equation}
\begin{aligned}
\delta_\lambda^{\text{DR}}&=2g^2(3+\xi)\epsilon^{-1}\eqncom\\
\delta_\lambda^{\ol{\text{DR}}}&=2g^2(3+\xi)\Bigl(\epsilon^{-1}-c_{\ol{\text{MS}}}\Bigr)\eqncom\\
\delta_\lambda^{\text{KS}}&=2g^2\Bigl[2+(3+\xi)\Bigl(1+\epsilon^{-1}-c_{\ol{\text{MS}}}\Bigr)\Bigr]\eqndot
\end{aligned}
\end{equation}

\subsubsection{A dimension \texorpdfstring{$\Delta^0=3$}{3} composite operator}\label{sec:comp_operator_renormalisation}
In this subsection, we evaluate the divergent contributions up to order $\gym^3$ that arise when the composite operator 
\begin{equation}\label{eq:example_operator}
\mathcal{O}=\frac{1}{N}\tr(\ol{\lambda}^{A}_{\dot{\alpha}}\ol{\lambda}^{B\dot{\beta}})
=\frac{1}{2N}\bigl(
\ol{\lambda}^{Aa}_{\dot{\alpha}}\ol{\lambda}^{B\dot \beta a}-\ol{\lambda}^{B\dot \beta a}\ol{\lambda}^{Aa}_{\dot{\alpha}}
\bigr)
\end{equation}
is inserted into particular correlation functions. To exemplify the diagonal renormalisation as well as operator mixing in \NfSYMt, we determine the one-loop counterterms that enter the two correlation functions
\begin{equation}
\begin{aligned}\label{eq:comp_O_correlation_func}
S^{\ol{\lambda}\,\ol{\lambda}}_{\mathcal{O}}&=
\vacl\T \ol{\lambda}^{\dot{\gamma}c_A}_A(-p)\ol{\lambda}^{c_B}_{B\dot{\delta}}(0)\mathcal{O}(p)\vac_{\complexi\mathcal{T}}
=
\frac{Z^{\ol\lambda\,\ol{\lambda}}_\cO}{Z^2_\lambda}
\settoheight{\eqoff}{$\times$}%
\setlength{\eqoff}{0.5\eqoff}%
\addtolength{\eqoff}{-3.75\unitlength}%
\raisebox{\eqoff}{
	\fmfframe(0,0)(-1,0){
		\begin{fmfchar*}(9,8.0)
		\fmfright{vop}
		\fmfforce{0 w, .9h}{v1}
		\fmfforce{0 w, .1h}{v3}
		\fmf{dashes,right=0.2,tension=0.6}{vop,v1}
		\fmf{dashes,left=0.2,tension=0.6}{vop,v3} 
		\fmffreeze
		\fmfposition
		\fmfipath{p[]}
		\fmfiset{p1}{vpath(__vop,__v1)}
		\fmfiset{p11}{subpath 2(0,length(p1)/5) of p1}
		\fmfiset{p12}{subpath 3(length(p1)/5,length(p1)) of p1}
		\fmfdraw
		\fmfiv{decor.shape=circle,decor.filled=full,decor.size=5}{vloc(__vop)}
		\end{fmfchar*}
	}
}
+
\settoheight{\eqoff}{$\times$}%
\setlength{\eqoff}{0.5\eqoff}%
\addtolength{\eqoff}{-3.75\unitlength}%
\raisebox{\eqoff}{
	\fmfframe(0,0)(-1,0){
		\begin{fmfchar*}(9,8.0)
		\fmfright{vop}
		\fmfforce{0 w, .9h}{v1}
		\fmfforce{0 w, .1h}{v3}
		\fmf{phantom,right=0.2,tension=0.6}{vop,v1}
		\fmf{phantom,left=0.2,tension=0.6}{vop,v3} 
		\fmffreeze
		\fmfposition
		\fmfipath{p[]}
		\fmfiset{p1}{vpath(__vop,__v1)}
		\fmfiset{p11}{subpath (0,3length(p1)/5) of p1}
		\fmfiset{p12}{subpath (3length(p1)/5,length(p1)) of p1}
		\fmfiset{p2}{vpath(__vop,__v3)}
		\fmfiset{p21}{subpath (0,3length(p2)/5) of p2}
		\fmfiset{p22}{subpath (3length(p2)/5,length(p2)) of p2}
		\fmfi{photon}{point 3length(p1)/5 of p1 -- point 3length(p2)/5 of p2}
		\fmfi{dashes}{p11}
		\fmfi{dashes}{p12}
		\fmfi{dashes}{p21}
		\fmfi{dashes}{p22}	
		\fmfdraw
		\fmfiv{decor.shape=circle,decor.filled=full,decor.size=5}{vloc(__vop)}
		\end{fmfchar*}
	}
}
+
2
\settoheight{\eqoff}{$\times$}%
\setlength{\eqoff}{0.5\eqoff}%
\addtolength{\eqoff}{-3.75\unitlength}%
\raisebox{\eqoff}{
	\fmfframe(0,0)(-1,0){
		\begin{fmfchar*}(9,8.0)
		\fmfright{vop}
		\fmfforce{0 w, .9h}{v1}
		\fmfforce{0 w, .1h}{v3}
		\fmf{phantom,right=0.2,tension=0.6}{vop,v1}
		\fmf{phantom,left=0.2,tension=0.6}{vop,v3} 
		\fmffreeze
		\fmfposition
		\fmfipath{p[]}
		\fmfiset{p1}{vpath(__vop,__v1)}
		\fmfiset{p11}{subpath (0,2length(p1)/5) of p1}
		\fmfiset{p12}{subpath (3length(p1)/5,length(p1)) of p1}
		\fmfiset{p2}{vpath(__vop,__v3)}
		\fmfiset{p21}{subpath (0,3length(p2)/5) of p2}
		\fmfiset{p22}{subpath (3length(p2)/5,length(p2)) of p2}
		\fmfi{dashes}{p11}
		\fmfi{dashes}{p12}
		\fmfi{dashes}{p2}
		\fmfdraw
		\fmfiv{decor.shape=circle,decor.size=5 thin,decor.filled=shaded}{point length(p1)/2 of p1}
		\fmfiv{decor.shape=circle,decor.filled=full,decor.size=5}{vloc(__vop)}
		\end{fmfchar*}
	}
}
+\order{\gym^4}
\\
S^{\varphi\varphi\varphi}_{\mathcal{O}}&=
\vacl\T \varphi^{ic_i}(-p)\varphi^{kc_k}(0) \varphi^{jc_j}(0)\mathcal{O}(p)\vac_{\complexi\mathcal{T}}
=
\frac{\mu^{\epsilon}Z^{\varphi\varphi\varphi}_{\cO}}{Z_\varphi^3}
\settoheight{\eqoff}{$\times$}%
\setlength{\eqoff}{0.5\eqoff}%
\addtolength{\eqoff}{-3.75\unitlength}%
\raisebox{\eqoff}{
	\fmfframe(0,0)(-1,0){
		\begin{fmfchar*}(9,8.0)
		\fmfright{vop}
		\fmfforce{0 w, .9h}{v1}
		\fmfforce{0 w, .1h}{v3}
		\fmfforce{0 w, .5h}{v2}
		\fmfforce{0.4 w, .5h}{v4}
		\fmf{plain,right=0.2,tension=0.6}{vop,v1}
		\fmf{plain,left=0.2,tension=0.6}{vop,v3} 
		\fmf{plain}{v2,vop}
		\fmffreeze
		\fmfposition
		\fmfdraw
		\fmfiv{decor.shape=circle,decor.filled=full,decor.size=5}{vloc(__vop)}
		\end{fmfchar*}
	}
}
+
\settoheight{\eqoff}{$\times$}%
\setlength{\eqoff}{0.5\eqoff}%
\addtolength{\eqoff}{-3.75\unitlength}%
\raisebox{\eqoff}{
	\fmfframe(0,0)(-1,0){
		\begin{fmfchar*}(9,8.0)
		\fmfright{vop}
		\fmfforce{0 w, .9h}{v1}
		\fmfforce{0 w, .1h}{v3}
		\fmfforce{0 w, .5h}{v2}
		\fmfforce{0.4 w, .5h}{v4}
		\fmf{phantom,right=0.2,tension=0.6}{vop,v1}
		\fmf{phantom,left=0.2,tension=0.6}{vop,v3} 
		\fmf{plain}{v2,v4}
		\fmffreeze
		\fmfposition
		\fmfipath{p[]}
		\fmfiset{p1}{vpath(__vop,__v1)}
		\fmfiset{p11}{subpath (0,3length(p1)/5) of p1}
		\fmfiset{p12}{subpath (3length(p1)/5,length(p1)) of p1}
		\fmfiset{p2}{vpath(__vop,__v3)}
		\fmfiset{p21}{subpath (0,3length(p2)/5) of p2}
		\fmfiset{p22}{subpath (3length(p2)/5,length(p2)) of p2}
		\fmfi{dashes}{point 3length(p1)/5 of p1 -- point 3length(p2)/5 of p2}
		\fmfi{dashes}{p11}
		\fmfi{plain}{p12}
		\fmfi{dashes}{p21}
		\fmfi{plain}{p22}	
		\fmfdraw
		\fmfiv{decor.shape=circle,decor.filled=full,decor.size=5}{vloc(__vop)}
		\end{fmfchar*}
	}
}
+\order{\gym^4}
\eqncom
\end{aligned}
\end{equation}
where we schematically depicted the contributing one-loop diagrams. The 1PI renormalisation constants stem from the elementary fields as well as the two elements of the complete operator renormalisation constant matrix $\cZ_{\cO}$, like in \eqref{eq:renormalised_operators}. Note that the final state fields in these two correlation functions are not gauge-invariant composite operators in contrast to the discussion on composite operator renormalisation in \subsecref{sec:composite-operator-insertionsN4}. We can, however, combine the final state fields to form a second gauge-invariant composite operator and determine all contributions to $\cZ_\cO$ via the corresponding two-point functions as in \eqref{eq:comp_op_phi3_g2}. Here, we refrain from this procedure to save one and respectively two loop orders in the calculation analogous to the calculation in \eqref{eq:phi2_renormalisation} at the expense of having nongauge-invariant objects in the calculation. We use the notational conventions from the calculation of the fermionic self-energy above to evaluate all possible transitions.

For the diagonal correlation function $S^{\ol{\lambda}\,\ol{\lambda}}_{\mathcal{O}}$, there are two divergent one-loop contributions that need to be cancelled by counterterms. The first is the 1PI contribution
\begin{equation}
\begin{aligned}\label{eq:comp_op1_id_contribution}
\bigl(S^{\ol\lambda\,\ol{\lambda}}_{\mathcal{O}}\bigr)_{\text{1PI}}&=
\Kop\Biggl[
\Biggl(
\settoheight{\eqoff}{$\times$}%
\setlength{\eqoff}{0.5\eqoff}%
\addtolength{\eqoff}{-8.\unitlength}%
\raisebox{\eqoff}{
	\fmfframe(1,.5)(2,1){
		\begin{fmfchar*}(20,15)
		\fmfright{vop}
		\fmfforce{0 w, 1h}{v1}
		\fmfforce{0 w, 0h}{v3}
		\fmfforce{0.34 w,0.8 h}{vc1}
		\fmfforce{0.34 w,0.2 h}{vc3}
		\fmf{dashes_ar}{vc1,v1}
		\fmf{dashes_ar}{vc3,v3}
		\fmf{dashes_ar,right=0.2,tension=0.6}{vop,vc1}
		\fmf{dashes_ar,left=0.2,tension=0.6}{vop,vc3}
		\fmf{photon}{vc1,vc3}
		\fmffreeze
		\fmfposition
		\fmfiv{label=$\scriptstyle A \dot\alpha a$,l.angle=90,l.dist=0.22w}{vloc(__vop)}
		\fmfiv{label=$\scriptstyle B \dot \beta a$,l.angle=-90,l.dist=0.22w}{vloc(__vop)}
		\fmfiv{label=$\scriptstyle -pA\dot{\gamma} c_A$,l.a=20,l.dist=0.1w}{vloc(__v1)}
		\fmfiv{label=$\scriptstyle 0 B\dot{\delta} c_B$,l.a=-20,l.dist=0.1w}{vloc(__v3)}
		\fmfiv{decor.shape=circle,decor.filled=full,decor.size=5}{vloc(__vop)}
		\fmfiv{label=$\scriptstyle p$,l.a=0,l.dist=0.1w}{vloc(__vop)}
		\end{fmfchar*}
	}
}
\Biggr)_{\text{1PI}}
\Biggr]\\
&=\text{nscf}(v_A,\mathcal{O},v_B)\Kop\Bigl[
\eta^{\mu_3\mu_4}\hat{I}_{(2,1)}^{\mu_2,\mu_1}(p)-(1-\xi)\hat{I}_{(3,1)}^{\mu_2\mu_3\mu_4,\mu_1}(p)\Bigr]
(T_{\mu_4\mu_1})^{\dot{\gamma}}_{\phan{\alpha}\dot\alpha}(T_{\mu_2\mu_3})^{\dot{\beta}}_{\phan{\alpha}\dot \delta}
\\
&=-\frac{2g^2}{\epsilon N}\bigl(
(1-\xi)\delta^{\dot{\gamma}}_{\phan{\alpha}\dot\alpha}\delta^{\dot{\beta}}_{\phan{\alpha}\dot \delta}
-2\delta^{\dot{\beta}}_{\phan{\alpha}\dot\alpha}\delta^{\dot{\gamma}}_{\phan{\alpha}\dot \delta}
\bigr)\Bigl((c_Ac_B)-\frac{1}{N}(c_A)(c_B)\Bigr)
\eqncom
\end{aligned}
\end{equation}
where the construction of the tensors $T$ follows by starting from the outgoing fermion line to the upper left and applying rule eight of the Feynman rules in \appref{sec:the-actual-feynman-rules}. The last line is obtained using 
\begin{equation}
\begin{aligned}\label{eq:K_integral_1}
\text{nscf}(v_A,\mathcal{O},v_B)&=2\complexi\gym^2N\Bigl(\frac{(c_Ac_B)}{N}-\frac{(c_A)(c_B)}{N^2}\Bigr)\eqncom\\
\Kop\bigl[\hat{I}^{\mu_2\mu_3\mu_4,\mu_1}_{(3,1)}(p)\bigr]&=
\frac{-\complexi}{(4\pi)^2}\frac{S(\hat\eta^{\mu_1\mu_2}\hat\eta^{\mu_3\mu_4})}{24\epsilon}\eqncom\\
\Kop\bigl[I^{\mu_2,\mu_1}_{(2,1)}(p)\bigr]
&=
\frac{-\complexi}{(4\pi)^2}\frac{\hat\eta^{\mu_1\mu_2}}{4\epsilon}\eqndot
\end{aligned}
\end{equation}
Since both integrals are IR convergent\footnote{The IR finiteness can be seen in Euclidean space by counting of loop momentum powers \cite{kleinert2001critical}, see also \appref{sec:UV_and_IR_div}.}, it is sufficient to extract the entire pole part of the integrals via the operator $\Kop$. The second divergent contribution to $S^{\ol\lambda\,\ol\lambda}_{\mathcal{O}}$ stems from the one-loop insertion of the fermionic self-energy \eqref{eq:fermion_SE_res} on one of the external legs of the operator. It is a connected non-1PI contribution ($\ol{\text{1PI}}$) and given by
\begin{equation}
\begin{aligned}\label{eq:SE_insertion_to_O}
\bigl(S_{\text{B}\,\mathcal{O}}^{\ol\lambda\,\ol\lambda}\bigr)_{\ol{\text{1PI}}}
&=\Kop\Biggl[
\Biggl(
\settoheight{\eqoff}{$\times$}%
\setlength{\eqoff}{0.5\eqoff}%
\addtolength{\eqoff}{-8.\unitlength}%
\raisebox{\eqoff}{
	\fmfframe(4,0)(2,1){
		\begin{fmfchar*}(15,15)
		\fmfright{vop}
		\fmfforce{0 w, .8h}{v1}
		\fmfforce{0 w, .2h}{v3}
		\fmf{phantom,right=0.2,tension=0.6}{vop,v1}
		\fmf{dashes_ar,left=0.2,tension=0.6,foreground=(0.65,,0.65,,0.65)}{vop,v3}
		\fmffreeze
		\fmfposition
		\fmfipath{p[]}
		\fmfiset{p1}{vpath(__vop,__v1)}
		\fmfiset{p11}{subpath 2(0,length(p1)/5) of p1}
		\fmfiset{p12}{subpath 3(length(p1)/5,length(p1)) of p1}
		\fmfiv{label=$\scriptstyle A \dot\alpha a$,l.angle=90,l.dist=0.22w}{vloc(__vop)}
		\fmfiv{label=$\scriptstyle B \dot \beta a$,l.angle=-90,l.dist=0.22w}{vloc(__vop)}
		\fmfiv{label=$\scriptstyle -pA\dot{\gamma} c_A$,l.a=70,l.dist=0.07w}{vloc(__v1)}
		\fmfiv{label=$\scriptstyle 0B\dot{\delta} c_B$,l.a=-70,l.dist=0.07w}{vloc(__v3)}
		\fmfiv{label=$\scriptstyle p$,l.a=0,l.dist=0.1w}{vloc(__vop)}
		\fmfi{dashes_ar}{p11}
		\fmfi{dashes_ar,foreground=(0.65,,0.65,,0.65)}{p12}
		\fmfdraw
		\fmfiv{decor.shape=circle,decor.size=10 thin,decor.filled=shaded}{point length(p1)/2 of p1}
		\fmfiv{decor.shape=circle,decor.filled=full,decor.size=5}{vloc(__vop)}
		\end{fmfchar*}
	}
}
+
\settoheight{\eqoff}{$\times$}%
\setlength{\eqoff}{0.5\eqoff}%
\addtolength{\eqoff}{-8.\unitlength}%
\raisebox{\eqoff}{
	\fmfframe(4,0)(2,1){
		\begin{fmfchar*}(15,15)
		\fmfright{vop}
		\fmfforce{0 w, .8h}{v1}
		\fmfforce{0 w, .2h}{v3}
		\fmf{phantom,left=0.2,tension=0.6}{vop,v3}
		\fmf{dashes_ar,right=0.2,tension=0.6,foreground=(0.65,,0.65,,0.65)}{vop,v1}
		\fmffreeze
		\fmfposition
		\fmfipath{p[]}
		\fmfiset{p1}{vpath(__vop,__v3)}
		\fmfiset{p11}{subpath (0,2length(p1)/5) of p1}
		\fmfiset{p12}{subpath (3length(p1)/5,length(p1)) of p1}
		\fmfiv{label=$\scriptstyle A \dot\alpha a$,l.angle=90,l.dist=0.22w}{vloc(__vop)}
		\fmfiv{label=$\scriptstyle B \dot \beta a$,l.angle=-90,l.dist=0.22w}{vloc(__vop)}
		\fmfiv{label=$\scriptstyle 0A\dot{\gamma} c_A$,l.a=70,l.dist=0.07w}{vloc(__v1)}
		\fmfiv{label=$\scriptstyle -pB\dot{\delta} c_B$,l.a=-70,l.dist=0.07w}{vloc(__v3)}
		\fmfiv{label=$\scriptstyle p$,l.a=0,l.dist=0.1w}{vloc(__vop)}
		\fmfi{dashes_ar}{p11}
		\fmfi{dashes_ar,foreground=(0.65,,0.65,,0.65)}{p12}
		\fmfdraw
		\fmfiv{decor.shape=circle,decor.size=10 thin,decor.filled=shaded}{point length(p1)/2 of p1}
		\fmfiv{decor.shape=circle,decor.filled=full,decor.size=5}{vloc(__vop)}
		\end{fmfchar*}
	}
}
\Biggr)_{\ol{\text{1PI}}}
\Biggr]\\
&=
-\frac{4g^2(3+\xi)}{\epsilon N}
\delta^{\dot{\gamma}}_{\phan{\alpha}\dot{\alpha}}\delta^{\dot{\beta}}_{\phan{\alpha}\dot{\delta}}\Bigl((c_Ac_B)-\frac{1}{N}(c_A)(c_B)\Bigr)\eqncom
\end{aligned}
\end{equation}
where the $\ol{\text{1PI}}$ operation on diagrams only amputates the external free propagators, which are grey shaded in the above diagrams. The divergent contributions \eqref{eq:comp_op1_id_contribution} and \eqref{eq:SE_insertion_to_O} are renormalised by the external fermion leg renormalisation constant $Z_\lambda^{-1}$ and the complete renormalisation matrix element $\cZ_{\cO}^{\ol{\lambda}\,\ol{\lambda}}=Z_{\cO}^{\ol{\lambda}\,\ol{\lambda}}Z_\lambda^{-1}$ in \eqref{eq:comp_O_correlation_func}. The one-loop contribution to the latter must absorb the following divergence\footnote{The colour structure guarantees that the operator does not mix with two \U{1} operators in this process, since the corresponding contraction vanishes. In a two-point function of $\cO$ with its conjugate, the colour factor drops out since $\frac{1}{N^2}(c_Ac_B)\bigl((c_Ac_B)-\frac 1N(c_A)(c_B)\bigr)=1-N^{-2}$.}
\begin{equation}
\bigl(S_{\text{B}\,\mathcal{O}}^{\ol\lambda\,\ol\lambda}\bigr)_{\text{1PI}}+\frac 12\bigl(S_{\text{B}\,\mathcal{O}}^{\ol\lambda\,\ol\lambda}\bigr)_{\ol{\text{1PI}}}=
-\frac{4g^2 }{\epsilon}\bigl(
2\delta^{\dot{\gamma}}_{\phan{\alpha}\dot\alpha}\delta^{\dot{\beta}}_{\phan{\alpha}\dot \delta}
-\delta^{\dot{\beta}}_{\phan{\alpha}\dot\alpha}\delta^{\dot{\gamma}}_{\phan{\alpha}\dot \delta}
\bigr)\frac 1N\Bigl((c_Ac_B)-\frac{1}{N}(c_A)(c_B)\Bigr)
\end{equation}
and from this, the diagonal entry of the renormalisation constant $\mathcal{Z}_{\mathcal{O}}$ can be constructed for given choices of the spinor indices. As discussed in \subsecref{sec:composite-operator-insertionsN4}, the gauge dependence drops out of the complete renormalisation matrix element and hence also the corresponding anomalous dimension matrix element $\gamma_{\mathcal{O}}$ is independent of $\xi$. The latter can be determined from this expression as in \eqref{eq:RGE_comp_operators} or directly \eqref{eq:anomalous_dim_matrix}.

Let us now calculate the a non-vanishing contribution to the correlation function $S_{\mathcal{O}}^{\varphi\varphi\varphi}$ with three different scalar field flavours\footnote{We restrict the scalar field flavours here to avoid contributions from the following range-$2$ diagram with final state radiations 
	\begin{equation}
	\settoheight{\eqoff}{$\times$}%
	\setlength{\eqoff}{0.5\eqoff}%
	\addtolength{\eqoff}{-6.\unitlength}%
	\raisebox{\eqoff}{
		\fmfframe(2,0)(2,0){
			\begin{fmfchar*}(15,10)
			\fmfright{vop}
			\fmfforce{0 w, 1h}{v1}
			\fmfforce{0 w, 0.5h}{v2}
			\fmfforce{0 w, 0.2h}{v3}
			\fmfforce{0.66 w,0.8 h}{vc1}
			\fmfforce{0.33 w,0.8 h}{vc2}
			\fmfforce{0.66 w,0.2 h}{vc3}
			\fmf{photon}{vc1,vc2}
			\fmf{plain}{vc2,v1}
			\fmf{plain}{vc2,v2}
			\fmf{plain}{vc3,v3}
			\fmf{dashes_ar,right=0.2,tension=0.6}{vop,vc1}
			\fmf{dashes_ar,left=0.2,tension=0.6}{vop,vc3}
			\fmf{dashes_ar}{vc1,vc3}
			\fmffreeze
			\fmfposition
			\fmfiv{label=$\scriptstyle A \dot\alpha $,l.angle=90,l.dist=0.22w}{vloc(__vop)}
			\fmfiv{label=$\scriptstyle B \dot \beta $,l.angle=-90,l.dist=0.22w}{vloc(__vop)}
			\fmfiv{label=$\scriptstyle -p $,l.a=30,l.dist=0.1w}{vloc(__vc2)}
			\fmfiv{label=$\scriptstyle -p i $,l.a=50,l.dist=-0.01w}{vloc(__v1)}
			\fmfiv{label=$\scriptstyle 0 i $,l.a=-20,l.dist=0.1w}{vloc(__v2)}
			\fmfiv{label=$\scriptstyle 0 j $,l.a=-40,l.dist=0.1w}{vloc(__v3)}
			\fmfiv{decor.shape=circle,decor.filled=full,decor.size=5}{vloc(__vop)}
			\fmfiv{label=$\scriptstyle p$,l.a=0,l.dist=0.1w}{vloc(__vop)}
			\end{fmfchar*}
		}
	}
	\eqndot
	\end{equation}	
} %
$i\neq j\neq k$. It changes the length of the operator $\mathcal{O}$ and starts at order $\gym^3$ with the contribution
\begin{equation}
\begin{aligned}\label{eq:lambda2_phi3}
S_{\mathcal{O}}^{\varphi\varphi\varphi}&=
\Kop\Biggl[\Biggl(
\settoheight{\eqoff}{$\times$}%
\setlength{\eqoff}{0.5\eqoff}%
\addtolength{\eqoff}{-8.\unitlength}%
\raisebox{\eqoff}{
	\fmfframe(2,1)(2,0){
		\begin{fmfchar*}(20,15)
		\fmfright{vop}
		\fmfforce{0 w, 1h}{v1}
		\fmfforce{0 w, 0.5h}{v2}
		\fmfforce{0 w, 0h}{v3}
		\fmfforce{0.34 w,0.8 h}{vc1}
		\fmfforce{0.34 w,0.5 h}{vc2}
		\fmfforce{0.34 w,0.2 h}{vc3}
		\fmf{plain}{vc1,v1}
		\fmf{plain}{vc2,v2}
		\fmf{plain}{vc3,v3}
		\fmf{dashes_ar,right=0.2,tension=0.6}{vop,vc1}
		\fmf{dashes_ar,left=0.2,tension=0.6}{vop,vc3}
		\fmf{dashes_ar}{vc2,vc1}
		\fmf{dashes_ar}{vc2,vc3}
		\fmffreeze
		\fmfposition
		\fmfiv{label=$\scriptstyle A \dot\alpha a$,l.angle=90,l.dist=0.22w}{vloc(__vop)}
		\fmfiv{label=$\scriptstyle B \dot \beta a$,l.angle=-90,l.dist=0.22w}{vloc(__vop)}
		\fmfiv{label=$\scriptstyle -pi c_i$,l.a=20,l.dist=0.1w}{vloc(__v1)}
		\fmfiv{label=$\scriptstyle 0k c_k$,l.a=-70,l.dist=0.02w}{vloc(__v2)}
		\fmfiv{label=$\scriptstyle 0j c_j$,l.a=-20,l.dist=0.1w}{vloc(__v3)}
		\fmfiv{decor.shape=circle,decor.filled=full,decor.size=5}{vloc(__vop)}
		\fmfiv{label=$\scriptstyle p$,l.a=0,l.dist=0.1w}{vloc(__vop)}
		\end{fmfchar*}
	}
}
\Biggr)_{\text{1PI}}
\Biggr]\\
&=\text{nscf}(v_i,\mathcal{O},v_j,v_k)
\text{K}\bigl[\hat{I}^{\mu_2\mu_3\mu_4,\mu_1}_{(3,1)}(p)\bigr]
\varepsilon_{\dot{\alpha}\dot{\gamma}}\varepsilon^{\dot{\beta}\dot{\delta}}
(T_{\mu_1\mu_4\mu_3\mu_2})^{\dot \gamma}_{\phan{\alpha}\dot{\delta}}\\
&=\frac{4\pi \complexi g^{3}\mu^\epsilon}{\epsilon} 
\frac{(c_i[c_j,c_k])}{N^{\frac 32}}(\bar\Sigma^{ikj})^{AB}\,\delta^{\dot{\beta}}_{\phan{\alpha}\dot{\alpha}}\eqncom
\end{aligned}
\end{equation}
where the integral was already given in \eqref{eq:K_integral_1} and the remaining factors stem from
\begin{equation}
\text{Rest}(v_i,\mathcal{O},v_k,v_j)=\gym^3 \mu^\epsilon(\bar{\Sigma}^i)^{AC}(\Sigma^{k})_{CD}(\bar{\Sigma}^j)^{DB}(c_i[c_j,c_k])
\equiv \mu^\epsilon(\gym^2N)^{\frac 32}(\bar\Sigma^{ikj})^{AB}\frac{(c_i[c_j,c_k])}{N^{\frac 32}}\eqndot
\end{equation}
The divergent contribution in \eqref{eq:lambda2_phi3} must be renormalised by the 1PI renormalisation matrix element $\mu^{\epsilon}Z^{\varphi\varphi\varphi}_{\cO}$ and hence we find that the operator $\mathcal{O}$ mixes non-trivially under renormalisation with a composite operator containing three real scalars in \NfSYMt. This process starts at order $g^3\sim\lambda^{\frac 32}$ in agreement with the fact that length-changing processes do not occur at order $\lambda$ in \NfSYMt. 

Note that the divergent contributions to the correlation functions \eqref{eq:comp_O_correlation_func} that we discussed in this subsection give two entries of the renormalisation matrix \eqref{eq:anomalous_dim_matrix}. In principle, also the operators $\tr(\varphi^i\varphi^j\varphi^k)$ and $\tr(\bar{\cF}_{\dot{\alpha}\dot{\beta}}\varphi)$ can have the same quantum numbers\footnote{These quantum numbers to order $g^3$ are its classical scaling dimension $\Delta_{\mathcal{O}}^0=3$, its $R$-symmetry Cartan charges characterised through $A$ and $B$ and the $\splbar{2}$-charge characterised through $\dot{\alpha}$ and $\dot{\beta}$.} as $\cO$ for certain choices of the spinor and flavour indices in \eqref{eq:example_operator}. For these choices, there are further entries in the complete renormalisation matrix $\cZ_\cO$.

\section{The \tHooft limit}\label{sec:tHooft_limit}
So far in this chapter, we discussed general aspects of renormalised field theories. As mentioned in the introduction, it is particularly interesting to investigate \NfSYMt and its deformations in the \tHooft limit where the parent theory was found to be integrable at one-loop order. Originally, the \tHooft limit was proposed in \cite{'tHooft:1973jz} to separate Feynman diagrams according to their topology: in \emph{gauge-invariant} correlation functions only planar diagrams survive. In these correlation functions, diagrams are planar if all their (closed) colour lines can be drawn in a plane without intersecting lines. For gauge groups \UN and \SUN, the \tHooft limit is given by simultaneously taking the colour degrees of freedom $N\to \infty$ and the Yang-Mills coupling $\gym\to 0$, so that the product $\lambda=\gym^2N$ remains fixed. 

To determine which terms survive in the \tHooft limit in a perturbative calculation, we sort all contributions according to the powers of $N$ that are generated by their respective colour structures. In a diagrammatic expansion, these powers are most conveniently determined in the colour-space double-line notation, see \cite{'tHooft:1973jz} or \cite[\chap{80}]{Srednicki:2007} for a textbook presentation. In this notation, fields carry the two fundamental matrix indices of the colour generators they are contracted with. For gauge groups \UN and \SUN, a field $X$ in the adjoint representation takes the form
\begin{equation}\label{eq:matrix_fields}
X^i{}_j=\sum_{a=s}^{N^2-1}X^a(\T^a)^i{}_j\eqncom\qquad 
s=\begin{cases}
1&\SUN\\
0&\UN
\end{cases}
\end{equation}
and it is connected in diagrams via its gauge group indices $i$ and $j$. The colour part of a propagator that connects an adjoint field with its conjugate is given by
\begin{equation}\label{eq:double_line_propagator}
\sum_{a=\colors}^{N^2-1}(\T^a)^i{}_j(\T^a)^k{}_l
=\delta^i_l\delta^k_j-\frac{\colors}{N}\delta^i_j\delta^k_l
=
\settoheight{\eqoff}{$\times$}%
\setlength{\eqoff}{0.5\eqoff}%
\addtolength{\eqoff}{-2\unitlength}%
\raisebox{\eqoff}{ 
	\fmfframe(1,2)(1,2){%
		\begin{fmfgraph*}(15,5)
		\fmfstraight
		\fmfpen{10}
		\fmfforce{0w,0h}{i}
		\fmfforce{1w,0h}{o}
		\fmf{plain_n}{i,o}
		\fmffreeze
		\fmfdraw
		\fmfpen{8}
		\fmf{iplain_n,foreground=white}{i,o}
		\fmfdraw
		\fmf{leftrightarrows,foreground=black}{i,o}
		\fmfdraw
		\fmfiv{l=$\scriptstyle i$,l.a=90,l.d=7}{vloc(__i)}
		\fmfiv{l=$\scriptstyle j$,l.a=-90,l.d=7}{vloc(__i)}
		\fmfiv{l=$\scriptstyle l$,l.a=90,l.d=7}{vloc(__o)}
		\fmfiv{l=$\scriptstyle k$,l.a=-90,l.d=7}{vloc(__o)}
		\fmfdraw
		\end{fmfgraph*}%
	}
}
\,
{}-{}\frac{\colors}{N}{}\,%
\settoheight{\eqoff}{$\times$}%
\setlength{\eqoff}{0.5\eqoff}%
\addtolength{\eqoff}{-2\unitlength}%
\raisebox{\eqoff}{ 
	\fmfframe(1,2)(1,2){%
		\begin{fmfgraph*}(15,5)
		\fmfstraight
		\fmfpen{10}
		\fmfforce{0w,0h}{i}
		\fmfforce{1w,0h}{o}
		\fmf{interrupted_plain_n}{i,o}
		\fmffreeze
		\fmfdraw
		\fmfpen{8}
		\fmf{iinterrupted_plain_n,foreground=white}{i,o}
		\fmfdraw
		\fmf{leftrightarrows_interruptedold,foreground=black}{i,o}
		\fmfdraw
		\fmfiv{l=$\scriptstyle i$,l.a=90,l.d=7}{vloc(__i)}
		\fmfiv{l=$\scriptstyle j$,l.a=-90,l.d=7}{vloc(__i)}
		\fmfiv{l=$\scriptstyle l$,l.a=90,l.d=7}{vloc(__o)}
		\fmfiv{l=$\scriptstyle k$,l.a=-90,l.d=7}{vloc(__o)}
		\fmfdraw
		\end{fmfgraph*}%
	}
}
\eqncom
\end{equation}
where the colour flow is directed from upper to lower fundamental indices. The $\frac{s}{N}$-term guarantees that the propagator in the \SUN theory is traceless. When we connect two \U{1} modes which have the colour generator $(\T^0)^i{}_j=\frac{1}{\sqrt{N}} (\one)^i{}_j$ by this type of propagator we find
\begin{equation}\label{eq:U1_mode_vanishes}
\frac{1}{\sqrt{N}^2}\Bigl(\,
\underbrace{%
	\settoheight{\eqoff}{$\times$}%
	\setlength{\eqoff}{0.5\eqoff}%
	\addtolength{\eqoff}{-4.5\unitlength}%
	\raisebox{\eqoff}{ 
		\fmfframe(3,2)(3,2){%
			\begin{fmfgraph*}(20,5)
			\fmfstraight
			\fmfpen{10}
			\fmfleft{i}
			\fmfright{o}
			\fmf{plain_n}{i,o}
			\fmfdraw
			\fmfpen{8}
			\fmf{iplain_n,foreground=white}{i,o}
			\fmfdraw
			\fmf{leftrightarrows,foreground=black}{i,o}
			\fmfdraw
			\fmfpen{1}
			\fmfcmd{z3=(vloc(__i)+(-10,0)); draw (vloc(__i)+(0,-4.5)){dir 180}...z3...(vloc(__i)+(0,4.5)){dir 0} withcolor (0.65,0.65,0.65);}
			\fmfcmd{z4=(vloc(__o)+(10,0)); draw (vloc(__o)+(0,-4.5)){dir 0}...z4...(vloc(__o)+(0,4.5)){dir 180} withcolor (0.65,0.65,0.65);}
			\fmfiv{l=$\scriptstyle i$,l.a=90,l.d=7}{vloc(__i)}
			\fmfiv{l=$\scriptstyle j$,l.a=-90,l.d=7}{vloc(__i)}
			\fmfiv{l=$\scriptstyle l$,l.a=90,l.d=7}{vloc(__o)}
			\fmfiv{l=$\scriptstyle k$,l.a=-90,l.d=7}{vloc(__o)}
			\fmfdraw
			\end{fmfgraph*}%
		}
	}%
}_{N^1}\,%
{}-{}\frac{\colors}{N}\,%
\underbrace{%
	\settoheight{\eqoff}{$\times$}%
	\setlength{\eqoff}{0.5\eqoff}%
	\addtolength{\eqoff}{-4.5\unitlength}%
	\raisebox{\eqoff}{ 
		\fmfframe(3,2)(3,2){%
			\begin{fmfgraph*}(20,5)
			\fmfstraight
			\fmfpen{10}
			\fmfleft{i}
			\fmfright{o}
			\fmf{interrupted_plain_n}{i,o}
			\fmffreeze
			\fmfdraw
			\fmfpen{8}
			\fmf{iinterrupted_plain_n,foreground=white}{i,o}
			\fmfdraw
			\fmf{leftrightarrows_interruptedold,foreground=black}{i,o}
			\fmfdraw 
			\fmfpen{1}
			\fmfcmd{z3=(vloc(__i)+(-10,0)); draw (vloc(__i)+(0,-4.5)){dir 180}...z3...(vloc(__i)+(0,4.5)){dir 0} withcolor (0.65,0.65,0.65);}
			\fmfcmd{z4=(vloc(__o)+(10,0)); draw (vloc(__o)+(0,-4.5)){dir 0}...z4...(vloc(__o)+(0,4.5)){dir 180} withcolor (0.65,0.65,0.65);}
			\fmfiv{l=$\scriptstyle i$,l.a=90,l.d=7}{vloc(__i)}
			\fmfiv{l=$\scriptstyle j$,l.a=-90,l.d=7}{vloc(__i)}
			\fmfiv{l=$\scriptstyle l$,l.a=90,l.d=7}{vloc(__o)}
			\fmfiv{l=$\scriptstyle k$,l.a=-90,l.d=7}{vloc(__o)}
			\fmfdraw
			\end{fmfgraph*}%
		}
	}%
}_{N^2}\,
\Bigr)
{}={}\left(1-\colors\right)\eqncom
\end{equation}
where we used that each closed fundamental colour-line loop corresponds to $\delta^i{}_i$ and yields a factor of $N$. We see that \eqref{eq:U1_mode_vanishes} vanishes for $s=1$ in the \SUN theory but is non-vanishing in the \UN theory with $s=0$. This is a first encounter of a contribution which must be present in the \tHooft limit despite its naive suppression factor\footnote{The naive idea to discard the $\frac{s}{N}$-term fails since propagators are not gauge invariant quantities without open colour lines and hence power counting in $N$ is meaningless.}  of $N^{-1}$. Apart from propagators, there are three- and four-point interactions in the actions of \NfSYMt and its deformations. In colour space, each planar single-trace part of an interaction with $n$ external legs is represented by a connected planar double-line graph with $n$ open double lines. For example the colour-space part of the scalar propagator, fermion-scalar and scalar-scalar interactions are given by
\begin{equation}\label{eq:double_line_building_blocs}
\begin{aligned}
\tr\bigl(\phi\ol{\phi}\,)
&\hat{=}
\Bigl(
\settoheight{\eqoff}{$\times$}%
\setlength{\eqoff}{0.5\eqoff}%
\addtolength{\eqoff}{-2\unitlength}%
\raisebox{\eqoff}{ 
	\fmfframe(1,2)(1,2){%
		\begin{fmfgraph*}(10,5)
		\fmfstraight
		\fmfpen{6}
		\fmfforce{0w,0h}{i}
		\fmfforce{1w,0h}{o}
		\fmf{plain_n}{i,o}
		\fmffreeze
		\fmfdraw
		\fmfpen{4}
		\fmf{iplain_n,foreground=white}{i,o}
		\fmfdraw
		\fmfpen{1}
		\fmfdraw
		\fmfdraw
		\end{fmfgraph*}%
	}
}
{}-{}\frac{\colors}{N}{}%
\settoheight{\eqoff}{$\times$}%
\setlength{\eqoff}{0.5\eqoff}%
\addtolength{\eqoff}{-2\unitlength}%
\raisebox{\eqoff}{ 
	\fmfframe(1,2)(1,2){%
		\begin{fmfgraph*}(10,5)
		\fmfstraight
		\fmfpen{6}
		\fmfforce{0w,0h}{i}
		\fmfforce{1w,0h}{o}
		\fmf{interrupted_plain_n}{i,o}
		\fmffreeze
		\fmfdraw
		\fmfpen{4}
		\fmf{iinterrupted_plain_n,foreground=white}{i,o}
		\fmfdraw
		\fmfpen{1}
		\fmfdraw
		\fmfdraw
		\end{fmfgraph*}%
	}
}
\Bigr)
\eqncom&
\gym\tr\bigl(\lambda\phi\lambda\bigr)
&\hat{=} \hspace{.2cm}
\ifpdf
\settoheight{\eqoff}{$+$}%
\setlength{\eqoff}{0.5\eqoff}%
\addtolength{\eqoff}{-1.8\unit}%
\raisebox{\eqoff}{%
	\includegraphics[angle={0},scale=0.04,trim=1cm 0cm 4cm 0]{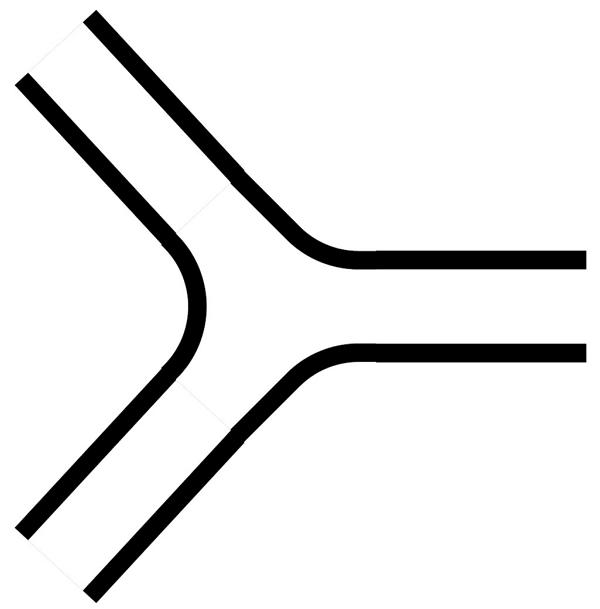} 
}\quad
\else
\scalebox{1}{
	\settoheight{\eqoff}{$+$}%
	\setlength{\eqoff}{0.5\eqoff}%
	\addtolength{\eqoff}{-2.3\unit}%
	\raisebox{\eqoff}{%
		\rotatebox{0}{
			\begin{pspicture}(-2.5,-2.5)(2.5,2.5)
			\recthreevertex{.2}{0}{0}
			\psset{linecolor=black,doubleline=true}
			\psline(-2,2.1)(-0.8,0.8)
			\psline(2.3,0)(0.6,0)
			\psline(-2,-2.1)(-0.8,-0.8)
			\psset{linecolor=black,doubleline=false,linestyle=solid}
			\end{pspicture}
		}
	}
}
\fi
\hspace{-.3cm}
\eqncom&
\\
\gym^2\tr\bigl(\phi\phi\ol{\phi}\,\ol{\phi}\,\bigr)
&\hat{=}  \hspace{.2cm}
\ifpdf
\settoheight{\eqoff}{$+$}%
\setlength{\eqoff}{0.5\eqoff}%
\addtolength{\eqoff}{-1.8\unit}%
\raisebox{\eqoff}{%
	\includegraphics[angle={0},scale=0.04,trim=2cm 0 2cm 0]{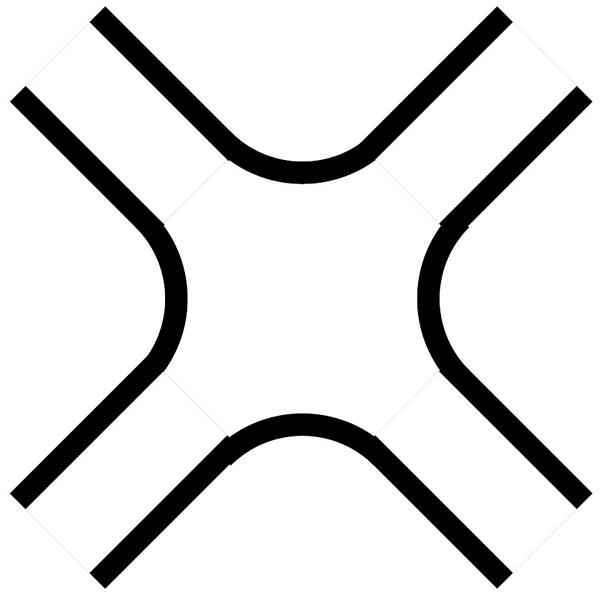} 
}
\else
\scalebox{1}{
	\settoheight{\eqoff}{$+$}%
	\setlength{\eqoff}{0.5\eqoff}%
	\addtolength{\eqoff}{0.5\unit}%
	\raisebox{\eqoff}{%
		\rotatebox{-45}{
			\begin{pspicture}(-2,-2)(2,2)
			\fourvertex{0.22}{0}{0}
			\psset{linecolor=black,doubleline=true}
			\psline(0,2.3)(0,1)
			\psline(0,-2.3)(0,-1)
			\psline(2.3,0)(1,0)
			\psline(-2.3,0)(-1,0)
			\end{pspicture}
		}
	}
}
\fi
\eqncom&
\frac{\gym^2}{N}\tr\bigl(\phi\phi\bigr)\tr\bigl(\ol{\phi}\,\ol{\phi}\,\bigr)
&\hat{=}  \hspace{.2cm}
\ifpdf
\settoheight{\eqoff}{$+$}%
\setlength{\eqoff}{0.5\eqoff}%
\addtolength{\eqoff}{-1.8\unit}%
\raisebox{\eqoff}{%
	\includegraphics[angle={0},scale=0.04,trim=2cm 0 2cm 0]{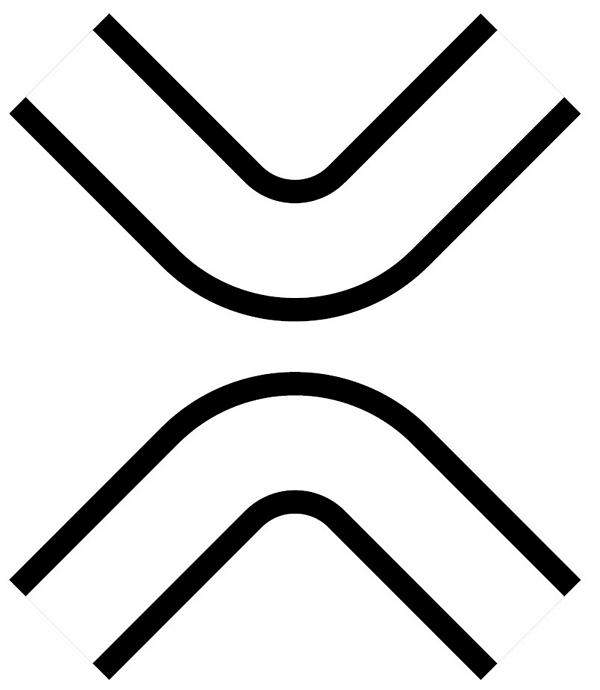}
}
\else
\scalebox{1}{
	\settoheight{\eqoff}{$+$}%
	\setlength{\eqoff}{0.5\eqoff}%
	\addtolength{\eqoff}{-2\unit}%
	\raisebox{\eqoff}{%
		\begin{pspicture}(-2,-2)(2,2)
		\psset{linecolor=black,doubleline=true,linestyle=solid}
		\psline[linearc=\linearc](-1.5,1.8)(0,0.3)(1.5,1.8)
		\psline[linearc=\linearc](-1.5,-1.8)(0,-0.3)(1.5,-1.8)
		\end{pspicture}
	}
}
\fi
\eqncom
\end{aligned}
\end{equation}
where we dropped the colour line arrows and suppressed all flavour and spinor indices.

In analogy to the fundamental interactions in  \eqref{eq:double_line_building_blocs}, we can define length-$L$ single-trace composite operators in colour space as 
\begin{equation}\label{eq:composite_operator_doubleline}
\mathcal{O}_{\text{i}}(x)=
\ifpdf
\settoheight{\eqoff}{$+$}%
\setlength{\eqoff}{0.5\eqoff}%
\addtolength{\eqoff}{-2\unit}%
\raisebox{\eqoff}{%
	\includegraphics[angle={0},scale=0.05,trim=2cm 4cm 2cm 0]{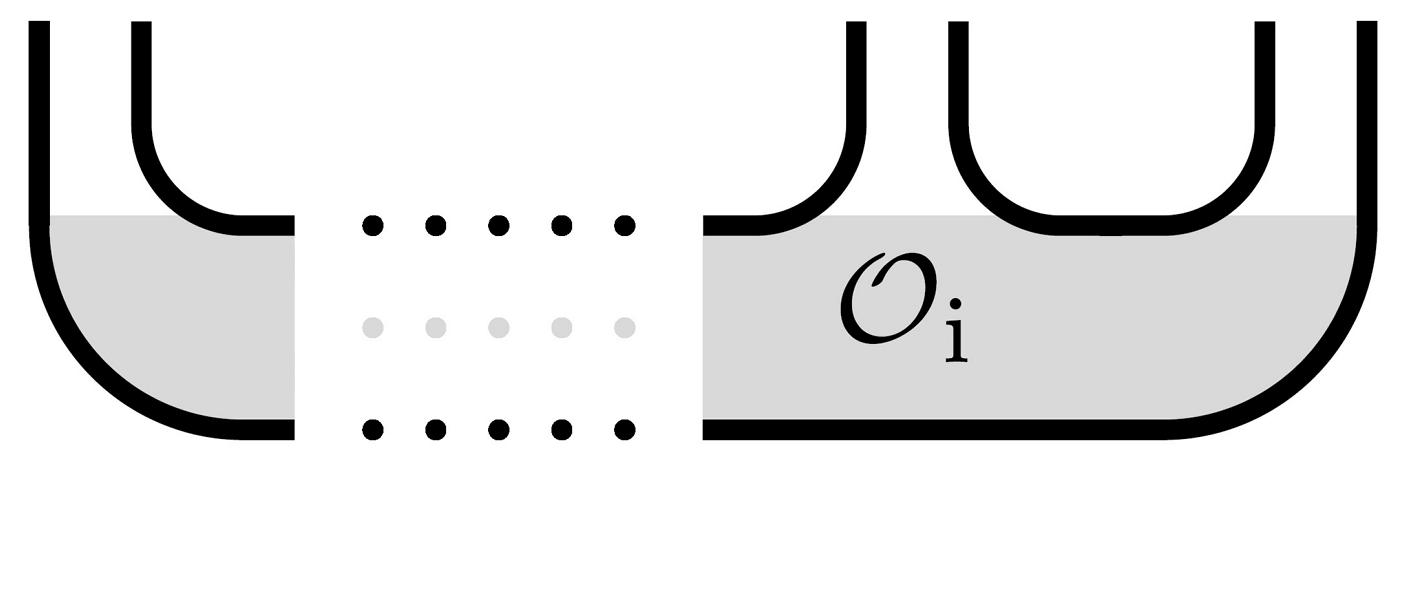}
}
\else
\settoheight{\eqoff}{$+$}%
\setlength{\eqoff}{0.5\eqoff}%
\addtolength{\eqoff}{1.5\unit}%
\raisebox{\eqoff}{%
	\scalebox{0.8}{
		\rotatebox{-90}{
			\begin{pspicture}(-1,-.5)(2,9)
			\urinsert{2}{9}
			\orvertex{2}{6}
			\psline[linestyle=dotted](.5,4)(.5,2)
			\psline[linestyle=dotted,linecolor=ogray](1.25,4)(1.25,2)
			\psline[linestyle=dotted](2,4)(2,2)
			\drinsert{2}{0}
			\rput{90}(1.1,6){$\scriptstyle \mathcal{O}_{\text{i}}$}
			\end{pspicture}
		}
	}
}
\fi
=\frac{1}{\sqrt{N^{L}}}\tr\bigl(\mathcal{A}_1\mathcal{A}_2\dots \mathcal{A}_L\bigr)\eqncom
\end{equation}
where the composite operator $\mathcal{O}_{\text{i}}(x)$ agrees with \eqref{eq:operator_as_tensorproduct} up to the flavour normalisation constant which depends on the explicit field-flavours in the operator, see e.g.\ \cite{Minahan:2010js}. The colour normalisation ensures that the planar leading term\footnote{The planar term in a free two-point correlation function is unique since the involved composite operators are graded cyclically invariant.} in two-point correlation functions scales as $N^0$, since
\begin{equation}\label{eq:operator_normalisation}
\vacl\T\mathcal{O}_{\text{i}}\ol{\mathcal{O}}_{\text{i}}\vac_{\text{free}} =\hspace{0.1cm}
\ifpdf
\settoheight{\eqoff}{$+$}%
\setlength{\eqoff}{0.5\eqoff}%
\addtolength{\eqoff}{-3.\unit}%
\raisebox{\eqoff}{%
	\includegraphics[angle={0},scale=0.08,trim=1cm 0cm 1cm 0]{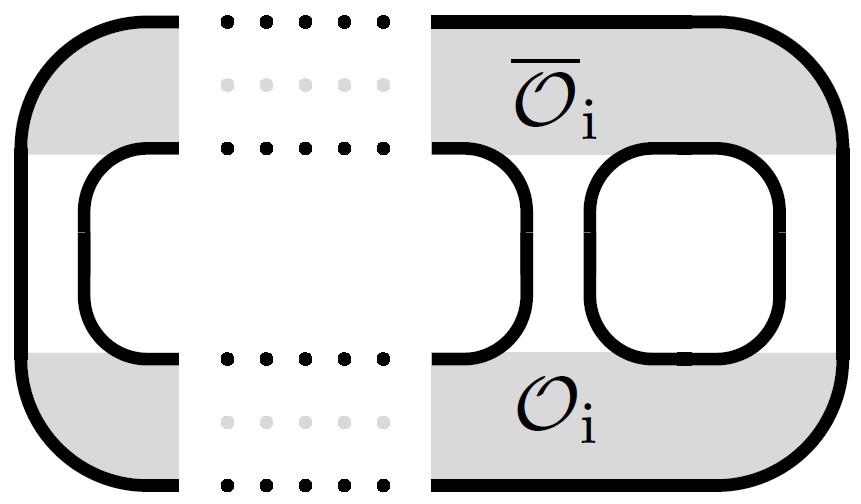}
}
\else
\settoheight{\eqoff}{$+$}%
\setlength{\eqoff}{0.5\eqoff}%
\addtolength{\eqoff}{.5\unit}%
\raisebox{\eqoff}{%
	\scalebox{0.8}{
		\rotatebox{-90}{
			\begin{pspicture}(-1,-.5)(2,9)
			\ulinsert{-3.5}{9}
			\ulinsert{-3.5}{9}
			\olvertex{-3.5}{6}
			\psline[linestyle=dotted](-2,4)(-2,2)
			\psline[linestyle=dotted,linecolor=ogray](-2.75,4)(-2.75,2)
			\psline[linestyle=dotted](-3.5,4)(-3.5,2)
			\dlinsert{-3.5}{0}
			\rput{90}(-2.6,6){$\scriptstyle \ol{\mathcal{O}}_{\text{i}}$}
			\urinsert{2}{9}
			\orvertex{2}{6}
			\psline[linestyle=dotted](.5,4)(.5,2)
			\psline[linestyle=dotted,linecolor=ogray](1.25,4)(1.25,2)
			\psline[linestyle=dotted](2,4)(2,2)
			\drinsert{2}{0}
			\rput{90}(1.1,6){$\scriptstyle \mathcal{O}_{\text{i}}$}
			\end{pspicture}
		}
	}
}
\fi
+\mathcal{O}\Bigl(\frac{1}{N^2}\Bigr)
= 
\frac{\delta^{i_1}{}_{i_1}\dots \delta^{i_L}{}_{i_L}}{N^L}+\mathcal{O}\Bigl(\frac{1}{N^2}\Bigr)=1+\mathcal{O}\Bigl(\frac{1}{N^2}\Bigr)\eqncom
\end{equation}
where the suppressed terms stem from $n$ non-planar double-line contractions in the two-point function and each such term comes with a suppression factor of $N^{-2n}$. We can repeat this exercise for the splitting of an initial single-trace operator into $n$ final ones and find that such processes are suppressed by a factor of at least $N^{-(n-1)}$ and hence the mixture of composite operators with different numbers of colour traces is suppressed in the \tHooft limit.

When we allow for interactions, we can use single-trace vertices like the ones displayed in \eqref{eq:double_line_building_blocs} to calculate the $N$- and $\gym$-power in a gauge-invariant correlation function. We find that each length-preserving ($\Delta L=0$) and length-changing ($\Delta L=1$) interaction yields a factor of $N\gym^2=\lambda$ and $N^{\frac 12}\gym=\lambda^{\frac 12}$, respectively. In a single-line representation these statements can be depicted for two-point functions in the \tHooft limit as 
\begin{equation}
\begin{aligned}\label{eq:tHooft_limit_transitions}
&
\settoheight{\eqoff}{$\times$}%
\setlength{\eqoff}{0.5\eqoff}%
\addtolength{\eqoff}{-7\unitlength}%
\raisebox{\eqoff}{ 
	\fmfframe(1,2)(1,2){%
		\begin{fmfgraph*}(20,10)
		\fmfstraight
		\fmfforce{0.00w,0h}{i0}
		\fmfforce{0.11w,0h}{i1}
		\fmfforce{0.22w,0h}{i2}
		\fmfforce{0.33w,0h}{i3}
		\fmfforce{0.44w,0h}{i4}
		\fmfforce{0.55w,0h}{i5}
		\fmfforce{0.66w,0h}{i6}
		\fmfforce{0.77w,0h}{i7}
		\fmfforce{0.88w,0h}{i8}
		\fmfforce{1.00w,0h}{i9}
		\fmfforce{0.00w,1h}{f0}
		\fmfforce{0.11w,1h}{f1}
		\fmfforce{0.22w,1h}{f2}
		\fmfforce{0.33w,1h}{f3}
		\fmfforce{0.44w,1h}{f4}
		\fmfforce{0.55w,1h}{f5}
		\fmfforce{0.66w,1h}{f6}
		\fmfforce{0.77w,1h}{f7}
		\fmfforce{0.88w,1h}{f8}
		\fmfforce{1.00w,1h}{f9}
		\fmf{plain}{i0,f0}
		\fmf{plain}{i1,f1}
		\fmf{plain}{i8,f8}
		\fmf{plain}{i9,f9}
		\fmfforce{0.44w,0.5h}{d1}
		\fmfforce{0.55w,0.5h}{d2}
		\fmfdraw
		\fmfpen{1}
		\fmfset{dot_len}{2mm}
		\fmf{dots}{d1,d2}
		\fmfdraw
		\fmf{plain,width=1.3mm,foreground=(0.65,,0.65,,0.65)}{i0,i9}
		\fmf{plain,width=1.3mm,foreground=(0.65,,0.65,,0.65)}{f0,f9}
		\end{fmfgraph*}%
	}
}
&&\propto 1
\eqncom
&
\settoheight{\eqoff}{$\times$}%
\setlength{\eqoff}{0.5\eqoff}%
\addtolength{\eqoff}{-7\unitlength}%
\raisebox{\eqoff}{ 
	\fmfframe(1,2)(1,2){%
		\begin{fmfgraph*}(20,10)
		\fmfstraight
		\fmfforce{0.00w,0h}{i0}
		\fmfforce{0.11w,0h}{i1}
		\fmfforce{0.22w,0h}{i2}
		\fmfforce{0.33w,0h}{i3}
		\fmfforce{0.44w,0h}{i4}
		\fmfforce{0.55w,0h}{i5}
		\fmfforce{0.66w,0h}{i6}
		\fmfforce{0.77w,0h}{i7}
		\fmfforce{0.88w,0h}{i8}
		\fmfforce{1.00w,0h}{i9}
		\fmfforce{0.00w,1h}{f0}
		\fmfforce{0.11w,1h}{f1}
		\fmfforce{0.22w,1h}{f2}
		\fmfforce{0.33w,1h}{f3}
		\fmfforce{0.44w,1h}{f4}
		\fmfforce{0.55w,1h}{f5}
		\fmfforce{0.66w,1h}{f6}
		\fmfforce{0.77w,1h}{f7}
		\fmfforce{0.88w,1h}{f8}
		\fmfforce{1.00w,1h}{f9}
		\fmf{plain}{i0,f0}
		\fmf{plain}{i9,f9}
		\fmf{phantom}{i3,f6}
		\fmf{phantom}{i6,f3}
		\fmffreeze
		\fmfposition
		\fmfipath{p[]}
		\fmfiset{p1}{vpath(__i3,__f6)}
		\fmfiset{p2}{vpath(__i6,__f3)}
		\fmfi{plain}{p1}
		\fmfi{plain}{p2}
		\fmfdraw
		\fmfiv{decor.shape=circle,decor.size=15 thin,decor.filled=empty,label={\scriptsize s.t.},label.dist=-0.12w}{point length(p1)/2 of p1}
		\fmfdraw
		\fmfforce{0.15w,0.5h}{d1}
		\fmfforce{0.26w,0.5h}{d2}
		\fmfforce{0.73w,0.5h}{d3}
		\fmfforce{0.84w,0.5h}{d4}
		\fmfpen{1}
		\fmfset{dot_len}{2mm}
		\fmf{dots}{d1,d2}
		\fmf{dots}{d3,d4}
		\fmfdraw
		\fmf{plain,width=1.3mm,foreground=(0.65,,0.65,,0.65)}{i0,i9}
		\fmf{plain,width=1.3mm,foreground=(0.65,,0.65,,0.65)}{f0,f9}
		\end{fmfgraph*}%
	}
}
&&\propto \lambda
\eqncom\qquad
&
\settoheight{\eqoff}{$\times$}%
\setlength{\eqoff}{0.5\eqoff}%
\addtolength{\eqoff}{-7\unitlength}%
\raisebox{\eqoff}{ 
	\fmfframe(1,2)(1,2){%
		\begin{fmfgraph*}(20,10)
		\fmfstraight
		\fmfforce{0.00w,0h}{i0}
		\fmfforce{0.11w,0h}{i1}
		\fmfforce{0.22w,0h}{i2}
		\fmfforce{0.33w,0h}{i3}
		\fmfforce{0.44w,0h}{i4}
		\fmfforce{0.55w,0h}{i5}
		\fmfforce{0.66w,0h}{i6}
		\fmfforce{0.77w,0h}{i7}
		\fmfforce{0.88w,0h}{i8}
		\fmfforce{1.00w,0h}{i9}
		\fmfforce{0.00w,1h}{f0}
		\fmfforce{0.11w,1h}{f1}
		\fmfforce{0.22w,1h}{f2}
		\fmfforce{0.33w,1h}{f3}
		\fmfforce{0.44w,1h}{f4}
		\fmfforce{0.55w,1h}{f5}
		\fmfforce{0.66w,1h}{f6}
		\fmfforce{0.77w,1h}{f7}
		\fmfforce{0.88w,1h}{f8}
		\fmfforce{1.00w,1h}{f9}
		\fmfforce{0.5w,0.5h}{m}
		\fmfforce{0.5w,1h}{f10}
		\fmf{plain}{m,f10}
		\fmf{plain}{i0,f0}
		\fmf{plain}{i9,f9}
		\fmf{phantom}{i3,f6}
		\fmf{phantom}{i6,f3}
		\fmffreeze
		\fmfposition
		\fmfipath{p[]}
		\fmfiset{p1}{vpath(__i3,__f6)}
		\fmfiset{p11}{subpath (0,length(p1)/2) of p1}
		\fmfiset{p2}{vpath(__i6,__f3)}
		\fmfiset{p21}{subpath (0,length(p2)/2) of p2}
		\fmfi{plain}{p11}
		\fmfi{plain}{p21}
		\fmfdraw
		\fmfiv{decor.shape=circle,decor.size=15 thin,decor.filled=empty,label={\scriptsize s.t.},label.dist=-0.12w}{point length(p1)/2 of p1}
		\fmfdraw
		\fmfforce{0.15w,0.5h}{d1}
		\fmfforce{0.26w,0.5h}{d2}
		\fmfforce{0.73w,0.5h}{d3}
		\fmfforce{0.84w,0.5h}{d4}
		\fmfpen{1}
		\fmfset{dot_len}{2mm}
		\fmf{dots}{d1,d2}
		\fmf{dots}{d3,d4}
		\fmfdraw
		\fmf{plain,width=1.3mm,foreground=(0.65,,0.65,,0.65)}{i0,i9}
		\fmf{plain,width=1.3mm,foreground=(0.65,,0.65,,0.65)}{f0,f9}
		\end{fmfgraph*}%
	}
}
&&\propto \sqrt{\lambda}
\eqncom\qquad
&
\end{aligned}
\end{equation}
where each white blobs represents one single-trace interaction of order $\ell=1$ or $\ell=2$ in the coupling constant $\gym$. At generic order $\gym^\ell$, we can also connect multiple single-trace interactions to form a connected interaction kernel which is attached through $r_i$ and $r_f$ external legs to the initial and final composite operators and has a total number of $r_i+r_f\leq \ell+2$ external legs. In the asymptotic regime, i.e.\ where the initial and final operator lengths exceeds the respective interaction ranges $r_i$ and $r_f$, relation \eqref{eq:tHooft_limit_transitions} can straightforwardly be generalised: the blobs are promoted to single-trace interaction kernels with $\ell$ powers of $\gym$. Such diagrams with planar interaction kernels contribute to correlation functions at order $\lambda^\frac{\ell}{2}$ in the \tHooft coupling and non-planar interaction kernels are suppressed with at least $N^{-2}$.

\subsection{Finite-size effects}\label{sec:finite-size-effects}
So far, we discussed asymptotic contributions to two-point correlation functions in which the lengths of the initial and final composite operators exceed the respective interaction ranges $r_i$ and $r_f$ of all interaction kernels. In this subsection, we discuss the so-called finite-size effects that start to occur when either the initial or the final operator length meets the interaction range $r_i$ or $r_f$ of the interaction kernel. 
 
\subsubsection{Wrapping}
First, we have the wrapping effect\footnote{In the context of \AdSCFT integrability, wrapping was discussed in \cite{Ambjorn:2005wa}.} which was systematically analysed in \cite{Sieg:2005kd}. It starts to occur in two-point correlation functions involving an initial length-$L_i$ and a final length-$L_f$ composite operator at loop order $K\geq \frac 12(L_i+L_f)$, when the interactions contain $L_i+L_f$ or more power of $\gym$. At this loop order, it is possible to build a connected chain of interactions that entirely wraps around the initial or final composite operator. The interaction kernel of such a process appears to be non-planar when drawn. A simple three-loop example of this in colour space is\footnote{We do not choose particular field flavours but with the interactions from \eqref{eq:deformed_action_complex_scalars2} it is clear that the drawn diagrams exist in general.}
\begin{equation}
\label{eq:wrapping_example}
\ifpdf
\settoheight{\eqoff}{$+$}%
\setlength{\eqoff}{0.5\eqoff}%
\addtolength{\eqoff}{-3.5\unit}%
\raisebox{\eqoff}{%
	\includegraphics[angle={0},scale=0.08,trim=0cm 0cm 0cm 0]{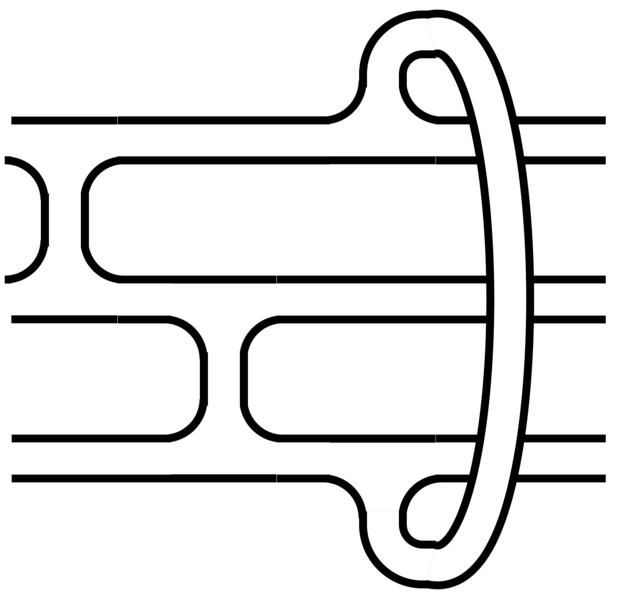}
}
\else
\scalebox{1}{
	\settoheight{\eqoff}{$+$}%
	\setlength{\eqoff}{0.5\eqoff}%
	\addtolength{\eqoff}{-5\unit}%
	\raisebox{\eqoff}{%
		\begin{pspicture}(-8.,-2.)(9.,8.)
		\threevertex{-3}{6}{-90}
		\threevertex{3}{6}{90}
		\threevertex{-3}{3}{90}
		\threevertex{0}{3}{-90}
		\threevertex{0}{0}{90}
		\threevertex{3}{0}{-90}
		\psset{linecolor=black,doubleline=true,linestyle=solid}
		\psline(-2,6)(2,6)
		\psline(4,6)(7,6)
		\psline(-2,3)(-1,3)
		\psline(1,3)(7,3)
		\psline(-4,0)(-1,0)
		\psline(1,0)(2,0)
		\psline(4,0)(7,0)
		\psline(-3,5)(-3,4)
		\psline(0,2)(0,1)
		\psline[linearc=\linearc](3,7)(3,8)(4,8)
		\psline[linearc=\linearc](3,-1)(3,-2)(4,-2)
		\psbezier(4,8)(6,8.5)(6,-2.5)(4,-2)
		\end{pspicture}
	}
}
\fi
\qquad
\eqncom\qquad
\ifpdf
\settoheight{\eqoff}{$+$}%
\setlength{\eqoff}{0.5\eqoff}%
\addtolength{\eqoff}{-3.5\unit}%
\raisebox{\eqoff}{%
	\includegraphics[angle={0},scale=0.08,trim=0cm 0cm 0cm 0]{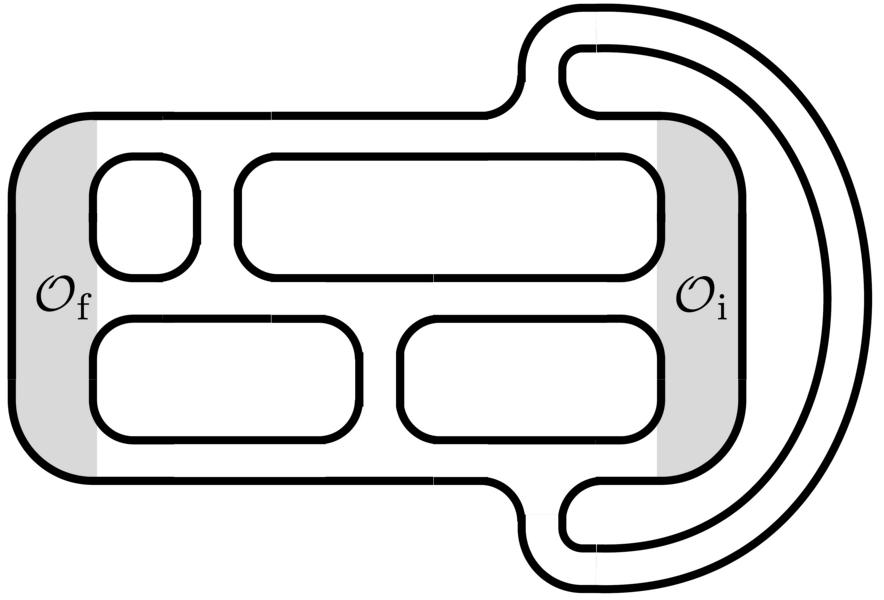}
}
\else
\scalebox{1}{
	\settoheight{\eqoff}{$+$}%
	\setlength{\eqoff}{0.5\eqoff}%
	\addtolength{\eqoff}{-5\unit}%
	\raisebox{\eqoff}{%
		\begin{pspicture}(-8.,-2.)(9.,8.)
		\ulinsert{-7}{6}
		\olvertex{-7}{3}
		\dlinsert{-7}{0}
		\rput[r](-5.5,3){$\scriptstyle \mathcal{O}_{\text{f}}$}
		\urinsert{7}{6}
		\orvertex{7}{3}
		\drinsert{7}{0}
		\rput[r](6.8,3){$\scriptstyle \mathcal{O}_{\text{i}}$}
		\threevertex{-3}{6}{-90}
		\threevertex{3}{6}{90}
		\threevertex{-3}{3}{90}
		\threevertex{0}{3}{-90}
		\threevertex{0}{0}{90}
		\threevertex{3}{0}{-90}
		\psset{linecolor=black,doubleline=true,linestyle=solid}
		\psline(-2,6)(2,6)
		\psline(-2,3)(-1,3)
		\psline(1,3)(4,3)
		\psline(-4,0)(-1,0)
		\psline(1,0)(2,0)
		\psline(-3,5)(-3,4)
		\psline(0,2)(0,1)
		\psline[linearc=\linearc](3,7)(3,8)(4,8)
		\psline[linearc=\linearc](3,-1)(3,-2)(4,-2)
		\psbezier(4,8)(10,8)(10,-2)(4,-2)
		\end{pspicture}
	}
}
\fi
\propto \gym^6 N^3=\lambda^3
\eqncom
\end{equation}
where the left hand expression is not a gauge invariant correlation function and hence the \tHooft limit cannot be taken. Wrapping diagrams that also involve four-point interactions can be drawn in a similar fashion. A general wrapping diagram for correlation functions of two operators with lengths $L_{\text{i}}$ and $L_{\text{f}}$ exists for loop orders $K\geq\frac 12(L_{\text{i}}+L_{\text{f}})$ and can be drawn as
\begin{equation}\label{eq:general_wrapping}
\ifpdf
\settoheight{\eqoff}{$+$}%
\setlength{\eqoff}{0.5\eqoff}%
\addtolength{\eqoff}{-8.5\unit}%
\raisebox{\eqoff}{%
	\includegraphics[angle={0},scale=0.2,trim=0cm 0cm 0cm 0]{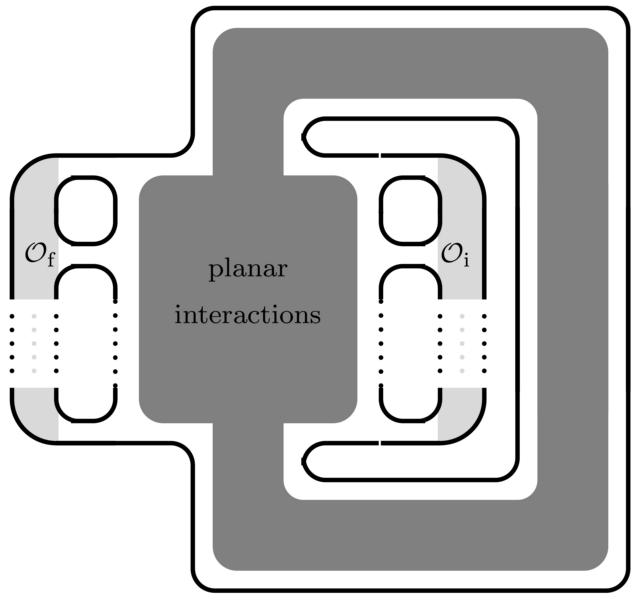}
}
\else
\scalebox{0.8}{
	\settoheight{\eqoff}{$+$}%
	\setlength{\eqoff}{0.5\eqoff}%
	\addtolength{\eqoff}{-10.5\unit}%
	\raisebox{\eqoff}{%
		\begin{pspicture}(-0.5,-6)(21.5,15)
		\ulinsert{0}{9}
		\olvertex{0}{6}
		\psline[linestyle=dotted](0,4)(0,2)
		\psline[linestyle=dotted,linecolor=ogray](0.75,4)(0.75,2)
		\psline[linestyle=dotted](1.5,4)(1.5,2)
		\dlinsert{0}{0}
		\urinsert{16}{9}
		\orvertex{16}{6}
		\psline[linestyle=dotted](14.5,4)(14.5,2)
		\psline[linestyle=dotted,linecolor=ogray](15.25,4)(15.25,2)
		\psline[linestyle=dotted](16,4)(16,2)
		\drinsert{16}{0}
		\rput[r](1.5,6){$\scriptstyle \mathcal{O}_{\text{f}}$}
		\rput[l](14.5,6){$\scriptstyle \mathcal{O}_{\text{i}}$}
		%
		\uinex{2}{9}
		\iinex{2}{6}
		\dinex{2}{0}
		\setlength{\xa}{4\unit}
		\addtolength{\xa}{-1\dlinewidth}
		\setlength{\ya}{9\unit}
		\addtolength{\ya}{0.5\dlinewidth}
		\setlength{\xb}{6.5\unit}
		\addtolength{\xb}{-0.5\dlinewidth}
		\setlength{\yb}{14\unit}
		\addtolength{\yb}{0.5\dlinewidth}
		\setlength{\xc}{9.5\unit}
		\addtolength{\xc}{0.5\dlinewidth}
		\setlength{\yc}{11\unit}
		\addtolength{\yc}{-0.5\dlinewidth}
		\setlength{\xd}{12\unit}
		\addtolength{\xd}{1.5\dlinewidth}
		\setlength{\yd}{0\unit}
		\addtolength{\yd}{-0.5\dlinewidth}
		\setlength{\xe}{17.5\unit}
		\addtolength{\xe}{-0.5\dlinewidth}
		\setlength{\ye}{-2\unit}
		\addtolength{\ye}{0.5\dlinewidth}
		\setlength{\xf}{20.5\unit}
		\addtolength{\xf}{0.5\dlinewidth}
		\setlength{\yf}{-5\unit}
		\addtolength{\yf}{-0.5\dlinewidth}
		\setlength{\yg}{8\unit}
		\addtolength{\yg}{-0.5\dlinewidth}
		\setlength{\yh}{1\unit}
		\addtolength{\yh}{0.5\dlinewidth}
		\psline[liftpen=1,linearc=\linearc](\xa,\ya)(\xb,\ya)(\xb,\yb)(\xf,\yb)(\xf,\yg)
		\psline[liftpen=1,linearc=\linearc](\xf,\yg)(\xf,\yh)
		\psline[liftpen=1,linearc=\linearc](\xf,\yh)(\xf,\yf)(\xb,\yf)(\xb,\yd)(\xa,\yd)
		\psline[liftpen=1,linearc=\linearc](\xd,\ya)(\xc,\ya)(\xc,\yc)(\xe,\yc)(\xe,\yg)
		\psline[liftpen=1,linearc=\linearc](\xe,\yg)(\xe,\yh)
		\psline[liftpen=1,linearc=\linearc](\xe,\yh)(\xe,\ye)(\xc,\ye)(\xc,\yd)(\xd,\yd)
		%
		%
		\psline[linestyle=dotted](3.5,4.5)(3.5,1.5)
		\doutex{14}{0}
		\ioutex{14}{6}
		\uoutex{14}{9}
		\psline[linestyle=dotted](12.5,4.5)(12.5,1.5)
		\setlength{\xa}{4\unit}
		\addtolength{\xa}{0.5\dlinewidth}
		\setlength{\ya}{9\unit}
		\addtolength{\ya}{-0.5\dlinewidth}
		\setlength{\xb}{6.5\unit}
		\addtolength{\xb}{0.5\dlinewidth}
		\setlength{\yb}{14\unit}
		\addtolength{\yb}{-0.5\dlinewidth}
		\setlength{\xc}{9.5\unit}
		\addtolength{\xc}{-0.5\dlinewidth}
		\setlength{\yc}{11\unit}
		\addtolength{\yc}{0.5\dlinewidth}
		\setlength{\xd}{12\unit}
		\addtolength{\xd}{-0.5\dlinewidth}
		\setlength{\yd}{0\unit}
		\addtolength{\yd}{0.5\dlinewidth}
		\setlength{\xe}{17.5\unit}
		\addtolength{\xe}{0.5\dlinewidth}
		\setlength{\ye}{-2\unit}
		\addtolength{\ye}{-0.5\dlinewidth}
		\setlength{\xf}{20.5\unit}
		\addtolength{\xf}{-0.5\dlinewidth}
		\setlength{\yf}{-5\unit}
		\addtolength{\yf}{0.5\dlinewidth}
		\setlength{\yg}{8\unit}
		\addtolength{\yg}{0.5\dlinewidth}
		\setlength{\yh}{1\unit}
		\addtolength{\yh}{-0.5\dlinewidth}
		%
		%
		\pscustom[linecolor=gray,fillstyle=solid,fillcolor=gray,linearc=\linearc]{%
			\psline[liftpen=1,linearc=\linearc](\xb,\ya)(\xd,\ya)(\xd,\yd)(\xa,\yd)(\xa,\ya)(\xb,\ya)
			\psline[liftpen=2,linearc=\linearc](\xb,\ya)(\xb,\yb)(\xf,\yb)(\xf,\yg)
			\psline[liftpen=1](\xf,\yg)(\xe,\yg)
			\psline[liftpen=1,linearc=\linearc](\xe,\yg)(\xe,\yc)(\xc,\yc)(\xc,\ya)
			\psline[liftpen=2,linearc=\linearc](\xb,\yd)(\xb,\yf)(\xf,\yf)(\xf,\yh)
			\psline[liftpen=1](\xf,\yh)(\xe,\yh)
			\psline[liftpen=1,linearc=\linearc](\xe,\yh)(\xe,\ye)(\xc,\ye)(\xc,\yd)
		}
		\pscustom[linecolor=gray,fillstyle=solid,fillcolor=gray]{%
			\psline[liftpen=1](\xe,\yg)(\xf,\yg)(\xf,\yh)
			\psline[liftpen=2](\xf,\yh)(\xe,\yh)(\xe,\yg)
		}
		\rput(8,5.5){$\scriptstyle \text{planar}$}
		\rput(8,4){$\scriptstyle \text{interactions}$}
		\end{pspicture}
	}
}
\fi
\propto \lambda^{\frac 12(L_{\text{i}}+L_{\text{f}})}
\eqncom
\end{equation}
where the dark grey area yields only planar interactions and $L_{\text{i}}+L_{\text{f}}$ powers $\gym$. The grey loop that wraps around the operator $\cO_i$ is also called wrapping loop. Since these wrapping diagrams have no suppression factor in $N$, they do contribute in the \tHooft limit.

\subsubsection{Prewrapping or multi-trace interactions}
The other finite-size effect which influences gauge invariant correlation functions in the \tHooft limit is prewrapping, which was introduced in \cite{Fokken:2013mza}. This effect subsumes all contributions from multi-trace interactions that start to occur one loop order prior to the wrapping contributions.
Such interactions contribute in the \tHooft limit if each single-trace factor of the multi-trace interaction is entirely contracted with an external single-trace composite operator -- despite the naive suppression factor in $N$ for multi-trace interactions in the actions in \subsecref{sec:multi-trace-parts-of-the-action}.

In two-point correlation functions involving single-trace composite operators of lengths $L_{\text{i}}$ and $L_{\text{f}}$, prewrapping can occur if each operator is connected to the single-trace factor of a double-trace interaction in a $K\geq \frac12(L_{\text{i}}+L_{\text{f}})-1$ loop process.
For example, two scalar $L=2$ composite operators can interact via the double-trace coupling displayed in \eqref{eq:double_line_building_blocs} in the following correlation function\footnote{Here the free correlation function is evaluated so that no additional interaction can be added to the one explicitly included.}
\begin{equation}\label{eq:O2_prewrapping}
\vacl\T\mathcal{O}_{2}\ol{\mathcal{O}}_{2}\Bigl(\frac{\gym^2}{N}\tr\bigl(\ol{\phi}\,\ol{\phi}\bigr)\tr\bigl(\phi\phi\bigr)\Bigr)\vac_{\text{free}} = \hspace{.1cm}
\ifpdf
\settoheight{\eqoff}{$+$}%
\setlength{\eqoff}{0.5\eqoff}%
\addtolength{\eqoff}{-2.2\unit}%
\raisebox{\eqoff}{%
	\includegraphics[angle={0},scale=0.08,trim=1cm 3cm 1cm 0]{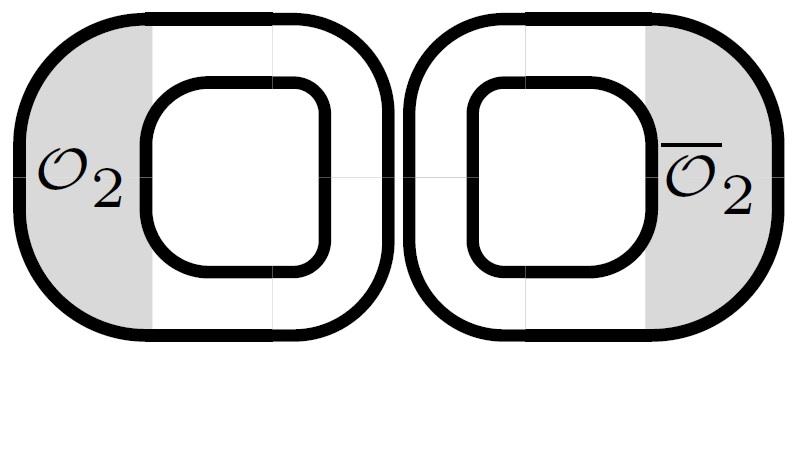}
}
\else
\settoheight{\eqoff}{$+$}%
\setlength{\eqoff}{0.5\eqoff}%
\addtolength{\eqoff}{-2\unit}%
\raisebox{\eqoff}{%
	\scalebox{0.8}{
		\rotatebox{0}{
			\begin{pspicture}(0,-1)(9,3.5)
			\ulinsert{0}{3}
			\dlinsert{0}{0}
			\rput{0}(.75,1.5){$\scriptscriptstyle \mathcal{O}_{2}$}
			\urinsert{9}{3}
			\drinsert{9}{0}
			\rput{0}(8.2,1.5){$\scriptscriptstyle \ol{\mathcal{O}}_{2}$}
			\psset{linecolor=black,doubleline=true,linestyle=solid}
			\psline[linearc=\linearc](3.,3)(4.,3)(4.,1.5)
			\psline[linearc=\linearc](3.,0)(4.,0)(4.,1.5)
			\psline[linearc=\linearc](6.,0)(5.,0)(5.,1.5)
			\psline[linearc=\linearc](6.,3)(5.,3)(5.,1.5)
			\end{pspicture}
		}
	}
}
\fi 
\propto 
\frac{\gym^2}{N}
\frac{\delta^{i_1}{}_{i_1}\dots \delta^{i_4}{}_{i_4}}{N^{2}}=\lambda \eqncom
\end{equation}
where the four colour loops generate an $N$-enhancement that absorbs the $\frac 1N$ suppression factor of the coupling, similar to the calculation in \eqref{eq:composite_operator_doubleline}. Note that this mechanism also applies, when the initial and final composite operator are both fused into a single line. In this case, the colour part of the propagator for that single connecting line is given in \eqref{eq:double_line_building_blocs} and for length-$L_i$ and $L_f$ operators we find the s-channel type prewrapping contribution\footnote{The overall power of the \tHooft coupling is determined in analogy to those of the length-changing processes depicted in \eqref{eq:tHooft_limit_transitions}.}
\begin{equation}\label{eq: generic prewrapping diagram}
\ifpdf
\settoheight{\eqoff}{$+$}%
\setlength{\eqoff}{0.5\eqoff}%
\addtolength{\eqoff}{-4\unit}%
\raisebox{\eqoff}{%
	\includegraphics[angle={0},scale=0.09,trim=0cm 0cm 0cm 0]{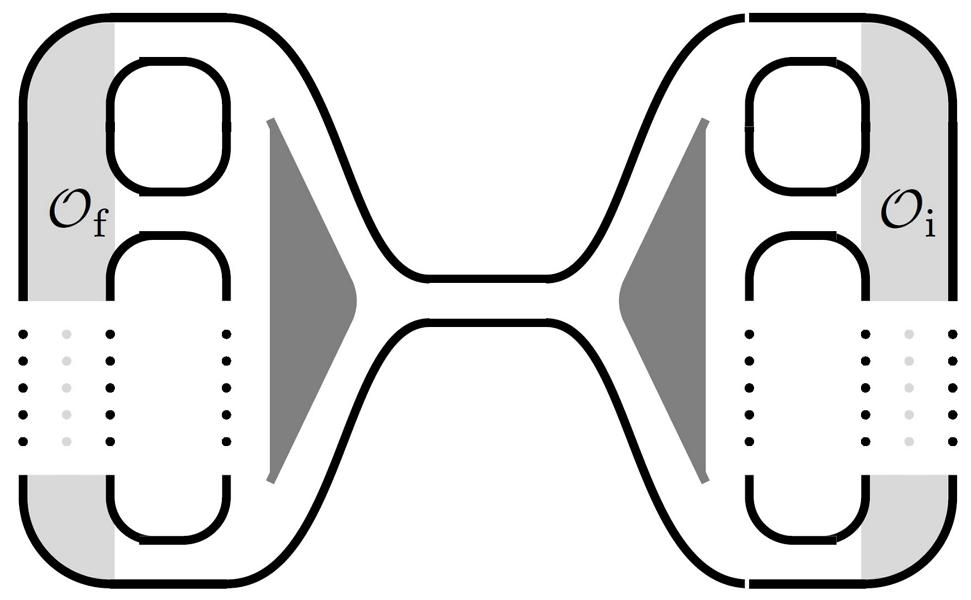}
}
\else
\settoheight{\eqoff}{$+$}%
\setlength{\eqoff}{1.5\eqoff}%
\addtolength{\eqoff}{-7\unit}%
\raisebox{\eqoff}{ 
	\rotatebox{0}{%
		\begin{pspicture}(0,-1.5)(16,10.5)
		\ulinsert{0}{9}
		\olvertex{0}{6}
		\psline[linestyle=dotted,linewidth=\blacklinew](0,4)(0,2)
		\psline[linestyle=dotted,linecolor=ogray,linewidth=\blacklinew](0.75,4)(0.75,2)
		\psline[linestyle=dotted,linewidth=\blacklinew](1.5,4)(1.5,2)
		\dlinsert{0}{0}
		\urinsert{16}{9}
		\orvertex{16}{6}
		\psline[linestyle=dotted,linewidth=\blacklinew](14.5,4)(14.5,2)
		\psline[linestyle=dotted,linecolor=ogray,linewidth=\blacklinew](15.25,4)(15.25,2)
		\psline[linestyle=dotted,linewidth=\blacklinew](16,4)(16,2)
		\drinsert{16}{0}
		\uinex{2}{9}
		\iinex{2}{6}
		\dinex{2}{0}
		\rput[r](1.5,6){$\scriptstyle \mathcal{O}_{\text{f}}$}
		\rput[r](15.8,6){$\scriptstyle \mathcal{O}_{\text{i}}$}
		\setlength{\ya}{9\unit}
		\addtolength{\ya}{0.5\dlinewidth}
		\setlength{\yb}{0\unit}
		\addtolength{\yb}{-0.5\dlinewidth}
		\setlength{\xc}{7.0\unit}
		\setlength{\yc}{4.5\unit}
		\addtolength{\yc}{-0.5\dlinewidth}
		\setlength{\xd}{9.0\unit}
		\setlength{\yd}{4.5\unit}
		\addtolength{\yd}{0.5\dlinewidth}
		\psbezier[linewidth=\blacklinew](3.5,\ya)(5.5,\ya)(5.5,\yd)(\xc,\yd)
		\psbezier[linewidth=\blacklinew](3.5,\yb)(5.5,\yb)(5.5,\yc)(\xc,\yc)
		\psline[linestyle=dotted,linewidth=\blacklinew](3.5,4)(3.5,2)
		\psbezier[linewidth=\blacklinew](12.5,\ya)(10.5,\ya)(10.5,\yd)(\xd,\yd)
		\psbezier[linewidth=\blacklinew](12.5,\yb)(10.5,\yb)(10.5,\yc)(\xd,\yc)
		\doutex{14}{0}
		\ioutex{14}{6}
		\uoutex{14}{9}
		\psline[linestyle=dotted,linewidth=\blacklinew](12.5,4)(12.5,2)
		\setlength{\xa}{3.5\unit}
		\addtolength{\xa}{\dlinewidth}
		\setlength{\xb}{6.5\unit}
		\addtolength{\xb}{-\dlinewidth}
		\setlength{\ya}{8\unit}
		\addtolength{\ya}{-0.5\dlinewidth}
		\setlength{\yb}{1\unit}
		\addtolength{\yb}{0.5\dlinewidth}
		\psline[linecolor=gray,fillstyle=solid,fillcolor=gray,linearc=\linearc](\xa,\ya)(\xb,4.5)(\xa,\yb)
		\setlength{\xa}{12.5\unit}
		\addtolength{\xa}{-\dlinewidth}
		\setlength{\xb}{9.5\unit}
		\addtolength{\xb}{\dlinewidth}
		\setlength{\ya}{8\unit}
		\addtolength{\ya}{-0.5\dlinewidth}
		\setlength{\yb}{1\unit}
		\addtolength{\yb}{0.5\dlinewidth}
		\psline[linecolor=gray,fillstyle=solid,fillcolor=gray,linearc=\linearc](\xa,\ya)(\xb,4.5)(\xa,\yb)
		\psset{linecolor=black,linewidth=\blacklinew}
		\psline[doubleline=true,doublesep=4.\blacklinew](7.0,4.5)(9.0,4.5)
		\end{pspicture}
	}
}%
\fi
\,{}-{}\displaystyle\frac{\colors}{N}\, %
\ifpdf
\settoheight{\eqoff}{$+$}%
\setlength{\eqoff}{0.5\eqoff}%
\addtolength{\eqoff}{-4\unit}%
\raisebox{\eqoff}{%
	\includegraphics[angle={0},scale=0.09,trim=0cm 0cm 0cm 0]{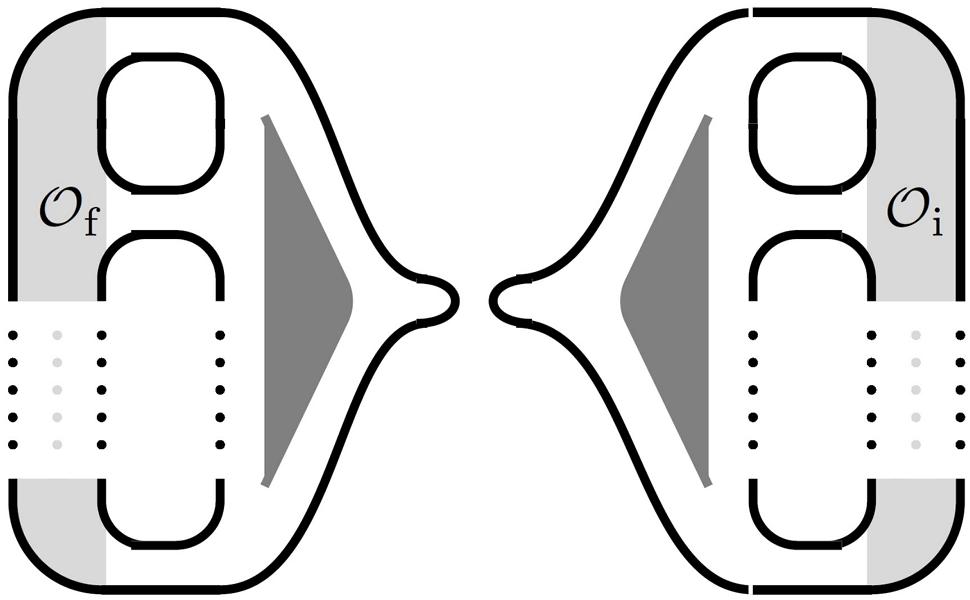}
}
\else
\settoheight{\eqoff}{$+$}%
\setlength{\eqoff}{1.5\eqoff}%
\addtolength{\eqoff}{-7\unit}%
\raisebox{\eqoff}{ 
	\rotatebox{0}{%
		\begin{pspicture}(0,-1.5)(16,10.5)
		\ulinsert{0}{9}
		\olvertex{0}{6}
		\psline[linestyle=dotted,linewidth=\blacklinew](0,4)(0,2)
		\psline[linestyle=dotted,linecolor=ogray,linewidth=\blacklinew](0.75,4)(0.75,2)
		\psline[linestyle=dotted,linewidth=\blacklinew](1.5,4)(1.5,2)
		\dlinsert{0}{0}
		\urinsert{16}{9}
		\orvertex{16}{6}
		\psline[linestyle=dotted,linewidth=\blacklinew](14.5,4)(14.5,2)
		\psline[linestyle=dotted,linecolor=ogray,linewidth=\blacklinew](15.25,4)(15.25,2)
		\psline[linestyle=dotted,linewidth=\blacklinew](16,4)(16,2)
		\drinsert{16}{0}
		\uinex{2}{9}
		\iinex{2}{6}
		\dinex{2}{0}
		\rput[r](1.5,6){$\scriptstyle \mathcal{O}_{\text{f}}$}
		\rput[r](15.8,6){$\scriptstyle \mathcal{O}_{\text{i}}$}
		\setlength{\ya}{9\unit}
		\addtolength{\ya}{0.5\dlinewidth}
		\setlength{\yb}{0\unit}
		\addtolength{\yb}{-0.5\dlinewidth}
		\setlength{\xc}{7.0\unit}
		\setlength{\yc}{4.5\unit}
		\addtolength{\yc}{-0.5\dlinewidth}
		\setlength{\xd}{8.5\unit}
		\setlength{\yd}{4.5\unit}
		\addtolength{\yd}{0.5\dlinewidth}
		\psbezier[linewidth=\blacklinew](3.5,\ya)(5.5,\ya)(5.5,\yd)(\xc,\yd)
		\psbezier[linewidth=\blacklinew](3.5,\yb)(5.5,\yb)(5.5,\yc)(\xc,\yc)
		\psline[linestyle=dotted,linewidth=\blacklinew](3.5,4)(3.5,2)
		\psbezier[linewidth=\blacklinew](12.5,\ya)(10.5,\ya)(10.5,\yd)(\xd,\yd)
		\psbezier[linewidth=\blacklinew](12.5,\yb)(10.5,\yb)(10.5,\yc)(\xd,\yc)
		\doutex{14}{0}
		\ioutex{14}{6}
		\uoutex{14}{9}
		\psline[linestyle=dotted,linewidth=\blacklinew](12.5,4)(12.5,2)
		\setlength{\xa}{3.5\unit}
		\addtolength{\xa}{\dlinewidth}
		\setlength{\xb}{6.5\unit}
		\addtolength{\xb}{-\dlinewidth}
		\setlength{\ya}{8\unit}
		\addtolength{\ya}{-0.5\dlinewidth}
		\setlength{\yb}{1\unit}
		\addtolength{\yb}{0.5\dlinewidth}
		\psline[linecolor=gray,fillstyle=solid,fillcolor=gray,linearc=\linearc](\xa,\ya)(\xb,4.5)(\xa,\yb)
		\setlength{\xa}{12.5\unit}
		\addtolength{\xa}{-\dlinewidth}
		\setlength{\xb}{9.5\unit}
		\addtolength{\xb}{\dlinewidth}
		\setlength{\ya}{8\unit}
		\addtolength{\ya}{-0.5\dlinewidth}
		\setlength{\yb}{1\unit}
		\addtolength{\yb}{0.5\dlinewidth}
		\psline[linecolor=gray,fillstyle=solid,fillcolor=gray,linearc=\linearc](\xa,\ya)(\xb,4.5)(\xa,\yb)
		\psset{linecolor=black,linewidth=\blacklinew}
		\uoneprop{8}{4.5}{0}
		\end{pspicture}
	}
}
\fi
\,{}\propto{}
\,\,\,(1-\colors)\lambda^{\frac 12(L_{\text{i}}+L_{\text{i}})-1}\eqncom
\end{equation}
where the dark grey areas encode planar interactions as in \eqref{eq:general_wrapping}. Hence, in any theory with gauge group \SUN where $s=1$ the s-channel-type processes vanish. Analogously, we can fuse the initial and final composite operator into each single-trace factor of any of the possible double-trace couplings given in \subsecref{sec:multi-trace-parts-of-the-action} and such a process contributes at the same loop order to the two-point correlation function.

\section{The (asymptotic) planar one-loop dilatation generator}\label{sec:the-quantum-dilatation-operator-on-composite-operators}
In \subsecref{sec:composite-operator-insertionsN4} we discussed how the renormalisation of composite operators in a quantised \CFT induces anomalous scaling dimensions that supplement the operators' classical scaling dimensions and alters the action of the dilatation generator \eqref{eq:quantum_D} in the interacting theory. In the \tHooft limit introduced in the previous section, the anomalous dimensions only receive planar contributions. In this section, we use this to simplify the structure of the one-loop dilatation generator\footnote{We restrict to the one-loop case here, since we will not need the higher-loop generalisations in \chapref{chap:applications}.} of \NfSYMt and discuss which parts of this structure can be carried over to the $\beta$- and $\gamma_i$-deformation.

The dilatation generator \eqref{eq:quantum_D} yields the perturbative corrections in $\gym$ to the anomalous dimension when it acts on a composite operator $\cO_i$ and these corrections were constructed from the complete renormalisation constant matrix of $\cO_i$ in \eqref{eq:anomalous_dim_matrix}. At one-loop order in \NfSYMt, all possible quantum corrections to $\cO_i$ stem from diagrams with a maximal interaction range of $r_i=2$, see the first line in \eqref{eq:comp_O_correlation_func} for an explicit example. In the \tHooft limit, the only contributing range $r_i=2$ diagrams connect two adjacent legs of the composite operator via a suitable interaction kernel. We can hence combine all range $r_i=2$ and $r_i=1$ one-loop contributions into a one-loop dilatation generator density $(\mathfrak{D}^{\cN=4}_2)^{\cA_3\cA_4}_{\cA_1\cA_2}$ which transforms two adjacent initial fields $\cA_1$ and $\cA_2$ of the composite operator into two final ones $\cA_3$ and $\cA_4$ and weighs the transition with the corresponding contribution to the anomalous dimension matrix element. For a composite operator of the form \eqref{eq:operator_as_tensorproduct}, the action of the one-loop dilatation generator can then be written analogously to \eqref{eq:symmetry_generator_on_comp_operators} as
\begin{equation}\label{eq:Dila_N4_one_loop}
[D^{\cN=4}_2,\mathcal{O}(x)]=-\complexi\sum_{i=1}^L \sum_{\cF_j\in \cA}
(\mathfrak{D}^{\cN=4}_2)^{\cF_1\cF_2}_{\cA_i\cA_{i+1}}\,\frac{1}{\cN}
\tr\bigl(
\cA_1\dots\cA_{i-1}\cF_1\cF_2\cA_{i+1}\dots\cA_L\bigr)\eqncom
\end{equation}
where $i+L$ is identified with $i$ and the sum over $\cF_j$ accounts for all possible transitions from the two initial fields to the two final ones. Note that $\mathfrak{D}^{\cN=4}_2$ is generically length-independent but for $L=1$ operators in the \UN theory it becomes explicitly length-dependent and the wrapping corrections discussed in \subsecref{sec:finite-size-effects} have to be included. So far, this representation of the one-loop dilatation generator is simply a repackaging of the prescription given in \subsecref{sec:composite-operator-insertionsN4} into a form suitable to the spin-chain description of composite operators in \NfSYMt. However, the dilatation generator density was completely determined in \cite{Beisert:2003jj} and in \appref{app:harmonic_action} we give $\mathfrak{D}^{\cN=4}_2$ explicitly in terms of occupation numbers in the oscillator representation. Using this representation, the planar one-loop dilatation generator of \NfSYMt with gauge group \SUN is completly determined via \eqref{eq:Dila_N4_one_loop}.

For the $\beta$- and $\gamma_i$-deformation, the question arises, if we can use \eqref{eq:Dila_N4_one_loop} to determine their respective dilatation generators as well and in the remainder of this section we follow the presentation of \cite{Fokken:2013mza} to address this question. In \secref{sec:The_deformations}, the deformations were obtained from \NfSYMt by turning the internal space to a non-commutative space with a Moyal-like $\ast$-product and by including all multi-trace interactions compatible with the remaining symmetries. In the asymptotic regime (which at one loop includes all operators with lengths $L\geq 3$), we know from \eqref{eq: generic prewrapping diagram} that the latter interactions cannot contribute. In this regime, the deformation-dependence of anomalous dimensions is determined entirely by the deformed elementary interactions that enter the weight factor in \eqref{eq:Dila_N4_one_loop}. Hence, if we can determine the deformation phase of a given one-loop contribution to the anomalous dimensions entirely from the fields that enter and leave the dilatation generator density, we can relate the deformed densities to the undeformed density $(\mathfrak{D}^{\cN=4}_2)^{\cA_3\cA_4}_{\cA_1\cA_2}$. To achieve this, we adapt the theorem of \cite{Filk96} for noncommutative field theories to the $\beta$- and $\gamma_i$-deformation: a planar single-trace Feynman diagram with $n$ entering fields in the deformed theory is given by the analogous diagram in the undeformed theory, multiplied by a phase that is determined purely by the order and the \su{4} Cartan charges $(q^1,q^2,q^3)$ of the $n$ incoming fields. For colour-ordered amplitudes an explicit formulation of this theorem in the  $\beta$-deformation was given in \cite{Khoze05} and when we take the $n=2R$ incoming fields from the alphabet \eqref{eq: alphabet} to enter the diagram in cyclic order, the relation reads
\begin{equation}\label{diagrel}
\begin{aligned}
\ifpdf
\settoheight{\eqoff}{$+$}%
\setlength{\eqoff}{0.5\eqoff}%
\addtolength{\eqoff}{-4.5\unit}%
\raisebox{\eqoff}{%
	\includegraphics[angle={0},scale=0.11,trim=0cm 0cm 0cm 0]{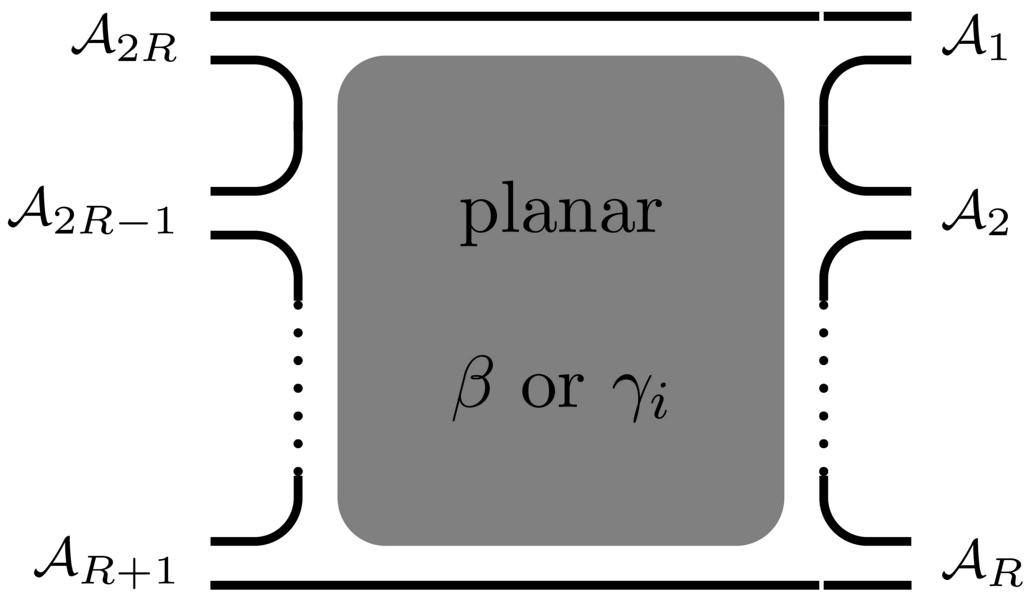}
}
\else
\settoheight{\eqoff}{$+$}%
\setlength{\eqoff}{0.5\eqoff}%
\addtolength{\eqoff}{-5.5\unit}%
\raisebox{\eqoff}{%
	\begin{pspicture}(-2.5,-1)(17.5,10)
	\rput[r](1.5,9){$\scriptstyle \mathcal{A}_{2R}$}
	\rput[r](1.5,6){$\scriptstyle \mathcal{A}_{2R-1}$}
	\rput[r](1.5,0){$\scriptstyle \mathcal{A}_{R+1}$}
	\rput[l](14.5,0){$\scriptstyle \mathcal{A}_{R}$}
	\rput[l](14.5,6){$\scriptstyle \mathcal{A}_{2}$}
	\rput[l](14.5,9){$\scriptstyle \mathcal{A}_{1}$}
	\uinex{2}{9}
	\iinex{2}{6}
	\dinex{2}{0}
	\setlength{\ya}{9\unit}
	\addtolength{\ya}{0.5\dlinewidth}
	\setlength{\yb}{0\unit}
	\addtolength{\yb}{-0.5\dlinewidth}
	\setlength{\xc}{7.5\unit}
	\setlength{\yc}{4.5\unit}
	\addtolength{\yc}{-0.5\dlinewidth}
	\setlength{\xd}{8.5\unit}
	\setlength{\yd}{4.5\unit}
	\addtolength{\yd}{0.5\dlinewidth}
	\psline(3.5,\ya)(12.5,\ya)
	\psline[linestyle=dotted](3.5,4.5)(3.5,1.5)
	\psline(12.5,\yb)(3.5,\yb)
	\doutex{14}{0}
	\ioutex{14}{6}
	\uoutex{14}{9}
	\psline[linestyle=dotted](12.5,4.5)(12.5,1.5)
	\setlength{\xa}{3.5\unit}
	\addtolength{\xa}{\dlinewidth}
	\setlength{\xb}{12.5\unit}
	\addtolength{\xb}{-\dlinewidth}
	\setlength{\ya}{9\unit}
	\addtolength{\ya}{-0.5\dlinewidth}
	\setlength{\yb}{0\unit}
	\addtolength{\yb}{0.5\dlinewidth}
	\pscustom[linecolor=gray,fillstyle=solid,fillcolor=gray,linearc=\linearc]{%
		\psline(\xa,4.5)(\xa,\ya)(\xb,\ya)(\xb,4.5)
		\psline[liftpen=2](\xb,4.5)(\xb,\yb)(\xa,\yb)(\xa,4.5)
	}
	\rput(8,6){planar}
	\rput(8,3){$\beta$ or $\gamma_i$}
	\end{pspicture}
}
\hspace{-.2cm}
\fi
&=
\ifpdf
\settoheight{\eqoff}{$+$}%
\setlength{\eqoff}{0.5\eqoff}%
\addtolength{\eqoff}{-4.5\unit}%
\raisebox{\eqoff}{%
	\includegraphics[angle={0},scale=0.11,trim=0cm 0cm 0cm 0]{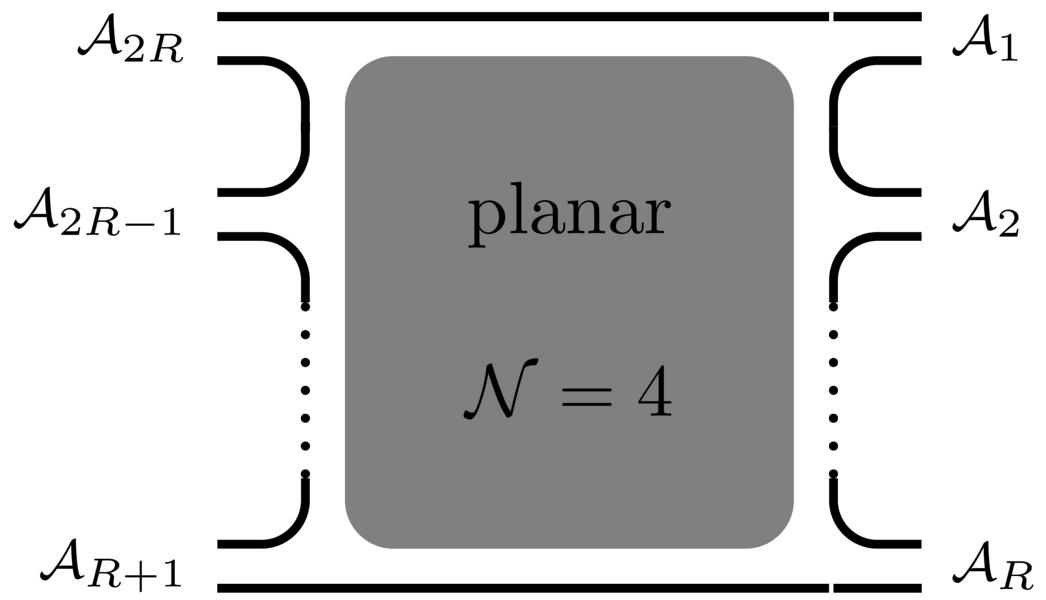}
}
\else
\settoheight{\eqoff}{$+$}%
\setlength{\eqoff}{0.5\eqoff}%
\addtolength{\eqoff}{-5.5\unit}%
\raisebox{\eqoff}{%
	\begin{pspicture}(-2.5,-1)(17.5,10)
	\rput[r](1.5,9){$\scriptstyle \mathcal{A}_{2R}$}
	\rput[r](1.5,6){$\scriptstyle \mathcal{A}_{2R-1}$}
	\rput[r](1.5,0){$\scriptstyle \mathcal{A}_{R+1}$}
	\rput[l](14.5,0){$\scriptstyle \mathcal{A}_{R}$}
	\rput[l](14.5,6){$\scriptstyle \mathcal{A}_{2}$}
	\rput[l](14.5,9){$\scriptstyle \mathcal{A}_{1}$}
	\uinex{2}{9}
	\iinex{2}{6}
	\dinex{2}{0}
	\setlength{\ya}{9\unit}
	\addtolength{\ya}{0.5\dlinewidth}
	\setlength{\yb}{0\unit}
	\addtolength{\yb}{-0.5\dlinewidth}
	\setlength{\xc}{7.5\unit}
	\setlength{\yc}{4.5\unit}
	\addtolength{\yc}{-0.5\dlinewidth}
	\setlength{\xd}{8.5\unit}
	\setlength{\yd}{4.5\unit}
	\addtolength{\yd}{0.5\dlinewidth}
	\psline(3.5,\ya)(12.5,\ya)
	\psline[linestyle=dotted](3.5,4.5)(3.5,1.5)
	\psline(12.5,\yb)(3.5,\yb)
	\doutex{14}{0}
	\ioutex{14}{6}
	\uoutex{14}{9}
	\psline[linestyle=dotted](12.5,4.5)(12.5,1.5)
	\setlength{\xa}{3.5\unit}
	\addtolength{\xa}{\dlinewidth}
	\setlength{\xb}{12.5\unit}
	\addtolength{\xb}{-\dlinewidth}
	\setlength{\ya}{9\unit}
	\addtolength{\ya}{-0.5\dlinewidth}
	\setlength{\yb}{0\unit}
	\addtolength{\yb}{0.5\dlinewidth}
	\pscustom[linecolor=gray,fillstyle=solid,fillcolor=gray,linearc=\linearc]{%
		\psline(\xa,4.5)(\xa,\ya)(\xb,\ya)(\xb,4.5)
		\psline[liftpen=2](\xb,4.5)(\xb,\yb)(\xa,\yb)(\xa,4.5)
	}
	\rput(8,6){planar}
	\rput(8,3){$\mathcal{N}=4$}
	\end{pspicture}
}
\fi
\,\Phi(\mathcal{A}_{1}\ast \mathcal{A}_{2}\ast\dots
\ast \mathcal{A}_{2R})
\eqncom
\end{aligned}
\end{equation}
where the operator $\Phi$ extracts the phase generated from the $\ast$-products which were defined in \eqref{eq:starproduct}. In the asymptotic regime, the order in which the initial and final fields in the dilatation generator density $\mathfrak{D}_2$ in \eqref{eq:Dila_N4_one_loop} appear is fixed. Therefore, we can determine the deformation phase from the elementary interactions from \eqref{diagrel} with $R=2$ in this case.
In \cite{Beisert:2005if}, the asymptotic one-loop dilatation operator density of the $\gamma_i$-deformation was constructed from the undeformed one using this method. Restricting to the asymptotic regime with operator lengths $L\geq3$ their result can be written as
\begin{equation}\label{eq: deformation of D_2}
\begin{aligned}
(\diladensity_{2}^{L\geq 3})_{\mathcal{A}_i\mathcal{A}_j}^{\mathcal{A}_k\mathcal{A}_l}&= \Phi(\mathcal{A}_k\cstar \mathcal{A}_l\cstar \mathcal{A}_j\cstar \mathcal{A}_i)(\diladensity_{2}^{ \cN=4 })_{\mathcal{A}_i\mathcal{A}_j}^{\mathcal{A}_k\mathcal{A}_l}
=\e^{\frac{\complexi}{2} (\mathbf{q}_{\mathcal{A}_k} \wedge \mathbf{q}_{\mathcal{A}_l}- \mathbf{q}_{\mathcal{A}_i} \wedge \mathbf{q}_{\mathcal{A}_j})}(\diladensity_{2}^{\cN=4})_{\mathcal{A}_i\mathcal{A}_j}^{\mathcal{A}_k\mathcal{A}_l} \eqncom
\end{aligned}
\end{equation}
where we used that all interactions in \eqref{eq:deformed_action_complex_scalars2} have vanishing total $(q^1,q^2,q^3)$ charge. 

For operator lengths $L\leq 2$ the above relation cannot be used\footnote{For $R\geq2$, we have prewrapping contributions which originate from multi-trace couplings and for $R>2$ we also have wrapping contributions, which are non-planar in an interaction kernel picture, \cf \eqref{eq:wrapping_example}.} to obtain the dilatation generator density in the deformed theories since the finite-size corrections discussed in \secref{sec:finite-size-effects} in general spoil the applicability of \eqref{diagrel}.
Subtleties how the finite-size corrections can be implemented in this equation are discussed in \secref{sec:beta_paper} for the $\beta$-deformation. For gauge group \UN, the corrections include wrapping contributions to $L=1$ operators and prewrapping contributions from additional double-trace couplings. For gauge group \SUN, the prewrapping contributions in $L=2$ states have to be included.

\chapter{Applications}\label{chap:applications}
In this chapter, we employ the renormalisation program reviewed in \chapref{chap:Renormalisation} to analyse the structure of the $\beta$- and $\gamma_i$-deformation introduced in \chapref{chap:The_models} in the \tHooft limit. This chapter is based on my publications \cite{Fokken:2013aea,Fokken:2014soa,Fokken:2013mza,Fokken:2014moa}. 

In the first section, we calculate the complete one-loop renormalisation constant of the coupling $Q_{\text{F}\,ii}^{ii}$ in the \tHooft limit. This is one of the additional double-trace couplings of the $\gamma_i$-deformation and was introduced in \eqref{eq:dtc}. We find that the complete one-loop counterterm $\mathfrak{d}_{Q_{\text{F}}}$ does not vanish, which shows the explicit breaking of conformal invariance in the $\gamma_i$-deformation at first loop order in the \tHooft limit.
	
In the second section, we use the close kinship of the $\gamma_i$-deformation to \NfSYMt to construct the leading wrapping correction to the spin-chain vacuum state $\sim\tr\bigl(\phi_i^L\bigr)$ for generic $L\geq 2$. For $L=2$ we calculate the two-loop anomalous dimension which is only finite if prewrapping contributions are included. The finite renormalisation-scheme dependent result directly demonstrates the non-conformality of the $\gamma_i$-deformation in the spectrum of composite operators.

In the third section, we derive all planar $L=2$ and $L=1$ one-loop finite-size corrections to construct the complete one-loop dilatation operator of the $\beta$-deformation. For this deformation, we show that a large class of composite operators is never affected by the deformation. For deformation-dependent operators, we classify which states may potentially be affected by prewrapping.

In the fourth section, we use the planar one-loop dilatation operator\footnote{In case of the conformal $\beta$-deformation with gauge group \SUN, we employ the dilatation operator derived in \secref{sec:beta_paper} and for the remaining theories we choose a particular one-loop dilatation operator with tree-level multi-trace couplings set to zero.} of the $\beta$- and $\gamma_i$-deformation to construct the planar one-loop thermal partition function of both theories on the space \RxSt. We do this by generalising the \Polya-theoretic approach that was used to find the respective partition function in the undeformed theory. From the partition function, we calculate the phase transition temperature, where the low-energy description of the theory in terms of colour neutral composite operators breaks down.

\newpage
\section{Non-conformality of the \texorpdfstring{$\gamma_i$}{gamma-i}-deformation}\label{sec:non-conformal_double_trace_coupling}
In this section, we employ the techniques discussed in \chapref{chap:Renormalisation} to show that the $\gamma_i$-deformation as introduced in \secref{sec:The_deformations} breaks conformal invariance in one loop quantum corrections. This section is based on \cite{Fokken:2013aea} and we will present the derivation and main results here including implications for the \AdSCFTc. We postpone the discussion of questions concerning integrability to \chapref{chap:Conclusion_outlook}, since they are also related to \secref{sec:cake} and \ref{sec:beta_paper}.

The $\gamma_i$-deformation, as we introduced it in \secref{sec:The_deformations}, is a non-supersymmetric deformation of the \NfSYM parent theory. This gauge theory together with its string-theory dual are supposed to be a non-supersymmetric example of the \AdSCFTc. On the string-theory side the three consecutive TsT transformations do not alter the $\AdS{5}$ part of the string background, as mentioned in \chapref{chap:Introduction}. Since the isometry group of the $\AdS{5}$ part is \SO{4,2}, which on the gauge-theory side corresponds to the  conformal symmetry, it is tempting to assume that the $\gamma_i$-deformation is also conformally invariant. While this statement is trivially true for the classical theory, see \subsecref{sec:conf_symmetry_N4}, it is premature to assume its validity in the quantum theory. In \chapref{chap:Renormalisation}, we have seen that the scaling symmetry as part of the conformal symmetries is in general influenced by renormalisation and it only remains exact if the $\beta$-functions of all elementary interactions vanish. This vanishing delicately\footnote{For a non-abelian gauge theory with $n_{\text{F}}$ and $n_{\text{S}}$ adjoint Weyl fermions and respectively complex scalars, the coupling's $\beta$-function for example is proportional to $\beta\sim g^3N(11-2n_{\text{F}}-n_{\text{S}})+\order{g^5}\stackrel{\mathcal{N}=4}{=}\order{g^5}$, \cite[\chap{78}]{Srednicki:2007}.} depends on the fine tuning between fermionic and bosonic \dof In deforming the $\text{S}^5$ part of \NfSYMt to obtain the $\gamma_i$-deformation, the relations between fermions and scalars are indeed altered and hence conformal invariance in the deformed theory must explicitly be checked for all couplings.

In the context of non-supersymmetric orbifold theories running $\beta$-functions of double-trace couplings without fix-points have already been found in \cite{Dymarsky:2005uh} and in \cite{Dymarsky:2005nc} these findings amounted to the no-go theorem: no non-supersymmetric orbifold exists which has a perturbatively accessible fix-points for all couplings in the \tHooft limit. This theorem does not exclude isolated Banks-Zaks fix-points \cite{Banks:1981nn}, where the two-loop corrections to the $\beta$-function cancel the one-loop corrections at a real and perturbatively accessible value of the coupling constant. On the string-theory side of the orbifold theories, the running double-trace coupling appears to correspond to the emergence of tachyonic modes in the twisted sectors, see \cite{Dymarsky:2005nc} and \cite{Armoni:2003va} for similar findings concerning non-commutative field theories. On the gauge-theory side, the running of the double-trace couplings can be attributed to dynamical symmetry breaking \cite{Pomoni:2008de}.

The analysis in the orbifold setup pushes forward the question whether the $\gamma_i$-de\-for\-ma\-tion also generates running couplings without fix-points which would spoil the conformal invariance. In this context, it is important to note that the proof of finiteness of the $\gamma_i$-deformation in \cite{Ananth:2006ac,Ananth:2007px} is incomplete. The non-renormalisation of couplings in the \tHooft limit can only be inherited from the parent \NfSYMt for single-trace couplings\footnote{The proof of finiteness of the single-trace couplings closely follows the proof in the undeformed theory \cite{Mandelstam:1982cb,Brink:1982wv}.}. In \secref{sec:The_deformations}, we have, however, seen that there are additional multi-trace couplings compatible with the symmetries of the $\gamma_i$-deformation. For these multi-trace couplings no inheritance principle exists \cite{Bershadsky:1998mb,Bershadsky:1998cb} and hence the vanishing of their $\beta$-functions is not guaranteed. In addition, the non-renormalisation proofs cannot be applied for these couplings, since they are restricted to planar single-trace diagrams without external states. Therefore, situations in which multi-trace couplings contribute in the \tHooft limit are not covered in the finiteness proofs of \cite{Ananth:2006ac,Ananth:2007px}, see \subsecref{sec:finite-size-effects}. 

In this section, we explicitly calculate the one-loop correction to the double-trace couplings
\begin{equation}\label{eq:running_dt_coupling}
-\frac{\gym^2}{2N}Q^{ii}_{\text{F}\,ii}\tr(\phi_i\phi_i)\tr(\bar\phi^i\bar\phi^i)\eqncom
\end{equation}
with fixed $i\in\{1,2,3\}$, which was given in \eqref{eq:dtc} as part of the multi-trace couplings in the $\gamma_i$-deformation. For gauge groups \UN and \SUN we calculate its $\beta$-function up to the first order correction in the effective planar coupling constant $g$ and find\footnote{To relate our $\beta$-function to the one in \cite{Fokken:2013aea,Fokken:2014soa}, the coupling constants are rescaled as $\gym=2^{-\frac12}\hat{g}_{\text{YM}}$ and $Q_{\text{F}\,ii}^{ii}=4\hat{Q}_{\text{F}\,ii}^{ii}$, where hatted quantities are the ones of \cite{Fokken:2013aea,Fokken:2014soa}. Furthermore, we have $\sin^2\gamma_1^+\sin^2\gamma_1^-=\frac 14(\cos \gamma_{2}-\cos\gamma_{3})^2$ and $\beta_{Q_{\text{F}\,ii}^{ii}}=4\hat{\beta}_{Q_{\text{F}\,ii}^{ii}}$, since it scales as $Q_{\text{F}\,ii}^{ii}$.}
\begin{equation}
\beta_{Q_{\text{F}\,ii}^{ii}}=2g^2\big(64\sin^2\gamma_i^+\sin^2\gamma_i^-
+(Q^{ii}_{\text{F}\,ii})^2\big)\eqndot
\end{equation}
Since this expression is non-vanishing for real deformation angles $\gamma_i^\pm$ and tree-level couplings $Q^{ii}_{\text{F}\,ii}$, the conformal invariance of the $\gamma_i$-deformation is broken by perturbative quantum corrections. This breaking of conformal invariance will in general reappear in the anomalous dimensions of the composite operators\footnote{In terms of oscillator occupation numbers \eqref{eq:occupation_numbers}, the composite operator with charge $q^1=2$ is for example characterised by $A_{\mathcal{O}_L}=(0,0,0,0,2,0,0,2)$, where we used \tabref{tab: su(4) charges}.} $\mathcal{O}_L$ with vanishing $\spl{2}$ and $\splbar{2}$ Lorentz charges and a total $\su{4}$ Cartan charges $\mathbf{q}_{\mathcal{O}_L}=2\mathbf{e}_i$, where $\mathbf{e}_i$ is the $i^{\text{th}}$ unit vector. Such an operator with length $L$ can be fused planarly to a length-$2$ operator involving two scalars $\phi_i$ in an $(L-2)$-loop length-changing process, analogously to the fusion in \eqref{eq: generic prewrapping diagram}. When this fused composite operator is connected via the coupling \eqref{eq:running_dt_coupling} in a two-point function $\vacl\T \ol{\mathcal{O}}_L\mathcal{O}_L\vac$, the resulting diagram is a generalisation of \eqref{eq:O2_prewrapping} for length-$L$ operators. These types of diagrams contribute in the \tHooft limit at order $\lambda^{L-1}$ and hence, the non-conformality of the coupling \eqref{eq:running_dt_coupling} will start to affect the corresponding anomalous dimension of the composite operator at this order.

\subsection{One-loop renormalisation of \texorpdfstring{$Q_{\text{F}\,ii}^{ii}$}{QF}}
In this subsection, we calculate the one-loop renormalisation constant to the double-trace coupling \eqref{eq:running_dt_coupling}, using the Feynman rules of \appref{app:Feynman_rules} and the \ttt{FokkenFeynPackage} described in \ref{sec:Feynman_rules_Mathematica}. We employ the general renormalisation program introduced in \chapref{chap:Renormalisation} and refer to \secref{sec:calc_Greens_functions} for our notational conventions.

The relevant counterterms entering the renormalised coupling constants are
\begin{equation}
\mathcal{Z}_{Q_{\text{F}}}Q_{\text{F}\,ii}^{ii}=(1+\mathfrak{d}_{Q_{\text{F}}})Q_{\text{F}\,ii}^{ii}=\frac{Z_{Q_{\text{F}}}}{Z_\phi^2}Q_{\text{F}\,ii}^{ii}
=(1+\delta^{(1)}_{Q_{\text{F}}}+2\delta^{(1)}_\phi)Q_{\text{F}\,ii}^{ii}+\order{\gym^4}\eqncom
\end{equation}
where we expressed the connected renormalisation constant $\mathcal{Z}_{Q_{\text{F}}}$ in terms of 1PI constants of the vertex $Z_{Q_{\text{F}}}=1+\delta_{Q_{\text{F}}}$ and of the external legs $Z_\phi=1-\delta_\phi$, see \subsecref{sec:N4_renormalised} and \appref{app:Feynman_rules} for details. While the one-loop 1PI renormalisation constant of the scalar fields (wave function renormalisation) is calculated separately in \appref{app:oneloopse}, we calculate the 1PI renormalisation constant of the coupling $Q_{\text{F}}$ here and construct the connected one-loop  counterterm $\mathfrak{d}_{Q_{\text{F}}}^{(1)}=\delta^{(1)}_{Q_{\text{F}}}+2\delta^{(1)}_\phi$. The 1PI counterterm $\delta^{(1)}_{Q_{\text{F}}}$ can be determined from the one-loop coefficient of the reduced correlation function of renormalised fields
\begin{equation}
\begin{aligned}\label{eq:gamma_i_matrix_element}
0&=
\Kop\Bigl[\Bigl(
\vacl\T\phi^a_i(p)\phi^b_{i}(0)\ol{\phi}^{ic}(-p)\ol{\phi}^{id}(0)\vac^{(1)}_{\complexi\mathcal{T}}
\Bigr)_{\text{1PI}}
\Bigr]
=
\Kop\Biggl[
\Biggl(
\settoheight{\eqoff}{$\times$}%
\setlength{\eqoff}{0.5\eqoff}%
\addtolength{\eqoff}{-6\unitlength}%
\raisebox{\eqoff}{%
	\fmfframe(2,1)(0,1){%
		\begin{fmfchar*}(10,10)
		\fmfforce{0 w,1 h}{v1}
		\fmfforce{1 w,1 h}{v2}
		\fmfforce{1 w,0 h}{v3}
		\fmfforce{0 w,0 h}{v4}
		\fmf{phantom}{vc,v1}
		\fmf{phantom}{vc,v2}
		\fmf{phantom}{vc,v3}
		\fmf{phantom}{vc,v4}
		\fmffreeze
		\fmfposition
		\fmfipath{p[]}
		\fmfiset{p1}{vpath(__vc,__v1)}
		\fmfiset{p11}{subpath (length(p1)/4,length(p1)) of p1}
		\fmfiset{p2}{vpath(__vc,__v2)}
		\fmfiset{p21}{subpath (length(p2)/4,length(p2)) of p2}
		\fmfiset{p3}{vpath(__vc,__v3)}
		\fmfiset{p31}{subpath (length(p3)/4,length(p3)) of p3}
		\fmfiset{p4}{vpath(__vc,__v4)}
		\fmfiset{p41}{subpath (length(p4)/4,length(p4)) of p4}
		\fmfi{plain_ar}{p11}
		\fmfi{plain_ar}{p21}
		\fmfi{plain_rar}{p31}
		\fmfi{plain_rar}{p41}
		\fmfiv{label=$\scriptstyle i a$,label.angle=-20,label.dist=6}{vloc(__v1)}
		\fmfiv{label=$\scriptstyle i b$,label.angle=-100,label.dist=6}{vloc(__v2)}
		\fmfiv{label=$\scriptstyle i c$,label.angle=160,label.dist=6}{vloc(__v3)}
		\fmfiv{label=$\scriptstyle i d$,label.angle=80,label.dist=6}{vloc(__v4)}
		\fmfiv{decor.shape=circle,decor.filled=shaded,decor.size=10thin}{vloc(__vc)}
		\end{fmfchar*}
	}
}
+
\settoheight{\eqoff}{$\times$}%
\setlength{\eqoff}{0.5\eqoff}%
\addtolength{\eqoff}{-6\unitlength}%
\raisebox{\eqoff}{%
	\fmfframe(2,1)(0,1){%
		\begin{fmfchar*}(10,10)
		\fmfforce{0 w,1 h}{v1}
		\fmfforce{1 w,1 h}{v2}
		\fmfforce{1 w,0 h}{v3}
		\fmfforce{0 w,0 h}{v4}
		\fmf{plain_ar}{vc,v1}
		\fmf{plain_ar}{vc,v2}
		\fmf{plain_rar}{vc,v3}
		\fmf{plain_rar}{vc,v4}
		\fmffreeze
		\fmfposition
		\fmfipath{p[]}
		\fmfiset{p1}{vpath(__vc,__v1)}
		\fmfiset{p11}{subpath (length(p1)/4,length(p1)) of p1}
		\fmfiset{p2}{vpath(__vc,__v2)}
		\fmfiset{p21}{subpath (length(p2)/4,length(p2)) of p2}
		\fmfiset{p3}{vpath(__vc,__v3)}
		\fmfiset{p31}{subpath (length(p3)/4,length(p3)) of p3}
		\fmfiset{p4}{vpath(__vc,__v4)}
		\fmfiset{p41}{subpath (length(p4)/4,length(p4)) of p4}
		\fmfiv{label=$\scriptstyle i a$,label.angle=-20,label.dist=6}{vloc(__v1)}
		\fmfiv{label=$\scriptstyle i b$,label.angle=-100,label.dist=6}{vloc(__v2)}
		\fmfiv{label=$\scriptstyle i c$,label.angle=160,label.dist=4}{vloc(__v3)}
		\fmfiv{label=$\scriptstyle i d$,label.angle=80,label.dist=6}{vloc(__v4)}
		\fmfiv{decor.shape=hexacross,decor.size=7thin}{vloc(__vc)}
		\fmfiv{label=$\scriptscriptstyle Q_{\text{F}}$,l.a=-80,l.dist=4}{vloc(__vc)}
		\end{fmfchar*}
	}
}
\Biggr)_{\text{1PI}}
\Biggr]
\\
&=-2\complexi\frac{\gym^2\mu^{2\epsilon}}{N}(ab)(cd)\Kop\Bigl[V^{(1)}_{Q_{\text{F}}\text{B}}(p,0,-p)
+\delta^{(1)}_{Q_{\text{F}}}Q_{\text{F}\,ii}^{ii}
\Bigr]
\eqncom
\end{aligned}
\end{equation}
where the identical field flavours are $i\in\{1,2,3\}$, the operator $\Kop$ extracts the divergence rendering the expression zero and we have projected to the double-trace contribution in the final equation. The first diagram contains all 1PI graphs that involve only bare quantities and the second diagram gives the corresponding 1PI counterterm of the coupling $Q_{\text{F}}$, in analogy to the situation in \eqref{eq:renormalised_phi3_vertex}. In the following, we evaluate $V^{(1)}_{Q_{\text{F}}\text{B}}$ in the dimensional reduction scheme in $D=4-2\epsilon$ dimensions and determine the 1PI counterterm so that \eqref{eq:gamma_i_matrix_element} vanishes. We then construct the connected counterterm $\mathfrak{d}_{Q_{\text{F}}}$ and use it to calculate the one-loop $\beta$-function of the coupling \eqref{eq:running_dt_coupling}.

\subsubsection{Gauge group \texorpdfstring{\SUN}{SU(N)}}
To calculate $V^{(1)}_{Q_{\text{F}}\text{B}}$ in the $\gamma_i$-deformation with gauge group \SUN, we have to add all one-loop diagrams with the coupling-tensor structure of \eqref{eq:gamma_i_matrix_element} that can be constructed from the single-trace action \eqref{eq:deformed_action_complex_scalars2} and the double-trace action \eqref{eq:dtc}. We can reduce the number of contributions, by exploiting that the counterterm of $\delta_{Q_{\text{F}}}$ vanishes in the parent \NfSYMt\footnote{Strictly speaking, the connected counterterm $\mathfrak{d}_{Q_{\text{F}}}$ vanishes in \NfSYMt. However, since any divergent contributions from the external field renormalisation is multiplied by $Q_{\text{F}}$ in the calculation of $\mathfrak{d}_{Q_{\text{F}}}$, they vanish in the limit $Q_{\text{F}}\rightarrow 0$.}, which allows us to calculate only those Feynman diagrams that depend on the deformation angles $\gamma_i^\pm$ and/or the double-trace couplings $Q_{\text{F}}$ and $Q_{\text{D}}$ explicitly. The remaining contributions can be reconstructed by enforcing that the divergent contributions to $V^{(1)}_{Q_{\text{F}}\text{B}}$ vanishes in the limit of vanishing deformation parameters $Q_{\text{F}}=Q_{\text{D}}=\gamma_i^\pm=0$.

The only vertices that are altered in the $\gamma_i$-deformation are cubic and quartic vertices involving only matter fields, i.e.\ fermionic and scalar fields. All 1PI one-loop diagrams that contain such couplings and contribute to $V^{(1)}_{Q_{\text{F}}\text{B}}$ are shown in \figref{fig:dtd}. These diagrams can be evaluated using the Feynman rules of \appref{app:Feynman_rules} and the \ttt{Mathematica} package \ttt{FokkenFeynPackage}. For generic gauge $\xi$ and fixed $i\in\{1,2,3\}$, we find the divergent parts in terms of the coupling tensors given in \eqref{eq:coupling_tensors_gammai} to be
\begin{subequations}\label{diagres}
\begin{equation}
\begin{aligned}\label{diagres1}
\subref{ss4s4}=\subref{us4s4}
&=\frac 12\gym^4\mu^{2\epsilon}\Kop[\hat{I}_{(1,1)}(p)]\sum_{r=1}^3F^{ir}_{ri}F^{ir}_{ri}(ab)(cd)
\eqncom\\
\Rop_|[\subref{ss4s4}]
=\Rop_|[\subref{us4s4}]
&=\frac 12\gym^4\mu^{2\epsilon}\Kop[\hat{I}_{(1,1)}(p)]\sum_{r=1}^3F^{ri}_{ir}F^{ri}_{ir}(ab)(cd)
\eqncom\\
\subref{ts4s4}
&= \frac 12
\gym^4\mu^{2\epsilon}\Kop[\hat{I}_{(1,1)}(p)]\sum_{r,s=1}^3(Q^{ii}_{\text{F}\,sr}+Q^{ii}_{\text{F}\,rs})(Q^{sr}_{\text{F}\,ii}+Q^{rs}_{\text{F}\,ii})(ab)(cd)
\eqncom\\
\end{aligned}
\end{equation}
\begin{equation}
\begin{aligned}\label{diagres2}
\subref{tDs4s4}
&=-4\gym^4\mu^{2\epsilon}\Kop[\hat{I}_{(1,1)}(p)]Q^{ii}_{\text{F}\,ii}(ab)(cd) 
\eqncom\\
\subref{s4gg2}=\subref{s4gg3}
&=2\xi \gym^4\mu^{2\epsilon}\Kop\bigl[\hat{I}_{(2,1)}(p)\bigr]Q^{ii}_{\text{F}\,ii}(ab)(cd) 
\eqncom\\
\subref{fbox1}=\subref{fbox2}
&=-2\gym^4\mu^{2\epsilon}\Kop[\hat{I}_{(1,1)}(p)]
\big[\tr\big((\rho^{\dagger i})^{\T}(\tilde\rho^{\dagger i})^{\T}\tilde\rho_i\rho_i\big)
+\tr\big((\tilde\rho^{\dagger i})^{\T}(\rho^{\dagger i})^{\T}\rho_i\tilde\rho_i\big)\big]{\scriptstyle(ab)(cd)}
\eqncom\\
\subref{fbox3}&=
-2\gym^4\mu^{2\epsilon}\Kop[\hat{I}_{(1,1)}(p)]
\big[
\tr\big((\rho^{\dagger i})^{\T}\rho_i(\rho^{\dagger i})^{\T}\rho_i\big)
+\tr\big((\tilde\rho^{\dagger i})^{\T}\tilde\rho_i(\tilde\rho^{\dagger i})^{\T}\tilde\rho_i\big)\big]{\scriptstyle(ab)(cd)}
\eqndot
\end{aligned}
\end{equation}
\end{subequations}
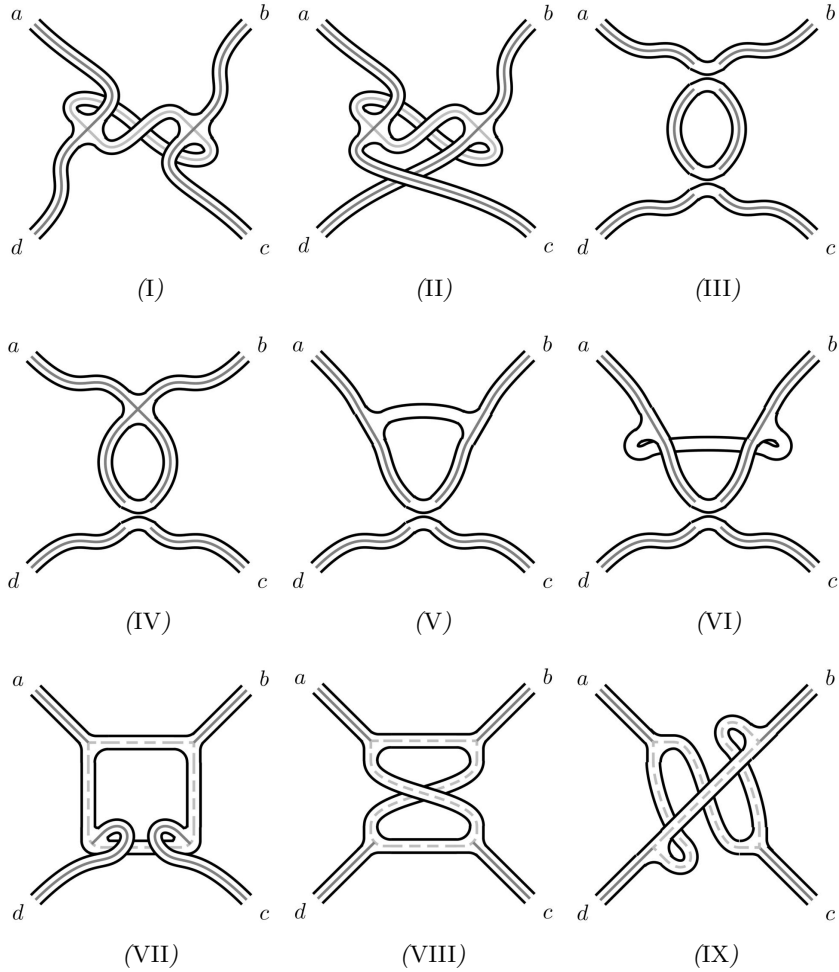
\begin{figure}[!b]
	\begin{center}
		\subfloat[\label{ss4s4}]{%
			\ifpdf
			\settoheight{\eqoff}{$+$}%
			\setlength{\eqoff}{0.5\eqoff}%
			\addtolength{\eqoff}{-0\unit}%
			\raisebox{\eqoff}{%
				\includegraphics[angle={0},scale=0.16,trim=0cm 0cm 0cm 0]{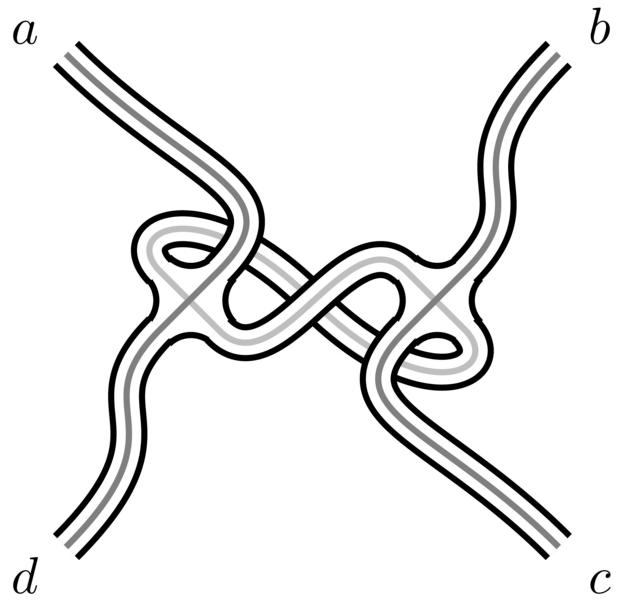}
			}
			\else
			\scalebox{.8}{
				\begin{pspicture}(-7.5,-7.5)(7.5,7.5)
				\fourvertex{-3}{0}{45}
				\fourvertex{3}{0}{45}
				\setlength{\xa}{1.5\unit}
				\addtolength{\xa}{-0.5\doublesep}
				\addtolength{\xa}{-\linew}
				\setlength{\xb}{4.5\unit}
				\addtolength{\xb}{0.5\doublesep}
				\addtolength{\xb}{\linew}
				\setlength{\xc}{7.5\unit}
				\addtolength{\xc}{-0.5\doublesep}
				\addtolength{\xc}{-\linew}
				\setlength{\xd}{10.5\unit}
				\addtolength{\xd}{0.5\doublesep}
				\addtolength{\xd}{\linew}
				\setlength{\ya}{6\unit}
				\addtolength{\ya}{0.5\doublesep}
				\addtolength{\ya}{\linew}
				\setlength{\yb}{3\unit}
				\addtolength{\yb}{-0.5\doublesep}
				\addtolength{\yb}{-\linew}
				\psset{linecolor=black,doubleline=true}
				\psbezier(-3.7071,0.7071)(-4.7071,1.7071)(-2.7071,2.7071)(0,0)
				\psbezier(0,0)(2.7071,-2.7071)(4.7071,-1.7071)(3.7071,-0.7071)
				\psset{linecolor=lightgray,doubleline=false}
				\psbezier(-3.7071,0.7071)(-4.7071,1.7071)(-2.7071,2.7071)(0,0)
				\psbezier(0,0)(2.7071,-2.7071)(4.7071,-1.7071)(3.7071,-0.7071)
				\psset{linecolor=black,doubleline=true}
				\psbezier(-2.2921,-0.7071)(-0.87868,-2.2071)(0.87868,2.12132)(2.2929,0.7071)
				\psset{linecolor=lightgray,doubleline=false}
				\psbezier(2.2929,0.7071)(0.87868,2.12132)(-0.87868,-2.2071)(-2.2921,-0.7071)
				\psline(3.7071,-0.7071)(2.2929,0.7071)
				\psline(-2.2921,-0.7071)(-3.7071,0.7071)
				\psset{linecolor=black,doubleline=true}
				\rput[B](-7,6.25){$a$}
				\psbezier(-2.2921,0.7071)(-0.2921,2.7071)(-3,3)(-6,6)
				\rput[B](7,6.25){$b$}
				\psbezier(3.7071,0.7071)(5.7071,2.7071)(3,3)(6,6)
				\rput[B](7,-7.125){$c$}
				\psbezier(6,-6)(3,-3)(0.2921,-2.7071)(2.2921,-0.7071)
				\rput[B](-7,-7.125){$d$}
				\psbezier(-6,-6)(-3,-3)(-5.7071,-2.7071)(-3.7071,-0.7071)
				\psset{linecolor=gray,doubleline=false}
				\psbezier(-6,-6)(-3,-3)(-5.7071,-2.7071)(-3.7071,-0.7071)
				\psline(-3.7071,-0.7071)(-2.2921,0.7071)
				\psbezier(-2.2921,0.7071)(-0.2921,2.7071)(-3,3)(-6,6)
				\psbezier(6,-6)(3,-3)(0.2921,-2.7071)(2.2921,-0.7071)
				\psline(2.2921,-0.7071)(3.7071,0.7071)
				\psbezier(3.7071,0.7071)(5.7071,2.7071)(3,3)(6,6)
				\end{pspicture}
			}
			\fi
		}
		\subfloat[\label{us4s4}]{%
			\ifpdf
			\settoheight{\eqoff}{$+$}%
			\setlength{\eqoff}{0.5\eqoff}%
			\addtolength{\eqoff}{-0\unit}%
			\raisebox{\eqoff}{%
				\includegraphics[angle={0},scale=0.16,trim=0cm 0cm 0cm 0]{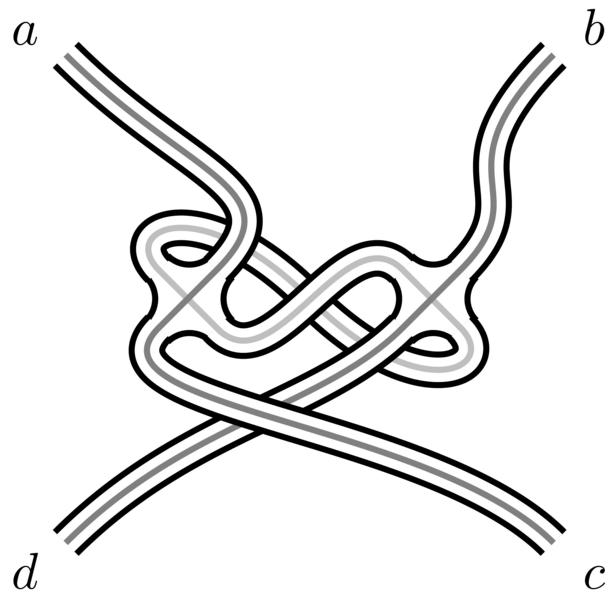}
			}
			\else
			\scalebox{.8}{
				\begin{pspicture}(-7.5,-7.5)(7.5,7.5)
				\fourvertex{-3}{0}{45}
				\fourvertex{3}{0}{45}
				\setlength{\xa}{1.5\unit}
				\addtolength{\xa}{-0.5\doublesep}
				\addtolength{\xa}{-\linew}
				\setlength{\xb}{4.5\unit}
				\addtolength{\xb}{0.5\doublesep}
				\addtolength{\xb}{\linew}
				\setlength{\xc}{7.5\unit}
				\addtolength{\xc}{-0.5\doublesep}
				\addtolength{\xc}{-\linew}
				\setlength{\xd}{10.5\unit}
				\addtolength{\xd}{0.5\doublesep}
				\addtolength{\xd}{\linew}
				\setlength{\ya}{6\unit}
				\addtolength{\ya}{0.5\doublesep}
				\addtolength{\ya}{\linew}
				\setlength{\yb}{3\unit}
				\addtolength{\yb}{-0.5\doublesep}
				\addtolength{\yb}{-\linew}
				\psset{linecolor=black,doubleline=true}
				\psbezier(-3.7071,0.7071)(-4.7071,1.7071)(-2.7071,2.7071)(0,0)
				\psbezier(0,0)(2.7071,-2.7071)(4.7071,-1.7071)(3.7071,-0.7071)
				\psset{linecolor=lightgray,doubleline=false}
				\psbezier(-3.7071,0.7071)(-4.7071,1.7071)(-2.7071,2.7071)(0,0)
				\psbezier(0,0)(2.7071,-2.7071)(4.7071,-1.7071)(3.7071,-0.7071)
				\psset{linecolor=black,doubleline=true}
				\psbezier(-2.2921,-0.7071)(-0.87868,-2.2071)(0.87868,2.12132)(2.2929,0.7071)
				\psset{linecolor=lightgray,doubleline=false}
				\psbezier(2.2929,0.7071)(0.87868,2.12132)(-0.87868,-2.2071)(-2.2921,-0.7071)
				\psline(3.7071,-0.7071)(2.2929,0.7071)
				\psline(-2.2921,-0.7071)(-3.7071,0.7071)
				\psset{linecolor=black,doubleline=true}
				\rput[B](-7,6.25){$a$}
				\psbezier(-2.2921,0.7071)(-0.2921,2.7071)(-3,3)(-6,6)
				\rput[B](7,6.25){$b$}
				\psbezier(3.7071,0.7071)(5.7071,2.7071)(3,3)(6,6)
				\rput[B](-7,-7.125){$d$}
				\psbezier(-6,-6)(-3,-3)(0.2921,-2.7071)(2.2921,-0.7071)
				\psset{linecolor=gray,doubleline=false}
				\psbezier(-6,-6)(-3,-3)(0.2921,-2.7071)(2.2921,-0.7071)
				\psline(2.2921,-0.7071)(3.7071,0.7071)
				\psbezier(3.7071,0.7071)(5.7071,2.7071)(3,3)(6,6)
				\psset{linecolor=black,doubleline=true}
				\rput[B](7,-7.125){$c$}
				\psbezier(6,-6)(3,-3)(-5.7071,-2.7071)(-3.7071,-0.7071)
				\psset{linecolor=gray,doubleline=false}
				\psbezier(6,-6)(3,-3)(-5.7071,-2.7071)(-3.7071,-0.7071)
				\psline(-3.7071,-0.7071)(-2.2921,0.7071)
				\psbezier(-2.2921,0.7071)(-0.2921,2.7071)(-3,3)(-6,6)
				\end{pspicture}
			}
			\fi
		}
		\subfloat[\label{ts4s4}]{%
			\ifpdf
			\settoheight{\eqoff}{$+$}%
			\setlength{\eqoff}{0.5\eqoff}%
			\addtolength{\eqoff}{-0\unit}%
			\raisebox{\eqoff}{%
				\includegraphics[angle={0},scale=0.16,trim=0cm 0cm 0cm 0]{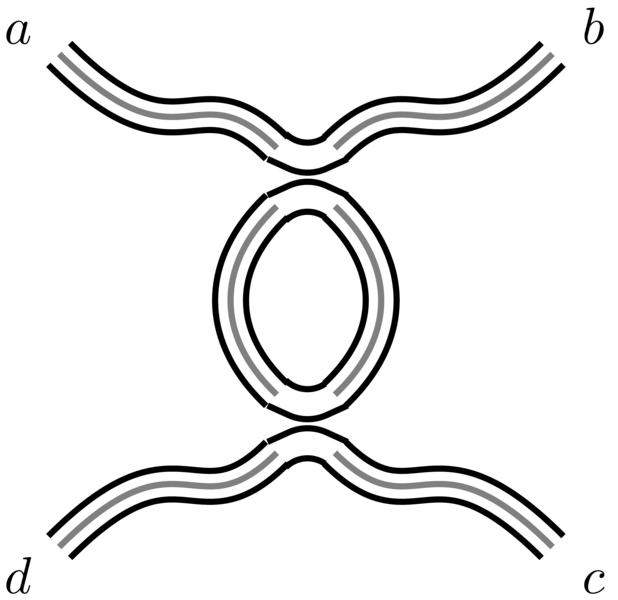}
			}
			\else
			\scalebox{.8}{
				\begin{pspicture}(-7.5,-7.5)(7.5,7.5)
				\fourvertexdbltr{0}{-3}{45}
				\fourvertexdbltr{0}{3}{45}
				\setlength{\xa}{1.5\unit}
				\addtolength{\xa}{-0.5\doublesep}
				\addtolength{\xa}{-\linew}
				\setlength{\xb}{4.5\unit}
				\addtolength{\xb}{0.5\doublesep}
				\addtolength{\xb}{\linew}
				\setlength{\xc}{7.5\unit}
				\addtolength{\xc}{-0.5\doublesep}
				\addtolength{\xc}{-\linew}
				\setlength{\xd}{10.5\unit}
				\addtolength{\xd}{0.5\doublesep}
				\addtolength{\xd}{\linew}
				\setlength{\ya}{6\unit}
				\addtolength{\ya}{0.5\doublesep}
				\addtolength{\ya}{\linew}
				\setlength{\yb}{3\unit}
				\addtolength{\yb}{-0.5\doublesep}
				\addtolength{\yb}{-\linew}
				\psset{doubleline=true}
				\psbezier(-0.7071,-2.2921)(-2.2071,-0.87868)(-2.2071,0.87868)(-0.7071,2.2929)
				\psbezier(0.7071,2.2929)(2.2071,0.87868)(2.2071,-0.87868)(0.7071,-2.2921)
				\rput[B](-7,6.25){$a$}
				\psbezier(-0.7071,3.7071)(-2.7071,5.7071)(-3,3)(-6,6)
				\rput[B](7,6.25){$b$}
				\psbezier(0.7071,3.7071)(2.7071,5.7071)(3,3)(6,6)
				\rput[B](7,-7.125){$c$}
				\psbezier(6,-6)(3,-3)(2.7071,-5.7071)(0.7071,-3.7071)
				\rput[B](-7,-7.125){$d$}
				\psbezier(-0.7071,-3.7071)(-2.7071,-5.7071)(-3,-3)(-6,-6)
				\psset{linecolor=gray,doubleline=false}
				\psbezier(-6,-6)(-3,-3)(-2.7071,-5.7071)(-0.7071,-3.7071)
				\psbezier(6,-6)(3,-3)(2.7071,-5.7071)(0.7071,-3.7071)
				\psbezier(-0.7071,3.7071)(-2.7071,5.7071)(-3,3)(-6,6)
				\psbezier(0.7071,3.7071)(2.7071,5.7071)(3,3)(6,6)
				\psbezier(-0.7071,-2.2921)(-2.2071,-0.87868)(-2.2071,0.87868)(-0.7071,2.2929)
				\psbezier(0.7071,-2.2921)(2.2071,-0.87868)(2.2071,0.87868)(0.7071,2.2929)
				\end{pspicture}
			}
			\fi
		}
		
		\subfloat[\label{tDs4s4}]{%
			\ifpdf
			\settoheight{\eqoff}{$+$}%
			\setlength{\eqoff}{0.5\eqoff}%
			\addtolength{\eqoff}{-0\unit}%
			\raisebox{\eqoff}{%
				\includegraphics[angle={0},scale=0.16,trim=0cm 0cm 0cm 0]{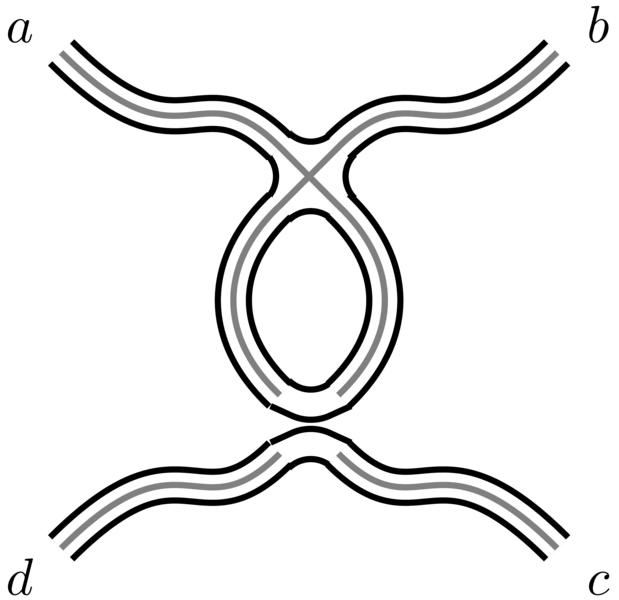}
			}
			\else
			\scalebox{.8}{
				\begin{pspicture}(-7.5,-7.5)(7.5,7.5)
				\fourvertexdbltr{0}{-3}{45}
				\fourvertex{0}{3}{45}
				\setlength{\xa}{1.5\unit}
				\addtolength{\xa}{-0.5\doublesep}
				\addtolength{\xa}{-\linew}
				\setlength{\xb}{4.5\unit}
				\addtolength{\xb}{0.5\doublesep}
				\addtolength{\xb}{\linew}
				\setlength{\xc}{7.5\unit}
				\addtolength{\xc}{-0.5\doublesep}
				\addtolength{\xc}{-\linew}
				\setlength{\xd}{10.5\unit}
				\addtolength{\xd}{0.5\doublesep}
				\addtolength{\xd}{\linew}
				\setlength{\ya}{6\unit}
				\addtolength{\ya}{0.5\doublesep}
				\addtolength{\ya}{\linew}
				\setlength{\yb}{3\unit}
				\addtolength{\yb}{-0.5\doublesep}
				\addtolength{\yb}{-\linew}
				\psset{doubleline=true}
				\psbezier(-0.7071,-2.2921)(-2.2071,-0.87868)(-2.2071,0.87868)(-0.7071,2.2929)
				\psbezier(0.7071,2.2929)(2.2071,0.87868)(2.2071,-0.87868)(0.7071,-2.2921)
				\rput[B](-7,6.25){$a$}
				\psbezier(-0.7071,3.7071)(-2.7071,5.7071)(-3,3)(-6,6)
				\rput[B](7,6.25){$b$}
				\psbezier(0.7071,3.7071)(2.7071,5.7071)(3,3)(6,6)
				\rput[B](7,-7.125){$c$}
				\psbezier(6,-6)(3,-3)(2.7071,-5.7071)(0.7071,-3.7071)
				\rput[B](-7,-7.125){$d$}
				\psbezier(-0.7071,-3.7071)(-2.7071,-5.7071)(-3,-3)(-6,-6)
				\psset{linecolor=gray,doubleline=false}
				\psbezier(-6,-6)(-3,-3)(-2.7071,-5.7071)(-0.7071,-3.7071)
				\psline(-0.7071,2.2929)(0.7071,3.7071)
				\psline(0.7071,2.2929)(-0.7071,3.7071)
				\psbezier(6,-6)(3,-3)(2.7071,-5.7071)(0.7071,-3.7071)
				\psbezier(-0.7071,3.7071)(-2.7071,5.7071)(-3,3)(-6,6)
				\psbezier(0.7071,3.7071)(2.7071,5.7071)(3,3)(6,6)
				\psbezier(-0.7071,-2.2921)(-2.2071,-0.87868)(-2.2071,0.87868)(-0.7071,2.2929)
				\psbezier(0.7071,-2.2921)(2.2071,-0.87868)(2.2071,0.87868)(0.7071,2.2929)
				\end{pspicture}
			}
			\fi
		}
		\subfloat[\label{s4gg2}]{%
			\ifpdf
			\settoheight{\eqoff}{$+$}%
			\setlength{\eqoff}{0.5\eqoff}%
			\addtolength{\eqoff}{-0\unit}%
			\raisebox{\eqoff}{%
				\includegraphics[angle={0},scale=0.16,trim=0cm 0cm 0cm 0]{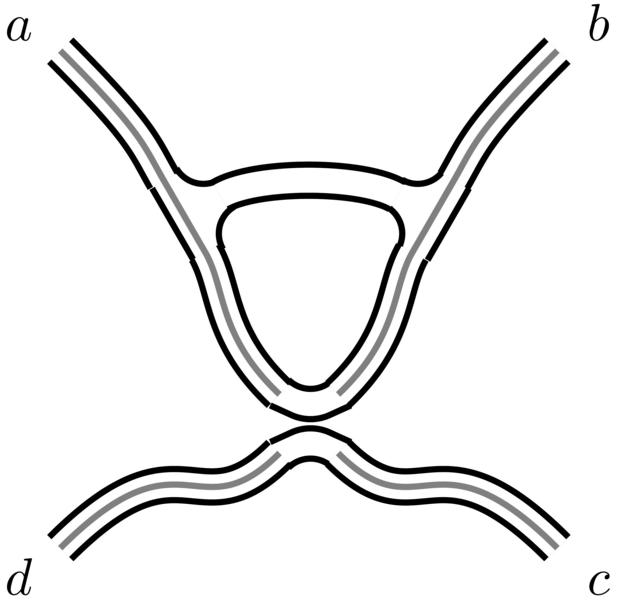}
			}
			\else
			\scalebox{.8}{
				\begin{pspicture}(-7.5,-7.5)(7.5,7.5)
				\fourvertexdbltr{0}{-3}{45}
				\threevertex{-3}{2}{30}
				\threevertex{3}{2}{150}
				\setlength{\xa}{1.5\unit}
				\addtolength{\xa}{-0.5\doublesep}
				\addtolength{\xa}{-\linew}
				\setlength{\xb}{4.5\unit}
				\addtolength{\xb}{0.5\doublesep}
				\addtolength{\xb}{\linew}
				\setlength{\xc}{7.5\unit}
				\addtolength{\xc}{-0.5\doublesep}
				\addtolength{\xc}{-\linew}
				\setlength{\xd}{10.5\unit}
				\addtolength{\xd}{0.5\doublesep}
				\addtolength{\xd}{\linew}
				\setlength{\ya}{6\unit}
				\addtolength{\ya}{0.5\doublesep}
				\addtolength{\ya}{\linew}
				\setlength{\yb}{3\unit}
				\addtolength{\yb}{-0.5\doublesep}
				\addtolength{\yb}{-\linew}
				\psset{doubleline=true}
				\psbezier(-2.134,2.5)(-1.268,3)(1.268,3)(2.134,2.5)
				\rput[B](-7,6.25){$a$}
				\psbezier(-0.7071,-2.2921)(-2.2071,-0.87868)(-2,0.268)(-2.5,1.134)
				\psbezier(-3.5,2.866)(-4,3.732)(-4,4)(-6,6)
				\rput[B](7,6.25){$b$}
				\psbezier(0.7071,-2.2921)(2.2071,-0.87868)(2,0.268)(2.5,1.134)
				\psbezier(3.5,2.866)(4,3.732)(4,4)(6,6)
				\rput[B](7,-7.125){$c$}
				\psbezier(6,-6)(3,-3)(2.7071,-5.7071)(0.7071,-3.7071)
				\rput[B](-7,-7.125){$d$}
				\psbezier(-0.7071,-3.7071)(-2.7071,-5.7071)(-3,-3)(-6,-6)
				\psset{linecolor=gray,doubleline=false}
				\psbezier(-6,-6)(-3,-3)(-2.7071,-5.7071)(-0.7071,-3.7071)
				\psbezier(6,-6)(3,-3)(2.7071,-5.7071)(0.7071,-3.7071)
				\psbezier(-0.7071,-2.2921)(-2.2071,-0.87868)(-2,0.268)(-2.5,1.134)
				\psline(-2.5,1.134)(-3.5,2.866)
				\psbezier(-3.5,2.866)(-4,3.732)(-4,4)(-6,6)
				\psbezier(0.7071,-2.2921)(2.2071,-0.87868)(2,0.268)(2.5,1.134)
				\psline(2.5,1.134)(3.5,2.866)
				\psbezier(3.5,2.866)(4,3.732)(4,4)(6,6)
				\end{pspicture}
			}
			\fi
		}
		\subfloat[\label{s4gg3}]{%
			\ifpdf
			\settoheight{\eqoff}{$+$}%
			\setlength{\eqoff}{0.5\eqoff}%
			\addtolength{\eqoff}{-0\unit}%
			\raisebox{\eqoff}{%
				\includegraphics[angle={0},scale=0.16,trim=0cm 0cm 0cm 0]{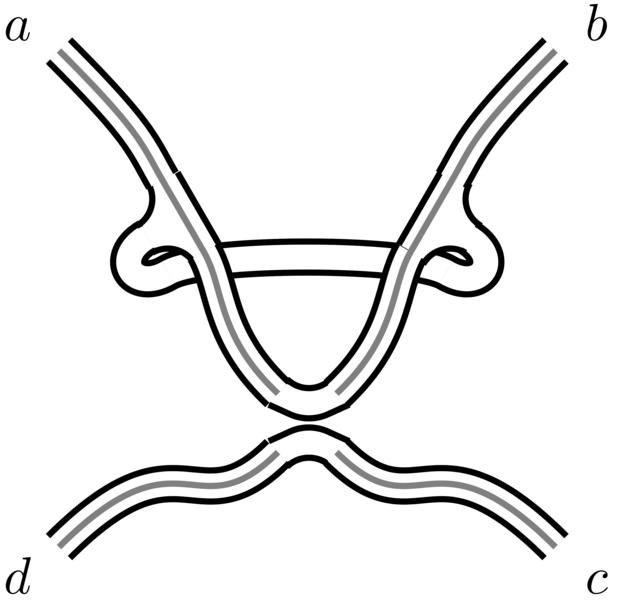}
			}
			\else
			\scalebox{.8}{
				\begin{pspicture}(-7.5,-7.5)(7.5,7.5)
				\psset{doubleline=true}
				\psbezier(-3.866,1.5)(-4.732,1)(-4.232,0.134)(-3.366,0.634)
				\psbezier(-3.366,0.634)(-2.5,1.134)(2.5,1.134)(3.366,0.634)
				\psbezier(3.866,1.5)(4.732,1)(4.232,0.134)(3.366,0.634)
				\fourvertexdbltr{0}{-3}{45}
				\threevertex{-3}{2}{210}
				\threevertex{3}{2}{-30}
				\setlength{\xa}{1.5\unit}
				\addtolength{\xa}{-0.5\doublesep}
				\addtolength{\xa}{-\linew}
				\setlength{\xb}{4.5\unit}
				\addtolength{\xb}{0.5\doublesep}
				\addtolength{\xb}{\linew}
				\setlength{\xc}{7.5\unit}
				\addtolength{\xc}{-0.5\doublesep}
				\addtolength{\xc}{-\linew}
				\setlength{\xd}{10.5\unit}
				\addtolength{\xd}{0.5\doublesep}
				\addtolength{\xd}{\linew}
				\setlength{\ya}{6\unit}
				\addtolength{\ya}{0.5\doublesep}
				\addtolength{\ya}{\linew}
				\setlength{\yb}{3\unit}
				\addtolength{\yb}{-0.5\doublesep}
				\addtolength{\yb}{-\linew}
				\psset{doubleline=true}
				\rput[B](-7,6.25){$a$}
				\psbezier(-0.7071,-2.2921)(-2.2071,-0.87868)(-2,0.268)(-2.5,1.134)
				\psbezier(-3.5,2.866)(-4,3.732)(-4,4)(-6,6)
				\rput[B](7,6.25){$b$}
				\psbezier(0.7071,-2.2921)(2.2071,-0.87868)(2,0.268)(2.5,1.134)
				\psbezier(3.5,2.866)(4,3.732)(4,4)(6,6)
				\rput[B](7,-7.125){$c$}
				\psbezier(6,-6)(3,-3)(2.7071,-5.7071)(0.7071,-3.7071)
				\rput[B](-7,-7.125){$d$}
				\psbezier(-0.7071,-3.7071)(-2.7071,-5.7071)(-3,-3)(-6,-6)
				\psset{linecolor=gray,doubleline=false}
				\psbezier(-6,-6)(-3,-3)(-2.7071,-5.7071)(-0.7071,-3.7071)
				\psbezier(6,-6)(3,-3)(2.7071,-5.7071)(0.7071,-3.7071)
				\psbezier(-0.7071,-2.2921)(-2.2071,-0.87868)(-2,0.268)(-2.5,1.134)
				\psline(-2.5,1.134)(-3.5,2.866)
				\psbezier(-3.5,2.866)(-4,3.732)(-4,4)(-6,6)
				\psbezier(0.7071,-2.2921)(2.2071,-0.87868)(2,0.268)(2.5,1.134)
				\psline(2.5,1.134)(3.5,2.866)
				\psbezier(3.5,2.866)(4,3.732)(4,4)(6,6)
				\end{pspicture}
			}
			\fi
		}
		
		\subfloat[\label{fbox1}]{%
			\ifpdf
			\settoheight{\eqoff}{$+$}%
			\setlength{\eqoff}{0.5\eqoff}%
			\addtolength{\eqoff}{-0\unit}%
			\raisebox{\eqoff}{%
				\includegraphics[angle={0},scale=0.16,trim=0cm 0cm 0cm 0]{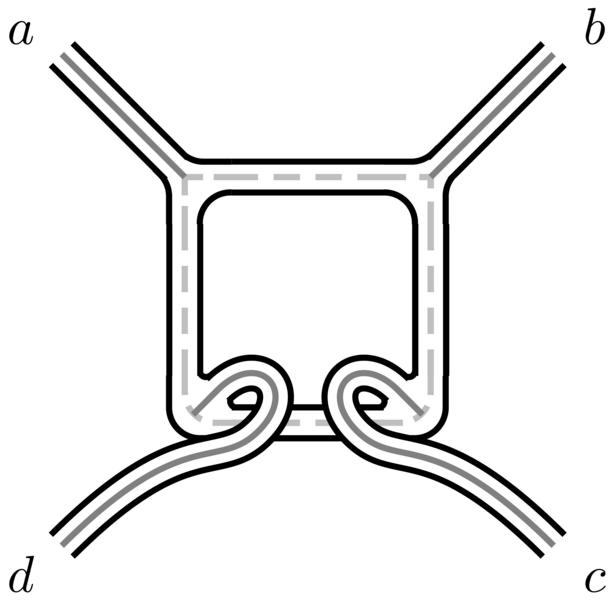}
			}
			\else
			\scalebox{.8}{
				\begin{pspicture}(-8.5,-8.5)(8.5,8.5)
				\recthreevertex{-3}{3}{135}
				\recthreevertex{3}{3}{45}
				\recthreevertexoneside{3}{-3}{-45}
				\recthreevertexoneside{-3}{-3}{-135}
				\psset{linecolor=black,doubleline=true}
				\psline(-1.85,3)(1.85,3)
				\psline(3,1.85)(3,-1.85)
				\psline(1.85,-3)(-1.85,-3)
				\psline(-3,-1.85)(-3,1.85)
				\psset{linecolor=lightgray,doubleline=false,linestyle=dashed}
				\psline(-3,1.85)(-3,3)
				\psline(-3,3)(-1.85,3)
				\psline(-1.85,3)(1.85,3)
				\psline(1.85,3)(3,3)(3,1.85)
				\psline(3,1.85)(3,-1.85)
				\psline[linearc=\linearc](3,-1.85)(3,-3)(1.85,-3)
				\psline(1.85,-3)(-1.85,-3)
				\psline[linearc=\linearc](-1.85,-3)(-3,-3)(-3,-1.85)
				\psline(-3,-1.85)(-3,1.85)
				\psset{linecolor=black,doubleline=true,linestyle=solid}
				\rput[B](-7,6.25){$a$}
				\psline(-3.2,3.2)(-6,6)
				\rput[B](7,6.25){$b$}
				\psline(3.2,3.2)(6,6)
				\rput[B](7,-7.125){$c$}
				\psbezier(2.2,-2.2)(1.2,-1.2)(0.2,-2.2)(1.2,-3.2)
				\psbezier(1.2,-3.2)(2.2,-4.2)(3,-3)(6,-6)
				\rput[B](-7,-7.125){$d$}
				\psbezier(-2.2,-2.2)(-1.2,-1.2)(-0.2,-2.2)(-1.2,-3.2)
				\psbezier(-1.2,-3.2)(-2.2,-4.2)(-3,-3)(-6,-6)
				\psset{linecolor=gray,doubleline=false}
				\psline(-3,3)(-3.2,3.2)
				\psline(-3.2,3.2)(-6,6)
				\psline(3,3)(3.2,3.2)
				\psline(3.2,3.2)(6,6)
				\psline(2.8,-2.8)(2.2,-2.2)
				\psbezier(2.2,-2.2)(1.2,-1.2)(0.2,-2.2)(1.2,-3.2)
				\psbezier(1.2,-3.2)(2.2,-4.2)(3,-3)(6,-6)
				\psline(-2.8,-2.8)(-2.2,-2.2)
				\psbezier(-2.2,-2.2)(-1.2,-1.2)(-0.2,-2.2)(-1.2,-3.2)
				\psbezier(-1.2,-3.2)(-2.2,-4.2)(-3,-3)(-6,-6)
				\end{pspicture}
			}
			\fi
		}
		\subfloat[\label{fbox2}]{%
			\ifpdf
			\settoheight{\eqoff}{$+$}%
			\setlength{\eqoff}{0.5\eqoff}%
			\addtolength{\eqoff}{-0\unit}%
			\raisebox{\eqoff}{%
				\includegraphics[angle={0},scale=0.16,trim=0cm 0cm 0cm 0]{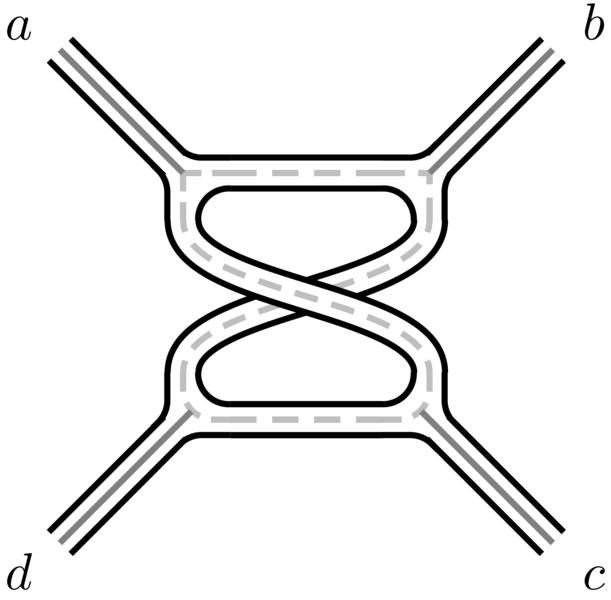}
			}
			\else
			\scalebox{.8}{
				\begin{pspicture}(-8.5,-8.5)(8.5,8.5)
				\recthreevertex{-3}{3}{135}
				\recthreevertex{3}{3}{45}
				\recthreevertex{3}{-3}{-45}
				\recthreevertex{-3}{-3}{-135}
				\psset{linecolor=black,doubleline=true}
				\psline(-1.85,3)(1.85,3)
				\psbezier(3,1.85)(3,0)(-3,0)(-3,-1.85)
				\psline(-1.85,-3)(1.85,-3)
				\psset{linecolor=lightgray,doubleline=false,linestyle=dashed}
				\psline(-3,1.85)(-3,3)
				\psline(-3,3)(-1.85,3)
				\psline(-1.85,3)(1.85,3)
				\psline(1.85,3)(3,3)(3,1.85)
				\psbezier(3,1.85)(3,0)(-3,0)(-3,-1.85)
				\psline[linearc=\linearc](-3,-1.85)(-3,-3)(-1.85,-3)
				\psline(-1.85,-3)(1.85,-3)
				\psline[linearc=\linearc](1.85,-3)(3,-3)(3,-1.85)
				\psset{linecolor=black,doubleline=true,linestyle=solid}
				\psbezier(3,-1.85)(3,0)(-3,0)(-3,1.85)
				\psset{linecolor=lightgray,doubleline=false,linestyle=dashed}
				\psbezier(3,-1.85)(3,0)(-3,0)(-3,1.85)
				\psset{linecolor=black,doubleline=true,linestyle=solid}
				\rput[B](-7,6.25){$a$}
				\psline(-3.2,3.2)(-6,6)
				\rput[B](7,6.25){$b$}
				\psline(3.2,3.2)(6,6)
				\rput[B](7,-7.125){$c$}
				\psline(-3.2,-3.2)(-6,-6)
				\rput[B](-7,-7.125){$d$}
				\psline(3.2,-3.2)(6,-6)
				\psset{linecolor=gray,doubleline=false}
				\psline(-3,3)(-3.2,3.2)
				\psline(-3.2,3.2)(-6,6)
				\psline(3,3)(3.2,3.2)
				\psline(3.2,3.2)(6,6)
				\psline(2.8,-2.8)(2.2,-2.2)
				\psline(-3.2,-3.2)(-6,-6)
				\psline(-2.8,-2.8)(-2.2,-2.2)
				\psline(3.2,-3.2)(6,-6)
				\end{pspicture}
			}
			\fi
		}
		\subfloat[\label{fbox3}]{%
			\ifpdf
			\settoheight{\eqoff}{$+$}%
			\setlength{\eqoff}{0.5\eqoff}%
			\addtolength{\eqoff}{-0\unit}%
			\raisebox{\eqoff}{%
				\includegraphics[angle={0},scale=0.16,trim=0cm 0cm 0cm 0]{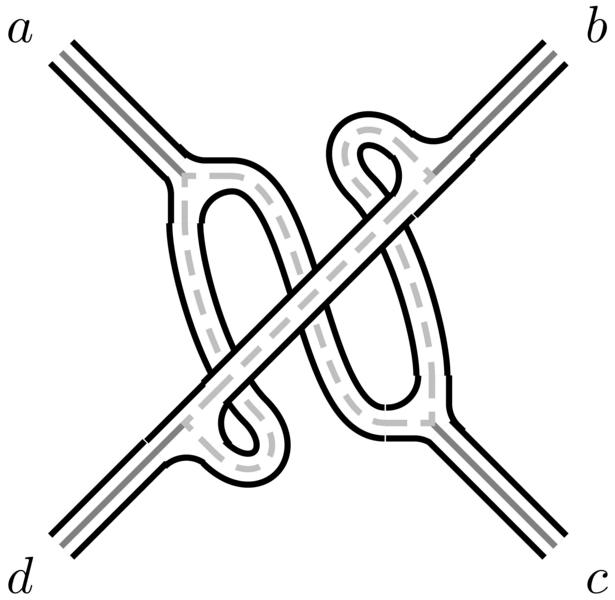}
			}
			\else
			\scalebox{.8}{
				\begin{pspicture}(-8.5,-8.5)(8.5,8.5)
				\recthreevertex{-3}{3}{135}
				\threevertex{3}{3}{135}
				\recthreevertex{3}{-3}{-45}
				\threevertex{-3}{-3}{-45}
				\psset{linecolor=black,doubleline=true}
				\psbezier(-1.85,3)(0,3)(0,-3)(1.85,-3)
				\psbezier(3,-1.85)(3,-0.85)(2.2929,1.7071)(1.2929,2.7071)
				\psbezier(1.2929,2.7071)(0.2929,3.7071)(1.2929,4.7071)(2.2929,3.7071)
				\psbezier(-2.2929,-3.7071)(-1.2929,-4.7071)(-0.2929,-3.7071)(-1.2929,-2.7071)
				\psbezier(-1.2929,-2.7071)(-2.2929,-1.7071)(-3,0.85)(-3,1.85)
				\rput[B](-7,6.25){$a$}
				\psline(-3.2,3.2)(-6,6)
				\rput[B](7,6.25){$b$}
				\psline(3.7071,3.7071)(6,6)
				\rput[B](7,-7.125){$c$}
				\psline(3.2,-3.2)(6,-6)
				\rput[B](-7,-7.125){$d$}
				\psline(-3.7071,-3.7071)(-6,-6)
				\psset{linecolor=gray,doubleline=false}
				\psline(-3,3)(-3.2,3.2)
				\psline(-3.2,3.2)(-6,6)
				\psline(3,3)(3.7071,3.7071)
				\psline(3.7071,3.7071)(6,6)
				\psline(3,-3)(3.2,-3.2)
				\psline(3.2,-3.2)(6,-6)
				\psline(-3,-3)(-3.7071,-3.7071)
				\psline(-3.7071,-3.7071)(-6,-6)
				\psset{linecolor=lightgray,doubleline=false,linestyle=dashed}
				\psline(-3,1.85)(-3,3)(-1.85,3)
				\psbezier(-1.85,3)(0,3)(0,-3)(1.85,-3)
				\psline(1.85,-3)(3,-3)(3,-1.85)
				\psbezier(3,-1.85)(3,-0.85)(2.2929,1.7071)(1.2929,2.7071)
				\psbezier(1.2929,2.7071)(0.2929,3.7071)(1.2929,4.7071)(2.2929,3.7071)
				\psline(2.2929,3.7071)(3,3)(2.2929,2.2929)
				\psline(-2.2929,-2.2929)(-3,-3)(-2.2929,-3.7071)
				\psbezier(-2.2929,-3.7071)(-1.2929,-4.7071)(-0.2929,-3.7071)(-1.2929,-2.7071)
				\psbezier(-1.2929,-2.7071)(-2.2929,-1.7071)(-3,0.85)(-3,1.85)
				\psset{linecolor=black,doubleline=true,linestyle=solid}
				\psline(-2.2929,-2.2929)(2.2929,2.2929)
				\psset{linecolor=lightgray,doubleline=false,linestyle=dashed}
				\psline(-2.2929,-2.2929)(2.2929,2.2929)
				\end{pspicture}
			}
			\fi
		}
		\caption{
			Complete list of 1PI diagrams (up to conjugation) that contribute to 			$\phi_i^{a}\phi_i^{b}\bar\phi^{ic}\bar\phi^{id}\big(ab\big)\big(cd\big)$ and deviate from the undeformed ones in \NfSYMt. The diagrams are given in the double-line notation introduced in \secref{sec:tHooft_limit}. A central plain or dashed line indicates the scalar or fermionic flavour, respectively and the flavour-neutral gauge fields appear without central line. Using the vertex classification of \eqref{eq:cubic_vertices} and \eqref{eq:quartic_vertices}, the diagrams involve \protect\subref{ss4s4}, \protect\subref{us4s4}: two single-trace $F$-tensor vertices; 
			\protect\subref{ts4s4}: two double-trace $Q_{\text{F}}$-tensor vertices; 
			\protect\subref{tDs4s4}: one double-trace $Q_{\text{F}}$-tensor and one single-trace $V^{\text{D}}$ vertex; 
			\protect\subref{s4gg2}, \protect\subref{s4gg3}: one double-trace $Q_{\text{F}}$-tensor with gauge field exchange; 
			\protect\subref{fbox1}, \protect\subref{fbox2}, \protect\subref{fbox3}: a fermion box with four 
			Yukawa-type interactions. 
		}\label{fig:dtd}
	\end{center}
\end{figure}
Note that we did not use the full vertices from the Feynman rules, but only considered the trace structure explicitly displayed in \figref{fig:dtd}. The operator $\Rop_|$ reflects a diagram at the vertical axis and restores the original order of the external labels and legs. In the special kinematics chosen in \eqref{eq:gamma_i_matrix_element}, the spacetime integral in each contribution can be evaluated\footnote{For more general techniques to evaluate the pole parts of momentum space integrals, see \appref{app:Evaluating_Feynman_integrals}.} exactly and its UV divergence in $D=4-2\epsilon$ dimensions is extracted from the pole part of the integral \eqref{eq:G_function} as
\begin{equation}\label{eq:def_dep_contributions}
\Kop[\hat{I}_{(\alpha,\beta)}(p)]=\WR^{-1}\Bigl(\complexi\mu^{2\epsilon}\Kop\bigl[
\int\frac{\de^D\bar l}{(2\pi)^D}\frac{1}{\bar l^{2\alpha}(\bar p-\bar l)^{2\beta}}
\bigr]\Bigr)\eqncom
\end{equation}
where the bar indicates momenta in Euclidean space and the operator $\WR^{-1}$ realises the inverse Wick rotation back to Minkowski space. The two integrals that occur in \eqref{diagres} have the pole parts $\Kop[\hat{I}_{(1,1)}(p)]=-\Kop[\hat{I}_{(2,1)}(p)]=\frac{\complexi}{(4\pi)^2\epsilon}$ and with these the sums of the UV divergences of each row in \figref{fig:dtd} is given by 
\begin{equation}
\begin{aligned}\label{dtdsubsum}
(1+\Rop_|)[\,\subref{ss4s4}+\subref{us4s4}\,]+\subref{ts4s4}
&=2\big(16\cos 2\gamma_i^+\cos 2\gamma_i^- 
+(Q^{ii}_{\text{F}\,ii})^2\big)\frac{\complexi\gym^4\mu^{2\epsilon}\big(ab\big)\big(cd\big)}{(4\pi)^2\epsilon}
\eqncom\\
2\bigl[\subref{tDs4s4}+\subref{s4gg2}+\subref{s4gg3}\bigr]
&=-8(1+\xi)Q^{ii}_{\text{F}\,ii}\frac{\complexi\gym^4\mu^{2\epsilon}\big(ab\big)\big(cd\big)}{(4\pi)^2\epsilon} 
\eqncom\\
(1+\Rop_|)\subref{fbox3}
&=-32(\cos 2\gamma_i^++\cos 2\gamma_i^-)
\frac{\complexi\gym^4\mu^{2\epsilon}\big(ab\big)\big(cd\big)}{(4\pi)^2\epsilon}
\eqncom
\end{aligned}
\end{equation}
where we used the flavour-tensor identity \eqref{FFsum} and the conservation of the $\mathfrak{u}(1)^{\times 3}$ Cartan charges in the first and flavour-tensor identity \eqref{l4rhotraces} in the last line. The additional factors of two for the contributions from \subref{tDs4s4}, \subref{s4gg2} and \subref{s4gg3} arise since these diagrams give the same result when the upper and lower vertices are interchanged. From the deformation-dependent contributions in \eqref{eq:def_dep_contributions}, we can now reconstruct the neglected ones by enforcing that their sum vanishes when we set $\gamma_i^\pm=Q^{ii}_{\text{F}\,ii}=0$. We find that the neglected terms yield a total contribution of $32 \gym^4\mu^{2\epsilon}\Kop[\hat{I}_{(1,1)}(p)]\big(ab\big)\big(cd\big)$. Adding this to the sum of \eqref{dtdsubsum} and removing the tensor structure displayed in \eqref{eq:gamma_i_matrix_element}, we find the divergent one-loop 1PI contribution
\begin{equation}
\begin{aligned}\label{deltaQ}
\Kop\bigl[V^{(1)}_{Q_{\text{F}}\text{B}}(p,0,-p)\bigr]=-\delta_{Q_{\text{F}}}^{(1)}Q^{ii}_{\text{F}\,ii}
=-\frac{\gym^2N}{(4\pi)^2\epsilon}\big(64\sin^2\gamma_i^+\sin^2\gamma_i^-
+(Q^{ii}_{\text{F}\,ii})^2-4(1+\xi)Q^{ii}_{\text{F}\,ii}\big)
\eqncom
\end{aligned}
\end{equation}
where the one-loop 1PI counterterm is fixed\footnote{Defining the counterterm like this amounts to choosing the dimensional reduction scheme with minimal subtraction where only the pole parts are subtracted, see \appref{subsec:The_renormalisation_procedure}.} so that \eqref{eq:gamma_i_matrix_element} is free of divergences.

For the non-1PI contributions to the connected renormalisation constant of $\mathfrak{d}_{Q_{\text{F}}}$ we need the one-loop scalar self-energy counterterm 
\begin{equation}\label{eq:delta_phi_1}
\delta_{\phi}^{(1)}=
-\Kop\Big[\Bigl(
\settoheight{\eqoff}{$\times$}%
\setlength{\eqoff}{0.5\eqoff}%
\addtolength{\eqoff}{-5.\unitlength}%
\raisebox{\eqoff}{%
	\fmfframe(0,1)(-2,1){%
		\begin{fmfchar*}(15,7.5)
		\fmfforce{0 w,0.5 h}{v1}
		\fmfforce{1 w,0.5 h}{v2}
		\fmf{phantom}{v1,v2}
		\fmffreeze
		\fmfposition
		\fmfipath{p[]}
		\fmfiset{p1}{vpath(__v1,__v2)}
		\fmfiset{p11}{subpath (0,2length(p1)/5) of p1}
		\fmfiset{p12}{subpath (3length(p1)/5,length(p1)) of p1}
		\fmfi{plain_ar}{p11}
		\fmfi{plain_ar}{p12}
		\fmfiv{label=$\scriptstyle i a$,label.angle=-60,l.dist=2}{vloc(__v1)}
		\fmfiv{label=$\scriptstyle j b$,label.angle=-120,l.dist=2}{vloc(__v2)}
		\fmfiv{decor.shape=circle,decor.filled=shaded,decor.size=11thin}{point length(p1)/2 of p1}
		\end{fmfchar*}
	}
}
\Bigr)_{\text{am}}
\Bigr]\Bigr|_{ip^2\delta^j_i(ab)}
=2(1+\xi)\frac{\gym^2N}{(4\pi)^2\epsilon}\eqncom
\end{equation}
which is calculated in \appref{app:oneloopse} and where we used the vertical bar to the right projects out the overall tensor structure. In the renormalised coupling $Q_{\text{F}}$ this counterterm renormalises contributions of the form
\begin{equation}
0=\Kop\Biggl[
\settoheight{\eqoff}{$\times$}%
\setlength{\eqoff}{0.5\eqoff}%
\addtolength{\eqoff}{-6\unitlength}%
\raisebox{\eqoff}{%
	\fmfframe(2,1)(0,1){%
		\begin{fmfchar*}(10,10)
		\fmfforce{0 w,1 h}{v1}
		\fmfforce{1 w,1 h}{v2}
		\fmfforce{1 w,0 h}{v3}
		\fmfforce{0 w,0 h}{v4}
		\fmf{phantom}{v1,vc}
		\fmf{plain_ar}{vc,v2}
		\fmf{plain_rar}{vc,v3}
		\fmf{plain_ar}{v4,vc}
		\fmffreeze
		\fmfposition
		\fmfiv{label=$\scriptstyle i a$,label.angle=-10,label.dist=6}{vloc(__v1)}
		\fmfiv{label=$\scriptstyle i b$,label.angle=-100,label.dist=6}{vloc(__v2)}
		\fmfiv{label=$\scriptstyle i c$,label.angle=160,label.dist=4}{vloc(__v3)}
		\fmfiv{label=$\scriptstyle i d$,label.angle=80,label.dist=6}{vloc(__v4)}
		\fmfiv{label=$\scriptscriptstyle Q_{\text{F}}$,l.a=-80,l.dist=4}{vloc(__vc)}
		\fmffreeze
		\fmfposition
		\fmfipath{p[]}
		\fmfiset{p1}{vpath(__vc,__v1)}
		\fmfiset{p11}{subpath (0,length(p1)/2) of p1}
		\fmfiset{p12}{subpath (length(p1)/2,length(p1)) of p1}
		\fmfi{plain_srarrow}{p11}
		\fmfi{plain_srarrow}{p12}
		\fmfiv{decor.shape=circle,decor.filled=shaded,decor.size=5thin}{point length(p1)/2 of p1}
		\end{fmfchar*}
	}
}
+
\settoheight{\eqoff}{$\times$}%
\setlength{\eqoff}{0.5\eqoff}%
\addtolength{\eqoff}{-6\unitlength}%
\raisebox{\eqoff}{%
	\fmfframe(2,1)(0,1){%
		\begin{fmfchar*}(10,10)
		\fmfforce{0 w,1 h}{v1}
		\fmfforce{1 w,1 h}{v2}
		\fmfforce{1 w,0 h}{v3}
		\fmfforce{0 w,0 h}{v4}
		\fmf{phantom}{v1,vc}
		\fmf{plain_ar}{vc,v2}
		\fmf{plain_rar}{vc,v3}
		\fmf{plain_ar}{v4,vc}
		\fmffreeze
		\fmfposition
		\fmfiv{label=$\scriptstyle i a$,label.angle=-10,label.dist=6}{vloc(__v1)}
		\fmfiv{label=$\scriptstyle i b$,label.angle=-100,label.dist=6}{vloc(__v2)}
		\fmfiv{label=$\scriptstyle i c$,label.angle=160,label.dist=4}{vloc(__v3)}
		\fmfiv{label=$\scriptstyle i d$,label.angle=80,label.dist=6}{vloc(__v4)}
		\fmfiv{label=$\scriptscriptstyle Q_{\text{F}}$,l.a=-80,l.dist=4}{vloc(__vc)}
		\fmffreeze
		\fmfposition
		\fmfipath{p[]}
		\fmfiset{p1}{vpath(__vc,__v1)}
		\fmfiset{p11}{subpath (0,length(p1)/2) of p1}
		\fmfiset{p12}{subpath (length(p1)/2,length(p1)) of p1}
		\fmfi{plain_srarrow}{p11}
		\fmfi{plain_srarrow}{p12}
		\fmfiv{decor.shape=hexacross,decor.size=7thin}{point length(p1)/2 of p1}
		\end{fmfchar*}
	}
}\Biggr]
\eqndot
\end{equation}
Combining \eqref{deltaQ} and \eqref{eq:delta_phi_1} yields the connected one-loop counterterm as
\begin{equation}\label{eq:full_ct_QF}
\mathfrak{d}_{Q_{\text{F}}}^{(1)}Q^{ii}_{\text{F}\,ii}=(\delta_{Q_{\text{F}}}^{(1)}+2\delta_\phi^{(1)})Q^{ii}_{\text{F}\,ii}
=
\frac{g^2}{\epsilon}\big(64\sin^2\gamma_i^+\sin^2\gamma_i^-
+(Q^{ii}_{\text{F}\,ii})^2\big)\eqncom
\end{equation}
where we wrote the result in terms of the effective planar coupling $g^2=\frac{\gym^2N}{(4\pi)^2}$. We see that the gauge-dependence drops out as required for observables. In the unpublished work \cite{DymarskyRoiban} this counterterm was also obtained and our results agree.\footnote{We thank Radu Roiban for communication on this point.} The one-loop anomalous dimension and $\beta$-function of the couplings $Q^{ii}_{\text{F}\,ii}$ are now obtained using \eqref{eq:renormalised_coupling_mu_dep} and \eqref{eq:beta_function} and we find
\begin{equation}\label{eq:beta_function_QF}
\beta^{(1)}_{Q_{\text{F}}}=Q^{ii}_{\text{F}\,ii}\gamma_{Q_{\text{F}}}^{(1)}
=Q^{ii}_{\text{F}\,ii}\frac{-\mu}{1+\mathfrak{d}_{Q_{\text{F}}}^{(1)}}\frac{\de\,\bigl(1+\mathfrak{d}_{Q_{\text{F}}}^{(1)}\bigr)}{\de \mu}
=2g^2\big(64\sin^2\gamma_i^+\sin^2\gamma_i^-
+(Q^{ii}_{\text{F}\,ii})^2\big)\eqncom
\end{equation}
where we used \eqref{eq:gYM} to for the $\mu$-dependence of $\gym$. This one-loop $\beta$-function is indeed non-vanishing for real parameters $\gamma_i^\pm$ and $Q^{ii}_{\text{F}\,ii}$ and hence the \SUN $\gamma_i$-deformation is not conformally invariant in the \tHooft limit.

\subsubsection{Gauge group \texorpdfstring{\UN}{U(N)}}
In the previous paragraph, we saw that the coupling $Q_{\text{F}}$ has a non-vanishing $\beta$-function, rendering the $\gamma_i$-deformation with gauge group \SUN non-conformal. For the \UN theory, there is still the hope that the additional couplings displayed in \subsecref{sec:multi-trace-parts-of-the-action} contribute to the renormalisation of $Q_{\text{F}}$ and can be tuned to render $\beta^{(1)}_{Q_{\text{F}}}$ zero. This is, however, not possible, as we show now.

Additional diagrams that contribute to the renormalisation of \eqref{eq:running_dt_coupling} in the \UN theory are of the form displayed in \figref{fig:dtd} but with a one or more vertices replaced by the higher-trace vertices given in \eqref{eq:3_trace_action} and \eqref{eq:4_trace_action}. In addition, the external double-trace structure must be kept in all diagrams. All diagrams that can be drawn like this are suppressed in the \tHooft limit, since the higher-trace vertices introduce additional suppression factors $\frac 1N$ from the coupling constant but cannot increase the number of internal colour loops. Hence, in the \tHooft limit at one-loop level the $\beta$-function of \eqref{eq:running_dt_coupling} is the same in the $\gamma_i$-deformation with gauge group \UN and it is given in \eqref{eq:beta_function_QF}.

\subsection{Immediate implications for the \texorpdfstring{\AdSCFTc}{AdS/CFT correspondence}}

We found that the $\gamma_i$-deformations with either gauge group \UN or \SUN has a positive $\beta$-function \eqref{eq:beta_function_QF} in the \tHooft limit for generic real deformation parameters $\gamma_i^\pm$ and $Q^{ii}_{\text{F}\,ii}$. In the absence further fix-points, the positivity of the $\beta$-function implies that the coupling increases at higher energy scales. Hence, the $\gamma_i$-deformations are not conformally invariant and potentially strongly coupled in the high energy regime, which raises questions concerning the \AdSCFTc of the $\gamma_i$-deformation. 

Most bluntly put, it is possible that the \AdSCFTc does not hold in the setup of the non-supersymmetric and non-conformal $\gamma_i$-deformation. In lack of a working proof of the correspondence in any setup, it is hard to verify this scenario. If the breakdown of the \AdSCFTc in the $\gamma_i$-deformed setup could, however, be shown, the question arises whether this breakdown is related to the lack of supersymmetry or the lack of conformality. In the other scenario, where the \AdSCFTc holds even for the $\gamma_i$-deformation, we have three possible implications from the non-conformality:
\begin{enumerate}
	\item The background of the dual string-theory is destabilised by the emergence of closed string tachyons related to the non-conformal multi-trace couplings, similar to the findings in a non-supersymmetric orbifold setup \cite{Dymarsky:2005nc}. Indeed, tachyons were found in the $\gamma_i$-deformed flat space \cite{Spradlin:2005sv}, but an exact connection to the instabilities of the $\gamma_i$-deformation still needs to be established.\footnote{We thank Radu Roiban for this comment.}
	\item 
	The string theory background introduced in \cite{Frolov:2005dj} receives perturbative string corrections that alter the $\AdS{5}$ part, so that the \SO{2,4} symmetry is dynamically broken. This scenario mimics the gauge theory situation, where the tree-level conformality is broken by perturbative corrections. 
	\item If the string theory background of \cite{Frolov:2005dj} is exact, it is possible that the corresponding gauge theory dual is not the $\gamma_i$-deformation. However, our definition of the $\gamma_i$-deformation includes all immediate candidates and in this scenario the \CFT dual would either include exotic couplings or even lack a Lagrangian description with the field content of \NfSYMt all together.
	\item Finally, the deformation angles $\gamma_i^\pm$ which we treated as independent may in fact be functions of the coupling $\gym$ with the relation $\gamma_i^-=\order{g}$, so that the $\gamma_i$-deformation can be viewed as a perturbation around the fix-point of the $\beta$-deformation. An analogous scenario appears in the ABJM and ABJ correspondences \cite{Aharony:2008ug,Aharony:2008gk} and in the interpolating quiver gauge theory \cite{Gadde:2009dj,Gadde:2010zi,Liendo:2011xb}. In these cases, finite functions of the couplings were found in \cite{Minahan:2009aq,Minahan:2009wg,Leoni:2010tb} and \cite{Pomoni:2011jj,Pomoni:2013poa,Mitev:2014yba,Mitev:2015oty}, respectively. In general, this scenario is hard to exclude, since non-vanishing terms in the $\beta$-functions can always be postponed to higher loop orders by adjusting the deformation angles in \eqref{eq:beta_function_QF}.
\end{enumerate}
To put our understanding of the \AdSCFTc on a firmer ground, it is crucial to determine which of the above scenarios is correct. In particular, it would be interesting to compute the one-loop corrections to the string background. On the side of the $\gamma_i$-deformation, it is also interesting to determine the two-loop contribution to the three double-trace couplings in \eqref{eq:running_dt_coupling} and determine whether they exhibit Banks-Zaks fix-points \cite{Banks:1981nn}.

\section{Ground-state energies at leading wrapping order in the \texorpdfstring{$\gamma_i$}{gamma-i}-deformation}\label{sec:cake}
In this section, which is based on \cite{Fokken:2014soa}, we test the widely believed inherited integrability\footnote{Here, we assume that the undeformed \NfSYMt is integrable \cite{Beisert:2003,Beisert:2003tq} and only discuss whether the deformation preserves the claimed integrability.} in the $\gamma_i$-deformation (introduced in \secref{sec:The_deformations}) by calculating the leading wrapping corrections to the anomalous dimensions of the length-$L$ integrability groundstate
\begin{equation}\label{eq:osci_vacuum}
\mathcal{O}_L=N^{-\frac{L}{2}}\tr\bigl(\phi_i^L\bigr)\eqncom\qquad i\in\{1,2,3\}\eqncom
\end{equation}
in the \SUN theory. While we confirm the integrability-based results for states that are not affected by prewrapping, we also calculate the renormalisation-scheme dependent anomalous dimension of the simplest example state $\mathcal{O}_2$ which is not accessible with the current integrability-based method. This state has the quantum numbers of $\tr\bigl(\phi_i\phi_i\bigr)$ and is hence subject to prewrapping corrections which stem from contributions that involve the running coupling \eqref{eq:running_dt_coupling}. Analogously, {\it all} length-$L$ states with the same quantum numbers are potentially affected by prewrapping and their anomalous dimensions was not yet derived from current integrability-based methods.

In \secref{sec:the-quantum-dilatation-operator-on-composite-operators}, we discussed how the asymptotic one-loop dilatation operator in the $\gamma_i$-deformation can be obtained from the undeformed one \eqref{eq: deformation of D_2}. This result, as well as the deformed gravity background \cite{Frolov:2005dj}, are compatible \cite{Roiban:2003dw,Beisert:2005if} with the integrability-based Bethe-ansatz techniques of the original \AdSCFTc\footnote{In the $\gamma_i$-deformation the deformation parameters are incorporated in the asymptotic Bethe-ansatz via twisted boundary conditions \cite{Beisert:2005if}. This procedure can be derived from a twisted S-matrix \cite{Ahn:2010ws} or a twisted	transfer matrix \cite{Gromov:2010dy} which corresponds to operational twisted boundary conditions \cite{Arutyunov:2010gu}.}, see the review collection \cite{Beisert:2010jr}. This is consistent with the finding that single-trace operators in the deformed theory inherit their properties from the respective single-trace operators in the parent theory \cite{Mauri:2005pa,Ananth:2006ac,AKS07}. Beyond the asymptotic regime, the finite-size effects of \subsecref{sec:finite-size-effects}, i.e.\ non-planar wrapping and multi-trace prewrapping interaction kernels, contribute to the anomalous dimensions. Hence, integrability may be spoiled at loop orders $K\geq L-1$ and further tests are required. In the closely related $\beta$-deformation, such a test was performed for the so-called single-impurity operators, which correspond to the single-magnon states in the integrability approach and differ from \eqref{eq:osci_vacuum} by replacing one of the $L$ chiral scalars by $\phi_j$ with $j\neq i$. The anomalous dimensions of such single-impurity operators vanish in \NfSYMt but render non-vanishing results in both deformations for lengths $L\geq 3$. In the $\beta$-deformation, the asymptotic contributions to their anomalous dimensions are obtained from the dispersion relation of the twisted Bethe ansatz \cite{Beisert:2005if}. These integrability-based findings can be verified by incorporating the modifications for the $\beta$-deformation \cite{Fiamberti:2008sm} into the explicit field-theoretic three-loop calculation \cite{Sieg:2010tz} and the all-loop argument of \cite{Gross:2002su}. For $L\geq 3$, also the occurring finite-size wrapping corrections which start at $K=L$ loops have been calculated. The eleven loop Feynman diagram calculation of  \cite{Fiamberti:2008sn} was reproduced by the integrability-based approach in \cite{Gunnesson:2009nn} for $\beta=\frac{1}{2}$ and in \cite{Gromov:2010dy}, \cite{Arutyunov:2010gu}, and \cite{Kazakov:2015efa} for generic $\beta$, based on L\"uscher corrections, Y-system, TBA equations, and the QSC approach, respectively. At $L=2$, the anomalous dimension of the single-magnon state vanishes to all loops\footnote{See \cite{Freedman:2005cg} and \cite{Penati:2005hp} for explicit one- and two-loop calculations, respectively.} in the conformal \SUN theory \cite{Fokken:2013mza} and it is non-vanishing at one loop in the non-conformal \UN theory without multi-trace couplings \cite{Freedman:2005cg}. On the integrability side which also does not incorporate contributions from multi-trace interactions, the present TBA and QSC result of \cite{Arutyunov:2010gu} and respectively \cite{Kazakov:2015efa} for the anomalous dimension of the $L=2$ states diverges\footnote{Such a divergence was encountered earlier in the expressions for the ground-state energy of the TBA \cite{Frolov:2009in}. In \cite{deLeeuw:2012hp}, it was found that the divergent ground-state energy vanishes in the undeformed theory when a regulating twist is introduced in the $\text{AdS}_5$ directions. This regularisation extends to the ground state of the supersymmetric deformations \cite{FrolovPC}.} which clearly cannot be correct.

In the $\gamma_i$-deformation, integrability beyond the asymptotic level can be tested already for the groundstate \eqref{eq:osci_vacuum}. While these states are still protected in the $\beta$-deformation, they have non-vanishing anomalous dimensions in the $\gamma_i$-deformation which solely stem from finite-size corrections. For $L\geq 3$, they were determined in \cite{Ahn:2011xq} from the integrability-based approach up to double-wrapping order $K=2L$. For $L=2$, this approach leads to a divergent anomalous dimension, similar to the discussed divergence of $L=2$ states in the $\beta$-deformation. From the field-theoretic perspective this is expected, since the state $\mathcal{O}_2=N^{-1}\tr\bigl(\phi_i\phi_i\bigr)$ can couple in the \tHooft limit to the running double-trace coupling \eqref{eq:running_dt_coupling} and processes originating form multi-trace interactions are not captured by the integrability-based approach.  

As mentioned in the beginning of this section, we will determine the planar anomalous dimensions of the states in \eqref{eq:osci_vacuum} in the $\gamma_i$-deformation with gauge group \SUN at critical wrapping order $K=L$ from a purely Feynman-diagrammatic approach. For lengths $L\geq 3$, they are given in terms of the Riemann-$\zeta$ function as
\begin{equation}
\begin{aligned}
\gamma_{\mathcal{O}_L}&=-32(2g^{2})^L\sin^2\frac{L\gamma_i^+}{2}\sin^2\frac{L\gamma_i^-}{2}
\binom{2L-2}{L-1}\zeta(2L-3)
\eqncom
\end{aligned}
\end{equation}
which reproduces\footnote{Our deformation angles and the coupling constant must be rescaled to match their conventions.} the leading-order expression of \cite{Ahn:2011xq}. For the $L=2$ state we calculate the non-divergent planar anomalous dimension
\begin{equation}\label{gammaO2res}
\begin{aligned}
\gamma_{\mathcal{O}_2}&=2g^2\bigl(Q_{\text{F}\,ii}^{ii}-\frac{\varrho}{2}\,\beta_{Q_{\text{F}\,ii}^{ii}}\bigr)
-32(2g^2)^2\sin^2\gamma_i^+\sin^2\gamma_i^- 
\eqndot
\end{aligned}
\end{equation}
At one loop, it receives contributions from the free tree-level coupling $Q_{\text{F}\,ii}^{ii}$ which entirely originate from prewrapping. At two loops, we have the $\gamma_i^\pm$-dependent term which entirely originates from wrapping diagrams and the one-loop $\beta$-function from the running double-trace coupling \eqref{eq:running_dt_coupling}. Since the $\beta$-function, which we calculated in \eqref{eq:beta_function_QF}, is non-vanishing, the two-loop anomalous dimension becomes dependent on the chosen renormalisation scheme: different choices of $\varrho$ characterise different schemes. In the DR scheme\footnote{For other renormalisation schemes like the kinematic subtraction scheme, the renormalisation scheme dependence may differ from the one displayed in \eqref{gammaO2res}.} introduced in \appref{sec:Renormalisation_schemes} and used in the following subsections, we have $\varrho=0$ and in the $\ol{\text{DR}}$ scheme we have $\varrho=c_{\MSbar}=\gammaE-\ln4\pi$. The renormalisation-scheme dependence implies that the anomalous dimension \eqref{gammaO2res} is non-observable, which is an explicit consequence of the $\gamma_i$-deformation's non-conformality. Implications of this renormalisation-scheme dependent anomalous dimension for integrability-based methods will be discussed in \chapref{chap:Conclusion_outlook}.

The non-conformality of the $\gamma_i$-deformation found in \secref{sec:non-conformal_double_trace_coupling} has immediate consequences for the correlation functions of composite operators in the \tHooft limit, as exemplified in \eqref{gammaO2res}. The limit, as it was originally proposed in \cite{'tHooft:1971fh}, must be applied to all contributions in correlation functions. Hence, from the perspective of two-point functions of composite operators in \eqref{eq:2_3_pt_function}, conformality in the \tHooft is broken, if any gauge-invariant local composite operator receives renormalisation-scheme dependent contributions to the anomalous dimensions. This is in sharp contrast to the phrasing of \cite[version 2]{Jin:2013baa}, where the author claims that the $\gamma_i$-deformation would be `conformally invariant in the planar limit'. It is in equally sharp contrast to the implicit definition\footnote{The implicit definition of `conformality in the \tHooft limit' that is used in \cite[version 1]{Kazakov:2015efa} excludes any operator for which the coupling \eqref{eq:running_dt_coupling} may contribute in a planar Feynman diagram from the theory all together \cite{Kazakovcommunication}. When the planar limit is defined like this, it preserves the conformality of the parent \NfSYMt by construction but severe constraints are put on the composite operators that are allowed.} used in \cite[version 1]{Kazakov:2015efa}, where the $\gamma_i$-deformation is claimed to preserve conformality in the $N\rightarrow\infty$ limit. In either of these two publications, the \tHooft limit as given in \cite{'tHooft:1971fh} cannot be applied to any correlation function of the $\gamma_i$-deformation in which the coupling \eqref{eq:running_dt_coupling} can contribute.

\subsection{Identifying deformation-dependent diagrams}
Instead of calculating all Feynman diagrams that contribute to the UV renormalisation constant $\mathcal{Z}_{\mathcal{O}_L}$ and hence the anomalous dimension $\gamma_{\mathcal{O}_L}$ at $K=L$ loops, we employ a trick similar to the one in the previous section. We exploit the kinship of the $\gamma_i$-deformation to \NfSYMt and only calculate those contributions that differ from their counterparts in the undeformed theory. All remaining contributions are then reconstructed using that the operator $\mathcal{O}_L$ does not receive quantum corrections in \NfSYMt or the $\beta$-deformation. In this subsection, we develop the tools to identify diagrams that differ in the $\gamma_i$-deformation from the ones in the undeformed theory and contribute in the \tHooft limit.

In the \tHooft limit, the connected $K$-loop order renormalisation constant $\mathcal{Z}_{\mathcal{O}_L}^{(K)}$ absorbs all\footnote{The only exceptions are divergences that arise when $q\rightarrow p$ in the following equation. In this case, we should consider the operator $\tilde\cO_{\text{B}}$ composed of $\mathcal{O}_{\text{B}L}$ and its conjugate instead of the individual operators. Then the occurring divergence is cancelled by the normal-ordering of the new operator since we have $\vacl \tilde{\cO_{\text{B}}(p)}\vac=0$, compare \subsecref{sec:Normal_ordering}.} planar UV-divergent contributions that arise in the $K$-loop two-point function of the bare operator $\mathcal{O}_{\text{B}L}$ and its conjugate, so that the $K^{\text{th}}$-order momentum space correlation function of renormalised operators
\begin{equation}\label{eq:2point_correlator_OL}
\vacl\T \ol{\mathcal{O}}_L(p)\mathcal{O}_L(q)\vac^{(K)}=\bigl(\mathcal{Z}_{\mathcal{O}_L}^{(K)}\bigr)^{2}
\vacl\T\ol{\mathcal{O}}_{\text{B}L}(p)\mathcal{O}_{\text{B}L}(q)\vac^{(K)}
\end{equation}
is finite. Any contribution to this correlation function that depends on the deformation parameters of the $\gamma_i$-deformation receives this dependence from the elementary interactions in the $K$-loop interaction kernel. Hence, to determine which contribution to \eqref{eq:2point_correlator_OL} is deformation-dependent, it is sufficient to determine the deformation-dependence of the respective interaction kernels. These kernels are obtained by cutting both composite operators out of a given Feynman diagram. The cutting lets the originally connected diagram fall apart into $c$ connected pieces of elementary interactions which together form the interaction kernel of the respective diagram. For loop orders $K<L-2$, where neither wrapping nor prewrapping can occur, all the connected pieces $c$ are planar single-trace diagrams of elementary interactions and hence we can determine their dependence on the deformation parameters $\gamma_i^\pm$ by employing relation \eqref{diagrel}. Since the external fields in each connected piece are chiral scalars $\phi_i$ or their conjugates $\bar{\phi}^i$, the $\ast$-product reduces to an ordinary product and their deformation dependence vanishes, compare \eqref{diagrel} and \subsecref{sec:deformation}. Therefore, in the asymptotic regime where $K<L-2$, all contributing interaction kernels are independent of deformation parameters. All their contributions to \eqref{eq:2point_correlator_OL} vanish like they do in the undeformed theory.

For loop orders $K\geq L-1$, prewrapping corrections contribute in \eqref{eq:2point_correlator_OL}. Since they originate from $N$-enhanced multi-trace contributions, we cannot employ \eqref{diagrel} to determine their deformation-dependence. From \subsecref{sec:finite-size-effects} we know that the $N$-enhancement occurs when a double-trace coupling from \subsecref{sec:multi-trace-parts-of-the-action} is connected to two single-trace operators that are the respective conjugates of each single-trace term in the coupling, as happens e.g.\ in \eqref{eq:O2_prewrapping}. This also extends to the double-trace part of an \SUN propagator depicted in \eqref{eq:double_line_propagator}, which is simply a two-point vertex with no coupling dependence. Luckily, for the operators $\mathcal{O}_L$ there are very few possible prewrapping contributions. Their \su{4} Cartan charges are $\mathbf{q}_{\mathcal{O}_L}=L\mathbf{e}_i$ and hence, for $L\geq 3$, they cannot be fused into any of the single-trace factors of the multi-trace couplings displayed in \subsecref{sec:multi-trace-parts-of-the-action}. For $L=2$, our operator allows for a single possible source for prewrapping contributions, which is the coupling $Q_{\text{F}}$ in \eqref{eq:dtc}. 

For loop orders $K\geq L$, also wrapping corrections contribute in \eqref{eq:2point_correlator_OL}. As discussed in \subsecref{sec:finite-size-effects}, they originate from non-planar interaction kernels and hence we cannot use \eqref{diagrel} for these corrections either. However, for the operators $\mathcal{O}_L$, we can at least sort all occurring interaction kernels for wrapping contributions into a class that contains all diagrams depending on $\gamma_i^\pm$ and a second class that contains all diagrams independent of $\gamma_i^\pm$. This decomposition reads\footnote{The open field lines to the right and left in each diagram are the points where the initial and final state operators have been cut out, respectively.}
\begin{equation}\label{wrapdiagdecomp}
\ifpdf
\settoheight{\eqoff}{$+$}%
\setlength{\eqoff}{0.5\eqoff}%
\addtolength{\eqoff}{-7\unit}%
\raisebox{\eqoff}{%
	\includegraphics[angle={0},scale=0.16,trim=0cm 0cm 0cm 0]{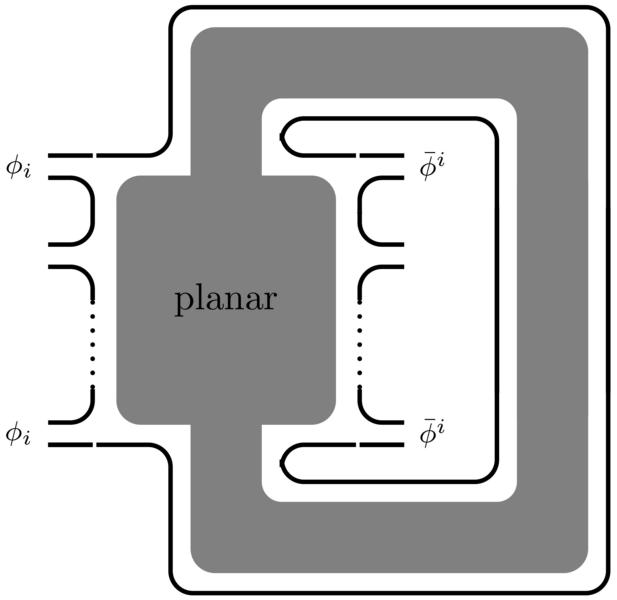}
}
\else
\scalebox{0.8}{
\settoheight{\eqoff}{$+$}%
\setlength{\eqoff}{0.5\eqoff}%
\addtolength{\eqoff}{-10.5\unit}%
\raisebox{\eqoff}{%
	\begin{pspicture}(-0.5,-6)(21.5,15)
	%
	\rput[r](1.5,9){$\scriptstyle \phi_i$}
	\rput[r](1.5,0){$\scriptstyle \phi_i$}
	\rput[l](14.5,0){$\scriptstyle \bar\phi^i$}
	\rput[l](14.5,9){$\scriptstyle \bar\phi^i$}
	%
	\uinex{2}{9}
	\iinex{2}{6}
	\dinex{2}{0}
	\setlength{\xa}{4\unit}
	\addtolength{\xa}{-0.5\dlinewidth}
	\setlength{\ya}{9\unit}
	\addtolength{\ya}{0.5\dlinewidth}
	\setlength{\xb}{6.5\unit}
	\addtolength{\xb}{-0.5\dlinewidth}
	\setlength{\yb}{14\unit}
	\addtolength{\yb}{0.5\dlinewidth}
	\setlength{\xc}{9.5\unit}
	\addtolength{\xc}{0.5\dlinewidth}
	\setlength{\yc}{11\unit}
	\addtolength{\yc}{-0.5\dlinewidth}
	\setlength{\xd}{12\unit}
	\addtolength{\xd}{0.5\dlinewidth}
	\setlength{\yd}{0\unit}
	\addtolength{\yd}{-0.5\dlinewidth}
	\setlength{\xe}{17.5\unit}
	\addtolength{\xe}{-0.5\dlinewidth}
	\setlength{\ye}{-2\unit}
	\addtolength{\ye}{0.5\dlinewidth}
	\setlength{\xf}{20.5\unit}
	\addtolength{\xf}{0.5\dlinewidth}
	\setlength{\yf}{-5\unit}
	\addtolength{\yf}{-0.5\dlinewidth}
	\setlength{\yg}{8\unit}
	\addtolength{\yg}{-0.5\dlinewidth}
	\setlength{\yh}{1\unit}
	\addtolength{\yh}{0.5\dlinewidth}
	\psline[liftpen=1,linearc=\linearc](\xa,\ya)(\xb,\ya)(\xb,\yb)(\xf,\yb)(\xf,\yg)
	\psline[liftpen=1,linearc=\linearc](\xf,\yg)(\xf,\yh)
	\psline[liftpen=1,linearc=\linearc](\xf,\yh)(\xf,\yf)(\xb,\yf)(\xb,\yd)(\xa,\yd)
	\psline[liftpen=1,linearc=\linearc](\xd,\ya)(\xc,\ya)(\xc,\yc)(\xe,\yc)(\xe,\yg)
	\psline[liftpen=1,linearc=\linearc](\xe,\yg)(\xe,\yh)
	\psline[liftpen=1,linearc=\linearc](\xe,\yh)(\xe,\ye)(\xc,\ye)(\xc,\yd)(\xd,\yd)
	%
	%
	\psline[linestyle=dotted](3.5,4.5)(3.5,1.5)
	\doutex{14}{0}
	\ioutex{14}{6}
	\uoutex{14}{9}
	\psline[linestyle=dotted](12.5,4.5)(12.5,1.5)
	\setlength{\xa}{4\unit}
	\addtolength{\xa}{0.5\dlinewidth}
	\setlength{\ya}{9\unit}
	\addtolength{\ya}{-0.5\dlinewidth}
	\setlength{\xb}{6.5\unit}
	\addtolength{\xb}{0.5\dlinewidth}
	\setlength{\yb}{14\unit}
	\addtolength{\yb}{-0.5\dlinewidth}
	\setlength{\xc}{9.5\unit}
	\addtolength{\xc}{-0.5\dlinewidth}
	\setlength{\yc}{11\unit}
	\addtolength{\yc}{0.5\dlinewidth}
	\setlength{\xd}{12\unit}
	\addtolength{\xd}{-0.5\dlinewidth}
	\setlength{\yd}{0\unit}
	\addtolength{\yd}{0.5\dlinewidth}
	\setlength{\xe}{17.5\unit}
	\addtolength{\xe}{0.5\dlinewidth}
	\setlength{\ye}{-2\unit}
	\addtolength{\ye}{-0.5\dlinewidth}
	\setlength{\xf}{20.5\unit}
	\addtolength{\xf}{-0.5\dlinewidth}
	\setlength{\yf}{-5\unit}
	\addtolength{\yf}{0.5\dlinewidth}
	\setlength{\yg}{8\unit}
	\addtolength{\yg}{0.5\dlinewidth}
	\setlength{\yh}{1\unit}
	\addtolength{\yh}{-0.5\dlinewidth}
	%
	%
	\pscustom[linecolor=gray,fillstyle=solid,fillcolor=gray,linearc=\linearc]{%
		\psline[liftpen=1,linearc=\linearc](\xb,\ya)(\xd,\ya)(\xd,\yd)(\xa,\yd)(\xa,\ya)(\xb,\ya)
		\psline[liftpen=2,linearc=\linearc](\xb,\ya)(\xb,\yb)(\xf,\yb)(\xf,\yg)
		\psline[liftpen=1](\xf,\yg)(\xe,\yg)
		\psline[liftpen=1,linearc=\linearc](\xe,\yg)(\xe,\yc)(\xc,\yc)(\xc,\ya)
		\psline[liftpen=2,linearc=\linearc](\xb,\yd)(\xb,\yf)(\xf,\yf)(\xf,\yh)
		\psline[liftpen=1](\xf,\yh)(\xe,\yh)
		\psline[liftpen=1,linearc=\linearc](\xe,\yh)(\xe,\ye)(\xc,\ye)(\xc,\yd)
	}
	\pscustom[linecolor=gray,fillstyle=solid,fillcolor=gray]{%
		\psline[liftpen=1](\xe,\yg)(\xf,\yg)(\xf,\yh)
		\psline[liftpen=2](\xf,\yh)(\xe,\yh)(\xe,\yg)
	}
	\rput(8,4.5){$\text{planar}$}
	\end{pspicture}
}
}
\fi
\,=
\ifpdf
\settoheight{\eqoff}{$+$}%
\setlength{\eqoff}{0.5\eqoff}%
\addtolength{\eqoff}{-7\unit}%
\raisebox{\eqoff}{%
	\includegraphics[angle={0},scale=0.16,trim=0cm 0cm 0cm 0]{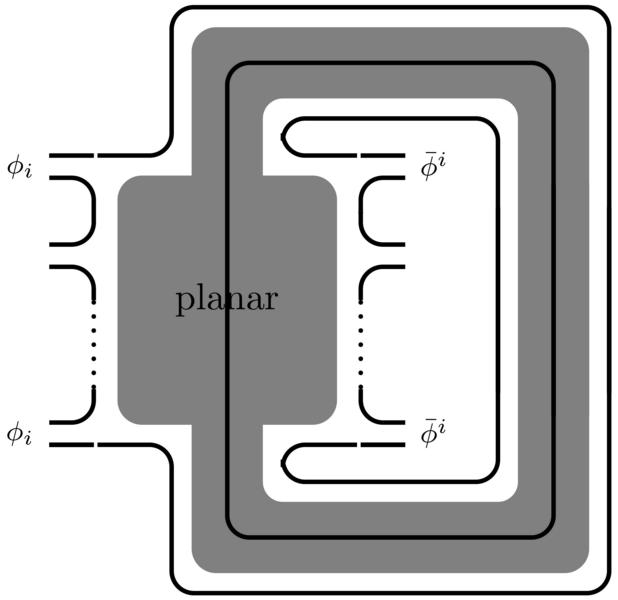}
}
\else
\scalebox{0.8}{
\settoheight{\eqoff}{$+$}%
\setlength{\eqoff}{0.5\eqoff}%
\addtolength{\eqoff}{-10.5\unit}%
\raisebox{\eqoff}{%
	\begin{pspicture}(-0.5,-6)(21.5,15)
	%
	\rput[r](1.5,9){$\scriptstyle \phi_i$}
	\rput[r](1.5,0){$\scriptstyle \phi_i$}
	\rput[l](14.5,0){$\scriptstyle \bar\phi^i$}
	\rput[l](14.5,9){$\scriptstyle \bar\phi^i$}
	%
	\uinex{2}{9}
	\iinex{2}{6}
	\dinex{2}{0}
	\setlength{\xa}{4\unit}
	\addtolength{\xa}{-0.5\dlinewidth}
	\setlength{\ya}{9\unit}
	\addtolength{\ya}{0.5\dlinewidth}
	\setlength{\xb}{6.5\unit}
	\addtolength{\xb}{-0.5\dlinewidth}
	\setlength{\yb}{14\unit}
	\addtolength{\yb}{0.5\dlinewidth}
	\setlength{\xc}{9.5\unit}
	\addtolength{\xc}{0.5\dlinewidth}
	\setlength{\yc}{11\unit}
	\addtolength{\yc}{-0.5\dlinewidth}
	\setlength{\xd}{12\unit}
	\addtolength{\xd}{0.5\dlinewidth}
	\setlength{\yd}{0\unit}
	\addtolength{\yd}{-0.5\dlinewidth}
	\setlength{\xe}{17.5\unit}
	\addtolength{\xe}{-0.5\dlinewidth}
	\setlength{\ye}{-2\unit}
	\addtolength{\ye}{0.5\dlinewidth}
	\setlength{\xf}{20.5\unit}
	\addtolength{\xf}{0.5\dlinewidth}
	\setlength{\yf}{-5\unit}
	\addtolength{\yf}{-0.5\dlinewidth}
	\setlength{\yg}{8\unit}
	\addtolength{\yg}{-0.5\dlinewidth}
	\setlength{\yh}{1\unit}
	\addtolength{\yh}{0.5\dlinewidth}
	\psline[liftpen=1,linearc=\linearc](\xa,\ya)(\xb,\ya)(\xb,\yb)(\xf,\yb)(\xf,\yg)
	\psline[liftpen=1,linearc=\linearc](\xf,\yg)(\xf,\yh)
	\psline[liftpen=1,linearc=\linearc](\xf,\yh)(\xf,\yf)(\xb,\yf)(\xb,\yd)(\xa,\yd)
	\psline[liftpen=1,linearc=\linearc](\xd,\ya)(\xc,\ya)(\xc,\yc)(\xe,\yc)(\xe,\yg)
	\psline[liftpen=1,linearc=\linearc](\xe,\yg)(\xe,\yh)
	\psline[liftpen=1,linearc=\linearc](\xe,\yh)(\xe,\ye)(\xc,\ye)(\xc,\yd)(\xd,\yd)
	%
	%
	\psline[linestyle=dotted](3.5,4.5)(3.5,1.5)
	\doutex{14}{0}
	\ioutex{14}{6}
	\uoutex{14}{9}
	\psline[linestyle=dotted](12.5,4.5)(12.5,1.5)
	\setlength{\xa}{4\unit}
	\addtolength{\xa}{0.5\dlinewidth}
	\setlength{\ya}{9\unit}
	\addtolength{\ya}{-0.5\dlinewidth}
	\setlength{\xb}{6.5\unit}
	\addtolength{\xb}{0.5\dlinewidth}
	\setlength{\yb}{14\unit}
	\addtolength{\yb}{-0.5\dlinewidth}
	\setlength{\xc}{9.5\unit}
	\addtolength{\xc}{-0.5\dlinewidth}
	\setlength{\yc}{11\unit}
	\addtolength{\yc}{0.5\dlinewidth}
	\setlength{\xd}{12\unit}
	\addtolength{\xd}{-0.5\dlinewidth}
	\setlength{\yd}{0\unit}
	\addtolength{\yd}{0.5\dlinewidth}
	\setlength{\xe}{17.5\unit}
	\addtolength{\xe}{0.5\dlinewidth}
	\setlength{\ye}{-2\unit}
	\addtolength{\ye}{-0.5\dlinewidth}
	\setlength{\xf}{20.5\unit}
	\addtolength{\xf}{-0.5\dlinewidth}
	\setlength{\yf}{-5\unit}
	\addtolength{\yf}{0.5\dlinewidth}
	\setlength{\yg}{8\unit}
	\addtolength{\yg}{0.5\dlinewidth}
	\setlength{\yh}{1\unit}
	\addtolength{\yh}{-0.5\dlinewidth}
	%
	%
	\pscustom[linecolor=gray,fillstyle=solid,fillcolor=gray,linearc=\linearc]{%
		\psline[liftpen=1,linearc=\linearc](\xb,\ya)(\xd,\ya)(\xd,\yd)(\xa,\yd)(\xa,\ya)(\xb,\ya)
		\psline[liftpen=2,linearc=\linearc](\xb,\ya)(\xb,\yb)(\xf,\yb)(\xf,\yg)
		\psline[liftpen=1](\xf,\yg)(\xe,\yg)
		\psline[liftpen=1,linearc=\linearc](\xe,\yg)(\xe,\yc)(\xc,\yc)(\xc,\ya)
		\psline[liftpen=2,linearc=\linearc](\xb,\yd)(\xb,\yf)(\xf,\yf)(\xf,\yh)
		\psline[liftpen=1](\xf,\yh)(\xe,\yh)
		\psline[liftpen=1,linearc=\linearc](\xe,\yh)(\xe,\ye)(\xc,\ye)(\xc,\yd)
	}
	
	\pscustom[linecolor=gray,fillstyle=solid,fillcolor=gray]{%
		\psline[liftpen=1](\xe,\yg)(\xf,\yg)(\xf,\yh)
		\psline[liftpen=2](\xf,\yh)(\xe,\yh)(\xe,\yg)
	}
	\setlength{\xb}{8\unit}
	\setlength{\yb}{12.5\unit}
	\setlength{\xc}{19\unit}
	\setlength{\yc}{-3.5\unit}
	\setlength{\yg}{8\unit}
	\addtolength{\yg}{0.5\dlinewidth}
	\setlength{\yh}{1\unit}
	\addtolength{\yh}{-0.5\dlinewidth}
	\psline[liftpen=2,linearc=\linearc](\xc,\yh)(\xc,\yc)(\xb,\yc)(\xb,\yb)(\xc,\yb)
	(\xc,\yg)
	\psline[liftpen=2,linearc=\linearc](\xc,\yg)(\xc,\yh)
	\rput(8,4.5){$\text{planar}$}
	\end{pspicture}
}
}
\fi
\,+
\ifpdf
\settoheight{\eqoff}{$+$}%
\setlength{\eqoff}{0.5\eqoff}%
\addtolength{\eqoff}{-7\unit}%
\raisebox{\eqoff}{%
	\includegraphics[angle={0},scale=0.16,trim=0cm 0cm 0cm 0]{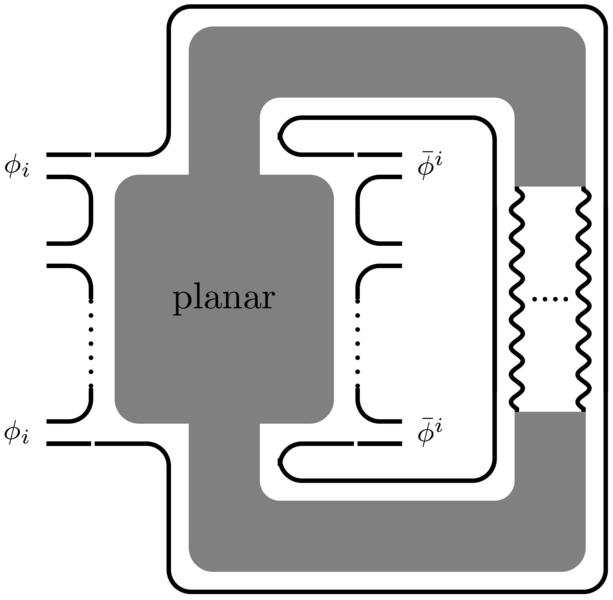}
}
\else
\scalebox{0.8}{
\settoheight{\eqoff}{$+$}%
\setlength{\eqoff}{0.5\eqoff}%
\addtolength{\eqoff}{-10.5\unit}%
\raisebox{\eqoff}{%
	\begin{pspicture}(-0.5,-6)(21.5,15)
	%
	\rput[r](1.5,9){$\scriptstyle \phi_i$}
	\rput[r](1.5,0){$\scriptstyle \phi_i$}
	\rput[l](14.5,0){$\scriptstyle \bar\phi^i$}
	\rput[l](14.5,9){$\scriptstyle \bar\phi^i$}
	%
	\uinex{2}{9}
	\iinex{2}{6}
	\dinex{2}{0}
	\setlength{\xa}{4\unit}
	\addtolength{\xa}{-0.5\dlinewidth}
	\setlength{\ya}{9\unit}
	\addtolength{\ya}{0.5\dlinewidth}
	\setlength{\xb}{6.5\unit}
	\addtolength{\xb}{-0.5\dlinewidth}
	\setlength{\yb}{14\unit}
	\addtolength{\yb}{0.5\dlinewidth}
	\setlength{\xc}{9.5\unit}
	\addtolength{\xc}{0.5\dlinewidth}
	\setlength{\yc}{11\unit}
	\addtolength{\yc}{-0.5\dlinewidth}
	\setlength{\xd}{12\unit}
	\addtolength{\xd}{0.5\dlinewidth}
	\setlength{\yd}{0\unit}
	\addtolength{\yd}{-0.5\dlinewidth}
	\setlength{\xe}{17.5\unit}
	\addtolength{\xe}{-0.5\dlinewidth}
	\setlength{\ye}{-2\unit}
	\addtolength{\ye}{0.5\dlinewidth}
	\setlength{\xf}{20.5\unit}
	\addtolength{\xf}{0.5\dlinewidth}
	\setlength{\yf}{-5\unit}
	\addtolength{\yf}{-0.5\dlinewidth}
	\setlength{\yg}{8\unit}
	\addtolength{\yg}{-0.5\dlinewidth}
	\setlength{\yh}{1\unit}
	\addtolength{\yh}{0.5\dlinewidth}
	\psline[liftpen=1,linearc=\linearc](\xa,\ya)(\xb,\ya)(\xb,\yb)(\xf,\yb)(\xf,\yg)
	\psline[liftpen=1,linearc=\linearc](\xf,\yg)(\xf,\yh)
	\psline[liftpen=1,linearc=\linearc](\xf,\yh)(\xf,\yf)(\xb,\yf)(\xb,\yd)(\xa,\yd)
	\psline[liftpen=1,linearc=\linearc](\xd,\ya)(\xc,\ya)(\xc,\yc)(\xe,\yc)(\xe,\yg)
	\psline[liftpen=1,linearc=\linearc](\xe,\yg)(\xe,\yh)
	\psline[liftpen=1,linearc=\linearc](\xe,\yh)(\xe,\ye)(\xc,\ye)(\xc,\yd)(\xd,\yd)
	%
	%
	\psline[linestyle=dotted](3.5,4.5)(3.5,1.5)
	\doutex{14}{0}
	\ioutex{14}{6}
	\uoutex{14}{9}
	\psline[linestyle=dotted](12.5,4.5)(12.5,1.5)
	\setlength{\xa}{4\unit}
	\addtolength{\xa}{0.5\dlinewidth}
	\setlength{\ya}{9\unit}
	\addtolength{\ya}{-0.5\dlinewidth}
	\setlength{\xb}{6.5\unit}
	\addtolength{\xb}{0.5\dlinewidth}
	\setlength{\yb}{14\unit}
	\addtolength{\yb}{-0.5\dlinewidth}
	\setlength{\xc}{9.5\unit}
	\addtolength{\xc}{-0.5\dlinewidth}
	\setlength{\yc}{11\unit}
	\addtolength{\yc}{0.5\dlinewidth}
	\setlength{\xd}{12\unit}
	\addtolength{\xd}{-0.5\dlinewidth}
	\setlength{\yd}{0\unit}
	\addtolength{\yd}{0.5\dlinewidth}
	\setlength{\xe}{17.5\unit}
	\addtolength{\xe}{0.5\dlinewidth}
	\setlength{\ye}{-2\unit}
	\addtolength{\ye}{-0.5\dlinewidth}
	\setlength{\xf}{20.5\unit}
	\addtolength{\xf}{-0.5\dlinewidth}
	\setlength{\yf}{-5\unit}
	\addtolength{\yf}{0.5\dlinewidth}
	\setlength{\yg}{8\unit}
	\addtolength{\yg}{0.5\dlinewidth}
	\setlength{\yh}{1\unit}
	\addtolength{\yh}{-0.5\dlinewidth}
	%
	%
	\pscustom[linecolor=gray,fillstyle=solid,fillcolor=gray,linearc=\linearc]{%
		\psline[liftpen=1,linearc=\linearc](\xb,\ya)(\xd,\ya)(\xd,\yd)(\xa,\yd)(\xa,\ya)(\xb,\ya)
		\psline[liftpen=2,linearc=\linearc](\xb,\ya)(\xb,\yb)(\xf,\yb)(\xf,\yg)
		\psline[liftpen=1](\xf,\yg)(\xe,\yg)
		\psline[liftpen=1,linearc=\linearc](\xe,\yg)(\xe,\yc)(\xc,\yc)(\xc,\ya)
		\psline[liftpen=2,linearc=\linearc](\xb,\yd)(\xb,\yf)(\xf,\yf)(\xf,\yh)
		\psline[liftpen=1](\xf,\yh)(\xe,\yh)
		\psline[liftpen=1,linearc=\linearc](\xe,\yh)(\xe,\ye)(\xc,\ye)(\xc,\yd)
	}
	%
	%
	\addtolength{\yh}{0.5\linew}
	\addtolength{\yg}{-0.5\linew}
	\pssin[periods=8,coilarm=0.1,amplitude=0.2](\xe,\yg)(\xe,\yh)
	\pssin[periods=8,coilarm=0.1,amplitude=0.2](\xf,\yg)(\xf,\yh)
	\newlength{\yaaa}
	\setlength{\yaaa}{0.5\yh}
	\addtolength{\yaaa}{0.5\yg}
	\addtolength{\xe}{0.7\dlinewidth}
	\addtolength{\xf}{-0.7\dlinewidth}
	\psline[linestyle=dotted,dotsep=1.5pt](\xe,\yaaa)(\xf,\yaaa)
	\newlength{\xaaa}
	\setlength{\xaaa}{0.5\xe}
	\addtolength{\xaaa}{0.5\xf}
	%
	\rput(8,4.5){$\text{planar}$}
	\end{pspicture}
}
}
\fi
\eqncom
\end{equation}
where planar indicates, that no additional non-planarity is contained in the gray shaded interaction area. The diagrams in the first class on the \rhs have at least one closed wrapping loop that is entirely made from fermionic or scalar fields and which is indicated by the solid flavour line. In the remaining wrapping diagrams, each wrapping line contains at least one segment that is made of a gauge field propagator, which is indicated by the wiggly lines in the rightmost diagram in \eqref{wrapdiagdecomp}. To see that all diagrams in the second class of \eqref{wrapdiagdecomp} are undeformed, we remove all gauge field propagators in the wrapping loop and replace the vertices they are connected to, according to
\begin{equation}\label{vertexreplacement} 
\setlength{\unit}{0.4cm}
\psset{xunit=\unit,yunit=\unit,runit=\unit}
\setlength{\dlinewidth}{0.75\unit}
\setlength{\linew}{1pt}
\setlength{\doublesep}{\dlinewidth}
\addtolength{\doublesep}{-\linew}
\psset{doublesep=\doublesep}
\psset{linewidth=\linew}
\setlength{\auxlen}{-0.2929\dlinewidth}
\addtolength{\auxlen}{\unit}
\setlength{\linearc}{0.75\unit}
\ifpdf
\settoheight{\eqoff}{$+$}%
\setlength{\eqoff}{0.5\eqoff}%
\addtolength{\eqoff}{-1.3\unit}%
\raisebox{\eqoff}{%
	\includegraphics[angle={0},scale=0.055,trim=0cm 0cm 0cm 0]{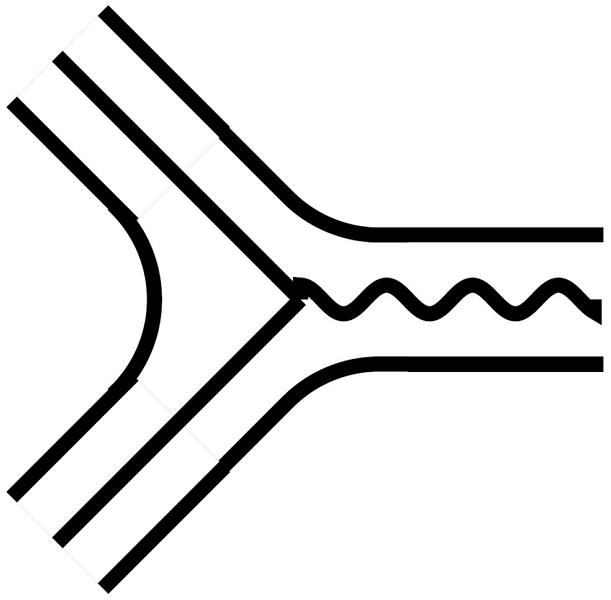}
}
\else
\settoheight{\eqoff}{$+$}%
\setlength{\eqoff}{0.5\eqoff}%
\addtolength{\eqoff}{-2\unit}%
\raisebox{\eqoff}{%
	\begin{pspicture}(-2,-2)(2,2)
	\recthreevertex{0}{0}{0}
	\psset{doubleline=true}
	\psline(-1.4142,1.4142)(-0.7071,0.7071)
	\psline(-0.7071,-0.7071)(-1.4142,-1.4142)
	\psline(0.7071,0)(1.7071,0)
	\psset{doubleline=false}
	\psline(-1.4142,1.4142)(0,0)
	\psline(0,0)(-1.4142,-1.4142)
	\pscoil[coilwidth=0.1666,coilheight=3,coilarm=0,coilaspect=0]{-}(0,0)(1.7071,0)
	\end{pspicture}
}
\fi
\quad\eqncom\quad
\ifpdf
\settoheight{\eqoff}{$+$}%
\setlength{\eqoff}{0.5\eqoff}%
\addtolength{\eqoff}{-1.3\unit}%
\raisebox{\eqoff}{%
	\includegraphics[angle={0},scale=0.055,trim=0cm 0cm 0cm 0]{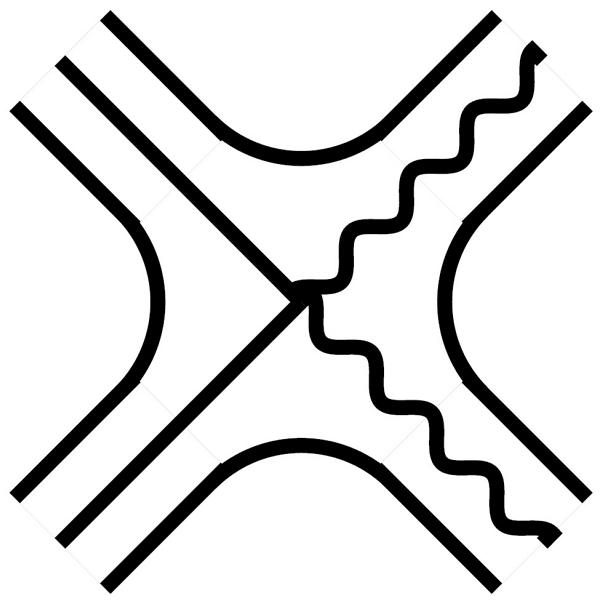}
}
\else
\settoheight{\eqoff}{$+$}%
\setlength{\eqoff}{0.5\eqoff}%
\addtolength{\eqoff}{-2\unit}%
\raisebox{\eqoff}{%
	\begin{pspicture}(-2,-2)(2,2)
	\fourvertex{0}{0}{45}
	\psset{doubleline=true}
	\psline(-1.4142,1.4142)(-0.7071,0.7071)
	\psline(1.4142,1.4142)(0.7071,0.7071)
	\psline(1.4142,-1.4142)(0.7071,-0.7071)
	\psline(-0.7071,-0.7071)(-1.4142,-1.4142)
	\psset{doubleline=false}
	\psline(-1.4142,1.4142)(0,0)
	\psline(0,0)(-1.4142,-1.4142)
	\pscoil[coilwidth=0.1666,coilheight=3,coilarm=0,coilaspect=0]{-}(0,0)(1.4142,1.4142)
	\pscoil[coilwidth=0.1666,coilheight=3,coilarm=0,coilaspect=0]{-}(0,0)(1.4142,-1.4142)
	\end{pspicture}
}
\fi
\quad
\longrightarrow
\quad
\ifpdf
\settoheight{\eqoff}{$+$}%
\setlength{\eqoff}{0.5\eqoff}%
\addtolength{\eqoff}{-1.3\unit}%
\raisebox{\eqoff}{%
	\includegraphics[angle={0},scale=0.04,trim=1cm 0cm 6cm 0]{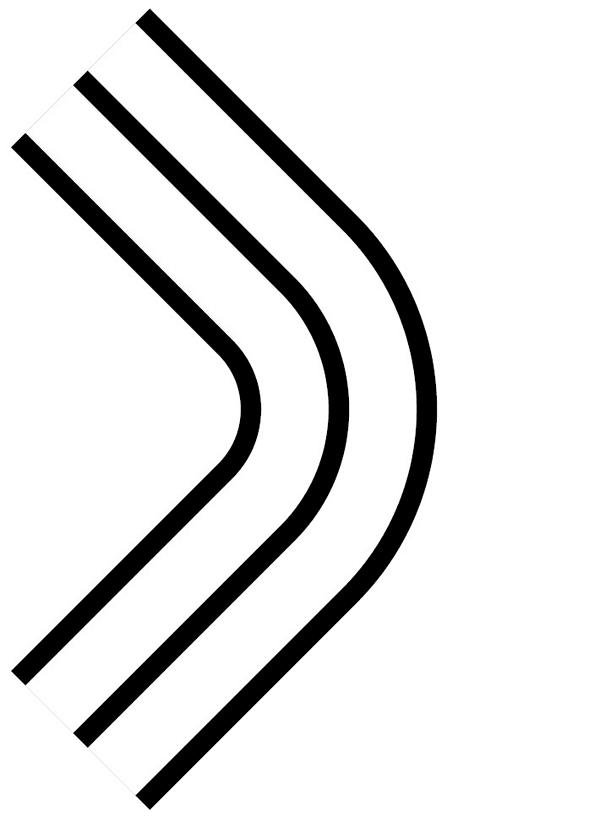}
}
\else
\settoheight{\eqoff}{$+$}%
\setlength{\eqoff}{0.5\eqoff}%
\addtolength{\eqoff}{-2\unit}%
\raisebox{\eqoff}{%
	\begin{pspicture}(-2,-2)(0.2921,2)
	\psset{doubleline=true}
	\psline[linearc=\linearc](-1.4142,1.4142)(0,0)(-1.4142,-1.4142)
	\psset{doubleline=false}
	\psline[linearc=\linearc](-1.4142,1.4142)(0,0)(-1.4142,-1.4142)
	\end{pspicture}}
\fi
\eqndot
\end{equation}
Here, the solid line stands for any fermionic or scalar flavour line as in \eqref{wrapdiagdecomp}. The resulting diagram has the same deformation-dependence as the original diagram, since all gauge field interactions are undeformed, compare \eqref{eq:deformed_action_complex_scalars2}. In addition, this removal procedure renders the resulting diagram\footnote{Eventually, the resulting diagram will be disconnected due to the removal of some interactions. In this case, we apply \eqref{diagrel} to each disconnected component, like in the asymptotic case discussed earlier.} planar and hence we can determine its dependence on $\gamma_i^\pm$ using \eqref{diagrel}. We find, in analogy to the $K<L-2$ case, that all such diagrams are independent of $\gamma_i^\pm$, which renders all diagrams in the second class of \eqref{wrapdiagdecomp} to be independent of $\gamma_i^\pm$ as well.

To sum it all up, the deformation-dependent diagrams that contribute to \eqref{eq:2point_correlator_OL} at order $K=L$ are for $L>2$ the wrapping diagrams with a closed wrapping loop entirely made of fermionic or scalar fields. For $L=2$, in addition the prewrapping diagrams that involve the coupling $Q_{\text{F}\,ii}^{ii}$ displayed in \eqref{eq:deformed_action_complex_scalars2} are deformation-dependent. These are the only diagrams necessary to construct the renormalisation constant $\mathcal{Z}_{\mathcal{O}_L}$ at leading wrapping order $K=L$.

\subsection{Finite-size corrections to the ground state}
In this subsection, we determine the connected $K=L$ loop UV renormalisation constant $\mathcal{Z}_{\mathcal{O}_L}$ of the length-$L$ groundstate $\mathcal{O}_L$ to leading wrapping order. From the discussion in the previous subsection, we know that $\mathcal{Z}_{\mathcal{O}_L}=1$ for loop orders $K\leq L-2$. At order $K\geq 1$, the prewrapping contributions that involve the coupling $Q_{\text{F}}$ occur for the state $\mathcal{O}_2$. Finally, at order $K=L$, we have the single-wrapping contributions from diagrams with a purely fermionic or scalar wrapping line. Since the operator $\mathcal{O}_L$ is not renormalised in the undeformed \NfSYMt, the mentioned prewrapping and wrapping contributions to $\mathcal{Z}_{\mathcal{O}_L}$ are the only ones that have to be calculated explicitly. We can reconstruct all remaining contributions by enforcing that the renormalisation constant is one when all deformation parameters are set to zero.

The complete renormalisation constant is built from the involved 1PI renormalisation constants as $\mathcal{Z}_{\mathcal{O}_L}=Z_{\mathcal{O}_L}Z_\phi^{-L/2}$. From the wave function renormalisation constant $Z_\phi$, which renormalises the external legs of $\mathcal{O}_L$, in the \SUN theory only the undeformed parts contribute\footnote{The constant $Z_\phi$ renormalises the scalar propagator, which can be seen as an $L=2$ operator. Its planar single-trace contribution is undeformed, since \eqref{diagrel} applies. Its deformed double-trace contribution can, however, only contribute in processes where two $L=1$ traces are connected to it, compare \eqref{eq:U1_mode_vanishes}. These states are absent in the \SUN theory.} in the \tHooft limit in this process. We do not calculate their contributions but reconstruct them later. It is therefore sufficient to calculate the 1PI renormalisation constant $Z_{\mathcal{O}_L}$, which is most conveniently extracted from the reduced 1PI momentum space correlation functions
\begin{equation}
\begin{aligned}\label{eq:wrapping_correlation_funct}
0&=\Kop\Bigl[\Bigl(\vacl\T\ol{\phi}^{i a_1}\dots\ol{\phi}^{i a_L}\mathcal{O}_L\vac_{\complexi \cT}\Bigr)_{\text{1PI}}\Bigr]
=\Kop\Bigl[Z_{\mathcal{O}_L}\Bigl(\vacl\T \ol{\phi}_{\text{B}}^{i a_1}\dots\ol{\phi}_{\text{B}}^{i a_L}\mathcal{O}_{\text{B}\,L}\vac_{\complexi \cT}\Bigr)_{\text{1PI}}\Bigr]\eqncom
\end{aligned}
\end{equation}
with $i\in\{1,2,3\}$, all fields evaluated at zero momentum with $p_i=0$ and the final state $\ol{\mathcal{O}}_L(0)$ being cut out. The operator $\Kop$ extracts the pole of the logarithmically divergent expression in the DR scheme in $D=4-2\epsilon$ dimensional Minkowski space. This correlation function is free of IR divergences\footnote{This becomes clear when choosing an explicit Euclidean-space loop momentum configuration in \eqref{PL}.}, even in the special configuration where all external kinematics are set to zero. The contributions to $Z_{\mathcal{O}_L}=1+\delta_{\mathcal{O}_L}$ are determined by evaluating the UV divergent correlation function and adjusting $\delta_{\mathcal{O}_L}$ to render \eqref{eq:wrapping_correlation_funct} true.

\subsubsection{Lengths \texorpdfstring{$L\geq 3$}{L>=3}}
For $L\geq 3$, prewrapping contributions are absent due to the particular state we are interested in. All deformation-dependent diagrams are hence $(K=L)$-loop diagrams with a single wrapping loop made of fermions or scalars. All contributing diagrams can be constructed using the Feynman rules of \appref{app:Feynman_rules}, the \ttt{Mathematica} package introduced in \appref{sec:Feynman_rules_Mathematica} and the tensor identities given in \appref{sec:Coupling_tensor_identities}. We find
\begin{subequations}\label{wrapLdiags}
\begin{equation}\label{wrapLdiags1}
\begin{aligned}
S(L)=
\settoheight{\eqoff}{$\times$}%
\setlength{\eqoff}{0.5\eqoff}%
\addtolength{\eqoff}{-16\unitlength}%
\raisebox{\eqoff}{%
	\fmfframe(7,1)(3,1){%
		\begin{fmfchar*}(30,30)
		\fmfleft{in}
		\fmfright{out}
		\fmf{phantom}{in,v5}
		\fmf{phantom}{out,v2}
		\fmf{phantom}{in,va5}
		\fmf{phantom}{out,va2}
		\fmfforce{(0,0.5h)}{in}
		\fmfforce{(w,0.5h)}{out}
		\fmfpoly{phantom}{v1,v6,v5,v4,v3,v2}
		\fmffixed{(0.75w,0)}{v5,v2}
		\fmfpoly{phantom}{va1,va6,va5,va4,va3,va2}
		\fmffixed{(w,0)}{va5,va2}
		\fmf{phantom}{vc,v1}
		\fmf{phantom}{vc,v4}
		\fmffreeze
		\fmf{plain_ar,left=0}{v1,v2}
		\fmf{plain_ar,left=0}{v2,v3}
		\fmf{plain_ar,left=0}{v3,v4}
		\fmf{dots,left=0}{v4,v5}
		\fmf{plain_ar,left=0}{v5,v6}
		\fmf{plain_ar,left=0}{v6,v1}
		\fmf{plain_rar}{vc,v1}
		\fmf{plain_rar}{vc,v2}
		\fmf{plain_rar}{vc,v3}
		\fmf{dots}{vc,v4}
		\fmf{plain_rar}{vc,v5}
		\fmf{plain_rar}{vc,v6}
		\fmf{plain_ar}{va1,v1}
		\fmf{plain_ar}{va2,v2}
		\fmf{plain_ar}{va3,v3}
		\fmf{dots}{va4,v4}
		\fmf{plain_ar}{va5,v5}
		\fmf{plain_ar}{va6,v6}
		\fmffreeze
		%
		\fmfv{l=$\scriptscriptstyle 1$,l.dist=2}{va1}
		\fmfv{l=$\scriptscriptstyle 2$,l.dist=2}{va2}
		\fmfv{l=$\scriptscriptstyle 3$,l.dist=2}{va3}
		\fmfv{l=$\scriptscriptstyle {L-1}$,l.dist=2}{va5}
		\fmfv{l=$\scriptscriptstyle L$,l.dist=2}{va6}
		\fmfiv{decor.shape=circle,decor.filled=full,decor.size=5}{vloc(__vc)}
		\end{fmfchar*}}}
&=\lambda^Lc(L)\sum_{j=1}^3 (\hat Q_{ij}^{ji})^L \frac{P_L}{\complexi^L}
\\[-1.5\baselineskip]
&=(2\lambda)^Lc(L)
\Big(2\e^{-iL\gamma_i^-}\cos{L\gamma_i^+}+\frac{1}{2^L}\Big) 
\frac{P_L}{\complexi^L}
\eqncom\\
\bar{S}(L)=
\settoheight{\eqoff}{$\times$}%
\setlength{\eqoff}{0.5\eqoff}%
\addtolength{\eqoff}{-16\unitlength}%
\raisebox{\eqoff}{%
	\fmfframe(7,1)(3,1){%
		\begin{fmfchar*}(30,30)
		\fmfleft{in}
		\fmfright{out}
		\fmf{phantom}{in,v5}
		\fmf{phantom}{out,v2}
		\fmf{phantom}{in,va5}
		\fmf{phantom}{out,va2}
		\fmfforce{(0,0.5h)}{in}
		\fmfforce{(w,0.5h)}{out}
		\fmfpoly{phantom}{v1,v6,v5,v4,v3,v2}
		\fmffixed{(0.75w,0)}{v5,v2}
		\fmfpoly{phantom}{va1,va6,va5,va4,va3,va2}
		\fmffixed{(w,0)}{va5,va2}
		\fmf{phantom}{vc,v1}
		\fmf{phantom}{vc,v4}
		\fmffreeze
		\fmf{plain_rar,left=0}{v1,v2}
		\fmf{plain_rar,left=0}{v2,v3}
		\fmf{plain_rar,left=0}{v3,v4}
		\fmf{dots,left=0}{v4,v5}
		\fmf{plain_rar,left=0}{v5,v6}
		\fmf{plain_rar,left=0}{v6,v1}
		\fmf{plain_rar}{vc,v1}
		\fmf{plain_rar}{vc,v2}
		\fmf{plain_rar}{vc,v3}
		\fmf{dots}{vc,v4}
		\fmf{plain_rar}{vc,v5}
		\fmf{plain_rar}{vc,v6}
		\fmf{plain_ar}{va1,v1}
		\fmf{plain_ar}{va2,v2}
		\fmf{plain_ar}{va3,v3}
		\fmf{dots}{va4,v4}
		\fmf{plain_ar}{va5,v5}
		\fmf{plain_ar}{va6,v6}
		\fmffreeze
		%
		\fmfv{l=$\scriptscriptstyle 1$,l.dist=2}{va1}
		\fmfv{l=$\scriptscriptstyle 2$,l.dist=2}{va2}
		\fmfv{l=$\scriptscriptstyle 3$,l.dist=2}{va3}
		\fmfv{l=$\scriptscriptstyle {L-1}$,l.dist=2}{va5}
		\fmfv{l=$\scriptscriptstyle L$,l.dist=2}{va6}
		\fmfiv{decor.shape=circle,decor.filled=full,decor.size=5}{vloc(__vc)}
		\end{fmfchar*}}}
&=\lambda^Lc(L)\sum_{j=1}^3 (\hat Q_{ji}^{ij})^L \frac{P_L}{\complexi^L}
\\[-1.5\baselineskip]
&=(2\lambda)^Lc(L)
\Big(2\e^{iL\gamma_i^-}\cos{L\gamma_i^+}+\frac{1}{2^L}\Big)\frac{P_L}{\complexi^L}\eqncom
\end{aligned}
\end{equation}
\begin{equation}\label{wrapLdiags2}
\begin{aligned}
%
F(L)=
\settoheight{\eqoff}{$\times$}%
\setlength{\eqoff}{0.5\eqoff}%
\addtolength{\eqoff}{-16\unitlength}%
\raisebox{\eqoff}{%
	\fmfframe(7,1)(3,1){%
		\begin{fmfchar*}(30,30)
		\fmfleft{in}
		\fmfright{out1}
		\fmf{phantom}{in,v10}
		\fmf{phantom}{out,v4}
		\fmf{phantom}{in,va5}
		\fmf{phantom}{out,va2}
		\fmfforce{(0,0.5h)}{in}
		\fmfforce{(w,0.5h)}{out}
		\fmfpoly{phantom}{v1,v12,v11,v10,v9,v8,v7,v6,v5,v4,v3,v2}
		\fmffixed{(0.75w,0)}{v10,v4}
		\fmfpoly{phantom}{va1,va6,va5,va4,va3,va2}
		\fmffixed{(w,0)}{va5,va2}
		\fmf{phantom}{vc,v1}
		\fmf{phantom}{vc,v7}
		\fmffreeze
		\fmf{dashes_ar,left=0}{v1,v2}
		\fmf{dashes_rar,left=0}{v2,v3}
		\fmf{dashes_ar,left=0}{v3,v4}
		\fmf{dashes_rar,left=0}{v4,v5}
		\fmf{dashes_ar,left=0}{v5,v6}
		\fmf{dots,left=0}{v6,v7}
		\fmf{dots,left=0}{v7,v8}
		\fmf{dashes_rar,left=0}{v8,v9}
		\fmf{dashes_ar,left=0}{v9,v10}
		\fmf{dashes_rar,left=0}{v10,v11}
		\fmf{dashes_ar,left=0}{v11,v12}
		\fmf{dashes_rar,left=0}{v12,v1}
		\fmf{plain_rar}{vc,v1}
		\fmf{plain_rar}{vc,v3}
		\fmf{plain_rar}{vc,v5}
		\fmf{dots}{vc,v7}
		\fmf{plain_rar}{vc,v9}
		\fmf{plain_rar}{vc,v11}
		\fmf{plain_ar}{va1,v2}
		\fmf{plain_ar}{va2,v4}
		\fmf{plain_ar}{va3,v6}
		\fmf{dots}{va4,v8}
		\fmf{plain_ar}{va5,v10}
		\fmf{plain_ar}{va6,v12}
		\fmffreeze
		%
		\fmfv{l=$\scriptscriptstyle 1$,l.dist=2}{va1}
		\fmfv{l=$\scriptscriptstyle 2$,l.dist=2}{va2}
		\fmfv{l=$\scriptscriptstyle 3$,l.dist=2}{va3}
		\fmfv{l=$\scriptscriptstyle {L-1}$,l.dist=2}{va5}
		\fmfv{l=$\scriptscriptstyle L$,l.dist=2}{va6}
		\fmfiv{decor.shape=circle,decor.filled=full,decor.size=5}{vloc(__vc)}
		\end{fmfchar*}}}
&=-2(-\lambda)^Lc(L)\tr\big[((\rho^{\dagger\,i})(\rho_{i})^{\T})^L\big]\frac{P_L}{\complexi^L}
\\[-1.5\baselineskip]&
=-4(2\lambda)^Lc(L)\cos L\gamma_i^+\frac{P_L}{\complexi^L}\eqncom\\ 
\tilde F(L)
=
\settoheight{\eqoff}{$\times$}%
\setlength{\eqoff}{0.5\eqoff}%
\addtolength{\eqoff}{-16\unitlength}%
\raisebox{\eqoff}{%
	\fmfframe(7,1)(3,1){%
		\begin{fmfchar*}(30,30)
		\fmfleft{in}
		\fmfright{out1}
		\fmf{phantom}{in,v10}
		\fmf{phantom}{out,v4}
		\fmf{phantom}{in,va5}
		\fmf{phantom}{out,va2}
		\fmfforce{(0,0.5h)}{in}
		\fmfforce{(w,0.5h)}{out}
		\fmfpoly{phantom}{v1,v12,v11,v10,v9,v8,v7,v6,v5,v4,v3,v2}
		\fmffixed{(0.75w,0)}{v10,v4}
		\fmfpoly{phantom}{va1,va6,va5,va4,va3,va2}
		\fmffixed{(w,0)}{va5,va2}
		\fmf{phantom}{vc,v1}
		\fmf{phantom}{vc,v7}
		\fmffreeze
		\fmf{dashes_rar,left=0}{v1,v2}
		\fmf{dashes_ar,left=0}{v2,v3}
		\fmf{dashes_rar,left=0}{v3,v4}
		\fmf{dashes_ar,left=0}{v4,v5}
		\fmf{dashes_rar,left=0}{v5,v6}
		\fmf{dots,left=0}{v6,v7}
		\fmf{dots,left=0}{v7,v8}
		\fmf{dashes_ar,left=0}{v8,v9}
		\fmf{dashes_rar,left=0}{v9,v10}
		\fmf{dashes_ar,left=0}{v10,v11}
		\fmf{dashes_rar,left=0}{v11,v12}
		\fmf{dashes_ar,left=0}{v12,v1}
		\fmf{plain_rar}{vc,v1}
		\fmf{plain_rar}{vc,v3}
		\fmf{plain_rar}{vc,v5}
		\fmf{dots}{vc,v7}
		\fmf{plain_rar}{vc,v9}
		\fmf{plain_rar}{vc,v11}
		\fmf{plain_ar}{va1,v2}
		\fmf{plain_ar}{va2,v4}
		\fmf{plain_ar}{va3,v6}
		\fmf{dots}{va4,v8}
		\fmf{plain_ar}{va5,v10}
		\fmf{plain_ar}{va6,v12}
		\fmffreeze
		%
		\fmfv{l=$\scriptscriptstyle 1$,l.dist=2}{va1}
		\fmfv{l=$\scriptscriptstyle 2$,l.dist=2}{va2}
		\fmfv{l=$\scriptscriptstyle 3$,l.dist=2}{va3}
		\fmfv{l=$\scriptscriptstyle {L-1}$,l.dist=2}{va5}
		\fmfv{l=$\scriptscriptstyle L$,l.dist=2}{va6}
		\fmfiv{decor.shape=circle,decor.filled=full,decor.size=5}{vloc(__vc)}
		\end{fmfchar*}}}
&=-2(-\lambda)^Lc(L)
\tr[((\tilde\rho^{\dagger\,i})(\tilde\rho_{i})^{\T})^L]\frac{P_L}{\complexi^L} 
\\[-1.5\baselineskip]&
=-4(2\lambda)^Lc(L)\cos L\gamma_i^-\frac{P_L}{\complexi^L} 
\eqncom
\end{aligned}
\end{equation}
\end{subequations}
where the composite operator with vanishing momentum $p=0$ is drawn as the central black dot. 
 The scalar coupling tensors $\hat{Q}$ are given in \eqref{eq:Qhat} and they are related to the usual $F$-tensors in \eqref{eq:F_to_Qhat}. The colour factor $c(L)=N^{-\frac L2}(a_1a_2\dots a_L)$ has to be separated off for the calculation of $Z_{\cO_L}$ since it is just the colour structure of the tree-level composite operator\footnote{When we close the open colour lines into the conjugate composite operator $\ol{\mathcal{O}}_L$, the complete colour structure becomes one, as in \eqref{eq:operator_normalisation}.}, compare \eqref{eq:composite_operator_doubleline}. All four diagrams depend on the scalar `cake' integral\footnote{For a connection of this integral to knot theory see \cite{Broadhurst:1996kc}.} $P_L$. Its diagrammatic integral representation, which is introduced in \appref{app:Evaluating_Feynman_integrals} and its UV divergence $\mathcal{P}_L$ given in \cite{Broadhurst:1985vq} are\footnote{The additional factor of $\complexi^L$ occurs since we work in Minkowski instead of Euclidean space, \cf \eqref{eq:wick_rotation_integral}.}
\begin{equation}\label{PL}
\complexi^{-L}P_L=
\settoheight{\eqoff}{$\times$}%
\setlength{\eqoff}{0.5\eqoff}%
\addtolength{\eqoff}{-8\unitlength}%
\raisebox{\eqoff}{%
	\fmfframe(3,-2)(0,-2){%
		\begin{fmfchar*}(20,20)
		\fmfleft{in}
		\fmfright{out1}
		\fmf{phantom}{in,v1}
		\fmf{phantom}{out,v2}
		\fmfforce{(0,0.5h)}{in}
		\fmfforce{(w,0.5h)}{out}
		\fmfpoly{phantom}{v1,va4,va3,v2,va2,va1}
		\fmffixed{(0.75w,0)}{v1,v2}
		\fmf{phantom}{vc,v1}
		\fmf{plain}{vc,v2}
		\fmffreeze
		\fmf{plain,left=0.25}{v1,va1}
		\fmf{plain,left=0.25}{va1,va2}
		\fmf{plain,left=0.25}{va2,v2}
		\fmf{plain,left=0.25}{v2,va3}
		\fmf{plain,left=0.25}{va3,va4}
		\fmf{dots,left=0.25}{va4,v1}
		\fmf{plain}{vc,va1}
		\fmf{plain}{vc,va2}
		\fmf{plain}{vc,va3}
		\fmf{dots}{vc,va4}
		\fmf{plain}{vc,v1}
		\fmffreeze
		\fmfv{l=$\scriptscriptstyle L$,l.dist=3}{va1}
		\fmfv{l=$\scriptscriptstyle 1$,l.dist=3}{va2}
		\fmfv{l=$\scriptscriptstyle 2$,l.dist=3}{v2}
		\fmfv{l=$\scriptscriptstyle 3$,l.dist=3}{va3}
		\fmfv{l=$\scriptscriptstyle L-1$,l.dist=3}{v1}
		\fmfv{decor.shape=circle,decor.filled=full,	decor.size=2thick}{vc,va1,va2,v2,va3,va4,v1}
		\end{fmfchar*}}}
\eqncom\qquad
\mathcal{P}_L
=\Kop[\complexi^{-L}P_L]
=\frac{1}{(4\pi)^{2L}}\frac{1}{\epsilon}
\frac{1}{L}\binom{2L-2}{L-1}\zeta(2L-3)
\eqncom
\end{equation}
where the dimension is assumed to be $D=4-2\epsilon$. Note that this integral has no subdivergences and therefore its overall UV divergence is given by a simple pole in the regulator. 

In the diagrams $F(L)$ and $\tilde F(L)$ in \eqref{wrapLdiags2}, the vertices which connect to the central composite operator always alternate with vertices that connect to external scalar fields. In principle, we could draw many more diagrams with a fermionic wrapping loop in which some of the vertices that connect to external fields are adjacent. However, the planar parts of all these diagrams vanish, which can be seen from the corresponding flavour tensor contraction. For two adjacent external scalars, they are given by either of the two possibilities
\begin{equation}
\begin{aligned}
\settoheight{\eqoff}{$\times$}%
\setlength{\eqoff}{0.5\eqoff}%
\addtolength{\eqoff}{-5\unitlength}%
\raisebox{\eqoff}{%
	\fmfframe(4,2)(4,2){%
		\begin{fmfchar*}(24,6)
		\fmfleft{vl}
		\fmfright{vr}
		\fmfforce{(0,0)}{vl}
		\fmfforce{(w,0)}{vr}
		\fmfforce{(0.333w,h)}{vi1}
		\fmfforce{(0.666w,h)}{vi2}
		\fmfforce{(0.333w,0)}{vc1}
		\fmfforce{(0.666w,0)}{vc2}
		\fmf{dashes_ar}{vl,vc1}
		\fmf{dashes_rar}{vc1,vc2}
		\fmf{dashes_ar}{vc2,vr}
		\fmffreeze
		\fmfposition
		\fmf{plain_ar}{vi1,vc1}
		\fmf{plain_ar}{vi2,vc2}
		\fmfiv{label=$\scriptstyle A$,l.dist=2}{vloc(__vl)}
		\fmfiv{label=$\scriptstyle i$,l.dist=2}{vloc(__vi1)}
		\fmfiv{label=$\scriptstyle i$,l.dist=2}{vloc(__vi2)}
		\fmfiv{label=$\scriptstyle B$,l.dist=2}{vloc(__vr)}
		\end{fmfchar*}}}
&\propto
\sum_{C=1}^4\rho^{i\,CA}\tilde \rho^i_{BC}
=0\eqncom\qquad
\settoheight{\eqoff}{$\times$}%
\setlength{\eqoff}{0.5\eqoff}%
\addtolength{\eqoff}{-5\unitlength}%
\raisebox{\eqoff}{%
	\fmfframe(4,2)(4,2){%
		\begin{fmfchar*}(24,6)
		\fmfleft{vl}
		\fmfright{vr}
		\fmfforce{(0,0)}{vl}
		\fmfforce{(w,0)}{vr}
		\fmfforce{(0.333w,h)}{vi1}
		\fmfforce{(0.666w,h)}{vi2}
		\fmfforce{(0.333w,0)}{vc1}
		\fmfforce{(0.666w,0)}{vc2}
		\fmf{dashes_rar}{vl,vc1}
		\fmf{dashes_ar}{vc1,vc2}
		\fmf{dashes_rar}{vc2,vr}
		\fmffreeze
		\fmfposition
		\fmf{plain_ar}{vi1,vc1}
		\fmf{plain_ar}{vi2,vc2}
		\fmfiv{label=$\scriptstyle A$,l.dist=2}{vloc(__vl)}
		\fmfiv{label=$\scriptstyle i$,l.dist=2}{vloc(__vi1)}
		\fmfiv{label=$\scriptstyle i$,l.dist=2}{vloc(__vi2)}
		\fmfiv{label=$\scriptstyle B$,l.dist=2}{vloc(__vr)}
		\end{fmfchar*}}}
&\propto
\sum_{C=1}^4\tilde \rho^i_{CA}\rho^{i\,BC}
=0\eqncom
\end{aligned}
\end{equation}
where the coupling tensors are given in \eqref{eq:coupling_tensors_gammai} and $i$ is fixed.

Expanding the 1PI renormalisation constant in \eqref{eq:wrapping_correlation_funct} to leading order in $\lambda$, we find the deformation-dependent $L$-loop contribution to the 1PI counterterm $(\delta_{\mathcal{O}_L}^{(L)})_{\text{def}}$. It is given by the negative sum of the diagrams displayed in \eqref{wrapLdiags}, contracted with the colour structure\footnote{This colour structure contraction simply erases the $N^{-\frac L2}$ suppression factor.} of the conjugate operator $\ol{\mathcal{O}}_L$, rendering the result
\begin{equation}\label{deltaZOLdef}
\begin{aligned}
(\delta^{(L)}_{\mathcal{O}_L})_{\text{def}}
&=
-\Kop[S(L)+\bar{S}(L)+F(L)+\tilde F(L)]\\
&=4(2\lambda)^L\Big(\cos L\gamma_i^++\cos L\gamma_i^--\cos L\gamma_i^+\cos L\gamma_i^--\frac{1}{2^{L+1}}\Big)\mathcal{P}_L
\eqndot
\end{aligned}
\end{equation}
In the full renormalisation constant $\mathcal{Z}_{\mathcal{O}_L}=1+\mathfrak{d}_{\mathcal{O}_L}$, this term occurs linearly at $L$ loops, compare \eqref{eq:comp_op_renormalisation}. Since \eqref{deltaZOLdef} is the only deformation-dependent contribution at $L$ loops, we can write it as $(\mathfrak{d}_{\mathcal{O}_L}^{(L)})_{\text{def}}=(\delta^{(L)}_{\mathcal{O}_L})_{\text{def}}$
and reconstruct the full $L$-loop counterterm by enforcing that it vanishes in the limit $\gamma_i^\pm\rightarrow 0$. For the undeformed contribution this procedure yields $(\mathfrak{d}_{\mathcal{O}_L}^{(L)})_{\ol{\text{def}}}
=-(\delta^{(L)}_{\mathcal{O}_L})_{\text{def}}\big|_{\gamma_i^\pm=0}$ and for the $L$-loop renormalisation constant we have
\begin{equation}\label{ZL}
\begin{aligned}
\mathcal{Z}_{\mathcal{O}_L}^{(L)}
&=
1+(\mathfrak{d}_{\mathcal{O}_L}^{(L)})_{\text{def}}+(\mathfrak{d}_{\mathcal{O}_L}^{(L)})_{\ol{\text{def}}}
&=1-16(2\lambda)^{L}\sin^2\frac{L\gamma_i^+}{2}\sin^2\frac{L\gamma_i^-}{2}\mathcal{P}_L\eqndot
\end{aligned}
\end{equation}
Note that the non-trivial renormalisation vanishes not only for the undeformed theory with $\gamma_i^\pm=0$, but also in the $\beta$-deformation, where $\gamma_i^+=\beta$ and $\gamma_i^-=0$. This vanishing provides us with an immediate consistency test, since $\mathcal{O}_L$ is not renormalised in the $\beta$-deformation. Expressing the renormalised \tHooft coupling constant in terms of the bare one as $\lambda=\mu^{-2\epsilon}\lambda_{\text{B}}$, we can calculate the anomalous dimension of $\mathcal{O}_{L\geq 3}$. Using \eqref{eq:anomalous_dim_matrix}, we find
\begin{equation}
\begin{aligned}\label{gammaOL}
\gamma_{\mathcal{O}_L}&=-32 (2g^2)^{L}\sin^2\frac{L\gamma_i^+}{2}\sin^2\frac{L\gamma_i^-}{2}\binom{2L-2}{L-1}\zeta(2L-3)\eqncom
\end{aligned}
\end{equation}
where we absorbed a factor of $(4\pi)^{-2}$ from \eqref{PL} into the effective planar coupling constant $g^2$. Up to a rescaling of the coupling constant and the deformation angles, this result agrees with the one found from the integrability-based calculation in \cite[equation (5.5)]{Ahn:2011xq}.

\subsubsection{Length \texorpdfstring{$L=2$}{L=2}}
For $L=2$, we have the deformation-dependent wrapping $(K=L=2)$-loop diagrams and, in addition, the deformation-dependent prewrapping contributions from $(K=1)$ and $(K=2)$-loop diagrams. The wrapping diagrams were constructed in \eqref{wrapLdiags}, but at two loops their pole part is not given by the scalar integral \eqref{PL} any more. At this low loop order the integral $P_L$ is IR-divergent if the composite operator is inserted with vanishing momentum. In order to extract only the UV divergence, we set the external momentum of the operator to $p$ and extract it at one of the two external legs, which removes the IR divergence from the $P_2$ integral. The remaining UV divergence is then given by 
\begin{equation}\label{eq:divergence_at_L2}
\mathcal{P}_2\,\rightarrow\,
\complexi^{-2}\Kop\Bigl[
\scalebox{0.6}{
	\settoheight{\eqoff}{$\times$}%
	\setlength{\eqoff}{0.5\eqoff}%
	\addtolength{\eqoff}{-6.75\unitlength}%
	\raisebox{\eqoff}{
		\fmfframe(-2,0)(-2,0){
			\begin{fmfchar*}(15,15.0)
			\fmfforce{0.35w,1h}{vout}
			\fmfforce{0.35w,0h}{vin}
			\fmfforce{0.5 w, 0.2h}{v1}
			\fmfforce{0.5 w, 0.8h}{v2}
			\fmfv{decor.shape=circle,decor.filled=full,decor.size=2thick}{v1,v2}
			\fmf{plain}{v2,vout}
			\fmf{plain}{vin,v1}
			\fmf{plain,left}{v1,v2,v1}
			\fmffreeze
			\fmf{plain}{v1,v2}
			\fmffreeze
			\fmfposition
			\fmfipath{p[]}
			\fmfiset{p1}{vpath(__v2,__v1)}
			\fmfiv{decor.shape=circle,decor.filled=full,decor.size=2thick}{point length(p1)/2 of p1}
			\fmfiv{label=$p$,decoration.angle=-90,l.a=0,l.dist=0.1w}{point length(p1)/2 of p1}
			\fmfiv{label=$p$,l.a=0,l.dist=0.1w}{vloc(__vout)}
			\end{fmfchar*}
		}
	}
}
\Bigr]
=
\Kop\bigl[\complexi^{-2}\hat{I}_{(1,1)}(p)\hat{I}_{(1+\epsilon,1)}(p)\bigr]
=\frac{-1}{(4\pi)^4}\Biggl(\frac{1}{2\epsilon^{2}}+\frac{ \frac 52-c_{\ol{\text{MS}}}+\log\frac{\mu^2}{p^2}}{\epsilon}\Biggr)
\eqncom
\end{equation}
where the occurring integrals are given in \eqref{eq:def_dep_contributions}, the constant is $c_{\ol{\text{MS}}}=\gammaE-\log 4\pi$ and $\mu$ is the renormalisation scale. The last equality is obtained using the techniques from \appref{app:Evaluating_Feynman_integrals}. The divergent two-loop contribution from wrapping interactions $(C_{\mathcal{O}_2}^{(2)})_{\text{wrap}}$ is therefore given by one minus the result of \eqref{ZL} with $\mathcal{P}_2$ replaced according to \eqref{eq:divergence_at_L2} and it reads
\begin{equation}\label{eq:2l_wrapping}
(C_{\mathcal{O}_2}^{(2)})_{\text{wrap}}
=
64\lambda^{2}\sin^2\gamma_i^+\sin^2\gamma_i^-
\Kop\bigl[\complexi^{-2}\hat{I}_{(1,1)}(p)\hat{I}_{(1+\epsilon,1)}(p)\bigr]\eqndot
\end{equation}
In contrast to the complete wrapping counterterm \eqref{ZL} at $L\geq 3$, this divergent contribution cannot be absorbed into local connected counterterms since it still contains a non-local pole $\propto\frac{1}{\epsilon} \log\frac{\mu^2}{p^2}$ which originates from subdivergence contributions that have not yet been accounted for. When we draw the $L=2$ wrapping diagrams from \eqref{wrapLdiags} via a one-loop vertex insertion as
\begin{equation}
\settoheight{\eqoff}{$\times$}%
\setlength{\eqoff}{0.5\eqoff}%
\addtolength{\eqoff}{-6.0\unitlength}%
\raisebox{\eqoff}{%
	\fmfframe(1,-1.25)(.5,1.25){%
		\begin{fmfchar*}(12,9)
		\fmfforce{(1w,0.5h)}{vo}
		\fmfforce{(0.8,1h)}{vl}
		\fmfforce{(0.8,0h)}{vr}
		\fmfforce{(0.3w,0.5h)}{vc}
		\fmf{phantom}{vl,vc}
		\fmf{phantom}{vr,vc}
		\fmf{phantom,left=0.75,tag=1}{vo,vc}
		\fmf{phantom,right=0.75,tag=2}{vo,vc}
		\fmffreeze
		\fmfposition
		\fmfipath{p[]}
		\fmfiset{p1}{vpath[1](__vo,__vc)}
		\fmfiset{p11}{subpath (0,4length(p1)/5) of p1}
		\fmfiset{p2}{vpath[2](__vo,__vc)}
		\fmfiset{p21}{subpath (0,4length(p2)/5) of p2}
		\fmfiset{p3}{vpath(__vl,__vc)}
		\fmfiset{p31}{subpath (0,4length(p3)/5) of p3}
		\fmfiset{p4}{vpath(__vr,__vc)}
		\fmfiset{p41}{subpath (0,4length(p4)/5) of p4}
		\fmfi{plain_rar,left=0.75}{p11}
		\fmfi{plain_rar,right=0.75}{p21}
		\fmfi{plain_ar}{p31}
		\fmfi{plain_ar}{p41}
		\fmfiv{decor.shape=circle,decor.filled=empty,decor.size=11thin}{vloc(__vc)}
		\fmfdraw
		\fmfiv{decor.shape=circle,decor.filled=shaded,decor.size=11thin}{vloc(__vc)}
		\fmfiv{decor.shape=circle,decor.filled=full,decor.size=5}{vloc(__vo)}
		\end{fmfchar*}
	}
}\eqncom
\end{equation}
it is clear that we also have to take the one-loop vertex renormalisation from the shaded area into account. 

All additional deformation-dependent contributions that can be constructed at two loops are prewrapping contributions with at least one power of the double-trace coupling $Q_{\text{F}\,ii}^{ii}$ whose one-loop anomalous dimension was calculated in \secref{sec:non-conformal_double_trace_coupling}. To construct all relevant contributions, we need all one-loop contributions that appear as subdiagrams in a two loop calculation of \eqref{eq:wrapping_correlation_funct} up to finite order in the regulator $\epsilon$. These one-loop diagrams can be constructed via the Feynman rules of \appref{app:Feynman_rules} and the \ttt{Mathematica} package \ttt{FokkenFeynPackage} introduced in \appref{sec:Feynman_rules_Mathematica}. The 1PI diagrams are given by 
\begin{equation}
\label{prewrapdiag}
\begin{aligned}
\settoheight{\eqoff}{$\times$}%
\setlength{\eqoff}{0.5\eqoff}%
\addtolength{\eqoff}{-6.0\unitlength}%
\raisebox{\eqoff}{%
	\fmfframe(2,-1.25)(1,1.25){%
		\begin{fmfchar*}(12,9)
		\fmfforce{(1w,0.5h)}{vo}
		\fmfforce{(0.8,1h)}{vl}
		\fmfforce{(0.8,0h)}{vr}
		\fmffreeze
		\fmfposition
		\fmfforce{(0.3w,0.5h)}{vc}
		\fmf{plain_rar,left=0.75}{vo,vc}
		\fmf{plain_rar,right=0.75}{vo,vc}
		\fmf{plain_rar}{vc,vl}
		\fmf{plain_rar}{vc,vr}
		\fmfiv{label=$\scriptstyle Q_{\text{F}}$,l.a=180,l.dist=5}{vloc(__vc)}
		\fmfiv{decor.shape=circle,decor.filled=full,decor.size=5}{vloc(__vo)}
		\end{fmfchar*}
	}
}
&=\complexi \lambda Q_{\text{F}\,ii}^{ii}\hat{I}_{(1,1)}(p)\frac{(ab)}{N}
\eqncom\\
\settoheight{\eqoff}{$\times$}%
\setlength{\eqoff}{0.5\eqoff}%
\addtolength{\eqoff}{-6.\unitlength}%
\raisebox{\eqoff}{
	\fmfframe(0,.5)(2,1){
		\begin{fmfchar*}(12,9)
		\fmfright{vop}
		\fmfforce{0 w, 1h}{v1}
		\fmfforce{0 w, 0h}{v3}
		\fmfforce{0.34 w,0.8 h}{vc1}
		\fmfforce{0.34 w,0.2 h}{vc3}
		\fmf{plain_rar}{vc1,v1}
		\fmf{plain_rar}{vc3,v3}
		\fmf{plain_rar,right=0.2,tension=0.6}{vop,vc1}
		\fmf{plain_rar,left=0.2,tension=0.6}{vop,vc3}
		\fmf{photon}{vc1,vc3}
		\fmffreeze
		\fmfposition
		\fmfiv{decor.shape=circle,decor.filled=full,decor.size=5}{vloc(__vop)}
		\end{fmfchar*}
	}
}
&=2\complexi \lambda\xi p^2 \hat{I}_{(2,1)}(p)\frac{(ab)}{N}
\eqncom
\qquad
\settoheight{\eqoff}{$\times$}%
\setlength{\eqoff}{0.5\eqoff}%
\addtolength{\eqoff}{-6.0\unitlength}%
\raisebox{\eqoff}{%
	\fmfframe(1,-1.25)(1,1.25){%
		\begin{fmfchar*}(12,9)
		\fmfforce{(1w,0.5h)}{vo}
		\fmfforce{(0.8,1h)}{vl}
		\fmfforce{(0.8,0h)}{vr}
		\fmffreeze
		\fmfposition
		\fmfforce{(0.3w,0.5h)}{vc}
		\fmf{plain_rar,left=0.75}{vo,vc}
		\fmf{plain_rar,right=0.75}{vo,vc}
		\fmf{plain_rar}{vc,vl}
		\fmf{plain_rar}{vc,vr}
		\fmfiv{decor.shape=circle,decor.filled=full,decor.size=5}{vloc(__vo)}
		\end{fmfchar*}
	}
}
=-2\complexi \lambda \hat{I}_{(1,1)}(p)\frac{(ab)}{N}
\eqncom\\
\settoheight{\eqoff}{$\times$}%
\setlength{\eqoff}{0.5\eqoff}%
\addtolength{\eqoff}{-5.\unitlength}%
\raisebox{\eqoff}{%
	\fmfframe(0,1)(-2,1){%
		\begin{fmfchar*}(15,7.5)
		\fmfforce{0 w,0.5 h}{v1}
		\fmfforce{1 w,0.5 h}{v2}
		\fmf{phantom}{v1,v2}
		\fmffreeze
		\fmfposition
		\fmfipath{p[]}
		\fmfiset{p1}{vpath(__v1,__v2)}
		\fmfiset{p11}{subpath (0,2length(p1)/5) of p1}
		\fmfiset{p12}{subpath (3length(p1)/5,length(p1)) of p1}
		\fmfi{plain_rar}{p11}
		\fmfi{plain_rar}{p12}
		\fmfiv{decor.shape=circle,decor.filled=shaded,decor.size=11thin}{point length(p1)/2 of p1}
		\end{fmfchar*}
	}
}
&=-2\lambda p^2\bigl(2\hat{I}_{(1,1)}(p)+ p^2 \hat{I}_{(2,1)}(p)(1-\xi)\bigr)(ab)
\eqncom
\end{aligned}
\end{equation}
where the occurrence of the double-trace coupling is explicitly indicated by the label $Q_{\text{F}}$ and we restricted to the diagrams with identical flavours on the external scalar lines. From the one-loop diagrams, the one-loop 1PI counterterms of the composite operator and the scalar field are easily extracted as the negative divergence of the respective diagrams with stripped off colour, flavour, and spacetime structure. We find
\begin{equation}\label{cttwopointv}
\begin{gathered}
\Bigl[
\settoheight{\eqoff}{$\times$}%
\setlength{\eqoff}{0.5\eqoff}%
\addtolength{\eqoff}{-5.5\unitlength}%
\raisebox{\eqoff}{
	\fmfframe(0,0)(4,1){
		\begin{fmfchar*}(12,9)
		\fmfright{vop}
		\fmfforce{0 w, .9h}{v1}
		\fmfforce{0 w, .1h}{v3}
		\fmf{plain_rar,right=0.2,tension=0.5}{vop,v1}
		\fmf{plain_rar,left=0.2,tension=0.5}{vop,v3}
		\fmffreeze
		\fmfposition
		\fmfiv{decor.shape=circle,decor.filled=full,decor.size=5}{vloc(__vop)}
		\fmfiv{decor.shape=hexacross,decor.size=12thin}{vloc(__vop)}
		\fmfiv{label=$\scriptstyle \text{def}$,l.a=0,l.dist=0.2w}{vloc(__vop)}
		\end{fmfchar*}
	}
}
\Bigr]_{\mathcal{O}_2}
=\bigl(\delta^{(1)}_{\mathcal{O}_2}\bigr)_{\text{def}}
=\frac{\lambda}{(4\pi)^2} \frac{Q_{\text{F}\,ii}^{ii}}{\epsilon}
\eqncom\quad
\Bigl[
\settoheight{\eqoff}{$\times$}%
\setlength{\eqoff}{0.5\eqoff}%
\addtolength{\eqoff}{-5.5\unitlength}%
\raisebox{\eqoff}{
	\fmfframe(0,0)(4,1){
		\begin{fmfchar*}(12,9)
		\fmfright{vop}
		\fmfforce{0 w, .9h}{v1}
		\fmfforce{0 w, .1h}{v3}
		\fmf{plain_rar,right=0.2,tension=0.5}{vop,v1}
		\fmf{plain_rar,left=0.2,tension=0.5}{vop,v3}
		\fmffreeze
		\fmfposition
		\fmfiv{decor.shape=circle,decor.filled=full,decor.size=5}{vloc(__vop)}
		\fmfiv{decor.shape=hexacross,decor.size=12thin}{vloc(__vop)}
		\fmfiv{label=$\scriptstyle \ol{\text{def}}$,l.a=0,l.dist=0.2w}{vloc(__vop)}
		\end{fmfchar*}
	}
}
\Bigr]_{\mathcal{O}_2}
=\bigl(\delta^{(1)}_{\mathcal{O}_2}\bigr)_{\ol{\text{def}}}
=\frac{-2\lambda}{(4\pi)^2}\frac{1+\xi}{\epsilon}
\eqncom\\
\bigl[\bigl(\twovertex{plain_rar}{}{plain_rar}{}\bigr)_{\text{1PI}}\bigr]_{\complexi p^2\delta^j_i(ab)}
{}={}
\delta^{(1)}_\phi=\frac{2\lambda}{(4\pi)^2}\frac{1+\xi}{\epsilon}\eqncom
\end{gathered}
\end{equation}
where we have split the contributions to the counterterm for the composite operator into deformation-dependent and deformation-independent ones. The complete one-loop counterterm can either be constructed from $\bigl(\delta^{(1)}_{\mathcal{O}_2}\bigr)_{\text{def}}$ alone, in analogy to the procedure in the $L\geq 3$ case, or from all above counterterms
\begin{equation}
\mathfrak{d}_{\mathcal{O}_2}^{(1)}=\delta_{\mathcal{O}_2}^{(1)}+\delta_\phi^{(1)}=\frac{\lambda}{(4\pi)^2}Q_{\text{F}\,ii}^{ii}\frac{1}{\epsilon}\eqndot
\end{equation}  

For the two-loop deformation-dependent prewrapping contributions, we construct the 1PI diagrams as
\begin{equation}
\begin{aligned}\label{eq:O2_2L_insertions}
\settoheight{\eqoff}{$\times$}%
\setlength{\eqoff}{0.5\eqoff}%
\addtolength{\eqoff}{-6.0\unitlength}%
\raisebox{\eqoff}{%
	\fmfframe(1,-1.25)(1,1.25){%
		\begin{fmfchar*}(12,9)
		\fmfforce{(1w,0.5h)}{vo}
		\fmfforce{(0.8,1h)}{vl}
		\fmfforce{(0.8,0h)}{vr}
		\fmffreeze
		\fmfposition
		\fmfforce{(0.2w,0.5h)}{vc2}
		\fmfforce{(0.6w,0.5h)}{vc1}
		\fmf{plain_rar,left=0.75}{vo,vc1}
		\fmf{plain_rar,right=0.75}{vo,vc1}
		\fmf{plain_rar,left=0.75}{vc1,vc2}
		\fmf{plain_rar,right=0.75}{vc1,vc2}
		\fmf{plain_rar}{vc2,vl}
		\fmf{plain_rar}{vc2,vr}
		\fmfiv{label=$\scriptstyle Q_{\text{F}}$,l.a=180,l.dist=5}{vloc(__vc2)}
		\fmfiv{decor.shape=circle,decor.filled=full,decor.size=5}{vloc(__vo)}
		\end{fmfchar*}
	}
}
&=
\settoheight{\eqoff}{$\times$}%
\setlength{\eqoff}{0.5\eqoff}%
\addtolength{\eqoff}{-6.0\unitlength}%
\raisebox{\eqoff}{%
	\fmfframe(1,-1.25)(1,1.25){%
		\begin{fmfchar*}(12,9)
		\fmfforce{(1w,0.5h)}{vo}
		\fmfforce{(0.8,1h)}{vl}
		\fmfforce{(0.8,0h)}{vr}
		\fmffreeze
		\fmfposition
		\fmfforce{(0.2w,0.5h)}{vc2}
		\fmfforce{(0.6w,0.5h)}{vc1}
		\fmf{plain_rar,left=0.75}{vo,vc1}
		\fmf{plain_rar,right=0.75}{vo,vc1}
		\fmf{plain_rar,left=0.75}{vc1,vc2}
		\fmf{plain_rar,right=0.75}{vc1,vc2}
		\fmf{plain_rar}{vc2,vl}
		\fmf{plain_rar}{vc2,vr}
		\fmfiv{label=$\scriptstyle Q_{\text{F}}$,l.a=90,l.dist=6}{vloc(__vc1)}
		\fmfiv{decor.shape=circle,decor.filled=full,decor.size=5}{vloc(__vo)}
		\end{fmfchar*}
	}
}
=2\lambda^2Q_{\text{F}\,ii}^{ii}\Kop\bigl[\hat{I}^2_{(1,1)}(p)\bigr]
\eqncom\qquad
\settoheight{\eqoff}{$\times$}%
\setlength{\eqoff}{0.5\eqoff}%
\addtolength{\eqoff}{-6.0\unitlength}%
\raisebox{\eqoff}{%
	\fmfframe(1,-1.25)(1,1.25){%
		\begin{fmfchar*}(12,9)
		\fmfforce{(1w,0.5h)}{vo}
		\fmfforce{(0.8,1h)}{vl}
		\fmfforce{(0.8,0h)}{vr}
		\fmffreeze
		\fmfposition
		\fmfforce{(0.2w,0.5h)}{vc2}
		\fmfforce{(0.6w,0.5h)}{vc1}
		\fmf{plain_rar,left=0.75}{vo,vc1}
		\fmf{plain_rar,right=0.75}{vo,vc1}
		\fmf{plain_rar,left=0.75}{vc1,vc2}
		\fmf{plain_rar,right=0.75}{vc1,vc2}
		\fmf{plain_rar}{vc2,vl}
		\fmf{plain_rar}{vc2,vr}
		\fmfiv{label=$\scriptstyle Q_{\text{F}}$,l.a=90,l.dist=6}{vloc(__vc1)}
		\fmfiv{label=$\scriptstyle Q_{\text{F}}$,l.a=180,l.dist=5}{vloc(__vc2)}
		\fmfiv{decor.shape=circle,decor.filled=full,decor.size=5}{vloc(__vo)}
		\end{fmfchar*}
	}
}
=-\lambda^2(Q_{\text{F}\,ii}^{ii})^2\Kop\bigl[\hat{I}_{(1,1)}^2(p)\bigr]
\eqncom\\
\settoheight{\eqoff}{$\times$}%
\setlength{\eqoff}{0.5\eqoff}%
\addtolength{\eqoff}{-6.\unitlength}%
\raisebox{\eqoff}{
	\fmfframe(1,-1.25)(1,1.25){
		\begin{fmfchar*}(12,9)
		\fmfright{vop}
		\fmfforce{0 w, 1h}{v1}
		\fmfforce{0 w, 0h}{v3}
		\fmfforce{0.6 w,0.5 h}{vc2}
		\fmf{phantom}{vc2,v1}
		\fmf{phantom}{vc2,v3}
		\fmf{plain_rar,left=0.75}{vop,vc2}
		\fmf{plain_rar,right=0.75}{vop,vc2}
		\fmffreeze
		\fmfposition
		\fmfipath{p[]}
		\fmfipair{w[]}
		\fmfiset{p1}{vpath(__vc2,__v1)}
		\fmfiset{p11}{subpath (0,length(p1)/2) of p1}
		\fmfiset{p12}{subpath (length(p1)/2,length(p1)) of p1}
		\fmfiset{w1}{point length(p1)/2 of p1}
		\fmfi{plain_rar}{p11}
		\fmfi{plain_rar}{p12}
		\fmfiset{p3}{vpath(__vc2,__v3)}
		\fmfiset{p31}{subpath (0,length(p3)/2) of p3}
		\fmfiset{p32}{subpath (length(p3)/2,length(p3)) of p3}
		\fmfiset{w3}{point length(p3)/2 of p3}
		\fmfi{plain_rar}{p31}
		\fmfi{plain_rar}{p32}
		\fmfforce{w1}{vc1}
		\fmfforce{w3}{vc3}
		\fmf{photon,right=0.75}{vc1,vc3}
		\fmfiv{decor.shape=circle,decor.filled=full,decor.size=5}{vloc(__vop)}
		\fmfiv{label=$\scriptstyle Q_{\text{F}}$,l.a=90,l.dist=6}{vloc(__vc2)}
		\end{fmfchar*}
	}
}
&=-2\lambda^2 \xi Q_{\text{F}\,ii}^{ii}\Kop\bigl[\hat{I}_{(1,1)}(p)p^2\hat{I}_{(1,2)}(p)\bigr]
\eqncom\\
\settoheight{\eqoff}{$\times$}%
\setlength{\eqoff}{0.5\eqoff}%
\addtolength{\eqoff}{-6.0\unitlength}%
\raisebox{\eqoff}{%
	\fmfframe(1,-1.25)(1,1.25){%
		\begin{fmfchar*}(12,9)
		\fmfforce{(1w,0.5h)}{vo}
		\fmfforce{(0.8,1h)}{vl}
		\fmfforce{(0.8,0h)}{vr}
		\fmf{plain_rar}{vc,vl}
		\fmf{plain_rar}{vc,vr}
		\fmffreeze
		\fmfposition
		\fmfforce{(0.3w,0.5h)}{vc}
		\fmf{phantom,left=0.75}{vo,vc}	
		\fmf{phantom,left=0.75}{vc,vo}	
		\fmffreeze
		\fmfposition
		\fmfipath{p[]}
		\fmfiset{p1}{vpath(__vo,__vc)}
		\fmfiset{p11}{subpath (0,length(p1)/2) of p1}
		\fmfiset{p12}{subpath (length(p1)/2,length(p1)) of p1}
		\fmfi{plain_rar}{p11}
		\fmfi{plain_rar}{p12}
		\fmfiset{p2}{vpath(__vc,__vo)}
		\fmfiset{p21}{subpath (0,length(p2)/2) of p2}
		\fmfiset{p22}{subpath (length(p2)/2,length(p2)) of p2}
		\fmfi{plain_ar}{p21}
		\fmfi{plain_ar}{p22}
		\fmfi{photon}{point length(p1)/2 of p1 -- point length(p2)/2 of p2}
		\fmfiv{label=$\scriptstyle Q_{\text{F}}$,l.a=180,l.dist=5}{vloc(__vc)}
		\fmfiv{decor.shape=circle,decor.filled=full,decor.size=5}{vloc(__vo)}
		\end{fmfchar*}
	}
}
&=-2\lambda^2Q_{\text{F}\,ii}^{ii}\Kop\bigl[\bigl(\hat{I}_{(1,1)}^2(p)-2\hat{I}_{(1+\epsilon,1)}(p)(2\hat{I}_{(1,1)}(p)+(1-\xi)p^2\hat{I}_{(1,2)}(p))\bigr)\bigr]
\eqncom\\
\settoheight{\eqoff}{$\times$}%
\setlength{\eqoff}{0.5\eqoff}%
\addtolength{\eqoff}{-6.0\unitlength}%
\raisebox{\eqoff}{%
	\fmfframe(1,-1.25)(1,1.25){%
		\begin{fmfchar*}(12,9)
		\fmfforce{(1w,0.5h)}{vo}
		\fmfforce{(0.8,1h)}{vl}
		\fmfforce{(0.8,0h)}{vr}
		\fmf{plain_rar}{vc,vl}
		\fmf{plain_rar}{vc,vr}
		\fmffreeze
		\fmfposition
		\fmfforce{(0.3w,0.5h)}{vc}
		\fmf{phantom,left=0.75}{vo,vc}		
		\fmffreeze
		\fmfposition
		\fmfipath{p[]}
		\fmfiset{p1}{vpath(__vo,__vc)}
		\fmfiset{p11}{subpath (0,length(p1)/2) of p1}
		\fmfiset{p12}{subpath (length(p1)/2,length(p1)) of p1}
		\fmfi{plain_rar}{p11}
		\fmfi{plain_rar}{p12}
		\fmfiv{decor.shape=circle,decor.filled=shaded,decor.size=8thin}{point length(p1)/2 of p1}
		\fmfiv{label=$\scriptstyle Q_{\text{F}}$,l.a=180,l.dist=5}{vloc(__vc)}
		\fmfiv{decor.shape=circle,decor.filled=full,decor.size=5}{vloc(__vo)}
		\fmf{plain_rar,right=0.75}{vo,vc}
		\end{fmfchar*}
	}
}
&=-4\lambda^2Q_{\text{F}\,ii}^{ii}\Kop\bigl[\hat{I}_{(1+\epsilon,1)}(p)(2\hat{I}_{(1,1)}(p)+(1-\xi)p^2\hat{I}_{(2,1)}(p))\bigr]
\eqncom
\end{aligned}
\end{equation}
where the divergent part of each integral\footnote{Note that we discarded contributions from the two-loop scalar master integral 
	$
	\settoheight{\eqoff}{$\times$}%
	\setlength{\eqoff}{0.5\eqoff}%
	\addtolength{\eqoff}{-2.8\unitlength}%
	\raisebox{\eqoff}{%
		\fmfframe(0,0)(-2.5,0){%
			\begin{fmfchar*}(8,5)
			\fmfforce{(1w,0.5h)}{vo}
			\fmfforce{(0w,0.5h)}{vl}
			\fmffreeze
			\fmfposition
			\fmfforce{(0.2w,0.5h)}{vcl}
			\fmfforce{(0.8w,0.5h)}{vcr}
			\fmf{plain}{vcl,vl}
			\fmf{plain}{vcr,vo}
			\fmf{phantom,left=0.75}{vcr,vcl}	
			\fmf{phantom,left=0.75}{vcl,vcr}	
			\fmffreeze
			\fmfposition
			\fmfipath{p[]}
			\fmfiset{p1}{vpath(__vcr,__vcl)}
			\fmfiset{p11}{subpath (0,length(p1)/2) of p1}
			\fmfiset{p12}{subpath (length(p1)/2,length(p1)) of p1}
			\fmfi{plain}{p11}
			\fmfi{plain}{p12}
			\fmfiset{p2}{vpath(__vcl,__vcr)}
			\fmfiset{p21}{subpath (0,length(p2)/2) of p2}
			\fmfiset{p22}{subpath (length(p2)/2,length(p2)) of p2}
			\fmfi{plain}{p21}
			\fmfi{plain}{p22}
			\fmfi{plain}{point length(p1)/2 of p1 -- point length(p2)/2 of p2}
			\fmfiv{decor.shape=circle,decor.filled=full,decor.size=2}{vloc(__vcr)}
			\fmfiv{decor.shape=circle,decor.filled=full,decor.size=2}{vloc(__vcl)}
			\fmfiv{decor.shape=circle,decor.filled=full,decor.size=2}{point length(p1)/2 of p1}
			\fmfiv{decor.shape=circle,decor.filled=full,decor.size=2}{point length(p2)/2 of p2}
			\end{fmfchar*}
		}
	}
	$ 
	in the third line. This is justified, since it is finite \cite{Chetyrkin:1980pr} and we are only interested in the divergent contributions.
} is obtained by iteratively evaluating a divergent subdiagram and inserting the result in the remaining one-loop diagram. All deformation-dependent 1PI diagrams that involve a one-loop counterterm read
\begin{equation}
\begin{aligned}\label{eq:O2_1L_CT_insertions}
\settoheight{\eqoff}{$\times$}%
\setlength{\eqoff}{0.5\eqoff}%
\addtolength{\eqoff}{-6.0\unitlength}%
\raisebox{\eqoff}{%
	\fmfframe(1,-1.25)(2.5,1.25){%
		\begin{fmfchar*}(12,9)
		\fmfforce{(1w,0.5h)}{vo}
		\fmfforce{(0.8,1h)}{vl}
		\fmfforce{(0.8,0h)}{vr}
		\fmf{plain_rar}{vc,vl}
		\fmf{plain_rar}{vc,vr}
		\fmffreeze
		\fmfposition
		\fmfforce{(0.3w,0.5h)}{vc}
		\fmf{phantom,left=0.75}{vo,vc}		
		\fmffreeze
		\fmfposition
		\fmfipath{p[]}
		\fmfiset{p1}{vpath(__vo,__vc)}
		\fmfiset{p11}{subpath (0,length(p1)/2) of p1}
		\fmfiset{p12}{subpath (length(p1)/2,length(p1)) of p1}
		\fmfi{plain_rar}{p11}
		\fmfi{plain_rar}{p12}
		\fmfiv{decor.shape=hexacross,decor.size=9thin}{point length(p1)/2 of p1}
		\fmfiv{label=$\scriptstyle Q_{\text{F}}$,l.a=180,l.dist=5}{vloc(__vc)}
		\fmfiv{decor.shape=circle,decor.filled=full,decor.size=5}{vloc(__vo)}
		\fmf{plain_rar,right=0.75}{vo,vc}
		\end{fmfchar*}
	}
}
&=2\complexi \lambda  Q_{\text{F}\,ii}^{ii}\Kop\bigl[\delta^{(1)}_\phi \hat{I}_{(1,1)}(p)\bigr]
\eqncom\qquad\quad\,
\settoheight{\eqoff}{$\times$}%
\setlength{\eqoff}{0.5\eqoff}%
\addtolength{\eqoff}{-6.0\unitlength}%
\raisebox{\eqoff}{%
	\fmfframe(1,-1.25)(2.5,1.25){%
		\begin{fmfchar*}(12,9)
		\fmfforce{(1w,0.5h)}{vo}
		\fmfforce{(0.8,1h)}{vl}
		\fmfforce{(0.8,0h)}{vr}
		\fmffreeze
		\fmfposition
		\fmfforce{(0.3w,0.5h)}{vc}
		\fmf{plain_rar,left=0.75}{vo,vc}
		\fmf{plain_rar,right=0.75}{vo,vc}
		\fmf{plain_rar}{vc,vl}
		\fmf{plain_rar}{vc,vr}
		\fmfiv{label=$\scriptstyle Q_{\text{F}}$,l.a=180,l.dist=5}{vloc(__vc)}
		\fmfiv{decor.shape=hexacross,decor.size=9thin}{vloc(__vc)}
		\fmfiv{decor.shape=circle,decor.filled=full,decor.size=5}{vloc(__vo)}
		\end{fmfchar*}
	}
}
=\complexi \lambda \Kop\bigl[ \delta^{(1)}_{Q_F}  Q_{\text{F}\,ii}^{ii}\hat{I}_{(1,1)}(p)\bigr]
\eqncom\\
\Bigl[
\settoheight{\eqoff}{$\times$}%
\setlength{\eqoff}{0.5\eqoff}%
\addtolength{\eqoff}{-6.0\unitlength}%
\raisebox{\eqoff}{%
	\fmfframe(2,-1.25)(1,1.25){%
		\begin{fmfchar*}(12,9)
		\fmfforce{(1w,0.5h)}{vo}
		\fmfforce{(0.8,1h)}{vl}
		\fmfforce{(0.8,0h)}{vr}
		\fmffreeze
		\fmfposition
		\fmfforce{(0.3w,0.5h)}{vc}
		\fmf{plain_rar,left=0.75}{vo,vc}
		\fmf{plain_rar,right=0.75}{vo,vc}
		\fmf{plain_rar}{vc,vl}
		\fmf{plain_rar}{vc,vr}
		\fmfiv{decor.shape=hexacross,decor.size=12thin}{vloc(__vo)}
		\fmfiv{decor.shape=circle,decor.filled=full,decor.size=5}{vloc(__vo)}
		\end{fmfchar*}
	}
}
\Bigr]_{\text{def}}
&=-2\complexi \lambda\Kop\bigl[\bigl(\delta^{(1)}_{\mathcal{O}_2}\bigr)_{\text{def}}\hat{I}_{(1,1)}(p)\bigr]
\eqncom\quad
\Bigl[
\settoheight{\eqoff}{$\times$}%
\setlength{\eqoff}{0.5\eqoff}%
\addtolength{\eqoff}{-6.\unitlength}%
\raisebox{\eqoff}{
	\fmfframe(-.5,.5)(1,1){
		\begin{fmfchar*}(12,9)
		\fmfright{vop}
		\fmfforce{0 w, 1h}{v1}
		\fmfforce{0 w, 0h}{v3}
		\fmfforce{0.34 w,0.8 h}{vc1}
		\fmfforce{0.34 w,0.2 h}{vc3}
		\fmf{plain_rar}{vc1,v1}
		\fmf{plain_rar}{vc3,v3}
		\fmf{plain_rar,right=0.2,tension=0.6}{vop,vc1}
		\fmf{plain_rar,left=0.2,tension=0.6}{vop,vc3}
		\fmf{photon}{vc1,vc3}
		\fmffreeze
		\fmfposition
		\fmfiv{decor.shape=circle,decor.filled=full,decor.size=5}{vloc(__vop)}
		\fmfiv{decor.shape=hexacross,decor.size=12thin}{vloc(__vop)}
		\end{fmfchar*}
	}
}
\Bigr]_{\text{def}}
=2\complexi \lambda \xi \Kop\bigl[\bigl(\delta^{(1)}_{\mathcal{O}_2}\bigr)_{\text{def}}p^2\hat{I}_{(2,1)}(p)\bigr]
\eqncom\\
\settoheight{\eqoff}{$\times$}%
\setlength{\eqoff}{0.5\eqoff}%
\addtolength{\eqoff}{-6.0\unitlength}%
\raisebox{\eqoff}{%
	\fmfframe(1,-1.25)(2.5,1.25){%
		\begin{fmfchar*}(12,9)
		\fmfforce{(1w,0.5h)}{vo}
		\fmfforce{(0.8,1h)}{vl}
		\fmfforce{(0.8,0h)}{vr}
		\fmffreeze
		\fmfposition
		\fmfforce{(0.3w,0.5h)}{vc}
		\fmf{plain_rar,left=0.75}{vo,vc}
		\fmf{plain_rar,right=0.75}{vo,vc}
		\fmf{plain_rar}{vc,vl}
		\fmf{plain_rar}{vc,vr}
		\fmfiv{label=$\scriptstyle Q_{\text{F}}$,l.a=180,l.dist=5}{vloc(__vc)}
		\fmfiv{decor.shape=hexacross,decor.size=12thin}{vloc(__vo)}
		\fmfiv{decor.shape=circle,decor.filled=full,decor.size=5}{vloc(__vo)}
		\end{fmfchar*}
	}
}
&=\complexi \lambda Q_{\text{F}\,ii}^{ii}\Kop\bigl[\bigl(\bigl(\delta^{(1)}_{\mathcal{O}_2}\bigr)_{\text{def}} +\bigl(\delta^{(1)}_{\mathcal{O}_2}\bigr)_{\ol{\text{def}}}\, \bigr)\hat{I}_{(1,1)}(p)\bigr] 
\eqncom
\end{aligned}
\end{equation}
where the counterterms are taken from \eqref{cttwopointv} and in case of the vertex renormalisation $Q_{\text{F}}$ from \eqref{deltaQ}. Combining \eqref{eq:2l_wrapping} with \eqref{eq:O2_2L_insertions} and \eqref{eq:O2_1L_CT_insertions}, we find the divergent 1PI two-loop contributions to the composite operator $\cO_2$ that stem wrapping diagrams as well as prewrapping contributions involving the double-trace coupling $Q_{\text{F}}$. They can be absorbed into the deformation-dependent two-loop 1PI counterterm of $\cO_2$ as 
\begin{equation}\label{deltaZO2dt1PI}
\begin{aligned}
(\delta_{\mathcal{O}_2}^{(2)})_{\text{def}}
=-(C_{\mathcal{O}_2}^{(2)})_{\text{def}}
&= \frac{g^4}{\epsilon^2}\left(32\sin^2\gamma_i^+\sin^2\gamma_i^-(1-\epsilon)+(Q_{\text{F}\,ii}^{ii})^2
-2Q_{\text{F}\,ii}^{ii}(1+\xi)
\right)\eqncom
\end{aligned}
\end{equation}
where we expressed the result in terms of the effective planar coupling. Here, the divergence can be absorbed into a counterterm, since the non-local divergence of \eqref{eq:2l_wrapping} has cancelled against the non-local divergence that occurs in the diagram containing the one-loop counterterm $\delta_{Q_{\text{F}}}^{(1)}$. Hence, the double-trace coupling is necessary for the consistency of the theory, even in the \tHooft limit. Upon expanding the full renormalisation constant $\mathcal{Z}_{\mathcal{O}_2}$ as it is defined in \eqref{eq:comp_op_renormalisation} to second order in the effective planar coupling, we find the full deformation-dependent two-loop counterterm to be 
\begin{equation}
\begin{aligned}\label{eq:def_dependent_d_O2}
(\mathfrak{d}_{\mathcal{O}_2}^{(2)})_{\text{def}}&=\Bigl(\delta_{\mathcal{O}_2}^{(2)}+\delta_{\phi}^{(2)}+\delta_{\mathcal{O}_2}^{(1)}\delta_{\phi}^{(1)}+(\delta_{\phi}^{(1)})^2\Bigr)_{\text{def}}=
(\delta_{\mathcal{O}_2}^{(2)})_{\text{def}}+(\delta_{\mathcal{O}_2}^{(1)})_{\text{def}}\delta_{\phi}^{(1)}
\\
&=
g^4\left(32\sin^2\gamma_i^+\sin^2\gamma_i^-\left(\frac{1}{\epsilon^2}-\frac{1}{\epsilon}\right)+(Q_{\text{F}\,ii}^{ii})^2\frac{1}{\epsilon^2}\right)
\eqndot
\end{aligned}
\end{equation}
Since the connected counterterm of $\cO_2$ vanishes in the undeformed \NfSYMt, we have $\mathfrak{d}_{\mathcal{O}_2}^{(2)}=(\mathfrak{d}_{\mathcal{O}_2}^{(2)})_{\text{def}}$ and the connected two-loop renormalisation constant becomes 
\begin{equation}\label{eq:Z_total_O2}
\mathcal{Z}_{\mathcal{O}_2}=1+\mathfrak{d}_{\mathcal{O}_2}^{(1)}+\mathfrak{d}_{\mathcal{O}_2}^{(2)}
=1+g^2Q_{\text{F}\,ii}^{ii}\frac 1\epsilon+
g^4\left(32\sin^2\gamma_i^+\sin^2\gamma_i^-\left(\frac{1}{\epsilon^2}-\frac{1}{\epsilon}\right)+(Q_{\text{F}\,ii}^{ii})^2\frac{1}{\epsilon^2}\right)\eqndot
\end{equation}
Using \eqref{eq:RGE_comp_operators}, we find the two-loop anomalous dimension in the DR scheme from this equation to be 
\begin{equation}\label{gammaO2}
\begin{aligned}
\gamma_{\mathcal{O}_2}
&=-\frac{1}{\mathcal{Z}_{\mathcal{O}_2}}\Big(\mu\parderiv{\mu}
+\beta_{Q_{\text{F}\,ii}^{ii}}\parderiv{Q_{\text{F}\,ii}^{ii}}\Big)\mathcal{Z}_{\mathcal{O}_2}
=2g^2Q_{\text{F}\,ii}^{ii}-32(2g^2)^2\sin^2\gamma_i^+\sin^2\gamma_i^-
+\order{g^6}\eqndot
\end{aligned}
\end{equation}

The anomalous dimension \eqref{gammaO2} depends on the $\beta$-function of the running coupling $Q_{\text{F}\,ii}^{ii}$ and is sensitive to the chosen renormalisation scheme, which we chose to be the DR scheme in the present calculation. For $\text{DR}_\varrho$ schemes\footnote{Such schemes include e.g.\  the $\ol{\text{DR}}$ scheme discussed in \subsecref{subsec:The_renormalisation_procedure} where the parameter is $\varrho=c_{\ol{\text{MS}}}=\gammaE-\log 4\pi$.} which are related to the DR scheme via a rescaling of the regularisation parameter $\mu\rightarrow \mu_\rho=\mu \e^{-\frac{\varrho}{2}}$, we can derive the renormalisation-scheme dependence directly. The renormalised effective planar coupling in the $\text{DR}_\varrho$ scheme is related to the one in the DR scheme as
\begin{equation}\label{eq:grho}
g_\varrho=\e^{\frac{\epsilon}{2}\varrho}g\eqncom
\end{equation} 
which follows from the exact Yang-Mills coupling relation $g_{\text{B}}=\mu^\epsilon g=\mu_\varrho^\epsilon g_\varrho$. For the coupling $Q_{\text{F}\,ii}^{ii}$ we also use the unique relation to the bare coupling. To get $\mathcal{Z}^\varrho_{Q^{ii}_{\text{F}\,ii}}$, we replace $g$ by $g_\varrho$ in all counterterms that contribute to its renormalisation \eqref{eq:full_ct_QF}. Upon reexpressing the result in terms of the couplings $g$ and $Q^{ii}_{\text{F}\,ii}$, we find\footnote{Note that the corresponding one-loop $\beta$-function from \eqref{eq:beta_function_QF} is independent of this scheme change. For the renormalisation-scheme independence of the one-loop $\beta$-function see e.g.\ \cite[\chap{7}]{Collins:1984xc}.}
\begin{equation}
\begin{aligned}\label{Qredef}
Q^{\varrho\,ii}_{\text{F}\,ii}=(\mathcal{Z}^\varrho_{Q^{ii}_{\text{F}\,ii}})^{-1}\mathcal{Z}_{Q^{ii}_{\text{F}\,ii}}Q^{ii}_{\text{F}\,ii}
=Q^{ii}_{\text{F}\,ii}-\frac{\varrho}{2}\beta_{Q_{\text{F}\,ii}^{ii}}+\mathcal{O}(g^4)
\eqncom
\end{aligned}
\end{equation}
where we dropped terms that vanish in the limit $\epsilon\to0$. The complete renormalisation constant $\mathcal{Z}_{\mathcal{O}_2}^\varrho$ in the $\text{DR}_\varrho$ scheme is then obtained by replacing all couplings in \eqref{eq:Z_total_O2} by the $\varrho$-dependent ones defined in \eqref{eq:grho} and \eqref{Qredef}. In this manner, we express the renormalisation constant in the scheme $\text{DR}_\varrho$ in terms of the quantities defined in the original DR scheme and hence we can calculate the anomalous dimension in the altered scheme using \eqref{gammaO2} as
\begin{equation}\label{gammaO2_rho}
\begin{aligned}
\gamma^\varrho_{\mathcal{O}_2}
&=\frac{1}{\mathcal{Z}^\varrho_{\mathcal{O}_2}}\Big(\mu\parderiv{\mu}
+\beta_{Q_{\text{F}\,ii}^{ii}}\parderiv{Q_{\text{F}\,ii}^{ii}}\Big)\mathcal{Z}^\varrho_{\mathcal{O}_2}
=\gamma_{\mathcal{O}_2}
-g^2\varrho\beta_{Q_{\text{F}\,ii}^{ii}}
+\order{g^6}\eqndot
\end{aligned}
\end{equation}
The respective one-loop anomalous dimension is renormalisation scheme-independent, since the $\varrho$-dependent term arises from the second derivative in \eqref{gammaO2_rho}, which enhances the power in $g$ through the multiplication with the $\beta$-function, see \cite[\chap{7}]{Collins:1984xc} for a general analysis.

\section{The complete one-loop dilatation operator of the planar \texorpdfstring{$\beta$}{beta}-deformation}\label{sec:beta_paper}
In this section, we determine the prewrapping and wrapping induced finite-size corrections to the asymptotic planar one-loop dilatation operator of the $\beta$-deformation given in \eqref{eq: deformation of D_2}. In doing so, we construct the complete planar one-loop dilatation operator for the conformal $\beta$-deformation with gauge group \SUN and of its non-conformal cousin with gauge group \UN. This section is based on \cite{Fokken:2013mza}.

In \secref{sec:The_deformations}, we introduced the $\beta$-deformation, presented its single-trace action in \eqref{eq:deformed_action_complex_scalars2} and for gauge group \SUN we also gave the double-trace part of the action in \eqref{eq:2_trace_action} in terms of elementary fields. The $\beta$-deformation with gauge group \SUN is an exactly marginal deformation with a single supersymmetry and it is part of the deformations classified in \cite{Leigh:1995ep}. In \cite{Mauri:2005pa}, it was shown for this theory that the coupling's one-loop finiteness constraint extends to higher loop orders and hence this theory is exactly superconformally invariant. The derivation employs the $\cN=1$ superspace formulation of the $\beta$-deformation which reads
\begin{equation}
\label{eq: superspace action}
\begin{aligned}
S &= \frac{1}{ 2 g^2_\YM}\int \! \de^4 x  \de^2 \theta \tr\left( W^\alpha W_\alpha\right) + \int \! \de^4 x \de^4 \theta \tr \big( \e^{- g_\YM V} \bar{\Phi}_i \e^{g_\YM V}\Phi^i\big) \\
& \phaneq {}+{} i g_\YM \int \! \de^4 x  \de^2 \theta \tr \big(\Phi_1\Phi_2\Phi_3 \e^{-i  \frac{\beta}{2}} 
{}-{} \Phi_1\Phi_3\Phi_2 \e^{i \frac{\beta}{2}} \big) + \text{h.c.} \eqncom
\end{aligned}
\end{equation}
with the vector superfield $V$, superfield strength $W_\alpha$ and the (anti-)chiral matter superfields $\ol{\Phi}_i$ and $\Phi^i$, respectively. For an introduction to the superfield formalism see \cite[\chap{2}]{GGRS83} and for a mapping between the super- and component-field formalism \cite{Siegnotes}. For gauge group \UN, the $\beta$-deformation is not conformally invariant, as was shown in \cite{Hollowood:2004ek} by the identification of a running double-trace coupling. In the IR regime, this running coupling assumes the non-vanishing IR fix-point value of the interaction given in \eqref{eq:2_trace_action} which belongs to the \SUN $\beta$-deformation. Hence, the \UN theory flows to the \SUN theory in the IR regime. Since the dilatation operator measures the anomalous dimensions of composite operators, its eigenvalues are only observables in theories with conformal symmetry. However, the effects of renormalisation-scheme dependence occur for loop orders $K\geq2$ and therefore we can still calculate renormalisation-scheme independent one-loop anomalous dimensions in non-conformal theories. In the previous section in \eqref{gammaO2_rho}, we saw an explicit example of this in the $\gamma_i$-deformation and in this section use this fact to calculate one-loop anomalous dimensions for the $\beta$-deformation with gauge group \UN.

In the integrability-based description the eigenvalues of the asymptotic dilatation operator can be constructed from the asymptotic Bethe ansatz for a given spin-chain state. In \cite{Beisert:2005if}, an appropriate asymptotic Bethe ansatz for the $\beta$-deformation was constructed from the corresponding ansatz in the undeformed \NfSYMt by introducing twists in the boundary conditions. Subsequently, also wrapping corrections in the $\beta$-deformation were discussed for the single-impurity states, which differ from the length-$L$ integrability vacuum by having a single excitation, e.g.\ $\tr\bigl(\phi_i^{L-1}\phi_j\bigr)$ with $i\neq j$. In the absence of prewrapping contributions (i.e.\ for $L\geq 3$ at one-loop level\footnote{See \cite{Arutyunov:2010gu,Bajnok:2010ud,Kazakov:2015efa} for higher-order results.}) the field-theory results of \cite{Fiamberti:2008sn} were reproduced in \cite{Gunnesson:2009nn} for $\beta=\frac{1}{2}$ and in \cite{Gromov:2010dy,Arutyunov:2010gu} for generic $\beta$. Since the gauge group sensitivity of the $\beta$-deformation is related to prewrapping contributions, these results are the same for gauge group \SUN and \UN. For the $L=2$ single-impurity state $\tr\bigl(\phi_i\phi_j\bigr)$, prewrapping contributions occur and with them a sensitivity to the gauge group \cite{FG05}: the anomalous dimension of this state vanishes for gauge group \SUN, while it is non-zero in the \UN case. The \UN anomalous dimension of this state is reproduced by the asymptotic dilatation operator constructed from \eqref{eq: deformation of D_2} and the corresponding asymptotic Bethe ansatz from \cite{Beisert:2005if}, as was noted in \cite{Frolov:2005iq}. However, the anomalous dimension for the prewrapping-dependent \SUN theory\footnote{This is the relevant theory for the \AdSCFTc, see \cite{Frolov:2005iq} for comments on the gauge group choice in the deformed \AdSCFTc.} cannot be reproduced. At two-loop level, the issue becomes even more prominent: in field-theoretic calculations, the two-loop anomalous dimension of $\tr\bigl(\phi_i\phi_j\bigr)$ vanishes in the \SUN theory \cite{Penati:2005hp}, while the corresponding integrability-based result diverges\footnote{In \cite{Frolov:2009in}, a similar divergence was encountered in the expressions for the integrability-vacuum state energy. If one is only interested in the vacuum state, the divergence can be regulated by introducing a twist in the \AdS{5} directions \cite{deLeeuw:2012hp}. This procedure is applicable for the vacuum state of the $\gamma_i$- and the $\beta$-deformation \cite{FrolovPC}.}  \cite{Arutyunov:2010gu,Kazakov:2015efa}.

In this section, we address the issues mentioned in the previous paragraph by analysing which states can be affected by prewrapping in the \tHooft limit. We generalise the Feynman diagram relation \eqref{diagrel} to be also applicable when states with vanishing \su{4} Cartan-charge components $Q^1=Q^2=0$ in the notation of \tabref{tab: su(4) charges} occur in the planar interaction kernel. An important consequence of this generalisation is that the anomalous dimensions and structure constants of $(Q^1,Q^2)$-neutral states are independent of the deformation parameter $\beta$ at all loop orders. For generic states, which cannot be included in \eqref{diagrel}, we show how prewrapping contributions can be incorporated into the definition of the planar asymptotic one-loop dilatation operator density \eqref{eq: deformation of D_2}. In the case of gauge group \SUN we find the complete version to be 
\begin{equation}\label{eq:density_intro}
(\diladensity_{2}^\beta)_{A_iA_j}^{A_kA_l}= \e^{\frac{\complexi}{2} (\mathbf{q}_{A_k} \wedge \mathbf{q}_{A_l}- \mathbf{q}_{A_i} \wedge \mathbf{q}_{A_j})} \rule[-0.96cm]{0.1mm}{1.415cm}\!{\phantom{|}}_{\substack{\\[0.2cm]
		\beta=0\text{ if }L=2\text{ and \phantom{...............}}\\ 
		(A_i,A_j,A_k,A_l\in\cA_\text{matter}\text{ or \phantom{)}}\\  
		\phantom{{}({}} A_i,A_j,A_k,A_l\in\ol{\cA}_\text{matter}) \phantom{\text{ or }}}}  (\diladensity_{2}^{\cN=4})_{A_iA_j}^{A_kA_l}
\eqncom
\end{equation}
where the fields $A_i$ are taken from the alphabet defined in \eqref{eq: alphabet} and the subalphabets of (anti-)chiral fields are given by
\begin{equation}
\begin{aligned}\label{eq:subalphabet_intro}
\cA_\text{matter}&=\{ \D^k\phi_i,\D^k\lambda_{i\alpha}\} \eqncom \qquad
\ol{\cA}_\text{matter}&=\{ \D^k\ol{\phi}^i, \D^k\ol{\lambda}^i_{\dot\alpha}\}\eqncom\qquad \forall i\in\{1,2,3\}
\eqndot
\end{aligned} 
\end{equation}
Our one-loop inclusion of prewrapping into the dilatation operator density is possible, since we can identify prewrapping affected contributions purely from the input fields of the density \eqref{eq:density_intro}. It is, however, not likely that this identification is still possible at higher loop orders. For gauge group \UN the one-loop finite-size corrections depend on the multi-trace couplings of the chosen theory displayed in \subsecref{sec:multi-trace-parts-of-the-action} and hence also the one-loop anomalous dimensions of $L=2$ and $L=1$ states depend on these couplings. If we choose a \UN $\beta$-deformation in which the tree-level values of all multi-trace couplings vanish, the asymptotic one loop dilatation operator density \eqref{eq: deformation of D_2} is correct for all states with length $L>1$. For the $L=1$ states the anomalous dimensions defining $g^2(\mathfrak{D}^\beta_{2,1})_i^f=\delta_i^f\bra{i}g^2\mathfrak{D}^\beta_{2,1}\ket{i}$ are
\begin{equation}
\begin{aligned}
\gamma_{\tr(\D^k\lambda_4)}^{(1)}= 
\gamma_{\tr(\D^k\cF)}^{(1)}= 
0\eqncom\qquad
\gamma_{\tr(\D^k\phi_i)}^{(1)}= 
\gamma_{\tr(\D^k\lambda_i)}^{(1)}= 
8g^2\sin^2\frac{\beta}{2}
\eqncom\qquad \forall i\in \{1,2,3\}\eqncom
\end{aligned}
\end{equation}
where the same equations hold for the conjugate elementary fields.

\subsection{\texorpdfstring{$(Q^1,Q^2)$}{(Q1,Q2)}-neutral states}\label{sec:Q1Q2_neutral_states}
The planar asymptotic one-loop dilatation operator density of the $\beta$-deformation was constructed using relation \eqref{diagrel} in \secref{sec:the-quantum-dilatation-operator-on-composite-operators}. Originally, this relation is a theorem from spacetime-noncommutative field theory which expresses deformed planar Feynman diagrams in terms of their undeformed counterparts times a phase factor which is determined by the order of the incoming momenta \cite{Filk96}. In this original setup, the planar interactions in \eqref{diagrel} may consist not only of elementary interactions but also of external states and the latter ones can be incorporated as additional interaction vertices at which momentum conservation holds. Hence, the deformation via a Moyal $\ast$-product is well defined also for external states, since the states phase factor is invariant under a cyclic relabelling of the external momenta.

In contrast to the above scenario, the noncommutative $\ast$-product in the $\beta$-deformation is distinctly different, since the ordering principle does not dictate the phase factor any more. In the $\beta$-deformation, fields are cyclically ordered by their colour arrangement, while the phase factors are determined by flavour via the two Cartan charges $(Q^1,Q^2)$ and therefore a single colour ordering may be accompanied by different phases, e.g. 
\begin{equation}
\tr\bigl(\phi_{j}\ast\phi_{k}\bigr)=\e^{-\frac{\complexi}{2}\beta\varepsilon_{ab3}Q^a_{\phi_j}Q^b_{\phi_k}}\tr\bigl(\phi_{j}\phi_{k}\bigr)
\qquad\text{and}\qquad
\tr\bigl(\phi_{k}\ast\phi_{j}\bigr)=\e^{\frac{\complexi}{2}\beta\varepsilon_{ab3}Q^a_{\phi_j}Q^b_{\phi_k}}\tr\bigl(\phi_{j}\phi_{k}\bigr)\eqncom
\end{equation}
see \subsecref{sec:deformation} for details. We postpone the discussion of $(Q^1,Q^2)$-charged traces to the following subsection and focus on traces with vanishing $Q^1=0=Q^2$ charge. We showed in \eqref{eq:star_product_trace_LN} for such traces that their cyclicity remains intact and the colour ordering principle also dictates the phase-dependence, like in the case of the spacetime-noncommutative field theory. Therefore, in the $\beta$-deformation, relation \eqref{diagrel} is valid for planar interactions composed of the charge-neutral single-trace interactions displayed in the action \eqref{eq:deformed_action_complex_scalars2}. In addition, it remains valid if we include external states whose single-trace factors individually have vanishing total $Q^1=Q^2=0$ charge. In this case the single-trace interactions as well as the states that enter the interaction kernel on the \lhs of \eqref{diagrel} are $\ast$-deformed.

The generalisation of relation \eqref{diagrel} to also include $(Q^1,Q^2)$-charge-neutral states in the interaction kernel has far reaching consequences. First, diagrams which only become planar when such states are included also fulfil \eqref{diagrel}, despite the fact that their subdiagrams of elementary interactions are non-planar. Diagrams in this class are for example the wrapping diagrams displayed in \eqref{eq:general_wrapping} when the two external states fulfil $Q^1=Q^2=0$. If we cut out the two external states, the remaining non-planar interaction kernel can be written as a double-trace diagram, see \cite{Sieg:2005kd}. Since we can apply relation \eqref{diagrel} in this special case, we find the double-trace relation
\begin{equation}\label{eq: relation for double-trace diagrams}
\begin{aligned}
\ifpdf
\settoheight{\eqoff}{$+$}%
\setlength{\eqoff}{0.5\eqoff}%
\addtolength{\eqoff}{-7\unit}%
\raisebox{\eqoff}{%
	\includegraphics[angle={0},scale=0.12,trim=0cm 0cm 0cm 0]{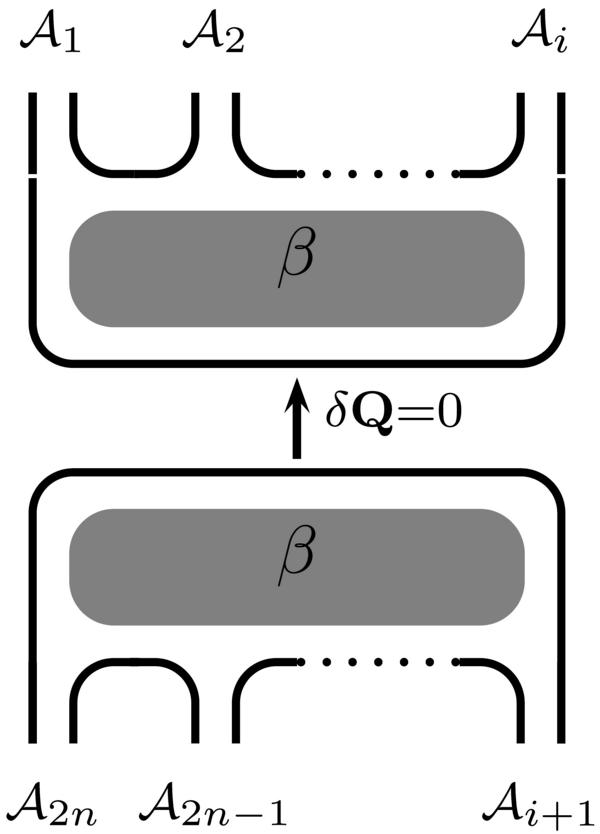}
}
\else
\settoheight{\eqoff}{$+$}%
\setlength{\eqoff}{0.5\eqoff}%
\addtolength{\eqoff}{-9\unit}%
\raisebox{\eqoff}{%
	\begin{pspicture}(-2,1)(11,-17)
	\rput(4.5,-8){%
		\rotatebox{90}{%
			\begin{pspicture}(-1,-2)(17,11)
			\uinex{2}{9}
			\iinex{2}{6}
			\dinex{2}{0}
			\setlength{\ya}{9\unit}
			\addtolength{\ya}{0.5\dlinewidth}
			\setlength{\yb}{0\unit}
			\addtolength{\yb}{-0.5\dlinewidth}
			\setlength{\xc}{7.0\unit}
			\setlength{\yc}{4.5\unit}
			\addtolength{\yc}{-0.5\dlinewidth}
			\setlength{\xd}{9.0\unit}
			\setlength{\yd}{4.5\unit}
			\addtolength{\yd}{0.5\dlinewidth}
			\psline[linearc=\linearc](3.5,\ya)(\xc,\ya)(\xc,\yb)(3.5,\yb)
			\psline[linestyle=dotted](3.5,4.5)(3.5,1.5)
			\psline[linearc=\linearc](12.5,\ya)(\xd,\ya)(\xd,\yb)(12.5,\yb)
			\doutex{14}{0}
			\ioutex{14}{6}
			\uoutex{14}{9}
			\psline[linestyle=dotted](12.5,4.5)(12.5,1.5)
			\setlength{\xa}{3.5\unit}
			\addtolength{\xa}{\dlinewidth}
			\setlength{\xb}{7.0\unit}
			\addtolength{\xb}{-\dlinewidth}
			\setlength{\ya}{9\unit}
			\addtolength{\ya}{-0.5\dlinewidth}
			\setlength{\yb}{0\unit}
			\addtolength{\yb}{0.5\dlinewidth}
			\pscustom[linecolor=gray,fillstyle=solid,fillcolor=gray,linearc=\linearc]{%
				\psline(\xa,4.5)(\xa,\ya)(\xb,\ya)(\xb,4.5)
				\psline[liftpen=2](\xb,4.5)(\xb,\yb)(\xa,\yb)(\xa,4.5)
			}
			\setlength{\xa}{12.5\unit}
			\addtolength{\xa}{-\dlinewidth}
			\setlength{\xb}{9.0\unit}
			\addtolength{\xb}{\dlinewidth}
			\setlength{\ya}{9\unit}
			\addtolength{\ya}{-0.5\dlinewidth}
			\setlength{\yb}{0\unit}
			\addtolength{\yb}{0.5\dlinewidth}
			\pscustom[linecolor=gray,fillstyle=solid,fillcolor=gray,linearc=\linearc]{%
				\psline(\xa,4.5)(\xa,\ya)(\xb,\ya)(\xb,4.5)
				\psline[liftpen=2](\xb,4.5)(\xb,\yb)(\xa,\yb)(\xa,4.5)
			}
			\pnode(7.25,4.5){left}
			\pnode(8.75,4.5){right}
			\ncline[arrows=->,linecolor=black]{left}{right}
			\end{pspicture}
		}
		\rput(-2,1.25){%
			\rput[b](-9,14.5){$\scriptstyle\mathcal{A}_{1}$}
			\rput[b](-6,14.5){$\scriptstyle\mathcal{A}_{2}$}
			\rput[b](0,14.5){$\scriptstyle\mathcal{A}_{i}$}
			\rput[t](0,1){$\scriptstyle\mathcal{A}_{i+1}$}
			\rput[t](-6,1){$\scriptstyle\mathcal{A}_{2n-1}$}
			\rput[t](-9,1){$\scriptstyle\mathcal{A}_{2n}$}}
	}
	\rput(4.5,-5.25){$\beta$}
	\rput(4.5,-10.75){$\beta$}
	\rput[l](5.0,-8.125){$\scriptstyle \delta \mathbf{Q} =0$}
	\end{pspicture}
}
\fi
\,
&=\,
\ifpdf
\settoheight{\eqoff}{$+$}%
\setlength{\eqoff}{0.5\eqoff}%
\addtolength{\eqoff}{-7\unit}%
\raisebox{\eqoff}{%
	\includegraphics[angle={0},scale=0.12,trim=0cm 0cm 0cm 0]{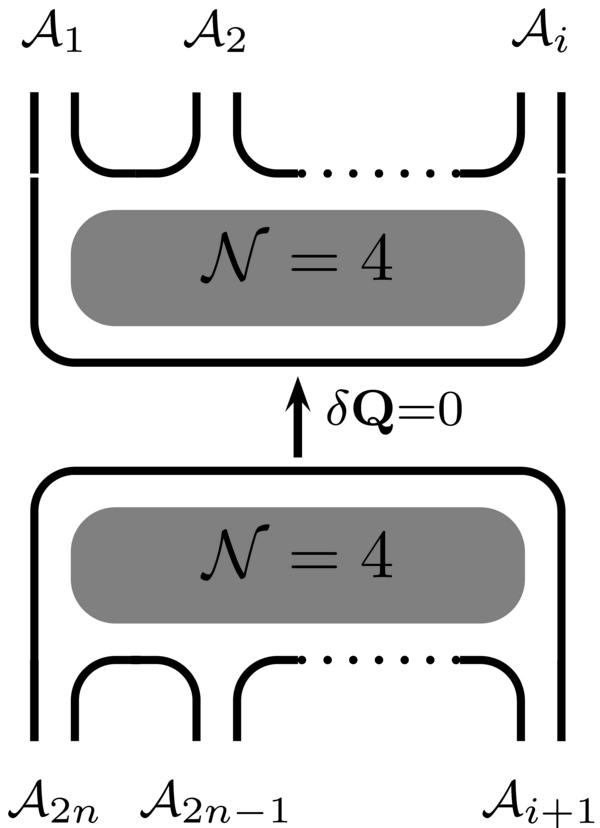}
}
\else
\settoheight{\eqoff}{$+$}%
\setlength{\eqoff}{0.5\eqoff}%
\addtolength{\eqoff}{-9\unit}%
\raisebox{\eqoff}{%
	\begin{pspicture}(-2,1)(11,-17)
	\rput(4.5,-8){%
		\rotatebox{90}{%
			\begin{pspicture}(-1,-2)(17,11)
			\uinex{2}{9}
			\iinex{2}{6}
			\dinex{2}{0}
			\setlength{\ya}{9\unit}
			\addtolength{\ya}{0.5\dlinewidth}
			\setlength{\yb}{0\unit}
			\addtolength{\yb}{-0.5\dlinewidth}
			\setlength{\xc}{7.0\unit}
			\setlength{\yc}{4.5\unit}
			\addtolength{\yc}{-0.5\dlinewidth}
			\setlength{\xd}{9.0\unit}
			\setlength{\yd}{4.5\unit}
			\addtolength{\yd}{0.5\dlinewidth}
			\psline[linearc=\linearc](3.5,\ya)(\xc,\ya)(\xc,\yb)(3.5,\yb)
			\psline[linestyle=dotted](3.5,4.5)(3.5,1.5)
			\psline[linearc=\linearc](12.5,\ya)(\xd,\ya)(\xd,\yb)(12.5,\yb)
			\doutex{14}{0}
			\ioutex{14}{6}
			\uoutex{14}{9}
			\psline[linestyle=dotted](12.5,4.5)(12.5,1.5)
			\setlength{\xa}{3.5\unit}
			\addtolength{\xa}{\dlinewidth}
			\setlength{\xb}{7.0\unit}
			\addtolength{\xb}{-\dlinewidth}
			\setlength{\ya}{9\unit}
			\addtolength{\ya}{-0.5\dlinewidth}
			\setlength{\yb}{0\unit}
			\addtolength{\yb}{0.5\dlinewidth}
			\pscustom[linecolor=gray,fillstyle=solid,fillcolor=gray,linearc=\linearc]{%
				\psline(\xa,4.5)(\xa,\ya)(\xb,\ya)(\xb,4.5)
				\psline[liftpen=2](\xb,4.5)(\xb,\yb)(\xa,\yb)(\xa,4.5)
			}
			\setlength{\xa}{12.5\unit}
			\addtolength{\xa}{-\dlinewidth}
			\setlength{\xb}{9.0\unit}
			\addtolength{\xb}{\dlinewidth}
			\setlength{\ya}{9\unit}
			\addtolength{\ya}{-0.5\dlinewidth}
			\setlength{\yb}{0\unit}
			\addtolength{\yb}{0.5\dlinewidth}
			\pscustom[linecolor=gray,fillstyle=solid,fillcolor=gray,linearc=\linearc]{%
				\psline(\xa,4.5)(\xa,\ya)(\xb,\ya)(\xb,4.5)
				\psline[liftpen=2](\xb,4.5)(\xb,\yb)(\xa,\yb)(\xa,4.5)
			}
			\pnode(7.25,4.5){left}
			\pnode(8.75,4.5){right}
			\ncline[arrows=->,linecolor=black]{left}{right}
			\end{pspicture}
		}
		\rput(-2,1.25){%
			\rput[b](-9,14.5){$\scriptstyle\mathcal{A}_{1}$}
			\rput[b](-6,14.5){$\scriptstyle\mathcal{A}_{2}$}
			\rput[b](0,14.5){$\scriptstyle\mathcal{A}_{i}$}
			\rput[t](0,1){$\scriptstyle\mathcal{A}_{i+1}$}
			\rput[t](-6,1){$\scriptstyle\mathcal{A}_{2n-1}$}
			\rput[t](-9,1){$\scriptstyle\mathcal{A}_{2n}$}}
	}
	\rput(4.5,-5.25){$\cN=4$}
	\rput(4.5,-10.75){$\cN=4$}
	\rput[l](5.0,-8.125){$\scriptstyle \delta \mathbf{Q} =0$}
	\end{pspicture}
}
\fi
\times
\,\,\,
\underbrace{\Phi\left(
	\mathcal{A}_{1}{}
	\ast{}\dots{}\ast{}\mathcal{A}_{i}
	\right)
	\Phi\left(
	\mathcal{A}_{i+1}{}
	\ast{}\dots{}\ast{}\mathcal{A}_{2n}
	\right)\rule[-0.25cm]{0pt}{0.5cm}}_{\rule[0pt]{0pt}{0.5cm}\displaystyle  \Phi\left(
	\mathcal{A}_{1}{}
	\ast{}\dots{}\ast{}\mathcal{A}_{i}{}\ast{}
	\mathcal{A}_{i+1}{}
	\ast{}\dots{}\ast{}\mathcal{A}_{2n}
	\right)}\eqncom
\end{aligned}
\end{equation}
where $0=\delta \mathbf{Q}=(\delta Q^1,\delta Q^2)$ denotes the vanishing charge flow between the separate traces and the grey-shaded regions represent arbitrary planar interactions. Second, any gauge-invariant correlation function of such states in the \tHooft limit is independent of the deformation parameter $\beta$, which follows from evaluating \eqref{diagrel} with zero external legs. Therefore, the anomalous dimensions and structure constants of states with total $Q^1=Q^2=0$ charge are independent of $\beta$ and in particular given by their \NfSYM counterparts. Immediate examples of such states are the Konishi state $N^{-1}\tr\bigl(\phi_i\ol{\phi}^i\bigr)$ and the chiral primary state
\begin{equation}
\mathcal{O}_j=N^{-\frac{3j}{2}}\tr\bigl(\phi_1^j\phi_2^j\phi_3^j\bigr)_{\ast}+\text{all permutations}\eqncom
\end{equation}
where the $\ast$ indicates that the state is non-trivially deformed. For the latter state the three-point functions $\langle \cO_j \cO_{j^\prime} \cO_{j^{\prime\prime}} \rangle$ were analysed at one-loop order in the planar gauge theory and at strong coupling in the Lunin-Maldacena background in \cite{David:2013oha}. They were found to be independent of $\beta$ and by the above arguments we can conclude that they are in fact independent of $\beta$ at arbitrary loop orders.\footnote{In \cite{Frolov:2005iq}, earlier arguments from \cite{Berenstein:2000ux,Berenstein:2000hy} were generalised for rational $\beta$ to show that the anomalous dimensions of the operator $\mathcal{O}_j$ vanishes.}

\subsection{\texorpdfstring{$(Q^1,Q^2)$}{(Q1,Q2)}-charged states}
Regardless of the findings in the previous subsection, relation \eqref{diagrel} cannot be generalised to include states with $(Q^1,Q^2)$-charged single-trace factors. For the planar one-loop dilatation operator in the asymptotic regime (for state lengths $L\geq 3$) this is not an issue, since neither non-planar wrapping nor multi-trace prewrapping contributions can occur, as was discussed in \secref{sec:the-quantum-dilatation-operator-on-composite-operators}. The absence of these effects was used to construct the asymptotic dilatation operator density \eqref{eq: deformation of D_2} as a deformed planar interaction kernel via \eqref{diagrel} in \secref{sec:the-quantum-dilatation-operator-on-composite-operators}. For the complete dilatation operator density, there are, however, additional prewrapping contributions that cannot be captured by \eqref{diagrel} and which can occur in length-preserving two-point functions of length-$L$ states at loop order $K=L-1$. 

From the general definition of prewrapping given in \subsecref{sec:finite-size-effects}, we can classify which states may receive prewrapping contributions in planar two-point correlation functions. In the \SUN $\beta$-deformation, prewrapping can occur when the state is fused either into one of the trace factors in the double-trace interaction of \eqref{eq:2_trace_action}, or into a single field line. Hence, the total \su{4} Cartan charge of the state must either be equal to $\pm(\mathbf{q}_{\phi_i}+\mathbf{q}_{\phi_j})$, or to one of the elementary fields with non-vanishing $(Q^1,Q^2)$-charge, \cf \tabref{tab: su(4) charges}. Applying this idea to the compact subsectors of the $\beta$-deformation, which were classified in \cite{Beisert:2003jj} for the undeformed theory, we find the potentially prewrapping affected subsectors given in \tabref{tab: prewrapping in compact subsectors}.
\begin{table}[htbp]
	\centering
	\caption{Prewrapping candidates in compact subsectors, where we omitted subsectors that are related to the above via the $\ZZ_3$ symmetry and/or charge conjugation. The spinor indices of fermions appear at the upper position in this table.}
	\label{tab: prewrapping in compact subsectors}
	\begin{tabular}{|c|l|l|}
		\hline
		Subsector & Fields & Prewrapping candidates \\ \hline
		$\SU2$ & $\phi_1,\phi_2$ & $\tr(\phi_1\phi_2)$ \\
		$\SU2$ & $\phi_1,\ol{\phi}_2$ & none \\ \hline
		$\U{1|1}$ & $\phi_1,\lambda_{4}^1$ & none \\
		$\U{1|1}$ & $\phi_1,\lambda_{1}^1$ & none \\
		$\U{1|1}$ & $\ol{\phi}^1,\lambda_{2}^1$ & none \\
		$\U{1|1}$ & $\ol{\phi}^1,\lambda_3^1$ & none \\ \hline
		$\U{1|2}$ & $\phi_1,\phi_2,\lambda_{4}^1$ & $\tr(\phi_1\phi_2)$ \\
		$\U{1|2}$ & $\ol{\phi}^2,\ol{\phi}^3,\lambda_{1}^1$ & $\tr(\ol{\phi}^2\ol{\phi}^3)$ \\
		$\U{1|2}$ & $\phi_1,\ol{\phi}^2,\lambda_{1}^1$ & none\\
		$\U{1|2}$ & $\phi_1,\ol{\phi}^3,\lambda_{1}^1$ & none \\ \hline
		$\U{1|3}$ & $\phi_1,\phi_2,\phi_3,\lambda_{4}^1$ & $\tr(\phi_1\phi_2)+\ZZ_3$ \\
		$\U{1|3}$ & $\phi_1,\ol{\phi}^2,\ol{\phi}^3,\lambda_{1}^1$ & $\tr(\ol{\phi}^2\ol{\phi}^3)$ \\ \hline
		$\SU{2|3}$ & $\phi_1,\phi_2,\phi_3,\lambda_{4}^1,\lambda_{4}^2$ & $\tr(\phi_1\phi_2)+\ZZ_3$ \\
		$\SU{2|3}$ & $\phi_1,\ol{\phi}^2,\ol{\phi}^3,\lambda_{1}^1,\lambda_{1}^2$ & $\tr(\ol{\phi}^2\ol{\phi}^3)$\\
		\hline 
	\end{tabular}
\end{table}
We see that all prewrapping candidates can be obtained from the $L=2$ single-impurity state\footnote{Note that the gauge group sensitivity of the state $\tr(\phi_2\phi_3)$ was already observed in \cite{FG05}.} $\tr(\phi_2\phi_3)$ with the charge conjugation operator and/or an operator realising the $\ZZ_3$ symmetry. The latter symmetry cyclically relabels the flavour indices of chiral scalars and fermions, as well as their conjugates. This relabelling leaves the action invariant, which can most easily be verified in the $\cN=1$ superspace formulation \eqref{eq: superspace action}. 

The noncompact subsectors can be classified analogously. In general, we find that in subsectors with restrictions on the flavour content no combination of $(Q^1,Q^2)$-charged fields exists whose total $(Q^1,Q^2)$-charge vanishes. Therefore, the criteria for prewrapping affected states mentioned in the beginning of this subsection can only be fulfilled by a finite number of field combinations in these subsectors. In the full theory, where the flavour content is unrestricted, this is of course not true and a large variety of prewrapping candidates exist, e.g.
\begin{equation}\label{eq: candidate}
\tr\Bigl(\phi_2\phi_3(\phi_1\ol{\phi}^1)^i(\phi_2\ol{\phi}^2)^j(\phi_3\ol{\phi}^3)^k(\lambda_1\ol{\lambda}^1)^l(\lambda_2\ol{\lambda}^2)^m(\lambda_3\ol{\lambda}^3)^n(\lambda_4\ol{\lambda}^4)^o\cF^p\ol{\cF}^q\Bigr)\eqncom
\end{equation}
with $i,j,k,l,m,n,o,p,q\in \NN_0$ and suppressed spinor indices and covariant derivatives. The structure of all occurring spinor indices has to be chosen such that all fields can in fact be fused into a single chiral field or the double-trace coupling \eqref{eq:2_trace_action}.

\subsection{The \texorpdfstring{\SUN}{SU(N)} dilatation operator}
In the previous two subsections, we classified which states are potentially affected by prewrapping in the \SUN $\beta$-deformation, depending on the state's overall $(Q^1,Q^2)$-charge. For the dilatation operator at loop order $K=1$, prewrapping can only occur for states with length $L=K+1=2$, which follows from the discussion in \subsecref{sec:finite-size-effects}. Wrapping contributions cannot occur at all, since these would require a state with length $L=K=1$, which is absent in the \SUN theory. Therefore, for the complete one-loop planar dilatation operator of the $\beta$-deformation with gauge group \SUN, we only need to determine the one-loop prewrapping contributions to all states.

Before analysing all possible states, we discuss prewrapping for the operator $\mathcal{O}=\tr(\lambda_1^\alpha\phi_2)$. We analyse how the asymptotic dilatation operator density is constructed from the UV renormalisation constant contributions in the undeformed theory and afterwards in the $\beta$-deformation. According to \eqref{eq:operator_as_spinchain}, the operator $\mathcal{O}$ is mapped to a cyclic spin-chain state as
\begin{equation}
\mathcal{O}\hat{=}
\frac{1}{N}\mathcal{P}_L\ket{\lambda_1^\alpha\phi_2}=\frac{1}{2N}\bigl(\ket{\lambda_1^\alpha\phi_2}+\ket{\phi_2\lambda_1^\alpha}\bigr)\eqncom
\end{equation}
where $\ket{\cdot}$ is a non-cyclic spin-chain state. The one-loop dilatation operator, given in \eqref{eq:Dila_N4_one_loop}, acts on a length $L=2$ state via the densities\footnote{Our Yang-Mills coupling $\gym$ is related to the one in \cite{Beisert:2003jj} via $\gym=2^{-\frac 12}\hat{g}_{\YM}$, hence the factor of two. The factor of $\complexi$ stems from our definition of the dilatation operator (or generator) in \eqref{eq:unitary_trafo_fields_2}.}
\begin{equation}
2\complexi g^2[D_{2},\mathcal{O}] =4 g^2\sum_{\cF_j\in \cA}
\Bigl(
(\mathfrak{D}_{2})^{\cF_1\cF_2}_{\lambda_1\phi_2}
+(\mathfrak{D}_{2})^{\cF_1\cF_2}_{\phi_2\lambda_1}
\Bigr)
\frac{1}{2N}\ket{\cF_1\cF_2}
\eqndot
\end{equation}
In the undeformed theory in the \tHooft limit, we obtain its elements from the planar part of the asymptotic expression  $(\diladensity_2^{\cN=4})_{\text{in}}^{\text{out}}=\bra{\text{out}}D^{\mathcal{N}=4}_{2}\ket{\text{in}}$, given in \eqref{eq: harminic action}. For the relevant non-cyclic states, it reads
\begin{equation}\label{eq: contributions}
\begin{aligned}
(\diladensity_2^{\cN=4})_{\lambda_1\phi_2}^{\lambda_1\phi_2}= +3\eqncom \quad
(\diladensity_2^{\cN=4})_{\lambda_1\phi_2}^{\phi_2\lambda_1}= -1\eqncom \quad
(\diladensity_2^{\cN=4})_{\phi_2\lambda_1}^{\phi_2\lambda_1}= +3\eqncom \quad
(\diladensity_2^{\cN=4})_{\phi_2\lambda_1}^{\lambda_1\phi_2}= -1 \eqndot
\end{aligned}
\end{equation}
In terms of Feynman diagrams, the density contributions arise from the counterterms of the following diagrams
\setlength{\fboxrule}{0pt} 
\begin{equation}
\begin{aligned}\label{eq:L2_state}
4g^2(\diladensity_2^{\cN=4})_{\lambda_1\phi_2}^{\lambda_1\phi_2}&=
\underbrace{\frac{1}{2}\fbox{\FDiagram[labelleftbottom=$\scriptstyle \lambda_1$,
		labelrightbottom=$\scriptstyle \phi_2$,
		labellefttop=$\scriptstyle \lambda_1$,
		labelrighttop=$\scriptstyle \phi_2$,
		leftSE,long,longup]{dashes_srarrow}{plain_srarrow}{}{}{}}}_{+g^2(3+\xi)}
\, + \,
\underbrace{\frac{1}{2}\FDiagram[
	labelleftbottom=$\scriptstyle \lambda_1$,
	labelrightbottom=$\scriptstyle \phi_2$,
	labellefttop=$\scriptstyle \lambda_1$,
	labelrighttop=$\scriptstyle \phi_2$,
	rightSE,long,longup]{dashes_srarrow}{plain_srarrow}{}{}{}}_{+g^2(1+\xi)}
\, + \,
\underbrace{\FDiagram[styleleftbottom=dashes,
	stylerightbottom=plain,
	stylemid=dashes,
	stylelefttop=dashes,
	stylerighttop=plain,
	labelleftbottom=$\scriptstyle \lambda_1$,
	labelrightbottom=$\scriptstyle \phi_2$,
	labellefttop=$\scriptstyle \lambda_1$,
	labelrighttop=$\scriptstyle \phi_2$,
	tchannel,long,longup]{dashes_srarrow}{plain_srarrow}{wiggly}{dashes_srarrow}{plain_srarrow}}_{-2\xi g^2}
\, + \,
\underbrace{\FDiagram[styleleftbottom=dashes,
	stylerightbottom=plain,
	stylemid=dashes,
	stylelefttop=dashes,
	stylerighttop=plain,
	labelleftbottom=$\scriptstyle \lambda_1$,
	labelrightbottom=$\scriptstyle \phi_2$,
	labellefttop=$\scriptstyle \lambda_1$,
	labelrighttop=$\scriptstyle \phi_2$,
	labelmid=$\scriptstyle \psi^3$,
	schannel,long,longup]{dashes_srarrow}{plain_srarrow}{dashes_sarrow}{dashes_srarrow}{plain_srarrow}}_{+2g^2} \eqncom \\
4g^2(\diladensity_2^{\cN=4})_{\lambda_1\phi_2}^{\phi_2\lambda_1}&=
\underbrace{\FDiagram[styleleftbottom=dashes,
	stylerightbottom=plain,
	stylemid=dashes,
	stylelefttop=dashes,
	stylerighttop=plain,
	labelleftbottom=$\scriptstyle \lambda_1$,
	labelrightbottom=$\scriptstyle \phi_2$,
	labellefttop=$\scriptstyle \phi_2$,
	labelrighttop=$\scriptstyle \lambda_1$,
	labelmid=$\scriptstyle \psi^3$,
	schannel,long,longup]{dashes_srarrow}{plain_srarrow}{dashes_sarrow}{plain_srarrow}{dashes_srarrow}}_{-2g^2}
\eqncom
\end{aligned}
\end{equation}
where the grey-shaded parts are cut out of each diagram and the shaded blobs represent one-loop self-energy insertions. We draw the operator insertion as an extended black line to emphasize that we only keep the terms that would also be present if the density is connected to a length $L\geq 3$ operator\footnote{In doing so, we explicitly avoid the occurrence of finite-size effects, \cf \eqref{sec:finite-size-effects}.}. The terms beneath each diagram are the respective numerical prefactors of the counterterms, which can be obtained using the Feynman rules from \appref{app:Feynman_rules} with colour structure fixed to give the planar contribution in a length $L\geq 3$ state. The last two equations in \eqref{eq: contributions} are obtained by reflecting all diagrams in \eqref{eq:L2_state} at the vertical axis, which leaves their counterterm contributions invariant. We find, that in \NfSYMt the s-channel contributions vanish, since the diagram in the second line of \eqref{eq:L2_state} cancels the last one in the first line. This cancellation exemplifies the situation discussed beneath \eqref{eq: generic prewrapping diagram}. In \NfSYMt, the cyclically symmetrised state renders a vanishing contribution when connected to an antisymmetric s-channel-type interaction. Hence, the asymptotic one-loop dilatation operator density gives the correct result, even when it is connected to an $L=2$ state.

In the $\beta$-deformation, the cancellation between the s-channel contributions in \eqref{eq:L2_state} does not occur, since our interactions are not antisymmetric any more. Using the result \eqref{eq: deformation of D_2} for the asymptotic dilatation operator density, the contributions in \eqref{eq: contributions} acquire the phases $1$, $\e^{\complexi\beta}$, $1$, and $\e^{\complexi\beta}$, respectively. From the Feynman-diagrammatic calculation in \eqref{eq:L2_state} we find the same result for asymptotic states. Hence, when combining the four s-channel contributions in the $\beta$-deformation, the automatic cancellation ceases to happen as
\begin{equation}
1-\e^{i \beta }+1-\e^{-i \beta }=4\sin^2\tfrac{\beta}{2} 
\eqndot
\end{equation}
However, we know from the colour space analysis, that there is no s-channel contribution in the \SUN theory and we have to set their contribution to zero by hand, giving an explicit example of prewrapping for the length $L=2$ state $\mathcal{O}$. In principle, this amounts to calculating the deformed one-loop s-channel Feynman diagrams of all $L=2$ non-cyclic states. In the supersymmetric $\beta$-deformation this is fortunately not necessary and we will present a short-cut for the implementation of prewrapping into the asymptotic dilatation operator density in the remainder of this subsection. The procedure can be verified in three steps: first, we find the tuples of input fields $(A_1,A_2,A_3,A_4)$ of $(\mathfrak{D}_2)_{A_2A_1}^{A_3A_4}$ for which the cancellation between s-channel contributions happens automatically, as in the undeformed theory. Secondly, we identify the tuples for which spurious s-channel contributions occur. These are removed by setting the deformation parameter $\beta$ to zero, which brings the diagrams back to the setup of the undeformed theory, where the cancellation is automatic. Third, we show that this undeforming procedure does not change any other contribution, so that the procedure can be applied to the sum of all contributions, i.e.\ for the full asymptotic dilatation operator density $(\mathfrak{D}_2)_{A_2A_1}^{A_3A_4}$.

In the $\beta$-deformation, only fields with non-vanishing $(Q^1,Q^2)$-charge are deformed, \cf \secref{sec:The_deformations}. In the $\cN=1$ superspace formulation, this means that only the matter superfields $\{\Phi_i, \ol{\Phi}^i\}$ in \eqref{eq: superspace action} are deformed. Interactions involving vector superfields $V$ are undeformed. In the component formulation, this translates to the fact that interactions involving gauge fields and/or the gluino $\lambda_4^\alpha$ and its conjugate $\ol{\lambda}^4_{\dot\alpha}$ are undeformed, while all remaining interactions may be deformed, as seen in \eqref{eq:deformed_action_complex_scalars2} and \eqref{eq:2_trace_action}. Contributions to s-channel diagrams in which one or both interaction vertices are undeformed cancel, as in the undeformed theory. The reason is that the combination of symmetric state with the commutator-type undeformed vertex in at least one of the interactions leads to the cancellation. Therefore, non-vanishing spurious s-channel contributions occur only in diagrams in which both interaction vertices are deformed. This implies that in deformed contributions the initial- and final-state fields all are of matter type. 

All one-loop s-channel diagrams with only matter fields are depicted in the first row of \tabref{tab: deformed possibilities}. In these interactions, two incoming matter fields $\{ \phi_i,\lambda_i^\alpha\}$ are transformed into two outgoing matter fields, or respectively, two incoming anti-matter fields $\{ \ol{\phi}^i,\ol{\lambda}^i_{\dot\alpha} \}$ are transformed into two outgoing anti-matter fields.\footnote{In the picture of the two-point function, these diagrams are connecting two matter fields $\{ \phi_i,\lambda_i^\alpha\}$ of an operator $\cO$ with two \emph{anti-matter} fields $\{ \ol{\phi}^i,\ol{\lambda}^i_{\dot\alpha} \}$ of a second operator $\bar\cO^\prime$, or, respectively, anti-matter fields in the former to matter fields in the latter.} All these spurious s-channel contributions can be removed by setting $\beta=0$ whenever such combinations of external fields occur. 
\begin{table}[tbp]
	\centering
	\caption{Asymptotic range $R=2$ one-loop diagrams with two deformed vertices that contribute to the renormalisation of an $L\geq 3$ state. Superfields are depicted by an oriented solid line, auxiliary fields by an oriented dotted line and component fields as in \appref{app:Feynman_rules}. The states with length $L\geq 3$ are drawn by bold horizontal lines. Twists mean a reflection along the vertical axis of only the upper half of the diagram. Scalar fields are treated on the same footing as the matter-type fermions, since the quartic vertices can be rewritten as cubic vertices with `propagating' auxiliary fields. Covariant derivatives are suppressed in the notation.}
	\label{tab: deformed possibilities}
	\begin{tabular}{|c|@{\quad}c@{\quad}|@{\quad}c@{\quad}|}
		\hline
		& in components & $\mathcal{N}=1$ \\\hline
		\begin{minipage}[c]{2.4cm}
			\vspace*{0.4\baselineskip}
			\centering {\bf s-channel} \\
			\vspace*{0.2\baselineskip}
			\begin{tabular}{@{}c@{}@{}l}
				$+${ } & vertical \& \\
				& horizontal \\ 
				& reflections \\
				$+${ } & twists
			\end{tabular}%
			\vspace*{0.2\baselineskip}
		\end{minipage}%
		& %
		$\FDiagram[labelleftbottom=$\scriptstyle \lambda_i$,
		labelrightbottom=$\scriptstyle \lambda_j$,
		labellefttop=$\scriptstyle \lambda_i$,
		labelrighttop=$\scriptstyle \lambda_j$,
		labelmid=$\scriptstyle \phi_k$,
		schannel,long,longup]{dashes_srarrow}{dashes_srarrow}{plain_sarrow}{dashes_srarrow}{dashes_srarrow}$ \,\,\,\,\,\, %
		$\FDiagram[labelleftbottom=$\scriptstyle \phi_i$,
		labelrightbottom=$\scriptstyle \lambda_j$,
		labellefttop=$\scriptstyle \phi_i$,
		labelrighttop=$\scriptstyle \lambda_j$,
		labelmid=$\scriptstyle \lambda_k$,
		schannel,long,longup]{plain_srarrow}{dashes_srarrow}{dashes_sarrow}{plain_srarrow}{dashes_srarrow}$ \,\,\,\,\,\, %
		$\FDiagram[labelleftbottom=$\scriptstyle \phi_i$,
		labelrightbottom=$\scriptstyle \phi_j$,
		labellefttop=$\scriptstyle \phi_i$,
		labelrighttop=$\scriptstyle \phi_j$,
		xchannel,long,longup]{plain_srarrow}{plain_srarrow}{}{plain_srarrow}{plain_srarrow}%
		\!\!=\!\!\FDiagram[
		labelleftbottom=$\scriptstyle \phi_i$,
		labelrightbottom=$\scriptstyle \phi_j$,
		labellefttop=$\scriptstyle \phi_i$,
		labelrighttop=$\scriptstyle \phi_j$,
		labelmid=$\scriptstyle F_k$,
		schannel,long,longup]{plain_srarrow}{plain_srarrow}{dots_sarrow}{plain_srarrow}{plain_srarrow}$ %
		& $\FDiagram[labelleftbottom=$\scriptstyle \varPhi_i$,
		labelrightbottom=$\scriptstyle \varPhi_j$,
		labellefttop=$\scriptstyle \varPhi_i$,
		labelrighttop=$\scriptstyle \varPhi_j$,
		labelmid=$\scriptstyle \varPhi_k$,
		schannel,long,longup]{plain_srarrow}{plain_srarrow}{plain_sarrow}{plain_srarrow}{plain_srarrow}$ \\ \hline
		\begin{minipage}[c]{2.4cm}
			\vspace*{0.4\baselineskip}
			\centering {\bf t-channel} \\
			\vspace*{0.2\baselineskip}
			\begin{tabular}{@{}c@{}@{}l}
				$+${ } & vertical \& \\
				& horizontal \\ 
				& reflections \\
				&
			\end{tabular}%
			\vspace*{0.2\baselineskip}
		\end{minipage}%
		& %
		$\FDiagram[labelleftbottom=$\scriptstyle \lambda_i$,
		labelrightbottom=$\scriptstyle \ol{\lambda}^j$,
		labellefttop=$\scriptstyle \ol{\lambda}^j$,
		labelrighttop=$\scriptstyle \lambda_i$,
		labelmid=$\scriptstyle \phi_k$,
		tchannel,long,longup]{dashes_srarrow}{dashes_sarrow}{plain_sarrow}{dashes_sarrow}{dashes_srarrow}$ \,\,\,\,\,\, %
		$\FDiagram[labelleftbottom=$\scriptstyle \phi_i$,
		labelrightbottom=$\scriptstyle \ol{\lambda}^j$,
		labellefttop=$\scriptstyle \ol{\lambda}^j$,
		labelrighttop=$\scriptstyle \phi_i$,
		labelmid=$\scriptstyle \ol{\lambda}^k$,
		tchannel,long,longup]{plain_srarrow}{dashes_sarrow}{dashes_sarrow}{dashes_sarrow}{plain_srarrow}$ \,\,\,\,\,\, %
		$\FDiagram[labelleftbottom=$\scriptstyle \phi_i$,
		labelrightbottom=$\scriptstyle \ol{\phi}^j$,
		labellefttop=$\scriptstyle \ol{\phi}^j$,
		labelrighttop=$\scriptstyle \phi_i$,
		xchannel,long,longup]{plain_srarrow}{plain_sarrow}{}{plain_sarrow}{plain_srarrow}%
		\!\!=\!\!\FDiagram[
		labelleftbottom=$\scriptstyle \phi_i$,
		labelrightbottom=$\scriptstyle \ol{\phi}^j$,
		labellefttop=$\scriptstyle \ol{\phi}^j$,
		labelrighttop=$\scriptstyle \phi_i$,
		labelmid=$\scriptstyle F^k$,
		tchannel,long,longup]{plain_srarrow}{plain_sarrow}{dots_sarrow}{plain_sarrow}{plain_srarrow}$ & %
		$\FDiagram[labelleftbottom=$\scriptstyle \varPhi_i$,
		labelrightbottom=$\scriptstyle \ol{\varPhi}^j$,
		labellefttop=$\scriptstyle \ol{\varPhi}^j$,
		labelrighttop=$\scriptstyle \varPhi_i$,
		labelmid=$\scriptstyle \ol{\varPhi}_k$,
		tchannel,long,longup]{plain_srarrow}{plain_sarrow}{plain_sarrow}{plain_sarrow}{plain_srarrow}$ \\\hline
	\end{tabular}
\end{table}

In the previous paragraph, we discussed how spurious s-channel contributions can be removed on the level of individual diagrams involving either four matter or four anti-matter fields. We now need to show that this procedure does not alter any non-s-channel contributions. Note that diagrams with one or two undeformed vertices are independent of $\beta$ and hence setting $\beta=0$ for such diagrams does not change their contribution. The only potentially affected diagrams necessarily have two deformed vertices. Apart from the already discussed s-channel diagrams, we have to analyse the t-channel- and self-energy-type diagrams. The t-channel-type diagrams are given in the second row of \tabref{tab: deformed possibilities} and we see that there are no such diagrams with only matter or only anti-matter fields. Hence, they are also not affected by the above procedure. Finally, we have self-energy-type contributions. Their subdiagrams of elementary interactions have range $R=1$, \cf \eqref{eq:L2_state}, and these subdiagrams are connected to an $L=2$ state. Therefore relation \eqref{diagrel} for the asymptotic dilatation operator is applicable\footnote{The $\ast$-product of a field with its own conjugate has a vanishing deformation phase.}, rendering their contributions independent of $\beta$ as well.
 
Since the interaction vertices only depend on the flavours of the involved elementary fields, the above analysis immediately lifts to diagrams with symmetrised covariant derivatives acting on some of the external fields. In particular, the analysis holds for all fields from the alphabet \eqref{eq: alphabet} that we used to build external states. We define the (anti-)matter subalphabets as
\begin{equation}
\begin{aligned}
\cA_\text{matter}&=\{ \D^k\phi_1, \D^k\phi_2, \D^k\phi_3, \D^k\lambda_{3\alpha}, \D^k\lambda_{2\alpha}, \D^k\lambda_{3\alpha} \} \eqncom \\
\ol{\cA}_\text{matter}&=\{ \D^k\ol{\phi}^1, \D^k\ol{\phi}^2, \D^k\ol{\phi}^3, \D^k\ol{\lambda}^1_{\dot\alpha}, \D^k\ol{\lambda}^2_{\dot\alpha},\D^k\ol{\lambda}^3_{\dot\alpha} \}
\eqndot
\end{aligned} 
\end{equation}
With these, we can give the complete one-loop dilatation operator \eqref{eq:Dila_N4_one_loop} of the planar $\beta$-deformation with gauge group \SUN in terms of the following density:
\begin{equation}
(\diladensity_2^\beta)_{A_iA_j}^{A_kA_l}= \e^{\frac{\complexi}{2} (\mathbf{q}_{A_k} \wedge \mathbf{q}_{A_l}- \mathbf{q}_{A_i} \wedge \mathbf{q}_{A_j})} \rule[-0.96cm]{0.1mm}{1.415cm}\!{\phantom{|}}_{\substack{\\[0.2cm]
		\beta=0\text{ if }L=2\text{ and \phantom{...............}}\\ 
		(A_i,A_j,A_k,A_l\in\cA_\text{matter}\text{ or \phantom{)}}\\  
		\phantom{{}({}} A_i,A_j,A_k,A_l\in\bar\cA_\text{matter}) \phantom{\text{ or }}}}  (\diladensity_2^{\cN=4})_{A_iA_j}^{A_kA_l}
\eqndot
\end{equation}
Note that the inclusion of the finite-size prewrapping effect induces an explicit length-dependence in the density.

\subsection{The \texorpdfstring{\UN}{U(N)} dilatation operator}\label{sec:UN_dila_beta}
For gauge group \UN, the one-loop finite-size corrections depend on the multi-trace couplings of the chosen theory displayed in \subsecref{sec:multi-trace-parts-of-the-action} and hence also the one-loop anomalous dimensions of $L=2$ and $L=1$ states depend on these couplings. We will not derive the general form of these anomalous dimensions, since they become renormalisation-scheme dependent at two-loop order, due to the non-conformality of the $\beta$-deformation with gauge group \UN, see \cite{FG05}. However, since the renormalisation-scheme dependence only starts at second loop order, we can still calculate the finite-size affected scheme-independent one-loop anomalous dimensions in our favourite $\beta$-deformation with this gauge group. For this theory, we choose that the tree-level values of all multi-trace couplings in \subsecref{sec:multi-trace-parts-of-the-action} vanish. In this case, the asymptotic one loop dilatation operator density \eqref{eq: deformation of D_2} is correct for all states with length $L>1$. For the $L=1$ states, which correspond to the \U{1} modes in the theory, we have to calculate the anomalous dimensions that enter the range $R=1$ part of the complete one-loop dilatation operator density $g^2(\mathfrak{D}^\beta_{2})_i^f=\delta_i^f\bra{i}g^2\mathfrak{D}^\beta_{2}\ket{i}$. The anomalous dimensions of (anti-)chiral scalars can be calculated from \eqref{eq:divergence_SEphi} using \eqref{eq:RGE_comp_operators}. Form the $\cN=1$ SUSY, this result also extends to the (anti-)chiral fermions. The anomalous dimensions of the gauge fields and gluinos vanish as in the undeformed theory since all contributing one-loop diagrams are undeformed. Combining these results we have
\begin{equation}
\begin{aligned}
\gamma_{\tr(\D^k\lambda_{4\alpha})}^{(1)}= 
\gamma_{\tr(\D^k\cF_{\alpha\beta})}^{(1)}= 
0\eqncom\qquad
\gamma_{\tr(\D^k\phi_i)}^{(1)}= 
\gamma_{\tr(\D^k\lambda_{i\alpha})}^{(1)}= 
8g^2\sin^2\frac{\beta}{2}
\eqncom\qquad \forall i\in \{1,2,3\}\eqncom
\end{aligned}
\end{equation}
where the same equations hold for the conjugate elementary fields.

\subsection{Immediate implications for the \texorpdfstring{\AdSCFTc}{AdS/CFT correspondence}}\label{sec:Implications_AdSCFT_beta}
Knowing the complete one-loop dilatation operator in the $\beta$-deformation with gauge group \UN and \SUN allows to compute the one-loop anomalous dimensions of all single-trace states. Explicit results and the involved calculatory steps are presented in \cite{Fokken:2013mza}, where the one-loop anomalous dimension of all superconformal primary states with classical scaling dimension $\Delta_0\leq 4.5$ are determined. There, it was found that the only multiplets affected by prewrapping at one loop are the ones characterised by the $L=2$ single-impurity highest-weight states, $\tr(\phi^i\phi^j)$ with $i\neq j$. In principle, multiplets whose conformal primaries are built from $\tr(\phi^i\phi^j)$ with $n$ covariant derivatives $\D_{1\dot{2}}$ acting on the individual scalars, can also be affected by prewrapping. However, it turns out that these multiplets have vanishing anomalous dimension, since the UV divergences in the occurring Feynman diagrams cancel, \cf \cite[\app{D}]{Fokken:2013mza}. For a better understanding of prewrapping in \CFTs, it would be desirable to understand why one-loop prewrapping candidates with $\Delta_0>2$ turn out to be not affected by prewrapping.

For the prewrapping affected single-impurity states $\tr(\phi^i\phi^j)$, it can be shown from an analysis in superspace that their anomalous dimension $\gamma_{\tr(\phi^i\phi^j)}=0$ is protected at all loop orders \cite{Fokken:2013mza}. In light of the \AdSCFTc, this state must therefore correspond to a supergravity mode (whose energy is not quantum-corrected) in the strong coupling regime. The calculation in \cite{Frolov:2005iq} indeed suggests that the mass of the corresponding mode receives no corrections. In this context, it would be interesting to understand how the $\beta$-deformation raises the mass of modes dual to non-prewrapping affected states, while keeping the $L=2$ single-impurity states at fixed masses.

Finally, the prewrapping effect in the $\beta$-deformation has important consequences for the integrability-based descriptions of this gauge theory. As these consequences are closely related to the ones in the non-supersymmetric $\gamma_i$-deformation, we postpone their discussion to \chapref{chap:Conclusion_outlook}.

\section{The thermal one-loop partition functions of the deformed theories}\label{sec:the-thermal-one-loop-partition-functions-of-the-deformed-theories}
In this section, we calculate the thermal one-loop partition functions of \NfSYMt and its deformations on the compactified spacetime \RxSt by the means of generalised \Polya theory. This section is based on \cite{Fokken:2014moa}.

For a general gauge theory, the confinement-deconfinement phase transition connecting the weak and strong coupling regimes of the theory is still not well understood, see e.g.\ \cite{PhysRevD.76.086003} for a discussion in the context of QCD. We will approach this questions for the $\beta$- and $\gamma_i$-deformation via the thermal partition function
\begin{equation}\label{eq: partition function intro}
\cZ(T)=\tr_{\RxSt}[e^{- H/T}]\eqncom
\end{equation}
where the Hamiltonian $H$ is given by the Wick-rotated action of the theory\footnote{For the connection between the path integral approach and statistical mechanics see e.g.\ \cite[\chap{2}]{Wipf:2013vp}.}, the temperature $T$ is measured in units of the Boltzmann constant and the trace sums over all admissible states (composite operators) on \RxSt. While the partition functions of \CFTs have a trivial temperature dependence in flat space, this is no longer the case on \RxSt, where the compactness of the space limits the number of possible (low energy) states \cite{Witten98Mar,AMMPR03}. For a gauge theory on \RxSt, the radius $R$ of the $\text{S}^3$ sphere combines with the flat space phase-transition scale $\Lambda$ to an effective dimensionless phase-transition scale $R \Lambda$. Therefore, we can tune $R$ so that the phase-transition becomes perturbatively accessible when $R\Lambda\ll 1$. Note that a sharp phase transition on the compact space $\text{S}^3$ is only observed in the \tHooft limit, where the number of colours $N$ serves as an order parameter in the partition function. Below the critical temperature\footnote{The critical temperature is also called Hagedorn temperature referring to \cite{Hagedorn65}.} $T_\text{H}$, the partition function \eqref{eq: partition function intro} scales as $N^0$ and above $T_\text{H}$ it scales as $N^2$, see \cite{Witten:1998qj}. In light of the \AdSCFTc, the low energy \dof of the system are colour-neutral composite operators and above the critical temperature, the system is described by the corresponding string theory dual \cite{Witten:1998qj}. In a direct path integral approach, the perturbative evaluation of \eqref{eq: partition function intro} amounts to calculating all vacuum diagrams of the theory up to a given order in the effective coupling, see \cite{AMMPR03,AMMPR05,AMR06,Mussel:2009uw}.
	
In a \CFT, we do not have to follow the direct path integral approach mentioned above. Instead, we can conformally map\footnote{In this mapping the \RxSt metric $\de s^2=\de t^2 +R^2\de \Omega_3$ is transformed via the coordinate change $t\rightarrow r=\e^{\complexi \frac{t}{R}}$ to the metric $(\de s^\prime)^2= \frac{R^2}{r^2}(-\de r^2 +r^2\de\Omega_3)$. The latter one is the metric of four-dimensional Minkowski space times a fixed ratio of the radii $R$ and $r$ of the $\text{S}^3$ and the decompactified $\text{S}^1$ sphere, respectively. See \cite{GL98,Witten:1998qj} for further information.} \RxSt to $\RR^{(3,1)}$ and express the partition function as
\begin{equation}\label{eq: partition function intro D}
\cZ(T)=\tr_{\RR^{(3,1)}}\left[x^D\right]\eqncom
\end{equation}
where $x=e^{- 1/RT}$ and $D$ is the dilatation operator \cite{Witten:1998qj}. Compared to the path integral approach, we save one loop order when using this formulation\footnote{The reason for this is that the anomalous dimensions as eigenvalues of the dilatation operator are obtained from the complete composite operator renormalisation constant, which can be extracted from a single operator insertion in a correlation function, as presented in \subsecref{sec:composite-operator-insertions}.} in perturbative calculations. The task to sum over all states on $\mathbb{R}^{(3,1)}$ now reduces to the enumeration of all graded cyclic spin-chains which can be thought of as necklaces. For the free \NfSYMt, this was done employing \Polya theory in \cite{Sundborg:1999ue}. In this approach, the single-site partition function $z(x)$, which enumerates all possible fields at one site, is employed to enumerate all composite states that can be built from it. It is given as a sum over all fields in the alphabet $\cA$ of the theory weighted by the eigenvalues of the classical dilatation-operator density $\mathfrak{D}_0$ as
\begin{equation}
z(x)=\sum_{\cA}\bra{\cA}x^{\mathfrak{D}_0}\ket{\cA}=\sum_{\cA}x^{\mathfrak{D}_0(\cA)}\eqndot
\end{equation}
In \cite{Spradlin:2004pp}, the first order correction in the \tHooft coupling to \eqref{eq: partition function intro D} was calculated via an extension of the \Polya-theoretic approach. In addition to $z(x)$, it also employs the two generalised expectation values\footnote{The function $F(\cdot)$ is the fermion number, which is zero or respectively one for bosonic or fermionic argument.}
\begin{align}
\label{eq: SV 3.10 intro}
\ev{ \mathfrak{D}_2^{L\ge3}(x) } &=
\sum_{A_1,A_2 \in \cal A} x^{(\mathfrak{D}_0)_{A_1}^{A_1}+(\mathfrak{D}_0)_{A_2}^{A_2}} 
(\mathfrak{D}_2^{L\ge3})^{ A_1A_2 }_{ A_1A_2 }
\eqncom \\
\label{eq: SV 3.11 intro}
\ev{ P\mathfrak{D}_2^{L\ge3}(w,y) }&=
\sum_{A_1,A_2 \in \cal A} (-1)^{F(A_1)F(A_2)} w^{(\mathfrak{D}_0)_{A_1}^{A_1}}y^{(\mathfrak{D}_0)_{A_2}^{A_2}}
(\mathfrak{D}_2^{L\ge3})^{ A_2A_1 }_{ A_1A_2 } \eqncom 
\end{align}
which also depend on the planar one-loop dilatation-operator density $\mathfrak{D}_2$ and employ the non-cyclic states given below \eqref{eq:occupation_numbers}.

The result from \cite{Sundborg:1999ue} for the free \NfSYMt is also correct for the $\beta$- and $\gamma_i$-deformation, since states in all these theories are constructed from the same alphabet given in \eqref{eq: alphabet}. The one-loop result, however, is not directly applicable for two reasons: first, it is not clear how the calculation in \cite{Spradlin:2004pp} can be adapted to treat also the deformations and second, the occurrence of prewrapping and wrapping finite-size corrections in the deformed one-loop calculations spoils any naive adaptation. As discussed in \secref{sec:non-conformal_double_trace_coupling} and \ref{sec:beta_paper}, only the $\beta$-deformation with gauge group \SUN is conformal. The $\beta$-deformation with gauge group \UN and the $\gamma_i$-deformation with either of those gauge groups are not conformally invariant, due to the running of multi-trace couplings introduced in \subsecref{sec:multi-trace-parts-of-the-action}. This non-conformality prevents us in principle to follow the partition function approach via \eqref{eq: partition function intro D}, which is only valid for \CFTs. However, since effects of non-conformality only appear at loop orders $K>1$ in this calculation, we can safely generalise the results of \cite{Spradlin:2004pp} to the $\beta$- and $\gamma_i$-deformation with either gauge group. For the non-conformal theories, that pose no restrictions on the occurring multi-trace structure, we only have to decide which of the multi-trace coupling from \subsecref{sec:multi-trace-parts-of-the-action} we want to include. For practical purposes, we choose these multi-trace couplings to vanish at tree-level. This choice corresponds to the proposed $\gamma_i$-deformation in \cite{Frolov:2005dj} and to the conventional \UN $\beta$-deformation, as it is obtained by deforming the \NfSYMt in the $\cN=1$ superspace formulation.

In this section, we rederive the calculation of \cite{Spradlin:2004pp} in a way that is also applicable in the deformed theories. That is, we compute the necessary ingredients for the one-loop partition function: $z(x)$, $\ev{P\mathfrak{D}_2^{L\geq 3}(w,y)}$, and $\ev{\mathfrak{D}^{L\geq 3}_2(x)}$ for the $\gamma_i$-deformation and obtain the results for the $\beta$-deformation and undeformed theory by taking the appropriate limits of the deformation angles $\gamma_i^\pm$. We calculate the occurring finite-size contributions $Z^{(1)}_{\text{f.s.c.}}(x)$, which are the prewrapping contributions in the \SUN $\beta$-deformation and the $L=1$ wrapping contributions for all deformed \UN theories. In addition to the asymptotic dilatation operator density $\mathfrak{D}_2^{L\geq 3}$ form \eqref{eq: deformation of D_2}, for the finite-size corrections we need the density $\mathfrak{D}_2^{L=2}$ given in \eqref{eq:density_intro} for $L=2$ states in the \SUN $\beta$-deformation and, in case of gauge group \UN, the density $\mathfrak{D}_2^{L=1}$ defined by anomalous dimensions that form the one-loop dilatation operator on $L=1$ states. The latter ones are given by\footnote{%
	The anomalous dimensions are obtained from the countertems that enter the definition \eqref{eq:RGE_comp_operators}. The one-loop counterterm of the scalar states $N^{-1/2}\tr(\phi_i)$ is obtained from \eqref{eq:divergence_SEphi}, like in \subsecref{sec:UN_dila_beta}. For the fermionic states $\tr(\lambda_{A\alpha})$, it can be obtained by deforming the fermionic self-energy \eqref{eq:fermion_SE_scalar_sol} and extracting the planar parts when the initial and final state are connected to it. The self-energy diagrams can be deformed by replacing the flavour and colour parts of the undeformed vertices in \eqref{eq:flavour_SElambda} by the deformed versions given in \eqref{eq:cubic_vertices}. The field strength states are not renormalised in the \tHooft limit, as in the undeformed theory.}
\begin{equation}\label{eq:U1_energies}
\begin{aligned}
\mathrlap{\gamma^{(1)}_{\tr(\D^k\phi_i)}}\hphantom{\gamma^{(1)}_{\tr(\D^k \cF_{\alpha\beta})}}&=
\mathrlap{\gamma^{(1)}_{\tr(\D^k\ol{\phi}^i)}}\hphantom{\gamma^{(1)}_{\tr(\D^k \bar\cF_{\dot\alpha\dot\beta})}}=
8g^2\Bigl(\sin^2\frac{\gamma^+_i}{2}+\sin^2\frac{\gamma^-_i}{2}\Bigr)\eqncom\\
\mathrlap{\gamma^{(1)}_{\tr(\D^k\lambda_{i\alpha})}}
\hphantom{\gamma^{(1)}_{\tr(\D^k \cF_{\alpha\beta})}}
&=
\mathrlap{\gamma^{(1)}_{\tr(\D^k\ol{\lambda}^i_{\dot{\alpha}})}}\hphantom{\gamma^{(1)}_{\tr(\D^k \bar\cF_{\dot\alpha\dot\beta})}}=
4g^2\Bigl(\sin^2\frac{\gamma^-_i}{2}+\sin^2\frac{\gamma^+_{i+1}}{2}+\sin^2\frac{\gamma^+_{i+2}}{2}\Bigr)\eqncom\\
\mathrlap{\gamma^{(1)}_{\tr(\D^k \lambda_{4\alpha})}}
\hphantom{\gamma^{(1)}_{\tr(\D^k \cF_{\alpha\beta})}}
&=
\mathrlap{\gamma^{(1)}_{\tr(\D^k\ol{\lambda}^4_{\dot{\alpha}})}}\hphantom{\gamma^{(1)}_{\tr(\D^k \bar\cF_{\dot\alpha\dot\beta})}}=
4g^2\Bigl(\sin^2\frac{\gamma^-_1}{2}+\sin^2\frac{\gamma^-_2}{2}+\sin^2\frac{\gamma^-_3}{2}\Bigr)\eqncom\\
\gamma^{(1)}_{\tr(\D^k \cF_{\alpha\beta})}
&=
\gamma^{(1)}_{\tr(\D^k \bar\cF_{\dot\alpha\dot\beta})}=0\eqncom
\end{aligned}
\end{equation}
where cyclic identification $i+3\sim i$ is understood.

From the thermal one-loop partition function of the deformed theories in the \tHooft limit we derive the one-loop correction to the phase-transition temperature to be 
\begin{equation}\label{eq:Hagedorn_intro}
T_\H(g)= T_\H\left(1+2 g^2+ \dots \right) \eqncom\qquad \text{with}\qquad
T_\H= \frac{1}{\ln(7+4\sqrt{3})}\frac 1R \eqndot
\end{equation}
In the calculation we find that the divergences of the thermal partition function at temperature $T_\H$ is entirely driven by those states with large lengths that are independent of the deformation parameters $\gamma_i^\pm$. Therefore, the critical temperature in \eqref{eq:Hagedorn_intro} stays the same as in the undeformed theory for the $\beta$- and $\gamma_i$-deformation at order $g^2$.

\subsection{Partition functions via \Polya theory}\label{sec:Partition functions via Polya}
Before going into the actual calculation, let us give a brief summary of the \Polya-theoretic approach of \cite{Spradlin:2004pp} to thermal one-loop partition functions. We follow their presentation and refer the reader there for details\footnote{Additional information regarding derivations may be found in \cite{thesis:Matthias}.}. In addition, we discuss the necessary adjustments for the deformations. We first discuss the approach in the context of single-trace states. Multi-trace states can be included afterwards by noting that the action of the dilatation operator $D$ on such states in the \tHooft limit is entirely determined by its action on single-trace states.

\subsubsection{The single-trace partition function}
The single-trace partition function is defined in analogy to \eqref{eq: partition function intro D} as trace over all single-trace operators:
\begin{equation}\label{eq: def single-trace partition function}
Z(x)=\tr_{\RR^{(3,1)}}\bigl[x^D\bigr]_{\text{s.t.}}\eqncom
\end{equation}
where s.t.\ restricts the trace to single-trace states only. Expanding the dilatation operator in the effective planar coupling $g$ as in \eqref{eq:quantum_D} yields the following expansion
\begin{equation}\label{eq: perturbative expansion of Z}
\begin{aligned}
Z(x,g) 
&= \tr\bigl[x^{D_0}\bigr]_{\text{s.t.}} + g^2\ln x \tr\bigl[x^{D_0} D_2\bigr]_{\text{s.t.}} + \cO(g^3) \\
&= Z^{(0)}(x) + g^2\ln x\,Z^{(1)}(x)+ \cO(g^3) \eqndot 
\end{aligned}
\end{equation}
Since the dilatation operator only starts to change the state's lengths at loop orders $K>1$, we can express the occurring traces in \eqref{eq: perturbative expansion of Z} as sums over traces with fixed lengths. Moreover, the trace over single-trace operators can be cast into a trace over spin-chains with fixed lengths using \eqref{eq:operator_as_spinchain}. 

We first evaluate the contribution from the free theory 
\begin{equation}\label{eq:free_st_partition_function}
Z^{(0)}(x)=\tr\bigl[x^{D_0}\bigr]_{\text{s.t.}}=\sum_{L=1+\colors}^\infty \tr_L[\cP x^{D_0}]\eqncom
\end{equation}
where we indicated the trace over a non-cyclic length-$L$ spin-chain state by $\tr_L$ and included the projector on graded cyclic states $\mathcal{P}$ from \eqref{eq:graded_projetor} explicitly. The sum over all state's lengths start at $L=1$ for gauge group \UN with $s=0$ and at $L=2$ for gauge group \SUN with $s=1$. To evaluate the trace in the last equality of \eqref{eq:free_st_partition_function} over cyclic states for a given length $L$, we note that such states correspond to necklaces with $L$ beads. Therefore, we can employ \Polya's enumeration theorem \cite{Polya37} which counts how many different necklaces of length $L$ can be built when each bead is taken from an alphabet $\cA$ and contributes with a weight $x=\e^{-1/RT}$. In this scenario, the single-site partition function takes the form
\begin{equation}\label{eq: single site partition function}
z(x)=\sum_{A\in\cA}x^{(\diladensity_0)_A^A}\eqncom
\end{equation}
where the classical dilatation operator density in the spin-chain representation is given in \eqref{eq: def classical dilatation op in osc language}. According to \cite{Polya37}, it enters the sum of all possible length-$L$ necklaces as
\begin{equation}\label{eq: Polya necklaces for fermions}
\tr_L[\cP x^{D_0}]= \frac{1}{L}\sum_{k \mid L} \EulerPhi(k) \left[ z(\omega^{k+1} x^k) \right]^{L/k} \eqncom
\end{equation}
where the sum runs over all divisors $k$ of $L$ and $\EulerPhi(k)$ is the Euler totient function. The formal quantity $\omega$ fulfils $\sqrt{\omega}=-1$ and was included to also account for the grading in fermionic necklaces\footnote{In this representation, we use that bosonic and fermionic fields have full and half-integer classical scaling dimensions, respectively.}, compare \eqref{eq:graded_shift_op}. Inserting \eqref{eq: Polya necklaces for fermions} into \eqref{eq:free_st_partition_function}, we find the single-trace partition function in the free theory
\begin{equation}\label{eq: free theory single-trace partition function}
Z^{(0)}(x) = \sum_{L=1+\colors}^\infty \tr_L[\cP x^{D_0}]
= - \colors z(x) - \sum_{k=1}^\infty \frac{\EulerPhi(k)}{k}
\ln[1 - z(\omega^{k+1} x^k)]\eqncom
\end{equation}
where we used that the following double sum over an arbitrary function $f$ can be written as
\begin{equation}\label{eq:sum_identity}
\sum_{L=1}^\infty \sum_{k|L}f(L,k)=\sum_{n,m=1}^\infty f(nm,m)\eqndot
\end{equation}

Let us now turn to the one-loop contribution in \eqref{eq: perturbative expansion of Z} of the interacting theory, which is given by 
\begin{equation}\label{eq:interacting_st_partition_function}
Z^{(1)}(x)=\tr\bigl[x^{D_0}D_2\bigr]_{\text{s.t.}}=\sum_{L=1+\colors}^\infty \tr_L[\cP x^{D_0}D_2]\eqndot
\end{equation}
In contrast to the undeformed theory, where the one-loop dilatation operator density is independent of the spin-chain's length, we have to include length-dependent contributions in the deformed theories explicitly. At one loop, they originate from the wrapping and prewrapping contributions for $L=1$ and $L=2$ spin-chains in theories with gauge groups \UN and \SUN, respectively. For spin-chains with $L\geq 3$, the planar one-loop dilatation operator density of the deformed theories is length-independent and we use the the result of \cite{Spradlin:2004pp} for the trace over a length-$L$ cyclic spin-chain state\footnote{In the second line, $(k,L)=1$ denotes that $k$ is relatively prime to $L$. The corresponding sum is related to the Euler totient function as $\EulerPhi(L)=\bigl(\sum_{k=0}^{L-1}1\bigr)_{(k,L)=1}$.}
\begin{equation}\label{eq: trL result}
\begin{aligned}
\tr_L[\cP x^{D_2} {D}_2^{L\ge3}]&= \sum_{m \mid L} \EulerPhi(m)\left[ z(\omega^{m+1}x^{m}) \right]^{L/m - 2}
\ev{ \mathfrak{D}_2^{L\ge3}(\omega^{m+1}x^{m})} \\
&\phan{=}+ \sum_{\substack{k=0 \\ (k,L) = 1}}^{L-1} \left[\ev{ P\mathfrak{D}_2^{L\ge3}(\omega^{L-k+1}x^{L-k}, \omega^{k+1}x^{k})} -  \frac{\ev{ \mathfrak{D}^{L\ge3}_2(\omega^{L+1}x^L)}}{z(\omega^{L+1}x^L)}\right] \eqndot
\end{aligned}
\end{equation}
The two generalised expectation values are given in terms of non-cyclic $L=2$ spin-chain states. They depend on the asymptotic dilatation operator density of the deformed theories from \eqref{eq: deformation of D_2}, which can be given in terms of the oscillator representation using \eqref{eq: harminic action}. The generalised expectation values take the explicit form\footnote{The term accounting for fermionic signs can also be written in terms of the formal quantity $\omega$ as $(-1)^{F(A_1)F(A_2)}=\omega^{2(\mathfrak{D}_0)_{A_1}^{A_1}(\mathfrak{D}_0)_{A_2}^{A_2}}$.} 
\begin{align}
\label{eq: SV 3.10}
\ev{ \mathfrak{D}_2^{L\ge3}(x) } &= \sum_{A_1,A_2 \in \cal A} x^{(\mathfrak{D}_0)_{A_1}^{A_1}+(\mathfrak{D}_0)_{A_2}^{A_2}} 
(\mathfrak{D}_2^{L\ge3})^{ A_1A_2 }_{ A_1A_2 }
\eqncom \\
\label{eq: SV 3.11}
\ev{ P\mathfrak{D}_2^{L\ge3}(w,y) }&=\sum_{A_1,A_2 \in \cal A} 
(-1)^{F(A_1)F(A_2)} 
w^{(\mathfrak{D}_0)_{A_1}^{A_1}}y^{(\mathfrak{D}_0)_{A_2}^{A_2}}
(\mathfrak{D}_2^{L\ge3})^{ A_2A_1 }_{ A_1A_2 } \eqncom
\end{align}
where the fermion number generator $F(\cdot)$ gives one for fermionic and zero for bosonic argument. It appears, since the order of the two fields in the outgoing state is reversed. For a detailed derivation of \eqref{eq: trL result}, we refer to \cite{Spradlin:2004pp}. For spin-chain states of length $L=2$ and $L=1$ we have the respective contributions
\begin{equation} 
\begin{aligned}
\label{eq: L=2 and L=1}
\tr_2[\cP x^{D_2} {D}_2^{L=2}]&= \ev{ \mathfrak{D}_2^{L=2}(x)} + \ev{ P\mathfrak{D}_2^{L=2}(x,x)} \eqncom\\
\tr_1[\cP x^{D_2} {D}_2^{L=1}] &= \sum_{A\in \cal A} x^{(\mathfrak{D}_0)_{A}^{A}}(\mathfrak{D}_2^{L=1})_A^A \eqncom
\end{aligned}
\end{equation}
with $\mathfrak{D}_2^{L=2}$ and $\mathfrak{D}_2^{L=1}$ given in \eqref{eq:density_intro} and \eqref{eq:U1_energies}, respectively. The complete one-loop correction to the single-trace partition function is obtained by inserting \eqref{eq: trL result} and \eqref{eq: L=2 and L=1} into \eqref{eq:interacting_st_partition_function} and summing over all admissible lengths. It takes the form
\begin{equation}\label{eq: single-trace one-loop} 
\begin{aligned}
Z^{(1)}(x)&= Z^{(1)}_{\text{f.s.c}}(x)+\sum_{L=2}^\infty \tr_L[\cP x^{D_0} D_2^{L\geq 3}]\\
&= Z^{(1)}_{\text{f.s.c}}(x)+ \sum_{n=1}^\infty\sum_{\substack{k=0 \\ (k,n)=1}}^{n-1}  \Bigg[\frac{\ev{\mathfrak{D}_2^{L\ge3}(\omega^{n+1}x^n)}}{1-z(\omega^{n+1}x^n)}  +\delta_{n\neq1} \ev{P\mathfrak{D}_2^{L\geq 3}(\omega^{n-k+1}x^{n-k},\omega^{k+1}x^{k})} \Bigg] \eqncom
\end{aligned}
\end{equation}
where the finite-size corrections are incorporated in 
\begin{equation}\label{eq:fsc}
Z^{(1)}_{\text{f.s.c}}(x)= (1-s) \tr_1[\cP x^{D_2} {D}_2^{L=1}] + \tr_2[\cP x^{D_2} {D}_2^{L=2}]-\tr_2[\cP x^{D_2} {D}_2^{L{\geq}3}]
\end{equation}
and we left the gauge group dependence of $\mathfrak{D}_2^{L=2}$ implicit.

\subsubsection{The multi-trace partition function}
We build multi-trace operators as products of single-trace operators that obey the correct statistics, i.e.\ Bose-Einstein statistic if the single-trace operator is bosonic and Fermi-Dirac statistic if it is fermionic. In the \tHooft limit, where the dilatation operator acts on multi-trace states $\mathcal{O}_{\text{m.t.}}$ by acting linearly on each of its respective bosonic or fermionic single-trace constituents, we can rewrite the sum over all multi-trace operators in terms of the single-trace partition function \eqref{eq: perturbative expansion of Z}. When we label each bosonic or fermionic single-trace operator defined via a suitable set of quantum numbers by $\mathcal{O}^{\text{B}}_{\text{s.t.}}$ or  $\mathcal{O}^{\text{F}}_{\text{s.t.}}$, respectively, the multi-trace partition function can be written as\footnote{The last equation is obtained by rewriting the bosonic and fermionic parts as individual exponentials $a=\e^{\log a}$, absorbing the respective products into the exponents and expanding the logarithms in a power series. Upon changing the order of summation the single-trace partition functions are obtained.}
\begin{equation}
\begin{aligned}\label{eq: def of multi-trace partition function omega}
\cZ(x,g)&=\sum_{\mathcal{O}_{\text{m.t.}}}x^{\diladensity(\mathcal{O}_{\text{m.t.}})}
=
\prod_{\mathcal{O}^{\text{B}}_{\text{s.t.}}}\frac{1}{1-x^{\diladensity(\mathcal{O}^{\text{B}}_{\text{s.t.}})}}
\prod_{\mathcal{O}^{\text{F}}_{\text{s.t.}}}\Bigl(1+x^{\diladensity(\mathcal{O}^{\text{F}}_{\text{s.t.}})}\Bigr)
=\exp\sum_{n=1}^\infty \frac{1}{n}Z(\omega^{n+1} x^n)  \eqncom
\end{aligned}
\end{equation}
where $\omega$ again accounts for signs occurring in the  expansion of the fermionic operator part. The perturbative expansion of the single-trace partition function \eqref{eq: perturbative expansion of Z} induces the following expansion in the multi-trace case
\begin{equation}
\begin{aligned}
\cZ(x,g)&=  \cZ^{(0)}(x)+ g^2 \ln x\, \cZ^{(1)}(x)  + \cO(g^3) \eqncom
\end{aligned}
\end{equation}
with
\begin{equation}
\begin{aligned}
\cZ^{(0)}(x) = \exp\sum_{n=1}^\infty \frac{1}{n}Z^{(0)}(\omega^{n+1} x^n) \eqncom \qquad
\cZ^{(1)}(x)= \cZ^{(0)}(x) \sum_{n=1}^\infty Z^{(1)}(\omega^{n+1} x^n) \eqndot
\end{aligned}
\end{equation}
In a final step, we can insert the single-trace expressions \eqref{eq: free theory single-trace partition function} and \eqref{eq: single-trace one-loop} into these equations. For the free multi-trace partition function we obtain
\begin{equation}
\begin{aligned}\label{eq:mt_partition_function_res}
\cZ^{(0)}(x)
&=\exp\Bigl[-\colors\sum_{n=1}^\infty \frac{1}{n}z(\omega^{n+1}x^n)\Bigr] \prod_{m=1}^\infty\frac{1}{1-z(\omega^{m+1}x^m)} \eqncom\\
\end{aligned}
\end{equation}
where we used \eqref{eq:sum_identity} from right to left and the identity $\sum_{k|L}\EulerPhi(k)=L$. The one-loop contribution can be simplified to\footnote{The second summand is obtained using the same steps as for \eqref{eq:mt_partition_function_res} including the defining identity $\EulerPhi(n)=\bigl(\sum_{k=0}^{n-1}1\bigr)_{(k,n)=1}$. For the third summand we used the identity \\
	$\sum_{n,L=1}^\infty\sum_{(k,L)=1}\delta_{L\neq 1}f(n(L-k),nk)=\sum_{a,b=1}^\infty f(a,b)$, which was proven in \cite{Spradlin:2004pp}.}
\begin{equation}\label{eq: first order of multi-trace partition function}
\begin{aligned}
\cZ^{(1)}(x)&= \cZ^{(0)}(x) \sum_{n=1}^\infty\Big[Z_{\text{f.s.c}}(\omega^{n+1} x^n) 
+ \frac{n}{1-z(\omega^{n+1}x^n)} \ev{\mathfrak{D}_2^{L\ge3}(\omega^{n+1}x^{n})}
\\ &\phan{{}={} \cZ^{(0)}(x) \Bigg[\sum_{n=1}^\infty Z_{\text{f.s.c}}(\omega^{n+1} x^n) {}}
+\sum_{m=1}^\infty
\ev{P\mathfrak{D}_2^{L\ge3}(\omega^{n+1}x^{n},\omega^{m+1}x^m)}\Big]\eqndot
\end{aligned}
\end{equation}

\subsection{Ingredients of the \Polya-theoretic approach}\label{sec:ingredients}
In this subsection, we compute the ingredients needed for the extended \Polya-theoretic method introduced in the previous subsection. These are the single-site partition function $z(x)$, the two generalised expectation values $\ev{P\mathfrak{D}_2^{L\geq 3}(w,y)}$ and $\ev{\mathfrak{D}^{L\geq 3}_2(x)}$ and the finite-size corrections $Z^{(1)}_{\text{f.s.c.}}(x)$. We focus on conveying the main ideas and results here and refer to the appendices \ref{app: PD2 calculation}, \ref{app: corrections}, and \ref{app: summation identities} for calculational details.

All ingredients can be computed in the oscillator representation of \appref{subsec:the-oscillator-representation}, using in particular the zero- and one-loop dilatation operator densities \eqref{eq: def classical dilatation op in osc language} and \eqref{eq: harminic action}. In this representation, the sum over all fields occurring in the alphabet theory \eqref{eq: alphabet} can be expressed in terms of sums over all oscillator occupation numbers \eqref{eq:occupation_numbers} as
\begin{equation}\label{eq:alphabet_sum}
\sum_{A_{i}\in \mathcal{A}}= \sum_{\akindsite[1]{i},\akindsite[2]{i}=0}^\infty \, \sum_{\bkindsite[\numberdot1]{i},\bkindsite[\numberdot2]{i}=0}^\infty \,
\sum_{\ckindsite[1]{i},\ckindsite[2]{i},\ckindsite[3]{i},\ckindsite[4]{i}=0}^1 
\delta_{C_{(i)}} \eqncom
\end{equation}
where the Kronecker-$\delta$ ensures that only combinations of oscillators occur whose central charge vanish, ensuring that the field is part of the alphabet. The central charge operator is defined in \eqref{eq:symmetry_generators_osci_LR} and can be written in terms of oscillator occupation numbers in analogy to the free dilatation operator density \eqref{eq: def classical dilatation op in osc language}.

\subsubsection{The single-site partition function}\label{sec:single-site partition function}
The single-site partition function depends on the field content of the theory alone and hence it is the same for \NfSYMt and its deformations, as discussed in \subsecref{sec:Building_blocs}. Using \eqref{eq:alphabet_sum} and \eqref{eq: def classical dilatation op in osc language} yields\footnote{To obtain the result, we perform the four finite sums and the sum over $a^1$ via the Kronecker-$\delta$. This restricts the sum over $a^2$ since $a^1\geq 0$ must remain valid. The sums over $b^\alpha$ are done using \eqref{eq:sum_order}.}
\begin{equation}\label{eq: single-bead partition function of nfsymt}
z(x)= \sum_{A\in \mathcal{A}}x^{(\mathfrak{D}_0)_A^A}
=\frac{2 \left(3-\sqrt{x}\right) x}{\left(1-\sqrt{x}\right)^3} \eqndot
\end{equation}
This result agrees with the ones of \cite{Sundborg:1999ue,Spradlin:2004pp,AMMPR03}.

\subsubsection{\texorpdfstring{The \expectationvalue $\ev{P\mathfrak{D}^{L\geq 3}_2(w,y)}$}{The expectation value <PD2(w,y)>}}\label{sec: PD2}
Next, we turn to the permuted expectation value of the asymptotic one-loop dilatation operator density. We obtain the explicit form of $ \ev{ P\mathfrak{D}^{L\geq 3}_2(w,y) } $ in the oscillator representation\footnote{Upon setting $A_{3}=A_{2}$ and $A_{4}=A_{1}$ in \eqref{eq: harminic action}, the product of fermionic signs in the last five lines simplifies to $(-1)^{\sum_{e=1}^4\sum_{l=1}^4c^e_{(1)}c^l_{(2)}+\ckind[e]}$.} by inserting \eqref{eq:alphabet_sum}, \eqref{eq: deformation of D_2} with \eqref{eq: harminic action} into \eqref{eq: SV 3.11}. This yields\footnote{Note that in the oscillator picture the fermion number operator takes the form $F(A)=\sum_{e=1}^4c^{e}$. Hence, the factor $(-1)^{F(A_1)F(A_2)}$ in \eqref{eq: SV 3.11} cancels the respective factor from the matrix element $(\diladensity_2^{\cN=4})_{A_1A_2}^{A_2A_1}$.}
\begin{align}
\ev{ P\mathfrak{D}^{L\geq 3}_2(w,y) } 
&=
\Biggl[\,
\prod_{i=1}^2\Bigg( \sum_{\akindsite[1]{i},\akindsite[2]{i}=0}^\infty \sum_{\bkindsite[\dot1]{i},\bkindsite[\dot2]{i}=0}^\infty
\sum_{\ckindsite[1]{i},\ckindsite[2]{i},\ckindsite[3]{i},\ckindsite[4]{i}=0}^1
\delta_{C_{(i)}} \Bigg)
\notag\\
&\phaneqtimes 
w^{\frac 12\left(2+\sum_{\alpha=1}^{2}a^\alpha_{(1)}+\sum_{\dot\alpha=\dot 1}^{\dot 2}b^{\dot\alpha}_{(1)}\right)}
y^{\frac 12\left(2+\sum_{\alpha=1}^{2}a^\alpha_{(2)}+\sum_{\dot\alpha=\dot 1}^{\dot 2}b^{\dot\alpha}_{(2)}\right)}
\notag\\ &\phaneqtimes
\e^{-\complexi \sum_{l,m=1}^4 \ckindsite[l]{1}\ckindsite[m]{2}\bq_{\lambda_{l}}\wedge\bq_{\lambda_{m}}}
 \notag\\
&\phaneqtimes \prod_{\alpha=1}^2\bigg(\sum_{\akind[\alpha]=0}^\infty \binom{\akindsite[\alpha]{1}}{\akind[\alpha]}\binom{\akindsite[\alpha]{2}}{\akind[\alpha]}\bigg)
\prod_{\alphadot=1}^2\bigg(\sum_{\bkind[\alphadot]=0}^\infty \binom{\bkindsite[\alphadot]{1}}{\bkind[\alphadot]}\binom{\bkindsite[\alphadot]{2}}{\bkind[\alphadot]}\bigg)
\notag\\
&\phaneqtimes \prod_{e=1}^4\bigg(\sum_{\ckind[e]=0}^1 \binom{\ckindsite[e]{1}}{{\ckind[e]}}\binom{\ckindsite[e]{2}}{\ckind[e]}(-1)^{\ckind[e]}\bigg)
\notag\\
&\phaneqtimes
c_{\mathrm{h}}\Bigl[\textstyle \sum_{i=1}^2(
\sum_{\alpha=1}^{2}a^\alpha_{(i)}+
\sum_{\dot\alpha=\dot1}^{\dot2}b^{\dot{\alpha}}_{(i)}+
\sum_{e=1}^{4}c^e_{(i)}),
\notag\\ 
&\phaneq\qquad\quad\,  
\textstyle \sum_{\alpha=1}^{2}(a^\alpha_{(1)}-a^\alpha)+
\sum_{\dot\alpha=\dot1}^{\dot2} (b^{\dot\alpha}_{(1)}-b^{\dot\alpha})+ \sum_{e=1}^{4}(c^e_{(1)}-c^e),
\notag\\
&\phaneq\qquad\quad\, 
\textstyle \sum_{\alpha=1}^{2}(a^\alpha_{(2)}-a^\alpha)+
\sum_{\dot\alpha=\dot1}^{\dot2} (b^{\dot\alpha}_{(2)}-b^{\dot\alpha})+ \sum_{e=1}^{4}(c^e_{(2)}-c^e)\Bigr]
\Biggr]
\eqncom\label{eq: first eqation of PD2}
\end{align}
where we have also used the antisymmetry of $\mathbf{q}_A\wedge\mathbf{q}_B$ defined in \eqref{eq: antisymmetric product}. 

The evaluation of the twelve infinite sums in \eqref{eq: first eqation of PD2} is a decisively complicated task in light of their entanglement via the central charge constraint and the coefficient $c_\text{h}$ in the harmonic action. We perform the sums in three steps, presented in detail in \appref{app: PD2 calculation}. First, we exploit that $c_{\mathrm{h}}(n,n_{12},n_{21})$ only depends on the total number of oscillators $n$ and the total number of oscillators that change sites $n_{ij}$ in a one-loop process. We can hence cut the number of infinite sums in half via summation identities\footnote{These are in particular the identities \eqref{eq:sum_order} and \eqref{eq:sum_binomial}.} for binomial coefficients. Second, we rewrite the coefficient $c_{\mathrm{h}}(n,n_{12},n_{21})$ in terms of the following integral representation\footnote{For a trigonometric version of this integral representation see \cite{Zwiebel:2007cpa}.}
\begin{equation}\label{def:Harmonic_action_integral}
\begin{aligned}
c_{\mathrm{h}}(n,n_{12},n_{21})&= \int_0^1 \measure{t}\Bigl(c^{\text{int}}(n,n_{12},n_{21})-t^{-1}\text{-pole}\Bigr) \eqncom\\
c^{\text{int}}(n,n_{12},n_{21})&=2(-1)^{1+n_{12}n_{21}} t^{\frac12(n_{12}+n_{21})-1} (1-t)^{\frac12(n-n_{12}-n_{21})} \eqncom
\end{aligned}
\end{equation}
where the $t^{-1}$-pole prescription denotes the subtraction of the $t^{-1}$-pole that occurs when all oscillators stay at their initial sites, i.e.\ $n_{12}=n_{21}=0$. In using this representation, we can reduce the entanglement of the remaining six infinite sums by defining
\begin{align}
\ev{ P \mathfrak{D}^{L\geq 3}_2(w,y) }&=\int_0^1\measure{t} \left(\ev{ P\mathfrak{D}^{L\geq 3}_2(w,y) }_{\text{int}}-\frac{1}{t}\text{-pole}\right)\eqncom\label{eq:integral_representation_PD2_2}
\intertext{with}
\ev{ P \mathfrak{D}^{L\geq 3}_2(w,y) }_{\text{int}}&=\left.\ev{ P \mathfrak{D}^{L\geq 3}_2(w,y)} \right|_{c_{\mathrm{h}}(n,n_{12},n_{21})\rightarrow c^{\text{int}}(n,n_{12},n_{21})}\eqndot\label{eq:integral_representation_PD2}
\end{align}
Third, we perform the remaining six infinite sums in $\ev{ P \mathfrak{D}^{L\geq 3}_2(w,y) }_{\text{int}}$. Two of these can be eliminated via the central charge constraint at sites one and two. To further disentangle the remaining four sums, we write the global summand as a product of differential and integral operators that act on simpler expressions and are independent of two of the summation variables. This allows us to perform two infinite sums and apply the operators afterwards. The remaining two summations then become feasible via the generating functions of Legendre polynomials. A  minimal example of this procedure is
\begin{equation}
\sum_{n=0}^\infty (n+1)x^{n}=  \sum_{n=0}^\infty \Diff{x}{} x^{n+1}=  \Diff{x}{} \sum_{n=0}^\infty x^{n+1}=  \Diff{x}{} \frac{x}{1-x}=\frac{1}{(1-x)^2} \eqncom
\end{equation}
where the feasible summation is the geometric series. 

After the above three steps, we arrive at the final result in the asymptotic regime
\begin{equation}
\begin{aligned}\label{eq:Result_PD2}
\ev{ P\mathfrak{D}^{L\geq 3}_2(w,y) } &=4\Biggl(
\frac{w y(1+w^{1/2})^2(1+y^{1/2})^2}{(1-w^{1/2})^2(1-y^{1/2})^2(w^{1/2}+y^{1/2})^2(1+w^{1/2}y^{1/2})^3}f_1(w,y)\\
&\phantom{{}={}4\Bigl(}+\frac{wy}{(1-w)^2(1-y)^2(1+w^{1/2}y^{1/2})(1-wy)}\sum_{i=1}^3f_2(w,y,\gamma^\pm_i)\\
&\phantom{{}={}4\Bigl(}+f_3(w,y)\ln\left[\frac{1-w}{1-w y}\right]\Biggr)\\
&\phantom{{}={}}+w\leftrightarrow y\eqncom
\end{aligned}
\end{equation}
where
\begin{align}
f_1(w,y)&=2-16w^{1/2}+7w+11w^{1/2}y^{1/2}-16wy^{1/2}+w^{3/2}y^{1/2}+3wy\eqncom\\
f_2(w,y,\gamma^\pm_i)&=
\Bigl(\sin^2\frac{\gamma^+_i}{2}+\sin^2\frac{\gamma^-_i}{2}\Bigr)
\Bigl(12w^{1/2}-4w^{3/2}+4wy^{1/2}-4w^{3/2}y^{1/2}-4w^{3/2}y \nonumber \\*
&\phantom{{}={}}-4w^2y-8w^2y^{3/2}+6w^{1/2}y^{1/2}+6wy-2w^{3/2}y^{3/2}-2w^2y^2\Bigr)
\nonumber \\*
&\phantom{{}={}}+4\sin^2\frac{\gamma^+_i+\gamma^-_i}{2}\Bigl(1+w^{1/2}y^{1/2}-wy-w^{3/2}y^{3/2}\Bigr)\eqncom\\
f_3(w,y)&=-\frac{w(w^{1/2}+3y^{1/2})}{(w^{1/2}+y^{1/2})^3}+\frac{2-6y^{1/2}}{(1-y^{1/2})^3}-\frac{1+3w^{1/2}y^{1/2}}{(1+w^{1/2}y^{1/2})^3}\eqndot
\label{eq:result_functions_last}
\end{align}
We obtain the respective result for the $\beta$-deformation in the asymptotic regime from \eqref{eq:Result_PD2} by setting $\gamma_i^+=\beta$ and $\gamma_i^-=0$. In the limit of vanishing deformation parameters $\gamma_i^\pm=0$, the second line of \eqref{eq:Result_PD2} drops out and the original result \cite{Spradlin:2004pp} for \NfSYMt is reproduced\footnote{Our conventions for $\mathfrak{D}_2$ differ by a factor of $4$ with respect to \cite{Spradlin:2004pp}, which induces the same factor in $\ev{ P\mathfrak{D}^{L\geq 3}_2(w,y)}$.}.

\subsubsection{\texorpdfstring{The expectation value $\ev{\mathfrak{D}^{L\geq 3}_2(x)}$}{The expectation value <D2(x)>}}\label{sec: D2}
For the generalised expectation value of the one-loop dilatation operator density $\ev{\mathfrak{D}^{L\geq 3}_2(x)}$ we can follow almost the same steps as for $\ev{P\mathfrak{D}^{L\geq 3}_2(x)}$. In order to apply the techniques of the previous paragraph, we only need to define
\begin{equation}
\ev{ \mathfrak{D}^{L\geq 3}_2(w,y)}\equiv\sum_{A_1,A_2 \in \cal A} w^{(\mathfrak{D}_0)_{A_1}^{A_1}}y^{(\mathfrak{D}_0)_{A_2}^{A_2}} (\mathfrak{D}^{L\geq 3}_2 )^{A_1A_2}_{A_1A_2}\eqncom
\end{equation}
which we force to reduce to the original definition \eqref{eq: SV 3.10}  for $w=y=x$. It follows from \eqref{eq: deformation of D_2} that the matrix element $(\mathfrak{D}^{L\geq 3}_2 )^{A_1A_2}_{A_1A_2}$ entering this expression is independent of the deformation parameters $\gamma_i^\pm$. This guarantees that also the finite-size corrected matrix element at length $L=2$ is independent of the deformation $(\mathfrak{D}^{L\geq 3}_2 )^{A_1A_2}_{A_1A_2}=(\mathfrak{D}^{L=2}_2 )^{A_1A_2}_{A_1A_2}=(\mathfrak{D}^{\mathcal{N}=4}_2 )^{A_1A_2}_{A_1A_2}$, see \subsecref{sec:Q1Q2_neutral_states} for details. Therefore, the generalised expectation value of all these matrix elements can be calculated with the same techniques, yielding 
\begin{equation}\label{eq:D2equalities}
\ev{\mathfrak{D}^{L\geq 3}_2(x)}=\ev{\mathfrak{D}^{L=2}_2(x)}=\ev{\mathfrak{D}^{\mathcal{N}=4}_2(x)}\eqndot
\end{equation}
For the full result we find\footnote{The matrix element $(\mathfrak{D}^{\mathcal{N}=4}_2 )^{A_1A_2}_{A_1A_2}$ can be obtained from \eqref{eq: harminic action} by setting $A_{3}=A_{1}$ and $A_{4}=A_{2}$. In addition, we have to shift the summation variables according to
	\begin{equation}
	\begin{aligned}
	\atkind[\alpha]=\akindsite[\alpha]{1}-a^\alpha\eqncom\qquad
	\btkind[\alphadot]=\bkindsite[\alphadot]{1}-b^{\alphadot}\eqncom\qquad
	\ctkind[e]=\ckindsite[e]{1}-c^e\eqncom
	\end{aligned}
	\end{equation}
	which amounts to summing over oscillators that hop from one site to the other instead of oscillators that stay at their original positions.}
\begin{equation}\label{eq: D_2 in full theory}
\ev{\mathfrak{D}^{L\geq 3}_2(x)}=4\left(\frac{(1 + \sqrt{x})^2 }{ (1 - \sqrt{x})^6}
\left[ - (1 - 4 \sqrt{x} + x)^2 \ln(1-x) - x (1 - 8 \sqrt{x} + 2 x)\right] \right)\eqncom
\end{equation}
which agrees with the result\footnote{Recall the proportionality factor of four between our convention for $\mathfrak{D}_2$ and the one of \cite{Spradlin:2004pp}.} in \cite{Spradlin:2004pp}. The latter was obtained by means of the representation theory of \PSLs44.

\subsubsection{\texorpdfstring{The finite-size contributions $Z^{(1)}_{\text{f.s.c.}}(x)$}{The finite size contributions Zfsc}}\label{sec:finite_size_contributions}
Finally, we construct the one-loop finite-size corrections that occur for spin-chains of length $L=2$ and $L=1$ and calculate $Z^{(1)}_{\text{f.s.c.}}(x)$ defined in \eqref{eq:fsc}. 

For the deformed theories with gauge group \UN and vanishing tree-level multi-trace couplings, the one-loop finite-size contributions stem from the wrapping-corrected one-loop anomalous dimensions of $L=1$ operators given in \eqref{eq:U1_energies}. According to \eqref{eq: L=2 and L=1}, we find their contributions to the partition function to be
\begin{equation} 
\begin{aligned}\label{eq: Z_corr2 gamma in full theory}
Z_{\text{f.s.c.}\,\UN}^{(1)}
(x,\gamma_i^\pm)&=\sum_{A\in\mathcal{A}}x^{(\mathfrak{D}_0)_{A}^{A}}(\mathfrak{D}_2^{L=1})_A^A
=8\sum_{i=1}^3\left(\sin^2\frac{\gamma_i^+}{2}+\sin^2\frac{\gamma_i^-}{2}\right)\frac{x-x^3+x^{\frac 32}-x^{\frac 52}}{(1-x)^4}\eqncom
\end{aligned} 
\end{equation}
which is obtained in a similar fashion as $z(x)$ in the beginning of this subsection. For the $\beta$-deformation with gauge group \SUN the $L=1$ wrapping corrections are absent but we have to account for the $L=2$ prewrapping contributions, i.e.\ instead of the asymptotic dilatation operator density \eqref{eq: deformation of D_2}, the finite-size corrected density \eqref{eq:density_intro} has to be taken for spin-chains with $L=2$. Inserting \eqref{eq: L=2 and L=1} into \eqref{eq:fsc} and using \eqref{eq:D2equalities}, we find
\begin{equation} 
\begin{aligned}\label{eq: Z_corr in full theory}
Z_{\text{f.s.c.}\,\SUN}^{(1)}(x,\beta)&=
\ev{P\mathfrak{D}^{L=2}_2(x,x)} -\ev{P\mathfrak{D}^{L\geq 3}_2(x,x)}
=-6\frac{(x+x^{\frac 32})^2}{(1-x)^4}\left(8\sin^2\frac{\beta}{2}\right)\eqncom
\end{aligned} 
\end{equation}
where the computation of $\ev{P\mathfrak{D}^{L=2}_2(x,x)} $ is presented in appendix \ref{app: corrections}. The complete finite-size contributions \eqref{eq:fsc} are given by 
\begin{equation}\label{eq:full_fsc}
Z_{\text{f.s.c.}}^{(1)}(x,\gamma_i^\pm)=(1-s)Z_{\text{f.s.c.}\,\UN}^{(1)}(x,\gamma_i^\pm)
+sZ_{\text{f.s.c.}\,\SUN}^{(1)}(x,\beta)\eqncom
\end{equation}
where $s$ is the gauge group identifier defined in \eqref{eq:matrix_fields}. To obtain the finite-size contribution for each deformation, the deformation parameters in the arguments of \eqref{eq:full_fsc} are adjusted as: $\gamma^-_i= 0$ and $\gamma^+_i= \beta$ for the $\beta$-deformation; and $\beta=0$ for the $\gamma_i$-deformation.

In case of the $\beta$-deformation, the finite-size corrections of \eqref{eq: Z_corr2 gamma in full theory} and \eqref{eq: Z_corr in full theory} can be understood directly from the anomalous dimensions and characters of the theorie's supermultiplets that were identified to be affected by the finite-size corrections in \cite{Fokken:2013mza}. For example \eqref{eq: Z_corr in full theory} subtracts the contributions from the six multiplets whose primary states are the prewrapping-affected $L=2$ single-impurity states discussed in \subsecref{sec:Implications_AdSCFT_beta}. From the asymptotic dilatation operator these states receive a falsely assigned anomalous dimension of $8g^2\sin^2\frac{\beta}{2}$, which is subtracted here. Analogous considerations apply to the $L=1$ states in the case of wrapping corrections.

\subsection{Partition function and Hagedorn temperature}\label{sec:Hagedorn temperature}
From the ingredients calculated in the previous subsection, we can construct the partition functions of the undeformed \NfSYMt as well as the $\beta$- and $\gamma_i$-deformation with gauge group \UN or \SUN. The final result for the multi-trace partition function is obtained upon inserting \eqref{eq:mt_partition_function_res}, \eqref{eq:Result_PD2}, \eqref{eq: D_2 in full theory}, and \eqref{eq:full_fsc} into \eqref{eq: first order of multi-trace partition function}. However, since the result does not render significant simplifications\footnote{When the partition function is expanded perturbatively, the occurring terms can be understood in terms of the single-trace operator multiplets and their anomalous dimensions. Up to classical scaling dimension $\Delta_0\leq 4.5$, the one-loop anomalous dimensions in case of the \SUN $\beta$-deformation were determined in \cite{Fokken:2013mza}.}, we do not show it explicitly. For gauge group \UN, the multi-trace partition function for the $\beta$- and $\gamma_i$-deformation up to $\mathcal{O}(x^6)=\mathcal{O}(\e^{-6/RT})$ can also be obtained by slightly generalising the result of \cite{Mussel:2009uw}, which was obtained in a direct two-loop Feynman-diagrammatic calculation on \RxSt. The necessary generalising steps were presented in \cite{Fokken:2014moa} and the multi-trace partition function calculated in this fashion agrees with our result up to the presented thermal weight order $\order{x^6}$.

In the beginning of this section, we discussed that the critical temperature $\THag$ in our theories separates the two phases where the multi-trace partition function scales as $N^0$ and $N^2$, respectively. Therefore, in the \tHooft limit where $N\rightarrow\infty$, we can compute $\THag$ by analysing where our low-temperature partition function diverges. For the free \NfSYMt with $\gym=0=g$ we see from \eqref{eq:mt_partition_function_res}, that this partition function only diverges when the single-site partition function \eqref{eq: single-bead partition function of nfsymt} turns to  $z(x_{\H})=1$. For the critical temperature in $x_\H=\e^{-1/RT_\H}$ we find
\begin{equation}\label{eq: zero-loop Hagedorn temperature}
x_\H=\frac{1}{7+4\sqrt{3}}\eqncom\qquad
T_\H= \frac{1}{\ln(7+4\sqrt{3})}\frac 1R \eqncom
\end{equation}
which was first calculated in \cite{Sundborg:1999ue}. This critical temperature is also valid in the free deformed theories, since they reduce to the free \NfSYMt in the limit of vanishing interactions.

For the interacting \NfSYMt, the order $g^2$ correction to the critical temperature was calculated in \cite{Spradlin:2004pp}. Close to the temperature $\THag$, the free partition function has a simple pole $\cZ(x) \sim \frac{C}{x_\H-x}$ for some constant $C$ and upon expanding around this pole in the interacting theory, we have
\begin{equation}\label{eq: hagedorn temperature expansion}
\frac{C}{x_\H+\delta x_\H-x}=\frac{C}{x_\H-x}\left[1-\frac{\delta x_\H}{x_\H-x}+\dots\right] \eqndot
\end{equation}
Hence, we find the order $g^2$ correction to the critical temperature from the double-pole contributions in \eqref{eq: first order of multi-trace partition function}. Unfortunately, we cannot perform the infinite sums in this equation and therefore we cannot conclusively determine $\delta x_\H$. What we can do, however, is follow the reasoning of \cite{Spradlin:2004pp} and compare the expansion \eqref{eq: hagedorn temperature expansion} to the multi-trace partition function \eqref{eq: first order of multi-trace partition function}. We see that the second term in \eqref{eq: first order of multi-trace partition function} yields a double-pole contribution for $n=1$. For the remaining terms and generic summation indices it is not entirely clear whether there are further double-pole contributions. Numerical studies\footnote{In the \su{2}, \so{6}, and \su{2|3} subsectors, where the summation of \eqref{eq: first order of multi-trace partition function} is possible, no additional double-poles appear, compare \cite{Fokken:2014moa,thesis:Matthias}.} suggest, however, that no such poles appear. Since $\ev{\mathfrak{D}^{L\geq 3}_2(x)}$ as well as $z(x)$ are not affected by the deformations, the only double-pole contribution that we found is also not affected. From the \Polya-theoretic perspective this appears to be reasonable. Contributions to the second term in \eqref{eq: first order of multi-trace partition function} involve the element $\ev{\mathfrak{D}_2}$, which acts as a weighted identity operator on the incoming fields and hence all states with non-vanishing anomalous dimensions contribute. In contrast, contributions to the third term in \eqref{eq: first order of multi-trace partition function} involve the element $\ev{P\mathfrak{D}_2}$, which acts as a weighted permutation operator on the incoming fields and hence only states that are invariant under such a permutation with non-vanishing anomalous dimensions contribute. Analogously, only the few terms that are prewrapping affected at one-loop order contribute to the first term in \eqref{eq: first order of multi-trace partition function}. In conclusion, the double-pole of \eqref{eq: first order of multi-trace partition function} is obtained from the residue of the aforementioned $n=1$ term and we obtain
\begin{equation}\label{eq:delta_Hagedorn}
\delta x_\H= - \lim_{x\rightarrow x_\H} \left[g^2 (x_\H-x) \ln x\frac{\ev{\mathfrak{D}^{L\geq 3}_2(x)}}{1-z(x)}\right]=-\frac{2}{3}g^2x_\H  \ln x_\H\ev{\mathfrak{D}^{L\geq 3}_2(x_\H)} 
=-2g^2x_\H  \ln x_\H
\eqndot
\end{equation}
From this, we find that the one-loop correction to the critical temperature in the $\beta$- and $\gamma_i$-deformation with gauge group \UN or \SUN is given by its value in the undeformed \NfSYMt found in \cite{Spradlin:2004pp}. Using \eqref{eq: zero-loop Hagedorn temperature}, it takes the explicit form
\begin{equation}
T_\H(g)= T_\H\left(1+2 g^2+ \dots \right) \eqncom\qquad g^2=\frac{\lambda}{(4\pi)^2}=\frac{Ng_\YM^2}{(4\pi)^2}\eqncom
\end{equation}
which is obtained from \eqref{eq:delta_Hagedorn} using $\frac{\delta T_\H}{T_\H}=-\frac{1}{\ln x_\H}\frac{\delta x_\H}{x_\H}= 2 g^2$. In light of the discussion above, we can also conjecture, that the critical temperature in the \SUN $\beta$-deformation is independent of the deformation parameter at all orders in the effective planar coupling constant, since the number of deformation-dependent states is always small, compared to the total number of states that contribute to the partition function.

\subsection{Immediate implications for the \texorpdfstring{\AdSCFTc}{AdS/CFT correspondence}}\label{sec:results_parti}

We found that the temperature of the deconfinement phase transition at order $g^2$ does not depend on the deformation parameters $\beta$ or $\gamma_i^\pm$, despite the fact that the corresponding partition function depends on them. In light of the \AdSCFTc, similar results were obtained for certain string theories in a one-parameter deformed background \cite{Gursoy06}. There, the critical temperature was also found to be undeformed while the partition function exhibited a non-trivial deformation-dependence. The extension of this analysis to the string-theory dual of the $\beta$-deformation was only successful in sectors which do not lead to non-trivial tests \cite{HMP07}. From the present evidence, it is, however, tempting to assume that the critical temperature in the $\beta$-deformation is the same as in the undeformed \NfSYMt for all values of the effective planar coupling $g$. Clearly, further investigations in this matter are necessary. 


\chapter{Summary, conclusion, and outlook}\label{chap:Conclusion_outlook}

\section{Summary and conclusion}
In this thesis, we investigated the properties of the $\beta$- and the $\gamma_i$-deformation at first order in the \tHooft limit and determined important observables in both theories. Of these deformations only the $\beta$-deformation with gauge group \SUN is conformal. The $\beta$-deformation with gauge group \UN as well as the $\gamma_i$-deformation with either of those gauge groups are not conformally invariant, due to the running of multi-trace couplings.

For the $\gamma_i$-deformation, we showed the non-conformality explicitly by identifying running double-trace couplings that cannot be neglected in the \tHooft limit. The breakdown of conformality in the \tHooft limit encouraged us to reinvestigate the integrability-based calculation of the ground-state energies -- or anomalous dimensions -- in the $\gamma_i$-deformation. Exploiting the close relation to the parent \NfSYMt, we constructed the full $L$-loop wrapping corrections to the length-$L$ ground-states entirely from Feynman-diagrammatic considerations. For $L\geq 3$, where the running double-trace couplings cannot contribute, we reproduced the integrability-based result. For $L=2$, where the integrability-based calculation leads to a divergent result, we calculated the finite and renormalisation-scheme dependent anomalous dimension. At this length, the running double-trace coupling that we calculated earlier contributes to the anomalous dimension of the state via prewrapping and induces a renormalisation-scheme dependence. In addition, this calculation exemplifies that the double-trace coupling must not be neglected in the \tHooft limit. If we ignore contributions from this coupling, the anomalous dimension of the $L=2$ state is divergent and persistent non-local poles appear in the Feynman-diagrammatic calculation.

For the $\beta$-deformation, we also investigated the occurrence of prewrapping. For composite operators whose total $\mathfrak{u}(1)^{\times 3}$ Cartan charge is only non-zero in the $R$-symmetry component, we found that the structure constants and anomalous dimensions are given by the respective quantities in the undeformed \NfSYMt. For the remaining operators, we analysed in which subsectors prewrapping may occur in the \tHooft limit and we derived a procedure that adds the correct one-loop prewrapping contributions to the asymptotic dilatation-operator density of the $\beta$-deformation. By also including the one-loop wrapping contributions for gauge group \UN, we constructed all one-loop finite-size corrections and presented the complete one-loop dilatation operator of the $\beta$-deformation in the \tHooft limit.

With the explicit one-loop dilatation operator of the $\beta$ and $\gamma_i$-deformation at hand, we constructed the planar one-loop thermal partition function of these theories on the space \RxSt. For this, we generalised the \Polya-theoretic approach of \cite{Spradlin:2004pp} and developed tools to perform the strongly entangled infinite sums in fairly general situations. From the one-loop thermal partition function, we calculated the order $g^2$ correction to the phase-transition temperature $\THag$ where the low-energy description of the system in terms of colour-neutral composite operator \dof breaks down. We found that this correction is the same as in the undeformed theory, since the deformation parameters only affect parts that render little contributions to the partition function on combinatorial grounds.

The finite-size corrections to anomalous dimension of composite operators raise the important question what prewrapping and the loss of conformality imply for the integrability-based description of gauge theories. The current integrability-based formalism \cite{Gromov:2013pga,Gromov:2014caa,Gromov:2015dfa} is problematic for length $L=2$ states. In the undeformed \NfSYMt and the supersymmetric $\beta$-deformation the energies of the ground-states $\tr\bigl(\phi_i\phi_i\bigr)$ diverge. To regulate these divergences and to find the correct vanishing of the ground-state energies, an additional twist in the \AdS{5} directions must be introduced. While being effective for these states, this ad hoc regularisation fails to work in the $\beta$-deformation already for the $L=2$ single-impurity states $\tr\bigl(\phi_i\phi_j\bigr)$ which also have divergent energies in the integrability-based approach. In case of the $\gamma_i$-deformation, even the energies of the $L=2$ ground-states $\tr\bigl(\phi_i\phi_i\bigr)$ render divergent results that cannot be regulated, in contrast to our finite and renormalisation-scheme dependent result \eqref{gammaO2res}. Based on our findings in the $\beta$- and $\gamma_i$-deformation, we can now formulate a test to find the limits of the integrability-based approach to \NfSYMt and its deformations in the \tHooft limit. First, a description that includes prewrapping in the framework of integrability must be found for the $L=2$ single-impurity states in the $\beta$-deformation with gauge group \SUN. This description must exist, if the theory is integrable as claimed. Second, this modified description of $L=2$ states should be applied to the $L=2$ ground-states in $\gamma_i$-deformation. If the modified integrability-based description still fails to give a finite result for these states, the divergence can be attributed to the breakdown of conformality in the $\gamma_i$-deformation. This would imply that integrability-based methods are limited to conformally invariant theories. If the modified integrability-based description, however, yields a finite result, it is quite possible that this modified description is implicitly tied to a particular choice of renormalisation scheme. In this case, the integrability-based description still gives the correct results in the particular renormalisation scheme for a $\gamma_i$-deformation with a certain structure of multi-trace interactions. The form of the renormalisation scheme and the multi-trace interactions could then be determined by comparing the newly found finite result for the two-loop energy of the state $\tr\bigl(\phi_i\phi_i\bigr)$ with our result \eqref{gammaO2res}. By adjusting the tree-level coupling $Q_{\text{F}}$ and the parameter characterising the renormalisation-scheme dependence $\varrho$ in our result, it would be possible to deduce properties of the renormalisation scheme that enters the definition modified methods of integrability.

All findings based on Feynman diagrammatic calculations were obtained using the unified framework presented in this thesis for \NfSYMt and its deformations as well as the renormalisation program suitable for the renormalisation of composite operators. Hence, this thesis constitutes an independent test of the calculations in \cite{Fokken:2013aea,Fokken:2014soa,Fokken:2013mza,Fokken:2014moa}. In addition, our framework is designed to be compatible with the conventions of \cite{Srednicki:2007,Collins:1984xc} to enhance the applicability in other contexts. We also provide the \ttt{Mathematica} package \ttt{FokkenFeynPackage} which uses the conventions introduced in this thesis. In this package, we distilled the Feynman rules for non-abelian gauge theories with scalars and Weyl fermions in the adjoint representation to an efficient tool for the evaluation of low-loop Feynman diagrams in \NfSYMt and its deformations.

\section{Outlook}
The findings presented in this thesis leave room for many interesting research directions in the future.

Very prominently, we have the proposed test of integrability-based methods. Its successful implementation would clarify the prerequisites for quantum integrability in \NfSYMt and its deformations as mentioned earlier in this chapter. If the best case scenario is realised in which integrability-based methods can be used to find renormalisation-scheme dependent anomalous dimensions in the $\gamma_i$-deformation, the question arises whether the powerful methods of integrability can also be extended to other non-conformal theories -- like \QCD.

Concerning prewrapping contributions, we clearly need further data, especially from higher-loop prewrapping candidate states in the $\beta$-deformation, to understand the general effects of prewrapping on anomalous dimensions and structure constants in \CFTs. For the conformal $\beta$-deformation, we discussed that only multiplets containing the single-impurity operators $\tr(\phi^i\phi^j)$ are affected by prewrapping at first loop order in the \tHooft limit. For other candidate states, one-loop prewrapping is absent due to a cancellation among Feynman diagram contributions and it is not yet clear whether this cancellation is a one-loop accident or related to a deeper principle that occurs also in higher-loop calculations. In addition, it is still an open problem how prewrapping appears on the string-theory side of the \AdSCFTc. On the gauge-theory side, the prewrapping-affected single-impurity state is protected to all orders in perturbation theory. Hence, it should correspond to a supergravity mode and the calculation in \cite{Frolov:2005iq} suggests that corrections to the mass of the corresponding mode indeed vanishes. It would be interesting to explicitly check how the energy of non-protected modes, e.g.\ the mode dual to the state $\tr(\phi^1\phi^2\phi^3)$, is increased by the $\beta$-deformation. This analysis hopefully helps to clarify the subtleties related to the choice of \UN or \SUN as gauge group on the string-theory side.

In a greater perspective, naturally the construction of the complete two-loop dilatation operator of \NfSYMt in the \tHooft limit would be very desirable. This task is highly complicated due to operator mixing and the correct identification of UV divergences in all contributing diagrams and for attempts in finding the two-loop dilatation operator see e.g.\  \cite{Belitsky:2005bu,Georgiou:2011xj,Loebbert:2015ova} and references therein. However, if this object becomes available, also the asymptotic dilatation operator of the $\beta$-deformation would follow and the question would arise anew, how wrapping and prewrapping must be implemented in the complete two-loop dilatation operator in the deformed theories. 

For the two-loop thermal partition function of \NfSYMt in the \tHooft limit, the two-loop dilatation operator also is the bottleneck input. If this object is at hand, the \Polya-theoretic methods of \cite{Spradlin:2004pp} still need to be generalised to second loop order. For the \su{2} sector of \NfSYMt, this generalisation was done in \cite{GRNS05} and it seems possible that these considerations can be generalised further to be also applicable in the full theory. If so, the computation of the partition function becomes a purely combinatorial exercise. Hopefully, the tools that we developed to evaluate the occurring entangled sums could be applied and/or modified to still render a closed expression for the two-loop thermal partition function. From the partition function, the second order correction to the critical temperature $\THag$ could be calculated, which would allow to determine the order of the confinement/deconfinement phase-transition in \NfSYMt on \RxSt, in analogy to the case of pure Yang-Mill theory studied in \cite{AMMPR05}. By employing the planar asymptotic two-loop dilatation operator of the $\beta$-deformation, the analogous calculation could be done in the deformed theory, at least asymptotically. This would then allow to test our conjecture, that $\THag$ is always independent of the deformation parameter $\beta$.

The detailed description of \NfSYMt and the renormalisation program that we gave in this thesis also opens the path to investigate the connection between quantum integrability and the perturbative structure of this theory. As discussed in the introduction, the anomalous dimensions of composite operators in \NfSYMt can efficiently be determined using the methods of integrability. It is, however, not yet clear how integrability emerges in the Feynman-diagrammatic description. This question can be approached in the Schwinger-Dyson formalism where the conformal invariance of the theory can be efficiently employed, see \cite{Kreimer:2005rw,EbrahimiFard:2005gx,Kreimer:2006ua,Kreimer:2009jt} for a modern formulation. In this description, all perturbative contributions are sorted in terms of primitive (subdivergence-free) Feynman diagrams and connected insertions of vertices and composite operators into such primitive diagrams. The exact conformality of the theory ensures that all connected vertex insertions in Feynman diagrams do not yield any additional divergent contributions since the corresponding connected renormalisation constant is $\cZ_v=1$. Therefore, in a correlation function with a composite operator insertion, the overall UV divergence is determined from the finite parts of the connected vertex functions and the numerical values of the primitive Feynman diagrams that contribute to the process. This also extends to the anomalous dimension of the composite operator since it is determined from the overall UV divergence of all Feynman diagrams with one composite operator insertion. As mentioned above, the anomalous dimensions can also be determined by the methods of integrability and therefore it should be possible to see the emergence of integrability by analysing the contributions to anomalous dimensions in the Dyson-Schwinger approach. Since the connected vertex functions must fulfil the supersymmetric Slavnov-Taylor identities of \NfSYMt, this emergence depends on these identities and the exact numerical values of the primitive Feynman diagrams. A starting point of such an investigation could be the construction of the two-loop Konishi anomalous dimension via the correlation function $\vacl\T \phi^a_k\ol{\phi}^{lb}\tr\bigl(\phi_i\ol{\phi}^i\bigr)\vac$ in the Dyson-Schwinger approach.

\ifpdf
\begin{figure}[!b]
	\begin{flushright}

\settoheight{\eqoff}{$+$}%
\setlength{\eqoff}{0.5\eqoff}%
\addtolength{\eqoff}{0\unit}%
\raisebox{\eqoff}{%
					\fbox{
	\includegraphics[angle={0},scale=0.06,trim=0cm 0cm 0cm 0]{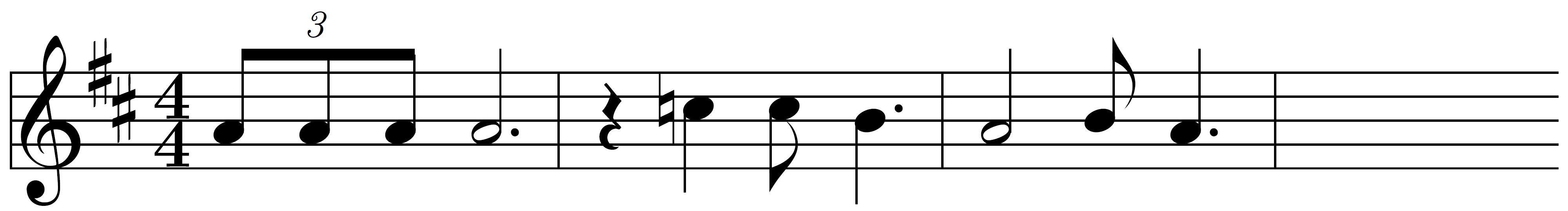}
					}
}
\end{flushright}
\end{figure}
\else
	\includegraphics[trim =4.1cm 29cm 0cm 7cm]{./jpggraphs/ThisIsTheEnd.ps}
\fi

\clearpage

\chapter*{Acknowledgements}
\markboth{Acknowledgements}{}
\addcontentsline{toc}{chapter}{Acknowledgements}%

It is a pleasure to thank my advisor Matthias Staudacher for accepting me as an outsider at the time in his group as a PhD candidate. I had the privilege to have very interesting and inspiring discussions with him as well as the members of his group.

I am also deeply grateful to Christoph Sieg for the dedication with which he fulfilled the task of being my coadvisor. He always made room for questions I had and encouraged me to learn many aspects of general physics and Feynman calculus in particular.

For the fruitful, inspiring, and at times challenging collaborations on the projects that I was involved in, I am much obliged to 
Christoph Sieg, David Meidinger, Gregor Richter, and Matthias Wilhelm.

I greatly benefited from the many opinions and expert views on various topics and for the many discussions I would like to thank Burkhard Eden, Ewa Felinska, Valentina Forini, Rouven Frassek, Sergey Frolov, Philip Hähnel, Nils Kanning, Yumi Ko, Laura Koster, Pedro Liendro, Tomasz \L{}ukowski, David Meidinger, Vladimir Mitev, Dhitiman Nandan, Brenda Penante, Jan Plefka, Elli Pomoni, Gregor Richter, Radu Roiban, Henning Samtleben, Volker Schomerus, Alessandro Sfondrini, Christoph Sieg, Vladimir Smirnov, Marcus Spradlin, Mat\-thi\-as Staudacher, Stijn van Tongeren, and Vitaly Velizhanin.

For reading the manuscript, I am grateful to Christoph Sieg, David Meidinger, Gregor Richter, and Julia Dierck.

I would also like to thank Matthias Staudacher for giving me the opportunity to see many places in the world that I would not have gotten to otherwise. In this regard I thank the Kavli Institute for the Physics and Mathematics of the Universe in Tokyo, and the Israel Institute for Advanced Studies of Jerusalem for warm hospitality during different stages of my PhD studies.

This thesis was supported through a PhD fellowship program by the \glqq Studienstiftung des deutschen Volkes\grqq,
by the DFG, SFB 647 \emph{Raum -- Zeit -- Materie. Analytische und Geometrische Strukturen}, the Marie Curie network GATIS (gatis.desy.eu) of the European Union's Seventh Framework Programme FP7/2007-2013/ under REA Grant Agreement No 317089 and the Marie Curie International Research Staff Exchange Network UNIFY (FP7-People-2010-IRSES under Grant Agreement No 269217),
and by the resolved support of my parents.

Ich danke Eike, Christine und Fokke für ihre unumstößliche Hilfe, Unterstützung und konstruktive Kritik in allen Lebenslagen.

Schließlich möchte ich Julia danken. Ohne sie wäre diese Arbeit und mein Leben im Allgemeinen nicht vollständig.



\chapter*{Appendix}

\addcontentsline{toc}{chapter}{Appendix}
\markboth{Appendix}{}
\numberwithin{equation}{section}
\setcounter{section}{0}
\renewcommand{\theHsection}{\Alph{section}}
\renewcommand{\thesection}{\Alph{section}}
\setcounter{table}{0}
\renewcommand{\thetable}{\Alph{section}.\arabic{table}}

\section{Conventions and list of used symbols and abbreviations}\label{sec:Conventions}
In this section, we present general conventions, identities, and in \tabref{tab:abbreviations} the list of abbreviations and symbol explanations used in this thesis.

\begin{description}
	\item[Einstein's summation convention] is implied in this thesis, unless explicitly stated otherwise. In a spacetime with metric $g$, it takes the form $a^\mu b_\mu=\sum_{\mu\nu}g_{\mu\nu}a^\mu b^\nu$.
	\item[Time orderd products] of operators $\T \mathcal{O}_1(t_1)\dots \mathcal{O}_n(t_n)$ are evaluated not as they are written, but such that operators at earlier times appear to the right of operators at later times \cite{PhysRev.80.268}, e.g.\ 
	\begin{equation}\label{eq:time_ordered_product}
	\T \mathcal{O}_1(t_1) \mathcal{O}_2(t_2)=\begin{cases}
	\mathcal{O}_1(t_1) \mathcal{O}_2(t_2)&t_1\geq t_2\eqncom\\
	\omega^{2\mathfrak{D}_0(\mathcal{O}_1)\mathfrak{D}_0(\mathcal{O}_2)}\mathcal{O}_2(t_2) \mathcal{O}_1(t_1)&t_1< t_2\eqncom
	\end{cases}
	\end{equation}
	where the prefactor in the second line involving $\omega$ is $(-1)$ if $\mathcal{O}_1$ and $\mathcal{O}_2$ are fermionic and $(+1)$ otherwise. Its explicit definition can be found below \eqref{eq:graded_Jacobi_identity}. When the time-ordering operator $\T$ appears in products of momentum-space objects, the pole structure of these objects must be chosen such that the  Fourier-transformed position-space expression is time ordered. This definition is compatible with the one given in \cite{Collins:1984xc}.
	\item[Fourier transforms] of fields are always taken in the convention 
	\begin{equation}\label{eq:Fourier_transformation}
	\tilde{\varphi}(k)=\int\de^d x\e^{-\complexi kx}\varphi(x)\eqncom\qquad
	\varphi(x)=\int\frac{\de^d k}{(2\pi)^{d}}\e^{\complexi k x}\tilde{\varphi}(k)\eqndot
	\end{equation}
	This definition of the Fourier transform agrees with the definitions in \cite{Srednicki:2007,Dreiner:2008tw} and yields the following realisation of the $\delta$-distribution $\delta^{(D)}(k)=\int \frac{\de^D x}{(2\pi)^{D}}\e^{\complexi k x}$.
	\item[Wick rotation] of a momentum-space vector $q$ is given by a counter-clockwise $90^{\circ}$ rotation in the complexified $q^0$-plane. This results in the transformations $q^0\Rightarrow \complexi \bar q_0$ and $q^j\Rightarrow  \bar q_j$ for $j\neq 0$ and Euclidean coordinates $\bar q$. The respective Wick rotation in coordinate space is chosen in the opposite direction in the complex plane, in accord with \cite[\chap{9}]{Peskin:1995ev}. Explicitly, we have for a position-space vector $x$ a clockwise $-90^{\circ}$ rotation in the complexified $x^0$-plane and the transformations $x^0\Rightarrow -\complexi \bar x_0$ and $x^j\Rightarrow  \bar x_j$ for $j\neq 0$. This choice guarantees that the Fourier transformation with the function $\e^{\complexi q x}$ is always a phase.
	
	For this choice of analytic continuation in momentum- and position-space, we use the following complex structure of the two-point function in four-dimensional Minkowski space
	\begin{equation}
	\vacl\T \cO(x)\ol{\cO}(y)\vac=\frac{1}{(\abs{x-y}^2+\complexi \epsilon)^{\Delta_{\cO}}}\eqncom\qquad
	\vacl\T \cO(p)\ol{\cO}(-p)\vac_{\complexi \cT}=\frac{1}{(p^2-\complexi \epsilon)^{\frac D2-\Delta_{\cO}}}\eqncom
	\end{equation}
	where $\epsilon>0$ is an infinitesimal parameter. See \appref{sec:the-fourier-transformation-of-the-free-two-point-function} for the relation between both correlators.
	\item[Colour space generators] of the groups \UN or \SUN fulfil the following relations
	\begin{equation}
	\tr\left(\T^a\right)=\sqrt{N}\delta^{a0}\eqncom\qquad
	\tr\left(\T^a \T^b\right)=\delta^{ab}\eqncom\qquad
	[\T^a,\T^b]=\complexi f^{abc}\T^c\eqndot
	\end{equation}
	\item[Traces over colour space generators] are often abbreviated as
	\begin{equation}\label{eq:colour_generator_trace}
	(a_1a_2\dots a_n)\equiv \tr\bigl[\T^{a_1}\T^{a_2}\dots \T^{a_n}\bigr]\eqndot
	\end{equation}
	\item[Signs in a Lagrangian] description appear in accord with the conventions of \cite{Srednicki:2007} with mostly plus metric $\eta_{\mu\nu}=\diag(-1,1,1,1)$, e.g.\ 
	\begin{equation}
	\mathcal{L}=-\frac 12 \partial^\mu\varphi\partial_\mu\varphi-\frac 12 m_{\varphi}^2\varphi^2
	+\complexi \ol{\Psi}\gamma^\mu\partial_\mu\Psi- m_{\Psi}\ol{\Psi}\Psi
	-\frac 14 F^{\mu\nu}F_{\mu\nu}-\frac 12 m^2_A A^\mu A_\mu\eqndot
	\end{equation}
	\item[Covariant derivatives] assume the form $\D_\mu=\partial_\mu-\complexi \gym A_\mu\eqndot$
	\item[The field strength tensor] is given by
	\begin{equation}\label{eq:F_mu_nu}
	F_{\mu\nu}=\frac{\complexi}{\gym}[\D_\mu,\D_\nu]=\D_\mu A_\nu-\D_\nu A_\mu\eqndot 
	\end{equation}
	\item[(Anti-)selfdual projectors] in $\RR^{(3,1)}$ take the explicit form
	\begin{equation}\label{eq:selfdual_projectors}
	\Pi_+^{\mu\nu\rho\gamma}=\frac 14(\eta^{\mu\rho}\eta^{\nu\gamma}-\eta^{\mu\gamma}\eta^{\nu\rho}+\complexi\varepsilon^{\mu\nu\rho\gamma})\eqncom\qquad
	\Pi_-^{\mu\nu\rho\gamma}=\frac 14(\eta^{\mu\rho}\eta^{\nu\gamma}-\eta^{\mu\gamma}\eta^{\nu\rho}-\complexi\varepsilon^{\mu\nu\rho\gamma})\eqncom
	\end{equation}
	and they have the usual properties of projectors, i.e.\ $\Pi_\pm^2=\Pi_\pm$, $\Pi_+\Pi_-=0$ and $(\Pi_++\Pi_-)=\text{id}$ on functions that are antisymmetric in two spacetime indices. 
	\item[The $\sigma^{\mu\nu}$ and $\bar\sigma^{\mu\nu}$] symbols are defined in terms of the $\sigma$- and $\bar{\sigma}$ matrices given in \eqref{eq:sigma_matrices} as
	\begin{equation}
	\begin{aligned}\label{eq:sigma_mu_nu}
	(\sigma^{\mu\nu})_\alpha^{\phan{\alpha}\beta}&=\frac{\complexi}{4}\bigl(\sigma^{\mu}\bar\sigma^{\nu}-\sigma^{\nu}\bar\sigma^{\mu}\bigr)_\alpha^{\phan{\alpha}\beta}
	=\frac{\complexi}{4}\bigl(\delta_{\alpha}^{\phan{\alpha}\rho}\delta^\beta_{\phan{\alpha}\omega}+\varepsilon_{\alpha\omega}\varepsilon^{\beta\rho}\bigr)(\sigma^\mu)_{\rho\dot{\alpha}}(\bar\sigma^\nu)^{\dot\alpha\omega}
	\eqncom\\
	(\bar\sigma^{\mu\nu})^{\dot\alpha}_{\phan{\alpha}\dot\beta}
	&=\frac{\complexi}{4}\bigl(\bar\sigma^{\mu}\sigma^{\nu}-\bar\sigma^{\nu}\sigma^{\mu}\bigr)^{\dot\alpha}_{\phan{\alpha}\dot\beta}
	=\frac{\complexi}{4}\bigl(\delta^{\dot\alpha}_{\phan{\alpha}\dot\rho}\delta_{\dot\beta}^{\phan{\alpha}\dot\omega}
	+\varepsilon_{\dot\beta\dot\rho}\varepsilon^{\dot\alpha\dot\omega}\bigr)
	(\bar\sigma^\mu)^{\dot\rho\alpha}(\sigma^\nu)_{\alpha\dot\omega}\eqndot
	\end{aligned}
	\end{equation}
	\item[Under hermitian conjugation], we have the following relations
	\begin{equation}
	\begin{aligned}\label{eq:sigma_hermitian_conjugation}
	\bigl((\sigma^{\mu\nu})_\alpha^{\phan{\alpha}\beta}\bigr)^\dagger&=(\bar\sigma^{\mu\nu})^{\dot\alpha}_{\phan{\alpha}\dot\beta}\eqncom\qquad
	\bigl((\sigma^\mu)_{\alpha\dot{\beta}}\bigr)^\dagger&=(\sigma^\mu)_{\beta\dot{\alpha}}\qquad
	\bigl((\bar\sigma_\mu)^{\dot\alpha\beta}\bigr)^\dagger&=(\bar\sigma_\mu)^{\dot\beta\alpha}\eqndot
	\end{aligned}
	\end{equation}
	\item[Identities for $\sigma^\mu$ and $\bar{\sigma}^\mu$ matrices], which are defined in \eqref{eq:sigma_matrices} and are commonly used throughout this thesis are
	\begin{equation}
	\begin{aligned}\label{eq:sigma_matrix_identities_2}
	(\sigma^\mu)_{\alpha\dot{\alpha}}(\sigma_\mu)_{\beta\dot{\beta}}&=-2\varepsilon_{\alpha\beta}\varepsilon_{\dot\alpha\dot\beta}\eqncom\\
	(\sigma^\mu)_{\alpha\dot{\alpha}}(\bar\sigma_\mu)^{\dot\beta\beta}&=-2\delta_{\alpha}^{\phan{\alpha}\beta}
	\delta^{\dot\beta}_{\phan{\alpha}\dot\alpha}\eqncom\\
	(\sigma_\mu)_{\alpha\dot{\alpha}}(\bar\sigma_\nu)^{\dot{\alpha}\alpha}&=-2\eta_{\mu\nu}\eqncom\\
	(\sigma_\mu)_{\alpha\dot{\alpha}}(\bar\sigma_\nu)^{\dot{\alpha}\beta}&=-\delta_{\alpha}^{\phan{\alpha}\beta}\eta_{\mu\nu}-2\complexi (\sigma_{\mu\nu})_{\alpha}^{\phan{\alpha}\beta}\eqncom\\
	(\bar\sigma_\mu)^{\dot{\alpha}\alpha}(\sigma_\nu)_{\alpha\dot\beta}&=-\delta^{\dot{\alpha}}_{\phan{\alpha}\dot\beta}\eta_{\mu\nu}-2\complexi (\bar\sigma_{\mu\nu})^{\dot{\alpha}}_{\phan{\alpha}\dot\beta}\eqncom\\
	(\bar\sigma_\mu)^{\dot{\alpha}\alpha}(\sigma_\nu)_{\beta\dot\beta}-
	(\bar\sigma_\nu)^{\dot{\alpha}\alpha}(\sigma_\mu)_{\beta\dot\beta}
	&=-2\complexi\Bigl(\delta^\alpha_\beta(\bar\sigma_{\mu\nu})^{\dot{\alpha}}_{\phan{\alpha}\dot{\beta}}
	-\delta^{\dot{\alpha}}_{\dot{\beta}}(\sigma_{\mu\nu})^{\phan{\alpha}\alpha}_{\beta}\Bigr)\eqndot
	\end{aligned}
	\end{equation}
	\item[Identities involving $\sigma^{\mu\nu}$ and $\bar\sigma^{\mu\nu}$]
	\begin{equation}
	\begin{aligned}\label{eq:sigma_munu_projector}
	2(\sigma^{\mu\nu})_\alpha^{\phan{\alpha}\beta}(\sigma^{\rho\gamma})_\beta^{\phan{\alpha}\alpha}&=
	\eta^{\mu\rho}\eta^{\nu\gamma}-\eta^{\mu\gamma}\eta^{\nu\rho}+\complexi\varepsilon^{\mu\nu\rho\gamma}\eqncom\\
	2(\bar\sigma^{\mu\nu})^{\dot\alpha}_{\phan{\alpha}\dot\beta}(\bar\sigma^{\rho\gamma})^{\dot\beta}_{\phan{\alpha}\dot\alpha}&=
	\eta^{\mu\rho}\eta^{\nu\gamma}-\eta^{\mu\gamma}\eta^{\nu\rho}-\complexi\varepsilon^{\mu\nu\rho\gamma}\eqncom\\
	(\sigma_{\mu\nu})_\alpha^{\phan{\alpha}\beta}
	(\sigma^{\mu\nu})_\gamma^{\phan{\alpha}\omega}
	&=\delta_\alpha^\omega\delta_\gamma^\beta+\varepsilon_{\alpha\gamma}\varepsilon^{\beta\omega}=
	2\delta_\alpha^\omega\delta_\gamma^\beta-\delta_\alpha^\beta\delta_\gamma^\omega\eqncom\\
	(\bar\sigma_{\mu\nu})^{\dot\beta}_{\phan{\alpha}\dot\alpha}
	(\bar\sigma^{\mu\nu})^{\dot\omega}_{\phan{\alpha}\dot\gamma}
	&=\delta^{\dot\omega}_{\dot\alpha}\delta^{\dot\beta}_{\dot\gamma}+\varepsilon_{\dot\alpha\dot\gamma}\varepsilon^{\dot\beta\dot\omega}=
	2\delta^{\dot\omega}_{\dot\alpha}\delta^{\dot\beta}_{\dot\gamma}-\delta^{\dot\beta}_{\dot\alpha}\delta^{\dot\omega}_{\dot\gamma}\eqncom\\
	(\sigma^{(\mu})_{\alpha\dot{\alpha}}(\sigma^{\nu\rho)})_{\beta}^{\phan{\alpha}\gamma}&=
	\varepsilon^{\mu\nu\rho\omega}(\sigma_\omega)_{\delta\dot{\alpha}}
	\Bigl(\delta^\gamma_\alpha\delta^\delta_\beta-\frac 12\delta_\alpha^\delta\delta^\gamma_\beta\Bigr)\eqncom\\
	(\sigma^{(\mu})_{\alpha\dot{\alpha}}(\bar\sigma^{\nu\rho)})^{\dot{\gamma}}_{\phan{\alpha}\dot\beta}
	&=
	\varepsilon^{\mu\nu\rho\omega}(\sigma_\omega)_{\alpha\dot{\delta}}
	\Bigl(\delta^{\dot{\gamma}}_{\dot{\alpha}}\delta^{\dot{\delta}}_{\dot{\beta}}-\frac 12\delta^{\dot{\gamma}}_{\dot{\beta}}\delta^{\dot{\delta}}_{\dot{\alpha}}\Bigr)\eqncom
	\end{aligned}
	\end{equation}	
	where the Minkowski space Levi-Civita symbol $\varepsilon^{0123}=-\varepsilon_{0123}=1$ was used. In the last two lines on the \lhs, the round parentheses in superscripts denote that the enclosed spacetime indices are cyclically symmetrised.
	\item[(Anti-)selfdual field strength tensors in spinorial indices] can be acquired from the definitions \eqref{eq:F_mu_nu} and \eqref{eq:sigma_mu_nu} and they read
	\begin{equation}
	\begin{aligned}
	\cF_{\alpha}^{\phan{\alpha}\beta}&=\frac{1}{2}F^{\mu\nu}(\sigma_{\mu\nu})_{\alpha}^{\phan{\alpha}\beta}
	=\frac{\complexi}{4}\bigl(\delta_{\alpha}^{\phan{\alpha}\rho}\delta^\beta_{\phan{\alpha}\omega}+\varepsilon_{\alpha\omega}\varepsilon^{\beta\rho}\bigr)\D_{\rho\dot{\alpha}}A^{\dot{\alpha}\omega}\eqncom\\
	\bar\cF^{\dot\alpha}_{\phan{\alpha}\dot\beta}&=\frac{1}{2}F^{\mu\nu}(\bar\sigma_{\mu\nu})^{\dot\alpha}_{\phan{\alpha}\dot\beta}
	=\frac{\complexi}{4}\bigl(\delta^{\dot\alpha}_{\phan{\alpha}\dot\rho}\delta_{\dot\beta}^{\phan{\alpha}\dot\omega}
	+\varepsilon_{\dot\beta\dot\rho}\varepsilon^{\dot\alpha\dot\omega}\bigr)\D^{\dot\rho\alpha}A_{\alpha\dot\omega}\eqncom\\
	\end{aligned}
	\end{equation}
	where the spinorial fields are defined as in \eqref{eq:vectorfield_spinorial}, e.g.\ $X_{\alpha\dot{\alpha}}=-\complexi(\sigma^\mu)_{\alpha\dot{\alpha}}X_\mu$.
	\item[The variation of the selfdual field strength] can be written as
	\begin{equation}
	\begin{aligned}\label{eq:selfdual_field_variation}
	\delta\cF_{\alpha\beta}&=
	\frac{1}{2}\varepsilon_{\beta\gamma}(\sigma^{\mu\nu})_\alpha^{\phan{\alpha}\gamma}\bigl[\partial^\mu\delta A_x^\nu-\partial^\nu\delta A_x^\mu+\gym f^{yzx}(\delta A^\mu_y A^\nu_z+A^\mu_y\delta A^\nu_z) \bigr]\T^x\\
	&=\frac{\complexi}{4}
	\bigl(\delta_{\alpha}^{\rho}\varepsilon_{\beta\omega}
	+\delta^{\rho}_\beta\varepsilon_{\alpha\omega}\bigr)
	\bigl(\partial_{\rho\dot{\alpha}}\delta A_x^{\dot{\alpha}\omega}+\gym f^{xyz} A_{y\rho\dot{\alpha}} \delta A_z^{\dot{\alpha}\omega} \bigr)T^x\\
	&=\frac{\complexi}{4}
	\bigl(\delta_{\alpha}^{\rho}\varepsilon_{\beta\omega}
	+\delta^{\rho}_\beta\varepsilon_{\alpha\omega}\bigr)
	\D_{\rho\dot{\alpha}}\delta A^{\dot{\alpha}\omega}
	=-\frac{\complexi}{4}
	\bigl(\delta_{\alpha}^{\rho}\delta_{\beta}^\omega
	+\delta^{\rho}_\beta\delta_{\alpha}^\omega\bigr)\varepsilon^{\dot{\alpha}\dot{\beta}}
	\D_{\rho\dot{\alpha}}\delta A_{\omega\dot{\beta}}\eqncom
	\end{aligned}
	\end{equation}
	where the covariant derivative in the last line acts as a commutator in colour space.
	\item[The pure gauge field term] in a Lagrangian description in spacetime component fields takes the explicit form
	\begin{equation}
	\begin{aligned}
	-\frac 14F^{\mu\nu}F_{\mu\nu}&=
	-\frac 14\Bigl(\partial^{[\mu}A^{\nu]}-\complexi \gym A^{[\mu}A^{\nu]}\Bigr)
	\Bigl(\partial_{[\mu}A_{\nu]}-\complexi \gym A_{[\mu}A_{\nu]}\Bigr)\\
	&=-\frac 14\Bigl(\partial^{[\mu}A^{\nu]}\partial_{[\mu}A_{\nu]}
	-2\complexi \gym \partial^{[\mu}A^{\nu]}A_{[\mu}A_{\nu]}
	-\gym^2 A^{[\mu}A^{\nu]}A_{[\mu}A_{\nu]}\Bigr)\\
	&=-\frac 12(\partial^\mu A^{a \nu}\partial_\mu A^a_{\nu}-\partial^\mu A^{a \nu}\partial_\nu A^a_{\mu})
	-\gym \partial^{\mu}A^{c \nu}A^a_{\mu}A^b_{\nu}f^{cab}\\
	&\phantom{=}{}
	-\frac{\gym^2}{4} A^{a \mu}A^{b \nu}A^c_{\mu}A^d_{\nu}f^{abe}f^{cde}\eqndot
	\end{aligned}
	\end{equation}
	In spinor indices we use \eqref{eq:sigma_munu_projector} to write it as
	\begin{equation}
	\begin{aligned}
	-\frac 14F^{\mu\nu}F_{\mu\nu}&=-\frac 18F^{\mu\nu}F^{\rho\sigma}(\eta_{\mu\rho}\eta_{\nu\sigma}-\eta_{\mu\sigma}\eta_{\nu\rho})\\
	&=-\frac 18F^{\mu\nu}F^{\rho\sigma}\bigl(
	(\sigma_{\mu\nu})_{\alpha}^{\phan{\alpha}\beta}(\sigma_{\rho\sigma})_{\beta}^{\phan{\alpha}\alpha}
	+(\bar\sigma_{\mu\nu})^{\dot\alpha}_{\phan{\alpha}\dot\beta}(\bar\sigma_{\rho\sigma})^{\dot\beta}_{\phan{\alpha}\dot\alpha}
	\bigr)\\
	&=-\frac 12\Bigl(
	\cF_{\alpha\beta}\cF^{\alpha\beta}
	+\bar{\cF}_{\dot{\alpha}\dot\beta}\bar{\cF}^{\dot{\alpha}\dot\beta}
	\Bigr)\eqndot
	\end{aligned}
	\end{equation}
\end{description}
\begin{table}[H]
	\caption{List of abbreviations and synonymous expressions.}
	\label{tab:abbreviations}
	\begin{tabularx}{\textwidth}{|c|X|}
		\hline
		abbreviation/symbol & synonymous expressions\\
		\hline
		&classical level; tree-level\\
		&\tHooft limit; planar limit -- see \secref{sec:tHooft_limit}\\
		\hline
		$\simeq$ & isomorphic\\
		$\otimes$ & tensor product\\
		$\times$ & direct product in representation-theoretic context; otherwise ordinary product\\
		$\floor{\cdot}$ & floor function\\
		$k|L$ & $k$ is a divisor of $L$\\
		$(m,L)$ & greatest common divisor of $m$ and $L$\\
		$\EulerPhi$ & Euler totient function\\
		$\RR^{(p,q)}$ & real vector space of dimension $p+q$ and a metric with $p$ and $q$ entries of $-1$ and $+1$, respectively\\
		1PI& one-particle irreducible (diagram); a connected diagram which does not separate into two disconnected ones if a single line is cut\\
		$\ol{\text{1PI}}$& non-1PI (diagram); a connected diagram which separates into two disconnected ones if a single line is cut\\
		$X_{\text{B}}$ & the subscript $B$ indicates that $X$ is the bare quantity which is not yet renormalised\\
		$X^{\text{M}}$ & the superscript $M$ indicates that $X$ is a spinor that fulfils the Majorana constraint -- see \appref{sec:spinor_in_various_dimensions}\\
		$X^{\pm}$ & the superscript $\pm$ indicates that $X$ is an eigen-spinor to the helicity projector $P^\pm$ -- see \appref{sec:spinor_in_various_dimensions}\\
		$\beta$-deformation & deformation of \NfSYMt with a single real parameter -- see \secref{sec:The_deformations}\\
		AdS & anti de-Sitter\\
		\AdSCFTc & correspondence between a superstring theory and a \CFT, proposed in \cite{Maldacena:1997re,Witten:1998qj,Witten:1998zw}\\
		\cf & confer\\
		\CFT & conformal field theory\\
		\dof & degrees of freedom\\
		DR & dimensional reduction (scheme) -- see \appref{sec:Renormalisation_schemes}\\
		$\ol{\text{DR}}$& modified dimensional reduction (scheme) -- see \appref{sec:Renormalisation_schemes}\\
		$\delta_X$  & 1PI counterterm of the quantity $X$ -- it is given by the negative sum of divergent 1PI contributions that involve $X$\\
		$\mathfrak{d}_X$  & connected counterterm of the quantity $X$ -- it is given by the negative sum of divergent connected contributions that involve $X$\\
		e.g. & `exemplum gratia'; for example\\
		$\mathbf{e}_i$ & unit vector in the $i^{\text{th}}$ direction\\
		\eom & equation of motion\\
		$\Gamma^M$, $\gamma^\mu$, $\rho^a$ & gamma matrices of $\RR^{(9,1)}$, $\RR^{(3,1)}$ and $\RR^{(6)}$, respectively -- \cf \appref{sec:Clifford_algebras_in_various_dimensions}\\
		$g=\sqrt{\lambda}(4\pi)^{-1}$ & effective planar coupling constant\\
	\end{tabularx}
\end{table}		
\setcounter{table}{-1} 
\begin{table}[H]
	\begin{tabularx}{\textwidth}{|c|X|}
		abbreviation/symbol & synonymous expressions\\
		\hline
		$\gym$ & coupling constant of the Yang-Mills gauge theory\\
		$\gamma_i$-deformation & deformation of \NfSYMt with three real parameters -- see \secref{sec:The_deformations}\\
		$G^{(\vec{A}\,)}(\vec{x})$& $n$-point function; correlation function with $n$ fields $A_i$ combined in the vector $\vec{A}$; Green's function with $n$ fields\\
		IBP & integration by parts methods for the evaluation of integrals -- see \cite{Chetyrkin:1981qh}\\
		i.e. & `id est'; that means\\
		IR & infrared\\
		$\Kop[\,\cdot\,]$& operator that extracts the divergent part (characterised in terms of some regulator) of its argument\\
		$\lambda=\gym^2N$ & \tHooft coupling constant\\
		\lhs & left-hand side\\
		MS & minimal subtraction (scheme) -- see \appref{sec:Renormalisation_schemes}\\
		$\ol{\text{MS}}$& modified minimal subtraction (scheme) -- see \secref{sec:Renormalisation_schemes}\\
		\NfSYMt & maximally supersymmetric gauge theory in four dimensions with fields in the adjoint representation of the gauge group \SUN or \UN, typically taken to live in Minkowski space\\
		$\mathcal{O}(x)$& composite operator; local gauge invariant composite operator composed of the elementary fields of the theory -- see \secref{sec:composite-operators}\\
		QCD & quantum chromo dynamics\\
		$\text{Q}_{d}\text{S}$ & quasi-$d$-dimensional space -- see \cite{Stockinger:2005gx}\\
		QFT & quantum field theory\\
		QSC & quantum spectral curve\\
		\RGE & renormalisation group equation\\
		\rhs & right-hand side\\
		$\sigma^\mu$, $\bar{\sigma}^\mu$, $\Sigma^a$, $\bar{\Sigma}^a$ & Weyl representation matrices and their conjugates in $\RR^{(3,1)}$ and $\RR^{(6)}$, respectively -- see \appref{sec:Clifford_algebras_in_various_dimensions}\\
		SYM & super Yang-Mills\\
		SUSY & supersymmetry\\
		$\T$ & time ordering symbol -- see \eqref{eq:time_ordered_product}\\
		TsT transformation & transformation of a string theory background via the consecutive application of a T-duality, a shift, and another T-duality transformation -- \cf \cite{Lunin:2005jy}\\
		UV & ultraviolet\\
		$Z_X$& 1PI renormalisation constant of the quantity $X$ -- see \eqref{eq:renormalisation_constants} and \secref{chap:Renormalisation}\\
		$\mathcal{Z}_X$ & connected renormalisation constant of the quantity $X$ -- see \eqref{eq:renormalisation_constants} and \secref{chap:Renormalisation}\\
		\hline
	\end{tabularx}
\end{table}

\section{Clifford algebras in various dimensions}
\label{sec:Clifford_algebras_in_various_dimensions}

In this section, we discuss some details of Clifford algebras that are needed in the main part of this work. We start with some general properties in \appref{sec:Clifford_algebras_general_properties} followed by a recursive construction of a $d$-dimensional Minkowski- or Euclidean-space Clifford algebra in \appref{sec:Clifford_algebra_construction}. In the remaining subsections, we present explicit matrix realisations in four, six, and ten dimensions. Up to notational adaptations we follow the presentation of \cite{Kugo1983} and refer the reader there for details. For a brief but comprehensive introduction to Clifford algebras in a physics context, we refer the reader to \cite{Park05, OFarrill, West:1998ey} and for a deeper discussion \cite{fulton1991representation, RauschdeTraubenberg:2005aa}.

\subsection{General properties}\label{sec:Clifford_algebras_general_properties}
A Clifford algebra $\clifford(d-t,t)$ in an even\footnote{For the construction of Clifford algebras in odd dimensions see \appref{sec:Clifford_algebra_construction}.} dimension $d$ is generated by the set of elements $\gamma^m$ that fulfil the relation 
\begin{equation}\label{eq:Clifford_algebra}
\{\gamma^m,\gamma^n\}\equiv \gamma^m\gamma^n+\gamma^n\gamma^m=-2\eta^{mn}\eqncom\qquad 0\leq m,n\leq d-1\eqncom
\end{equation}
where the flat metric $\eta^{mn}=\diag(-1,\dots,-1,+1,\dots,+1)$ has $d-t$ positive entries and $t$ negative ones. As we will be interested in Minkowski space $\mathbb{R}^{(d-1,1)}$ or Euclidean space $\mathbb{R}^{(d)}$, we will choose $\eta^{mn}=\diag(-1,-\xi^2,\dots,-\xi^2)$ with $\xi^2=\pm 1$. The Clifford algebra only has one irreducible representation with dimension greater than one. Its elements $\gamma^m$ can be realised explicitly as unitary $(2^{d/2}\times2^{d/2})$ matrices such that
\begin{equation}\label{eq:unitary_gamma_matrix}
\gamma^m(\gamma^m)^\dagger=\one_{2^{d/2}}\eqncom
\end{equation}
where the hermitian conjugate $(\gamma^m)^\dagger=(\gamma^m)^{*\T}$ is given by complex conjugation and transposition. 
We can combine \eqref{eq:Clifford_algebra} with \eqref{eq:unitary_gamma_matrix} to see how the $\gamma$-matrices transform under hermitian conjugation:
\begin{equation}\label{eq:gamma_hc}
	\left.\begin{array}{lclc}
		(\gamma^0)^\dagger&=&\gamma^0& \\
		(\gamma^m)^\dagger&=&\xi^2\gamma^m & m\neq 0
	\end{array}\right\}
	\quad
	\Leftrightarrow
	\quad
	\begin{cases}
		(\gamma^m)^\dagger=\gamma^m &\text{ for } \xi^2=+1\\
		(\gamma^m)^\dagger=\gamma^0\gamma^m\gamma^0 &\text{ for } \xi^2=-1\\		
	\end{cases}
	\eqncom
\end{equation}
where $\xi^2=+1$ corresponds to Euclidean space and $\xi^2=-1$ corresponds to Minkowski space.

Apart from the representation described above, there are two seemingly different representations that also fulfil \eqref{eq:Clifford_algebra} and \eqref{eq:unitary_gamma_matrix}. The first is the complex conjugate representation with elements $(\gamma^m)^*$ and the second is the transposed representation with elements $(\gamma^m)^{\T}$. Since there is only one irreducible representation of the Clifford algebra, these three representations must be related via two constant matrices\footnote{The matrix $C$ is often called charge-conjugation matrix, since it relates massless Dirac spinors to their counterparts with opposite (electric) charge.} as
\begin{equation}\label{eq:gamma_trafo_matrices}
(\gamma^m)^*= \pm B^{-1}\gamma^m B\eqncom\qquad
(\gamma^m)^{\T}= -C^{-1}\gamma^m C\eqndot
\end{equation}
Note that the sign in the second equation is chosen to be $(-1)$. Analogously, we could fix the sign in the first equation and these two conditions would determine the behaviour of $\gamma^m$ under hermitian conjugation. However, in explicit realisations we rather fix the hermitian conjugation properties of $\gamma^m$.

\subsection{Construction of a \texorpdfstring{$d$}{d}-dimensional Clifford algebra}\label{sec:Clifford_algebra_construction}
In this subsection, we will construct the generators of a Clifford algebra in $d+1$ and $d+2$ dimensions inductively from the $d$-dimensional generators. In a recursive construction of the $d$-dimensional Clifford algebra from a one-dimensional one\footnote{For $d=1$ we will take $\gamma_{(1)}=1$ in a recursive construction of higher dimensional Clifford algebras.}, this method automatically generates a Weyl representation in Euclidean ($\xi=1$) or Minkowski space ($\xi=\complexi$).

We take the Clifford algebra generators in even dimension $d$ to be given by the elements $\tilde\gamma^m$ that fulfil 
\begin{equation}\label{eq:Clifford_algebra_D_1}
\{\tilde\gamma^m,\tilde\gamma^n\}\equiv \tilde\gamma^m\tilde\gamma^n+\tilde\gamma^n\tilde\gamma^m=-2\eta^{mn}\onee{2^{(d/2)}}\eqncom\qquad 1\leq m,n\leq d\eqndot
\end{equation}
The metric takes the form $\eta^{mn}=\diag(-1,-\xi^2,\dots,-\xi^2)$. For the construction of the $(d+1)$- and $(d+2)$-dimensional Clifford algebra, we present the new elements for the higher dimensional algebras and then show that they fulfil \eqref{eq:Clifford_algebra_D_1}, i.e. $\gamma^m\gamma^n=-\gamma^n\gamma^m$ for $m\neq n$ and $\gamma^m\gamma^m=-\eta^{mm}\onee{2^{(d/2)}}$.

We first present the generators of the $(d+1)$-dimensional Clifford algebra by constructing the additional element $\tilde\gamma^{d+1}$ that also fulfils \eqref{eq:Clifford_algebra_D_1}: 
\begin{equation}\label{eq:gamma_D1}
\tilde\gamma^{d+1}=\complexi^{\frac d2}\prod_{j=1}^{d}\tilde{\gamma}^j=\complexi^{\frac d2}\tilde{\gamma}^1\tilde{\gamma}^2\dots\tilde{\gamma}^d\eqndot
\end{equation}
This object anticommutes with all the $\tilde\gamma^m$ since each $\tilde\gamma^m$ anticommutes with every but one factor in the product in \eqref{eq:gamma_D1}. Thus, commuting $\tilde\gamma^m$ past $\tilde\gamma^{d+1}$ generates a factor of $(-1)^{d-1}$ which is just $-1$ since we assumed $d$ to be even. The square of $\tilde\gamma^{d+1}$ evaluates to
\begin{equation}
\begin{aligned}
\tilde\gamma^{d+1}\tilde\gamma^{d+1}&=\complexi^d 
\tilde{\gamma}^1\prod_{j=2}^{d}\tilde{\gamma}^j\,
\tilde{\gamma}^1\prod_{k=2}^{d}\tilde{\gamma}^k
=\complexi^d 
(\tilde{\gamma}^1)^2(-1)^{d-1}
\tilde{\gamma}^2\prod_{j=3}^{d}\tilde{\gamma}^j\,
\tilde{\gamma}^2\prod_{k=3}^{d}\tilde{\gamma}^k\\
&=\complexi^d 
(\tilde{\gamma}^1)^2(-1)^{d-1}(\tilde{\gamma}^2)^2(-1)^{d-2}
\tilde{\gamma}^3\prod_{j=4}^{d}\tilde{\gamma}^j\,
\tilde{\gamma}^3\prod_{k=4}^{d}\tilde{\gamma}^k=\dots\\
&=\complexi^d (-1)^{\frac 12d(d-1)}\prod_{j=1}^{d}(\tilde{\gamma}^j)^2
=(\xi^2)^{d-1}\onee{d/2}
=\xi^2\onee{2^{(d/2)}}\eqncom
\end{aligned}
\end{equation}
where we used that $\xi^2=\pm1$ and again that $d$ is even. The dots mean that the $\tilde{\gamma}^m$ in the products are commuted until each doubly occuring pair stands next to each other. 
This concludes the $(d+1)$-dimensional case. The $(d+1)$-dimensional Clifford algebra is generated by the $d$-dimensional one supplemented with $\tilde\gamma^{d+1}$ defined in \eqref{eq:gamma_D1}.

In a second step, we construct the $(d+2)$-dimensional Clifford algebra generators in terms of the $(d+1)$-dimensional one. 
We take the $(d+2)$-dimensional elements $\gamma^m$ to be 
\begin{equation}\label{eq:Clifford_algebra_D2}
\gamma^m=
\begin{pmatrix}
0& \sigma^m\\
\bar{\sigma}^m& 0
\end{pmatrix}\eqncom\qquad
1\leq m\leq d+2\eqncom
\end{equation}
with
\begin{equation}\label{eq:sigma_matrices_general}
\sigma^m=\begin{cases}
\complexi \tilde{\gamma}^m& 1\leq m\leq d+1\\
-\xi \onee{d/2}& m=d+2
\end{cases}\eqncom\qquad
\bar\sigma^m=\begin{cases}
-\complexi \tilde{\gamma}^m& 1\leq m\leq d+1\\
-\xi \onee{d/2}& m=d+2
\end{cases}\eqndot
\end{equation}
For the first $(d+1)$ elements, we immediately find
\begin{equation}
\{\gamma^m ,\gamma^n\}=\begin{pmatrix}
\{\tilde\gamma^m ,\tilde\gamma^n\}&0\\
0&\{\tilde\gamma^m ,\tilde\gamma^n\}
\end{pmatrix}
=-2\eta^{mn}\onee{2^{(d+2)/2}}\eqncom\qquad
1\leq m,n\leq d+1\eqndot
\end{equation}
Finally, the last element squares to $\gamma^{d+2}\gamma^{d+2}=\xi^2\onee{2^{(d+2)/2}}$ and its anticommutation relation with the remaining elements is
\begin{equation}
\{\gamma^m ,\gamma^{d+2}\}=\begin{pmatrix}
-\complexi \xi[\tilde\gamma^m ,\onee{2^{(d/2)}}]&0\\
0&+\complexi \xi[\tilde\gamma^m ,\onee{2^{(d/2)}}]
\end{pmatrix}
=0\eqncom\qquad
1\leq m\leq d+1\eqndot
\end{equation}
This concludes the construction of the $(d+2)$-dimensional Clifford algebra generators from the $(d+1)$-dimensional ones.

From the structure in \eqref{eq:Clifford_algebra_D2} it is clear that the $(d+2)$-dimensional generators are in the Weyl representation and the $\gamma^m$ are hermitian if the $\tilde\gamma^m$ are. In addition, the Clifford algebra for $\gamma^m$ induces a similar relation for the $\sigma$ and $\bar{\sigma}$ matrices
\begin{equation}
\sigma^m\bar{\sigma}^n+\sigma^n\bar{\sigma}^m=-2\eta^{mn}\onee{2^{d/2}}\eqncom\qquad
\bar\sigma^m\sigma^n+\bar\sigma^n\sigma^m=-2\eta^{mn}\onee{2^{d/2}}\eqndot
\end{equation}

\subsection{Four-dimensional Minkowski Clifford algebra}\label{sec:4D_Clifford_algebra}
In this subsection, we will present the four-dimensional Minkowski-space generators of the Clifford algebra $\clifford(3,1)$ which we use in the main part. For compatibility reasons, we will not employ the construction presented in \appref{sec:Clifford_algebra_construction}, but instead follow the conventions of \cite{Srednicki:2007}.\footnote{Both representations are linked via a unitary transformation, as is necessary from the general discussion in \appref{sec:Clifford_algebras_general_properties}. Later in \appref{sec:6D_Clifford_algebra}, for the six-dimensional Clifford algebra we will use the construction of \appref{sec:Clifford_algebra_construction}.}

In four-dimensional Minkowski space with mostly plus metric $\eta_{\mu\nu}=\diag(-1,1,1,1)$ we work in the Weyl representation. We construct the $\gamma$-matrices from the $\sigma$-matrices that take values in $\so{1,3}\otimes\spl{2}\otimes\splbar{2}$ and have the explicit form
\begin{equation}\label{eq:sigma_matrices}
(\sigma^\mu)_{\alpha\dot{\beta}}=(\one,\sigma_1,\sigma_2,\sigma_3)_{\alpha\dot{\beta}}\eqncom\qquad
(\bar{\sigma}^\mu)^{\dot\alpha\beta}=(\one,-\sigma_1,-\sigma_2,-\sigma_3)_{\dot\alpha\beta}\eqncom
\end{equation}
where the Pauli matrices $\sigma_i$ are
\begin{equation}\label{eq:Pauli_matrices}
\one=
\begin{pmatrix}
1&0\\
0&1
\end{pmatrix}
\eqncom\qquad
\sigma_1=
\begin{pmatrix}
0&1\\
1&0
\end{pmatrix}
\eqncom\qquad
\sigma_2=
\begin{pmatrix}
0&-i\\
i&0
\end{pmatrix}
\eqncom\qquad
\sigma_3=
\begin{pmatrix}
1&0\\
0&-1
\end{pmatrix}
\eqndot
\end{equation}
The $\sigma$-matrices have the following properties under transposition and complex conjugation
\begin{equation}\label{eq:sigma_T_cc}
\begin{aligned}
\bigl((\sigma^\mu)_{\alpha\dot{\beta}}\bigr)^{\T}&=(\sigma^\mu)_{\dot{\beta}\alpha}
=(\one,\sigma_1,-\sigma_2,\sigma_3)_{\dot{\beta}\alpha}\eqncom\\
\bigl((\sigma^\mu)_{\alpha\dot{\beta}}\bigr)^{*}&=(\sigma^\mu)_{\dot{\alpha}\beta}
=(\one,\sigma_1,-\sigma_2,\sigma_3)_{\dot{\alpha}\beta}\eqncom
\end{aligned}
\end{equation}
and analogous relations hold for the $\bar{\sigma}$-matrices. Combining transposition and complex conjugation to hermitian conjugation, we have $(\sigma^\mu)^\dagger=\sigma^\mu$ and $(\bar\sigma^\mu)^\dagger=\bar\sigma^\mu$, which we denote in index-notation as
\begin{equation}\label{eq:sigma_hc}
\bigl((\sigma^\mu)_{\alpha\dot{\beta}}\bigr)^{\dagger}=(\sigma^\mu)_{\beta\dot{\alpha}}\eqncom\qquad
\bigl((\bar\sigma^\mu)^{\dot\alpha\beta}\bigr)^{\dagger}=(\bar\sigma^\mu)^{\dot\beta\alpha}\eqndot
\end{equation}
We can transform $\sigma^\mu$ into $\bar{\sigma}^\mu$ by raising the spinor indices with the $\varepsilon$-tensors defined in \secref{sec:Four-dimensional_Minkowski_spinors}
\begin{equation}
\varepsilon_{\alpha\beta}\varepsilon_{\dot\gamma\dot\beta}(\bar\sigma^\mu)^{\dot\beta\beta}
=
-\varepsilon_{\dot\gamma\dot\beta}(\bar\sigma^\mu)^{\dot\beta\beta}\varepsilon_{\beta\alpha}
=
(\sigma_2\cdot \bar\sigma^\mu\cdot\sigma_2)_{\dot{\gamma}\alpha}
=(\one,\sigma_1,\sigma_2,\sigma_3)_{\alpha\dot{\gamma}}\eqncom
\end{equation}
where the dot denotes an ordinary matrix multiplication. The additional sign in the third component of $\bar\sigma^\mu$ is absorbed into an index switch of the spinor indices in the last equality. In general, we find
\begin{equation}\label{eq:sigma_matrix_identities}
(\sigma^\mu)_{\alpha\dot{\alpha}}=\varepsilon_{\alpha\beta}\varepsilon_{\dot\alpha\dot\beta}(\bar\sigma^\mu)^{\dot\beta\beta}
\eqncom\qquad
(\bar\sigma^\mu)^{\dot\alpha\alpha}=\varepsilon^{\alpha\beta}\varepsilon^{\dot\alpha\dot\beta}(\sigma^\mu)_{\beta\dot\beta}\eqndot
\end{equation}

The $\gamma$-matrices in the Weyl representation can be expressed in terms of the $\sigma$-matrices, as
\begin{equation}\label{eq:gamma_Weyl}
\gamma^\mu=
\begin{pmatrix}
0&(\sigma^\mu)_{\alpha\dot{\beta}}\\
(\bar{\sigma}^\mu)^{\dot{\alpha}\beta}&0
\end{pmatrix}\eqncom\qquad
\gamma^5=\begin{pmatrix}
-\delta_{\alpha}^{\phan{\alpha}\beta}&0\\
0&+\delta^{\dot\alpha}_{\phan{\alpha}\dot\beta}
\end{pmatrix}
\end{equation}
where we have indicated the spinor-index structure only on the \rhs for notational reasons. In general we know that hermitian conjugation, transposition and complex conjugation of the $\gamma$-matrices can be realised via the transformation \eqref{eq:gamma_hc} and the two transformations in \eqref{eq:gamma_trafo_matrices}, respectively. With the position-choice for spinor indices in \eqref{eq:gamma_Weyl}, the transformations are realised via the three matrices 
\begin{equation}
\beta=
\begin{pmatrix}
0&\delta^{\dot{\alpha}}_{\phan{\dot{\alpha}}\dot{\beta}}\\
\delta_\alpha^{\phan{\alpha}\beta}&0
\end{pmatrix}\eqncom\qquad
C=
\begin{pmatrix}
\varepsilon_{\alpha\beta}&0\\
0&\varepsilon^{\dot{\alpha}\dot{\beta}}
\end{pmatrix}\eqncom\qquad
B=C\beta^{\T}=
\begin{pmatrix}
0&	\varepsilon_{\alpha\beta}\\
\varepsilon^{\dot\alpha\dot\beta}	&0
\end{pmatrix}\eqndot
\end{equation}
Note that numerically $\beta=\gamma^0$ and  we only need to introduce it to match the spinor-index structure of the $\gamma$-matrices. It is now straightforward to verify in this representation that\footnote{We chose to employ the same definition of $\beta$ as in \cite{Srednicki:2007} in order to have the same definition of a Dirac-conjugate spinor $\ol{\Psi}=\Psi^\dagger\beta$. In \eqref{eq:gamma_trafos_4d}, this choice fixes the transformation rules for $\beta$, $C$ and $B$.}
\begin{equation}\label{eq:gamma_trafos_4d}
(\gamma^\mu)^\dagger=\beta\gamma^\mu\beta^{-1}\eqncom\qquad
(\gamma^\mu)^{\T}=-C^{-1}\gamma^\mu C\eqncom\qquad
(\gamma^\mu)^*=-B^{-1}\gamma^\mu B\eqncom
\end{equation} 
where we used \eqref{eq:sigma_T_cc} and \eqref{eq:sigma_hc}.

\subsection{Six-dimensional Euclidean Clifford algebra}\label{sec:6D_Clifford_algebra}
In this subsection, we present the six-dimensional Euclidean-space Clifford algebra generators.

We follow the recursive construction presented in \appref{sec:Clifford_algebra_construction}, starting from the one-dimensional matrix $\gamma_{(1)}=1$, to get the six-dimensional Euclidean-space $\gamma$-matrices $\gamma_{(6)}^a$. Note that this construction gives the Clifford algebra $\clifford(0,6)$ with metric $\delta^{ab}=-\diag(1,\dots,1)$. Fortunately this is exactly the six-dimensional algebra we need for the construction of the ten-dimensional Minkowski-space algebra in \appref{sec:ten-dimensional_Minkowski_Clifford_algebra}. For later convenience and direct compatibility with \cite{Brink197777}, we take the $\gamma$-matrices to be $\rho^a=u^{-1} \gamma_{(6)}^a u$, where $u$ is a unitary transformation given in terms of the $2\times 2$ matrices \eqref{eq:Pauli_matrices} as $u=\diag (\onee{2},\complexi \sigma_3,-\sigma_2,\sigma_1)$. We find the explicit realisation
\begin{equation}\label{eq:6D_gamma}
\rho^a=\begin{pmatrix}
0& (\Sigma^a)_{AB}\\
(\bar\Sigma^a)^{AB}&0
\end{pmatrix}\eqncom\qquad
\rho^7=
\begin{pmatrix}
-\delta_A^{\phan{A}B}&0\\
0&\delta^A_{\phan{A}B}
\end{pmatrix}\eqncom\qquad
1\leq A,B\leq 4\eqncom
\end{equation}
where the indices $A,B\in\{1,2,3,4\}$ are fundamental indices of \su{4} and its conjugate\footnote{The isomorphism $\spin{6}\simeq\su{4}$ ensures that the Euclidean space indices $a$ can be related to the \su{4} indices.} and we suppressed these on the \lhs for notational reasons.
The constituent matrices $\delta$ are unit matrices and the $\Sigma^a$ are given in terms of the Pauli matrices as
\begin{equation}
\begin{aligned}\label{eq:Sigma_6}
\Sigma^1&=\bar\Sigma^1=-\sigma_2\otimes\sigma_3\eqncom&
\Sigma^2&=-\bar\Sigma^2=\complexi\sigma_2\otimes\onee{2}\eqncom&
\Sigma^3&=-\bar\Sigma^3=\complexi\sigma_1\otimes\sigma_2\eqncom&\\
\Sigma^4&=\bar\Sigma^4=-\sigma_2\otimes\sigma_1\eqncom&
\Sigma^5&=-\bar\Sigma^5=\complexi\sigma_3\otimes\sigma_2\eqncom&
\Sigma^6&=\bar\Sigma^6=-\onee{2}\otimes\sigma_2\eqndot&
\end{aligned}
\end{equation}
Note that our matrix index positions (upper indices at $\Sigma$ and lower indices at $\bar{\Sigma}$) are simply a choice. In this choice, an ordinary matrix product is realised when an upper index is contracted with a lower one and the indices can be raised or lowered with Kronecker-$\delta$. 

To find the explicit realisation of $(\rho^a)^\dagger$, we calculate the hermitian conjugate of \eqref{eq:Sigma_6}. Numerically, this yields $(\Sigma^a)^\dagger=\bar{\Sigma}^a$ and for fixed indices $A$ and $B$ we find
\begin{equation}\label{eq:Sigmabar_6}
\bigl((\Sigma^a)_{AB}\bigr)^*=\delta_{BC}(\bar\Sigma^a)^{CD}\delta_{DA}\eqncom\qquad
\bigl((\bar\Sigma^a)^{AB}\bigr)^*=\delta^{BC}(\Sigma^a)_{CD}\delta^{DA}\eqndot
\end{equation}
Combining \eqref{eq:Sigmabar_6} with \eqref{eq:6D_gamma}, we see that the $\rho^a$ are hermitian and in explicit index notation the hermitian conjugation can be realised as 
\begin{equation}\label{eq:rho6_hermitian_conjugation}
(\rho^a)^\dagger=\beta_{6}\rho^a\beta_{6}^{-1}\eqncom\qquad
\beta_6=\begin{pmatrix}\delta^{AB}&0\\
0&\delta_{AB}
\end{pmatrix}\eqncom
\end{equation}
where we introduced $\beta_6$ as the equivalent of $\gamma_0$ in six Euclidean dimensions. For the transposition, we have
\begin{equation}\label{eq:rho6_transpose}
(\rho^a)^{\T}=-(C_6)^{-1}\rho^a C_6=
\begin{pmatrix}
0& -(\bar{\Sigma}^a)^{AB}\\
 -(\Sigma^a)_{AB}&0
\end{pmatrix}\qquad\text{with}\qquad
C_6=\begin{pmatrix}
0& \delta_A^{\phan{A}B}\\
\delta^A_{\phan{A}B}&0
\end{pmatrix}\eqndot
\end{equation}
Note that this equation is just an explicit realisation of the second part of \eqref{eq:gamma_trafo_matrices}. The complex conjugation can also be expressed in terms of a unitary transformation by combining \eqref{eq:rho6_hermitian_conjugation} with \eqref{eq:rho6_transpose}:
\begin{equation}
(\rho^a)^* 
=(\beta_{6}\rho^a\beta_{6}^{-1})^{\T}=
-B_6^{-1}\rho^aB_6\eqncom\qquad
B_6=C_6\beta_6^{\T}=
\begin{pmatrix}
0 & \delta_{AB}\\
\delta^{AB}& 0
\end{pmatrix}\eqndot
\end{equation}
With this definition, we explicitly find
\begin{equation}\label{eq:rho6_cc}
(\rho^a)^{*}=-(B_6)^{-1}\rho^a B_6=
\begin{pmatrix}
0& -(\delta\cdot\bar{\Sigma}^a\cdot \delta)_{AB}\\
-(\delta\cdot\Sigma^a\cdot\delta)^{AB}&0
\end{pmatrix}\eqncom
\end{equation}
where we used the Kronecker-$\delta$ in $B_6$ to raise or lower the \su{4} spinor indices, see also \appref{sec:spinors-in-six-dimensional-euclidean-space}.

\subsection{Ten-dimensional Minkowski Clifford algebra}\label{sec:ten-dimensional_Minkowski_Clifford_algebra}
In this subsection, we profit from the previous two subsections and stitch the ten-dimensional Minkowski space Clifford algebra $\clifford(9,1)$ together from the four-dimensional one given in \appref{sec:4D_Clifford_algebra} and the six-dimensional one given in \appref{sec:6D_Clifford_algebra}.

We write the ten-dimensional $\Gamma$-matrices as a Kronecker product of the six-dimensional matrices defined in \eqref{eq:6D_gamma} and the four-dimensional ones defined in \eqref{eq:gamma_Weyl}:
\begin{equation}
\Gamma^m=\begin{cases}
\onee{8}\otimes \gamma^m	& 0\leq m\leq 3\\
 \rho^{m-3}\otimes \complexi\gamma^5& 4\leq m\leq 9
\end{cases}\eqncom\qquad
\Gamma^{11}= \complexi^{10/2}\prod_{i=0}^9\Gamma^{i}=\rho^7\otimes \complexi\gamma^5\eqncom
\end{equation}
where the unit matrix is to be understood as $\onee{8}=\diag(\delta_A^{\phan{A}B},\delta^A_{\phan{A}B})$ in the index conventions of \appref{sec:6D_Clifford_algebra}. A direct calculation using the definitions of $\gamma^\mu$ and $\rho^a$ immediately confirms that this choice of $\Gamma^m$ fulfils the Clifford algebra relation \eqref{eq:Clifford_algebra}. Analogously, the matrices that realise transposition, complex conjugation and hermitian conjugation are obtained from respective Kronecker products
\begin{equation}\label{eq:beta_B_C_10}
C_{10}=C_6\otimes C\eqncom\qquad
\beta_{10}=\beta_6\otimes\beta\eqncom\qquad
B_{10}=C_{10}\beta_{10}^{\T}=B_6\otimes B\eqncom
\end{equation}
where the lower dimensional matrices are defined in \appref{sec:4D_Clifford_algebra} and \ref{sec:6D_Clifford_algebra}. In these conventions the following relations hold
\begin{equation}
(\Gamma^m)^\dagger=\beta_{10}\Gamma^m\beta_{10}^{-1}\eqncom\qquad
(\Gamma^m)^*=-B_{10}^{-1}\Gamma^mB_{10}\eqncom\qquad
(\Gamma^m)^{\T}=-C^{-1}_{10}\Gamma^mC_{10}\eqndot
\end{equation}

\section{Spinors in various dimensions}\label{sec:spinor_in_various_dimensions}
In this appendix, we use the Clifford algebra introduced in \appref{sec:Clifford_algebras_in_various_dimensions} to determine the properties of the spinor representation. In particular, we will discuss six-dimensional spinors in Euclidean space and four- and ten-dimenional spinors in Minkowski space. This section follows the presentation of \cite{OFarrill} and further details are taken from \cite{Park05, West:1998ey, OFarrill, fulton1991representation, Deligne99quantumfields, RauschdeTraubenberg:2005aa, VanProeyen:1999ni}. Furthermore, we adopt the conventions of \cite{Srednicki:2007} in four-dimensional Minkowski space.

\subsection{General properties}\label{sec:spinors_general_properties}
Spinors in $d$-dimensional Euclidean or Minkowski space transform under the spinor representation \spin{d} or \spin{d-1,1}, respectively. This representation can be obtained as the irreducible representation of a subgroup of the Clifford algebra $\clifford(d-t,t)$. For certain $d$ the Clifford algebra decomposes into invariant subspaces and hence the spinors and the spinor representation can be restricted to these subspaces. The invariant subspaces occur in two different ways. First, in even dimensions there exists a Weyl representation which splits the original representation of $\gamma$-matrices into a plus- and a minus-chirality part. Second, depending on $d$ the $\gamma$-matrices can be restricted to be in a Majorana representation, i.e.\ they are purely real or purely imaginary. Accordingly, in the Majorana representation also the spinors can be restricted to be purely real or purely imaginary. The explicit realisations of the Weyl- and/or Majorana-condition will be discussed below.

The Clifford algebra is $2^d$ dimensional and consists of the elements
\begin{equation}
\clifford(d-t,t)=\{1,\gamma^m,\gamma^{m_1m_2},\dots,\gamma^{m_1\dots m_d}\}\eqncom
\end{equation}
where the multi-index elements can be chosen as totally antisymmetric products of the generators e.g.\ $\gamma^{m_1m_2}\sim (\gamma^{m_1}\gamma^{m_2}-\gamma^{m_2}\gamma^{m_1})$. Note that the Clifford algebra splits into an even and odd part under the identification $\gamma^m\rightarrow -\gamma^m$ as
\begin{equation}\label{eq:Clifford_even_odd}
\clifford(d-t,t)=\clifford(d-t,t)^{(\text{even})}\oplus \clifford(d-t,t)^{(\text{odd})}\eqncom
\end{equation}
where the even (odd) part contains products of even (odd) numbers of generators. The irreducible representation of $\clifford(d-t,t)^{(\text{even})}$ is called the spinor representation and its generators can be chosen as
\begin{equation}\label{eq:Lorentz_generators}
M^{mn}=\frac \complexi 4 [\gamma^m,\gamma^n]=\frac \complexi 4 (\gamma^m\gamma^n-\gamma^n\gamma^m)\eqndot
\end{equation}
These are also the generators of the special orthogonal group with the identity component\footnote{The subscript $0$ means that the group SO$_0(d-t,t)$ contains the identity element.} SO$_0(d-t,t)$, since the spin group Spin$(d-t,t)$ is a double cover of SO$_0(d-t,t)$ for $d\geq 2$. In particular, for $t=0$ these are the generators of SO$(d)$ and for $t=1$ they are the generators of the proper orthochronous Lorentz group SO$_0^+(d-1,1)$. When we take $\Lambda=1+\delta\omega$ to realise an SO$(d)$ or SO$_0^+(d-1,1)$ transformation, we can introduce a unitary operator that realises the corresponding spin transformation
\begin{equation}
\Uop(\Lambda)=\e^{-\frac \complexi 2\delta\omega_{mn}\mathbb{M}^{mn}}\eqncom\qquad
\end{equation}
where $\delta\omega_{mn}$ is an antisymmetric matrix that contains the real parameters of the transformation $\Lambda$ and $\mathbb{M}^{mn}$ are abstract hermitian representations of the generators $M^{mn}$.

A spinor $\lambda$ in even dimensions $d$ (Euclidean or Minkowski space) is a $2^{d/2}$-dimensional tuple that transforms under the spin group as
\begin{equation}\label{eq:spinor_trafo}
\Uop^{-1}(\Lambda)\,\lambda\, \Uop(\Lambda)
=\lambda+\delta\lambda+\order{\delta\omega^2}\eqncom\qquad
\delta\lambda=\frac \complexi 2 \delta\omega_{mn}[\mathbb{M}^{mn},\lambda]
=\frac \complexi 2 \delta\omega_{mn}M^{mn}\lambda\eqncom
\end{equation} 
where we have suppressed any spacial dependence of the spinor\footnote{The transformation of a space-time dependent spinor would assume the form\\
$\Uop^{-1}(\Lambda)\,\lambda(x)\, \Uop(\Lambda)=D(\Lambda)\lambda(\Lambda^{-1}x)$, where $D(\Lambda)$ realises the spin transformation, compare \appref{sec:conformal_algebra} for details.} by choosing $\lambda=\lambda(0)$. We can also define the conjugate $\bar{\lambda}$ by requiring that the contraction $\bar{\lambda}^a\lambda_a$ is invariant under spin transformations. From \eqref{eq:spinor_trafo}, we see that $\bar{\lambda}$ then must transform as
\begin{equation}\label{eq:Conjugate}
\delta\bar{\lambda}=\bar{\lambda}\Bigl( -\frac \complexi 2\delta\omega^{mn}M_{mn}\Bigr)\eqndot
\end{equation} 
Alternatively, we could also take the hermitian conjugate of \eqref{eq:spinor_trafo}. In Minkowski space this gives
\begin{equation}\label{eq:Dirac_conjugate_2}
\delta(\lambda^\dagger\gamma_0)= (\lambda^\dagger \gamma_0)\Bigl(-\frac \complexi 2\delta\omega^{mn}M_{mn}\Bigr)\eqncom
\end{equation} 
where we have used \eqref{eq:Clifford_algebra}, \eqref{eq:gamma_hc}, and \eqref{eq:Lorentz_generators} to obtain $(M_{mn})^\dagger=\gamma_0M_{mn}\gamma_0$. Combining \eqref{eq:Conjugate} and \eqref{eq:Dirac_conjugate_2} we define the Dirac conjugate
\begin{equation}\label{eq:Dirac_conjugate}
\bar{\lambda}^D\equiv\lambda^\dagger\gamma_0\eqndot
\end{equation}
In Euclidean space, we have a different hermitian conjugation relation given in \eqref{eq:gamma_hc}. However, in the cases we will be interested in we can choose the Euclidean $\gamma_m$ matrices to be anti-hermitian. This choice yields $(M_{mn})^\dagger=M_{mn}$ and thus the hermitian conjugate of \eqref{eq:spinor_trafo} in Euclidean space is
\begin{equation}\label{eq:Euclidean_conjugate_2}
\delta(\lambda^\dagger)= \lambda^\dagger\Bigl(-\frac \complexi 2\delta\omega^{mn}M_{mn}\Bigr)\eqndot
\end{equation} 
Combining \eqref{eq:Conjugate} with \eqref{eq:Euclidean_conjugate_2}, we define the Euclidean conjugate
\begin{equation}\label{eq:Euclidean_conjugate}
\bar{\lambda}^E\equiv\lambda^\dagger\gamma^E_0\eqncom\qquad\gamma^E_0=\onee{2^{d/2}}\eqndot
\end{equation}

In even dimensions there exists a Weyl representation and thus we know that \eqref{eq:Lorentz_generators} is reducible. We can employ the general construction of the Weyl representation in \appref{sec:Clifford_algebra_construction} and find that $M_{mn}$ splits into two $2^{d/2-1}$-dimensional invariant subspaces. Accordingly, the spinor splits into $\lambda=(\lambda^-,\lambda^+)^{\T}$ and transforms as
\begin{equation}\label{eq:spinor_trafo_Weyl}
\begin{pmatrix}
\delta\lambda^-_{a}\\
\delta\lambda^{+a}
\end{pmatrix}= \delta\omega^{mn}
\begin{pmatrix}
\frac \complexi 4(\sigma_m\bar{\sigma}_n-\sigma_n\bar\sigma_m)_a^{\phan{a}b}&0\\
0&\frac \complexi 4(\bar\sigma_m\sigma_n-\bar\sigma_n\sigma_m)^a_{\phan{a}b}
\end{pmatrix}
\begin{pmatrix}
\lambda^-_{b}\\
\lambda^{+b}
\end{pmatrix}
\eqncom
\end{equation} 
where $\sigma$ and $\bar{\sigma}$ can be constructed from the $d-1$ dimensional Clifford algebra, compare \eqref{eq:sigma_matrices_general}. The spinors $\lambda^-$ and $\lambda^+$ are called minus- and plus-chirality Weyl spinors, respectively and we can use $\gamma^{d+1}$ to define suitable Weyl projectors
\begin{equation}
P^\pm=\frac 12 \left(\onee{2^{d/2}}\pm\gamma^{d+1}\right)\eqndot
\end{equation}

In the beginning of this subsection, we mentioned that the $\gamma$-matrices in the Majorana representation can be chosen to be purely real, which renders the generators of the spinor representation $M_\tM^{mn}$ purely imaginary and the Majorana spinors $\lambda_\tM$ purely real. Under complex conjugation, we therefore have
\begin{equation}\label{eq:spinor_trafo_majorana}
(\delta\lambda_\tM)^*= \Bigl(-\frac \complexi 2\delta\omega_{mn}(M^{mn}_\tM)^*\Bigr)\lambda_\tM^*
=\Bigl(\frac \complexi 2\delta\omega_{mn}M^{mn}_\tM\Bigr)\lambda_\tM
=\delta\lambda_\tM\eqndot
\end{equation}
However, we do not want to work in the Majorana representation. In a general representation, we can show from \eqref{eq:gamma_trafo_matrices} that the generators conjugation can be written as $(M^{mn})^{*}=B^{-1}M^{mn}B$. So, in order to preserve the transformation property \eqref{eq:spinor_trafo_majorana}, we define the Majorana conjugate (often called charge conjugate) spinor $\lambda^C$ as
\begin{equation}\label{eq:Majorana_equation}
\lambda^C\equiv B\lambda^*= C(\bar{\lambda})^{\T}\eqncom
\end{equation}
where $\bar{\lambda}=\bar{\lambda}^D$ or $\bar{\lambda}=\bar{\lambda}^E$ given in \eqref{eq:Dirac_conjugate} or \eqref{eq:Euclidean_conjugate}, respectively. Note that we must choose $B B^*=\one$ in order to have $(\lambda^C)^C=\lambda$. If we now compute the complex conjugate of \eqref{eq:spinor_trafo}, we find
\begin{equation}
\delta\lambda^C= \Bigl(\frac \complexi 2\delta\omega_{mn}M^{mn}\Bigr)\lambda^C
\end{equation}
and we see that for $\lambda^C=\lambda$ the Majorana condition \eqref{eq:spinor_trafo_majorana} is fulfilled.

With the Weyl- and Majorana-condition, we have projections to two different subspaces. We can now ask whether we can define a combined projector onto a Majorana-Weyl subspace. In general, this is possible if the reality condition for Majorana spinors respects the Weyl projectors, i.e. if $B$ commutes with $P^\pm$ . In the cases of interest in this work, a Majorana-Weyl subspace only exists for ten-dimensional Minkowski space. A classification which types of spinors appear up to space-time dimension twelve with arbitrary signs in the space-time metric can be found in \cite{VanProeyen:1999ni, OFarrill}.

\subsection{Spinors in four-dimensional Minkowski space}
\label{sec:Four-dimensional_Minkowski_spinors}
In four-dimensional Minkowski space, we adopt the conventions of \cite{Srednicki:2007} for the manipulation of spinor indices. We will give a brief review on the definitions employed in this thesis and refer the reader to \cite[\chap{7A}]{tanedo2013flight} and in general to \cite{Srednicki:2007,Dreiner:2008tw} for further details.

Spinors in four-dimensional Minkowski space come in the two inequivalent fundamental complex two-dimensional representations of $\text{SL}(2,\CC)\simeq\text{Spin}(3,1)$. These are the minus- and plus-chirality Weyl spinors. They carry undotted and dotted greek indices, respectively and in addition the plus-chirality spinors carry a dagger that distinguishes them from minus-chirality ones in an index-free notation. Under hermitian conjugation, the Weyl spinors change their chirality and we explicitly have
\begin{equation}
\chi^{\dagger\dot \alpha}\equiv(\chi^\alpha)^\dagger\eqncom\qquad
\chi^\dagger_{\dot \alpha}\equiv(\chi_\alpha)^\dagger
\qquad\text{and}\qquad
(\chi^{\dagger\dot \alpha})^\dagger=\chi^{\alpha}\eqncom\qquad
(\chi^\dagger_{\dot \alpha})^\dagger=\chi_{\alpha}\eqndot
\end{equation}
Note that for fixed spinor indices hermitian and complex conjugation yield the same conjugate spinor. Spinor indices can be raised and lowered with the Lorentz-invariant antisymmetric tensors $\varepsilon=\pm \complexi \sigma_2$ with the explicit definition 
\begin{equation}\label{eq:epsilon_spinor_indices}
\varepsilon^{12}=\varepsilon^{\dot 1 \dot 2}=\varepsilon_{21}=\varepsilon_{\dot 2 \dot 1}=1\eqncom\quad
\varepsilon^{21}=\varepsilon^{\dot 2 \dot 1}=\varepsilon_{12}=\varepsilon_{\dot 1 \dot 2}=-1\eqndot
\end{equation}
Note that the $\varepsilon$-tensor with upper indices is the inverse of the one with lower indices and so they satisfy
\begin{equation}
\varepsilon_{\gamma\delta}\varepsilon^{\alpha\beta}=
\delta^{\phan{\alpha}\alpha}_{\delta}\delta^{\phan{\alpha}\beta}_{\gamma}-
\delta^{\phan{\alpha}\alpha}_{\gamma}\delta^{\phan{\alpha}\beta}_{\delta}\eqncom\qquad
\varepsilon^{\dot\alpha\dot\beta}\varepsilon_{\dot\gamma\dot\delta}=
\delta^{\dot\alpha}_{\phan{\alpha}\dot\delta}\delta^{\dot\beta}_{\phan{\alpha}\dot\gamma}-
\delta^{\dot\alpha}_{\phan{\alpha}\dot\gamma}\delta^{\dot\beta}_{\phan{\alpha}\dot\delta}\eqndot
\end{equation}
For rank one spinor fields this renders the explicit relations\footnote{%
	Note that higher rank spinor fields transform analogously, e.g. $A^{\alpha\beta}=\varepsilon^{\alpha\gamma}\varepsilon^{\beta\delta}A_{\gamma\delta}$. The only exception to this is the transformation of the $\varepsilon$-tensor itself, whose indices are transformed according to
	$\varepsilon^{\alpha\beta}=\varepsilon^{\alpha\gamma}\varepsilon_{\gamma\delta}\varepsilon^{\delta\beta}=(\complexi \sigma_2\cdot\sigma_2\cdot\sigma_2)^{\alpha\beta}=(\complexi \sigma_2)^{\alpha\beta}$.
}
\begin{equation}\label{eq:Weyl_spinor_raise_lower}
\chi_\alpha=\varepsilon_{\alpha\beta}\chi^\beta\eqncom\qquad
\chi^\alpha=\varepsilon^{\alpha\beta}\chi_\beta\eqncom\qquad
\chi^\dagger_{\dot{\alpha}}=\varepsilon_{\dot\alpha\dot\beta}\chi^{\dagger\dot{\beta}}\eqncom\qquad
\chi^{\dagger\dot{\alpha}}=\varepsilon^{\dot\alpha\dot\beta}\chi^{\dagger}_{\dot{\beta}}\eqndot
\end{equation}
We define the canonic product of two minus- and respectively plus-chirality spinors to be
\begin{equation}
\chi\xi\equiv\chi^\alpha \xi_\alpha=\varepsilon_{\alpha\beta}\chi^\alpha \xi^\beta\eqncom\qquad
\chi^\dagger \xi^\dagger\equiv\chi^\dagger_{\dot{\alpha}} \xi^{\dagger\dot\alpha}=\varepsilon_{\dot\alpha\dot\beta}\chi^{\dagger\dot{\beta}} \chi^{\dagger\dot{\alpha}}\eqndot
\end{equation}
With the $\sigma$-matrices from \appref{sec:4D_Clifford_algebra} we can also construct four-vectors from spinors of the form $\chi^\alpha\sigma^\mu_{\alpha\dot{\beta}}\xi^{\dagger\dot{\beta}}$ and $\chi^\dagger_{\dot{\alpha}}\bar\sigma^{\mu\dot\alpha\beta}\xi_\beta$. Under hermitian conjugation, these products transform as
\begin{equation}
\begin{aligned}
(\chi \xi)^\dagger&=(\varepsilon_{\alpha\beta})^\dagger(\xi^\beta)^\dagger(\chi^\alpha )^\dagger
=\varepsilon_{\dot\alpha\dot\beta}\xi^{\dagger\dot\beta}\chi^{\dagger\dot\alpha}=\xi^\dagger\chi^\dagger\eqncom\\
(\chi\sigma^\mu\xi^\dagger)^\dagger&=(\xi^{\dagger\dot{\beta}})^\dagger(\sigma^\mu_{\alpha\dot{\beta}})^\dagger(\chi^\alpha)^\dagger
=\xi^{\beta}\sigma^\mu_{\beta\dot{\alpha}}\chi^{\dagger\dot\alpha}
=\xi\sigma^\mu\chi^\dagger\eqncom\\
(\chi^\dagger\bar\sigma^\mu\xi)^\dagger&=(\xi_{\beta})^\dagger(\bar\sigma^{\mu\dot\alpha\beta})^\dagger(\chi^{\dagger}_{\dot\alpha})^\dagger
=\xi^{\dagger}_{\dot\beta}\bar\sigma^{\mu\dot\beta\alpha}\chi_{\alpha}
=\xi^{\dagger}\bar\sigma^\mu\chi\eqndot
\end{aligned}
\end{equation}

A general Dirac spinor is composed of a minus- and a plus-chirality Weyl spinor. In four dimensions the Dirac spinor and its conjugate take the form 
\begin{equation}
\Psi_4=\begin{pmatrix}
\chi_\alpha\\
\xi^{\dagger\dot{\alpha}}
\end{pmatrix}\eqncom\qquad
\ol{\Psi}_4=\Psi^\dagger\beta=(\xi^\alpha,\chi^\dagger_{\dot\alpha})
\end{equation}
and its minus- and plus-chirality contributions are obtained via the projectors\footnote{We suppressed the spinor indices on the \lhs in analogy to the situation in \eqref{eq:gamma_Weyl}.}
\begin{equation}
	P^-_{4}=\frac 12 \left(\onee{4}-\gamma^5\right)
	=\begin{pmatrix}
		\delta_\alpha^{\phan{\alpha}\beta} &0\\
		0&0
	\end{pmatrix}\eqncom\qquad
	P^+_{4}=\frac 12 \left(\onee{4}+\gamma^5\right)
	=\begin{pmatrix}
		0 &0\\
		0&\delta^{\dot\alpha}_{\phan{\alpha}\dot\beta}
	\end{pmatrix}\eqndot
\end{equation}

Finally, a Majorana spinor must be invariant under the transformation \eqref{eq:Majorana_equation}. The explicit realisation of $B$ in \appref{sec:4D_Clifford_algebra} restricts the form of a Majorana spinor in four dimensions to
\begin{equation}
\Psi_4^\tM=\begin{pmatrix}
\chi_\alpha\\
\chi^{\dagger\dot{\alpha}}
\end{pmatrix}\eqncom\qquad
\ol{\Psi_4^\tM}=(\Psi_4^\tM)^\dagger\beta 
=(\chi^\alpha,\chi^\dagger_{\dot\alpha})\eqndot
\end{equation}

\subsection{Spinors in six-dimensional Euclidean space}\label{sec:spinors-in-six-dimensional-euclidean-space}
We construct spinors in six-dimensional Euclidean space from the general considerations in  \appref{sec:spinors_general_properties} and the explicit Clifford algebra representation in \appref{sec:6D_Clifford_algebra}. Further details regarding general results in this subsection can be found in \cite{OFarrill,Deligne99quantumfields}.

As before, the spinor representation splits into a minus- and plus-chirality Weyl representation. In six dimensions, however, there is an isomorphism $\text{Spin}(6)\simeq\SU{4}$ under which the minus- and plus-chirality representations are mapped to the fundamental and anti-fundamental representation of \SU{4}, respectively. Therefore, the minus- and plus-chirality Weyl representations become dual to each other, as can be seen from \eqref{eq:Sigmabar_6} and we only need one set of (capital latin) indices to characterise spinors in six-dimensional Euclidean space. We choose plus-chirality spinors to carry a star that distinguishes them from minus-chirality ones in an index-free notation. Under hermitian conjugation, they turn into each other as
\begin{equation}
\psi^{* A}\equiv(\psi^A)^\dagger\eqncom\qquad
\psi^*_{A}\equiv(\psi_A)^\dagger
\qquad\text{and}\qquad
(\psi^{* A})^\dagger=\psi^{A}\eqncom\qquad
(\psi^*_{A})^\dagger=\psi_{A}\eqndot
\end{equation}
Note that for fixed spinor indices hermitian and complex conjugation yield the same conjugate spinor. Two Weyl spinors in the fundamental and anti-fundamental representations can be contracted as
\begin{equation}
\psi^*\chi=\psi^{*A}\chi_A=\psi^{*}_A\chi^A\eqncom\qquad
\psi\chi^*=\psi^{A}\chi^*_A=\psi_A\chi^{*A}\eqncom
\end{equation}
where the position of the \su{4} summation-indices does not matter since they are raised and lowered with Kronecker-$\delta$'s. Using the $\Sigma$-matrices defined in \appref{sec:6D_Clifford_algebra}, we can construct six-dimensional vectors from the Weyl spinors of the form $\psi^A\Sigma^a_{AB}\chi^B$ and $\psi_A\bar\Sigma^{aAB}\chi_B$. Under hermitian conjugation, these products transform as
\begin{equation}
\begin{aligned}
(\psi^*\chi)^\dagger&=(\chi^A)^\dagger(\psi^*_A)^\dagger=\chi^*\psi\eqncom\\
(\psi^A\Sigma^a_{AB}\chi^B)^\dagger&=(\chi^B)^\dagger(\Sigma^a_{AB})^\dagger(\psi^A)^\dagger
=\chi^{*}_B\bar\Sigma^{aBA}\psi^{*}_A\eqncom\\
(\psi_A\bar\Sigma^{aAB}\chi_B)^\dagger&=(\chi_B)^\dagger(\bar\Sigma^{aAB})^\dagger(\psi_A)^\dagger
=\chi^{*B}\Sigma^{a}_{BA}\psi^{*A}\eqncom\\
\end{aligned}
\end{equation}
where we used that $(\Sigma^a_{AB})^\dagger=(\Sigma^a_{AB})^*$ for fixed $A$ and $B$.

From these Weyl spinors the eight-dimensional Dirac spinor is constructed as
\begin{equation}
\Psi_6=\begin{pmatrix}
\psi_A\\ \chi^{* A}
\end{pmatrix}\eqncom\qquad
\ol{\Psi}_6=\Psi_6^\dagger\,\beta_6=(\psi^{* A},\chi_A)\eqndot
\end{equation}
and the projectors to left- and right-handed subspaces take the explicit form
\begin{equation}
	P^{-}_6=\frac 12 \left(\onee{8}-\rho^7\right)
	=\begin{pmatrix}
	\delta_A^{\phan{A}B} &0\\
	0&0
	\end{pmatrix}
\eqncom\qquad
	P^{+}_6=\frac 12 \left(\onee{8}+\rho^7\right)
	=\begin{pmatrix}
	0 &0\\
	0&\delta^{A}_{\phan{A}B}
	\end{pmatrix}
\eqndot
\end{equation}
By construction, $\Psi_6$ transforms under the minus- and plus-chirality Weyl representation. As the latter two are dual to each other in six-dimensional Euclidean space, $\Psi_6$ transforms under a reducible representation.

The Majorana representation, in constrast to the Dirac representation is irreducible in six-dimensional Euclidean space. A Majorana spinor must be invariant under the transformation \eqref{eq:Majorana_equation}. The explicit realisation of $B_6$ in \appref{sec:6D_Clifford_algebra} restricts the form of a Majoranor spinor in six Euclidean dimensions to
\begin{equation}
\Psi_6^\tM=\begin{pmatrix}
\psi_A\\
\psi^{*A}
\end{pmatrix}\eqncom\qquad
\ol{\Psi_6^\tM} =(\Psi_6^\tM)^\dagger\beta_6 
=(\psi^{* A},\psi_A)\eqndot
\end{equation}

\subsection{Spinors in ten-dimensional Minkowski space}\label{sec:Spinors_in_ten-dimensional_Minkowski_space}
For spinors in ten-dimensional Minkowski space, we can now combine the four-dimensional Minkwoski space spinors with the six-dimensional Euclidean space spinors. The ten-di\-men\-sio\-nal spinors fall into inequivalent representations, which are the minus- and plus-chirality Majorana-Weyl spinors. Further details regarding general results in this subsection can be found in \cite{OFarrill,RauschdeTraubenberg:2005aa,Deligne99quantumfields}.

In analogy to the construction in \appref{sec:ten-dimensional_Minkowski_Clifford_algebra}, we take the ten-dimensional spinors to be
\begin{equation}
\Psi_{10}=\Psi_6\otimes\Psi_4\eqncom\qquad
\ol{\Psi}_{10}=\ol{\Psi}_6\otimes\ol{\Psi}_4\eqndot
\end{equation}
The projectors to the minus- and plus-chirality subspaces can be expressed in terms of the projectors in six and four dimensions as
\begin{equation}\label{eq:Projectors_LR_10}
\begin{aligned}
	P^{-}_{10} 
		=P^-_{6}\otimes P^-_{4}+P^+_{6}\otimes P^+_{4}\eqncom\qquad
	P^+_{10} 
		=P^-_{6}\otimes P^+_{4}+P^+_{6}\otimes P^-_{4}	\eqndot
\end{aligned}
\end{equation}
Like before, a Majorana spinor must be invariant under the transformation in \eqref{eq:Majorana_equation}. Since the Majorana condition matrix takes the form $B_{10}=B_6\otimes B$, we can construct it as a Kronecker product of the lower dimensional Majorana spinors as
\begin{equation}\label{eq:Majorana_spinor_10}
\begin{aligned}
\Psi_{10}^\tM=\Psi_{6}^\tM\otimes\Psi_{4}^\tM
=\begin{pmatrix}
\psi_A\\
\psi^{* A}
\end{pmatrix}
\otimes
\begin{pmatrix}
\chi_\alpha\\
\chi^{\dagger\dot{\alpha}}
\end{pmatrix}\quad\eqncom\quad
\ol{\Psi_{10}^\tM}=\ol{\Psi_{6}^\tM}\otimes\ol{\Psi_{4}^\tM}
=(\psi^{* A},\psi_A)
\otimes
(\chi^\alpha,\chi^\dagger_{\dot\alpha})\eqndot
\end{aligned}
\end{equation}
In ten-dimensional Minkowski space, the Majorana- and Weyl-condition are compatible. Therefore, we combine \eqref{eq:Majorana_spinor_10} with \eqref{eq:Projectors_LR_10} to define Majorana-Weyl spinors as
\begin{equation}
\begin{aligned}
\Psi_{10}^{\tM -}&=
\begin{pmatrix}
\psi_A\\
0
\end{pmatrix}
\otimes
\begin{pmatrix}
\chi_\alpha\\
0
\end{pmatrix}+
\begin{pmatrix}
0\\
\psi^{*A}
\end{pmatrix}
\otimes
\begin{pmatrix}
0\\
\chi^{\dagger\dot\alpha}
\end{pmatrix}\eqncom\\ 
\ol{\Psi_{10}^{\tM -}}&=
(\psi^{*A},0)\otimes(0,\chi^{\dagger}_{\dot{\alpha}})+
(0,\psi_A)\otimes(\chi^\alpha,0)\eqncom\\
\Psi_{10}^{\tM +}&=
\begin{pmatrix}
0\\
\psi^{* A}
\end{pmatrix}
\otimes
\begin{pmatrix}
\chi_\alpha\\
0
\end{pmatrix}+
\begin{pmatrix}
\psi_{A}\\
0
\end{pmatrix}
\otimes
\begin{pmatrix}
0\\
\chi^{\dagger\dot\alpha}
\end{pmatrix}\eqncom\\ 
\ol{\Psi^{\tM+}_{10}}&=
(0,\psi_A)\otimes(0,\chi^{\dagger}_{\dot{\alpha}})+
(\psi^{*A},0)\otimes(\chi^\alpha,0)\eqndot
\end{aligned}
\end{equation}
To eliminate the redundancies of the Dirac-representation, let us define the eight-component Majorana-Weyl spinors
\begin{equation}
\begin{aligned}
\lambda_{A\alpha}=
\psi_A\otimes\chi_\alpha\eqncom\quad
\ol{\lambda}^{ A\dot\alpha}=
\psi^{*A}\otimes\chi^{\dagger\dot\alpha}\eqncom\quad
\lambda^A_{\alpha}=
\psi^{* A}
\otimes
\chi_\alpha\eqncom\quad
\ol{\lambda}_A^{\dot\alpha}=
\psi_{A}\otimes\chi^{\dagger\dot\alpha}\eqncom
\end{aligned}
\end{equation}
where the first two spinors have negative chirality and the last two ones have positive chirality in ten dimensions. Under hermitian conjugation, they transform into each other as
\begin{equation}
(\lambda_{A\alpha})^\dagger=\ol{\lambda}^A_{\dot\alpha}\eqncom\qquad
(\ol{\lambda}^{ A\dot\alpha})^\dagger=\lambda^\alpha_A\eqncom\qquad
(\lambda^A_{\alpha})^\dagger=\ol{\lambda}_{A\dot{\alpha}}\eqncom\qquad
(\ol{\lambda}_A^{\dot\alpha})^\dagger=\lambda^{A\alpha}\eqndot
\end{equation}

\section{Kaluza-Klein compactification}\label{app:Kaluza_Klein_compactification}
In this appendix, we present the Kaluza-Klein compactification of the non-abelian gauge fields and fermions in the action \eqref{eq:10d_action} from $4+q$ spacetime dimensions to $4$ spacetime dimensions and $q$ compact dimensions of radius $R$. The idea behind the Kaluza-Klein compactification is to have a physical system that stretches over more than the naive macroscopic dimensions. The extra dimensions are compactified on some manifold that is not accessible at low energies -- that is the scale $R$ of the manifold is small compared to distances which can be probed at low energies. The concept was first introduced in \cite{Nordstroem} and later applied to describe Einstein gravity together with Electro-Magnetism \cite{Kaluza, Klein}. While this original idea cannot be applied to describe natural phenomena up to the present moment, the concept of extra dimensions proofed very useful in the developed superstring theories \cite{Candelas198546, Yau}. The extra dimensions required by superstring theory can be compactified so that it appears as a lower-dimensional theory augmented with internal dimensions that are only accessible at the Planck scale. Despite the many interesting features of Kaluza-Klein compactifications in string theory, we will focus on the original and simple mechanism described in \cite{Scherk1974347, Witten1981412} and in the overview articles \cite{Cheng:2010pt, Duff19861, Witten_KK}. We will first discuss the gauge-field part and thereafter the fermion part of \eqref{eq:10d_action}. 

Let us start with a non-abelian gauge field $A^M$ in five dimensions composed of a four-dimensional gauge field $A^\mu$ and an additional component $A^5=\phi$. We would like to compactify the space as $\RR^{(4,1)}\rightarrow \RR^{(3,1)}\times S^1$ where the $S^1$ has radius $R$. The gauge field then becomes periodic in the fifth coordinate as $A^M(x^\mu,y)=A^M(x^\mu,y+2\pi R)$ and we can decompose it into Fourier modes along $y$:
\begin{equation}\label{eq:gauge_field_in_extra_dim}
A^\mu(x^\nu,y)=\sum_{n\in\ZZ}A_{(n)}^\mu(x^\nu)\frac{\e^{\frac{\complexi n y}{R}}}{\sqrt{2\pi R}}\eqncom\qquad
\phi(x^\nu,y)=\sum_{n\in\ZZ}\phi_{(n)}(x^\nu)\frac{\e^{\frac{\complexi n y}{R}}}{\sqrt{2\pi R}}\eqndot
\end{equation}
Note that $\bigl(A_{(n)}^M\bigr)^\dagger=A_{(-n)}^M$, since we take $A^M$ to be real-valued. The five-dimensional coupling constant $g_5$ has mass dimension $[g_5]=-\frac 12$ and we define the four-dimensional coupling constant as
\begin{equation}
g_4=\frac{g_5}{\sqrt{2\pi R}}\eqncom
\end{equation}
so that it has mass dimension $[g_4]=0$. The five-dimensional covariant derivative acts on a field in the Fourier representation given above as
\begin{equation}\label{eq:covariant_derivative_KK}
\D^A X=\sum_{n\in\ZZ}\D^A X_{(n)}\frac{\e^{\frac{\complexi n y}{R}}}{\sqrt{2\pi R}}\qquad\text{with}\quad
\begin{cases}
\D^\mu X_{(n)}=\bigl(\partial^\mu X_{(n)}-\complexi g_4\sum_{m\in\ZZ}[A^\mu_{(m)},X_{(n-m)}]\bigr)\\
\D^5X_{(n)}=\bigl(\frac{\complexi n}{R} X_{(n)}-\complexi g_4\sum_{m\in\ZZ}[\phi_{(m)},X_{(n-m)}]\bigr)
\end{cases}\hspace{-.4cm}\eqndot
\end{equation}
With these definitions, the field strength components are
\begin{equation}
F^{MN}=\frac{\complexi}{g_5}[\D^M,\D^N]=\sum_{n\in\ZZ}F^{MN}_{(n)}\frac{\e^{\frac{\complexi n y}{R}}}{\sqrt{2\pi R}}\eqncom
\end{equation}
with
\begin{align}
F^{\mu\nu}_{(n)}=\Bigl(\partial^\mu A^\nu_{(n)}-\partial^\nu A^\mu_{(n)}-\complexi g_4 \sum_{m\in\ZZ}[A^\mu_{(m)},A^\nu_{(n-m)}]\Bigr)\quad\eqncom\quad
F^{\mu5}_{(n)}=\Bigl(-\frac{\complexi n}{R}A^\mu_{(n)}+\D^\mu \phi_{(n)}
\Bigr)\eqndot
\end{align}
Note that we can lift the dependence on the kinetic part of $\phi_{(n)}$ by the following gauge transformation
\begin{equation}\label{eq:gauge_extra_dimensions}
A^\mu_{(n)}\rightarrow A^\mu_{(n)}+\frac{R}{\complexi n}D^\mu \phi_{(n)} \qquad \text{for } n\neq 0\eqndot
\end{equation}
Now we can calculate the five-dimensional action of a free non-abelian gauge field with one compactified dimension
\begin{equation}\label{eq:action_KK_gaugefield}
\begin{aligned}
S&= -\frac 14\int \de^4 x\de y\,\tr\bigl[F^{MN}F_{MN}\bigr]
= -\frac 14\int \de^4 x\de y\tr\bigl[F^{\mu\nu}F_{\mu\nu}+2F^{\mu 5}F_{\mu 5}\bigr]\\
&= -\frac 14\sum_{m,n\in\ZZ}\int \de^4 x\de y\,\frac{\e^{\frac{\complexi (n+m) y}{R}}}{2\pi R}
\tr\bigl[F^{\mu\nu}_{(m)}F_{(n)\mu\nu}+2F^{\mu 5}_{(m)}F_{(n)\mu 5}\bigr]\\
&= -\frac 14\sum_{n\in\ZZ}\int \de^4 x
\tr\bigl[F^{\mu\nu}_{(-n)}F_{(n)\mu\nu}+2F^{\mu 5}_{(-n)}F_{(n)\mu 5}\bigr]\\
&= -\frac 14\int \de^4 x\tr\Bigl[\sum_{n\in\ZZ}\Bigl(F^{\mu\nu}_{(-n)}F_{(n)\mu\nu}
+2\frac{n^2}{R^2}A^{\mu}_{(-n)}A_{(n)\mu}\Bigr)
+2\bigl(\D^{\mu}\phi_{(0)}\bigr)\bigl(\D_\mu\phi_{(0)}\bigr)\Bigr]\eqncom
\end{aligned}
\end{equation}
where we used the gauge transformation \eqref{eq:gauge_extra_dimensions} in the last equality. In the action with one compactified dimension, we find a tower of four-dimensional non-abelian gauge fields $A^\mu_{(n)}$ with masses $m_{A_{(n)}}=\frac{n^2}{R^2}$ and a massless scalar field $\phi_{(0)}$. All scalar fields $\phi_{(n\neq 0)}$ have vanished from the action and transformed into the longitudinal \dof of the massive gauge fields $A^\mu_{(n\neq 0)}$. The two massless fields $A^\mu_{(0)}$ and $\phi_{(0)}$ are coupled to all the remaining massive fields through the covariant derivative \eqref{eq:covariant_derivative_KK}. 

Let us take the limit when the size of the compactified dimension goes to zero, i.e.\ $R\rightarrow 0$. We find that all massive fields in \eqref{eq:action_KK_gaugefield} become infinitely heavy and hence do not contribute to any observable process as long as we only probe distances $d\gg R$. In this limit, we obtain the action of a non-abelian gauge field that minimally interacts with a real scalar field
\begin{equation}
\begin{aligned}
S&= \int \de^4 x\tr\Bigl[-\frac 14 F^{\mu\nu}F_{\mu\nu}
-\frac 12\bigl(\D^{\mu}\phi\bigr)\bigl(\D_\mu\phi\bigr)\Bigr]\eqncom
\end{aligned}
\end{equation}
where the covariant derivative is $\D^{\mu}X=\partial^\mu X-\complexi g_4[A^\mu,X]$, the field strength tensor is $F^{\mu\nu}=\frac{\complexi}{g^4}[\D^\mu,\D^\nu]$ and we have dropped the Fourier-mode index $(0)$. Note that we could have gotten this result by heuristically setting $\partial^5\rightarrow 0$ and dropping the integration over $y$ in the five-dimensional action.

After compactifying one dimension on a circle with radius $R$, let us generalise this result to compactify $q$ dimensions on a $q$-torus $\T^{q}=(S^1)^{\times q}$, that is on $q$ circles each with radius $R$. In the six-dimensional case with two compactified dimensions this was done in \cite{0954-3899-15-6-007}.
In principle, we can start from $(4+q)$ dimensions and iterate the above procedure until we reach four dimensions. We then have $q$ different Fourier-mode numbers ${\bf{n}}=(n_1,\dots,n_q)$ and scalar fields $\phi^i_{(\bf{n})}$ with $i=\{1,2,\dots,q\}$. Each of the $q$ scalar fields obtains a mass term of the form $m_{\phi_{(\bf{n})}^i}\sim\frac{\bf{n}^2}{R^2}$. So, if we focus again on the limit $R\rightarrow 0$ all massive modes vanish and we are left with the zero-mode action with ${\bf{n}}=(0,\dots,0)$. Like in the five-dimensional case, this action can simply be obtained by setting the partial derivatives with respect to the extra dimensions to zero in the $(4+q)$-dimensional action. For the field strength part of \eqref{eq:10d_action}, we obtain
\begin{equation}
\begin{aligned}\label{eq:Derivation_Kaluza_Klein_gaugefield}
S&= \int \de^4 x\tr\Bigl[-\frac 14 F^{\mu\nu}F_{\mu\nu}
-\frac 12\bigl(\D^{\mu}\phi^i\bigr)\bigl(\D_\mu\phi_i\bigr)
+\frac{g_4}{4}[\phi^i,\phi^j][\phi_i,\phi_j]
\Bigr]\eqncom
\end{aligned}
\end{equation}
where $i$ and $j$ run over the extra $q$ dimensions and we have suppressed the Fourier-mode indices again.

Having dealt with gauge fields, we will now investigate how fermions behave under the compactification procedure. The action of a massless Dirac fermion coupled to a non-abelian gauge field in $4+q$ dimensions, like in the action \eqref{eq:10d_action}, directly splits into a four-dimensional contribution with summation index $\mu$ and a $q$-dimensional remainder with summation index $A$ as
\begin{equation}\label{eq:action_Dirac_KK}
S=\int\de^{4+q}x \tr\left[\complexi\, \ol{\Psi}\Gamma^N\D_N\Psi\right]
=\int\de^{4+q}x \tr\Bigl[\ol{\Psi}\bigl(
\complexi\Gamma^\mu \D_\mu+
\complexi\Gamma^A \D_A\bigr)\Psi\Bigr]\eqndot
\end{equation}
After the compactification, the $q$-dimensional space is a torus $\T^{q}=(S^1)^{\times q}$ where each circle has radius $R$. The spectrum of the Dirac operator in a compact space is discrete and its eigenvalues $\lambda_{(n)}$ are either zero or of $\frac{n}{R}$ order\footnote{In principle, the fermion field can be Fourier-expanded in extra dimensions in analogy to \eqref{eq:gauge_field_in_extra_dim}. For a free fermion it is then clear that $\lambda_{(n)}$ is proportional to $\frac{n}{R}$.}. So, if we write \eqref{eq:action_Dirac_KK} in terms of eigenfunctions of $\complexi \Gamma^j \partial_j$, we can integrate out the extra dimensions and obtain
\begin{equation}
S=\int\de^{4}x \sum_{n\in\ZZ}\tr\Bigl[\ol{\psi}_{(n)}\bigl(
\complexi\Gamma^\mu \D_\mu+
\lambda_{(n)}\bigr)\psi_{(n)}
+ g_4 \sum_{m\in\ZZ}\ol{\psi}_{(n)}\bigl[\Gamma^j\phi_{(m)j},\psi_{(n-m)}\bigr]
\Bigr]\eqncom
\end{equation}
where $\psi_{(m)}$ and $\phi_{(m)j}$ are the Fourier-modes of the fermion field and of the $j^{\text{th}}$ gauge-field component, respectively. From dimensional analysis, we have the relations
\begin{equation}
[\psi_{(m)}]=\frac{3}{2}\eqncom\qquad
[\phi_{(m)j}]=1\eqncom\qquad
[g_{4+q}]=-\frac q2\eqncom\qquad
g_4=\frac{g_{4+q}}{(2\pi R)^{\frac q2}}\eqndot
\end{equation}
This is the action of $n$ massive Dirac fermions\footnote{Note that in a chiral representation the chiralities of the $4$- and the $q$-dimensional fermions are correlated, c.f.\ \cite{Witten_KK}. We choose to ignore this subtlety, as we explicitly work out all needed $\Gamma$-matrices in \appref{sec:Clifford_algebras_in_various_dimensions}.} in four dimensions with masses $|\lambda_{(n)}|$ coupled to $j$ scalar fields through the coupling tensors $\Gamma^j$. Like in the gauge-field case we now take the limit $R\rightarrow 0$, which sends all masses to infinity and thus effectively decouples all massive fields from the theory. We obtain
\begin{equation}\label{eq:Derivation_Kaluza_Klein_fermion}
S=\int\de^{4}x \tr\Bigl[\ol{\psi}\complexi\Gamma^\mu \D_\mu\psi
+ g_4 \ol{\psi}\bigl[\Gamma^j\phi_{j},\psi\bigr]
\Bigr]\eqncom
\end{equation}
where the covariant derivative acts as $\D^{\mu}X=\partial^\mu X-\complexi g_4[A^\mu,X]$ and we have dropped the Fourier-mode index $(0)$.

\section{The conformal algebra}\label{sec:conformal_algebra}
In this appendix, we review some aspects of the conformal algebra which generates the conformal group. We follow the detailed presentation of \cite[\chap{4}]{philippe1997conformal} and \cite{Mack1969174,CFT} and refer the reader there for further discussions.

\subsection{Conformal transformations of coordinates}
The conformal group can be seen as an extension of the Poincar\'e group by dilatations or scale transformations and the so-called special conformal transformations. Its generators are the ones of proper orthochronous Lorentz transformations $M_{\mu\nu}$, translations $P_\mu$, dilatations $D$, and special conformal transformations $K_\mu$. When these generators act on coordinates, they can be represented by the following differential operators
\begin{equation}\label{eq:conformal_generators}
\begin{aligned}
\mathcal{P}_\mu=\frac{1}{\complexi} \partial_\mu\eqncom\qquad
\mathcal{M}_{\mu\nu}=\complexi (x_\mu\partial_\nu-x_\nu\partial_\mu)\eqncom\qquad
\mathcal{D}=\complexi x_\mu\partial^\mu\eqncom\qquad
\mathcal{K}_\mu=\frac{\complexi}{2}\Bigl( x_\mu x_\nu\partial^\nu-\frac{x^2}{2}\partial_\mu\Bigr)\eqndot
\end{aligned}
\end{equation}
In $D$ dimensions, we have $D$ translations and special conformal translations, $\frac{1}{2}D(D-1)$ Lorentz transformations and one dilatation, yielding a total of $\frac{1}{2}(D+2)(D+1)$ generators of the conformal group. This is the exact number of generators that the algebra $\so{D,2}$ has, and indeed the conformal group in $D$-dimensional Minkowski space is isomorphic to the group $SO(D,2)$.
The commutation relations of the conformal algebra consist of the commutation relations of the Poincar\'e algebra
\begin{equation}\label{eq:Poincare_algebra}
\begin{aligned}
[\mathcal{P}_\mu,\mathcal{M}_{\nu\rho}]&=\complexi( \eta_{\mu\nu}\mathcal{P}_\rho-\eta_{\mu\rho}\mathcal{P}_\nu)\eqncom&\\
[\mathcal{M}_{\mu\nu},\mathcal{M}_{\rho\sigma}]&=
\complexi(\eta_{\mu\sigma}\mathcal{M}_{\nu\rho}
+\eta_{\nu\rho}\mathcal{M}_{\mu\sigma}
-\eta_{\mu\rho}\mathcal{M}_{\nu\sigma}
-\eta_{\nu\sigma}\mathcal{M}_{\mu\rho}
)\eqncom&
\end{aligned}
\end{equation}
supplemented with the additional commutation relations
\begin{equation}\label{eq:conformal_algebra}
\begin{aligned}
[\mathcal{K}_\mu,\mathcal{M}_{\nu\rho}]&=\complexi( \eta_{\mu\nu}\mathcal{K}_\rho-\eta_{\mu\rho}\mathcal{K}_\nu)\eqncom&
[\mathcal{K}_\mu,\mathcal{P}_\nu]&=\frac{\complexi}{2}(\eta_{\mu\nu}\mathcal{D}+ \mathcal{M}_{\mu\nu})\eqncom&\\
[\mathcal{D},\mathcal{P}_\mu]&=-\complexi \mathcal{P}_\mu\eqncom&
[\mathcal{D},\mathcal{K}_\mu]&=\complexi \mathcal{K}_\mu\eqndot&
\end{aligned}
\end{equation}

We can use the exponential map to realise macroscopic conformal transformations $ (C_g)^\mu_{\phan{\mu}\nu}$ of a point $x^\mu$ via unitary operators as
\begin{equation}\label{eq:unitary_trafo_coord}
\widehat{x^\mu}=U^{-1}_{(\alpha\cdot g)}x^\mu U_{(\alpha\cdot g)}\equiv (C_g)^\mu_{\phan{\mu}\nu}x^\nu\eqncom \qquad \text{with}\qquad
U_{(\alpha\cdot g)}=\e^{-\complexi \alpha\cdot g}\eqncom
\end{equation} 
where $g\in\{\mathcal{P}^\mu,\mathcal{K}^\mu,\mathcal{M}^{\mu\nu},\mathcal{D}\}$ is one of the conformal generators, $\alpha\cdot g$ denotes a suitable index contraction of the generator with the transformation parameter $\alpha$ and $C_g=C_g(\alpha)$ is a matrix representation of the transformation in coordinate space. For an infinitesimal transformation $C_g=1+\delta c$ we can linearise the transformation and obtain\footnote{The additional factor of $\frac 12$ in front of the parameter $\omega_{\nu\rho}$ is conventional and accounts for the antisymmetry of $\mathcal{M}^{\mu\nu}$ and $\omega_{\mu\nu}$.}
\begin{equation}
\begin{aligned}\label{eq:unitary_trafo_coord_infin}
M&:	&(x^\mu)^\prime-x^\mu&= 
\frac{\complexi}{2}\,\delta\omega_{\nu\rho}[\mathcal{M}^{\nu\rho},x^\mu]
&&=(\delta\omega^{\mu\nu})x_\nu\eqncom&\\
P&:	&(x^\mu)^\prime-x^\mu&= 
\complexi\, \delta a_\nu[\mathcal{P}^\nu,x^\mu]
&&=\left(\frac{\delta a^\mu x^\nu}{x^2}\right)x_\nu\eqncom&\\
K&:	&(x^\mu)^\prime-x^\mu&= 
\complexi\, \delta a_\nu[\mathcal{K}^\nu,x^\mu]
&&=\left(-\frac{\delta a_\rho x^\rho}{2}\eta^{\mu\nu}+\frac 14\delta a^\mu x^\nu\right)x_\nu\eqncom&\\
D&:	&(x^\mu)^\prime-x^\mu&= 
\complexi\, \delta s[\mathcal{D},x^\mu]
&&=(-\delta s\,\eta^{\mu\nu}) x_\nu\eqndot&
\end{aligned}
\end{equation}
While the special conformal transformation as above does not yield an intuitive interpretation, it becomes a simple translation, when it acts on the inverse of a vector $\complexi \delta a_\nu[\mathcal{K}^\nu, x^\mu x^{-2}]=\frac 14\delta a^\mu$. Note that the first term on the rightmost side in each line of \eqref{eq:unitary_trafo_coord_infin} is the linear contribution $\delta c$ of the matrix $C$. Combining \eqref{eq:unitary_trafo_coord_infin} with the exponentiation formula \eqref{eq:unitary_trafo_coord}, the macroscopic transformations can be obtained and they read
\begin{equation}
\begin{aligned}
P&:	&\widehat{x^\mu}&= 
x^\mu+a^\mu\eqncom
& M&:	&\widehat{x^\mu}&= 
\Lambda^\mu_{\phan{\mu}\nu}x^\nu\eqncom\\
K&:	&\widehat{x^\mu}&= 
\frac{x^\mu-a^\mu x^2}{1-2a\cdot x+a^2 x^2}\eqncom
& D&:	&\widehat{x^\mu}&= 
\lambda x^\mu\eqncom
\end{aligned}
\end{equation}
with the shift $a^\mu$, the Lorentz transformation $\Lambda^\mu_{\phan{\mu}\nu}$ and the scale $\lambda$.

\subsection{Conformal transformations of fields}\label{sec:conformal-transformations-of-fields}
Having the transformations of coordinates under the action of the conformal group, we still need to work out the transformations of fields in given Lorentz representations. For this, we loosely follow \cite{Mack:1969rr} adapted to our notation. Let us define the conformal generators $\mathfrak{g}=\{\mathfrak{P}^\mu,\mathfrak{K}^\mu,\mathfrak{M}^{\mu\nu},\mathfrak{D}\}$ that realise the conformal transformations only on fields. They fulfil the same commutation relations \eqref{eq:Poincare_algebra} and \eqref{eq:conformal_algebra} as the generators $g$. A conformal transformation of a field $f_A(x)$ in a given Lorentz representation characterised by the (multi-) index $A$ takes the form
\begin{equation}\label{eq:conf_trafo_fields}
\hat{f}_A(x)=U_{(\alpha\cdot\mathfrak{g})}^{-1}f_A(x) U_{(\alpha\cdot\mathfrak{g})}
=(R_{\mathfrak{g}})_A^{\phan{A}B}f_B\bigl(C^{-1}_gx\bigr)\eqncom
\end{equation}
in analogy to \eqref{eq:unitary_trafo_coord}. This time, the matrix $R_{\mathfrak{g}}=R_{\mathfrak{g}}(\alpha)$ realises the transformation of the Lorentz representation and $ C^{-1}_g= C^{-1}_g(\alpha)$ accounts for the coordinate transformation. Note that the induced transformation on coordinates is inverted compared to \eqref{eq:unitary_trafo_coord}. This guarantees that a conformal transformation of the coordinate system is cancelled by a simultaneous transformation of the field values as
\begin{equation}\label{eq:shift_invariance}
\hat{f}_A(x^\prime)=\hat{f}_A(C_gx)=(R_{\mathfrak{g}})_A^{\phan{A}B}f_B(C^{-1}_gC_gx)=(R_{\mathfrak{g}})_A^{\phan{A}B}f_B(x)\eqndot
\end{equation}
To find the matrix representation $R_{\mathfrak{g}}$, we exploit the translational invariance to express all generators at position $x$ in terms of their counterparts at the origin\footnote{That is to say we characterise fields in terms of little group transformations that leave the origin invariant, as was done in \cite{Mack:1969rr}.}. In particular, we choose a basis such that the generator of translations $\mathfrak{P}^\mu$ does not act on Lorentz representation indices, i.e.\ $[\mathfrak{P^\mu},f_A(x)]=[\mathcal{P}^\mu,f_A(x)]=-\complexi \partial^\mu f_A(x)$. A field at position $x+a$ is then related to the field at position $x$ as\footnote{Note that for a fixed position $\hat{x}$ the commutator is $[P^\mu,f_A(\hat{x})]=-\complexi \partial^\mu f_A(x)\bigr|_{x=\hat{x}}$.}
\begin{equation}\label{eq:translation_operator}
f_A(x+a)=U^{-1}_{(a\cdot\mathfrak{P})}f_A(x)U_{(a\cdot\mathfrak{P})}\eqndot
\end{equation}
Using this translational invariance, we can express the commutators of the remaining generators with the field $f_A(x)$ in terms of fields located at the origin $f_A(0)$ as
\begin{equation}
\begin{aligned}\label{eq:fprime_conformal}
\bigl(\hat{f}_A(x)-f_A(x)\bigr)_{\delta c^2=0}&=\complexi \delta c_{m}\bigl[\mathfrak{g}^m,f_A(x)\bigr]
=\complexi \delta c_{m}U^{-1}_{(x\cdot \mathfrak{P})}
\bigl[\tilde{\mathfrak{g}}^m,f_A(0)\bigr]U_{(x\cdot \mathfrak{P})}\eqncom
\end{aligned}
\end{equation}
where $m$ is a generic (multi-) index comprised of the free indices of $\mathfrak{g}$, and the hatted generators are $\tilde{\mathfrak{g}}^m=U_{(x\cdot \mathfrak{P})}\mathfrak{g}^mU^{-1}_{(x\cdot \mathfrak{P})}$. They can be expressed in terms of generators $\mathfrak{g}^m_0$ acting on fields at the origin and the differential operators \eqref{eq:conformal_generators} by using the Baker-Campbell-Hausdorff formula and the fundamental commutation relations \eqref{eq:Poincare_algebra} -- \eqref{eq:conformal_algebra}:
\begin{equation}\label{eq:transformed_conformal_generator}
\tilde{\mathfrak{g}}^m
=U_{(x\cdot \mathfrak{P})}\mathfrak{g}^mU^{-1}_{(x\cdot \mathfrak{P})}
=\sum_{n=0}^\infty\frac{(-\complexi)^n}{n!}x_{\nu_1}\dots x_{\nu_n}
[\mathfrak{P}^{\nu_1},[\mathfrak{P}^{\nu_2},\dots,[\mathfrak{P}^{\nu_n},\mathfrak{g}^m]\dots]
\end{equation}
The representation of conformal generators on fields $\mathfrak{g}_0^m$ commute with position space elements, for example $[\mathfrak{g}_0^m,x^\mu]=[\mathfrak{g}_0^m,P^\mu]=0$ and they have the following commutation relations with conformal primary fields at the origin\footnote{In principle, the commutator of the special conformal generator could be chosen to be $[\mathfrak{K}_0^{\mu},f(0)]=\kappa^\mu f(0)$, with $\mathfrak{K}_0^{\mu}$ nilpotent and $\kappa^\mu$ massless \cite{Mack:1969rr}.}
\begin{equation}
\begin{aligned}\label{eq:commutator_fields_conf_generators}
\bigl[\mathfrak{K}_0^{\mu},f_A(0)\bigr]&=0\eqncom&
\bigl[\mathfrak{M}_0^{\mu\nu},f_{A}(0)\bigr]&=(S^{\mu\nu})_A^{\phan{A}B}f_{B}(0)\eqncom&
\bigl[(\mathfrak{D}_0),f_{A}(0)\bigr]&=-\complexi\Delta f_{A}(0)\eqncom&
\end{aligned}
\end{equation}
where $\Delta$ is the scaling dimension of the field \cite{philippe1997conformal}. We build $S^{\mu\nu}$ in the trivial-, spinor-, and vector-representation by combining invariant tensors such that all labels and the commutation relations are preserved. Acting on scalars, fermions or gauge fields, we find
\begin{equation}
S^{\mu\nu}\phi=0\eqncom\qquad
S^{\mu\nu}\psi=\frac{\complexi}{4}[\gamma^\mu,\gamma^\nu]\psi\eqncom\qquad
S^{\mu\nu}A^\rho=-\complexi (\eta^{\mu\rho}A^\nu-\eta^{\nu\rho}A^\mu)\eqndot
\end{equation}
Further details on the construction of the spinor-representation can be found in \ref{sec:spinor_in_various_dimensions}. The action of the conformal generators in \eqref{eq:fprime_conformal} can now be determined in two steps. First, using \eqref{eq:transformed_conformal_generator} and \eqref{eq:commutator_fields_conf_generators}, evaluate the action of $\tilde{\mathfrak{g}}^m$ on $f_A(0)$. In a second step, commute the unitary operators in $U_{(x\cdot\mathfrak{P})}$ back to the fields $f_A(0)$. For the variation of fields $\delta f_A(x) =\hat{f}_A(x)-f_A(x)$, we find
\begin{equation}
\begin{aligned}\label{eq:unitary_trafo_fields}
M&:	&\delta f_A(x)& 
=\frac{\complexi\delta\omega_{\mu\nu}}{2}[\mathfrak{M}^{\mu\nu},f_A(x)]
\hspace{-0.3cm}&&
=\frac{\complexi\delta\omega_{\mu\nu}}{2}\bigl(
(S^{\mu\nu})_A^{\phan{A}B}
-\mathcal{M}^{\mu\nu}\delta_A^{B}\bigr)f_B(x)\eqncom&&\\
P&:	&\delta f_A(x)& 
=\complexi\delta a_{\mu}[\mathfrak{P}^{\mu},f_A(x)]
\hspace{-0.3cm}&&
=\complexi\,\delta a_{\mu}\mathcal{P}^{\mu}f_A(x)\eqncom&&\\
K&:	&\delta f_A(x)& 
=\complexi\,\delta a_{\mu}[\mathfrak{K}^{\mu},f_A(x)]
\hspace{-0.3cm}&&
=\complexi\delta a_{\mu}\bigl(
(\frac{\complexi}{2} x^\mu \Delta+\mathcal{K}^{\mu})\delta_A^{B}-\frac{x_\nu}{2} (S^{\mu\nu})_A^{\phan{A}B}
\bigr)f_B(x)\eqncom&&\\
D&:	&\delta f_A(x)& 
=\complexi\,\delta s[\mathfrak{D},f_B(x)]
\hspace{-0.3cm}&&
=\complexi\delta s\bigl(-\complexi\Delta- \mathcal{D}\bigr)f_A(x)\eqncom&&
\end{aligned}
\end{equation}
where the generators $\{\mathcal{M},\mathcal{P},\mathcal{K},\mathcal{D}\}$ were defined in \eqref{eq:conformal_generators}. The linear contribution to $R_{\mathfrak{g}}$ in \eqref{eq:conf_trafo_fields} can immediately be obtained from the rightmost side of each line in \eqref{eq:unitary_trafo_fields} by setting the occurring field to $f_A(x)=1$.

In principle, we can now combine \eqref{eq:unitary_trafo_fields} with the exponentiation formula \eqref{eq:conf_trafo_fields} to obtain macroscopic conformal transformations of the fields $f_A$. Let us split the $D$-dimensional coordinate transformation $C^\mu_{\phan{\mu}\nu}$ into a scale contribution $\e^{-\alpha}=(\det C)^{1/D}$ and an angle transformation $\tilde C^\mu_{\phan{\mu}\nu}=\e^{\alpha} C^\mu_{\phan{\mu}\nu}$. Then, a conformal transformation turns into \cite[\chap{4}]{philippe1997conformal}
\begin{equation}
\begin{aligned}
\widehat{x^\mu}&=\e^{-\alpha}\tilde C^\mu_{\phan{\mu}\nu}x^\nu\eqncom\\
\hat{f}_A(\hat{x})&=\e^{\alpha\Delta_{f_A}}L_A^{\phan{A}B}f_B(x)\eqncom
\end{aligned}
\end{equation}
where $L=L(\tilde C^\mu_{\phan{\mu}\nu})$ realises all conformal transformations of the Lorentz representation index $A$ except those of dilatations.

\section{Comparison of the field and the oscillator representation}\label{sec:comparison-of-field-and-oscillator-representation}
In \subsecref{subsec:the-oscillator-representation}, the mapping of the symmetry generators of \NfSYMt from the field to the oscillator representation was given. In this appendix, we show that the symmetry transformations of fields given in \eqref{eq:unitary_trafo_fields_2}, \eqref{eq:R_symmetry_on_fields}, \eqref{eq:SUSY_trafo_scalar}, and \eqref{eq:SUSY_trafo_psi} are compatible with the symmetry transformations as they follow from (anti-)commuting oscillators in the oscillator picture presented in \subsecref{subsec:the-oscillator-representation}. To free ourselves from induced coordinate transformations of symmetry generators in the field representation, we focus on the transformation of fields at the origin. Finally, we discuss the mutual commutation relations of the symmetry generators.

For the dilatation operator $\mathfrak{D}_0$ in \eqref{eq:symmetry_generators_osci_LR} it is easy to see that it reproduces the classical scaling dimensions of all fields in \eqref{eq:fields_osci}. Hence, it is compatible with the definition \eqref{eq:unitary_trafo_fields_2} if the mapping \eqref{eq:generators_in_osci_rep} is included.

The $R$-symmetry generators with their action on fields given in \eqref{eq:R_symmetry_on_fields} are already in the oscillator representation of \eqref{eq:symmetry_generators_osci_LR}.

Next, we turn to the Lorentz generators which were defined in terms of spinorial indices in \eqref{eq:generators_in_osci_rep}. Their action on a field with spinor index $\alpha$ is\footnote{We could also evaluate the action of the spinorial Lorentz generators on vector fields. After expressing the vector fields in terms of $\spl{2} \times \splbar{2}$ indices we can again employ the identities of \appref{sec:Conventions} to find the same relation.}
\begin{equation}
[\mathfrak{M}_\alpha^{\phan{\alpha}\beta},f_\gamma]=
\frac 12 (\sigma^{\mu\nu})_\alpha^{\phan{\alpha}\beta}[\mathfrak{M}_{\mu\nu},f_\gamma]=
\frac 12 (\sigma^{\mu\nu})_\alpha^{\phan{\alpha}\beta} (\sigma_{\mu\nu})_\gamma^{\phan{\alpha}\sigma}f_\sigma
=\delta_\gamma^\beta f_\alpha-\frac 12 \delta_\alpha^\beta f_\gamma\eqncom
\end{equation}
where we used \eqref{eq:sigma_munu_projector} to arrive at the last equality. Taking the hermitian conjugate of this relation gives the analogous relation for $\ol{\mathfrak{M}}$ and we see that the oscillator representation of $\mathfrak{L}$ and $\ol{\mathfrak{L}}$ are also compatible with the spinorial representation.

The central charge was not present in our discussion of the symmetries in \subsecref{sec:symmetries}, since we restricted to theories where it vanishes when acting on physical fields. Hence, $\mathfrak{C}$ must vanish on all fields, which is indeed the case as can be seen from \eqref{eq:symmetry_generators_osci_LR} and \eqref{eq:fields_osci}.

Next, we turn to translations and special conformal transformations. On fields at the origin $f_A(0)$, we confirm that the oscillator representation reproduces the symmetry transformation \eqref{eq:unitary_trafo_fields_2} when we include the mapping \eqref{eq:generators_in_osci_rep} and note that translations act as $[\mathfrak{P}_{\alpha\dot\alpha},X]=-\complexi(\sigma^\mu)_{\alpha\dot\alpha}\D_\mu X=\D_{\alpha\dot{\alpha}}X$. Slightly more interesting, we can also calculate the action of $\mathfrak{P}_{\beta\dot\beta}$ followed by the action of $\mathfrak{K}^{\dot\alpha\alpha}$ on a field, say $\lambda_{\gamma}(0)$. In the oscillator representation, this yields
\begin{equation}\label{eq:KP_on_lambda_osci}
\mathfrak{K}^{\dot\alpha\alpha}\mathfrak{P}_{\beta\dot\beta}\lambda_\gamma=\delta^{\dot{\alpha}}_{\dot{\beta}}
(\delta^\alpha_\beta\lambda_\omega+\delta^\alpha_\omega\lambda_\beta)\eqndot
\end{equation}
In the field representation, this transformation is given by
\begin{equation}
\begin{aligned}\label{eq:KP_on_lambda_field}
[\mathfrak{K}^{\dot\alpha\alpha},[\mathfrak{P}_{\beta\dot\beta},\lambda_\gamma(0)]]&=
-(\bar\sigma_\mu)^{\dot\alpha\alpha}(\sigma_\nu)_{\beta\dot\beta}
\bigl([\mathfrak{P}^\nu,[\mathfrak{K}^{\mu},\lambda_\gamma(0)]]+[[\mathfrak{K}^{\mu},\mathfrak{P}^\nu],\lambda_\gamma(0)]]\bigr)\\
&=\complexi\delta^{\dot{\alpha}}_{\dot{\beta}}\delta^\alpha_\beta (-\complexi\Delta_\lambda) \lambda_\gamma(0)
-\frac{\complexi}{2}(\bar\sigma_\mu)^{\dot\alpha\alpha}(\sigma_\nu)_{\beta\dot\beta}(\sigma^{\mu\nu})_\gamma^{\phan{\alpha}\delta}\lambda_\delta(0)\\
&=\delta^{\dot{\alpha}}_{\dot{\beta}}\delta^\alpha_\beta \Bigl(\Delta_\lambda-\frac 12\Bigr) \lambda_\gamma(0)
+\delta^{\dot{\alpha}}_{\dot{\beta}}\delta^\alpha_\gamma\lambda_\beta(0)\eqncom
\end{aligned}
\end{equation}
where we used the Jacobi identity in the first equality, the algebra relations \eqref{eq:unitary_trafo_fields_2} and \eqref{eq:conformal_algebra} in the second and third equality and the $\sigma$ matrix identities of \appref{sec:Conventions} to get to the last equality. At the classical level with $\Delta_\lambda=3/2$, we find agreement with the oscillator representation result \eqref{eq:KP_on_lambda_osci}. At the quantum level, however, coupling-dependent corrections to $\Delta_\lambda$ spoil the relation and corrections to the symmetry generators need to be taken into account.

Next, we have the action of the supersymmetry generators $\mathfrak{Q}^{A}_\alpha$ and $\ol{\mathfrak{Q}}_{A\dot{\alpha}}$ on fields. Using the fundamental commutation relations \eqref{eq:osci_commutation_relations}, it is fairly simple to deduce in the oscillator representation
\begin{equation}
\begin{aligned}\label{oscillator_variations}
\ol{\mathfrak{Q}}_{O\dot\omega}\cF_{\alpha\beta}&=\frac 12(\D_{\alpha\dot{\omega}}\lambda_{O\beta}+\D_{\beta\dot{\omega}}\lambda_{O\alpha})\eqncom &
\mathfrak{Q}^{O}_\omega\cF_{\alpha\beta}&=0\eqncom\\
\ol{\mathfrak{Q}}_{O\dot\omega}\lambda_{A\alpha}&=\D_{\alpha\dot{\omega}}\varphi_{OA}\eqncom &
\mathfrak{Q}^{O}_\omega\lambda_{A\alpha}&=\delta^O_A\cF_{\alpha\omega}\eqncom\\
\ol{\mathfrak{Q}}_{O\dot\omega}\varphi_{AB}&=-\varepsilon_{OABC}\ol{\lambda}^C_{\dot{\omega}}\eqncom &
\mathfrak{Q}^{O}_\omega\varphi_{AB}&=\delta^O_A\lambda_{B\omega}-\delta^O_B\lambda_{A\omega}\eqncom\\
\ol{\mathfrak{Q}}_{O\dot\omega}\ol{\lambda}^A_{\dot\alpha}&=\delta_O^A\bar{\cF}_{\dot\alpha\dot{\omega}}\eqncom &
\mathfrak{Q}^{O}_\omega\ol{\lambda}^A_{\dot\alpha}&=-\D_{\omega\dot{\alpha}}\varphi^{OA}\eqncom\\
\ol{\mathfrak{Q}}_{O\dot\omega}\bar{\cF}_{\dot\alpha\dot\beta}&=0\eqncom &
\mathfrak{Q}^{O}_\omega\bar{\cF}_{\dot\alpha\dot\beta}&=\frac 12(\D_{\omega\dot{\alpha}}\ol{\lambda}^O_{\dot\beta}+\D_{\omega\dot{\beta}}\ol{\lambda}_{O\dot\alpha})\eqndot
\end{aligned}
\end{equation}
and for the SUSY variation including the parameters $\complexi\delta_{\epsilon,\ol{\epsilon}}=\complexi\bigl(\epsilon^\omega_O\mathfrak{Q}^O_\omega+\ol{\mathfrak{Q}}_{O\dot\omega}\ol{\epsilon}^{O\dot\omega}\bigr)$ this yields
\begin{equation}
\begin{aligned}\label{oscillator_variations_with_parameters}
\delta_{\epsilon,\ol{\epsilon}}\cF_{\alpha\beta}&=\frac{\complexi}{2}(\D_{\alpha\dot{\omega}}\lambda_{O\beta}+\D_{\beta\dot{\omega}}\lambda_{O\alpha})\ol{\epsilon}^{O\dot\omega}\eqncom&
\delta_{\epsilon,\ol{\epsilon}}\bar{\cF}_{\dot\alpha\dot\beta}&=
\frac{\complexi}{2}\epsilon^\omega_O(\D_{\omega\dot{\alpha}}\ol{\lambda}^O_{\dot\beta}+\D_{\omega\dot{\beta}}\ol{\lambda}_{O\dot\alpha})\eqncom&
\\
\delta_{\epsilon,\ol{\epsilon}}\lambda_{A\alpha}&=
-\complexi\D_{\alpha\dot{\omega}}\varphi_{OA}\ol{\epsilon}^{O\dot\omega}+\complexi\epsilon^\omega_A\cF_{\alpha\omega}\eqncom&
\delta_{\epsilon,\ol{\epsilon}}\ol{\lambda}^A_{\dot\alpha}&=
-\complexi\epsilon^\omega_O\D_{\omega\dot{\alpha}}\varphi^{OA}
-\complexi\bar{\cF}_{\dot\alpha\dot{\omega}}\ol{\epsilon}^{A\dot\omega}\eqncom&\\
\delta_{\epsilon,\ol{\epsilon}}\varphi_{AB}&=\complexi\bigl(\epsilon^\omega_A\lambda_{B\omega}-\epsilon^\omega_B\lambda_{A\omega}
-\varepsilon_{OABC}\ol{\lambda}^C_{\dot{\omega}}\ol{\epsilon}^{O\dot\omega}\bigr)\eqncom&
&&
\end{aligned}
\end{equation}
where we used that the parameters $\epsilon$ and $\ol{\epsilon}$ are fermionic. Note that the terms in the right column are the hermitian conjugates\footnote{For this, keep in mind the hermitian conjugation properties: 	$(\lambda_{A\alpha})^\dagger=\ol{\lambda}^A_{\dot{\alpha}}$, $(\varphi_{AB})^\dagger=\varphi^{AB}$, $(\D_{\alpha\dot{\beta}})^\dagger=-\D_{\beta\dot{\alpha}}$, $(\D^{\dot\alpha\beta})^\dagger=-\D^{\dot\beta\alpha}$ and $(\cF_{\alpha\beta})^\dagger=\bar\cF_{\dot{\beta}\dot{\alpha}}$.} of the terms in the left column. Comparing the individual results in \eqref{oscillator_variations_with_parameters} to the respective results \eqref{eq:scalar_variations_spinorial}, \eqref{eq:SUSY_trafo_psi}, and \eqref{eq:fieldstrength_variation}, we find that the SUSY transformations in terms of spinor indices are indeed represented by the oscillator representation \eqref{eq:osci_commutation_relations} -- \eqref{eq:symmetry_generators_osci_QSP}. In the transformation involving $\mathfrak{Q}$ and $\lambda$, we ignored the $\gym$-dependent terms with two scalars that appear in \eqref{eq:SUSY_trafo_psi}. In the oscillator picture, such transitions are possible by inserting the identity in the form of $\one=\cosc^{\dagger}_1 \cosc^{\dagger}_2 \cosc^{\dagger}_3 \cosc^{\dagger}_4$ before acting with $\mathfrak{Q}$. The fermionic oscillators that remain after the action of $\mathfrak{Q}$ are then symmetrically distributed on two spin-chain sites, which gives the desired term involving two scalars. In fact, since this transition depends on $\gym$, it is not part of the tree-level oscillator algebra and we take the field representation to determine the numerical prefactor that this transition must have in the oscillator representation. An analogous observation can be made for the transformation involving $\ol{\mathfrak{Q}}$ and $\ol\lambda$.

Finally, for special conformal SUSY generators, let us focus on the example
\begin{equation}
\mathfrak{S}^\alpha_A\lambda_{B\beta}=\delta^\alpha_\beta\varphi_{AB}\eqncom
\end{equation}
which immediately follows from the oscillator picture definitions. In the field representation, we can use \eqref{eq:action_of_special_conf_generators} and find
\begin{equation}
\begin{aligned}
\{\mathfrak{S}^\alpha_A,\lambda_{B\beta}(0)\}
&=-\frac{1}{2}(\bar\sigma_\mu)^{\dot{\alpha}\alpha}\Bigl(
-\{\ol{\mathfrak{Q}}_{A\dot{\alpha}},[\mathfrak{K}^\mu,\lambda_{B\beta}(0)]\}
+[\mathfrak{K}^\mu,\{\ol{\mathfrak{Q}}_{A\dot{\alpha}},\lambda_{B\beta}(0)\}]
\Bigr)\\
&=-\frac{1}{2}(\bar\sigma_\mu)^{\dot{\alpha}\alpha}[\mathfrak{K}^\mu,\D_{\beta\dot{\alpha}}\varphi_{AB}]
=\frac 12
[\mathfrak{K}^{\dot{\alpha}\alpha},[\mathfrak{P}_{\beta\dot{\alpha}},\varphi_{AB}]]\\
&=\delta^\alpha_\beta\varphi_{AB}\eqncom
\end{aligned}
\end{equation}
where we used that $[\mathfrak{K}^\mu,f(0)]=0$ and the last equality is obtained as a slight alteration of \eqref{eq:KP_on_lambda_field}. We see that the oscillator transformation indeed coincides with the field transformation.

Having dealt with the action of the symmetry generators on fields, we need to confirm that the algebra generated by the generators in the oscillator representation agrees with the algebra defined in \subsecref{sec:symmetries}. Upon rewriting\footnote{For this, some identities of \appref{sec:Conventions} are necessary.} \eqref{eq:symmetry_algebra_1} -- \eqref{eq:symmetry_algebra_last} in terms of the spinorial generators defined in \eqref{eq:generators_in_osci_rep} we find agreement with the algebra generated by the oscillator representation generators. This algebra was presented in detail in \cite[Appendix D]{Beisert:2004ry} and naturally it is no coincidence but due to construction that we arrive at this representation.

\section{Derivation of Feynman rules}\label{app:Feynman_rules}
In this appendix, we present the momentum-space Feynman rules for renormalised non-abelian gauge theories with massless Weyl fermions and complex scalars in Minkowski space. The presentation in this appendix employs the path integral formalism for correlation functions introduced in \secref{sec:path_integral_approach} and the derived rules are applicable in the cases of \NfSYMt and its deformations. We follow the discussion of Feynman rules with Weyl fermions \cite{Dreiner:2008tw} and in particular the momentum space discussion in \cite{Dedes:2006ni} and refer the reader there for details. For a general introduction to Feynman rules, \QFTs, and the path integral approach compatible with our conventions see also \cite{Srednicki:2007}. The rules derived in this appendix were successfully tested against the Euclidean-space Feynman rules derived in \cite{Siegnotes} on which all Feynman diagrammatic calculations in \cite{Fokken:2013aea,Fokken:2014soa,Fokken:2013mza} are based. For the test, we employed the \ttt{Mathematica} package \ttt{FokkenFeynPackage} which we constructed for this thesis and that will be introduced in \appref{sec:Feynman_rules_Mathematica}.

\subsection{The action and general setting}\label{sec:the-action-and-general-setting}
We focus on a non-abelian gauge theory with a gauge field $A^\mu$, with $A$ massless Weyl fermions $\lambda_{A\alpha}$ and $j$ massless complex scalars $\phi_j$. All fields transform in the adjoint representation of the gauge group \UN or \SUN. The colour group generators fulfil
\begin{equation}\label{eq:colour_generators}
\begin{aligned}
\tr\bigl(\T^a\bigr)&=\sqrt{N}\delta^{a0}\eqncom\quad&
\tr\bigl(\T^a \T^b\bigr)&=\delta^{ab}\eqncom\\
[\T^a,\T^b]&=\complexi f^{abc}\T^c\eqncom\quad&
\sum_{a=s}^{N^2-1}(\T^a)^i_{\phan{i}j}(\T^a)^k_{\phan{k}l}&=\delta^i_l\delta^k_j-\frac{s}{N}\delta^i_j\delta^k_l\eqncom&
\end{aligned}
\end{equation}
where $s=0$ and $s=1$ for gauge group \UN and \SUN, respectively. We assume that the renormalised action in $(D=4-2\epsilon)$-dimensional Minkowski space has the form
\begin{equation}\label{eq:action_Feynman_rules}
S=\int\de^{D}x \left(\tr\bigl[\mathcal{L}_0+\mathcal{L}_{\text{g}}+\mathcal{L}_{\text{m}}+\mathcal{L}_{\text{ct}}\bigr]+\mathcal{L}_{\text{dt}}\right)\eqncom
\end{equation}
where the trace is taken over the colour group generators. The free Lagrange density, the gauge field, matter and double-trace interactions have the explicit form\footnote{%
	This action is transformed to the one in the conventions of \cite{Fokken:2013aea} by replacing the couplings and tensors displayed here according to
	\begin{equation}
	\begin{aligned}
	\gym&\rightarrow\frac{\gym}{\sqrt{2}}\eqncom&
	(\rho^j)^{AB}&\rightarrow\sqrt{2}(\rho_j)_{AB}\eqncom&
	(\tilde{\rho}_j)^{AB}&\rightarrow\sqrt{2}(\tilde{\rho}^{\dagger j})_{AB}\eqncom&
	F^{ij}_{lk}&\rightarrow 2F^{ij}_{lk}\eqncom&
	Q_{\text{F}lk}^{ij}&\rightarrow 4Q_{\text{F}lk}^{ij}\eqncom
	\end{aligned}
	\end{equation}
	where the tensors on the \rhs are the ones defined in \cite{Fokken:2013aea}.
}
\begin{align}
\label{eq:Lagrangian_Feynman_rules_1}
\mathcal{L}_{0}&{}={}\frac 12 A^{\mu}\bigl(\eta_{\mu\nu}\partial^2-(1-\xi^{-1})\partial_\mu \partial_\nu\bigr)A^{\nu}
+ \ol{c}\partial^2 c
+\ol{\phi}^{j}\partial^2 \phi_{j}
+\ol{\lambda}^{A}_{\dot{\alpha}}\complexi(\bar\sigma^\mu)^{\dot{\alpha}\beta}\partial_\mu\lambda_{A\beta}
\eqncom
\\
\label{eq:Lagrangian_Feynman_rules_2}
\mathcal{L}_{\text{ct}}&{}={}
-\frac 12\delta_{A}A^{\mu}\bigl(\eta_{\mu\nu}\partial^2-(1-\xi^{-1})\partial_\mu \partial_\nu\bigr)A^{\nu}
-\delta_{c}\ol{c}\partial^2 c
-\delta_{\phi}\ol{\phi}^{j}\partial^2 \phi_{j}
-\delta_{\lambda}\ol{\lambda}^{A}_{\dot{\alpha}}\complexi(\bar\sigma^\mu)^{\dot{\alpha}\beta}\partial_\mu\lambda_{A\beta}
\eqncom\\
\label{eq:Lagrangian_Feynman_rules_gluon}
\mathcal{L}_{\text{g}}&{}={}
		\complexi \gym \mu^{\epsilon}Z_{A^3}[A_{\mu},A_{\nu}]\partial^{\mu}A^{\nu}+\frac{\gym^2\mu^{2\epsilon}}{4}Z_{A^4}[A^{ \mu},A^{ \nu}][A_{\mu},A_{\nu}]
		+\complexi\gym \mu^{\epsilon}Z_{\ol cAc}\partial^\mu\ol{c}[A_\mu,c]
		\nonumber\\
		&\phan{{}={}}
		+\complexi\gym \mu^{\epsilon}Z_{\ol \phi A\phi} \bigl([\partial_\mu\ol{\phi}^{j},A^{\mu}]\phi_j-[\ol{\phi}^{j},A^{\mu}]\partial_\mu\phi_j\bigr)
		+Z_{\ol \phi A^2\phi}\gym^2\mu^{2\epsilon}[A^{\mu},\ol{\phi}^{j}][A_\mu ,\phi_{j}]
		\nonumber\\
		&\phan{{}={}}
		+\gym\mu^{\epsilon} Z_{\ol \lambda A\lambda} (\bar\sigma^\mu)^{\dot{\alpha}\beta}\,\ol{\lambda}^{A}_{\dot{\alpha}}[A_\mu,\lambda_{A\beta}]\eqncom
\end{align}
\begin{align}
\label{eq:Lagrangian_Feynman_rules_3}
\mathcal{L}_{\text{m}}&{}={}
\gym \mu^{\epsilon}Z_{\lambda\phi\lambda}\Bigl(
(\rho^j)^{AB}\lambda^\alpha_{B}\phi_j\lambda_{A\alpha}
+(\rho^\dagger_j)_{AB}\ol\lambda^{B}_{\dot\alpha}\ol{\phi}^j\ol\lambda^{A\dot\alpha}\Bigr)
\nonumber\\&\phan{{}={}}
+\gym \mu^{\epsilon}Z_{\lambda\ol\phi\lambda}\Bigl((\tilde\rho^\dagger_j)^{AB}\lambda^\alpha_{B}\ol{\phi}^{j}\lambda_{A\alpha}
+(\tilde\rho^j)_{AB}\ol\lambda^{B}_{\dot\alpha}\phi_{j}\ol\lambda^{A\dot\alpha}
\Bigr)
\nonumber\\
&\phan{{}={}}
+\gym^2\mu^{2\epsilon}Z_{\text{F}}F^{ij}_{lk}\phi_{i}\phi_{j}\ol{\phi}^{k}\ol{\phi}^{l}
-\frac{\gym^2\mu^{2\epsilon}}{2}Z_{\text{D}}\bigl[\phi_{j},\ol{\phi}^{j}\bigr]\bigl[\phi_{k},\ol{\phi}^{k}\bigr]\eqncom
\\
\label{eq:Lagrangian_Feynman_rules_4}
\mathcal{L}_{\text{dt}}&{}={}
-\frac{\gym^2\mu^{2\epsilon}}{2N}\left(
Z_{\text{Fdt}}Q_{\text{F}lk}^{ij}\tr\bigl[\phi_i\phi_j\bigr]\tr\bigl[\ol{\phi}^k\ol{\phi}^l]
+
Z_{\text{Ddt}}Q_{\text{D}jl}^{ik}\tr\bigl[\phi_i\ol{\phi}^j\bigr]\tr\bigl[\phi_k\ol{\phi}^l]
\right)\eqndot
\end{align}
We use capital Latin indices $A,B\in\{1,2,3,4\}$ to label the fermionic flavours, small Latin indices $i,j,k,l\in\{1,2,3\}$ to label the flavours of the complex scalars, greek letters $\mu,\nu\in\{0,1,2,3\}$ for spacetime indices and $\alpha,\beta,\dot\alpha,\dot\beta\in\{1,2\}$ for \spl{2} and \splbar{2} spinors, respectively. In addition, small Latin indices $a,b,...\in\{s,1,\dots,N^2-1\}$ are colour indices. If colour indices appear twice, they are summed over although we always write them as upper indices, e.g.\ $\lambda_{A\alpha}=\lambda_{A\alpha}^a\T^a=\sum_{a=s}^{N^2-1} \lambda_{A\alpha}^a\T^a$. The dimensions of fields in $D$-dimensional position space are
\begin{equation}
[\phi]=[A]=[c]=\frac 12(D-2)\eqncom\qquad
[\lambda]=\frac 12(D-1)\eqncom
\end{equation}
which ensures that the partial derivatives have exactly dimension $[\partial^\mu]=1$ in generic dimensions. The mostly plus spacetime metric is $\eta_{\mu\nu}$ and in four-dimensional Minkowski space we use the $\sigma$ and $\bar{\sigma}$ matrices defined in \appref{sec:4D_Clifford_algebra}. Finally, for a field $f$ and a coupling $g$, the 1PI renormalisation constants are\footnote{The relative sign in the field renormalisation constants was chosen such that both counterterms of fields and couplings are given by the negative sum of divergences. In \cite[\chap{14}]{Srednicki:2007}, the sign in front of $\delta_f$ is absorbed into the counterterm.}
\begin{equation}\label{eq:renormalisation_constants}
Z_f=1-\delta_f\eqncom \qquad 
Z_g=1+\delta_g\eqncom \qquad 
\delta_{f\text{ or }g}=-\bigl(\text{divergence of involved 1PI graphs}\bigr)\eqncom
\end{equation}
where the exact and real counterterm $\delta$ starts at order $\gym^2$ and ensures that all calculated quantities remain finite. The bare fields and parameters, which are labelled by a subscript B and live in $(d=4)$-dimensional Minkowski space are related to the renormalised ones via the renormalisation constants as
\begin{equation}
	f_\text{B}=Z_f^{\frac 12} f\eqncom\qquad
	g_{\text{B}}=\mathcal{Z}_gg=\frac{\mu^{\Delta^0_{g}(d)-\Delta^0_{g}(D)}Z_{g}}{(Z_{f_1}Z_{f_2}\dots Z_{f_n})^{\frac 12}}g\eqncom\quad
	Z_f=1-\delta_f\eqncom\quad
	\cZ_g=1+\mathfrak{d}_g\eqncom\quad
	Z_g=1+\delta_g\eqncom
\end{equation}
where the coupling constant $g$ appears in an interaction term with the $n$ fields $f$ as in \secref{sec:N4SYM_renormalisation}. The action is real if the coupling tensors have the following properties under complex conjugation
\begin{equation}
\begin{aligned}\label{eq:coupling_tensor_conjugation}
(\rho^\dagger_j)_{BA}&=\bigl[(\rho^j)^{AB}\bigr]^\ast\eqncom&
Q^{lk}_{\text{F}ij}&=\bigl[Q^{ij}_{\text{F}lk}\bigr]^\ast\eqncom&
F^{lk}_{ij}&=\bigl[F^{ij}_{lk}\bigr]^\ast\eqncom&\\
(\tilde\rho^j)_{BA}&=\bigl[(\tilde\rho^\dagger_j)^{AB}\bigr]^\ast\eqncom&
Q^{jl}_{\text{D}ik}&=\bigl[Q^{ik}_{\text{D}jl}\bigr]^\ast\eqndot&&&
\end{aligned}
\end{equation}
In \NfSYMt, we have three and four flavours for the complex scalars and Weyl fermions, respectively and the coupling tensors take the form
\begin{equation}
\begin{aligned}\label{eq:coupling_tensors_N4}
(\rho^j)^{AB}&=-\complexi \sqrt{2}\varepsilon^{jAB4}\eqncom&
Q^{lk}_{\text{F}ij}&=0\eqncom&
F^{ij}_{lk}&=2\bigl(\delta^i_k\delta^j_l-\delta^j_k\delta^i_l\bigr)\eqncom&\\
(\tilde\rho^\dagger_j)^{AB}&=\complexi\sqrt{2} (\delta_j^A\delta_4^B-\delta_j^B\delta_4^A)\eqncom&
Q^{jl}_{\text{D}ik}&=0\eqndot&&&
\end{aligned}
\end{equation}
In the $\gamma_i$-deformed theory with gauge group \SUN, the field content is the same as in the undeformed theory, but the matter coupling tensors change to
\begin{equation}
\begin{aligned}\label{eq:coupling_tensors_gammai}
(\rho^j)^{AB}&=-\complexi \sqrt{2}\varepsilon^{jAB4}\e^{\frac{\complexi}{2}\mathbf{q}_{\lambda_A}\wedge\mathbf{q}_{\lambda_B}}\eqncom&
Q^{ij}_{\text{F}lk}&=\text{free}\eqncom&
F^{ij}_{lk}&=2\bigl(\delta^i_k\delta^j_l\e^{\complexi\mathbf{q}_{\phi_i}\wedge\mathbf{q}_{\phi_j}}
-\delta^j_k\delta^i_l\bigr)\eqncom&
\\
(\tilde\rho^\dagger_j)^{AB}&=\complexi\sqrt{2} (\delta_j^A\delta_4^B-\delta_j^B\delta_4^A)\e^{\frac{\complexi}{2}\mathbf{q}_{\lambda_A}\wedge\mathbf{q}_{\lambda_B}}\eqncom&
Q^{jl}_{\text{D}ik}&=\text{free}\eqncom&
&&
\end{aligned}
\end{equation}
where $\mathbf{q}$ are the $\mathfrak{u}(1)^{\times 3}$ Cartan-charge vectors defined in \tabref{tab: su(4) charges}, the antisymmetric product $\wedge$ is defined in \eqref{eq: antisymmetric product} and free for $Q_{\text{F}}$ and $Q_{\text{D}}$ means that these couplings are free constants up to \su{4} charge conservation. In the conformally invariant $\beta$-deformation with gauge group \SUN, the two free double-trace couplings are fixed to 
\begin{equation}
Q^{\beta\,ij}_{\text{F}lk}=2 F^{ij}_{lk}\eqncom\qquad Q^{\beta\,jl}_{\text{D}ik}=0\eqndot
\end{equation}
If the gauge group is \UN, additional couplings to \U{1} modes are possible for all fields, but we refrain from giving these couplings explicitly.

\subsection{Propagators and the free theory}\label{sec:propagators-and-the-free-theory}
To derive the free propagators from the action \eqref{eq:action_Feynman_rules}, we rewrite the free action as an integral over momentum space rather than position space. We adopt the notation and conventions from \cite{Srednicki:2007}. In a Fourier transformation, defined in \eqref{eq:Fourier_transformation}, the partial derivatives turn into $\partial_\mu\rightarrow \complexi p_\mu$, where the momentum also has dimension $[p^\mu]=1$ regardless of the dimension $D$ of spacetime. For the free action we use
\begin{equation}
\begin{aligned}
\int\de^D xf(x)M^{-1}(x)g(x)
&{}={}\int\de^D x\de^D y \frac{\de^Dk}{(2\pi)^D}\e^{-\complexi k(x-y)}f(x)M^{-1}(y)g(y)
\\
&=\int\de^D x\de^D y\frac{\de^D p}{(2\pi)^D}\frac{\de^D k}{(2\pi)^D}\frac{\de^D l}{(2\pi)^D}
\e^{ilx}\e^{-ik(x-y)}\e^{ipy}
\widehat{f}(l)\widehat{M^{-1}g}(p)\\
&=\int\de^D k\de^D p\frac{\delta^{(D)}(k+p)}{(2\pi)^D}\hat{f}(k)\widehat{M^{-1}g}(p)\eqncom
\end{aligned}
\end{equation}
where we used hats to denote Fourier transformed objects and refrained from integrating $k$ for later convenience. Note, however, that we will drop this distinction of the Fourier transformed fields immediately as we will only work in momentum space. The operator $M^{-1}$ corresponds to the inverse of the free Feynman propagators which take the explicit form\footnote{The spacetime part of the inverse propagators follows from \eqref{eq:Lagrangian_Feynman_rules_1} and is for scalars and ghosts $\Delta^{-1}(-p)=p^2$, for fermions $(S^{-1}(p))^{\dot{\alpha}\beta}=(\bar{\sigma}_\mu)^{\dot{\alpha}\beta}p^\mu$, and for gauge fields $(\Delta^{-1}(-p))_{\mu\nu}=\eta_{\mu\nu}p^2-(1-\xi^{-1})p_\mu p_\nu$.}
\begin{equation}
\begin{aligned}\label{eq:free_propagators}
\Delta_i^j&=\frac{\delta_i^j}{p^2-\complexi\epsilon}\eqncom&
(S(p))^B_{A\alpha\dot{\beta}}&=\frac{-(\sigma_\mu)_{\alpha\dot{\beta}}p^\mu}{p^2-\complexi\epsilon}\delta_A^B\eqncom&
\Delta&=\frac{1}{p^2-\complexi\epsilon}\eqncom&
\Delta^{\mu\nu}&=\frac{\eta^{\mu\nu}-(1-\xi)\frac{p^\mu p^\nu}{p^2}}{p^2-\complexi\epsilon}\eqncom&
\end{aligned}
\end{equation}
for scalars, fermions, ghosts, and gauge fields, respectively. They have exact integer dimensions regardless of the dimension $D$ of spacetime, since we have $[p^\mu]=1$. The fermionic propagator with upper indices is obtained by raising the spinor indices with the invariant tensors $(S(p))^{B\dot{\alpha}\alpha}_{A}=\varepsilon^{\alpha\beta}\varepsilon^{\dot\alpha\dot\beta}(S(p))^B_{A\beta\dot{\beta}}$. We assume that the free path integral is normalised to
\begin{equation}\label{eq:free_path_integral}
\int\mathcal{D}\{\ol{\phi}\phi\ol{\lambda}\lambda\ol{c}c A\}\e^{
	\complexi\int\de^Dk\de^Dp
	\frac{\delta^{(D)}(k+p)}{(2\pi)^D}\tr\left[
		-\ol{\phi}\Delta^{-1}\phi
		-\ol{\lambda}S^{-1}\lambda
		-\ol{c}\Delta^{-1}c
		-\frac 12 A\Delta^{-1}A
		\right]
	}=1\eqncom
\end{equation}
where we suppressed all contracted indices, the leftmost fields in each term depend on $k$ and the inverse propagators and rightmost fields depend on $p$. The generating functional of the free theory is obtained from this path integral via a shift of the field variables that leaves the overall result unchanged:
\begin{equation}
\begin{aligned}
\phi_j&\rightarrow\phi_j-(2\pi)^D\Delta_j^i J_i\eqncom\quad&
\lambda_{A\alpha}&\rightarrow \lambda_{A\alpha}-(2\pi)^DS^B_{A\alpha\dot{\beta}}\ol\eta_B^{\dot\beta}\eqncom\quad&
c_j&\rightarrow c_j-(2\pi)^D\Delta \eta\eqncom\quad&
\\
\ol{\phi}^j&\rightarrow\ol{\phi}^j-(2\pi)^D\ol{J}^i\Delta_i^j\eqncom\quad&
\ol{\lambda}^A_{\dot{\alpha}}&\rightarrow \ol{\lambda}^A_{\dot{\alpha}}+(2\pi)^D\eta^{B\beta}S^A_{B\beta\dot{\alpha}}\eqncom\quad&
\ol{c}&\rightarrow\ol{c}-(2\pi)^D\ol{\eta}\Delta\eqncom\quad&
\\
A^\mu&\rightarrow A^\mu-(2\pi)^D\Delta^{\mu\nu} J_\nu\eqncom&&&&&
\end{aligned}
\end{equation}
where the momentum dependence is always $f(p)\rightarrow f(p)\pm(2\pi)^DM_f(p)J_f(p)$.
Note that the sources transform like the fields under the colour group of the theory, e.g.\ $\eta^{A\alpha}=\eta^{A\alpha a}\T^a$. Under these shifts, the path integral acquires a source Lagrange density for renormalised fields of the form
\begin{equation}\label{eq:source_Lagrangian}
\mathcal{L}_{\text{source}}=\bigl(\eta^{A\alpha}(k)\lambda_{A\alpha}(p)+\ol{J}^j(k)\phi_j(p)+\ol{\eta}(k)c(p)+\text{h.c.}\bigr)
+J^\mu(k) A_\mu(p)\eqndot
\end{equation}
Using $\mathcal{L}_0$ implicitly given in \eqref{eq:free_path_integral}, the generating functional $Z_0=Z_0\bigl[\{J,\eta\}\bigr]$ in terms of all sources $\{J,\eta\}$ becomes
\begin{equation}
\begin{aligned}\label{eq:generating_functional_app}
Z_0&=
\int\mathcal{D}\{\ol{\phi}\phi\ol{\lambda}\lambda\ol{c}c A\}
\e^{
	\complexi\int\de^D k\de^D p\,\delta^{(D)}(k+p)\tr [
	\frac{1}{(2\pi)^D}\mathcal{L}_0+\mathcal{L}_{\text{source}}
	]	
	}\\
&=\exp\Biggl(\complexi\int\de^D k\de^Dp(2\pi)^D\delta^{(D)}(k+p)
\\
&\phan{=\exp\Biggl(\complexi\int\de^D}\Bigl[
\ol{J}^{ia}\Delta_i^{jab} J^b_j
+\eta^{A\alpha a}S^{Bab}_{A\alpha\dot{\beta}}\ol{\eta}_B^{\dot{\beta}b}
+\ol{\eta}^a\Delta^{ab}\eta^b+\frac 12J^{\mu a}\Delta^{ab}_{\mu\nu}J^{\nu b}
\Bigr]
\Biggr)\eqndot
\end{aligned}
\end{equation}
In the last equality the leftmost sources in each term depend on $k$ and the propagators and rightmost sources depend on $p$. We can now generate a field with momentum $p$ by taking a functional derivative of $Z_0$ with respect its corresponding source with momentum $-p$. We demand that the functional derivatives obey $\frac{\delta}{\delta f(p)}f(k)=\delta^{(D)}(p-k)$ for scalar and fermionic functions $f$. We can hence generate a field by acting with a functional derivative with respect to its source on $\complexi S_{\text{source}}=\complexi\int\de^D k\de^Dl\,\delta^{(D)}(k+l)\tr\mathcal{L}_{\text{source}}$. The exact field replacements are
\begin{equation}
\begin{aligned}\label{eq:field_as_source_der}
\phi_j^a(p)&\rightarrow \frac{1}{\complexi}\frac{\delta}{\delta \ol{J}^{j a}(-p)}\eqncom&
\lambda_{A\alpha}^a(p)&\rightarrow \frac{1}{\complexi}\frac{\delta}{\delta \eta^{A\alpha a}(-p)}\eqncom&
c^a(p)&\rightarrow \frac{1}{\complexi}\frac{\delta}{\delta \eta^{a}(-p)}\eqncom&
\\
\ol{\phi}^{ja}(p)&\rightarrow \frac{1}{\complexi}\frac{\delta}{\delta J_j^{a}(-p)}\eqncom&
\ol\lambda^{Aa}_{\dot\alpha}(p)&\rightarrow \complexi \frac{\delta}{\delta \ol\eta_A^{\dot\alpha a}(-p)}\eqncom&
\ol{c}^{a}(p)&\rightarrow\complexi \frac{\delta}{\delta \ol\eta^{a}(-p)}\eqncom&
\\
A^{\mu a}(p)&\rightarrow \frac{1}{\complexi}\frac{\delta}{\delta J_\mu^{a}(-p)}\eqncom&&&&&
\end{aligned}
\end{equation}
where the additional sign for anti-fermionic fields arises from commuting the fermionic source derivative past the fermionic field in the Lagrange density. 

A time-ordered correlation function (so far of the free theory) is obtained from the generating functional $Z_0$ by taking functional derivatives with respect to the corresponding sources and setting all sources to zero afterwards:\footnote{The time ordering symbol means that the Fourier transformation of the momentum space expression is a time-ordered position space expression, compare the definition in \appref{sec:Conventions}.}
\begin{equation}\label{eq:Npoint_function}
\vacl\T f_n(p_n)\dots f_1(p_1)\vac=
\left[
\frac{\pm\complexi\delta}{\delta J_{f_n}(-p_n)}\dots
\frac{\pm\complexi\delta}{\delta J_{f_1}(-p_1)}Z_0\right]_{\{J,\eta\}=0}\eqncom
\end{equation}
where we have $+\complexi$ for $\ol{\lambda}$ and $\ol{c}$ and $-\complexi$ for the remaining fields and $J,\eta=0$ indicates that all sources are set to zero. In particular, using \eqref{eq:generating_functional_app}, we see that the two-point correlation function of a field $f$ is directly related to its propagator $M_f$ as
\begin{equation}\label{eq:two_point_correlator_free}
\vacl\T f(p_{\text{out}}) \ol{f}(p_{\text{in}})\vac
=(2\pi)^D\delta^{(D)}(p_{\text{in}}+p_{\text{out}})\,\frac{1}{\complexi}M_f(p_{\text{out}})\eqncom
\end{equation}
where we used the momentum conserving $\delta$-distribution to swap $-p_{\text{in}}$ for $p_{\text{out}}$ in the propagator. The momentum conserving factor of $(2\pi)^D\delta^{(D)}(p_{\text{in}}+p_{\text{out}})$ is no coincidence, but does occur in any connected $n$-point correlation function, as we have restricted ourselves to a theory with local interactions \eqref{eq:action_Feynman_rules}. We can therefore separate this factor off and focus on the calculation of the reduced correlation function $\complexi \mathcal{T}$ which in \eqref{eq:two_point_correlator_free} is given by $-\complexi M_f(p_{\text{out}})$. As indicated in \eqref{eq:two_point_correlator_free}, the calculation of $\complexi \mathcal{T}$ in the free theory simply yields the free propagator of the involved field times a factor of $-\complexi$. In Feynman diagrams, we hence associate the following Feynman propagators to the different line types:
\begin{equation}\label{eq:propagators_scalar}
\begin{aligned}
\propagatorR{photon}{\mu a}{\nu b}
&{}={}
\vacl \T A^{\mu a}(p)A^{\nu b}(-p)\vac_{\complexi\mathcal{T}}
&=&\frac{1}{\complexi}\frac{1}{p^2-\complexi \epsilon}\Big(\eta^{\mu\nu}-(1-\xi)\frac{p^\mu p^\nu}{p^2}\Big)\delta^{ab}
\eqncom\\
\propagatorR{plain_rar}{ia}{jb}&{}={}
\vacl\T \phi_i^{a}(p)\ol{\phi}^{jb}(-p)\vac_{\complexi\mathcal{T}}
&=&\frac{1}{\complexi}\frac{1}{p^2-\complexi \epsilon}\delta_i^j\delta^{ab}
\eqncom\\
\propagatorR{dots_rar}{a}{b}
&{}={}
\vacl\T c^a(p)\bar c^b(-p)\vac_{\complexi\mathcal{T}}
&=&\frac{1}{\complexi}\frac{1}{p^2-\complexi \epsilon}\delta^{ab}\eqncom
\end{aligned}
\end{equation}
where the subscript $\complexi\mathcal{T}$ at the correlators means that a factor of $(2\pi)^D\delta^{(D)}(p_{\text{in}}+p_{\text{out}})$ has been dropped and we have adopted the general index conventions described below \eqref{eq:Lagrangian_Feynman_rules_4}. The arrows on line types always flow from anti field to field and the momentum arrow below flows from $-p$ to $p$. Note that time flows from right to left in our conventions, to match the expressions within the correlation functions. For the propagators of Weyl fermions there is a slight inconvenience, as there are two types of propagators, i.e.\ $S_{\alpha\dot\alpha}$ and $S^{\dot{\alpha}\alpha}$. To get the correct expressions in correlation functions we rewrite the fermionic part in the generating functional
\begin{equation}
\begin{aligned}\label{eq:fermion_part_in_Z0}
\complexi S_{0\ol{\lambda}\lambda}&=\complexi\int\de^D k\de^Dp(2\pi)^D\delta^{(D)}(k+p)
\eta^{A\alpha a}(k)S^{Bab}_{A\alpha\dot{\beta}}(p)\ol{\eta}_B^{\dot{\beta}b}(p)\\
&=\complexi\int\de^D k\de^Dp(2\pi)^D\delta^{(D)}(k+p)
\ol{\eta}_{\dot{\beta}}^{Bb}(p)S^{A \dot{\beta}\alpha ba}_{B}(k)\eta_{A\alpha}^{ a}(k)\eqncom
\end{aligned}
\end{equation}
where we used the $\delta$-distribution in the last line to rewrite the propagator $-S(p)=S(-p)=S(k)$. Depending on the index position of fermions in the two-point function, we can hence use the first or the second line of \eqref{eq:fermion_part_in_Z0} in the generating functional. Using \eqref{eq:field_as_source_der} and \eqref{eq:Npoint_function}, the fermionic propagators are then given by
\begin{equation}
\begin{aligned}\label{eq:propagators_fermion}
\propagatorR{dashes_rar}{A\alpha a}{B\dot{\beta} b}&{}={}
\vacl\T\lambda^{a}_{A\alpha}(p)\ol\lambda_{\dot\beta}^{Bb}(-p)\vac_{\complexi\mathcal{T}}
&=&\frac{1}{\complexi}S^{Bab}_{A\alpha\dot{\beta}}(p)
&=&
\frac{1}{\complexi}\frac{-(\sigma^\mu)_{\alpha\dot\beta}p_\mu}{p^2-\complexi \epsilon}\delta_A^B\delta^{ab}
\eqncom\\
\propagator{dashes_ar}{B\dot{\beta} b}{A\alpha a}&{}={}
\vacl\T\ol\lambda_B^{\dot\beta b}(-p)\lambda^{A\alpha a}(p)\vac_{\complexi\mathcal{T}}
&=&\frac{1}{\complexi}S_B^{A\dot\beta\alpha ba}(-p)
&=&\frac{1}{\complexi}\frac{(\bar\sigma^\mu)^{\dot\beta\alpha}p_\mu}{p^2-\complexi \epsilon}\delta^A_B\delta^{ba}
\eqncom
\end{aligned}
\end{equation}
where it is crucial that the index position in the correlation function matches the index position in the resulting $\sigma$- or $\bar{\sigma}$-matrices to get the correct sign for the propagator.

\subsection{Interactions and the full theory}
To also account for interactions, we follow the general path integral procedure, compare e.g.\ \cite[\chap{9,45,71,72}]{Srednicki:2007} for details. We write the generating functional of the interacting theory as an interaction part in which all fields are replaced by functional derivatives with respect to the sources $\{J,\eta\}$ that acts on the free generating functional:
\begin{equation}
Z\bigl[\{J,\eta\}]=Z_{\text{int}}[\{\tfrac{\delta}{\delta J},\tfrac{\delta}{\delta J}\}] Z_0[\{J,\eta\}]\eqncom
\end{equation}
where the interaction part contains the Fourier transformed action of all interactions including the two-point counterterm interactions
\begin{equation}
Z_{\text{int}}[\{J,\eta\}]=\e^{\complexi\, \text{FT}\bigl(\int\de^D x\tr(
	\mathcal{L}_{\text{ct}}[\{J,\eta\}]+\mathcal{L}_g[\{J,\eta\}]+\mathcal{L}_m[\{J,\eta\}])
	+\mathcal{L}_{\text{dt}}[\{J,\eta\}]\bigr)
	}\eqndot
\end{equation}
The Lagrange densities are given in \eqref{eq:Lagrangian_Feynman_rules_2} -- \eqref{eq:Lagrangian_Feynman_rules_4} and all fields have been replaced by functional derivatives with respect to the corresponding sources as in \eqref{eq:field_as_source_der}. To get a feeling for contributions to $Z_{\text{int}}$, let us focus on a tree-level interaction involving two fermions and one scalar given by the first term in \eqref{eq:Lagrangian_Feynman_rules_3}. When we set the renormalisation constants to one for the moment and depict sources with filled dots, this interaction is represented by
\begin{equation}
\begin{aligned}
\settoheight{\eqoff}{$\times$}%
\setlength{\eqoff}{0.5\eqoff}%
\addtolength{\eqoff}{-9\unitlength}%
\raisebox{\eqoff}{%
	\fmfframe(5,1)(0,1){%
		\begin{fmfchar*}(14,14)
		\fmfforce{0 w,1 h}{v1}
		\fmfforce{1 w,0.5 h}{v2}
		\fmfforce{0 w,0 h}{v3}
		\fmfv{decor.shape=circle,decor.filled=full,decor.size=0.15w}{v1,v2,v3}
		\fmf{dashes_ar}{v1,vc}
		\fmf{plain_rar}{vc,v2}
		\fmf{dashes_ar}{v3,vc}
		\end{fmfchar*}
	}
}&=\complexi\,\text{FT}\left.\left(\gym(\rho^{j})^{AB}\delta^\alpha_\beta (bca)\int\de^D x 
\frac{\delta}{\complexi\delta \eta_\beta^{B b}(x)}
\frac{\delta}{\complexi\delta \ol{J}^{jc}(x)}
\frac{\delta}{\complexi\delta \eta^{A\alpha a}(x)}
\right)Z_0\right|_{\{J,\eta\}=0}
\\
&=
-\gym(\rho^{j})^{AB}\delta^\alpha_\beta (bca)\int\de^Dp^\prime\de^Dq^\prime\de^Dr^\prime\Bigl[
(2\pi)^D\delta^{(D)}(p^\prime+q^\prime+r^\prime)
\\
&\phan{=-\gym(\rho^{j})^{AB}{}}
S_{\alpha\dot{\alpha}}(p^\prime)\Delta(r^\prime)S^{\dot{\beta}\beta}(q^\prime)
\bigl(-\complexi \ol{\eta}^{\dot{\alpha}a}_{A}(p^\prime)\bigr)
\bigl(\complexi J^c_j(r^\prime)\bigr)
\bigl(-\complexi \ol{\eta}^{b}_{B\dot{\beta}}(q^\prime)\bigr)
\Bigr]\eqncom
\end{aligned}
\end{equation}
where we used the colour trace abbreviation $(a_1\dots a_n)=\tr(\T^{a_1}\dots \T^{a_n})$ and the propagators in the last line are given by the spacetime part of \eqref{eq:free_propagators} alone. Note that the source terms in the last line appear in the reversed order compared to the order of the functional derivatives in the first line. This reversed order guarantees that no additional signs appear when we calculate the corresponding three-point scattering matrix element using \eqref{eq:Npoint_function}:
\begin{equation}
\begin{aligned}\label{eq:ffs_test}
\vacl\T\ol{\lambda}_{\dot{\beta}}^{Bb}(q)\ol{\phi}^{jc}(r)\ol{\lambda}^{A\dot{\alpha}a}(p)\vac_{\complexi\mathcal{T}}&=
\Delta^{jcc^\prime}_{j^\prime}(-r)
S^{B\dot{\beta}\beta^\prime bb^\prime}_{B^\prime}(-q)
S_{A^\prime\alpha^\prime\dot{\alpha}}^{A aa^\prime}(-p)
\\&\phan{{}={}}\left(
-\gym\delta^{\alpha^\prime}_{\beta^\prime}\bigl[
(\rho^{j^\prime})^{A^\prime B^\prime}(c^\prime a^\prime b^\prime )+(\rho^{j^\prime})^{B^\prime A^\prime }(c^\prime b^\prime a^\prime )
\bigr]\right)\eqncom
\end{aligned}
\end{equation}
where we reinstated the complete propagators and used the cyclicity of the colour trace. We can use this result to find the vertex factor associated with this interaction in a Feynman diagram. If we had built this scattering matrix element directly from Feynman diagrams, each propagator in the first line would have come with an additional factor of $-\complexi$. Hence, the vertex that connects the three propagators is the second line of \eqref{eq:ffs_test} times a factor of $-\complexi$.

In principle, we can redo the above calculation for every interaction in order to get all vertex factors of our theory. The faster way, however, is to take $\complexi$ times the Fourier transform of the desired interaction given in \eqref{eq:action_Feynman_rules} with \eqref{eq:Lagrangian_Feynman_rules_2} -- \eqref{eq:Lagrangian_Feynman_rules_4}, drop the factor of $(2\pi)^D\delta^{(D)}(\sum_i p_i)$ and erase the fields by taking the corresponding functional derivatives with momentum arguments $(-p_i)$. For fermionic interactions, the order of the functional derivatives has to be reversed\footnote{Up to nomenclature, this procedure to derive interaction vertices in a non-abelian gauge theory reproduces the vertex factors that are given in \cite[\chap{72}]{Srednicki:2007}.} compared to the order in which the fields appear in \eqref{eq:Lagrangian_Feynman_rules_2} -- \eqref{eq:Lagrangian_Feynman_rules_3}, as we have seen in the above example. For all vertices we assume that the momenta leave the vertex and we label all $n$-valent vertices by the type of propagator that can be connected to each of the $n$ legs. First, we discuss the counterterms or two-point vertices. Since they have the same field structure as the kinetic term, it follows from the free path integral \eqref{eq:free_path_integral} that they are given by the negative inverse of the propagator which can be connected to it times the corresponding counterterm. Using \eqref{eq:propagators_scalar} and \eqref{eq:propagators_fermion}, we find the counterterm vertices
\begin{equation}\label{eq:counterterms}
\begin{gathered}
\begin{aligned}
\twovertex{photon}{-p\mu a}{photon}{p\nu b}&=\complexi \delta_A\Bigl(\eta_{\mu\nu} p^2 -(1-\xi^{-1})p^\mu p^\nu\Bigr)\delta^{ab}\eqncom\\
\twovertex{plain_rar}{-pia}{plain_rar}{pjb}&=\complexi \delta_\phi p^2\delta^j_i\delta^{ab}\eqncom\\
\twovertex{dots_rar}{-pa}{dots_rar}{pb}&=\complexi \delta_c p^2\delta^{ab}\eqncom\\
\twovertex{dashes_rar}{-pA\dot\alpha a}{dashes_rar}{pB\beta b}&=\complexi \delta_\lambda (\bar\sigma_\mu)^{\dot\alpha \beta} p_\mu\delta^B_A\delta^{ab}\eqncom\\
\twovertex{dashes_ar}{pA\alpha a}{dashes_ar}{-pB\dot\beta b}&=-\complexi \delta_\lambda(\sigma^\mu)_{\beta\dot\alpha} p^\mu\delta^A_B\delta^{ab}\eqncom
\end{aligned}
\end{gathered}
\end{equation}
where the entries at each leg label the indices that can be connected at this point and we did not show the 1PI projector on the graphs explicitly. Of course, we can also build counterterm propagators by simply connecting two propagators via the corresponding counterterm vertex. This procedure yields the ordinary propagators of \eqref{eq:propagators_scalar} and \eqref{eq:propagators_fermion} times a factor of the corresponding counterterm.\footnote{Alternatively, we could have derived the free propagators including a formal factor of $Z$. This would have resulted in the free propagators times a factor of $Z^{-1}=(1-\delta)^{-1}$. After expanding this factor in powers of the coupling constant, it yields at lowest order exactly the propagator times $\delta$ as we have found here.} In calculations, we often distinguish two-point vertices from propagators by explicitly stating that the former ones have amputated legs. Second, we turn to the three-point vertices. We label them starting in the lower left corner and going around the vertex in clockwise order. This procedure gives the following three-valent vertices\footnote{Our Minkowski space Feynman rules can be mapped to the Euclidean space Feynman rules given in \cite{Fokken:2013aea} by replacing our coupling constant $\gym$ with $\frac{\gym}{\sqrt{2}}$ and performing a Wick rotation to Euclidean space. This rotation effectively generates a factor $-\complexi$ and $\complexi$ in each of our vertices and propagators, respectively. Note, however, that we use a different convention to raise and lower spinor indices in this thesis.} %
\begin{subequations}\label{eq:cubic_vertices}
	\begin{equation}
	\begin{aligned}\label{eq:cubic_vertices1}
V_{AAA}
&=
\threevertexJ{photon}{$\scriptstyle q\nu b$}{photon}{$\scriptstyle r\gamma c$}{photon}{$\scriptstyle p\mu a$}
=-\complexi\gym \mu^{\epsilon}Z_{A^3}\bigl[
(q-r)_\mu\eta_{\nu\gamma}
+(r-p)_\nu\eta_{\gamma\mu}
+(p-q)_\gamma\eta_{\mu\nu}\bigr](c[a,b])
\eqncom\\
V_{\ol cA c}
&=
\threevertexJ{dots_rar}{$\scriptstyle q b$}{photon}{$\scriptstyle r \mu c$}{dots_ar}{$\scriptstyle p a$}
=
\complexi\gym \mu^{\epsilon}Z_{\ol cAc}q_\mu(c[a,b])
\eqncom\\
V_{\ol\phi A\phi}
&=
\threevertexJ{plain_rar}{$\scriptstyle q j b$}{photon}{$\scriptstyle r\mu c$}{plain_ar}{$\scriptstyle p i a$}
=
\threevertexJ{plain_ar}{$\scriptstyle q i b$}{photon}{$\scriptstyle r\mu c$}{plain_rar}{$\scriptstyle p j a$}
=
-\complexi\gym \mu^{\epsilon}Z_{\ol \phi A\phi}\delta^i_j(p_\mu-q_\mu)(c[a,b])
\eqncom\\
V_{\ol\lambda A\lambda}
&=
\threevertexJ{dashes_rar}{$\scriptstyle qB\dot\beta b$}{photon}{$\scriptstyle r\mu c$}{dashes_ar}{$\scriptstyle p A\alpha a$}
=
\begin{cases}
\complexi \gym\mu^{\epsilon} Z_{\ol\lambda A\lambda}(\bar{\sigma}_\mu)^{\dot{\beta}\alpha}\delta^A_B(c[a,b])\\
\complexi \gym \mu^{\epsilon}Z_{\ol\lambda A\lambda}(\sigma_\mu)_{\alpha\dot{\beta}}\delta^A_B(c[a,b])
\end{cases}
\eqncom\\
\end{aligned}
\end{equation}
\begin{equation}\label{eq:cubic_vertices2}
\begin{aligned}
V_{\lambda\phi\lambda}
&=
\threevertexJ{dashes_ar}{$\scriptstyle q B\beta b$}{plain_rar}{$\scriptstyle r j c$}{dashes_ar}{$\scriptstyle pA\alpha a$}
=
\complexi\gym \mu^{\epsilon}Z_{\lambda\phi\lambda}\delta^{\alpha}_{\beta}\bigl[
(\rho^{j})^{A B}(c a b )+(\rho^{j})^{B A }(c b a )
\bigr]
\eqncom\\
V_{\ol\lambda\,\ol\phi\,\ol\lambda}
&=
\threevertexJ{dashes_rar}{$\scriptstyle qB\dot\beta b$}{plain_ar}{$\scriptstyle r j c$}{dashes_rar}{$\scriptstyle p A\dot\alpha a$}
=
\complexi\gym \mu^{\epsilon}Z_{\lambda\phi\lambda}\delta^{\dot\beta}_{\dot\alpha}\bigl[
(\rho^\dagger_{j})_{A B}(c a b )+(\rho^\dagger_{j})_{B A }(c b a )
\bigr]
\eqncom\\
V_{\lambda\ol\phi\lambda}
&=
\threevertexJ{dashes_ar}{$\scriptstyle  qB\beta b$}{plain_ar}{$\scriptstyle r j c$}{dashes_ar}{$\scriptstyle p A\alpha a$}
=
\complexi\gym\mu^{\epsilon} Z_{\lambda\ol\phi\lambda}\delta_{\beta}^{\alpha}\bigl[
(\tilde\rho^\dagger_{j})^{A B}(c a b )+(\tilde\rho^\dagger_{j})^{B A }(c b a )
\bigr]
\eqncom\\
V_{\ol\lambda\phi\ol\lambda}
&=
\threevertexJ{dashes_rar}{$\scriptstyle qB\dot\beta b$}{plain_rar}{$\scriptstyle r j c$}{dashes_rar}{$\scriptstyle p A\dot\alpha a$}
=
\complexi\gym\mu^{\epsilon} Z_{\lambda\ol\phi\lambda}\delta^{\dot\beta}_{\dot\alpha}\bigl[
(\tilde\rho^{j})_{A B}(c a b )+(\tilde\rho^{j})_{B A }(c b a )
\bigr]
\eqndot
\end{aligned}
\end{equation}
\end{subequations}
Here, we combined the ordinary vertices with the counterterm vertices, since we included the 1PI renormalisation constants $Z_v=1+\delta_v$ for each vertex $v$. This is a slight abuse of notation, since we did not draw the
$
\settoheight{\eqoff}{$\times$}%
\setlength{\eqoff}{0.5\eqoff}%
\addtolength{\eqoff}{0.0\unitlength}%
\raisebox{\eqoff}{%
	\fmfframe(0,0)(-2,-1){%
		\begin{fmfchar*}(2,2)
		\fmfforce{0.5 w,0.5 h}{v1}
		\fmfv{decor.shape=hexacross,decor.size=7thin}{v1}
		\end{fmfchar*}
	}
}
$-decorated counterterm vertices in the diagrams. The uncorrected and counterterm vertices are, however, easily obtained from the above expressions by setting $Z_v=1$ and $Z_v=\delta_v$, respectively. Note that both couplings in the third line have the same sign since the sign from exchanging the colour indices between $\phi$ and $\ol{\phi}$ is compensated by the one from exchanging the corresponding momenta. Analogously, both versions in the fourth line have the same sign, since the sign from the colour index exchange is compensated by changing the $\sigma$-matrices as $\bar\sigma\leftrightarrow (-\sigma)$. Finally, the four-valent vertices follow analogously:
\begin{equation}\label{eq:quartic_vertices}
\begin{aligned}
V_{AAAA}
&=
\fourvertextwo{photon}{$\scriptstyle q\nu b$}{photon}{$\scriptstyle r\rho c$}{photon}{$\scriptstyle s\sigma d$}{photon}{$\scriptstyle p\mu a$}
&{}={}&
\complexi\gym^2\mu^{2\epsilon} Z_{A^4}\bigl[{}
+(\eta_{\mu\rho}\eta_{\nu\sigma}-\eta_{\mu\sigma}\eta_{\nu\rho})([a,b][c,d])
\\[-1.25\baselineskip]
&&&\phan{\complexi\gym^2 Z_{A^4}\bigl[{}}
+(\eta_{\mu\sigma}\eta_{\rho\nu}-\eta_{\mu\nu}\eta_{\rho\sigma})([a,c][d,b])
\\
&&&\phan{\complexi\gym^2 Z_{A^4}\bigl[{}}
+
(\eta_{\mu\nu}\eta_{\sigma\rho}-\eta_{\mu\rho}\eta_{\sigma\nu})([a,d][b,c])
\bigr]
\eqncom\\
V_{A\ol{\phi}A\phi}
&=
\fourvertextwo{plain_rar}{$\scriptstyle q i b$}{photon}{$\scriptstyle r\rho c$}{plain_rar}{$\scriptstyle s j d$}{photon}{$\scriptstyle p\mu a$}
&{}={}&
\complexi\gym^2\mu^{2\epsilon} Z_{\ol{\phi}A^2 \phi}\delta^j_i\eta_{\mu\rho}\bigl[([a,b][c,d])+([c,b][a,d])\bigr]
\eqncom\\
V^{\text{F}}_{\phi\phi\ol{\phi}\,\ol{\phi}}
&=
\fourvertextwo{plain_ar}{$\scriptstyle q j b$}{plain_ar}{$\scriptstyle rk c$}{plain_ar}{$\scriptstyle s l d$}{plain_ar}{$\scriptstyle p i a$}
&{}={}&
\complexi\gym^2\mu^{2\epsilon} \Bigl[
Z_{\text{F}}\bigl(F^{ij}_{lk}(abcd)
+F^{ji}_{lk}(bacd)
+F^{ij}_{kl}(abdc)
+F^{ji}_{lk}(badc)\bigr)
\\[-1.\baselineskip]
&&&\phan{\complexi\gym^2 \Bigl[}
-\frac{Z_{\text{Fdt}}}{N}\bigl(Q^{ij}_{\text{F}lk}+Q^{ij}_{\text{F}kl}\bigr)(ab)(cd)\Bigr]
\eqncom\\
V^{\text{D}}_{\phi\ol{\phi}\phi\ol\phi}
&=
\fourvertextwo{plain_rar}{$\scriptstyle q j b$}{plain_rar}{$\scriptstyle rk c$}{plain_ar}{$\scriptstyle s l d$}{plain_ar}{$\scriptstyle p i a$}
&{}={}&
-\complexi\gym^2\mu^{2\epsilon}\Bigl[
Z_{\text{D}}\bigl(\delta^i_j\delta^k_l([a,b][c,d])
+\delta^i_l\delta^k_j([a,d][c,b])\bigr)
\\[-1.\baselineskip]
&&&\phan{-\complexi\gym^2\Bigl[}
+\frac{Z_{\text{Ddt}}}{N}\bigl(Q^{ik}_{\text{D}jl}(ab)(cd)
+Q^{ki}_{\text{D}jl}(ad)(bc)\bigr)
\Bigr]
\eqndot
\end{aligned}
\end{equation}

\subsection{Feynman rules}\label{sec:the-actual-feynman-rules}
We will not include the explicit wave function structure of external fields in our Feynman rules, as their inclusion is more conveniently discussed from the perspective of canonical quantisation. However, our Feynman rules are compatible with the ones in \cite{Dreiner:2008tw} and hence external state wave functions can be included into our rules by simply adopting their rules for external states. In particular, using the representation of massless Weyl fermion $\xi_\alpha$ and $\xi^\dagger_{\dot{\alpha}}$ in \cite[\chap{3.1}]{Dreiner:2008tw}, allows to adopt the rules\footnote{Note that their time direction is reversed compared to ours.} for initial and final states in \cite[\chap{4.1, 4.4}]{Dreiner:2008tw}. We first give the Feynman rules for a single diagram with external propagators and time flow from right to left. The amputated version of a diagram is obtained by simply dropping the external propagators in the final expression. The drawing rules are:
\begin{itemize}
	\item[1.] Treat each incoming and outgoing field as an external vertex which is labelled by the field {\it momentum} $p_i$ and its non-trivial quantum numbers, i.e.\ its {\it flavour}, {\it colour}, {\it spacetime}, and $\spl{2}$- and $\splbar{2}$-{\it spinor} indices.
	\item[2.] Incoming $\spl{2}$ and $\splbar{2}$ indices must be placed at the lower and upper position, respectively.
	\item[3.] Ensure the overall momentum conservation by requiring that the sum of all external momenta vanishes $\sum_i p_i=0$.
	\item[4.] Connect all external fields with suitable propagators \eqref{eq:propagators_scalar} and \eqref{eq:propagators_fermion} and vertices \eqref{eq:cubic_vertices}, and \eqref{eq:quartic_vertices} to the desired diagram. This fixes the spinor-index position of outgoing fermions.
	\item[5.] Associate to each closed loop in the diagram a momentum $\ell_i$.
	\item[6.] By ensuring momentum conservation at each internal vertex, find the momentum of every internal propagator. Associate this momentum to the momentum arrow of the propagator (seen in the direction of the arrow).
	\item[7.] Label each internal vertex with the required indices. (This may be simplified by directly incorporating the Kronecker-$\delta$'s from the propagators.)
\end{itemize}
The rules to translate Feynman diagrams into Feynman integrals are:
\begin{itemize}
	\item[8.] For each incoming fermion field (to the right), follow its path through the diagram to an outgoing field on the left. Write down the factors of propagators and vertices in the order in which they appear on the path. The $\sigma$ and $\bar{\sigma}$ matrices must alternate along the path\footnote{A subproduct of the form $\dots(\sigma_\mu)_{\alpha\dot{\alpha}}(\bar\sigma_\nu)^{\dot{\alpha}\beta}\dots$ is admissible, while $\dots(\sigma_\mu)_{\alpha\dot{\alpha}}(\bar\sigma_\nu)^{\dot{\beta}\alpha}\dots$ is not.} and contracted indices must appear next to each other. For the vertex $V_{\ol\lambda A\lambda}$, this requirement determines whether the first line with $\sigma$ or second line with $\bar{\sigma}$ has to be chosen.
	\item[9.] For each closed fermion loop, pick one vertex as a starting point and follow the path of the loop in clockwise order. Beginning with the first propagator, write down the factors of fermion propagators and vertices in the order in which they appear along the path. Each fermion propagator that is transversed {\it in its arrow direction} gives the $\sigma$ version of the propagator and each fermion propagator that is transversed {\it against its arrow direction} gives the $\bar\sigma$ version, \cf \eqref{eq:propagators_fermion}. The value of the $V_{\ol\lambda A\lambda}$ vertices is determined as in rule 8.
	\item[10.] Write down the remaining factors of propagators and vertices that arise from scalars and gauge fields. For cubic vertices $V_{AAA}$ and $V_{\ol{\phi}A\phi}$, which are given with momenta leaving the vertex, it may be necessary to adjust the momentum factors to match the momentum directions in the diagram.
	\item[11.] Divide the resulting expression by the symmetry factor of the diagram. It accounts for exchange symmetry of internal vertices and propagators that leave the overall diagram unchanged and is hence closely related to the existence of identical particles, see e.g.\ \cite[\chap{10}]{Srednicki:2007} for further details.
	\item[12.] Multiply the resulting expression with a factor of $(-1)$ for each closed fermion loop\footnote{A fermion loop is closed when dashed lines form a closed path, regardless of the orientation of the contributing line segments.}.
	\item[13.] Integrate\footnote{To evaluate the integrals in praxis, we perform a Wick rotation to Euclidean space for each integral. This generates an additional factor of $\complexi$ for each loop, \cf \appref{subsec:Wick_rotation}.} over all loop momenta $\ell_i$ with a measure $\frac{\de^D \ell_i}{(2\pi)^D}$ over $D$-dimensional Minkowski space. Techniques for the evaluation of such integrals are discussed in \appref{sec:Renormalisation_schemes}.	
\end{itemize}
For physical processes usually several diagrams contribute. The last two rules determine which diagrams contribute and if there are relative signs between them
\begin{itemize}
	\item[14.] For a scattering matrix element $\complexi\mathcal{T}$ specified by a set of incoming and outgoing fields, all topologically distinct diagrams that can be generated by the rules 1 -- 7 contribute to $\complexi\mathcal{T}$. Up to a given maximal order $\mathcal{O}(\gym^{n_{\text{max}}})$, of all these diagrams only those with $\ell\leq \lceil \tfrac{n_{\text{max}}}{2}\rceil$ loops contribute to $\complexi\mathcal{T}$.
	\item[15.] If there are fermions in the external states, the overall sign of each diagram is determined as follows: take the ordering of external fermionic fields (in a right to left formula as derived by the rules 1 -- 13) in one of the contributing diagrams to be the canonical ordering. If the ordering of external fermionic fields in another diagram differs by an odd permutation, then there is a relative sign between both diagrams.
\end{itemize}
To get a little feeling how these rules work and for abbreviating notations see \secref{sec:calc_Greens_functions}.

\subsection{Feynman rules for real scalars}\label{subsec:Feynman_real_scalars}
The Feynman rules, as derived in the previous appendix are suitable for \NfSYMt and its deformations. However, if we are not interested in deformations that depend on the \su{4} Cartan charges of each field, it may be more convenient to work with the action \eqref{eq:N4_action_real_scalars} which contains real scalars instead of complex ones. For the Feynman rules the different action implies that the scalar propagator and scalar interactions have to be altered.

For the real scalar propagator, the only change is that the scalar lines are not directed any more in Feynman diagrams and we have
\begin{equation}
\propagatorR{plain}{ia}{jb}{}={}
\vacl\T\bigl[ \varphi_i^{a}(p)\varphi_j^{b}(-p)\bigr]\vac_{\complexi\mathcal{T}}
=\frac{1}{\complexi}\frac{1}{p^2-\complexi \epsilon}\delta_{ij}\delta^{ab}
\eqncom
\end{equation}
where the \so{6} flavour indices of the real scalars are $i,j\in\{1,\dots,6\}$. For the three- and four-valent vertices we follow the prescription beneath \eqref{eq:ffs_test} to derive the vertices from the interactions given in \eqref{eq:N4_action_real_scalars}. We find
\begin{subequations}\label{eq:cubic_vertices_real_scalar}
\begin{equation}\label{eq:cubic_vertices_real_scalar1}
\begin{aligned}
V_{\lambda\varphi\lambda}
&=
\threevertexJ{dashes_ar}{$\scriptstyle q B\beta b$}{plain}{$\scriptstyle r j c$}{dashes_ar}{$\scriptstyle pA\alpha a$}
=
-\gym \mu^{\epsilon}Z_{\lambda\varphi\lambda}(\bar\Sigma^{j})^{A B}\delta^{\alpha}_{\beta}(c [a, b])\eqncom\\
V_{\ol\lambda\varphi\ol\lambda}
&=
\threevertexJ{dashes_rar}{$\scriptstyle qB\dot\beta b$}{plain}{$\scriptstyle r j c$}{dashes_rar}{$\scriptstyle p A\dot\alpha a$}
=
\gym \mu^{\epsilon} Z_{\lambda\varphi\lambda}(\Sigma^{j})_{A B}\delta^{\dot\beta}_{\dot\alpha}(c [a, b])\eqncom\\
\end{aligned}
\end{equation}
\begin{equation}\label{eq:cubic_vertices_real_scalar2}
\begin{aligned}
V_{\varphi A\varphi}
&=
\threevertexJ{plain}{$\scriptstyle q j b$}{photon}{$\scriptstyle r\mu c$}{plain}{$\scriptstyle p i a$}
=
-\complexi\gym\mu^{\epsilon} Z_{\varphi A\varphi}\delta^i_j(p_\mu-q_\mu)(c[a,b])
\eqncom\\
V_{A\varphi A\varphi}
&=
\fourvertextwo{plain}{$\scriptstyle q i b$}{photon}{$\scriptstyle r\rho c$}{plain}{$\scriptstyle s j d$}{photon}{$\scriptstyle p\mu a$}
=
\complexi\gym^2\mu^{2\epsilon} Z_{\varphi^2 A^2}\delta_{ij}\eta_{\mu\rho}\bigl[([a,b][c,d])+([c,b][a,d])\bigr]
\eqncom\\
\end{aligned}
\end{equation}
\end{subequations}
For the four scalar interaction we could follow the same derivation. However, as we derived \NfSYMt from the ten-dimensional Yang-Mills theory in \secref{sec:N1SYM_10D_to4D}, we know that this interaction has the same form as the four gluon interaction, compare \eqref{eq:N4_action_real_scalars} and \eqref{eq:Lagrangian_Feynman_rules_gluon}. Therefore, the four scalar interaction is obtained from $V_{AAAA}$ in \eqref{eq:quartic_vertices} by replacing the metric $\eta$ by the $\so{6}$ metric $\delta$ everywhere.
\begin{equation}
\begin{aligned}
V_{\varphi\varphi\varphi\varphi}
&=
\fourvertextwo{plain}{$\scriptstyle qjb$}{plain}{$\scriptstyle rkc$}{plain}{$\scriptstyle sld$}{plain}{$\scriptstyle pia$}
&{}={}&
\complexi\gym^2\mu^{2\epsilon} Z_{\varphi^4}\bigl[{}
+(\delta_{ik}\delta_{jl}-\delta_{il}\delta_{jk})([a,b][c,d])
\\[-1.25\baselineskip]
&&&\phan{\complexi\gym^2 Z_{A^4}\bigl[{}}
+(\delta_{il}\delta_{k j}-\delta_{ij}\delta_{kl})([a,c][d,b])
\\
&&&\phan{\complexi\gym^2 Z_{A^4}\bigl[{}}
+
(\delta_{ij}\delta_{l k}-\delta_{ik}\delta_{l j})([a,d][b,c])
\bigr]\eqndot
\end{aligned}
\end{equation}

\section{Equations of motion and the Bianchi identity}\label{app:EOM_Bianchi}
To define the alphabet from which local gauge-invariant composite are built, it is necessary to lift redundancies which appear when the Bianchi identity or the \eom relate different composite operators to one another. In this appendix, we present the Bianchi identity as well as the classical \eom in our conventions compatible with our representations of classical and quantised theories.

First, the Bianchi identity in our conventions can be written as
\begin{equation}
\begin{aligned}
0&=\D^{(\mu}F^{\nu\rho)}=\D^{(\mu}_{\phan{+}}F^{\nu\rho)}_++\D^{(\mu}_{\phan{+}}F^{\nu\rho)}_-
=-\frac{\complexi}{2}\varepsilon^{\mu\nu\rho\omega}(\sigma_{\omega})_{\alpha\dot{\alpha}}
\bigl(\D^{\dot{\alpha}\beta}\cF_{\beta}^{\phan{\alpha}\alpha}+\D^{\dot{\beta}\alpha}\bar{\cF}^{\dot{\alpha}}_{\phan{\alpha}\dot{\beta}}\bigr)\eqncom
\end{aligned}
\end{equation}
where we used the (anti-)self-dual projectors \eqref{eq:selfdual_projectors}, the mapping to spinor indices \eqref{eq:vectorfield_spinorial} and \eqref{eq:selfdual_fieldstrength} and the identity \eqref{eq:sigma_munu_projector}. Since the tensors are non-vanishing, this implies that the sum of self-dual and anti-self-dual \eom in parentheses on the \rhs must vanish. 

Second, we have the \eom of all elementary fields, which relate different composite operators. The \eom of a field $X^{A\alpha a}$ with flavour spinor and colour index $A$, $\alpha$ and $a$, respectively follow from the Euler-Lagrange equation and reads in general
\begin{equation}
0=\frac{\delta S}{\delta X^{A\alpha a}} 
=\frac{\delta \mathcal{L}(x)}{\delta X^{A\alpha a}(x)}-\D_{\beta\dot{\beta}} \frac{\delta \mathcal{L}(x)}{\delta (\D_{\beta\dot{\beta}} X^{A\alpha a})}\eqndot
\end{equation}
Concretely, for the action \eqref{eq:deformed_action_complex_scalars2} or equivalently \eqref{eq:action_Feynman_rules} the \eom become for the single-trace part\footnote{For the last equation we used the fundamental definitions of \subsecref{sec:N4SYM_4D} and varied explicitly the action with respect to $\delta A_{\alpha\dot{\alpha}}$. For the final result we used several identities of \appref{sec:Conventions} and the Bianchi identity which connects the \eom of the self-dual and anti-self-dual components of the field strength.}
\begin{equation}
\begin{aligned}\label{eq:eoms}
\frac 12\D_{\alpha\dot{\alpha}}\D^{\dot{\alpha}\alpha}\ol{\phi}^{ja}&{}={}-\gym\Bigl((\rho^j)^{AB}\tr\bigl(\lambda_{B}^\alpha \T^a \lambda_{A\alpha}\bigr)+(\tilde{\rho}^j)_{AB}\tr\bigl(\ol{\lambda}^B_{\dot{\alpha}}\T^a\ol{\lambda}^{A\dot{\alpha}}\bigr)\Bigr)
\\
&\phan{{}={}}
-\gym^2\Bigl(F^{ij}_{lk}\tr\bigl(\phi_i\T^a\ol{\phi}^k\ol{\phi}^l\bigr)+F^{ji}_{lk}\tr\bigl(\T^a\phi_i\ol{\phi}^k\ol{\phi}^l\bigr)
-\tr\bigl(\bigl[\T^a,\ol{\phi}^j\bigr]\bigl[\phi_k,\ol{\phi}^k\bigr]\bigr)\Bigr)\eqncom
\\
\D^{\dot{\alpha}\alpha}\lambda_{A\alpha}^a
&{}={}
\gym\Bigl(
(\rho^\dagger_j)_{BA}\tr\bigl(\T^a\ol{\phi}^j\ol{\lambda}^{B\dot{\alpha}}\bigr)
+(\rho^\dagger_j)_{AB}\tr\bigl(\T^a\ol{\lambda}^{B\dot{\alpha}}\ol{\phi}^j\bigr)
\Bigr)\\
&\phan{{}={}\gym\Bigl(}
+(\tilde\rho^j)_{BA}\tr\bigl(\T^a\phi_j\ol{\lambda}^{B\dot{\alpha}}\bigr)
+(\tilde\rho^j)_{AB}\tr\bigl(\T^a\ol{\lambda}^{B\dot{\alpha}}\phi_j\bigr)
\Bigr)\eqncom\\
\D^{\dot{\alpha}\beta}\cF_\beta^{\phan{\alpha}\alpha}&=
\gym \tr\bigl(\T^a\bigl[\ol{\phi}^j,\D^{\dot{\alpha}\alpha}\phi_j]\bigr)
+\gym\tr\bigl(\T^a\bigl\{\lambda^\alpha_{A},\ol{\lambda}^{A\dot{\alpha}}\bigr\}\bigr)\eqncom
\end{aligned}
\end{equation}
and the remaining equations are obtained by hermitian conjugation (see \secref{sec:N4SYM_4D} for the respective transformation rules). In composite operators we can hence replace all occurrences of the terms on the \lhs of \eqref{eq:eoms} by the respective \rhs Note that these are only the classical \eom and the higher-order corrections\footnote{These corrections have to be taken into account from two-loop order on in perturbative calculations that involve composite operators.} which can be obtained from the effective action are missing. Moreover, note that the terms on the \lhs of \eqref{eq:eoms} all involve spinor indices which are contracted by the antisymmetric symbols $\varepsilon_{\alpha\beta}$ or $\varepsilon_{\dot{\alpha}\dot{\beta}}$. Therefore, we can prevent the occurrence of such terms in composite operators by symmetrising all $\spl{2}$ and $\splbar{2}$ spinor indices of covariant derivatives and the fields they act on. This prescription yields the alphabet \eqref{eq: alphabet} which we use to construct composite operators.

\section{The \texorpdfstring{{\tt FokkenFeynPackage}}{FokkenFeynPackage}}\label{sec:Feynman_rules_Mathematica}
In this appendix, we give a manual for the tool {\tt FokkenFeynPackage}, which implements the Feynman rules of \appref{app:Feynman_rules} for {\tt Mathematica} 8.0. The package will be uploaded together with the \ttt{arXiv} source file of this thesis and it allows to construct the integrands of arbitrary\footnote{Note that the lack of optimisation, probably restricts the applicability of the package to relatively low loop orders and numbers of external legs. The limits of applicability are, however, not conceptual but arise in the not optimised contraction of occurring interaction tensors.} Feynman integrals from a given Feynman diagram. For one- and two-loop propagator-type scalar integrals, it can also be used to construct the explicit result of the Feynman integrals in terms of $\Gamma$ functions and possible occurring IR divergence factors. We first describe general aspects of the package, including the explicit form of propagators and vertices, then we discuss how it can be used to solve Feynman diagrams in two examples and finally we include a list of additional variables and functions that are defined in this package. For further examples, see also \ttt{examples.nb} in the package source file, where most of the calculations that were done in \chapref{chap:applications} are presented.

The package {\tt FokkenFeynPackage} should be called with the command 
\begin{equation}
\begin{aligned}
&\ttt{SetDirectory["}\text{path}\ttt{"]}\\
&\ttt{<<FokkenFeynPackage`}
\end{aligned}
\end{equation}
where "path" stands for the directory in which the package is saved. This initialises all variables and functions. The package itself contains the following seven files:
\begin{itemize}
	\item {\tt FokkenFeynPackage.m} is the the packages main file, which initialises the remaining six files, when it is loaded in {\tt Mathematica}.
	\item {\tt Feynpar.m}, which is the self-contained Mathematica package\footnote{Note that the package {\tt Feynpar} contains some further function definitions and we refer the reader to the manual of {\tt Feynpar} for details.} of \cite{West1993286} which provides spacetime tensors with the usual contraction properties of indices.
	\item {\tt VariablesReplacements.m}, in which general variable names and replacement rules are stored.
	\item {\tt Tensors.m}, which contains all tensors that occur in the construction of Feynman integrals.
	\item {\tt Functions.m}, which contains all functions necessary to transform a given Feynman diagram into the appropriate Feynman integrals.
	\item {\tt Feynmanrules.m}, which contains the {\tt Mathematica} implementation of the propagators and vertices given in \appref{app:Feynman_rules}.
	\item {\tt solveUpto2loop.m}, which contains all functions to solve one- and two-loop Feynman integrals that depend on a single external scale.
\end{itemize}

In \tabref{tab:variables}, a list of all variables and parameters defined in the program is given and in \tabref{tab:function}, a list of all general rules and functions with a short description is given. In addition to these general definitions, the program contains the {\tt Mathematica} implementation for all vertices and propagators of the theory presented in \appref{app:Feynman_rules}.
All propagators are labelled with an initial \ttt{P}, followed by the field type that characterises it. The propagators of \eqref{eq:propagators_scalar} and \eqref{eq:propagators_fermion} can be called in {\tt Mathematica} via the functions
\begin{equation}\label{program:propagators}
\begin{aligned}
\propagatorR{photon}{n_1}{n_2}
&=\text{{\tt PA[n1,n2][p]}}\eqncom
&\propagatorR{dots_rar}{n_1}{n_2}
&=\text{{\tt Pc[n1,n2][p]}}\eqncom\\
\propagatorR{dashes_rar}{n_1}{n_2}
&=\text{{\tt P}}\lambda\lambda\text{{\tt b[n1,n2][p]}}\eqncom
&\propagator{dashes_ar}{n_1}{n_2}
&=\text{{\tt P}}\lambda\text{{\tt b}}\lambda\text{{\tt[n1,n2][p]}}
\eqncom\\
\propagatorR{plain_rar}{n_1}{n_2}
&=\text{{\tt P}}\phi\text{{\tt[n1,n2][p]}}
\eqncom
&&
\end{aligned}
\end{equation}
where the labels \ttt{n1} and \ttt{n2} characterise the connection points at which the propagators may be attached. The first and second Weyl fermion propagators in the second line correspond to the first and second line of \eqref{eq:propagators_fermion}, respectively. The positions of fermion and anti-fermion in the name correspond to the first and second label in the fermion propagator functions, e.g.\ the Weyl fermion propagator to the left in \eqref{program:propagators} has ingoing and outgoing fermions $\lambda$ and $\ol{\lambda}$ at labels $n_1$ and $n_2$, respectively. The counterterms of \eqref{eq:counterterms} are named with an initial \ttt{PCT} and otherwise exactly like the propagators and we have 
\begin{equation}\label{program:counterterms}
\begin{gathered}
\begin{aligned}
\twovertex{photon}{n_1}{photon}{n_2}&=\text{{\tt PCTA[n1,n2][p]}}\eqncom
&\twovertex{dots_rar}{n_1}{dots_rar}{n_2}&=\text{{\tt PCTc[n1,n2][p]}}\eqncom\\
\twovertex{dashes_rar}{n_1}{dashes_rar}{n_2}&=\text{{\tt PCT}}\lambda\lambda\text{{\tt b[n1,n2][p]}}\eqncom
&\twovertex{dashes_ar}{n_1}{dashes_ar}{n_2}&=\text{{\tt PCT}}\lambda\text{{\tt b}}\lambda\text{{\tt[n1,n2][p]}}\eqncom\\
\twovertex{plain_rar}{n_1}{plain_rar}{n_2}&=\text{{\tt PCT}}\phi\text{{\tt[n1,n2][p]}}\eqndot&&
\end{aligned}
\end{gathered}
\end{equation}
Note, however, that these objects are two-point vertices and hence the labels \ttt{n1} and \ttt{n2} are connection points to which propagators may be connected. Obviously, in such connections the flavour-charge flow must be conserved, meaning that only propagators with the correct flavour-arrow direction can be connected to a given point \ttt{n}. Three-point vertices are named with an initial \ttt{V} followed by the fields that enter the respective vertex. The order of appearing fields corresponds to the ordering of the indices that characterise the vertices' connection points. If the vertices depend on the momenta of connecting fields, the same rules apply as in \eqref{eq:cubic_vertices}. Namely, the momentum \ttt{pi} in a vertex function must be inserted such that \ttt{pi} leaves the vertex. We have the three-valent vertices
	\begin{equation}\label{program:cubic_vertices}
	\begin{aligned}
	\threevertexJ{photon}{$\scriptstyle p_3n_3$}{photon}{$\scriptstyle p_1n_1$}{photon}{$\scriptstyle p_2n_2$}
	&=\text{{\tt VAAA[n1,n2,n3][p1,p2,p3]}}\eqncom
	&\threevertexJ{dots_rar}{$\scriptstyle p_3n_3$}{photon}{$\scriptstyle n_1$}{dots_ar}{$\scriptstyle n_2$}
	&=\text{{\tt VAccb[n1,n2,n3][p3]}}\eqncom\\
	\threevertexJ{dashes_rar}{$\scriptstyle n_3$}{photon}{$\scriptstyle n_1$}{dashes_ar}{$\scriptstyle n_2$}
	&=
	\begin{cases}
	\text{{\tt VA}}\lambda\lambda\text{{\tt b}}\sigma{\text{\tt[n1,n2,n3]}}\\
	\text{{\tt VA}}\lambda\lambda\text{{\tt b}}\sigma\text{{\tt b[n1,n2,n3]}}
	\end{cases}
	\hspace{-.4cm}
	\eqncom
	&\threevertexJ{dashes_ar}{$\scriptstyle n_3$}{photon}{$\scriptstyle n_1$}{dashes_rar}{$\scriptstyle n_2$}
	&=
	\begin{cases}
	\text{{\tt VA}}\lambda\text{{\tt b}}\lambda\sigma{\text{\tt[n1,n2,n3]}}\\
	\text{{\tt VA}}\lambda\text{{\tt b}}\lambda\sigma\text{{\tt b[n1,n2,n3]}}
	\end{cases}
	\hspace{-.4cm}
	\eqncom\\
	\threevertexJ{dashes_ar}{$\scriptstyle n_3$}{plain_rar}{$\scriptstyle n_1$}{dashes_ar}{$\scriptstyle n_2$}
	&=
	\text{{\tt V}}\phi\lambda\lambda{\tt[n1,n2,n3]}\eqncom
	&\threevertexJ{dashes_rar}{$\scriptstyle n_3$}{plain_ar}{$\scriptstyle n_1$}{dashes_rar}{$\scriptstyle n_2$}
	&=
	\text{{\tt V}}\phi\text{{\tt b}}\lambda\text{{\tt b}}\lambda\text{{\tt b[n1,n2,n3]}}\eqncom\\
	\threevertexJ{dashes_ar}{$\scriptstyle  n_3$}{plain_ar}{$\scriptstyle n_1$}{dashes_ar}{$\scriptstyle n_2$}
	&=
	\text{{\tt V}}\phi\text{{\tt b}}\lambda\lambda\text{{\tt[n1,n2,n3]}}\eqncom
	&\threevertexJ{dashes_rar}{$\scriptstyle n_3$}{plain_rar}{$\scriptstyle n_1$}{dashes_rar}{$\scriptstyle n_2$}
	&=\text{{\tt V}}\phi\lambda\text{{\tt b}}\lambda\text{{\tt b[n1,n2,n3]}}\eqncom\\
	\threevertexJ{plain_ar}{$\scriptstyle p_3n_3$}{photon}{$\scriptstyle n_1$}{plain_rar}{$\scriptstyle p_2n_2$}
	&=\text{{\tt VA}}\phi\text{{\tt b}}\phi\text{{\tt[n1,n2,n3][p2,p3]}}\eqncom
	&&
	\end{aligned}
	\end{equation}
where we combined the ordinary vertices with the counterterm vertices, since we included the 1PI renormalisation constants $Z_v=1+\delta_v$ into each vertex $v$ in analogy to the situation in \eqref{eq:cubic_vertices}. Note that the fermion-gluon vertices distinguish between the version with a $\sigma$ and a $\bar{\sigma}$ matrix, as in \eqref{eq:cubic_vertices}. For quartic vertices, the same rules as for cubic vertices apply, except that we chose to label F- and D-term type scalar vertices with \ttt{VF} and \ttt{VD}, respectively. This gives the four-valent vertices as
\begin{equation}\label{program:quartic_vertices}
\begin{gathered}
\begin{aligned}
\fourvertextwo{photon}{$\scriptstyle n_2$}{photon}{$\scriptstyle n_3$}{photon}{$\scriptstyle n_4$}{photon}{$\scriptstyle n_1$}
&=\text{{\tt VAAAA[n1,n2,n3,n4]}}\eqncom
&\fourvertextwo{plain_rar}{$\scriptstyle n_2$}{photon}{$\scriptstyle n_3$}{plain_rar}{$\scriptstyle n_4$}{photon}{$\scriptstyle n_1$}
&=\text{{\tt VA}}\phi\text{{\tt bA}}\phi\text{{\tt[n1,n2,n3,n4]}}
\eqncom\\
\fourvertextwo{plain_ar}{$\scriptstyle n_2$}{plain_ar}{$\scriptstyle n_3$}{plain_ar}{$\scriptstyle n_4$}{plain_ar}{$\scriptstyle n_1$}
&=\text{{\tt VF[n1,n2,n3,n4]}}\eqncom
&\fourvertextwo{plain_rar}{$\scriptstyle n_2$}{plain_rar}{$\scriptstyle n_3$}{plain_ar}{$\scriptstyle n_4$}{plain_ar}{$\scriptstyle n_1$}
&=\text{{\tt VD[n1,n2,n3,n4]}}\eqndot
\end{aligned}
\end{gathered}
\end{equation}
Finally, to also accommodate real scalar fields in the case of \NfSYMt, we have the following {\tt Mathematica} implementation of the objects in \appref{subsec:Feynman_real_scalars}:
\begin{equation}\label{program:vertices_real_scalar}
\begin{aligned}
\propagatorR{plain}{n_1}{n_2}&=\text{{\tt P}}\varphi\text{{\tt[n1,n2][p]}}\eqncom
&\twovertex{plain}{n_1}{plain}{n_2}&=\text{{\tt PCT}}\varphi\text{{\tt[n1,n2][p]}}\eqncom\\
\threevertexJ{plain}{$\scriptstyle p_3n_3$}{photon}{$\scriptstyle n_1$}{plain}{$\scriptstyle p_2n_2$}
&=\text{{\tt VA}}\varphi\varphi\text{{\tt[n1,n2,n3][p2,p3]}}\eqncom&&\\
\threevertexJ{dashes_ar}{$\scriptstyle n_3$}{plain}{$\scriptstyle n_1$}{dashes_ar}{$\scriptstyle n_2$}
&=\text{{\tt V}}\varphi\lambda\lambda\text{{\tt[n1,n2,n3]}}\eqncom
&\threevertexJ{dashes_rar}{$\scriptstyle n_3$}{plain}{$\scriptstyle n_1$}{dashes_rar}{$\scriptstyle n_2$}
&=\text{{\tt V}}\varphi\lambda\text{{\tt b}}\lambda\text{{\tt b[n1,n2,n3]}}\eqncom\\
\fourvertextwo{plain}{$\scriptstyle n_2$}{photon}{$\scriptstyle n_3$}{plain}{$\scriptstyle n_4$}{photon}{$\scriptstyle n_1$}
&=
\text{{\tt VA}}\varphi\text{{\tt A}}\varphi\text{{\tt[n1,n2,n3,n4]}}\eqncom
&\fourvertextwo{plain}{$\scriptstyle n_2$}{plain}{$\scriptstyle n_3$}{plain}{$\scriptstyle n_4$}{plain}{$\scriptstyle n_1$}
&=\text{{\tt V}}\varphi\varphi\varphi\varphi\text{{\tt[n1,n2,n3,n4]}}\eqncom
\end{aligned}
\end{equation}
where the same rules apply as for \eqref{program:propagators} -- \eqref{program:quartic_vertices}. Within the \ttt{Mathematica} functions for propagators and vertices different types of indices are distinguished by having different heads and a list of all distinguished objects is given in \tabref{tab:index_types}.
\begin{table}[H]
	\centering
	\caption{Possible internal index types for a leg label $n1$.}
	\label{tab:index_types}
	\begin{tabular}{|c|c|}
		\hline
		index type & internal appearance\\
		\hline
		colour $\T^{\ttt{\scriptsize n1}}$& \ttt{trc}$\scriptstyle[\ttt{\scriptsize n1}]$\\
		fermionic flavour & \ttt{\scriptsize ff}$\scriptstyle[\ttt{\scriptsize n1}]$\\
		scalar flavour & $ \ttt{\scriptsize fs}\scriptstyle[\ttt{\scriptsize n1}]$\\
		spacetime index & $\scriptstyle \mu[\ttt{\scriptsize n1}]$\\
		$\spl{2}$ index & $\scriptstyle\alpha[\ttt{\scriptsize n1}]$\\
		$\splbar{2}$ index & $\scriptstyle\beta \ttt{\scriptsize d}[\ttt{\scriptsize n1}]$\\
		spacetime vector \ttt{p} in direction ${\scriptstyle \mu[\ttt{\scriptsize n1}]}$ & $\ttt{mom}[\ttt{p},{\scriptstyle\mu[\ttt{\scriptsize n1}]}]$\\
		\hline
	\end{tabular}
\end{table}
So, for example calling the function for the scalar gluon vertex, with scalar momenta \ttt{p2} and \ttt{p3} that leave the vertex, yields
\begin{equation}
\begin{aligned}
\ttt{VA}\phi\ttt{b}\phi\ttt{[n1,n2,n3][p2,p3]}&=
-\complexi \ttt{gYM}\,
\bigl(
\ttt{mom[p2,}\mu\ttt{[n1]]}
-\ttt{mom[p3,}\mu\ttt{[n1]]}
\bigr)
\\&\phan{{}={}}
\bigl(
\ttt{trc[c[\ttt{n1}],c[\ttt{n2}],c[\ttt{n3}]]}
-
\ttt{trc[c[\ttt{n1}],c[\ttt{n3}],c[\ttt{n2}]]}
\bigr)
\\&\phan{{}={}}
\delta\ttt{ss[fs[n2],fs[n3]]}\eqncom
\end{aligned}
\end{equation}
where the occurring functions are described in \tabref{tab:function}.

To exemplify how a Feynman diagram calculation with the {\tt FokkenFeynPackage} works, let us evaluate a one- and a two-loop propagator type diagram explicitly. All unexplained symbols are briefly described in \tabref{tab:variables} and \ref{tab:function}. First, we evaluate a contribution to the one-loop scalar self-energy, given by
\begin{equation}
\begin{aligned}
\text{{\tt diag1}}=
\Biggl(
\settoheight{\eqoff}{$\times$}%
\setlength{\eqoff}{0.5\eqoff}%
\addtolength{\eqoff}{-5.\unitlength}%
\raisebox{\eqoff}{
	\fmfframe(0,1)(0,0){
		\begin{fmfchar*}(20,7.5)
		\fmfforce{0 w, 0.5h}{vout}
		\fmfforce{1 w, 0.5h}{vin}
		\fmfforce{0.25 w, 0.5h}{v2}
		\fmfforce{0.75 w, 0.5h}{v1}
		\fmf{plain}{vin,v1}
		\fmf{plain}{v1,v2}
		\fmf{plain}{v2,vout}
		\fmf{photon,right,tension=0}{v1,v2}
		\fmffreeze
		\fmfposition
		\Marrow{a}{down}{bot}{"p2" infont "cmvtt10" scaled 0.6}{v1,v2}{9} 
		\Marrow{b}{up}{top}{"p1" infont "cmvtt10" scaled 0.6}{v1,v2}{18}
		\Marrow{c}{down}{bot}{"p0" infont "cmvtt10" scaled 0.6}{vin,v1}{9}
		\Marrow{d}{down}{bot}{"p0" infont "cmvtt10" scaled 0.6}{v2,vout}{9}
		\fmfiv{label={\tt\tiny 22},l.angle=65,l.dist=0.075w}{vloc(__v1)}
		\fmfiv{label={\tt\tiny in},l.angle=-50,l.dist=0.05w}{vloc(__v1)}
		\fmfiv{label={\tt\tiny 11},l.angle=-130,l.dist=0.05w}{vloc(__v1)}
		\fmfiv{label={\tt\tiny 21},l.angle=115,l.dist=0.075w}{vloc(__v2)}
		\fmfiv{label={\tt\tiny out},l.angle=-130,l.dist=0.05w}{vloc(__v2)}
		\fmfiv{label={\tt\tiny 12},l.angle=-50,l.dist=0.05w}{vloc(__v2)}
		\end{fmfchar*}
	}
}
\Biggr)_{\text{1PI}}
&{}={}
\text{{\tt VA}}\phi\text{{\tt b}}\phi\text{{\tt [22,in,11][-p0,p2]}}
\\[-0.7\baselineskip]
&\phan{{}={}}
\text{{\tt VA}}\phi\text{{\tt b}}\phi\text{{\tt [21,12,out][-p2,p0]}}
\\&\phan{{}={}}
\text{{\tt P}}\phi\text{{\tt [11,12][p2]}}
\\&\phan{{}={}}
\text{{\tt PA}}\text{{\tt [21,22][-p1]}}\eqncom
\end{aligned}
\end{equation}
where the order of vertices and propagators in the last equality is not important. Note that the signs in the momentum input\footnote{Signs in the momentum input in propagators ensure that the momentum flow corresponds to the flow from the first label to the second in the respective function. Since both propagators are bosonic and hence quadratic in momenta, these signs do not change the final result in this example.} in the vertices ensure that the vertex functions have momentum arguments that leave the vertex. We contract all occurring indices via
\begin{equation}
\begin{aligned}
\ttt{diag1a}&=\ttt{ContractIndices[diag1]}
\\
&=
2\ttt{gYMB}^2\ttt{NN}\,\frac{\ttt{p0}^2+2\ttt{p0}.\ttt{p2}+\ttt{p2}^2-(1-\xi)\frac{(\ttt{p0}.\ttt{p1}+\ttt{p1}.\ttt{p2})^2}{\ttt{p1}^2}}{\ttt{p1}^2\ttt{p2}^2}
\\
&\phan{{}={}}
\bigl(\ttt{trc[c[in],c[out]]}-\frac{1}{\ttt{NN}}\ttt{trc[c[in]]trc[c[out]]}\bigr)
\\
&\phan{{}={}}
\delta\ttt{s[fs[in],fs[out]]}\eqncom
\end{aligned}
\end{equation}
where we did not show the \ttt{FullSimplify} operation that is needed to obtain the above form. For integrals that do not involve contracted \spl{2,\CC} spinor indices, it is also possible to employ the function $\delta$\ttt{Replace[\expr]} instead of \ttt{ContractIndices[\expr]}. The colour and flavour indices in the final expression can be renamed via
\begin{equation}
\begin{aligned}\label{eq:diag1b}
\ttt{diag1b}&=\ttt{gTo}\lambda\ttt{[FinalIndex[diag1a]]}\\
&=2\lambda\,\mumu^{4-\ttt{DIM}}\,\frac{\ttt{p0}^2+2\ttt{p0}.\ttt{p2}+\ttt{p2}^2-(1-\xi)\frac{(\ttt{p0}.\ttt{p1}+\ttt{p1}.\ttt{p2})^2}{\ttt{p1}^2}}{\ttt{p1}^2\ttt{p2}^2}
\\
&\phan{{}={}}
\bigl(\ttt{trc[in,out]}-\frac{1}{\ttt{NN}}\ttt{trc[in]trc[out]}\bigr)
\delta\ttt{s[sin,sout]}\eqncom
\end{aligned}
\end{equation}
where we also replaced the bare Yang-Mills coupling constant by the renormalised \tHooft coupling $\lambda$. This was the last general step of the \ttt{FokkenFeynPackage}. In the present example of a propagator-type scalar one-loop integral, we can, however, proceed further. The contracted momenta \ttt{p0.p1} can be rewritten in terms of squares of irreducible momenta via \ttt{rulesMomentaL1[p0.p2]}$=\frac12(\ttt{p0}^2 - \ttt{p1}^2 + \ttt{p2}^2)$. Of course this replacement depends on the basis choice of irreducible momenta at a given loop order. Our explicit one- and two-loop choices are
\begin{equation}
\begin{aligned}\label{eq:momentum_parametrisation}
&\text{1 loop:}& \ttt{p0}& =p\eqncom& \ttt{p1}&=\ell\eqncom & \ttt{p2}&=p-\ell\eqncom& &&&&&&\\
&\text{2 loop:}& \ttt{p0}& =p\eqncom& \ttt{p1}&=\ell\eqncom & \ttt{p2}&=k\eqncom& \ttt{p3}&=p-k\eqncom&\ttt{p4}&=p-\ell\eqncom& \ttt{p5}&=k-\ell&
\end{aligned}
\end{equation} 
and the replacement function for two-loop momenta is \ttt{rulesMomentaL2[\expr]}. Coming back to \eqref{eq:diag1b}, we can generate the one-loop integrands in terms of scalar integrals as
\begin{equation}
\begin{aligned}\label{eq:diag1c}
\ttt{diag1c}&=\ttt{GenerateIntegrands[diag1b][1]}
\\&=
-2\lambda\,\mumu^{4-\ttt{DIM}}\,
\Bigl(\ttt{i[0,1]}-2\ttt{i[1,0]}-2\ttt{p}^2\ttt{i[1,1]}
\\
&\phan{{}={}-2\lambda\,\Bigl(}
+(1-\xi)\bigl(\ttt{i[2,-1]}-2\ttt{p}^2\ttt{i[2,0]}+\ttt{p}^4\ttt{i[2,1]}\bigr)
\Bigr)
\\
&\phan{{}={}}
\bigl(\ttt{trc[in,out]}-\frac{1}{\ttt{NN}}\ttt{trc[in]trc[out]}\bigr)
\delta\ttt{s[sin,sout]}\eqncom
\end{aligned}
\end{equation}
where the integrals \ttt{i}$[\alpha,\beta]$ are the Minkowski space analoga of the integrals $\bar{I}_{(\alpha,\beta)}(\bar p)$ given in  \eqref{eq:I_alpha_beta}. The scalar integrals can be solved explicitly in terms of $\Gamma$ functions via the function
\begin{equation}
\begin{aligned}\label{eq:diag1d}
\ttt{diag1d}&=\ttt{SolveIntegrals[diag1c][1]}
\\&=
-2\lambda\,\complexi \ttt{p}^2\left( \frac{4\pi\mumu^2}{\ttt{p}^2}\right)^{\frac 12(4-\ttt{DIM})}
\bigl(2+(1-\xi)(\ttt{DIM}-3)\bigr)
\frac{\csc[\frac{\pi}{2}\ttt{DIM}]}{2^{\ttt{DIM}+1}\sqrt{\pi}\,\Gamma[\frac{1}{2}(\ttt{DIM}-1)]}
\\
&\phan{{}={}}
\bigl(\ttt{trc[in,out]}-\frac{1}{\ttt{NN}}\ttt{trc[in]trc[out]}\bigr)
\delta\ttt{s[sin,sout]}
\\
&\phan{{}={}}
+(\text{IR divergence parts labelled by \ttt{iIR1}}[\num{1},\num{2}])
\eqncom
\end{aligned}
\end{equation}
where the IR divergence line appears since the program does not determine whether a given diagram contains persistent IR divergences. This task is left for the user and the program simply assigns a possible IR divergences to each integral occurring in \eqref{eq:diag1c}. At one loop, the IR divergence is characterised by $\ttt{iIR1[\num{1},\num{2}]}=(\ttt{i[\num{1},\num{2}]})_{\text{IR div.}}$. The present example is free of IR divergences\footnote{To see this, we use power counting method that was used to obtain the IR divergence in \eqref{eq:IR_divergent_diag}. In the present case we also have to rewrite a momentum vector in terms of its magnitude times a normalised direction component $\ell_\mu=\abs{\ell}\,\hat{\ell}_\mu$.}, so that we can set $\ttt{iIR1[\num{1},\num{2}]}\rightarrow 0$ in \eqref{eq:diag1d}. This concludes our one-loop example.

Next, let us turn to a slightly more complicated two-loop diagram. To exemplify how fermions must be treated we calculate the following contribution to the fermion two-loop self energy 
\begin{equation}
\begin{aligned}\label{eq:diag2}
\ttt{diag2}=
\settoheight{\eqoff}{$\times$}%
\setlength{\eqoff}{0.5\eqoff}%
\addtolength{\eqoff}{-7.5\unitlength}%
\Biggl(
\raisebox{\eqoff}{%
	\fbox{
		\fmfframe(0,0)(0,0){%
			\begin{fmfchar*}(30,15)
			\fmfforce{(.75w,0.5h)}{vo}
			\fmfforce{(0.w,0.5h)}{vl}
			\fmfforce{(1w,0.5h)}{vr}
			\fmf{dashes_ar}{vc,vl}
			\fmf{dashes_rar}{vo,vr}
			\fmffreeze
			\fmfposition
			\fmfforce{(0.25w,0.5h)}{vc}
			\fmf{phantom,left=0.75}{vo,vc}	
			\fmf{phantom,left=0.75}{vc,vo}
			\fmffreeze
			\fmfposition
			\fmfipath{p[]}
			\fmfiset{p1}{vpath(__vo,__vc)}
			\fmfiset{p11}{subpath (length(p1)/2,0) of p1}
			\fmfiset{p12}{subpath (length(p1),length(p1)/2) of p1}
			\fmfi{plain}{p11}
			\fmfi{dashes_ar}{p12}
			\fmfiset{p2}{vpath(__vc,__vo)}
			\fmfiset{p21}{subpath (0,length(p2)/2) of p2}
			\fmfiset{p22}{subpath (length(p2)/2,length(p2)) of p2}
			\fmfi{plain}{p21}
			\fmfi{dashes_ar}{p22}
			\fmfi{dashes_rar}{point length(p1)/2 of p1 -- point length(p2)/2 of p2}
			\fmfiv{label={\tt\tiny 31},l.angle=112,l.dist=0.045w}{vloc(__vc)}
			\fmfiv{label={\tt\tiny 22},l.angle=-112,l.dist=0.045w}{vloc(__vc)}
			\fmfiv{label={\tt\tiny out},l.angle=-60,l.dist=0.05w}{vloc(__vl)}
			\fmfiv{label={\tt\tiny 42},l.angle=68,l.dist=0.045w}{vloc(__vo)}
			\fmfiv{label={\tt\tiny 11},l.angle=-68,l.dist=0.045w}{vloc(__vo)}
			\fmfiv{label={\tt\tiny in},l.angle=-120,l.dist=0.05w}{vloc(__vr)}
			\fmfiv{label={\tt\tiny 51},l.angle=65,l.dist=0.045w}{point length(p1)/2 of p1}
			\fmfiv{label={\tt\tiny 12},l.angle=-40,l.dist=0.025w}{point length(p1)/2 of p1}
			\fmfiv{label={\tt\tiny 21},l.angle=-140,l.dist=0.025w}{point length(p1)/2 of p1}
			\fmfiv{label={\tt\tiny 52},l.angle=-65,l.dist=0.045w}{point length(p2)/2 of p2}
			\fmfiv{label={\tt\tiny 41},l.angle=40,l.dist=0.025w}{point length(p2)/2 of p2}
			\fmfiv{label={\tt\tiny 32},l.angle=140,l.dist=0.025w}{point length(p2)/2 of p2}
			\MarrowShifti{a}{(4,-4)}{bot}{"-p1" infont "cmvtt10" scaled 0.6}{p11} 
			\MarrowShifti{b}{(-4,-4)}{lft}{"-p2" infont "cmvtt10" scaled 0.6}{p12}
			\MarrowShifti{c}{(-4,4)}{lft}{"-p3" infont "cmvtt10" scaled 0.6}{p21}
			\MarrowShifti{d}{(4,4)}{top}{"-p4" infont "cmvtt10" scaled 0.6}{p22}
			\Marrowi{e}{left}{lft}{"p5" infont "cmvtt10" scaled 0.6}{point length(p2)/2 of p2 -- point length(p1)/2 of p1}{3}
			\MarrowShift{f}{(2,4)}{top}{"p0" infont "cmvtt10" scaled 0.6}{vr,vo}
			\MarrowShift{g}{(-2,4)}{top}{"p0" infont "cmvtt10" scaled 0.6}{vc,vl}
			\end{fmfchar*}
		}
	}
}
\Biggr)_{\text{1PI}}
&=
\ttt{V}\varphi\lambda\lambda\ttt{[11,42,in]}
\ttt{V}\varphi\lambda\lambda\ttt{[12,21,51]}
\\[-0.9\baselineskip]
&\phan{{}={}}
\ttt{V}\varphi\lambda\ttt{b}\lambda\ttt{b[31,22,out]}
\ttt{V}\varphi\lambda\ttt{b}\lambda\ttt{b[32,41,52]}
\\&\phan{{}={}}
\ttt{P}\varphi\ttt{[11,12][p1]}
\ttt{P}\varphi\ttt{[31,32][$-$p3]}
\\&\phan{{}={}}
\ttt{P}\lambda\ttt{b}\lambda\ttt{[41,42][$-$p4]}
\ttt{P}\lambda\lambda\ttt{b}\ttt{[51,52][$-$p5]}
\\&\phan{{}={}}
\ttt{P}\lambda\ttt{b}\lambda\ttt{[22,21][$-$p2]}
\eqncom
\end{aligned}
\end{equation}
where the two-loop momentum parametrisation of \eqref{eq:momentum_parametrisation} is chosen and the occurring signs ensure momentum conservation at each vertex. Since we have fermions in this diagram, we have to be careful when writing the associated propagators and vertices. Feynman rule 8 of \appref{sec:the-actual-feynman-rules} forces us to have alternating $\sigma$ and $\bar{\sigma}$ matrices along a path of fermion propagators in a Feynman diagram. This requirement dictates that the $\ttt{P}\lambda\ttt{b}\lambda$ and the $\ttt{P}\lambda\lambda\ttt{b}$ version of the fermion propagator have to alternate along the fermion path in the above diagram.\footnote{In diagrams involving the gauge field fermion interaction vertex the Feynman rule 8 of \appref{sec:the-actual-feynman-rules} forces us to choose the appropriate vertex from \eqref{program:cubic_vertices}.} Note also that we have to pick up a sign in the momentum input if we cross a propagator against the direction of its momentum arrow. In \eqref{eq:diag2}, this happens only in the first scalar propagator $\ttt{P}\varphi$, where the sign has no net influence since the function is quadratic in the momentum. For fermionic propagators, which are linear in the momentum, these signs are, however, important. As before, we proceed by contracting all indices and replacing the \tHooft coupling and indices in the final expression to obtain
\begin{equation}
\begin{aligned}\label{eq:diag2a}
\ttt{diag2a}&=
\ttt{gTo[}\lambda\ttt{FinalIndex[ContractIndices[diag2]]]}\\
&=48\complexi \lambda^2\, \mumu^{8-2\ttt{DIM}}\,
\frac{\ttt{mom[p2,\muind{3}]mom[p4,\muind{1}]mom[p5,\muind{2}]}}{\ttt{p1}^2\ttt{p2}^2\ttt{p3}^2\ttt{p4}^2\ttt{p5}^2}	
\\
&\phan{{}={}}
\ttt{T}\sigma\ttt{[}\{\ttt{\muind{3},\muind{2},\muind{1}}\}\ttt{][}\sigma\ttt{b,}{\scriptstyle\beta\ttt{d}[\ttt{out}]}\ttt{,}{\scriptstyle\alpha[\ttt{in}]}\ttt{]}
\\
&\phan{{}={}}
\bigl(\ttt{trc[in,out]}-\frac{1}{\ttt{NN}}\ttt{trc[in]trc[out]}\bigr)
\delta\ttt{4[fin,fout]}\eqndot
\end{aligned}
\end{equation}
The structure \ttt{T}$\sigma$ represents a product of alternating $\sigma$ and $\bar{\sigma}$ matrices whose open spacetime indices occur in the first argument and the initial and final open spinor indices appear in the third and fourth argument. It is generated in \ttt{ContractIndices} via the function $\sigma$\ttt{toOpenTrace} and in principle the following four structures are possible
\begin{equation}
\begin{aligned}
&\text{{\tt T}}\sigma[\{\mu 1,\dots,\mu n\}][\sigma,\alphaindex,\betadcindex]
&=&
(\sigma_{\mu_1})_{\alphaindex\dot{\gamma}_1}(\bar{\sigma}_{\mu_2})^{\dot{\gamma}_1\gamma_2}\dots(\sigma_{\mu_n})_{\gamma_{n-1}\betadcindex}
\eqncom&\\
&\text{{\tt T}}\sigma[\{\mu 1,\dots,\mu n\}][\sigma\text{{\tt b}},\betadcindex,\alphaindex]
&=&
(\bar\sigma_{\mu_1})^{\betadcindex\gamma_1}(\sigma_{\mu_2})_{\gamma_1\dot{\gamma}_2}\dots(\bar\sigma_{\mu_n})^{\dot{\gamma}_{n-1}\alphaindex}
\eqncom&\\
&\text{{\tt T}}\sigma[\{\mu 1,\dots,\mu n\}][\delta\delta,{\scriptstyle\alpha[1]},{\scriptstyle\alpha[2]}]
&=&
(\sigma_{\mu_1})_{{\scriptstyle\alpha[1]}\dot{\gamma}_1}(\bar{\sigma}_{\mu_2})^{\dot{\gamma}_1\gamma_2}
\dots(\bar{\sigma}_{\mu_n})^{\dot{\gamma}_{n-1}{\scriptstyle\alpha[2]}}
\eqncom&\\
&\text{{\tt T}}\sigma[\{\mu 1,\dots,\mu n\}][\delta\delta\text{{\tt d}},{\scriptstyle\beta}\text{{\scriptsize d}}{\scriptstyle[2]},{\scriptstyle\beta}\text{{\scriptsize d}}{\scriptstyle[2]}]
&=&
(\bar\sigma_{\mu_1})^{{\scriptstyle\beta}\text{{\scriptsize d}}{\scriptstyle[1]}\gamma_1}(\sigma_{\mu_2})_{\gamma_1\dot{\gamma}_2}\dots
(\sigma_{\mu_n})_{\gamma_{n-1}{\scriptstyle\beta}\text{{\scriptsize d}}{\scriptstyle[2]}}\eqndot&
\end{aligned}
\end{equation}
This concludes the general manipulations of \ttt{FokkenFeynPackage}, but like in the previous one-loop example, we can proceed further for propagator-type two-loop scalar integrals. To transform \eqref{eq:diag2a} into a scalar integral, we contract it with an external momentum vector $-\frac{\complexi}{2}p_{\alpha_{\text{in}}\dot{\beta}_{\text{out}}}$ which is realised in \ttt{Mathematica} as\footnote{For the mapping of momentum vector to the spinorial representation \cf \eqref{eq:vectorfield_spinorial}.}
\begin{equation}
\ttt{pext}=\ttt{toOpenTrace}\Bigl[-\frac 12 \sigma[\nu][{\scriptstyle\alpha[\ttt{in}]},{\scriptstyle\beta\ttt{d}[\ttt{out}]}]\ttt{mom[p0,\muind{\nu}]}\Bigr]\eqndot
\end{equation}
The two-loop integral with fully contracted spacetime indices is now obtained as 
\begin{equation}
\begin{aligned}
\ttt{diag2b}&=
\ttt{ContractIndices[pext diag2a]}\\
&=48\complexi \lambda^2\, \mumu^{8-2\ttt{DIM}}\,
\frac{
	(\ttt{p0}.\ttt{p5})\,(\ttt{p2}.\ttt{p4})
	-(\ttt{p0}.\ttt{p4})\,(\ttt{p2}.\ttt{p5})
	-(\ttt{p0}.\ttt{p2})\,(\ttt{p4}.\ttt{p5})
}{\ttt{p1}^2\ttt{p2}^2\ttt{p3}^2\ttt{p4}^2\ttt{p5}^2}	
\\
&\phan{{}={}}
\bigl(\ttt{trc[in,out]}-\frac{1}{\ttt{NN}}\ttt{trc[in]trc[out]}\bigr)
\delta\ttt{4[fin,fout]}\eqncom
\end{aligned}
\end{equation}
which can be written in terms of scalar two-loop integrals\footnote{The spacetime integrals are characterised in terms of weights of their irreducible loop momenta as $\ttt{i[n1,n2,n3,n4,n5]=}\ttt{p1}^{-2\ttt{n1}}\ttt{p2}^{-2\ttt{n2}}\ttt{p3}^{-2\ttt{n3}}\ttt{p4}^{-2\ttt{n4}}\ttt{p5}^{-2\ttt{n5}}$.} via
\begin{equation}
\begin{aligned}\label{eq:diag2c}
\ttt{diag2c}&=
\ttt{GenerateIntegrands[diag2b][2]}\\
&=24\complexi \lambda^2\, \mumu^{8-2\ttt{DIM}}\,
\\
&\phan{{}={}}
\bigl(
\ttt{i[0,1,0,1,1]}
-\ttt{i[1,0,1,0,1]}
-\ttt{p}^{2}\ttt{i[1,1,1,1,0]}
\bigr)
\\
&\phan{{}={}}
\bigl(\ttt{trc[in,out]}-\frac{1}{\ttt{NN}}\ttt{trc[in]trc[out]}\bigr)
\delta\ttt{4[fin,fout]}\eqndot
\end{aligned}
\end{equation}
Finally, the scalar propagator-type two-loop integrals can be solved analytically via the function \ttt{SolveIntegrals[\expr][2]}, which solves nested one-loop integrals directly and employs IBP relations of \cite{Chetyrkin:1981qh} in more complicated cases. The solution is given in terms of $\Gamma$ functions and unresolved possible IR divergences characterised by the function \ttt{iIR2[n1,n2,n3,n4,n5]}. The analysis whether IR divergences occur is left to the user, as in the one-loop example. In the present example, the first two integrals in \eqref{eq:diag2c}, which contain IR divergences, cancel and we find the final IR-finite solution
\begin{equation}
\begin{aligned}
\ttt{diag2d}&=
\ttt{SolveIntegrals[diag2c][2]}\\
&=6\lambda^2\,\complexi\ttt{p}^2
\left( \frac{4\pi\mumu^2}{\ttt{p}^2}\right)^{(4-\ttt{DIM})}
\frac{\csc[\frac{\pi}{2}\ttt{DIM}]^2}{2^{2\ttt{DIM}}\pi\Gamma[\frac 12(\ttt{DIM}-1)]}
\\
&\phan{{}={}}
\bigl(\ttt{trc[in,out]}-\frac{1}{\ttt{NN}}\ttt{trc[in]trc[out]}\bigr)
\delta\ttt{4[fin,fout]}
\eqncom
\end{aligned}
\end{equation}
where we suppressed the artificial IR divergence output \ttt{iIR2[\num{1},\num{2},\num{3},\num{4},\num{5}]}.

\begin{table}[H]
	\small
	\centering
	\caption{The variables and parameters in this table are defined in $1$: {\tt Feynpar.m}; $2$: {\tt VariablesReplacements.m}; $3$: {\tt Tensors.m}; $4$: {\tt Functions.m}; $5$: {\tt Feynmanrules.m}; and $6$: {\tt solveUpto2loop.m}.}
	\label{tab:variables}
	\begin{tabularx}{\textwidth}{|c|c|X|}
		\hline
		symbol &file& usage\\
		\hline
		\ttt{a}$.$\ttt{b} & & The multiplication operator $.$ connects a contracted pair of spacetime vetors, e.g.\ $a_\mu b^\mu\hat{=} \ttt{a}.\ttt{b} $.\\
		{\tt \alphaindex} &2& An \spl{2} index must be indicated via this head in calculations.\\
		{\tt $\beta$}&2& Deformation angle of the $\beta$-deformation.\\
		{\tt \betadcindex} &2&  An \splbar{2} index must be indicated via this head in calculations.\\
		{\tt \cindex} &2& A colour index must be indicated via this head in calculations.\\
		{\tt cmsb}&2& The constant appearing in the $\ol{\text{MS}}$ scheme {\tt cmsb}$=\gammaE-\log 4\pi$.\\
		{\tt $\delta$AA, $\delta$ccb, $\delta\varphi\varphi$, $\delta\phi\phi$b, $\delta\lambda\lambda$b} &5& The counterterms of gauge fields, ghosts, real scalars, complex scalars and Weyl fermions are given by these symbols.\\
		{\tt DIM} &2& spacetime dimension used in {\tt FokkenFeynPackage}\\
		{\tt dimension} &1& spacetime dimension used in {\tt Feynpar.m}; by default $\text{{\tt dimension}}=\text{{\tt DIM}}$.\\
		{\tt \ff} &2& A fermion flavour index must be indicated via this head in calculations.\\
		{\tt \fs} &2& A scalar flavour index must be indicated via this head in calculations.\\
		{\tt $\lambda$}&2& This is the \tHooft coupling constant $\lambda =\gym^2 N$.\\
		{\tt $\gamma[\_]$}, {\tt $\gamma$p$[\_]$}, {\tt $\gamma$m$[\_]$}&2& These are the deformation angles of the $\gamma_i$-deformation. The first variable gives the angles $\gamma_i$, the second $\gamma_i^+$ and the third $\gamma_i^-$, with $i\in\{1,2,3\}$.\\
		{\tt gp}&2&This is the effective planar coupling constant $g=\frac{\sqrt{\lambda}}{4\pi}$.\\
		{\tt gYMB}&2& This is the bare Yang-Mills coupling constant. In a \CFT in four-dimensional Minkowski space it is related to the renormalised coupling as $g_{\scriptscriptstyle{\text{YM}}\,\text{B}}=\mu^{\frac 12(4-D)}\gym=\mu^\epsilon\gym$.\\
		{\tt L}&2& This is the object $L=\log\frac{p^2}{\mumu^2}$.\\
		{\tt i$[\num{1},\num{2},\num{3},\num{4},\num{5}]$} &2& This function characterises a scalar one- or two-loop Feynman integral. The integral is given in terms of the weights of the two, or respectively five irreducible momenta, which occur as $p_i^{-2n_i}$ in the integrals.\\
	\end{tabularx}
\end{table}
\begin{table}[H]
	\small
	\centering
	\begin{tabularx}{\textwidth}{|c|c|X|}
		symbol &file& usage\\
		\hline
		{\tt iDL$[1,1,1,1,1]$}&2& This function gives the (unevaluated) two-loop dimensionless scalar master integral 
		{\tt iDL$[1,1,1,1,1]=$p$^{2(5-\text{{\tt DIM}})}$i$[1,1,1,1,1]$}.\\
		{\tt iIR1$[\num{1},\num{2}]$}&2& Encodes possible IR divergences that may have entered the calculation of a scalar one-loop integral in terms of powers of the irreducible one-loop momenta $p_1^{-2n_1}p_2^{-2n_2}$. Whether {\tt iIR1} occurs is not determined by {\tt FokkenFeynProgram} and must be checked by hand.\\
		{\tt iIR2$[\num{1},\num{2},\num{3},\num{4},\num{5}]$} &2& Encodes possible IR divergences that may have entered the calculation of a scalar two-loop integral in terms of powers of the five irreducible two-loop momenta $p_i^{-2n_i}$. Whether {\tt iIR1} occurs is not determined by {\tt FokkenFeynProgram} and must be checked by hand.\\
		{\tt \muindex} &2& A spacetime index must be indicated via this head in calculations.\\
		{\tt $\mumu$}&2& This is the renormalisation scale of Feynman diagrams.\\
		{\tt mom$[$p$\_,\muindex]$} &3& A momentum must be indicated via this structure in calculations. It can be generated via the function {\tt momentum}.\\
		{\tt NN}&2& This is the number of colours.\\
		{\tt N$\psi$}&2& This gives the number of fermionic flavours.\\
		{\tt N$\phi$}&2& This gives the number of scalar flavours.\\
		{\tt p}, {\tt p0}, {\tt p1}, {\tt p2}, {\tt p3}, {\tt p4}, {\tt p5}&2& These are the irreducible one- and two-loop momenta. For an external momentum $p$, and one-loop momentum $\ell$, they are given by: {\tt p0}$=p$, {\tt p1}$=\ell$, and {\tt p2}$=p-\ell$. For an external momentum $p$, and two-loop momenta $\ell$ and $k$, they are given by: {\tt p0}$=p$, {\tt p1}$=\ell$, {\tt p2}$=k$, {\tt p3}$=p-k$, {\tt p4}$=p-\ell$, and {\tt p5}$=k-\ell$.\\
		{\tt sun}&2& This is the gauge group identifier: it is $s=0$ for gauge group \UN and $s=1$ for gauge group \SUN.\\
		{\tt trc$[\cindex,\dots,\cindex]$} &3& A colour trace over colour generators $\T^{\cindex}$ must be indicated via this head in calculations.\\
		{\tt $\xi$}&2& This is the gauge parameter as it enters \eqref{eq:Lagrangian_Feynman_rules_2} and in particular the gauge field propagator \eqref{eq:propagators_scalar}.\\
		\hline
	\end{tabularx}
\end{table}
\setcounter{table}{2}
\begin{table}[H]
	\small
	\centering
	\caption{The functions in this table are defined in $1$: {\tt Feynpar.m}; $2$: {\tt VariablesReplacements.m}; $3$: {\tt Tensors.m}; $4$: {\tt Functions.m}; $5$: {\tt Feynmanrules.m}; and $6$: {\tt solveUpto2loop.m}. Free indices of special types are labelled according to $\muindex$: spacetime; $\ff$: fermionic flavour; $\fs$: scalar flavour. If more than one index-type is admissible in an input argument, it is indicated by $\_$ and real numerical inputs are labelled as $\numn$, $\num{1}$, $\num{2}$, ect. Finally, generic input expressions are given as the input $\expr$.}
	\label{tab:function}
	\begin{tabularx}{\textwidth}{|c|c|X|}
		\hline
		symbol & file & usage\\
		\hline
		{\tt $\beta$def$[\expr]$}&2 & This function replaces $\gamma$, $\gamma${\tt p} and  $\gamma${\tt m} with $\beta$ to obtain expressions in the $\beta$-deformation from the respective expressions in the $\gamma_i$-deformation.\\
		{\tt ContractIndices$[\expr]$} &4& This function contracts all doubly occurring indices.\\
		{\tt $\delta[\_,\_]$} &3& This is a generic unevaluated Kronecker-$\delta$.\\
		{\tt $\delta$4$[\_,\_]$} &3& This is the unevaluated Kronecker-$\delta$ in four Euclidean dimensions. It automatically contracts doubly occurring indices, e.g.\ $\delta${\tt4}$[a,b]f[b]=f[a]$. In calculations of \NfSYM quantities, it may appear in places where usually $\delta$\ttt{f}$[\_,\_]$ or  $\delta$\ttt{s}$[\_,\_]$ appears.\\
		{\tt $\delta$c$[\cindex,\cindex]$} &3& This is an unevaluated Kronecker-$\delta$ for colour indices with $\delta$c$[\text{{\scriptsize c}}{\scriptstyle[a]},\text{{\scriptsize c}}{\scriptstyle[a]}]=\text{{\tt NN}}$.\\
	\end{tabularx}
\end{table}
\begin{table}[H]
	\small
	\centering
	\begin{tabularx}{\textwidth}{|c|c|X|}
		symbol &file & usage\\
		\hline
		{\tt $\delta$f$[\ff,\ff]$} &3& This is an unevaluated Kronecker-$\delta$ for fermion-flavour indices with $\delta$f$[${\scriptsize f}${\scriptstyle[a]},${\scriptsize f}${\scriptstyle[a]}]=${\tt N}$\psi$.\\
		{\tt $\delta$ff$[\ff,\ff]$} &3& This is an unevaluated fermionic Kronecker-$\delta$ generated by the Feynman rules. It is replaced via $\delta$\ttt{Replace[\expr]} by {\tt $\delta$f$[\ff,\ff]$}.\\
		{\tt $\delta$s$[\fs,\fs]$} &3& This is an unevaluated Kronecker-$\delta$. For scalar-flavour indices with $\delta$f$[\text{{\scriptsize s}}{\scriptstyle[a]},\text{{\scriptsize s}}{\scriptstyle[a]}]=\text{{\tt N}}\phi$.\\
		{\tt $\delta$ss$[\fs,\fs]$} &3& This is an unevaluated scalar Kronecker-$\delta$ generated by the Feynman rules. It is replaced via $\delta$\ttt{Replace[\expr]} by {\tt $\delta$s$[\fs,\fs]$}.\\
		{\tt $\delta$Replace$[\expr]$} &4& This function realises index contractions of doubly occurring indices in Kronecker-$\delta$'s. All products involving {\tt $\delta$}, {\tt $\delta$c}, {\tt $\delta$f}, and {\tt $\delta$s} are reduced. If the summation indices in $\expr$ are not are not correctly assigned an error message is returned.\\
		{\tt e[\_,\_,\_,\_]} &1& This function originates from the \ttt{Feynpar} package and realises the (unevaluated) totally antisymmetric tensor in four-dimensional Minkowski space. Note, that it does not distinguish lower and upper indices and doubly occurring indices are assumed to be contracted.\\
		{\tt F$[\fs,\fs,\fs,\fs]$}&5& This is the abstract symbol $\text{{\tt F}}[i,j,k,l]=F^{ij}_{lk}$ that enters \eqref{eq:Lagrangian_Feynman_rules_3} and realises the four-scalar F-term-type coupling in the theory.\\
		{\tt FinalIndex$[\expr]$}&2& This function replaces flavour and colour indices in final expressions for simple further treatment in {\tt Mathematica}. For example, a fermion external and summation index labelled by {\tt in} and $1$, respectively are replaced according to {\scriptsize\tt ff}${\scriptstyle[1]}\rightarrow {\tt sum}1$ and {\scriptsize\tt ff}${\scriptstyle[in]}\rightarrow {\tt fin}$.\\
		{\tt g$[\muindex,\muindex]$} &1& This is the metric tensor in {\tt DIM} dimensions. It automatically contracts doubly occurring indices, e.g.\ {\tt g}$[{\scriptstyle \mu[1]},{\scriptstyle \mu[2]}]f[{\scriptstyle \mu[1]}]=f[{\scriptstyle \mu[2]}]$.\\
		{\tt GenerateIntegrands$[\expr][$l$\_]$} &4& For {\tt l}$\_\in\{1,2\}$, this function generates one- or two-loop integrands in expressions where all spacetime indices are contracted.\\
		{\tt G$[\num{1},\num{2},$DIM$\_]$} &6& This is the explicit realisation of the $G$-function defined in \eqref{eq:G_function}.\\
		{\tt Gln$[\numn,\num{1},\num{2},$DIM$\_]$}  &6& This is the explicit realisation of the $G_{(n)}$-function defined in \eqref{eq:Gn_definition}\\
		{\tt GG$[\num{1},\num{2}]$}&2& This is the unevaluated $G$-function of \eqref{eq:G_function} with weight arguments $\num{1}$ and $\num{2}$ in {\tt DIM} dimensions.\\
		{\tt GGn$[\numn,\num{1},\num{2}]$}&2& This is the unevaluated $G_{(n)}$-function of \eqref{eq:Gn_definition} with level $\numn$ and weight arguments $\num{1}$ and $\num{2}$ in {\tt DIM} dimensions.\\
		{\tt $\gamma$pmTo$\gamma[\expr]$}&2& This function maps $\expr$ from the representation in \eqref{eq:gamma_pm} with deformation angles $\gamma_i^\pm$ to the one in \eqref{eq: antisymmetric product} with deformation angles $\gamma_i$.\\
		{\tt $\gamma$To$\gamma$pm1$[\expr]$}, {\tt $\gamma$To$\gamma$pm2$[\expr]$}&2& These are two possible mappings from the representation in \eqref{eq: antisymmetric product} with deformation angles $\gamma_i$ to the representation \eqref{eq:gamma_pm} with deformation angles $\gamma_i^\pm$.\\
		{\tt gTo$\lambda[\expr]$} &2& This function maps {\tt gYM} to the \tHooft coupling $\lambda$ and introduces the $\ol{\text{MS}}$ constant {\tt cmsb}.\\
		{\tt IntegrandToWeights$[\expr]$} &4& This function rewrites a scalar integrand $\expr$ in terms of weights of the linearly independent one- or two-loop momenta, e.g.\  {\tt IntegrandToWeights$[c\,\prod_{i=0}^5p_i^{-2n_i}]=c\,p_0^{-2n_0}$i$[n_1,n_2,n_3,n_4,n_5]$} for a constant $c$ and external scale $p_0^2$.\\
	\end{tabularx}
\end{table}
\begin{table}[H]
	\small
	\centering
	\begin{tabularx}{\textwidth}{|c|c|X|}
		symbol &file& usage\\
		\hline
		{\tt intsolveL1}, {\tt intsolveL2} &6& These functions both have the input $[\expr]$. They solve one- and respectively two-loop integrals that were generated via {\tt IntegrandToWeights$[\expr]$}. The solution is given in terms of unevaluated G-functions and possible IR divergent contributions {\tt iIR1$[\expr]$} or {\tt iIR2$[\expr]$}.\\
		{\tt $\lambda$Togeff$[\expr]$}&2& This function replaces the \tHooft coupling $\lambda$ with the effective planar coupling constant {\tt gp}.\\
		{\tt M$\delta[\_,\_]$} &3&  This is an explicit (evaluated) realisation of a generic Kronecker-$\delta$.\\
		{\tt M$\epsilon 2$u}, {\tt M$\epsilon 2$d} &3& These functions ${\tt func}[\_,\_]$ take two \spl{2} or \splbar{2} indices as arguments and are the explicit realisations of \eqref{eq:epsilon_spinor_indices}. The first and second function raises and lowers \spl{2,\CC} spinor indices, respectively.\\
		{\tt M$\epsilon 4[\_,\_,\_,\_]$} &3& This function is the explicit realisation of the totally antisymmetric tensor in four Euclidean dimensions with upper or lower indices with $\varepsilon^{1234}=\varepsilon_{1234}=1$. It usually takes \ff or \fs indices as input.\\
		{\tt M$\epsilon 4$d}, {\tt M$\epsilon 4$u} &3&These two functions take spacetime arguments {\tt func}$[\muindex,\muindex,\muindex,\muindex]$ and are the explicit realisations of the totally antisymmetric tensor in four Minkowski dimensions with lower and respectively upper indices and the normalisation $\varepsilon^{1234}=-\varepsilon_{1234}=1$.\\
		{\tt M$\gamma[\muindex]$} &3& This function explicitly implements the $\gamma$-matrices of \eqref{eq:gamma_Weyl}.\\
		{\tt M$\sigma[\muindex]$}, {\tt M$\sigma$b$[\muindex]$}&3& These functions are the explicit matrix implementations of $(\sigma_\mu)_{\alpha\dot{\beta}}$ and $(\bar\sigma_\mu)^{\dot{\beta}\alpha}$ given in \eqref{eq:sigma_matrices}.\\
		{\tt M$\Sigma[\fs]$}, {\tt M$\Sigma$b$[\fs]$} &3& These functions are the explicit matrix implementations of $(\Sigma_j)_{AB}$ and $(\bar\Sigma_j)^{AB}$ given in \eqref{eq:Sigma_6}.\\
		{\tt M$\sigma\mu\nu[\muindex,\muindex]$}, {\tt M$\sigma$b$\mu\nu[\muindex,\muindex]$} &3& These functions are the explicit matrix implementations of $(\sigma_{\mu\nu})_{\alpha}{}^\beta$ and $(\bar\sigma_{\mu\nu})^{\dot{\alpha}}{}_{\dot{\beta}}$ given in \eqref{eq:sigma_mu_nu}.\\
		{\tt momentum$[$p$\_,\muindex]$} &4& This function generates a momentum vector {\tt mom}$[$p$\_,\muindex]$ which has magnitude {\tt p}$\_$ and direction \muindex. If a sum is inserted, the output is a sum of momenta, e.g.\ $\ttt{momentum}[\ttt{p}+\ttt{q},\mu]\hat{=}p_{\mu}+q_{\mu}$.\\
		{\tt pToL$[\expr]$} &2& This function replaces $\log[\frac{\text{{\tt p}}^2}{\mumu^2}]$ by {\tt L}.\\
		{\tt QD}, {\tt QF}&5& These two abstract symbols with input {\tt func}$[\fs,\fs,\fs,\fs]$ realise the four-scalar double-trace interactions that enter \eqref{eq:Lagrangian_Feynman_rules_4}. The argument order is {\tt QF}$[i,j,k,l]=Q_{\text{F}lk}^{ij}$ and analogously for {\tt QD}.\\
		{\tt $\rho$}, {\tt $\rho$d}, {\tt $\rho$t}, {\tt $\rho$td} &5& These abstract symbols with input {\tt func}$[\fs,\ff,\ff]$ realise the scalar-fermion interactions in \eqref{eq:Lagrangian_Feynman_rules_3} with $\rho\hat{=}\rho$, $\rho${\tt d}$\hat{=}\rho^\dagger$, $\rho${\tt t}$\hat{=}\tilde\rho$, $\rho${\tt td}$\hat{=}\tilde{\rho}^\dagger$ and canonical input, e.g.\ {\tt $\rho[$j$,$A$,$B$]=(\rho^j)^{AB}$}.\\
		{\tt rulesMomentaL1}, {\tt rulesMomentaL2}&2& These functions with argument {\tt func}$[\expr]$ replace contractions of irreducible one- and two-loop momenta by squares of irreducible momenta, e.g.\ the one-loop replacement $p_1\cdot p_2=\frac 12(p^2-p_1^2-p_2^2)$.\\
		{\tt rulesOpenIndices$[\expr]$}&2& This function rewrites the two-loop irreducible momenta $p_3^\mu$, $p_4^\mu$, and $p_5^\mu$ in terms of $p_0^\mu$, $p_1^\mu$, and $p_2^\mu$.\\
		{\tt $\sigma[\muindex][\alphaindex,\betadcindex]$} &5& This abstract symbol represents a general four-dimensional Minkowski space $\sigma$-matrix in the Weyl-representation, see \eqref{eq:sigma_matrices} for our explicit realisation.\\
	\end{tabularx}
\end{table}
\begin{table}[H]
	\small
	\centering
	\begin{tabularx}{\textwidth}{|c|c|X|}
		symbol &file& usage\\
		\hline
		{\tt $\sigma$b$[\muindex][\betadcindex,\alphaindex]$} &5& This abstract symbol represents a general four-dimensional Minkowski space $\bar\sigma$-matrix in the Weyl-representation, see \eqref{eq:sigma_matrices} for our explicit realisation.\\
		{\tt $\Sigma[\fs][\ff,\ff]$} &5& This abstract symbol represents a general six-dimensional Euclidean space $\Sigma$-matrix in the Weyl-representation, see \eqref{eq:Sigma_6} for our explicit realisation.\\
		{\tt $\Sigma$b$[\fs][\ff,\ff]$} &5& This abstract symbol represents a general six-dimensional Euclidean space $\bar\Sigma$-matrix in the Weyl-representation, see \eqref{eq:Sigma_6} for our explicit realisation.\\
		{\tt $\sigma$toOpenTrace$[\expr]$} &5& This function replaces the abstract $\sigma$ and $\bar{\sigma}$ symbols by the structure {\tt T}$\sigma$. In addition, the abstract $\Sigma$ and $\bar{\Sigma}$ symbols are replaced with the respective tensors {\tt T}$\Sigma$ and {\tt T}$\bar{\Sigma}$.\\
		{\tt SolveIntegrals$[\expr][$l$\_]$} &4& This function solves all {\tt l}$\_\in\{1,2\}$ loop integrals \expr that were generated by {\tt GenerateIntegrands$[\expr]$}.\\
		{\tt spacetimetensors$[\expr]$} &4& This function evaluates traces over \spl{2,\CC} spinor indices. Such traces involve alternating products of $\sigma$ and $\bar{\sigma}$ matrices and this function gives explicit results up to length eight.\\
		\ttt{T}$\epsilon$\ttt{4Ed}, \ttt{T}$\epsilon$\ttt{4Eu} & 3 & These two functions depend on \ttt{func}$[\_,\_,\_,\_]$ and realise the (unevaluated) totally antisymmetric symbol in four-dimensional Euclidean space. Note that also the properties displayed in \eqref{eq:identities_Sigma} are implemented.\\
		{\tt TFN4}, {\tt TF$\gamma$}, {\tt TFpm} &3& These tensors take input {\tt func}$[\fs,\fs,\fs,\fs]$ in analogy to {\tt F} and realise the explicit four-scalar F-term-type coupling of the undeformed, $\gamma_i$-deformed, and $\gamma_i^\pm$-deformed theory. Their definitions can be found in \eqref{eq:coupling_tensors_N4} and \eqref{eq:coupling_tensors_gammai}.\\
		{\tt ToCouplingTensors$\gamma$pm$[\expr]$}&2& This function replaces the abstract flavour couplings by the explicit coupling tensors of the $\gamma_i$-deformation with deformation angles $\gamma_i^\pm$.\\
		{\tt toFunction$[\expr]$} &6& This function evaluates the abstract {\tt GG} functions in terms of $\Gamma$-functions, provided that all arguments render finite expressions. The abstract {\tt GG} functions are introduced in {\tt intsolveL1} and  {\tt intsolveL2}.\\
		{\tt T$\rho$N4}, {\tt T$\rho$dN4}, {\tt T$\rho$tN4}, {\tt T$\rho$tdN4} &3& These tensors take input {\tt func}$[\fs,\ff,\ff]$ in analogy to the abstract symbols $\rho$, etc. and realise the explicit scalar-fermion interaction in the \NfSYMt. Their definitions can be found in \eqref{eq:coupling_tensors_N4}.\\
		{\tt T$\rho\gamma$}, {\tt T$\rho$d$\gamma$}, {\tt T$\rho$t$\gamma$}, {\tt T$\rho$td$\gamma$}  &3& These tensors take input {\tt func}$[\fs,\ff,\ff]$ in analogy to the abstract symbols $\rho$, etc. and realise the explicit scalar-fermion interaction in the $\gamma_i$-deformation with deformation angles $\gamma_i$. Their definitions can be found in \eqref{eq:coupling_tensors_gammai}.\\
		{\tt T$\rho$pm}, {\tt T$\rho$dpm}, {\tt T$\rho$tpm}, {\tt T$\rho$tdpm}  &3& These tensors take input {\tt func}$[\fs,\ff,\ff]$ in analogy to the abstract symbols $\rho$, etc. and realise the explicit scalar-fermion interaction in the $\gamma_i$-deformation with deformation angles $\gamma_i^\pm$. Their definitions can be found in \eqref{eq:coupling_tensors_gammai} and the deformation angles are given in \eqref{eq:gamma_pm}.\\
		{\tt TQhat}, {\tt TQhatpm} && These tensors take input {\tt func}$[\fs,\fs,\fs,\fs]$ and realise the explicit four-scalar coupling {\tt TQhat$[$i$,$j$,$k$,$l$]=\hat{Q}^{ij}_{lk}$} given in \eqref{eq:F_to_Qhat}. They give the coupling with deformation angles $\gamma_i$ and $\gamma_i^\pm$, respectively.\\
	\end{tabularx}
\end{table}
\begin{table}[H]
	\small
	\centering
	\begin{tabularx}{\textwidth}{|c|c|X|}
		symbol &file& usage\\
		\hline
		{\tt T$\sigma[$array$\_][$type$\_,\_,\_]$} &3& This structure encodes alternating products of $(\sigma_\mu)_{\alpha\dot{\beta}}$ and $(\bar{\sigma}_\nu)^{\dot{\beta}\gamma}$ matrices with spacetime indices encoded from left to right in {\tt array}$\_$. The variable {\tt type}$\_\in\{\delta\delta,\delta\delta$d$, \sigma, \sigma$b$\}$ characterises if the number of $\sigma$ and $\bar{\sigma}$ matrices is equal (first two) and with which type of matrix the product starts (first and third start with $\sigma$). Depending on the variable {\tt type}$\_$, the last two entries are either of the pairs $\{(\alphaindex,\alphaindex),(\betadcindex,\betadcindex),(\alphaindex,\betadcindex),(\betadcindex,\alphaindex)\}$. They characterise the left and right open spinor indices in the product.\\
		{\tt T$\Sigma$}, {\tt T$\Sigma$b} &3& These tensors take the input {\tt func}$[\fs,\ff,\ff]$ and realise the \NfSYMt $\Sigma$ identities given in \eqref{eq:identities_Sigma}.\\
		\hline
	\end{tabularx}
\end{table}

\section{Dimensional renormalisation schemes}\label{sec:Renormalisation_schemes}
In this appendix, we present the so-called $\text{MS}$, $\ol{\text{MS}}$, $\text{DR}$, and $\ol{\text{DR}}$ renormalisation schemes\footnote{These schemes were defined in \cite{'tHooft:1973mm}, \cite{breitenlohner1977,Bardeen:1978yd,Mertig:1995ny,Vogelsang:1995vh}, and \cite{Siegel:1979wq}, respectively. The $\ol{\text{DR}}$-scheme is defined analogously to the $\ol{\text{MS}}$-scheme.}. All these schemes regularise divergent contributions by the analytic continuation of the integer spacetime dimension $d$ to some real dimension $D=d-2\epsilon$, so that divergences are encoded in terms of the regulator $\epsilon$. The difference between the schemes occurs in the treatment of spinor structures and in the subtraction procedure that erases the occurring divergences from renormalised contributions. We will call a renormalisation scheme a procedure that contains both, a regularisation and a subtraction procedure. Such a scheme relates the bare action which depends on bare quantities $q_{\text{B}}$ (sources and fields) to a renormalised action in which all bare quantities are replaced renormalised ones and respective 1PI renormalisation constants $q_{\text{B}}=Z_q q$. The renormalisation constants contain the local counterterms\footnote{The locality of counterterms was proven in \cite{PhysRev.118.838}. See also \cite{hahn1968} for a refined version of the proof.} which subtract divergent parts from bare contributions. The general discussion is inspired by \cite[\chap{14,27}]{Srednicki:2007} and \cite{Collins:1984xc} and we refer the reader there for further details. 

Generically, expanding a non-renormalised interacting \QFT around its free solution, as done in \eqref{eq:Zj_perturbative}, yields divergent perturbative contributions. Using the Feynman rules of \appref{app:Feynman_rules}, these contributions which depend on external momenta $p_i$ and spacetime indices $\mu_{j}$ can be expressed in terms of an appropriate tensor structures\footnote{Here we only give the spacetime indices explicitly and suppress all remaining structures.} $T_{\nu_1\dots\nu_{\text{m}}}$ contracted with a spacetime tensor-integral $I^{\nu_1\dots\nu_{\text{m}}\mu_1\dots\mu_{\text{m}}}$. In $d$-dimensional Minkowski momentum space the tensor integral takes the form
\begin{equation}\label{eq:Feynman_integrals}
I^{\nu_1\dots\nu_{\text{m}}\mu_1\dots\mu_{\text{m}}}(\{p_i\})=
\int\left(\prod_{l=1}^{L}\frac{\de^{d}k_l}{(2\pi)^{d}}\right)\frac{N^{\nu_1\dots\nu_{\text{m}}\mu_1\dots\mu_{\text{m}}}(\{P_n\})}{\prod_{n}P_n^{2\alpha_n}}\eqncom
\end{equation}
where $P_n=P_n(\{k_l,p_i\})$ is a set of irreducible momenta, $k_l$ are the $L$ loop momenta and the numerator $N$ is a polynomial of momentum space tensors\footnote{We assume that the Fourier transform of all expressions can unambiguously be defined. We choose to work in momentum space, to avoid technical complications. For a modern discussion of Epstein-Glaser renormalisation in position space see e.g.\ \cite{Gracia-Bondia:2014zwa}.}. Depending on the integral measure and the occurring integrands, this integral may develop UV and/or IR divergences\footnote{For the identification of UV and IR divergences in concrete integrals see the discussion in \appref{sec:UV_and_IR_div}.} when irreducible momenta approach $\abs{P_n}\rightarrow\infty$ and/or $\abs{P_n}\rightarrow 0$, respectively. When these UV and IR divergences do not cancel among themselves respectively in a physical process, they encode that different shortcomings of our approach. Persistent IR divergences, on the one hand, occur when some contributions to a process are not yet accounted for, \cf \cite[\chap{26 -- 27}]{Srednicki:2007} for a detailed description of possible issues. UV divergences, on the other hand, have two possible explanations. Either, the theory only describes physical processes accurately at low energies and needs an ultraviolet completion to be valid at all energy scales, or they are part of the perturbative approach and we can define a renormalisation scheme that eliminates the UV divergent contributions in all observables. The second scenario is what we have for \NfSYMt and its deformations in flat Minkowski space. For these theories, we can fix renormalisation scheme at some energy scale $\mu$ and then determine the dependence of all couplings on $\mu$ via the renormalisation group flow.

The regularisation procedure makes formally divergent integral contributions finite by altering the theory using some type of regulator. For example, the integral \eqref{eq:G_function} may be divergent if all its parameters are integer valued but it becomes finite if some of the parameters are shifted by a real regulator $0<\epsilon\ll 1$. It is advantageous to choose a regulator that spoils as few symmetries of the original theory as possible, since this allows to adopt symmetry-based simplifications like the Slavnov-Taylor identities discussed in \subsecref{sec:N4SYM_renormalisation} in the regularised theory. Symmetries that are broken by the regulator lead to spurious finite contributions that have to be subtracted by the counterterms in the subtraction procedure in order to restore the symmetry for the renormalised theory \cite{'tHooft:1971fh,Stockinger:2006bi,Stockinger:2005gx}. For example, the renormalisation schemes that we discuss in this appendix do not preserve SUSY in all possible correlation functions and hence finite SUSY-restoring counterterms have to be added in the corresponding processes, see e.g.\ \cite{Stockinger:2006bi,Stockinger:2005gx} for details.

\subsection{Dimensional regularisation}\label{subsec:Regularisation}
The dimensional regularisation procedure is widely used to define regularised theories since it preserves most of the symmetries of quantum chromo dynamics \cite{breitenlohner1977}. Here, we summarise the findings of \cite{breitenlohner1977,Avdeev1983262,AVDEEV1981272,Stockinger:2005gx} with a focus on obstacles in the context of SUSY. For a detailed introduction to dimensional regularisation we refer to the literature, e.g.\ \cite[\chap{4}]{Collins:1984xc}.

It is possible to regularise divergent integrals by altering the measure of loop integrals, in particular the dimension $d\rightarrow D=d-2\epsilon$ of the integration space. In this dimensional regularisation procedure \cite{'tHooft:1972fi} the regularised theory lives in $D\in\RR$ dimensions, while the original theory lives in $d\in\NN$ dimensions. Divergent contributions in the latter are encoded in inverse powers of $0<\epsilon\ll 1$. In the integrals \eqref{eq:Feynman_integrals}, the measure is altered according to
\begin{equation}
\de\Omega_{d}\de r\, r^{d-1}\rightarrow
\de\Omega_{D}\de r\,r^{D-1}\eqncom
\end{equation}
where $\de\Omega_{d}$ gives the integration over a $d$-dimensional unit sphere. For scalar Feynman integrals $I(\{p_i\})$, this procedure is well defined and many techniques are available for the evaluation of regularised integrals, see e.g.\ \cite{Collins:1984xc,Smirnov:2004ym}. The parameters $\lambda^{(d)}_{\text{B}\,i}$ of the unregulated $d$-dimensional theory have classical scaling dimensions $\Delta^{d}_{\lambda_i}$ which were defined in \eqref{eq:dimension_source} and ensure that the corresponding action is dimensionless. In the regulated theory, the dimension changes to $D$ and hence the scaling dimensions of fields and couplings in principle must also change, so that the action may remain dimensionless. However, since we want to keep the scaling dimensions of parameters in $D$ dimensions fixed to the respective values in $d$ dimensions, we introduce a regularisation scale $\mu$ which absorbs the difference as
\begin{equation}\label{eq:mu_dependence_of_g}
\lambda^{(d)}_{\text{B}\,i}\rightarrow \lambda^{(D)}_{\text{B}\,i}= \mu^{\bigl(\Delta^{D}_{\lambda_i}-\Delta^{d}_{\lambda_i}\bigr)}\lambda_{\text{B}\,i}^{(d)}\eqndot
\end{equation}

For a consistent definition of the complete regularised theory, also all Lorentz covariants in $d$ dimensions must be generalised to Lorentz covariants in non-integer dimensions $D$, so that all tensor integrals $\hat{I}^{\nu_1\dots\mu_n}$ and the accompanying tensors $\hat{T}_{\nu_1\dots\nu_m}$ are well defined. In particular, the spinor representation and along with it $\gamma$-matrices and the representation of fermions must be generalised, see \appref{sec:spinor_in_various_dimensions} for details on the interdependences. An explicit discussion how this may be achieved was presented in \cite{breitenlohner1977}. From a conceptual point of view \cite{Avdeev1983262,AVDEEV1981272}, the $d$-dimensional space is promoted to a quasi-$d$-dimensional space ($\text{Q}_{d}\text{S}$) which is formally infinite-dimensional and has no antisymmetric tensor $\varepsilon_{\mu_1\mu_2\dots\mu_{d}}$. This space still retains the properties $\eta_{\mu\nu}\eta^{\mu\nu}=d$ for the metric and $\tr(\one)=2^{d/2}$ for the dimensionality of the $\gamma$-matrices from the four-dimensional space, see also \cite{Stockinger:2005gx}. It can be written as the direct sum $\text{Q}_{d}\text{S}=\text{Q}_{D}\text{S}\oplus\text{Q}_{2\epsilon}\text{S}$, so that the $d$-dimensional metric decomposes into 
\begin{equation}\label{eq:eta_in_dimreg}
\eta_{\mu\nu}=\hat{\eta}_{\mu\nu}+\tilde{\eta}_{\mu\nu}
\end{equation} 
on the $D$- and the $2\epsilon$-dimensional quasi-subspaces, respectively. The metrics on the quasi-subspaces have the projector properties
\begin{equation}
\begin{aligned}\label{eq:dim_reg_projectors}
\hat\eta_{\mu\nu}\hat\eta^\nu_{\phan{\mu}\rho}
&=\hat{\eta}_{\mu\rho}\eqncom\qquad
\tilde\eta_{\mu\nu}\tilde\eta^\nu_{\phan{\mu}\rho}
=\tilde{\eta}_{\mu\rho}\eqncom\qquad
\hat\eta_{\mu\nu}\tilde\eta^\nu_{\phan{\mu}\rho}
=0\eqncom\qquad
\tilde\eta^\mu_{\phan{\mu}\mu}
&=d-D\eqncom\qquad
\hat\eta^\mu_{\phan{\mu}\mu}
=D
\end{aligned}
\end{equation}
and can be used to construct all Lorentz covariants in the appropriate subspaces, e.g.\ $\hat{\gamma}_\mu=\hat{\eta}_{\mu\nu}\gamma^\nu$. In particular, the Clifford algebra defined by $\{\gamma_\mu,\gamma_\nu\}=-2\eta_{\mu\nu}\one$ induces analogous relations in $\text{Q}_{D}\text{S}$ and $\text{Q}_{2\epsilon}\text{S}$.

This construction is not possible for objects whose properties manifestly require integer dimensions like Fierz identities, the $d$-dimensional Levi-Civita tensor $\varepsilon_{\mu_1\dots\mu_{d}}$ and $\gamma_{d+1}$. The Fierz identities can be abandoned in the regularised theory, as we do not explicitly use them in the construction of the theory. For the Levi-Civita tensor, which may appear in Feynman diagrams, we wish to define a $D$-dimensional analogon. Following \cite{breitenlohner1977}, we start with the $d$-dimensional identity
\begin{equation}\label{eq:Levi_Civita_identity}
\varepsilon_{\mu_1\dots\mu_{d}}\varepsilon_{\nu_1\dots\nu_{d}}=
\begin{vmatrix}
\eta_{\mu_1\nu_1}& \cdots&\eta_{\mu_1\nu_{d}}\\
\vdots&&\vdots\\
\eta_{\mu_{d}\nu_1}& \cdots&\eta_{\mu_{d}\nu_{d}}
\end{vmatrix}\eqncom
\end{equation}
replace the metric tensors on the \rhs according to \eqref{eq:eta_in_dimreg} and take this equation as the defining identity of the $D$-dimensional Levi-Civita tensor. We also like to generalise $\gamma_{d+1}$ to a symbol $\hat{\gamma}_{D+1}$, since we explicitly used it in the construction of Weyl fermions in our theories in \chapref{chap:The_models}. For the generalisation, note that in $d$-dimensional $\gamma_{d+1}$ can be fixed entirely via the two relations\footnote{To derive the first identity, note that $\gamma_{d+1}$ may be written as $\gamma_{d+1}=\complexi^{\floor{\frac d2}}\frac{1}{d!}\varepsilon_{12\dots d}\varepsilon^{\nu_1\dots\nu_d}\gamma_{\nu_1}\dots\gamma_{\nu_d}$, in the notation of \appref{sec:Clifford_algebra_construction}.}
\begin{align}
\tr\bigl(\gamma_{d+1}\gamma_{\mu_1}\dots\gamma_{\mu_{d}}\bigr)&=c^*\varepsilon_{\mu_1\dots\mu_{d}}
\tr\bigl(\onee{2^{d/2}}\bigr)\eqncom\label{eq:trace_relation}\\
\{\gamma_{d+1},\gamma_\mu\}&=0\eqncom\label{eq:anti_commutativity}
\end{align}
where the constant $c^*$ depends on the definitions of the $d$-dimensional Clifford algebra, compare \appref{sec:Clifford_algebra_construction}. In the $D$-dimensional generalisation, these relations are not mutually compatible any more. If we want to keep a notion of Weyl fermions in the $D$-dimensional theory, we must keep the anti-commutativity relation \eqref{eq:anti_commutativity}. However, enforcing this relation implies for a trace involving $\hat\gamma_{D+1}$ and $(d+2)$ $\hat{\gamma}$-matrices\footnote{This identity is obtained by commuting $\hat{\gamma}_\alpha$ around the trace $\hat{\eta}^{\alpha\beta}\tr\bigl[\hat{\gamma}_\alpha\hat{\gamma}_\beta\hat{\gamma}_{D+1}\hat{\gamma}_{\mu_1}\dots\hat{\gamma}_{\mu_{d}}\bigr]$, with $\mu_i\neq\mu_j$ and the Clifford algebra relation $\{\hat{\gamma}_\mu,\hat{\gamma}_\nu\}=-2\hat{\eta}_{\mu\nu}\one$.}
\begin{equation}
0=-2(D-d)\tr\bigl(\hat\gamma_{D+1}\hat\gamma_{\mu_1}\dots\hat\gamma_{\mu_{d}}\bigr)\eqndot
\end{equation}
Since $D-d=-2\epsilon$ is non-vanishing, the trace must vanish for this choice of $\hat{\gamma}_{D+1}$ and therefore we cannot construct a smooth limit $D\rightarrow d$ to the trace relation \eqref{eq:trace_relation}. From the algebraic point of view, some sort of obstacle is to be expected since Weyl fermions only exist in even dimensions, see \appref{sec:spinor_in_various_dimensions} and the explicit construction in \appref{sec:Clifford_algebra_construction}. The second possibility is to keep the trace relation. This implicitly forces $\hat{\gamma}_{D+1}$ back to the $d$-dimensional one and we define
\begin{equation}
\hat{\gamma}_{D+1}=\gamma_{d+1}=\complexi^{\floor{\frac{d}{2}}}\gamma_1\gamma_2\dots \gamma_{d}\eqndot
\end{equation}
For the anti-commutativity relation \eqref{eq:anti_commutativity} this choice induces\footnote{The sign difference compared to the relation in \cite{breitenlohner1977} originates from a relative sign in the definition $\tilde{\eta}_\mu^{\phan{\mu}\mu}=d-D$.}
\begin{equation}
\{\gamma_{d+1},\hat{\gamma}_\mu\}=\{\gamma_{d+1},\gamma_\mu-\tilde{\gamma}_\mu\}=-2\gamma_{d+1}\tilde{\gamma}_\mu\eqncom
\end{equation}
with $\tilde{\gamma}_\mu\in\text{Q}_{2\epsilon}\text{S}$ as before and we used that objects in different subspaces commute \cite{breitenlohner1977}. Note that the \rhs encodes the breaking of the Weyl symmetry. This can be seen explicitly by writing the action $\mathcal{L}=\complexi\ol{\psi} \Gamma^\mu \D_\mu P_-^2\psi$ in terms of left-handed Weyl spinors $\psi^{-}=P_- \psi=\frac 12(\one-\gamma_{d+1})\psi$, once in $d\in 2\NN$ and once in $D\in \RR$ dimensions:
\begin{equation}
\begin{aligned}\label{eq:example_Weyl_breaking}
\mathcal{L}_{d}&=\complexi\ol{\psi}\, \gamma^\mu P_-^2 \D_\mu\psi=\complexi\ol{\psi^-}\, \gamma^\mu \D_\mu \psi^-\eqncom\\
\mathcal{L}_{D}&=\complexi\ol{\psi}\, \hat{\gamma}^\mu P_-^2 \D_\mu\psi=\complexi\ol{\psi^-}\, \hat{\gamma}^\mu \D_\mu \psi^-
+2\complexi\ol{\psi}\, \gamma_{d+1}\tilde\gamma^\mu \D_\mu \psi\eqndot
\end{aligned}
\end{equation}
So, we see an explicit example of a regularised theory which does not preserve the symmetries of the original one. As a consequence, additional finite renormalisation constants must be included in the renormalisation procedure to restore the left-handedness of fermions in the renormalised theory.

\subsection{Dimensional reduction}
The purpose of the dimensional reduction procedure is to adapt the concept of dimensional regularisation, so that it manifestly preserves supersymmetry in addition to Lorentz and gauge symmetry. For a general introduction to dimensional reduction\footnote{For an explicit comparison of the $\ol{\text{MS}}$ and $\ol{\text{DR}}$ scheme in QCD amplitude calculations see \cite{Bern:2002zk,Kunszt:1993sd}.} we also refer to \cite{Avdeev1983262,AVDEEV1981272,Jack:1997sr,Stockinger:2005gx,Stoeckinger2006250}. We adopt all notational conventions from the previous subsection and specialise to the $(d=4)$-dimensional case.

In \cite{SIEGEL1979193}, dimensional regularisation was slightly modified to give the dimensional reduction procedure which should allow for a manifestly SUSY invariant regularisation of theories. Dimensional regularisation is not SUSY invariant, since it treats bosonic and fermionic degrees of freedom manifestly different, as exemplified in the previous subsection. The idea in \cite{SIEGEL1979193} was to keep all tensorial structures in exact $d=4$ dimensions while analytically continuing momenta and positions to $D=4-2\epsilon\in\RR$ dimensions and define suitable projectors \eqref{eq:dim_reg_projectors} so that all products of tensors in $d$ and $D$ dimensions are well defined. In this construction, supersymmetry is naively preserved since all structures on which it relies are not touched by the regularisation. However, it was shown in \cite{Siegel:1980qs,AVDEEV1982317,Avdeev1983262} that the original dimensional reduction procedure leads to inconsistencies. Very prominently this can be seen by taking the $d$-dimensional identity \eqref{eq:Levi_Civita_identity} and projecting it to the subspace combinations $\tilde{\varepsilon}^{\mu_1\dots\mu_4}\tilde{\varepsilon}_{\mu_1\dots\mu_4}$, $\hat{\varepsilon}^{\mu_1\dots\mu_4}\hat{\varepsilon}_{\mu_1\dots\mu_4}$ and $\hat{\varepsilon}^{\mu_1\dots\mu_4}\tilde{\varepsilon}_{\nu_1\dots\nu_4}$. Upon building combinations of the results one arrives at $0=D(D-1)^2(D-2)^2(D-3)^2(D-4)$, which forces the dimensional continuation in $D=4-2\epsilon$ back to $D=d$ dimensions, see \cite{AVDEEV1982317} for details.

It is possible to define a dimensional reduction scheme without introducing mathematical inconsistencies \cite{Stockinger:2005gx}. In fact, the inconsistencies in the dimensional reduction procedure can be traced back to enforcing the non-mutually compatible relations \eqref{eq:trace_relation} and \eqref{eq:anti_commutativity} or to the application of truly four-dimensional identities (like Fierz identities) in the dimensionally reduced theory. Therefore, we may use dimensional reduction without introducing inconsistencies by applying the rules: 
\begin{itemize}
	\item Momenta within Feynman integrals are kept in $d=4$ dimensions.
	\item The integration measure in Feynman integrals is altered to $D=4-2\epsilon$ dimensions. For the evaluation of these integrals see \appref{app:Evaluating_Feynman_integrals}.
	\item All remaining tensors, $\gamma$-matrices, etc.\ are kept in $d=4$ dimensions.
	\item Abandon Fierz identities in the regularised theory.
	\item Abandon \eqref{eq:Levi_Civita_identity} in the regularised theory.
	\item Use the projectors \eqref{eq:dim_reg_projectors} to contract indices between $d$- and $D$-dimensional objects.
	\item Index counting is not possible in the regularised theory. For example, the indices $\mu_1,\dots,\mu_5$ may all be different, since we work in $\text{Q}_4\text{S}$. Only in truly four dimensions at least two of the indices are equal.
\end{itemize}
For $\hat{\gamma}_5$ use one of the two options:
\begin{itemize}
	\item either $\hat{\gamma}_5$ is completely antisymmetric: $\{\hat{\gamma}_5,\gamma_\mu\}=0$. This implies that the trace of any number of $\hat{\gamma}_\mu$ matrices with one $\hat{\gamma}_5$ vanishes,	
	\item or $\hat{\gamma}_5=\gamma_5$ is the four-dimensional one. This implies that the anticommutator yields $\{\hat\gamma_5,\hat\gamma_\mu\}=-2\gamma_5\tilde{\gamma}_\mu$.
\end{itemize}

While these rules lead to a consistently defined regularised theory, it is not guaranteed that this theory preserves supersymmetry. In fact, if the four-dimensional $\gamma_5$ is chosen, as in \eqref{eq:example_Weyl_breaking}, supersymmetry is broken already at the one-loop level \cite{Stockinger:2005gx}. Therefore, additional counterterms must be included in the renormalisation procedure to restore the symmetries falsely broken by the regularisation. If the first option with a completely antisymmetric $\hat{\gamma}_5$ is chosen, it is not immediately clear if the regularisation procedure preserves supersymmetry. In many low-loop checks like \cite{Capper1980479} and the corresponding references within \cite{Jack:1997sr,Stockinger:2006bi}, this scheme turned out to preserve supersymmetry. However, from the construction of $\hat{\gamma}_5$ it is clear that results in the regularised theory may differ from the unregulated theory when we combine four or more $\hat{\gamma}$-matrices with $\hat{\gamma}_5$ in a trace. Four-fermion operators in Feynman diagrams may for example induces such traces in physical processes. Whether SUSY breaking terms do appear in a given physical process can be investigated using the so-called quantum action principle presented in \cite{Stockinger:2005gx}.

\subsection{The subtraction procedure and renormalisation schemes}\label{subsec:The_renormalisation_procedure}

After the regularisation, former divergences of integrals appear as terms that become infinite when the regulator $\epsilon$ is removed. Those terms are removed by introducing a type of subtraction procedure. This subtraction renders the regularised theory finite in the limit $D\rightarrow d$ where we remove $\epsilon$ to obtain the renormalised theory in $d$ dimensions. As mentioned in \subsecref{subsec:Regularisation}, the subtraction procedure must also eliminate finite symmetry-violating terms that arise if a symmetry-violating regularisation procedure was chosen. The determination of such terms is in general complicated, since the finite parts of Feynman diagrams are far less restricted than the divergent contributions. Only after the subtraction procedure is completed and the theory is transformed to the renormalised theory in $d$-dimensional Minkowski space, the abandoned Fierz identities and \eqref{eq:Levi_Civita_identity} can be used again.

Apart from the requirements mentioned in the previous paragraph, the choice of $\Lambda$ which defines the subtraction procedure is not unique. The minimal subtraction procedure only eliminates the occurring divergences in $\epsilon$. Combining this subtraction with the  dimensional regularisation or dimensional reduction procedure gives the minimal subtraction (MS) scheme \cite{'tHooft:1973mm} or the dimensional reduction (DR) scheme, respectively. If in addition to the divergences, factors of $c_{\ol{\text{MS}}}=\log 4\pi-\gammaE$ which arise in the expansion of $\Gamma$-functions are subtracted, the renormalisation schemes are called modified minimal subtraction ($\ol{\text{MS}}$) scheme \cite{Bardeen:1978yd} or modified dimensional reduction ($\ol{\text{DR}}$) scheme.

More bluntly, we could assume that our theory is only valid in some low energy regime and incorporate this in the calculation by subtracting all contributions from energies higher than a chosen cutoff scale $\Lambda_{\text{max}}$. 

Finally, in the kinematic subtraction procedure, we enforce that $n$-point functions do not receive any perturbative corrections at a chosen energy scale $\mu$. For a divergent integral $I(p^2)$ depending on an external scale $p^2$ this procedure amounts to subtracting the same integral evaluated at scale $\mu$. Hence, the divergent integral is replaced by $I(p^2)-I(\mu)$, which clearly vanishes at $p^2=\mu$. As the overall divergence of our integrals is independent of the external kinematics, it is also clear that this renormalisation procedure is finite when the regulator is removed.

\section{Evaluating Feynman integrals}\label{app:Evaluating_Feynman_integrals}
In this appendix, we discuss the techniques that are necessary to evaluate integrals over $D$-dimensional Minkowski space perturbatively. In general, we will Wick rotate -- i.e.\ analytically continue in the time direction -- a given Minkowski space integral so that we arrive at a Euclidean space integral. The concepts needed to do this are well presented in \cite[\chap{14}]{Srednicki:2007} and we refer the reader there for further details. The one-loop Euclidean space integrals can then be solved using the results of \cite{Vladimirov:1979zm,Chetyrkin:1980pr}. For higher-loop integration by parts (IBP) techniques can be used to significantly reduce the complexity of a given integral. However, as we do not explicitly need the IBP relations in this thesis, we refer the reader to \cite{Chetyrkin:1981qh} for a comprehensive discussion of the needed concepts. For the extraction of divergent contributions of integrals it may become necessary to renormalise IR divergences and we briefly introduce the idea behind it following the presentation of \cite{kleinert2001critical}.

\subsection{Wick rotation and Euclidean space integrals}\label{subsec:Wick_rotation}
First, we transform a given one-loop integral from Minkowski to Euclidean space. The generalisation to higher loops then follows immediately.

The one-loop Minkowski space integrals that we wish to transform have the form\footnote{As we are interested in models with massless fields, masses $m_i$ are absent in the denominator.}
\begin{equation}\label{eq:example_integral}
\int\frac{\de^Dl}{(2\pi)^D}\frac{P(\{l^\mu\})}{(l^{2}-\complexi \epsilon)\prod_{i=1}^{n-1}((p_i-l)^{2}-\complexi \epsilon)}\eqncom
\end{equation}
where the $p_i$ are external momenta, $P$ is a finite polynomial in the loop momentum. The infinitesimal parameter $\epsilon$ appears since we work with Feynman propagators and the position of the poles of the integrand fixes the integration contour in the complex plane. Higher loop integrals are concatenations of this structure and conceptually they can be treated analogously. 
We will now use Feynman's parametrisation formula to rewrite the denominator of the integrand such that we can analytically continue\footnote{For the analytic continuation we must not move our integration contour over any singularities. As the numerator does not introduce any new poles, we can ignore it for this part.} the time direction of the loop integral. Then we reverse the parametrisation procedure to obtain the Euclidean space version of \eqref{eq:example_integral}, which can be solved by the means of \cite{Vladimirov:1979zm,Chetyrkin:1980pr,Chetyrkin:1981qh}. We use the integration measure for Feynman parameters
\begin{equation}
\int\de F_n=(n-1)!\int_0^1\de x_1\dots\de x_n\delta^{(n)}(x_1+\dots+x_n-1)
\end{equation}
to rewrite the denominator of \eqref{eq:example_integral} as
\begin{equation}
\begin{aligned}\label{eq:Feynman_parameterised}
\frac{1}{l^{2}-\complexi \epsilon}\prod_{i=1}^{n-1}\frac{1}{(p_i-l)^{2}-\complexi \epsilon}&=
(n-1)!\int_0^1\de x_1\dots\de x_{n-1}
\Biggl[l^2+\sum_{i=1}^{n-1}x_i\bigl((p_i-l)^2-l^2\bigr)-\complexi\epsilon\Biggr]^{-n}\\
&=
(n-1)!\int_0^1\de x_1\dots\de x_{n-1}
\left[q^2+C-\complexi\epsilon\right]^{-n}\eqndot
\end{aligned}
\end{equation}
In the second line, we introduced the shifted loop momentum $q=l-\sum_{i=1}^{n-1}x_ip_i$ and the loop-momentum independent term $C=\sum_{i=1}^{n-1}x_i p_i\bigl(p_i-\sum_{j=1}^{n-1}x_jp_j\bigr)$. Note that the shift to the integration variable $q$ does not change the integral measure in \eqref{eq:example_integral}. From the last equality in \eqref{eq:Feynman_parameterised} we see that the integrand has poles in the $q^0$-plane for $q^0=\pm (\sqrt{|\vec{q}|^2+C}-\complexi\tilde{\epsilon})$ for some $\tilde{\epsilon}$. When the integrand vanishes fast enough for $\abs{q^0}\rightarrow\infty$, the integral exists and we can analytically continue the $q^0$-axis counter-clockwise by $90^{\circ}$ since we do not encounter any poles in this Wick rotation. The Wick-rotated Euclidean space integral is obtained by substituting the Minkowski space variable $q$ by its Euclidean counterpart $\bar q$ as\footnote{Following \cite[\chap{9}]{Peskin:1995ev}, this choice of Wick rotating to Euclidean momentum space via $q^0\Rightarrow\complexi \bar q^d$ implies that the respective Wick rotation to Euclidean position space is obtained by replacing $x^0\Rightarrow\complexi \bar x^d$ and $x^j\Rightarrow\bar x^j$.} 
\begin{equation}
q^0\Rightarrow\complexi \bar q^d\eqncom\qquad
q^j\Rightarrow\bar q^j\eqncom \qquad
q^2\Rightarrow\bar q^2\eqncom\qquad
\de^Dq\Rightarrow\complexi \de^D\bar q\eqndot
\end{equation}
Under this Wick rotation, the integral \eqref{eq:example_integral} turns into
\begin{equation}
\begin{aligned}
\complexi(n-1)!\int_0^1\de x_1\dots\de x_{n-1}\int\frac{\de^D\bar q}{(2\pi)^D}\frac{P(\{\bar l^\mu\})}{\bar q^2 +\bar C}\eqncom
\end{aligned}
\end{equation}
where we set the infinitesimal parameter $\epsilon$ to zero as this does no longer interfere with the integration contour. Upon reversing the Feynman parametrisation\footnote{This is possible, as the parametrisation does not depend on the signature of the integration space.}, we find the Wick rotated integral
\begin{equation}\label{eq:wick_rotation_integral}
\text{WR}\Biggl[\int\frac{\de^Dl}{(2\pi)^D}\frac{P(\{l^\mu\})}{(l^{2}-\complexi \epsilon)\prod_{i=1}^{n-1}((p_i-l)^{2}-\complexi \epsilon)}\Biggr]=
\complexi\int\frac{\de^D\bar l}{(2\pi)^D}\frac{P(\{\bar l^\mu\})}{\bar l^{2}\prod_{i=1}^{n-1}(\bar p_i-\bar l)^{2}}\eqndot
\end{equation}

\subsection{Ultraviolet and infrared divergences in scalar integrals}\label{sec:UV_and_IR_div}
Having the Euclidean space integrals, the question is how to solve them. In this subsection we will restrict ourselves to scalar integrals, i.e.\ integrals with a trivial numerator polynomial. We will introduce a graphical notation for the integrals and discuss issues of definiteness as well as ultraviolet and infrared divergences.

The integral \eqref{eq:wick_rotation_integral} may be ill defined in a given dimension and for certain integrands. To circumvent this problem, we will work in dimensional regularisation introduced in \appref{sec:Renormalisation_schemes} with spacetime dimension $D=4-2\epsilon$, where the real parameter $\epsilon\ll 1$ regularises the occurring divergences of the integrals. Restricting ourselves to the so-called propagator integrals with a trivial numerator and only one external scale, i.e.\ $\bar p_i=0$ for $i\neq 1$, the integral \eqref{eq:wick_rotation_integral} is solved exactly by \eqref{eq:G_function}. This result is, however, not yet satisfactory as it is not clear what to make of the various poles that occur for particular combinations of $\alpha$, $\beta$, and $D$. Before we address this issue, let us introduce an integral diagram representation for the integral in \eqref{eq:G_function} with Euclidean $(d=4)$-dimensional integrand momenta $\bar{l}$ and an integration measure over $D$-dimensional Euclidean space. We choose the representation
\begin{equation}\label{eq:I_alpha_beta}
\hat{I}_{(\alpha,\beta)}(\bar p)\equiv
\settoheight{\eqoff}{$\times$}%
\setlength{\eqoff}{0.5\eqoff}%
\addtolength{\eqoff}{-2.75\unitlength}%
\raisebox{\eqoff}{
	\fmfframe(0,0)(-1,0){
		\begin{fmfchar*}(15,5.0)
		\fmfleft{vout}
		\fmfright{vin}
		\fmfforce{0.15 w, 0.5h}{v2}
		\fmfforce{0.85 w, 0.5h}{v1}
		\fmfv{decor.shape=circle,decor.filled=full,
			decor.size=2thick}{v1,v2}
		\fmf{plain}{v2,vout}
		\fmf{plain}{vin,v1}
		\fmf{phantom,left=0.6}{v1,v2,v1}
		\fmffreeze
		\fmfposition
		\fmfipath{p[]}
		\fmfipair{v[]}
		\fmfiset{p1}{vpath(__v1,__v2)}
		\fmfiset{p11}{subpath (0,length(p1)/2) of p1}
		\fmfiset{p12}{subpath (length(p1)/2,1) of p1}
		\fmfiv{label=$\scriptscriptstyle \alpha$,label.angle=-90,label.dist=2}{point length(p1)/2 of p1}
		\fmfiset{p2}{vpath(__v2,__v1)}
		\fmfiset{p21}{subpath (0,length(p2)/2) of p2}
		\fmfiset{p22}{subpath (length(p2)/2,1) of p2}
		\fmfiv{label=$\scriptscriptstyle \beta$,label.angle=90,label.dist=2}{point length(p2)/2 of p2}
		\fmfi{plain}{p1}
		\fmfi{plain}{p2}
		\end{fmfchar*}
	}
}
\equiv
\int\frac{\de^D\bar l}{(2\pi)^D}\frac{1}{\bar l^{2\alpha}(\bar p-\bar l)^{2\beta}}
=\int\frac{\de^D\bar l}{(2\pi)^D}\frac{1}{\bar l^{2\beta}(\bar p-\bar l)^{2\alpha}}\eqncom
\end{equation}
where each factor (propagator) in the integrand is represented by a line that connects two dots and is labelled by the propagator power, if this power is not one. Each external scale is represented by an open line segment and we enforce momentum conservation\footnote{Note that a diagram with $n$ external lines depends only on $n-1$ external momenta due to the momentum conservation.} at each vertex. 

For $\beta=0$, the integral becomes a tadpole-type integral which is independent of the external scale $\bar p$. While it is highly divergent from dimensional analysis, in the dimensional regularisation scheme it evaluates to\footnote{This is in fact advantageous, as tadpole-type diagrams must vanish in physical processes if the involved fields $f$ have vanishing vacuum expectation value, i.e.\ $\vacl f(x)\vac=0$. In the path integral formalism, the subtraction of tadpole-type diagrams in the renormalisation procedure is presented in \cite[\chap{9}]{Srednicki:2007} in detail.}
\begin{equation}\label{eq:tadpole_type_diagram}
\hat{I}_{(\alpha,0)}=
\settoheight{\eqoff}{$\times$}%
\setlength{\eqoff}{0.5\eqoff}%
\addtolength{\eqoff}{-2.75\unitlength}%
\raisebox{\eqoff}{
	\fmfframe(0,0)(-1,0){
		\begin{fmfchar*}(6,5.0)
		\fmftop{vout}
		\fmfbottom{vin}
		\fmf{plain,left}{vin,vout,vin}
		\fmffreeze
		\fmfposition
		\fmfiv{label=$\scriptscriptstyle \alpha$,l.angle=-90,l.dist=2}{vloc(__vin)}
		\end{fmfchar*}
	}
}
=
\int\frac{\de^D \bar l}{(2\pi)^{D}}\frac{1}{\bar l^{2\alpha}}
=0\eqndot
\end{equation}
The reason for this naive contradiction is that this integral is not well defined in four dimensions and as we analytically continue the spacetime dimension from four to $D$ dimensions, we introduce a choice how to evaluate the Gaussian integral in this spacetime. Following the prescription\footnote{Note that the spacetime dimension $2\omega$ in \cite{Capper:1974,RevModPhys:47849} must be complex. In contrast to this prescription, we chose to work with integrands that are analytically continued into the complex plane and with a real spacetime dimension $D$ in \eqref{eq:wick_rotation_integral}. However, both prescriptions are compatible.} of \cite{Capper:1974,RevModPhys:47849} to redefine the Gaussian integral and evaluating the integral afterwards, we find the above result.

In more general situations with $\alpha,\beta\neq 0$, the $G$ function in \eqref{eq:G_function} can directly be expanded in terms of the regulator $\epsilon$. Depending on the choices of $\alpha$ and $\beta$ this expansion may yield poles in $\epsilon$ which either originate from the IR ($\abs{\bar l}\rightarrow 0$) or the UV regime ($\abs{\bar l}\rightarrow\infty$). For the interpretation of the two types of divergences see \secref{sec:Renormalisation_schemes}. If we are interested in low energy phenomena, we can employ the methods of renormalisation \cite{PhysRev.118.838} to cancel the ultraviolet divergences. However, our regulator does not distinguish between UV and IR divergences and hence we have to identify the type of divergence for each integral to decide whether it has to cancel with other contributions or if we have to renormalise it in a UV renormalisation procedure. This identification can be done by counting powers of the loop momenta as we are in Euclidean space.

Let us discuss the simplest cases of UV and IR divergences to get acquainted with the calculatory method and the graphical notation of both divergences. Close to four dimensions with $D=4-2\epsilon$, the simplest UV divergent example of \eqref{eq:G_function} is the integral with $\alpha=1$ and $\beta=1$, given by
\begin{equation}
\settoheight{\eqoff}{$\times$}%
\setlength{\eqoff}{0.5\eqoff}%
\addtolength{\eqoff}{-2.75\unitlength}%
\raisebox{\eqoff}{
	\fmfframe(0,0)(-1,0){
		\begin{fmfchar*}(15,5.0)
		\fmfleft{vout}
		\fmfright{vin}
		\fmfforce{0.15 w, 0.5h}{v2}
		\fmfforce{0.85 w, 0.5h}{v1}
		\fmfv{decor.shape=circle,decor.filled=full,
			decor.size=2thick}{v1,v2}
		\fmf{plain}{v2,vout}
		\fmf{plain}{vin,v1}
		\fmf{plain,left=0.6}{v1,v2,v1}
		\fmffreeze
		\fmfposition
		\end{fmfchar*}
	}
}
=
\int\frac{\de^D \bar l}{(2\pi)^D}\frac{1}{\bar l^{2}(\bar p-\bar l)^2}
\eqndot
\end{equation}
As the loop momentum approaches zero, the integrand scales as $\abs{\bar l}^{1}\de \abs{\bar l}$ and as it approaches infinity it scales\footnote{This is of course only true if the external scale does not vanish.} as $\abs{l}^{-1}\de\abs{\bar l}$. Hence, the integral is divergent in the ultraviolet with the divergence 
\begin{equation}\label{eq:UV_divergence_example}
\Kop\bigl(
\settoheight{\eqoff}{$\times$}%
\setlength{\eqoff}{0.5\eqoff}%
\addtolength{\eqoff}{-2.75\unitlength}%
\raisebox{\eqoff}{
	\fmfframe(0,0)(-1,0){
		\begin{fmfchar*}(15,5.0)
		\fmfleft{vout}
		\fmfright{vin}
		\fmfforce{0.15 w, 0.5h}{v2}
		\fmfforce{0.85 w, 0.5h}{v1}
		\fmfv{decor.shape=circle,decor.filled=full,
			decor.size=2thick}{v1,v2}
		\fmf{plain}{v2,vout}
		\fmf{plain}{vin,v1}
		\fmf{plain,left=0.6}{v1,v2,v1}
		\fmffreeze
		\fmfposition
		\end{fmfchar*}
	}
}
\bigr)
\otimes
\settoheight{\eqoff}{$\times$}%
\setlength{\eqoff}{0.5\eqoff}%
\addtolength{\eqoff}{-2.75\unitlength}%
\raisebox{\eqoff}{
	\fmfframe(0,0)(-1,0){
		\begin{fmfchar*}(5.0,5.0)
		\fmfleft{vout}
		\fmfright{vin}
		\fmfforce{0.5 w, 0.5h}{v2}
		\fmfv{decor.shape=circle,decor.filled=full,
			decor.size=2thick}{v1}
		\fmf{plain}{vin,v1,vout}
		\fmffreeze
		\fmfposition
		\end{fmfchar*}
	}
}
\equiv
\Kop\left(\frac{G(1,1)}{(4\pi)^{2-\epsilon}}\right)\frac{1}{\bar p^{2\epsilon}}
=\frac{1}{(4\pi)^2\epsilon}\,\frac{1}{\bar p^{2\epsilon}}
\eqncom
\end{equation}
where the operator $\Kop$ extracts the divergence in the regulator $\epsilon$. As this UV divergence is generated by the whole loop integrand, the cograph to the right of $\otimes$ is given by shrinking the UV divergent (sub-)graph to a point. For further details on the procedure to isolate UV divergences in integral diagrams \cf \cite[\chap{11}]{kleinert2001critical}. On the integral level, the operator $\Kop(\cdot)$ gives the divergent part of its argument and $(\bar p)^{-2\epsilon}$ corresponds to the cograph. The simplest IR divergent example of \eqref{eq:G_function} is the integral with $\alpha=2$ and $\beta=1$ and is represented by the integral diagram
\begin{equation}\label{eq:IR_divergent_diag}
\settoheight{\eqoff}{$\times$}%
\setlength{\eqoff}{0.5\eqoff}%
\addtolength{\eqoff}{-2.75\unitlength}%
\raisebox{\eqoff}{
	\fmfframe(0,0)(-1,0){
		\begin{fmfchar*}(15,5.0)
		\fmfleft{vout}
		\fmfright{vin}
		\fmfforce{0.15 w, 0.5h}{v2}
		\fmfforce{0.85 w, 0.5h}{v1}
		\fmfv{decor.shape=circle,decor.filled=full,
			decor.size=2thick}{v1,v2}
		\fmf{plain}{v2,vout}
		\fmf{plain}{vin,v1}
		\fmf{plain,left=0.6}{v1,v2,v1}
		\fmffreeze
		\fmfposition
		\fmfipath{p[]}
		\fmfiset{p1}{vpath(__v1,__v2)}
		\fmfiv{label=$\scriptscriptstyle 2$,label.angle=-90,label.dist=2}{point length(p1)/2 of p1}
		\end{fmfchar*}
	}
}
\equiv
\int\frac{\de^D \bar l}{(2\pi)^D}\frac{1}{\bar l^{4}(\bar p-\bar l)^2}\eqndot
\end{equation}
As the loop momentum approaches zero the integrand scales as $\abs{\bar l}^{-1}\de \abs{\bar l}$ and as it approaches infinity it scales as $\abs{l}^{-3}\de\abs{\bar l}$. Hence, the integral is divergent in the infrared with the divergence
\begin{equation}
\Kop\bigl(
\settoheight{\eqoff}{$\times$}%
\setlength{\eqoff}{0.5\eqoff}%
\addtolength{\eqoff}{-2.75\unitlength}%
\raisebox{\eqoff}{
	\fmfframe(0,0)(-1,0){
		\begin{fmfchar*}(15,5.0)
		\fmfleft{vout}
		\fmfright{vin}
		\fmfforce{0.15 w, 0.5h}{v2}
		\fmfforce{0.85 w, 0.5h}{v1}
		\fmfv{decor.shape=circle,decor.filled=empty,
			decor.size=2thick}{v1,v2}
		\fmf{plain,left=0.6}{v1,v2}
		\fmffreeze
		\fmfposition
		\fmfipath{p[]}
		\fmfiset{p1}{vpath(__v1,__v2)}
		\fmfiv{label=$\scriptscriptstyle 2$,label.angle=-90,label.dist=2}{point length(p1)/2 of p1}
		\end{fmfchar*}
	}
}
\bigr)
\otimes
\settoheight{\eqoff}{$\times$}%
\setlength{\eqoff}{0.5\eqoff}%
\addtolength{\eqoff}{-2.75\unitlength}%
\raisebox{\eqoff}{
	\fmfframe(0,0)(-1,0){
		\begin{fmfchar*}(15,5.0)
		\fmfleft{vout}
		\fmfright{vin}
		\fmfforce{0.15 w, 0.5h}{v2}
		\fmfforce{0.85 w, 0.5h}{v1}
		\fmfv{decor.shape=circle,decor.filled=empty,
			decor.size=2thick}{v1,v2}
		\fmf{plain}{v2,vout}
		\fmf{plain}{vin,v1}
		\fmf{plain,left=0.6}{v2,v1}
		\fmffreeze
		\fmfposition
		\end{fmfchar*}
	}
}
\equiv
\Kop\left(\frac{G(2,1)}{(4\pi)^{2-\epsilon}}\right)\frac{1}{\bar p^{2(1+\epsilon)}}
=
-\frac{1}{(4\pi)^2\epsilon}\frac{1}{\bar p^{2(1+\epsilon)}}\eqncom
\end{equation}
where we adopted the graphical notation for IR divergences from \cite[\chap{12}]{kleinert2001critical}, in which the IR (sub-)divergent part is cut out of the original graph. The gluing points are marked with white dots, which otherwise have the same properties as black dots and the rules for integral diagrams still apply. Essentially, the IR divergence in the original integral stems from the term $\bar l^{-4}$ and not the whole loop. Therefore, the divergence can be isolated by cutting this part out\footnote{The IR divergence is nevertheless determined by the full diagram with $G(2,1)$, since we need the term $(\bar p-\bar l)^{-2}$ to make the integral UV convergent.} of the diagram and not shrinking it to a point. In the cograph, the cutting is implemented via the replacement $\bar{l}^{-4}\rightarrow \delta^{(D)}(\bar l)$. For a detailed and pedagogical introduction to the isolation of IR divergences see \cite[\chap{12}]{kleinert2001critical}.

When we turn to more complicated scalar integrals that depend on multiple scales $p_i$ or are of higher loop order, we only have analytic results for special subclasses of generic integrals. However, if the integrals are logarithmically UV divergent, we can still determine their UV divergence with relatively little effort. A logarithmic UV divergence on the one hand is independent of any scale, see e.g.\ \eqref{eq:UV_divergence_example}, and hence we can choose a special combination of the external scales that simplifies the evaluation of the integral. IR divergences on the other hand are generated by some lines of the integral diagram alone and they vanish if an external momentum flows through that line. Therefore, if we choose a special combination of external scales we may accidentally introduce spurious IR divergences into the integral which we consequently have to subtract. For example, we cannot solve the integral \footnote{This integral can be solved using the techniques presented in \cite{Bern:1993kr}.}
\begin{equation}\label{eq:integral_3_vertex}
\settoheight{\eqoff}{$\times$}%
\setlength{\eqoff}{0.5\eqoff}%
\addtolength{\eqoff}{-5.75\unitlength}%
\raisebox{\eqoff}{
	\fmfframe(0,0)(-1,0){
		\begin{fmfchar*}(15,9.0)
		\fmfforce{0w,0.6h}{vout}
		\fmfforce{1w,0.6h}{vin}
		\fmfforce{0.5w,0h}{vdown}
		\fmfforce{0.15 w, 0.6h}{v2}
		\fmfforce{0.85 w, 0.6h}{v1}
		\fmfv{decor.shape=circle,decor.filled=full,decor.size=2thick}{v1,v2}
		\fmf{plain}{v2,vout}
		\fmf{plain}{vin,v1}
		\fmf{plain,left=0.6}{v1,v2,v1}
		\fmffreeze
		\fmfposition
		\fmfipath{p[]}
		\fmfiset{p1}{vpath(__v1,__v2)}
		\fmfiv{decor.shape=circle,decor.filled=full,decor.size=2thick}{point length(p1)/2 of p1}
		\fmfi{plain}{point length(p1)/2 of p1 -- vloc(__vdown)}
		\end{fmfchar*}
	}
}
=\int \frac{\de^D \bar l}{(2\pi)^D}\frac{1}{\bar l^2(\bar p-\bar l)^2(\bar q+\bar l)^2}
\end{equation}
directly but it is clearly IR convergent for $D>2$. When we choose the special momentum configuration $\bar p\neq 0$ and $\bar q=0$ to find its UV divergence in $D=4-2\epsilon$ dimensions, the integral turns into \eqref{eq:IR_divergent_diag}, which is directly solvable but also has an IR divergence. So we have to subtract this spurious IR divergence, whose origin lies in the special kinematic configuration\footnote{Physically, this configuration describes the collinear limit in which one particle in a three-body interaction scatters with zero momentum transfer. As the interacting particles are massless, it is possible in this limit to radiate off infinitely many particles from a two-body interaction. This missing contribution would render the physical process IR finite if there are not further contributions missing, \cf \cite[\chap{26 -- 27}]{Srednicki:2007}.}. To see how the UV divergence extraction of a multi-scale diagram works, let us follow an example of \cite[\chap{8,9,12}]{kleinert2001critical}. We calculate a two-loop contribution to the four-point interaction in $\phi^4$-theory in two ways: first at generic non-vanishing external scales and then at a special scale configuration where the integrals can be solved analytically as concatenations of \eqref{eq:G_function}. Starting with the first approach, we want to calculate the divergence of
\begin{equation}\label{eq:phi_4_example}
\scalebox{0.6}{
	\settoheight{\eqoff}{$\times$}%
	\setlength{\eqoff}{0.4\eqoff}%
	\addtolength{\eqoff}{-6.75\unitlength}%
	\raisebox{\eqoff}{
		\fmfframe(-2,0)(-2,0){
			\begin{fmfchar*}(15,15.0)
			\fmfleft{vout}
			\fmfright{vin}
			\fmfforce{0.4w,1.0h}{vt1}
			\fmfforce{0.6w,1.0h}{vt2}
			\fmfforce{0.15 w, 0.5h}{v2}
			\fmfforce{0.85 w, 0.5h}{v1}
			\fmfv{decor.shape=circle,decor.filled=full,decor.size=2thick}{v1,v2}
			\fmf{plain}{v2,vout}
			\fmf{plain}{vin,v1}
			\fmf{plain,left}{v1,v2,v1}
			\fmffreeze
			\fmf{plain}{v1,v2}
			\fmffreeze
			\fmfposition
			\fmfipath{p[]}
			\fmfiset{p1}{vpath(__v2,__v1)}
			\fmfiv{decor.shape=circle,decor.filled=full,decor.size=2thick}{point length(p1)/2 of p1}
			\fmfi{plain}{point length(p1)/2 of p1 -- vloc(__vt1)}
			\fmfi{plain}{point length(p1)/2 of p1 -- vloc(__vt2)}
			\end{fmfchar*}
		}
	}
}
=\mu^{4\epsilon}\int\frac{\de^D\bar l}{(2\pi)^D}\frac{\de^D\bar k}{(2\pi)^D}\frac{1}{(\bar p-\bar l)^2(\bar q+\bar k)^2\bar{k}^2(\bar k-\bar l)^2}\eqncom
\end{equation}
where $p$ and $q$ enter through the left and upper vertex, respectively and the scale $\mu$ is introduced to render the integral dimensionless in $D=4-2\epsilon$ dimensions. Note that we slightly abused the pictorial representation to the left by not multiplying it with the $\mu$ factor. This integral is free of IR divergences for $p\neq0\neq q$ but it does contain a UV subdivergence in the lower bubble, which corresponds to the integration over $l$. As in the previous example, it is complicated to evaluate this two-scale integral in general but its pole part can be extracted relatively directly
\begin{equation}
\Kop\Bigl[
\scalebox{0.6}{
	\settoheight{\eqoff}{$\times$}%
	\setlength{\eqoff}{0.4\eqoff}%
	\addtolength{\eqoff}{-6.75\unitlength}%
	\raisebox{\eqoff}{
		\fmfframe(-2,0)(-2,0){
			\begin{fmfchar*}(15,15.0)
			\fmfleft{vout}
			\fmfright{vin}
			\fmfforce{0.4w,1.0h}{vt1}
			\fmfforce{0.6w,1.0h}{vt2}
			\fmfforce{0.15 w, 0.5h}{v2}
			\fmfforce{0.85 w, 0.5h}{v1}
			\fmfv{decor.shape=circle,decor.filled=full,decor.size=2thick}{v1,v2}
			\fmf{plain}{v2,vout}
			\fmf{plain}{vin,v1}
			\fmf{plain,left}{v1,v2,v1}
			\fmffreeze
			\fmf{plain}{v1,v2}
			\fmffreeze
			\fmfposition
			\fmfipath{p[]}
			\fmfiset{p1}{vpath(__v2,__v1)}
			\fmfiv{decor.shape=circle,decor.filled=full,decor.size=2thick}{point length(p1)/2 of p1}
			\fmfi{plain}{point length(p1)/2 of p1 -- vloc(__vt1)}
			\fmfi{plain}{point length(p1)/2 of p1 -- vloc(__vt2)}
			\end{fmfchar*}
		}
	}
}
\Bigr]
=
\frac{1}{(4\pi)^4}\left[\frac{1}{2\epsilon^2}+\frac{1}{\epsilon}\left(\frac 52- c_{\ol{\text{MS}}}
+\log\frac{\mu^2}{\bar{p}^2+\bar{q}^2} \right)\right]\eqncom
\end{equation}
with the same momentum configuration as in \eqref{eq:phi_4_example}, $c_{\ol{\text{MS}}}=\gammaE-\log 4\pi$, and an explicit derivation given in\footnote{Note that \cite[\chap{8}]{kleinert2001critical} works in $D=4-\epsilon$, whereas we work in $D=4-2\epsilon$ dimensions.} \cite[\chap{8}]{kleinert2001critical}. The subdivergence in the integral gives rise to the non-local pole and including the one-loop vertex counterterm insertion 
\begin{equation}\label{eq:subdivergence_UV_ren}
\frac{1}{g_{\phi}^2}\Kop\Bigl[
\scalebox{0.6}{
	\settoheight{\eqoff}{$\times$}%
	\setlength{\eqoff}{0.5\eqoff}%
	\addtolength{\eqoff}{-6.75\unitlength}%
	\raisebox{\eqoff}{
		\fmfframe(-2,0)(-2,0){
			\begin{fmfchar*}(15,15.0)
			\fmfforce{.35w,0h}{vout1}
			\fmfforce{0.65w,0h}{vout2}
			\fmfforce{0.35w,1h}{vin1}
			\fmfforce{0.65w,1h}{vin2}
			\fmfforce{0.5 w, 0.2h}{v2}
			\fmfforce{0.5 w, 0.8h}{v1}
			\fmfv{decor.shape=hexacross,decor.size=10 thin}{v2}
			\fmf{plain}{vout1,v2,vout2}
			\fmf{plain}{vin1,v1,vin2}
			\fmf{plain,left=0.7}{v1,v2,v1}
			\end{fmfchar*}
		}
	}
}
\Bigr]
=
\Kop\Bigl[
\Kop\bigl(
\scalebox{0.6}{
	\settoheight{\eqoff}{$\times$}%
	\setlength{\eqoff}{0.5\eqoff}%
	\addtolength{\eqoff}{-6.75\unitlength}%
	\raisebox{\eqoff}{
		\fmfframe(-2,0)(-2,0){
			\begin{fmfchar*}(15,15.0)
			\fmfforce{0w,0.6h}{vout1}
			\fmfforce{0w,0.4h}{vout2}
			\fmfforce{1w,0.6h}{vin1}
			\fmfforce{1w,0.4h}{vin2}
			\fmfforce{0.15 w, 0.5h}{v2}
			\fmfforce{0.85 w, 0.5h}{v1}
			\fmfv{decor.shape=circle,decor.filled=full,decor.size=2thick}{v1,v2}
			\fmf{plain}{vout1,v2,vout2}
			\fmf{plain}{vin1,v1,vin2}
			\fmf{plain,left}{v1,v2,v1}
			\end{fmfchar*}
		}
	}
}
\bigr)
\otimes
\scalebox{0.6}{
	\settoheight{\eqoff}{$\times$}%
	\setlength{\eqoff}{0.5\eqoff}%
	\addtolength{\eqoff}{-6.75\unitlength}%
	\raisebox{\eqoff}{
		\fmfframe(-2,0)(-2,0){
			\begin{fmfchar*}(15,15.0)
			\fmfforce{.4w,0h}{vout1}
			\fmfforce{0.6w,0h}{vout2}
			\fmfforce{0.4w,1h}{vin1}
			\fmfforce{0.6w,1h}{vin2}
			\fmfforce{0.5 w, 0.15h}{v2}
			\fmfforce{0.5 w, 0.85h}{v1}
			\fmfv{decor.shape=circle,decor.filled=full,decor.size=2thick}{v1,v2}
			\fmf{plain}{vout1,v2,vout2}
			\fmf{plain}{vin1,v1,vin2}
			\fmf{plain,left}{v1,v2,v1}
			\end{fmfchar*}
		}
	}
}
\Bigr]
=\frac{-1}{(4\pi)^2\epsilon}\frac{1}{(4\pi)^2}\left(
\frac{1}{\epsilon}+2-c_{\ol{\text{MS}}}
+\log\frac{\mu^2}{\bar{p}^2+\bar{q}^2}\right)
\end{equation}
renders the sum of both contributions free of non-local poles. Note that the leftmost diagram is not an integral diagram but a Feynman diagram in $\phi^4$-theory with coupling constant $g_{\phi}$ and a one-loop counterterm insertion at the lower vertex. Combining the latter two contributions, the overall UV divergence of \eqref{eq:phi_4_example} is given by
\begin{equation}\label{eq:overall_divergence_UV_ren}
\Kop \bar R_{\text{UV}}\Bigl[
\scalebox{0.6}{
	\settoheight{\eqoff}{$\times$}%
	\setlength{\eqoff}{0.4\eqoff}%
	\addtolength{\eqoff}{-6.75\unitlength}%
	\raisebox{\eqoff}{
		\fmfframe(-2,0)(-2,0){
			\begin{fmfchar*}(15,15.0)
			\fmfleft{vout}
			\fmfright{vin}
			\fmfforce{0.4w,1.0h}{vt1}
			\fmfforce{0.6w,1.0h}{vt2}
			\fmfforce{0.15 w, 0.5h}{v2}
			\fmfforce{0.85 w, 0.5h}{v1}
			\fmfv{decor.shape=circle,decor.filled=full,decor.size=2thick}{v1,v2}
			\fmf{plain}{v2,vout}
			\fmf{plain}{vin,v1}
			\fmf{plain,left}{v1,v2,v1}
			\fmffreeze
			\fmf{plain}{v1,v2}
			\fmffreeze
			\fmfposition
			\fmfipath{p[]}
			\fmfiset{p1}{vpath(__v2,__v1)}
			\fmfiv{decor.shape=circle,decor.filled=full,decor.size=2thick}{point length(p1)/2 of p1}
			\fmfi{plain}{point length(p1)/2 of p1 -- vloc(__vt1)}
			\fmfi{plain}{point length(p1)/2 of p1 -- vloc(__vt2)}
			\end{fmfchar*}
		}
	}
}
\Bigr]
=
\Kop\Bigl[
\scalebox{0.6}{
	\settoheight{\eqoff}{$\times$}%
	\setlength{\eqoff}{0.4\eqoff}%
	\addtolength{\eqoff}{-6.75\unitlength}%
	\raisebox{\eqoff}{
		\fmfframe(-2,0)(-2,0){
			\begin{fmfchar*}(15,15.0)
			\fmfleft{vout}
			\fmfright{vin}
			\fmfforce{0.4w,1.0h}{vt1}
			\fmfforce{0.6w,1.0h}{vt2}
			\fmfforce{0.15 w, 0.5h}{v2}
			\fmfforce{0.85 w, 0.5h}{v1}
			\fmfv{decor.shape=circle,decor.filled=full,decor.size=2thick}{v1,v2}
			\fmf{plain}{v2,vout}
			\fmf{plain}{vin,v1}
			\fmf{plain,left}{v1,v2,v1}
			\fmffreeze
			\fmf{plain}{v1,v2}
			\fmffreeze
			\fmfposition
			\fmfipath{p[]}
			\fmfiset{p1}{vpath(__v2,__v1)}
			\fmfiv{decor.shape=circle,decor.filled=full,decor.size=2thick}{point length(p1)/2 of p1}
			\fmfi{plain}{point length(p1)/2 of p1 -- vloc(__vt1)}
			\fmfi{plain}{point length(p1)/2 of p1 -- vloc(__vt2)}
			\end{fmfchar*}
		}
	}
}
+
\Kop\bigl(
\scalebox{0.6}{
	\settoheight{\eqoff}{$\times$}%
	\setlength{\eqoff}{0.5\eqoff}%
	\addtolength{\eqoff}{-6.75\unitlength}%
	\raisebox{\eqoff}{
		\fmfframe(-2,0)(-2,0){
			\begin{fmfchar*}(15,15.0)
			\fmfforce{0w,0.6h}{vout1}
			\fmfforce{0w,0.4h}{vout2}
			\fmfforce{1w,0.6h}{vin1}
			\fmfforce{1w,0.4h}{vin2}
			\fmfforce{0.15 w, 0.5h}{v2}
			\fmfforce{0.85 w, 0.5h}{v1}
			\fmfv{decor.shape=circle,decor.filled=full,decor.size=2thick}{v1,v2}
			\fmf{plain}{vout1,v2,vout2}
			\fmf{plain}{vin1,v1,vin2}
			\fmf{plain,left}{v1,v2,v1}
			\end{fmfchar*}
		}
	}
}
\bigr)
\otimes
\scalebox{0.6}{
	\settoheight{\eqoff}{$\times$}%
	\setlength{\eqoff}{0.5\eqoff}%
	\addtolength{\eqoff}{-6.75\unitlength}%
	\raisebox{\eqoff}{
		\fmfframe(-2,0)(-2,0){
			\begin{fmfchar*}(15,15.0)
			\fmfforce{.4w,0h}{vout1}
			\fmfforce{0.6w,0h}{vout2}
			\fmfforce{0.4w,1h}{vin1}
			\fmfforce{0.6w,1h}{vin2}
			\fmfforce{0.5 w, 0.15h}{v2}
			\fmfforce{0.5 w, 0.85h}{v1}
			\fmfv{decor.shape=circle,decor.filled=full,decor.size=2thick}{v1,v2}
			\fmf{plain}{vout1,v2,vout2}
			\fmf{plain}{vin1,v1,vin2}
			\fmf{plain,left}{v1,v2,v1}
			\end{fmfchar*}
		}
	}
}
\Bigr]
=\frac{1}{(4\pi)^4}\left[-\frac{1}{2\epsilon^2}+\frac{1}{2\epsilon}\right]\eqncom
\end{equation}
where we used the $\bar R_{\text{UV}}$ operation as presented in \cite[\chap{11}]{kleinert2001critical} to recursively subtract all UV subdivergences from a given diagram. Keep in mind that the $\bar R_{\text{UV}}$ operation is a pure integral operation that subtracts all UV subdivergences regardless whether these subtractions do exist in the physical theory that generated the integral. From the perspective of a renormalisable field theory, the subtraction of the UV subdivergence in \eqref{eq:overall_divergence_UV_ren} occurs as the Feynman diagram \eqref{eq:subdivergence_UV_ren} generates it. Let us now recalculate the overall divergence of \eqref{eq:phi_4_example} at the special kinematical point where the external momentum $q$ entering the upper vertex in \eqref{eq:phi_4_example} is set to zero. For the integral, this amounts to
\begin{equation}
\scalebox{0.6}{
	\settoheight{\eqoff}{$\times$}%
	\setlength{\eqoff}{0.5\eqoff}%
	\addtolength{\eqoff}{-6.75\unitlength}%
	\raisebox{\eqoff}{
		\fmfframe(-2,0)(-2,0){
			\begin{fmfchar*}(15,15.0)
			\fmfleft{vout}
			\fmfright{vin}
			\fmfforce{0.15 w, 0.5h}{v2}
			\fmfforce{0.85 w, 0.5h}{v1}
			\fmfv{decor.shape=circle,decor.filled=full,decor.size=2thick}{v1,v2}
			\fmf{plain}{v2,vout}
			\fmf{plain}{vin,v1}
			\fmf{plain,left}{v1,v2,v1}
			\fmffreeze
			\fmf{plain}{v1,v2}
			\fmffreeze
			\fmfposition
			\fmfipath{p[]}
			\fmfiset{p1}{vpath(__v2,__v1)}
			\fmfiv{label=$\scriptstyle 2$,label.angle=90,label.dist=2}{point length(p1)/2 of p1}
			\end{fmfchar*}
		}
	}
}
=\int\frac{\de^D\bar l}{(2\pi)^D}\frac{\de^D\bar k}{(2\pi)^D}\frac{\bigl(\mu^{2(2-D/2)}\bigr)^2}{\bar {k}^4(\bar k-\bar l)^2(\bar p-\bar l)^2}
=\frac{1}{(4\pi)^D}G(2,1)G(3-\tfrac{D}{2},1)\left(\frac{\mu^2}{\bar{p}^2}\right)^{4-D}
\eqncom
\end{equation}
which we solved analytically using \eqref{eq:G_function}. In addition to the UV subdivergence from the lower bubble, the absence of the second scale now introduces an IR divergence in the upper bubble. The pole part of the whole diagram is
\begin{equation}\label{eq:polepart_UVIR_ren}
\Kop\Bigl[
\scalebox{0.6}{
	\settoheight{\eqoff}{$\times$}%
	\setlength{\eqoff}{0.5\eqoff}%
	\addtolength{\eqoff}{-6.75\unitlength}%
	\raisebox{\eqoff}{
		\fmfframe(-2,0)(-2,0){
			\begin{fmfchar*}(15,15.0)
			\fmfleft{vout}
			\fmfright{vin}
			\fmfforce{0.15 w, 0.5h}{v2}
			\fmfforce{0.85 w, 0.5h}{v1}
			\fmfv{decor.shape=circle,decor.filled=full,decor.size=2thick}{v1,v2}
			\fmf{plain}{v2,vout}
			\fmf{plain}{vin,v1}
			\fmf{plain,left}{v1,v2,v1}
			\fmffreeze
			\fmf{plain}{v1,v2}
			\fmffreeze
			\fmfposition
			\fmfipath{p[]}
			\fmfiset{p1}{vpath(__v2,__v1)}
			\fmfiv{label=$\scriptstyle 2$,label.angle=90,label.dist=2}{point length(p1)/2 of p1}
			\end{fmfchar*}
		}
	}
}
\Bigr]
=\frac{1}{(4\pi)^4}\left[-\frac{1}{2\epsilon^2}+\frac{1}{\epsilon}\left(-\frac 32+c_{\ol{\text{MS}}}-\log\frac{\mu^2}{\bar{p}^2}\right)\right]\eqncom
\end{equation}
where we have to subtract by hand the additional IR subdivergence contribution
\begin{equation}\label{eq:subdivergence1_UVIR_ren}
\Kop\Bigl[\Kop\bigl(
\scalebox{0.7}{
	\settoheight{\eqoff}{$\times$}%
	\setlength{\eqoff}{0.5\eqoff}%
	\addtolength{\eqoff}{0.5\unitlength}%
	\raisebox{\eqoff}{
		\fmfframe(-2,0)(-2,0){
			\begin{fmfchar*}(6,2.0)
			\fmfforce{0 w, 0h}{vout}
			\fmfforce{1 w, 0h}{vin}
			\fmf{plain}{vout,v1,vin}
			\fmfv{decor.shape=circle,decor.filled=empty,decor.size=2thick}{vout,vin}
			\fmfv{label=$\scriptstyle 2$,label.angle=-90,label.dist=2}{v1}
			\end{fmfchar*}
		}
	}
}
\bigr)
\,{}\otimes{}\,
\scalebox{0.6}{
	\settoheight{\eqoff}{$\times$}%
	\setlength{\eqoff}{0.5\eqoff}%
	\addtolength{\eqoff}{-5.75\unitlength}%
	\raisebox{\eqoff}{
		\fmfframe(-2,0)(-2,0){
			\begin{fmfchar*}(15,15.0)
			\fmfleft{vout}
			\fmfright{vin}
			\fmfforce{0.15 w, 0.5h}{v2}
			\fmfforce{0.85 w, 0.5h}{v1}
			\fmfv{decor.shape=circle,decor.filled=empty,decor.size=2thick}{v1,v2}
			\fmf{plain}{v2,vout}
			\fmf{plain}{vin,v1}
			\fmf{plain,left=1}{v1,v2}
			\fmffreeze
			\fmf{plain}{v1,v2}
			\end{fmfchar*}
		}
	}
}
\Bigr]
=\frac{1}{(4\pi)^2\epsilon}\frac{1}{(4\pi)^2}\left(\frac{1}{\epsilon}+2-c_{\ol{\text{MS}}}+\log\frac{\mu^2}{\bar{p}^2}\right)\eqndot
\end{equation}
Comparing the sum of \eqref{eq:polepart_UVIR_ren} and \eqref{eq:subdivergence1_UVIR_ren} with \eqref{eq:overall_divergence_UV_ren}, we find that we correctly reproduce the $\epsilon^{-1}$ pole but fail to get the correct result for the $\epsilon^{-2}$ pole. The reason for this mismatch is that we accidentally subtracted a UV divergent contribution when we subtract the IR subdivergence. This can be seen from \eqref{eq:subdivergence1_UVIR_ren}, where the cograph contains a pure UV divergence. So we have to also include a factor of 
\begin{equation}
\Kop\bigl(
\scalebox{0.7}{
	\settoheight{\eqoff}{$\times$}%
	\setlength{\eqoff}{0.5\eqoff}%
	\addtolength{\eqoff}{0.5\unitlength}%
	\raisebox{\eqoff}{
		\fmfframe(-2,0)(-2,0){
			\begin{fmfchar*}(6,2.0)
			\fmfforce{0 w, 0h}{vout}
			\fmfforce{1 w, 0h}{vin}
			\fmf{plain}{vout,v1,vin}
			\fmfv{decor.shape=circle,decor.filled=empty,decor.size=2thick}{vout,vin}
			\fmfv{label=$\scriptstyle 2$,label.angle=-90,label.dist=2}{v1}
			\end{fmfchar*}
		}
	}
}
\bigr)
\,
\Kop\bigl(
\scalebox{0.6}{
	\settoheight{\eqoff}{$\times$}%
	\setlength{\eqoff}{0.5\eqoff}%
	\addtolength{\eqoff}{-6.75\unitlength}%
	\raisebox{\eqoff}{
		\fmfframe(-2,0)(-2,0){
			\begin{fmfchar*}(15,15.0)
			\fmfforce{0w,0.6h}{vout1}
			\fmfforce{0w,0.4h}{vout2}
			\fmfforce{1w,0.6h}{vin1}
			\fmfforce{1w,0.4h}{vin2}
			\fmfforce{0.15 w, 0.5h}{v2}
			\fmfforce{0.85 w, 0.5h}{v1}
			\fmfv{decor.shape=circle,decor.filled=full,decor.size=2thick}{v1,v2}
			\fmf{plain}{vout1,v2,vout2}
			\fmf{plain}{vin1,v1,vin2}
			\fmf{plain,left}{v1,v2,v1}
			\end{fmfchar*}
		}
	}	
}
\bigr)
=\frac{1}{(4\pi)^2\epsilon}\frac{-1}{(4\pi)^2\epsilon}
\end{equation}
to remedy this mistake and reproduce \eqref{eq:overall_divergence_UV_ren}. It is possible to formalise this procedure of subtracting UV- and IR-subdivergences from any given scalar integral by introducing the $\bar R_{\text{IR}}$ operation which, in analogy to $\bar R_{\text{UV}}$, subtracts the IR subdivergences of a diagram. Combining both subtractions as $\bar R^*=\bar R_{\text{IR}}\bar R_{\text{UV}}$ leads to the so-called $\bar R^*$-operation \cite{CHETYRKIN1982340,Chetyrkin:1984xa,Chetyrkin:1985} which iteratively subtracts the UV and afterwards IR subdivergences.\footnote{For the example discussed here, the order in which IR and UV subdivergences are subtracted does not matter. However, in generic diagrams the different momentum dependence of UV and IR divergences dictates that the UV divergences have to be subtracted before the IR divergences \cite[\chap{12}]{kleinert2001critical}.} The present example can then be expressed as
\begin{equation}
\begin{aligned}
\Kop \bar R^*\Bigl(
\scalebox{0.55}{
	\settoheight{\eqoff}{$\times$}%
	\setlength{\eqoff}{0.5\eqoff}%
	\addtolength{\eqoff}{-6.75\unitlength}%
	\raisebox{\eqoff}{
		\fmfframe(-2,0)(-2,0){
			\begin{fmfchar*}(15,15.0)
			\fmfleft{vout}
			\fmfright{vin}
			\fmfforce{0.15 w, 0.5h}{v2}
			\fmfforce{0.85 w, 0.5h}{v1}
			\fmfv{decor.shape=circle,decor.filled=full,decor.size=2thick}{v1,v2}
			\fmf{plain}{v2,vout}
			\fmf{plain}{vin,v1}
			\fmf{plain,left}{v1,v2,v1}
			\fmffreeze
			\fmf{plain}{v1,v2}
			\fmffreeze
			\fmfposition
			\fmfipath{p[]}
			\fmfiset{p1}{vpath(__v2,__v1)}
			\fmfiv{label=$\scriptstyle 2$,label.angle=90,label.dist=2}{point length(p1)/2 of p1}
			\end{fmfchar*}
		}
	}
}
\Bigr)
&=
\Kop \bar R_{\text{IR}}\Bigl[
\scalebox{0.55}{
	\settoheight{\eqoff}{$\times$}%
	\setlength{\eqoff}{0.5\eqoff}%
	\addtolength{\eqoff}{-6.75\unitlength}%
	\raisebox{\eqoff}{
		\fmfframe(-2,0)(-2,0){
			\begin{fmfchar*}(15,15.0)
			\fmfleft{vout}
			\fmfright{vin}
			\fmfforce{0.15 w, 0.5h}{v2}
			\fmfforce{0.85 w, 0.5h}{v1}
			\fmfv{decor.shape=circle,decor.filled=full,decor.size=2thick}{v1,v2}
			\fmf{plain}{v2,vout}
			\fmf{plain}{vin,v1}
			\fmf{plain,left}{v1,v2,v1}
			\fmffreeze
			\fmf{plain}{v1,v2}
			\fmffreeze
			\fmfposition
			\fmfipath{p[]}
			\fmfiset{p1}{vpath(__v2,__v1)}
			\fmfiv{label=$\scriptstyle 2$,label.angle=90,label.dist=2}{point length(p1)/2 of p1}
			\end{fmfchar*}
		}
	}	
}
+\Kop\bigl(
\scalebox{0.55}{
	\settoheight{\eqoff}{$\times$}%
	\setlength{\eqoff}{0.5\eqoff}%
	\addtolength{\eqoff}{-6.75\unitlength}%
	\raisebox{\eqoff}{
		\fmfframe(-2,0)(-2,0){
			\begin{fmfchar*}(15,15.0)
			\fmfforce{0w,0.6h}{vout1}
			\fmfforce{0w,0.4h}{vout2}
			\fmfforce{1w,0.6h}{vin1}
			\fmfforce{1w,0.4h}{vin2}
			\fmfforce{0.15 w, 0.5h}{v2}
			\fmfforce{0.85 w, 0.5h}{v1}
			\fmfv{decor.shape=circle,decor.filled=full,decor.size=2thick}{v1,v2}
			\fmf{plain}{vout1,v2,vout2}
			\fmf{plain}{vin1,v1,vin2}
			\fmf{plain,left}{v1,v2,v1}
			\end{fmfchar*}
		}
	}
}
\bigr)
\,{}\otimes{}\,
\scalebox{0.5}{
	\settoheight{\eqoff}{$\times$}%
	\setlength{\eqoff}{0.5\eqoff}%
	\addtolength{\eqoff}{-2.75\unitlength}%
	\raisebox{\eqoff}{
		\fmfframe(-2,0)(-2,0){
			\begin{fmfchar*}(10,10.0)
			\fmfforce{0w,0h}{vout}
			\fmfforce{1w,0h}{vin}
			\fmf{plain}{vout,v1,vin}
			\fmf{plain,right,tension=0.4}{v1,v1}
			\fmfv{decor.shape=circle,decor.filled=full,decor.size=2thick}{v1}
			\fmffreeze
			\fmfposition
			\fmfipath{p[]}
			\fmfiset{p1}{vpath(__v1,__v1)}
			\fmfiv{label=$\scriptstyle 2$,label.angle=90,label.dist=2}{point length(p1)/2 of p1}
			\end{fmfchar*}
		}
	}
}
\Bigr]
\\
&=
\Kop\Bigl[
\scalebox{0.55}{
	\settoheight{\eqoff}{$\times$}%
	\setlength{\eqoff}{0.5\eqoff}%
	\addtolength{\eqoff}{-6.75\unitlength}%
	\raisebox{\eqoff}{
		\fmfframe(-2,0)(-2,0){
			\begin{fmfchar*}(15,15.0)
			\fmfleft{vout}
			\fmfright{vin}
			\fmfforce{0.15 w, 0.5h}{v2}
			\fmfforce{0.85 w, 0.5h}{v1}
			\fmfv{decor.shape=circle,decor.filled=full,decor.size=2thick}{v1,v2}
			\fmf{plain}{v2,vout}
			\fmf{plain}{vin,v1}
			\fmf{plain,left}{v1,v2,v1}
			\fmffreeze
			\fmf{plain}{v1,v2}
			\fmffreeze
			\fmfposition
			\fmfipath{p[]}
			\fmfiset{p1}{vpath(__v2,__v1)}
			\fmfiv{label=$\scriptstyle 2$,label.angle=90,label.dist=2}{point length(p1)/2 of p1}
			\end{fmfchar*}
		}
	}	
}
+\Kop\bigl(
\scalebox{0.55}{
	\settoheight{\eqoff}{$\times$}%
	\setlength{\eqoff}{0.5\eqoff}%
	\addtolength{\eqoff}{-6.75\unitlength}%
	\raisebox{\eqoff}{
		\fmfframe(-2,0)(-2,0){
			\begin{fmfchar*}(15,15.0)
			\fmfforce{0w,0.6h}{vout1}
			\fmfforce{0w,0.4h}{vout2}
			\fmfforce{1w,0.6h}{vin1}
			\fmfforce{1w,0.4h}{vin2}
			\fmfforce{0.15 w, 0.5h}{v2}
			\fmfforce{0.85 w, 0.5h}{v1}
			\fmfv{decor.shape=circle,decor.filled=full,decor.size=2thick}{v1,v2}
			\fmf{plain}{vout1,v2,vout2}
			\fmf{plain}{vin1,v1,vin2}
			\fmf{plain,left}{v1,v2,v1}
			\end{fmfchar*}
		}
	}
}
\bigr)
\,{}\otimes{}\,
\scalebox{0.5}{
	\settoheight{\eqoff}{$\times$}%
	\setlength{\eqoff}{0.5\eqoff}%
	\addtolength{\eqoff}{-2.75\unitlength}%
	\raisebox{\eqoff}{
		\fmfframe(-2,0)(-2,0){
			\begin{fmfchar*}(10,10.0)
			\fmfforce{0w,0h}{vout}
			\fmfforce{1w,0h}{vin}
			\fmf{plain}{vout,v1,vin}
			\fmf{plain,right,tension=0.4}{v1,v1}
			\fmfv{decor.shape=circle,decor.filled=full,decor.size=2thick}{v1}
			\fmffreeze
			\fmfposition
			\fmfipath{p[]}
			\fmfiset{p1}{vpath(__v1,__v1)}
			\fmfiv{label=$\scriptstyle 2$,label.angle=90,label.dist=2}{point length(p1)/2 of p1}
			\end{fmfchar*}
		}
	}
}
+\Kop\bigl(
\scalebox{0.7}{
	\settoheight{\eqoff}{$\times$}%
	\setlength{\eqoff}{0.5\eqoff}%
	\addtolength{\eqoff}{0.5\unitlength}%
	\raisebox{\eqoff}{
		\fmfframe(-2,0)(-2,0){
			\begin{fmfchar*}(6,2.0)
			\fmfforce{0 w, 0h}{vout}
			\fmfforce{1 w, 0h}{vin}
			\fmf{plain}{vout,v1,vin}
			\fmfv{decor.shape=circle,decor.filled=empty,decor.size=2thick}{vout,vin}
			\fmfv{label=$\scriptstyle 2$,label.angle=90,label.dist=2}{v1}
			\end{fmfchar*}
		}
	}
}
\bigr)
\,{}\otimes{}\,
\scalebox{0.55}{
	\settoheight{\eqoff}{$\times$}%
	\setlength{\eqoff}{0.5\eqoff}%
	\addtolength{\eqoff}{-5.75\unitlength}%
	\raisebox{\eqoff}{
		\fmfframe(-2,0)(-2,0){
			\begin{fmfchar*}(15,15.0)
			\fmfleft{vout}
			\fmfright{vin}
			\fmfforce{0.15 w, 0.5h}{v2}
			\fmfforce{0.85 w, 0.5h}{v1}
			\fmfv{decor.shape=circle,decor.filled=empty,decor.size=2thick}{v1,v2}
			\fmf{plain}{v2,vout}
			\fmf{plain}{vin,v1}
			\fmf{plain,left=1}{v1,v2}
			\fmffreeze
			\fmf{plain}{v1,v2}
			\end{fmfchar*}
		}
	}
}
+\Kop\bigl(
\scalebox{0.55}{
	\settoheight{\eqoff}{$\times$}%
	\setlength{\eqoff}{0.5\eqoff}%
	\addtolength{\eqoff}{-6.75\unitlength}%
	\raisebox{\eqoff}{
		\fmfframe(-2,0)(-2,0){
			\begin{fmfchar*}(15,15.0)
			\fmfforce{0w,0.6h}{vout1}
			\fmfforce{0w,0.4h}{vout2}
			\fmfforce{1w,0.6h}{vin1}
			\fmfforce{1w,0.4h}{vin2}
			\fmfforce{0.15 w, 0.5h}{v2}
			\fmfforce{0.85 w, 0.5h}{v1}
			\fmfv{decor.shape=circle,decor.filled=full,decor.size=2thick}{v1,v2}
			\fmf{plain}{vout1,v2,vout2}
			\fmf{plain}{vin1,v1,vin2}
			\fmf{plain,left}{v1,v2,v1}
			\end{fmfchar*}
		}
	}	
}
\bigr)
\Kop\bigl(
\scalebox{0.7}{
	\settoheight{\eqoff}{$\times$}%
	\setlength{\eqoff}{0.5\eqoff}%
	\addtolength{\eqoff}{0.5\unitlength}%
	\raisebox{\eqoff}{
		\fmfframe(-2,0)(-2,0){
			\begin{fmfchar*}(6,2.0)
			\fmfforce{0 w, 0h}{vout}
			\fmfforce{1 w, 0h}{vin}
			\fmf{plain}{vout,v1,vin}
			\fmfv{decor.shape=circle,decor.filled=empty,decor.size=2thick}{vout,vin}
			\fmfv{label=$\scriptstyle 2$,label.angle=90,label.dist=2}{v1}
			\end{fmfchar*}
		}
	}
}
\bigr)
\Bigr]
\\
&=
\Kop\bar R_{\text{UV}}\Bigl(
\scalebox{0.55}{
	\settoheight{\eqoff}{$\times$}%
	\setlength{\eqoff}{0.4\eqoff}%
	\addtolength{\eqoff}{-6.75\unitlength}%
	\raisebox{\eqoff}{
		\fmfframe(-2,0)(-2,0){
			\begin{fmfchar*}(15,15.0)
			\fmfleft{vout}
			\fmfright{vin}
			\fmfforce{0.4w,1.0h}{vt1}
			\fmfforce{0.6w,1.0h}{vt2}
			\fmfforce{0.15 w, 0.5h}{v2}
			\fmfforce{0.85 w, 0.5h}{v1}
			\fmfv{decor.shape=circle,decor.filled=full,decor.size=2thick}{v1,v2}
			\fmf{plain}{v2,vout}
			\fmf{plain}{vin,v1}
			\fmf{plain,left}{v1,v2,v1}
			\fmffreeze
			\fmf{plain}{v1,v2}
			\fmffreeze
			\fmfposition
			\fmfipath{p[]}
			\fmfiset{p1}{vpath(__v2,__v1)}
			\fmfiv{decor.shape=circle,decor.filled=full,decor.size=2thick}{point length(p1)/2 of p1}
			\fmfi{plain}{point length(p1)/2 of p1 -- vloc(__vt1)}
			\fmfi{plain}{point length(p1)/2 of p1 -- vloc(__vt2)}
			\end{fmfchar*}
		}
	}
}
\Bigr)\eqncom
\end{aligned}
\end{equation}
where the third and fourth term in the second equality renormalise the IR divergences of the first and second term, respectively. Note that the second term in the second equality on the \rhs evaluates to zero under $\Kop$ as the cograph is of tadpole-type, compare \eqref{eq:tadpole_type_diagram}. However, it is necessary to keep tadpole-type diagrams in the evaluation of the $\bar R$ operations, as we have to associate the IR divergence
$\Kop\bigl(
\scalebox{0.4}{
	\settoheight{\eqoff}{$\times$}%
	\setlength{\eqoff}{0.5\eqoff}%
	\addtolength{\eqoff}{-2.75\unitlength}%
	\raisebox{\eqoff}{
		\fmfframe(-2,0)(-2,0){
			\begin{fmfchar*}(10,10.0)
			\fmfforce{0w,0h}{vout}
			\fmfforce{1w,0h}{vin}
			\fmf{plain}{vout,v1,vin}
			\fmf{plain,right,tension=0.4}{v1,v1}
			\fmfv{decor.shape=circle,decor.filled=full,decor.size=2thick}{v1}
			\fmffreeze
			\fmfposition
			\fmfipath{p[]}
			\fmfiset{p1}{vpath(__v1,__v1)}
			\fmfiv{label=$ 2$,label.angle=90,label.dist=2}{point length(p1)/2 of p1}
			\end{fmfchar*}
		}
	}
}\bigr)
=
\Kop\bigl(
\scalebox{0.7}{
	\settoheight{\eqoff}{$\times$}%
	\setlength{\eqoff}{0.5\eqoff}%
	\addtolength{\eqoff}{0.5\unitlength}%
	\raisebox{\eqoff}{
		\fmfframe(-2,0)(-2,0){
			\begin{fmfchar*}(6,2.0)
			\fmfforce{0 w, 0h}{vout}
			\fmfforce{1 w, 0h}{vin}
			\fmf{plain}{vout,v1,vin}
			\fmfv{decor.shape=circle,decor.filled=empty,decor.size=2thick}{vout,vin}
			\fmfv{label=$\scriptstyle 2$,label.angle=90,label.dist=2}{v1}
			\end{fmfchar*}
		}
	}
}
\bigr)
$
with these types of diagrams in order to get the correct higher pole cancellations. Only after the $\bar R^*$ operation is completed, we set tadpole-type diagrams to zero\footnote{Keeping tadpole-type diagrams during the evaluation of the $\bar R^*$ operation is sensible, as the $\bar R^*$ operator is defined to act purely on integrals, regardless of the physical context in which the integral appears. Setting tadpole-type diagrams to zero after the $\bar R^*$ operation is completed invokes the physical argument that all these diagrams must vanish in a physical process after renormalisation if all vacuum expectation values of physical fields vanish, \cf \cite[\chap{9}]{Srednicki:2007}.}. This concludes our minimal example how to extract the UV divergence of a multi-scale integral by the means of choosing a special kinematical point and renormalising the occurring spurious IR divergences.

\subsection{Tensor integrals}\label{sec:tensor-integrals}
After the scalar integrals, we now turn to one-scale tensor integrals, i.e.\ integrals of the form \eqref{eq:G_function} but with a non-trivial numerator polynomial. For the evaluation of these integrals in an MS or DR related renormalisation scheme discussed in \appref{sec:Renormalisation_schemes}, it is important to notice that they are generated from the Feynman rules in \appref{app:Feynman_rules} in a way that the integrand momenta live in $d=4$ dimensions, while the integral measure lives in $D=d-2\epsilon$ dimensions. This leads to a peculiar structure of the solution of tensor integrals.

We can generate tensor integrals from \eqref{eq:G_function} by taking derivatives with respect to the external momentum of the \lhs and \rhs of the equation. For a rank one tensor this yields
\begin{equation}\label{eq:rank_1_G_function}
\settoheight{\eqoff}{$\times$}%
\setlength{\eqoff}{0.5\eqoff}%
\addtolength{\eqoff}{-2.75\unitlength}%
\raisebox{\eqoff}{
	\fmfframe(0,0)(-1,0){
		\begin{fmfchar*}(15,5.0)
		\fmfleft{vout}
		\fmfright{vin}
		\fmfforce{0.15 w, 0.5h}{v2}
		\fmfforce{0.85 w, 0.5h}{v1}
		\fmfv{decor.shape=circle,decor.filled=full,
			decor.size=2thick}{v1,v2}
		\fmf{plain}{v2,vout}
		\fmf{plain}{vin,v1}
		\fmf{phantom,left=0.6}{v1,v2,v1}
		\fmffreeze
		\fmfposition
		\fmfipath{p[]}
		\fmfipair{vv[]}
		\fmfiset{p1}{vpath(__v1,__v2)}
		\fmfiset{p11}{subpath (length(p1)/2,0) of p1}
		\fmfiset{p12}{subpath (length(p1)/2,length(p1)) of p1}
		\fmfiv{label=$\scriptscriptstyle \beta$,label.angle=-115,label.dist=4}{point 3length(p1)/4 of p1}
		\fmfiset{p2}{vpath(__v2,__v1)}
		\fmfiset{p21}{subpath (0,length(p2)/2) of p2}
		\fmfiset{p22}{subpath (length(p2)/2,length(p2)) of p2}
		\fmfiv{label=$\scriptscriptstyle \alpha$,label.angle=90,label.dist=2}{point length(p2)/2 of p2}
		\fmfi{derplain,right=0.6,label.side=right,label.dist=2,label=$\scriptscriptstyle \mu$}{p11}
		\fmfi{plain}{p12}
		\fmfi{plain}{p2}
		\end{fmfchar*}
	}
}
\equiv
\int\frac{\de^D\bar l}{(2\pi)^D}\frac{(\bar p-\bar l)^\mu}{\bar l^{2\alpha}(\bar p-\bar l)^{2\beta}}
=\frac{G(\alpha,\beta-1)}{(4\pi)^{\frac{D}{2}}}\frac{\alpha+\beta-1-\frac{D}{2}}{\beta-1}\frac{\bar p^\mu}{\bar p^{2(\alpha+\beta-\frac{D}{2})}}\eqncom
\end{equation}
where the arrow in the diagram represents a numerator momentum on that line with index $\mu$ and all occurring factors have been reshuffled to the \rhs As mentioned, the Euclidean momenta live in $d$ dimensions while the integration is taken over the $D$-dimensional space. We can also generalise this calculation to produce higher rank tensor integrals, provided that we are only interested in traceless symmetric tensors. In this case, each derivative that acts on the integral \eqref{eq:rank_1_G_function} yields zero when acting on the numerator monomial and hence we only need to act with $n$ derivatives consecutively on the denominator part of the integral. After a shift of the integration variable, we find
\begin{equation}\label{eq:G_function_traceless_symmetric}
\hat{I}^{(\mu_1\dots\mu_n)}_{(\alpha,\beta)}(\bar p)=
\settoheight{\eqoff}{$\times$}%
\setlength{\eqoff}{0.5\eqoff}%
\addtolength{\eqoff}{-2.75\unitlength}%
\raisebox{\eqoff}{
	\fmfframe(2,0)(-1,0){
		\begin{fmfchar*}(15,5.0)
		\fmfleft{vout}
		\fmfright{vin}
		\fmfforce{0.15 w, 0.5h}{v2}
		\fmfforce{0.85 w, 0.5h}{v1}
		\fmfv{decor.shape=circle,decor.filled=full,
			decor.size=2thick}{v1,v2}
		\fmf{plain}{v2,vout}
		\fmf{plain}{vin,v1}
		\fmf{phantom,left=0.6}{v1,v2,v1}
		\fmffreeze
		\fmfposition
		\fmfipath{p[]}
		\fmfipair{vv[]}
		\fmfiset{p1}{vpath(__v1,__v2)}
		\fmfiset{p11}{subpath (length(p1)/2,0) of p1}
		\fmfiset{p12}{subpath (length(p1),length(p1)/2) of p1}
		\fmfiv{label=$\scriptscriptstyle \alpha$,label.angle=-55,label.dist=4}{point length(p1)/4 of p1}
		\fmfiset{p2}{vpath(__v2,__v1)}
		\fmfiset{p21}{subpath (0,length(p2)/2) of p2}
		\fmfiset{p22}{subpath (length(p2)/2,length(p2)) of p2}
		\fmfiv{label=$\scriptscriptstyle \beta$,label.angle=90,label.dist=2}{point length(p2)/2 of p2}
		\fmfi{derplain,right=0.6,label.dist=2,label.angle=-115,label=$\scriptscriptstyle (\mu_1\dots\mu_n)$}{p12}
		\fmfi{plain}{p11}
		\fmfi{plain}{p2}
		\end{fmfchar*}
	}
}
\equiv
\int\frac{\de^D\bar l}{(2\pi)^D}\frac{\bar l^{(\mu_1\mu_2\dots \mu_n)}}{\bar l^{2\alpha}(\bar p-\bar l)^{2\beta}}
=\frac{G_{(n)}(\alpha,\beta)}{(4\pi)^{\frac{D}{2}}}\frac{\bar p^{(\mu_1\mu_2\dots\mu_n)}}{\bar p^{2(\alpha+\beta-\frac{D}{2})}}\eqncom
\end{equation}
where the parentheses in the diagram indicate that the numerator momenta are symmetrised and traceless with respect to the spacetime indices and all factors are combined into the $G$-function of rank $n$:
\begin{equation}\label{eq:Gn_definition}
G_{(n)}(\alpha,\beta)=G(\alpha-n,\beta)\prod_{i=1}^{n}\frac{\alpha-i+\beta-\frac{D}{2}}{\alpha-i}\eqndot
\end{equation}
The tensor $p^{(\mu_1\mu_2\dots \mu_n)}$ is a traceless symmetric product of the momenta $p^{\mu_1}$ -- $p^{\mu_n}$ which all live in $d$ dimensions. The momenta are, however, traceless in the $D$-dimensional space, since the $D$-dimensional integration projects occurring spacetime tensors to the $(D=d-2\epsilon)$-dimensional subspace, see \appref{sec:Renormalisation_schemes} and the references therein. The tensor can be written as\footnote{Compared to the representation in \cite{Kotikov:1995cw}, in our representation the alternating sign is hidden in the Pochhamer symbol, as can be seen using the identity $(-n)_t=(-1)^t(n-t+1)_t$.}
\begin{equation}\label{eq:p_traceless}
p^{(\mu_1\mu_2\dots \mu_n)}=
\sum_{t=0}^{\floor{\frac{n}{2}}}
\frac{p^{2t}}{2^t(2-\frac D2-n)_t}
S\bigl(\hat{\eta}^{\mu_1\mu_2}\hat{\eta}^{\mu_3\mu_4}\dots \hat{\eta}^{\mu_{2t-1}\mu_{2t}}\times p^{\mu_{2t+1}}\dots p^{\mu_n}\bigr)\eqncom
\end{equation}
where $(x)_n$ is the Pochhammer symbol, $\hat{\eta}$ the metric tensor in $D$ dimensions with $\hat{\eta}_{\mu\nu}\hat{\eta}^{\mu\nu}=D$ as in \appref{subsec:Regularisation} and the operator $S$ symmetrises the product of metrics and momenta with respect to the indices $\mu_j$. Note that the action of $S$ depends on its argument, for example $\hat{\eta}^{\mu_1\mu_2}$ is already symmetric under index exchange and the indices are not exchanged by the symmetrisation operator\footnote{When $S$ acts on $n$ indices which are grouped into $j$ different metric tensors and a symmetric remainder function $f_{\text{s}}$, the symmetrisation procedure via $S$ yields $(2j-1)!!\begin{pmatrix}n\\2j\end{pmatrix}$ different terms.}. We can also invert this relation and express a product of $n$ momenta in terms of traces and traceless symmetric parts as
\begin{equation}\label{eq:p_inptraceless}
p^{\mu_1}p^{\mu_2}\dots p^{\mu_n}=
\sum_{t=0}^{\floor{\frac{n}{2}}}
\frac{p^{2t}}{2^{t}(\frac D2-2t+n)_t}
S\bigl(\hat{\eta}^{\mu_1\mu_2}\hat{\eta}^{\mu_3\mu_4}\dots \hat{\eta}^{\mu_{2t-1}\mu_{2t}}\times p^{(\mu_{2t+1}\dots \mu_n)}\bigr)\eqndot
\end{equation}
Note that we provide an explicit implementation of \eqref{eq:p_traceless} and \eqref{eq:p_inptraceless} for {\tt Mathematica} in the package {\tt FokkenMomentumTensors.m}, which can be found in the \ttt{arXiv} source file of this thesis.
Using the last relation and \eqref{eq:G_function_traceless_symmetric}, we find the one scale integral with a monomial of $n$ momenta in the numerator
\begin{equation}
\begin{aligned}\label{eq:G_function_rank_n}
\hat{I}^{\mu_1\dots\mu_n}_{(\alpha,\beta)}(\bar p)&=
\settoheight{\eqoff}{$\times$}%
\setlength{\eqoff}{0.5\eqoff}%
\addtolength{\eqoff}{-2.75\unitlength}%
\raisebox{\eqoff}{
	\fmfframe(2,0)(-1,0){
		\begin{fmfchar*}(15,5.0)
		\fmfleft{vout}
		\fmfright{vin}
		\fmfforce{0.15 w, 0.5h}{v2}
		\fmfforce{0.85 w, 0.5h}{v1}
		\fmfv{decor.shape=circle,decor.filled=full,
			decor.size=2thick}{v1,v2}
		\fmf{plain}{v2,vout}
		\fmf{plain}{vin,v1}
		\fmf{phantom,left=0.6}{v1,v2,v1}
		\fmffreeze
		\fmfposition
		\fmfipath{p[]}
		\fmfipair{vv[]}
		\fmfiset{p1}{vpath(__v1,__v2)}
		\fmfiset{p11}{subpath (length(p1)/2,0) of p1}
		\fmfiset{p12}{subpath (length(p1),length(p1)/2) of p1}
		\fmfiv{label=$\scriptscriptstyle \alpha$,label.angle=-55,label.dist=4}{point length(p1)/4 of p1}
		\fmfiset{p2}{vpath(__v2,__v1)}
		\fmfiset{p21}{subpath (0,length(p2)/2) of p2}
		\fmfiset{p22}{subpath (length(p2)/2,length(p2)) of p2}
		\fmfiv{label=$\scriptscriptstyle \beta$,label.angle=90,label.dist=2}{point length(p2)/2 of p2}
		\fmfi{derplain,right=0.6,label.dist=2,label.angle=-115,label=$\scriptscriptstyle \mu_1\dots\mu_n$}{p12}
		\fmfi{plain}{p11}
		\fmfi{plain}{p2}
		\end{fmfchar*}
	}
}
\equiv
\int\frac{\de^D\bar l}{(2\pi)^D}\frac{\bar l^{\mu_1}\bar l^{\mu_2}\dots \bar l^{\mu_n}}{\bar l^{2\alpha}(\bar p-\bar l)^{2\beta}}\\
&=
\frac{1}{(4\pi)^{\frac{D}{2}}}
\sum_{t=0}^{\floor{\frac{n}{2}}}
\frac{G_{(n-2t)}(\alpha-t,\beta)}{2^{t}(\frac D2-2t+n)_t}
\frac{ S\bigl(\hat{\eta}^{\mu_1\mu_2}\dots \hat{\eta}^{\mu_{2t-1}\mu_{2t}}\times \bar p^{(\mu_{2t+1}\dots \mu_n)}\bigr)}{\bar p^{2(\alpha-t+\beta-\frac{D}{2})}}\eqndot
\end{aligned}
\end{equation}

Let us now briefly discuss the possibility of numerator momenta on more than one internal line. In the one-loop case discussed above, we can rewrite a tensor in the momentum $(\bar p -\bar l)$ by expanding it. The result is a sum of tensors depending on the loop momentum $\bar l$ times appropriate factors of the external momentum $\bar p$. We can then simply apply \eqref{eq:G_function_traceless_symmetric} and \eqref{eq:G_function_rank_n} to solve the loop-momentum dependent tensor integral. Diagrammatically, this amounts to shifting the numerator momentum arrows through the diagram while keeping track of the relative signs. A minimal example of this procedure is
\begin{equation}
\begin{aligned}
\int\frac{\de^D\bar l}{(2\pi)^D}\frac{\bar l^{\mu}(\bar p-\bar l)^{\nu}}{\bar l^{2}(\bar p-\bar l)^{2}}
&=
\left(\int\frac{\de^D\bar l}{(2\pi)^D}\frac{\bar l^{\mu}}{\bar l^{2}(\bar p-\bar l)^{2}}\right)\bar p^{\nu}
&-&\int\frac{\de^D\bar l}{(2\pi)^D}\frac{\bar l^{\mu}\bar l^{\nu}}{\bar l^{2}(\bar p-\bar l)^{2}}&
\\[0.5\baselineskip]
\settoheight{\eqoff}{$\times$}%
\setlength{\eqoff}{0.5\eqoff}%
\addtolength{\eqoff}{-2.75\unitlength}%
\raisebox{\eqoff}{
	\fmfframe(0,0)(-1,0){
		\begin{fmfchar*}(15,5.0)
		\fmfleft{vout}
		\fmfright{vin}
		\fmfforce{0.15 w, 0.5h}{v2}
		\fmfforce{0.85 w, 0.5h}{v1}
		\fmfv{decor.shape=circle,decor.filled=full,
			decor.size=2thick}{v1,v2}
		\fmf{plain}{v2,vout}
		\fmf{plain}{vin,v1}
		\fmf{phantom,left=0.6}{v1,v2,v1}
		\fmffreeze
		\fmfposition
		\fmfipath{p[]}
		\fmfipair{vv[]}
		\fmfiset{p1}{vpath(__v1,__v2)}
		\fmfiset{p13}{subpath (length(p1),0) of p1}
		\fmfiset{p12}{subpath (length(p1),length(p1)/2) of p1}
		\fmfi{plain}{p1}
		\fmfiset{p2}{vpath(__v2,__v1)}
		\fmfiset{p21}{subpath (0,length(p2)/2) of p2}
		\fmfiset{p22}{subpath (length(p2)/2,length(p2)) of p2}
		\fmfi{plain}{p2}
		\fmfi{derplain,right=0.6,label.dist=4,label.angle=-115,label=$\scriptscriptstyle \nu$}{p13}
		\fmfi{derplain,right=0.6,label.dist=4,label.angle=-115,label=$\scriptscriptstyle \mu$}{p2}
		\end{fmfchar*}
	}
}\quad
&=
\qquad\quad
\settoheight{\eqoff}{$\times$}%
\setlength{\eqoff}{0.5\eqoff}%
\addtolength{\eqoff}{-2.75\unitlength}%
\raisebox{\eqoff}{
	\fmfframe(0,0)(-1,0){
		\begin{fmfchar*}(15,5.0)
		\fmfleft{vout}
		\fmfright{vin}
		\fmfforce{0.15 w, 0.5h}{v2}
		\fmfforce{0.65 w, 0.5h}{v1}
		\fmfv{decor.shape=circle,decor.filled=full,
			decor.size=2thick}{v1,v2}
		\fmf{plain}{v2,vout}
		\fmf{plain}{vin,v1}
		\fmf{phantom,left=0.6}{v1,v2,v1}
		\fmffreeze
		\fmfposition
		\fmfipath{p[]}
		\fmfipair{vv[]}
		\fmfiset{p1}{vpath(__v1,__v2)}
		\fmfiset{p13}{subpath (length(p1),0) of p1}
		\fmfiset{p12}{subpath (length(p1),length(p1)/2) of p1}
		\fmfi{plain}{p1}
		\fmfiset{p2}{vpath(__v2,__v1)}
		\fmfiset{p21}{subpath (0,length(p2)/2) of p2}
		\fmfiset{p22}{subpath (length(p2)/2,length(p2)) of p2}
		\fmfi{plain}{p2}
		\fmfi{derplain,right=0.6,label.dist=4,label=$\scriptscriptstyle \mu$}{p2}
		\fmfiset{p3}{vpath(__v1,__vin)}
		\fmfiset{p31}{subpath (length(p3),0) of p3}
		\fmfi{derplain,left=0.6,label.dist=4,label=$\scriptscriptstyle \nu$}{p31}
		\end{fmfchar*}
	}
}
&-&
\qquad
\settoheight{\eqoff}{$\times$}%
\setlength{\eqoff}{0.5\eqoff}%
\addtolength{\eqoff}{-2.75\unitlength}%
\raisebox{\eqoff}{
	\fmfframe(0,0)(-1,0){
		\begin{fmfchar*}(15,5.0)
		\fmfleft{vout}
		\fmfright{vin}
		\fmfforce{0.15 w, 0.5h}{v2}
		\fmfforce{0.85 w, 0.5h}{v1}
		\fmfv{decor.shape=circle,decor.filled=full,
			decor.size=2thick}{v1,v2}
		\fmf{plain}{v2,vout}
		\fmf{plain}{vin,v1}
		\fmf{phantom,left=0.6}{v1,v2,v1}
		\fmffreeze
		\fmfposition
		\fmfipath{p[]}
		\fmfipair{vv[]}
		\fmfiset{p1}{vpath(__v1,__v2)}
		\fmfiset{p13}{subpath (length(p1),0) of p1}
		\fmfiset{p12}{subpath (length(p1),length(p1)/2) of p1}
		\fmfi{plain}{p1}
		\fmfiset{p2}{vpath(__v2,__v1)}
		\fmfiset{p21}{subpath (0,length(p2)/2) of p2}
		\fmfiset{p22}{subpath (length(p2)/2,length(p2)) of p2}
		\fmfi{plain}{p2}
		\fmfi{derplain,right=0.6,label.dist=4,label.angle=-115,label=$\scriptscriptstyle \mu\nu$}{p2}
		\end{fmfchar*}
	}
}
&\eqndot 
\end{aligned}
\end{equation}
In higher loop examples, this procedure becomes more involved but ultimately it is possible to reduce a given numerator polynomial to a set of irreducible numerators whose explicit structure depend on the type of integral we are interested in.

Finally, if we have contracted numerator momenta on two different lines within a diagram, we can use the completion of squares to rewrite the contracted pairs in terms of squares of the denominator momenta or the irreducible numerator momenta. The simplest example of this is
\begin{equation}
\begin{aligned}
\settoheight{\eqoff}{$\times$}%
\setlength{\eqoff}{0.5\eqoff}%
\addtolength{\eqoff}{-2.75\unitlength}%
\raisebox{\eqoff}{
	\fmfframe(0,0)(-1,0){
		\begin{fmfchar*}(15,5.0)
		\fmfleft{vout}
		\fmfright{vin}
		\fmfforce{0.15 w, 0.5h}{v2}
		\fmfforce{0.85 w, 0.5h}{v1}
		\fmfv{decor.shape=circle,decor.filled=full,
			decor.size=2thick}{v1,v2}
		\fmf{plain}{v2,vout}
		\fmf{plain}{vin,v1}
		\fmf{phantom,left=0.6}{v1,v2,v1}
		\fmffreeze
		\fmfposition
		\fmfipath{p[]}
		\fmfipair{vv[]}
		\fmfiset{p1}{vpath(__v1,__v2)}
		\fmfiset{p11}{subpath (length(p1)/2,0) of p1}
		\fmfiset{p12}{subpath (length(p1),length(p1)/2) of p1}
		\fmfiv{
			label=$\scriptscriptstyle \alpha$,label.angle=-55,label.dist=2}{point length(p1)/4 of p1}
		\fmfi{plain}{p1}
		\fmfiset{p2}{vpath(__v2,__v1)}
		\fmfiset{p21}{subpath (0,length(p2)/2) of p2}
		\fmfiset{p22}{subpath (length(p2)/2,length(p2)) of p2}
		\fmfiv{
			label=$\scriptscriptstyle \beta$,label.angle=55,label.dist=2}{point 3length(p2)/4 of p2}
		\fmfi{plain}{p2}
		\fmfi{derplain,right=0.6,label.dist=4,label.angle=-115,label=$\scriptscriptstyle \mu$}{p12}
		\fmfi{derplain,right=0.6,label.dist=4,label.angle=-115,label=$\scriptscriptstyle \mu$}{p21}
		\end{fmfchar*}
	}
}
&=
\int\frac{\de^D\bar l}{(2\pi)^D}\frac{\bar l_\mu (\bar p-\bar l)^\mu}{\bar l^{2\alpha}(\bar p-\bar l)^{2\beta}}
=\frac 12\int\frac{\de^D\bar l}{(2\pi)^D}\frac{\bar p^2-\bar l^2-(\bar p-\bar l)^2}{\bar l^{2\alpha}(\bar p-\bar l)^{2\beta}}
\\[0.5\baselineskip]
&=
\frac 12\Bigl(
\settoheight{\eqoff}{$\times$}%
\setlength{\eqoff}{0.5\eqoff}%
\addtolength{\eqoff}{-2.75\unitlength}%
\raisebox{\eqoff}{
	\fmfframe(0,0)(-1,0){
		\begin{fmfchar*}(15,5.0)
		\fmfleft{vout}
		\fmfright{vin}
		\fmfforce{0.15 w, 0.5h}{v2}
		\fmfforce{0.65 w, 0.5h}{v1}
		\fmfv{decor.shape=circle,decor.filled=full,
			decor.size=2thick}{v1,v2}
		\fmf{plain}{v2,vout}
		\fmf{plain}{vin,v1}
		\fmf{phantom,left=0.6}{v1,v2,v1}
		\fmffreeze
		\fmfposition
		\fmfipath{p[]}
		\fmfipair{v[]}
		\fmfiset{p1}{vpath(__v1,__v2)}
		\fmfiset{p11}{subpath (0,length(p1)/2) of p1}
		\fmfiset{p12}{subpath (length(p1)/2,1) of p1}
		\fmfiv{
			label=$\scriptscriptstyle \alpha$,label.angle=-90,label.dist=2}{point length(p1)/2 of p1}
		\fmfi{plain}{p1}
		\fmfiset{p2}{vpath(__v2,__v1)}
		\fmfiset{p21}{subpath (0,length(p2)/2) of p2}
		\fmfiset{p22}{subpath (length(p2)/2,1) of p2}
		\fmfiv{
			label=$\scriptscriptstyle \beta$,label.angle=90,label.dist=2}{point length(p2)/2 of p2}
		\fmfi{plain}{p2}
		\fmfiset{p3}{vpath(__v1,__vin)}
		\fmfiv{decor.shape=circle,decor.filled=full,decor.size=2thick}{point length(p3)/2 of p3}
		\end{fmfchar*}
	}
}
-\settoheight{\eqoff}{$\times$}%
\setlength{\eqoff}{0.5\eqoff}%
\addtolength{\eqoff}{-2.75\unitlength}%
\raisebox{\eqoff}{
	\fmfframe(0,0)(-1,0){
		\begin{fmfchar*}(15,5.0)
		\fmfleft{vout}
		\fmfright{vin}
		\fmfforce{0.15 w, 0.5h}{v2}
		\fmfforce{0.85 w, 0.5h}{v1}
		\fmfv{decor.shape=circle,decor.filled=full,
			decor.size=2thick}{v1,v2}
		\fmf{plain}{v2,vout}
		\fmf{plain}{vin,v1}
		\fmf{phantom,left=0.6}{v1,v2,v1}
		\fmffreeze
		\fmfposition
		\fmfipath{p[]}
		\fmfipair{v[]}
		\fmfiset{p1}{vpath(__v1,__v2)}
		\fmfiset{p11}{subpath (0,length(p1)/2) of p1}
		\fmfiset{p12}{subpath (length(p1)/2,1) of p1}
		\fmfiv{
			label=$\scriptscriptstyle \alpha-1$,label.angle=-90,label.dist=2}{point length(p1)/2 of p1}
		\fmfi{plain}{p1}
		\fmfiset{p2}{vpath(__v2,__v1)}
		\fmfiset{p21}{subpath (0,length(p2)/2) of p2}
		\fmfiset{p22}{subpath (length(p2)/2,1) of p2}
		\fmfiv{
			label=$\scriptscriptstyle \beta$,label.angle=90,label.dist=2}{point length(p2)/2 of p2}
		\fmfi{plain}{p2}
		\end{fmfchar*}
	}
}
-
\settoheight{\eqoff}{$\times$}%
\setlength{\eqoff}{0.5\eqoff}%
\addtolength{\eqoff}{-2.75\unitlength}%
\raisebox{\eqoff}{
	\fmfframe(0,0)(-1,0){
		\begin{fmfchar*}(15,5.0)
		\fmfleft{vout}
		\fmfright{vin}
		\fmfforce{0.15 w, 0.5h}{v2}
		\fmfforce{0.85 w, 0.5h}{v1}
		\fmfv{decor.shape=circle,decor.filled=full,
			decor.size=2thick}{v1,v2}
		\fmf{plain}{v2,vout}
		\fmf{plain}{vin,v1}
		\fmf{phantom,left=0.6}{v1,v2,v1}
		\fmffreeze
		\fmfposition
		\fmfipath{p[]}
		\fmfipair{v[]}
		\fmfiset{p1}{vpath(__v1,__v2)}
		\fmfiset{p11}{subpath (0,length(p1)/2) of p1}
		\fmfiset{p12}{subpath (length(p1)/2,1) of p1}
		\fmfiv{
			label=$\scriptscriptstyle \alpha$,label.angle=-90,label.dist=2}{point length(p1)/2 of p1}
		\fmfi{plain}{p1}
		\fmfiset{p2}{vpath(__v2,__v1)}
		\fmfiset{p21}{subpath (0,length(p2)/2) of p2}
		\fmfiset{p22}{subpath (length(p2)/2,1) of p2}
		\fmfiv{
			label=$\scriptscriptstyle \beta-1$,label.angle=90,label.dist=2}{point length(p2)/2 of p2}
		\fmfi{plain}{p2}
		\end{fmfchar*}
	}
}
\Bigr)\eqndot
\end{aligned}
\end{equation}

\subsection{Products of \texorpdfstring{$\sigma$-matrices}{sigma-deformation}}\label{subsec:products_of_sigma}
Apart from the (tensor-) integrals, also products of $\gamma$-matrices enter the calculation of Feynman diagrams, whenever fermions are involved. When we employ the Feynman rules of \appref{app:Feynman_rules} where the fermions are in the Weyl representation, these products turn into alternating products of $\sigma$- and $\bar{\sigma}$-matrices. Here, we provide the identities necessary to reduce such alternating products in strictly four dimensional spacetime. That is to say, the reduction in this appendix is valid in any regularisation procedure that leaves the spacetime strictly in $D=4$ dimensions and in the DR regularisation procedure with anticommuting $\hat{\gamma}_5$, see \subsecref{subsec:Regularisation} for details. For differences in the dimensional regularisation procedure, where the concept of Weyl fermions breaks down, see also \subsecref{subsec:Regularisation}.

To evaluate tensors of the form given in \eqref{eq:sigma_matrix_product}, we give a set of identities which suffices to reduce any given product that contains only $\sigma$- and $\bar{\sigma}$-matrices in the beginning. We first need the product of a $\sigma$- and a $\bar \sigma$-matrix
\begin{equation}
\begin{aligned}\label{eq:sigma_product}
(\sigma_\mu)_{\alpha\dot{\alpha}}(\bar\sigma_\nu)^{\dot{\alpha}\beta}
&=-\delta_{\alpha}^{\phan{\alpha}\beta}\eta_{\mu\nu}-2\complexi (\sigma_{\mu\nu})_{\alpha}^{\phan{\alpha}\beta}\eqncom\\
(\bar\sigma_\mu)^{\dot{\alpha}\alpha}(\sigma_\nu)_{\alpha\dot\beta}
&=-\delta^{\dot{\alpha}}_{\phan{\alpha}\dot\beta}\eta_{\mu\nu}-2\complexi
(\bar\sigma_{\mu\nu})^{\dot{\alpha}}_{\phan{\alpha}\dot\beta}\eqncom
\end{aligned}
\end{equation}
where the antisymmetric products $\sigma_{\mu\nu}$ and $\bar\sigma_{\mu\nu}$ are explicitly given in \eqref{eq:sigma_mu_nu}.
Next, we need all possible products of of $\sigma$-matrices and these new objects:
\begin{equation}
\begin{aligned}
(\sigma_{\mu_1\mu_2})_\alpha^{\phan{\alpha}\gamma}(\sigma_{\mu_3})_{\gamma\dot{\alpha}}&=
\frac{\complexi}{2}\bigl[\eta_{\mu_1\mu_3}(\sigma_{\mu_2})_{\alpha\dot{\alpha}}
-\eta_{\mu_2\mu_3}(\sigma_{\mu_1})_{\alpha\dot{\alpha}}+\complexi\varepsilon_{\mu_1\mu_2\mu_3\nu}(\sigma^\nu)_{\alpha\dot{\alpha}}\bigr]\eqncom\\
(\bar\sigma_{\mu_3})^{\dot\alpha\gamma}(\sigma_{\mu_1\mu_2})_\gamma^{\phan{\alpha}\alpha}&=
-\frac{\complexi}{2}\bigl[\eta_{\mu_1\mu_3}(\bar\sigma_{\mu_2})^{\dot\alpha\alpha}
-\eta_{\mu_2\mu_3}(\bar\sigma_{\mu_1})^{\dot\alpha\alpha}+\complexi\varepsilon_{\mu_1\mu_2\mu_3\nu}(\bar\sigma^\nu)^{\dot\alpha\alpha}\bigr]\eqncom\\
(\bar\sigma_{\mu_1\mu_2})^{\dot\alpha}_{\phan{\alpha}\dot{\gamma}}(\bar\sigma_{\mu_3})^{\dot\gamma\alpha}&=
\frac{\complexi}{2}\bigl[\eta_{\mu_1\mu_3}(\bar\sigma_{\mu_2})^{\dot\alpha\alpha}
-\eta_{\mu_2\mu_3}(\bar\sigma_{\mu_1})^{\dot\alpha\alpha}-\complexi\varepsilon_{\mu_1\mu_2\mu_3\nu}(\bar\sigma^\nu)^{\dot\alpha\alpha}\bigr]\eqncom\\
(\sigma_{\mu_3})_{\alpha\dot{\gamma}}(\bar\sigma_{\mu_1\mu_2})^{\dot\gamma}_{\phan{\alpha}\dot{\alpha}}&=
-\frac{\complexi}{2}\bigl[\eta_{\mu_1\mu_3}(\sigma_{\mu_2})_{\alpha\dot{\alpha}}
-\eta_{\mu_2\mu_3}(\sigma_{\mu_1})_{\alpha\dot{\alpha}}-\complexi\varepsilon_{\mu_1\mu_2\mu_3\nu}(\sigma^\nu)_{\alpha\dot{\alpha}}\bigr]\eqncom\\
(\sigma_{\mu_1\mu_2})_\alpha^{\phan{\alpha}\gamma}(\sigma_{\mu_3\mu_4})_\gamma^{\phan{\alpha}\beta}&=
\frac{1}{4}\delta_\alpha^{\phan{\alpha}\beta}\bigl[\eta_{\mu_1\mu_3}\eta_{\mu_2\mu_4}-\eta_{\mu_1\mu_4}\eta_{\mu_2\mu_3}+\complexi\varepsilon_{\mu_1\mu_2\mu_3\mu_4}\bigr]\\
&\phan{=}+\frac{\complexi}{2}\bigl[
\eta_{\mu_1\mu_3}(\sigma_{\mu_2\mu_4})_\alpha^{\phan{\alpha}\beta}
+\eta_{\mu_1\mu_4}(\sigma_{\mu_3\mu_2})_\alpha^{\phan{\alpha}\beta}
-(\mu_1\leftrightarrow\mu_2)
\bigr]\eqncom\\
(\bar\sigma_{\mu_1\mu_2})^{\dot{\alpha}}_{\phan{\alpha}\dot\gamma}(\bar\sigma_{\mu_3\mu_4})^{\dot{\gamma}}_{\phan{\alpha}\dot\beta}&=
\frac{1}{4}\delta^{\dot\alpha}_{\phan{\alpha}\dot\beta}\bigl[\eta_{\mu_1\mu_3}\eta_{\mu_2\mu_4}-\eta_{\mu_1\mu_4}\eta_{\mu_2\mu_3}-\complexi\varepsilon_{\mu_1\mu_2\mu_3\mu_4}\bigr]\\
&\phan{=}+\frac{\complexi}{2}\bigl[
\eta_{\mu_1\mu_3}(\bar\sigma_{\mu_2\mu_4})^{\dot{\alpha}}_{\phan{\alpha}\dot\beta}
+\eta_{\mu_1\mu_4}(\bar\sigma_{\mu_3\mu_2})^{\dot{\alpha}}_{\phan{\alpha}\dot\beta}
-(\mu_1\leftrightarrow\mu_2)
\bigr]\eqndot
\end{aligned}
\end{equation}
Finally, the all products that contain a Levi-Civita tensor can be reduced using 
\begin{equation}
\begin{aligned}
\varepsilon_{\mu_1\mu_2\nu\rho}(\sigma^{\nu\rho})_\alpha^{\phan{\alpha}\beta}&=-2\complexi (\sigma_{\mu_1\mu_2})_\alpha^{\phan{\alpha}\beta}\eqncom\\
\varepsilon_{\mu_1\mu_2\nu\rho}(\bar\sigma^{\nu\rho})^{\dot{\alpha}}_{\phan{\alpha}\dot\beta}&=2\complexi (\bar\sigma_{\mu_1\mu_2})^{\dot{\alpha}}_{\phan{\alpha}\dot\beta}\eqncom\\
\varepsilon_{\mu_1\mu_2\mu_3\nu}(\sigma_{\mu_4}{}^{\nu})_\alpha^{\phan{\alpha}\beta}&=-\complexi\bigl[
\eta_{\mu_1\mu_4}(\sigma_{\mu_2\mu_3})_\alpha^{\phan{\alpha}\beta}
-\eta_{\mu_2\mu_4}(\sigma_{\mu_1\mu_3})_\alpha^{\phan{\alpha}\beta}
+\eta_{\mu_3\mu_4}(\sigma_{\mu_1\mu_2})_\alpha^{\phan{\alpha}\beta}
\bigr] \eqncom\\
\varepsilon_{\mu_1\mu_2\mu_3\nu}(\bar\sigma_{\mu_4}{}^{\nu})^{\dot{\alpha}}_{\phan{\alpha}\dot\beta}&=\complexi\bigl[
\eta_{\mu_1\mu_4}(\bar\sigma_{\mu_2\mu_3})^{\dot{\alpha}}_{\phan{\alpha}\dot\beta}
-\eta_{\mu_2\mu_4}(\bar\sigma_{\mu_1\mu_3})^{\dot{\alpha}}_{\phan{\alpha}\dot\beta}
+\eta_{\mu_3\mu_4}(\bar\sigma_{\mu_1\mu_2})^{\dot{\alpha}}_{\phan{\alpha}\dot\beta}
\bigr] \eqncom\\
\varepsilon^{\mu\nu\rho\sigma}\varepsilon_{\alpha\beta\gamma\sigma}&=
-\delta^\mu_{(\alpha}\delta^\nu_\beta\delta^\rho_{\gamma)}+\delta^\mu_{(\beta}\delta^\nu_\alpha\delta^\rho_{\gamma)}\eqncom\\
\varepsilon^{\mu\nu\rho\sigma}\varepsilon_{\alpha\beta\rho\sigma}&=-2\delta^\mu_{[\alpha}\delta^\nu_{\beta]}\eqncom\\
\varepsilon^{\mu\nu\rho\sigma}\varepsilon_{\alpha\nu\rho\sigma}&=-6\delta^\mu_\alpha\eqncom
\end{aligned}
\end{equation}
where the parentheses around indices indicate that the indices are cyclically symmetrised and the brackets indicate that the indices are antisymmetrised. 

Since we reduce the products of $\sigma$-matrices in $D=4$ dimensions, all indices are raised and lowered with the Minkowski space metric $\eta$. Reductions in Feynman diagrams that yield a non-vanishing contribution involving $\varepsilon_{\mu_1\mu_2\mu_3\mu_4}$ must be treated with caution, as we work in the DR scheme in a quasi-4-dimensional space where this tensor does not exist, see \subsecref{subsec:Regularisation} for details.

\section{Fourier transformation of the free two-point function}\label{sec:the-fourier-transformation-of-the-free-two-point-function}
In this appendix, we present the Fourier transformation of the free two-point correlation function. For simplicity, we restrict to the purely scalar state $\cO_L(x)=(L!)^{-\frac 12}\tr(\varphi^L)$ of the $\varphi^3$-theory that we discussed in \secref{sec:phi3_theory}. A complete set of Feynman rules for this theory compatible with our conventions is given in \cite[\chap{9}]{Srednicki:2007}.

In $D$-dimensional Minkowski configuration space, the free two-point function is given by \eqref{eq:2pt_func_classical} and explicitly reads
\begin{equation}
\vacl\T\mathcal{O}_L(x)\mathcal{O}_L(0)\vac=\frac{1}{(x^2+\complexi \epsilon)^{\Delta^0_{\cO_L}}}\eqncom
\end{equation}
where the classical scaling dimension the composite operator $\Delta^0_{\cO_L}=\frac L2(D-2)$ is the sum of its constituent classical scaling dimensions and these were given in \eqref{eq:phi3_dimensions}. To Fourier transform this expression, we Wick rotate to Euclidean space, perform the transformation and then perform an inverse Wick rotation back to Minkowski space. For a two-point function with generic classical scaling dimension $\Delta$, we obtain
\begin{equation}
\begin{aligned}\label{eq:Fourier-transform}
\int\de^D x\frac{\e^{-\complexi p x}}{(x^2+\complexi \epsilon)^{\Delta}}
&=\complexi \WR^{-1}
\int_0^\infty\de \abs{\bar x} \abs{\bar{x}}^{D-1}\int_0^\pi\de\bar\vartheta \sin^{D-2}(\bar\vartheta)\int \de\bar \Omega_{D-1}\frac{\e^{-\complexi \abs{\bar p}\abs{\bar x}\cos \bar\vartheta}}{\abs{\bar x}^{2\Delta}}
\\
&=\complexi (2\pi)^{\frac D2}\int_0^\infty\de u u^{\frac D2-2\Delta}J_{\frac{D-2}{2}}(u)
\WR^{-1}\frac{1}{\bar p^{2(\frac D2-\Delta)}}
\\
&=\complexi 2^{D-2\Delta}\pi^{\frac D2}\frac{\Gamma(\frac D2-\Delta)}{\Gamma(\Delta)}
\frac{1}{(p^2-\complexi \epsilon)^{(\frac D2-\Delta)}}\eqncom
\end{aligned}
\end{equation}
where barred quantities live in Euclidean space and we substituted $u=\abs{\bar x}\abs{\bar p}$. The first kind Bessel function $J_n(x)$ occurs in the integration over $\bar \vartheta$ and the integration over the $(D-1)$-dimensional unit sphere is obtained from $\int \de \bar{\Omega}_{D}=2\pi^{\frac D2}\Gamma^{-1}(\frac D2)$. The divergence that occurs in this Fourier transform for $2\Delta\geq D$ originates from the integration over the origin and arises when the operators in the position space two-point function approach a coincident point.

In \eqref{eq:free_2pt_comp_op} we calculated the free reduced momentum space two-point function for the operator $\cO_2=\frac{1}{2!} \varphi^2$ explicitly from Feynman diagrams. There, we found that it is proportional to $\vacl\T \cO_2(p)\cO_2(-p)\vac_{\complexi\cT}|_{g=0}\sim \Gamma(2-\frac D2)$, which matches the divergence of \eqref{eq:Fourier-transform} for $\Delta_{\cO_2}=D-2$. We can also generalise the calculation of \eqref{eq:free_2pt_comp_op} to composite operators with generic lengths $\cO_{L+1}$ with classical scaling dimension $\Delta_{\cO_{L+1}}=\frac{L+1}{2}(D-2)$. In this case, all $L$ loop integrals can be performed consecutively and we find
\begin{equation}
\begin{aligned}
\vacl\T \cO_{L+1}(p)\cO_{L+1}(-p)\vac_{\complexi\cT}|_{g=0}
&=
\settoheight{\eqoff}{$\times$}%
\setlength{\eqoff}{0.5\eqoff}%
\addtolength{\eqoff}{-2.75\unitlength}%
\raisebox{\eqoff}{
	\fmfframe(0,0)(-1,0){
		\begin{fmfchar*}(10,5.0)
		\fmfforce{0 w, 0.5h}{v2}
		\fmfforce{1 w, 0.5h}{v1}
		\fmfv{decor.shape=circle,decor.filled=full,
			decor.size=2thick}{v1,v2}
		\fmf{plain,left=0.6}{v1,v2,v1}
		\fmf{phantom,left=0.2,tag=1}{v1,v2,v1}
		\fmffreeze
		\fmfposition
		\fmfipath{p[]}
		\fmfipair{v[]}
		\fmfiset{p1}{vpath1(__v1,__v2)}
		\fmfiset{p2}{vpath1(__v2,__v1)}
		\fmfi{dots}{point length(p1)/2 of p1 -- point length(p2)/2 of p2}
		\end{fmfchar*}
	}
}
=
\frac{\WR^{-1}}{\complexi L!(2\pi)^{LD}} \int 
\frac{\de^D \bar l_1\dots \de^D \bar l_L}{\bar l_1^2\prod_{j=1}^{L-1}(\bar l_j-\bar l_{j+1})^2(\bar p-\bar l_L)^2}
\\
&=
\frac{\Gamma^{L+1}(\frac D2-1)\Gamma(L+1-\frac{DL}{2})}{\complexi L!(4\pi)^{\frac{LD}{2}}\Gamma(\frac{L+1}{2}(D-2))}
\frac{1}{(p^{2}-\complexi\epsilon)^{(L+1-\frac{LD}{2})}}
\eqncom
\end{aligned}
\end{equation}
where the integrals can be solved consecutively in terms of $G$-functions resulting in a telescope product of $\Gamma$ functions. This time, the divergence stems from the second $\Gamma$ function in the numerator and comparing with \eqref{eq:Fourier-transform}, we find that this is precisely the divergence that occurs for a scalar length-$(L+1)$ operator in the Fourier transformation.

\section{The harmonic action}\label{app:harmonic_action}
In this appendix, we present the zero- and one-loop dilatation operator density of \NfSYMt in the oscillator representation.

In \subsecref{sec:Building_blocs}, we presented the mapping of length-$L$ single-trace composite operators to cyclic spin-chain states with $L$-sites. The cyclic spin-chain states are constructed from non-cyclic states that are characterised in terms of a vector $\ket{A}$ of oscillator occupation numbers $A_i$ at each site.

We can write the action of the planar zero and one-loop dilatation operator of \eqref{eq:D_cl_on_O} and \eqref{eq:Dila_N4_one_loop} in the oscillator representation in terms of range $R\leq 2$ dilatation operator densities $\mathfrak{D}_{0}$ and $\mathfrak{D}_{2}$ which act on these non-cyclic states. For \NfSYMt and its deformations the zero-loop density follows immediately from \eqref{eq:symmetry_generators_osci_LR}:
\begin{equation}\label{eq: def classical dilatation op in osc language}
(\mathfrak{D}_{0})_{A_i}^{A_j}=\bra{A_j}D_{0}\ket{A_i}=\delta_{A_i}^{A_j}\Bigl(1+\frac12\sum_{\alpha=1}^{2}\akindsite[\alpha]{i}
+\frac12\sum_{\dot\alpha=1}^{2}\bkindsite[\dot{\alpha}]{i}\Bigr)\eqndot
\end{equation}
The one-loop density of the parent \NfSYMt was derived in \cite{Beisert:2003jj}, and in \cite{Fokken:2013mza} it was written explicitly in terms of the occupation numbers \eqref{eq:occupation_numbers} of the two incoming fields $A_1$, $A_2$ and outgoing fields $A_3$, $A_4$ at sites $i$ and $i+1$ as
\begin{align}\label{eq: harminic action}
(\diladensity^{\cN=4}_{2})_{A_{1}A_{2}}^{A_{3}A_{4}}&=
\bra{A_3A_4}\mathfrak{D}_2^{\cN=4}\ket{A_1A_2}
\notag\\
&=
\Biggl[
\prod_{\alpha=1}^2\Biggl(\sum_{\akind[\alpha]=\maxset{\akindsite[\alpha]{3} - \akindsite[\alpha]{2},
		0}}^{\minset{\akindsite[\alpha]{1}, \akindsite[\alpha]{3}}}
\binom{\akindsite[\alpha]{1}}{\akind[\alpha]}
\binom{\akindsite[\alpha]{2}}{\akindsite[\alpha]{3} - \akind[\alpha]} \Biggr)
\notag\\
&\phaneq \prod_{\alphadot=1}^{2}\Biggl(\sum_{\bkind[\alphadot]=\maxset{\bkindsite[\alphadot]{3} - \bkindsite[\alphadot]{2},
		0}}^{\minset{\bkindsite[\alphadot]{1}, \bkindsite[\alphadot]{3}}}
\binom{\bkindsite[\alphadot]{1}}{\bkind[\alphadot]}
\binom{\bkindsite[\alphadot]{2}}{\bkindsite[\alphadot]{3} - \bkind[\alphadot]}\Biggr) 
\notag\\
&\phaneq \prod_{a=1}^4\Biggl(\sum_{\ckind[a]=\maxset{\ckindsite[a]{3} - \ckindsite[a]{2},
		0}}^{\minset{\ckindsite[a]{1}, \ckindsite[a]{3}}}
\binom{\ckindsite[a]{1}}{\ckind[a]}
\binom{\ckindsite[a]{2}}{\ckindsite[a]{3} - \ckind[a]}\Biggr)
\notag\\
& \phaneq c_{\text{h}}\Big[{\textstyle \sum_{i=1}^2 \big(\sum_{\beta=1}^2 \akindsite[\beta]{i}+\sum_{\betadot=1}^{2} \bkindsite[\betadot]{i}+\sum_{B=1}^4 \ckindsite[B]{i}\big)}
,\notag\\
&\phaneq \hphantom{c\Big[} {\textstyle \sum_{\beta=1}^2 (\akindsite[\beta]{1}-\akind[\beta])+\sum_{\betadot=1}^{2} (\bkindsite[\betadot]{1}-\bkind[\betadot])+\sum_{B=1}^4 (\ckindsite[B]{1}-\ckind[B])}
,\notag\\
&\phaneq\hphantom{c\Big[}{\textstyle \sum_{\beta=1}^2 (\akindsite[\beta]{3}-\akind[\beta])+\sum_{\betadot=1}^{2} (\bkindsite[\betadot]{3}-\bkind[\betadot])+\sum_{B=1}^4 (\ckindsite[B]{3}-\ckind[B]) }
\Big]
\displaybreak\notag\\
&\phaneq(-1)^{(\ckindsite[1]{1} - \ckind[1]
	+ \ckindsite[2]{1} - \ckind[2]
	+ \ckindsite[3]{1} - \ckind[3]
	+ \ckindsite[4]{1} - \ckind[4])
	(\ckindsite[1]{3} - \ckind[1]
	+ \ckindsite[2]{3} - \ckind[2]
	+ \ckindsite[3]{3} - \ckind[3]
	+ \ckindsite[4]{3} - \ckind[4])}
\notag\\
&\phaneq(-1)^{( \ckind[2]
	+ \ckind[3]
	+ \ckind[4])
	(\ckindsite[1]{1}+\ckindsite[1]{3})}
\notag\\
&\phaneq(-1)^{(\ckindsite[1]{2} - \ckindsite[1]{3} + \ckind[1]
	+ \ckind[3]
	+ \ckind[4])
	(\ckindsite[2]{1}+\ckindsite[2]{3})}
\notag\\
&\phaneq(-1)^{(\ckindsite[1]{2} - \ckindsite[1]{3} + \ckind[1]
	+\ckindsite[2]{2} - \ckindsite[2]{3} + \ckind[2]
	+ \ckind[4])
	(\ckindsite[3]{1}+\ckindsite[3]{3})}
\notag\\
&\phaneq(-1)^{(\ckindsite[1]{2} - \ckindsite[1]{3} + \ckind[1]
	+\ckindsite[2]{2} - \ckindsite[2]{3} + \ckind[2]
	+\ckindsite[3]{2} - \ckindsite[3]{3} + \ckind[3])
	(\ckindsite[4]{1} + \ckindsite[4]{3})} 
\Biggr]
\eqncom
\end{align}
where the occupation numbers fulfil $A_{1}+A_{2}=A_{3}+A_{4}$ and the indices $a^\alpha$, $b^\alpha$ and $c^a$ characterise how many oscillators stay at their initial sites. The coefficient $c_{\text{h}}[n,n_{12},n_{21}]$ gives a weight to a particular transition depending on the total number of oscillators $n=n_1+n_2$ and the number $n_{ij}$ of oscillators that hop from site $i$ to $j$ and vice versa. It is called the harmonic action and it is given in terms of the harmonic numbers $h(k)=\sum_{i=1}^k\frac{1}{i}$ and the Euler gamma function as 
\begin{equation}\label{eq: harmonic action coefficient}
c_{\text{h}}[n,n_{12},n_{21}]=\begin{cases}2 h(\frac12 n) & \text{if }n_{12}=n_{21}=0 \eqncom\\
2(-1)^{1+n_{12}n_{21}} 
\frac{\Gamma(\frac12(n_{12}+n_{21}))\Gamma(1+\frac12(n-n_{12}-n_{21}))}{\Gamma(1+\frac12
	n)} & \text{else.}
\end{cases}
\end{equation}

\section{Scalar one-loop self-energy}
\label{app:oneloopse}

Using the Feynman rules \ref{app:Feynman_rules} and relations \eqref{rhorhotr}, the one-loop self-energy contributions to the scalar propagators in the $\gamma_i$-deformation are given by
\begin{equation}
\begin{aligned}\label{sephi}
\Bigl(\swfone[
\fmfiv{label=$\scriptstyle i a$,label.angle=-60,l.dist=2}{vloc(__v1)}
\fmfiv{label=$\scriptstyle j b$,label.angle=-120,l.dist=2}{vloc(__v2)}
]{plain_rar}{dashes_rar,left=1}{dashes_ar,left=1}\Bigr)_{\text{am}}
&=-4\lambda  p^2 \hat{I}_{(1,1)}(p)\delta_i^j
\Big[\big(ab\big)-\cos\gamma_i^-\frac 1N\big(a\big)\big(b\big)\Big]
\eqncom\\
\Bigl(\swfone[
\fmfiv{label=$\scriptstyle i a$,label.angle=-60,l.dist=2}{vloc(__v1)}
\fmfiv{label=$\scriptstyle j b$,label.angle=-120,l.dist=2}{vloc(__v2)}
]{plain_rar}{dashes_ar,left=1}{dashes_rar,left=1}
\Bigr)_{\text{am}}
&=
-4\lambda   p^2 \hat{I}_{(1,1)}(p)\delta_i^j
\Big[\big(ab\big)-\cos\gamma_i^+\frac 1N\big(a\big)\big(b\big)\Big]
\eqncom\\
\Bigl(\swfone[
\fmfiv{label=$\scriptstyle i a$,label.angle=-60,l.dist=2}{vloc(__v1)}
\fmfiv{label=$\scriptstyle j b$,label.angle=-120,l.dist=2}{vloc(__v2)}
]{plain_rar}{photon,left=1}{plain_ar}
\Bigr)_{\text{am}}
&=2 \lambda  p^2\bigl(2\hat{I}_{(1,1)}(p)- p^2 \hat{I}_{(2,1)}(p)(1-\xi)\bigr)\delta_i^j
\Big[\big(ab\big)-\frac 1N\big(a\big)\big(b\big)\Big]
\eqncom
\end{aligned}
\end{equation}
where the external momentum is $p$ and the integrals are the Minkowski space versions of \eqref{eq:G_function}. Note that these diagrams are derived from the Feynman rules and hence they include all possible insertions of the occurring vertices. In particular, also the version where one of the external legs ends inside the loop. In the theory with gauge group \SUN, the full self-energy contribution to the propagator and the counterterm subtracting the respective UV divergence in the DR-scheme read
\begin{equation}
\begin{aligned}
\Bigl(
\settoheight{\eqoff}{$\times$}%
\setlength{\eqoff}{0.5\eqoff}%
\addtolength{\eqoff}{-5.\unitlength}%
\raisebox{\eqoff}{%
	\fmfframe(0,1)(-2,1){%
		\begin{fmfchar*}(15,7.5)
		\fmfforce{0 w,0.5 h}{v1}
		\fmfforce{1 w,0.5 h}{v2}
		\fmf{phantom}{v1,v2}
		\fmffreeze
		\fmfposition
		\fmfipath{p[]}
		\fmfiset{p1}{vpath(__v1,__v2)}
		\fmfiset{p11}{subpath (0,2length(p1)/5) of p1}
		\fmfiset{p12}{subpath (3length(p1)/5,length(p1)) of p1}
		\fmfi{plain_ar}{p11}
		\fmfi{plain_ar}{p12}
		\fmfiv{label=$\scriptstyle i a$,label.angle=-60,l.dist=2}{vloc(__v1)}
		\fmfiv{label=$\scriptstyle j b$,label.angle=-120,l.dist=2}{vloc(__v2)}
		\fmfiv{decor.shape=circle,decor.filled=shaded,decor.size=11thin}{point length(p1)/2 of p1}
		\end{fmfchar*}
	}
}
\Bigr)_{\text{am}}	
&=
-2\lambda p^2\bigl(2\hat{I}_{(1,1)}(p)+ p^2 \hat{I}_{(2,1)}(p)(1-\xi)\bigr)\delta^j_i(ab)
\eqncom\\
\delta_\phi^{(1)}&=
-\Kop\Bigl[
\Bigl(
\settoheight{\eqoff}{$\times$}%
\setlength{\eqoff}{0.5\eqoff}%
\addtolength{\eqoff}{-5.\unitlength}%
\raisebox{\eqoff}{%
	\fmfframe(0,1)(-2,1){%
		\begin{fmfchar*}(15,7.5)
		\fmfforce{0 w,0.5 h}{v1}
		\fmfforce{1 w,0.5 h}{v2}
		\fmf{phantom}{v1,v2}
		\fmffreeze
		\fmfposition
		\fmfipath{p[]}
		\fmfiset{p1}{vpath(__v1,__v2)}
		\fmfiset{p11}{subpath (0,2length(p1)/5) of p1}
		\fmfiset{p12}{subpath (3length(p1)/5,length(p1)) of p1}
		\fmfi{plain_ar}{p11}
		\fmfi{plain_ar}{p12}
		\fmfiv{label=$\scriptstyle i a$,label.angle=-60,l.dist=2}{vloc(__v1)}
		\fmfiv{label=$\scriptstyle j b$,label.angle=-120,l.dist=2}{vloc(__v2)}
		\fmfiv{decor.shape=circle,decor.filled=shaded,decor.size=11thin}{point length(p1)/2 of p1}
		\end{fmfchar*}
	}
}
\Bigr)_{\text{am}}
\Bigr]_{\complexi p^2 \delta^j_i(ab)}
=
\frac{2\lambda (1+\xi)}{(4\pi)^2}\frac{1}{\epsilon}=2g^2 (1+\xi)\frac{1}{\epsilon}\eqncom
\end{aligned}
\end{equation}
and it is the one-loop contribution to the respective counterterm in \eqref{eq:counterterms}. If we keep the $\frac 1N$ corrections, as is necessary for the contributions to \U1 modes, the divergent part of \eqref{sephi} reads
\begin{equation}
\begin{aligned}\label{eq:divergence_SEphi}
\Kop\Big[\Bigl(
\settoheight{\eqoff}{$\times$}%
\setlength{\eqoff}{0.5\eqoff}%
\addtolength{\eqoff}{-5.\unitlength}%
\raisebox{\eqoff}{%
	\fmfframe(0,1)(-2,1){%
		\begin{fmfchar*}(15,7.5)
		\fmfforce{0 w,0.5 h}{v1}
		\fmfforce{1 w,0.5 h}{v2}
		\fmf{phantom}{v1,v2}
		\fmffreeze
		\fmfposition
		\fmfipath{p[]}
		\fmfiset{p1}{vpath(__v1,__v2)}
		\fmfiset{p11}{subpath (0,2length(p1)/5) of p1}
		\fmfiset{p12}{subpath (3length(p1)/5,length(p1)) of p1}
		\fmfi{plain_ar}{p11}
		\fmfi{plain_ar}{p12}
		\fmfiv{label=$\scriptstyle i a$,label.angle=-60,l.dist=2}{vloc(__v1)}
		\fmfiv{label=$\scriptstyle j b$,label.angle=-120,l.dist=2}{vloc(__v2)}
		\fmfiv{decor.shape=circle,decor.filled=shaded,decor.size=11thin}{point length(p1)/2 of p1}
		\end{fmfchar*}
	}
}
\Bigr)_{\text{am}}	
\Big]
&=
-2\frac{\complexi p^2g^2\delta^j_i}{\epsilon}\Bigl((1+\xi)(ab)
-\Bigl((1+\xi)+4\sin^2\frac{\gamma_i^+}{2}+4\sin^2\frac{\gamma_i^-}{2}\Bigr)\frac{(a)(b)}{N}\Bigr)\eqndot
\end{aligned}
\end{equation}

\section{Coupling tensor identities for the \texorpdfstring{$\gamma_i$}{gamma}-de\-for\-ma\-tion}\label{sec:Coupling_tensor_identities}
For the evaluation of the Feynman diagrams in \secref{sec:non-conformal_double_trace_coupling} and \secref{sec:cake}, several contraction identities for the various interaction tensors in the action \eqref{eq:action_Feynman_rules} are needed. In this appendix, we give a complete list of the used identities, all of which are easily derived by the means of {\tt{Mathematica}}. In this appendix, we restrict the indices to assume the values $i,j,k=1,2,3$ and $A,B,C=1,2,3,4$.

For \secref{sec:non-conformal_double_trace_coupling}, we introduce the transverse Kronecker-$\delta$
\begin{equation}
\tau_A^B=\delta_A^i\delta_i^B=\delta_A^B-\delta_A^4\delta_4^B\eqncom
\end{equation}
and we write the following contractions of Yukawa-type couplings as
\begin{equation}
\begin{aligned}\label{eq:rho_relation1}
\bigl(\rho^{i}\rho^{\dagger}_j\bigr)^A{}_{B}=
(\rho^{i})^{AC}(\rho^{\dagger}_j)_{CB}
&=2\Bigl(\delta^i_j\tau^A_B-\delta^A_j\delta^i_B\e^{\frac{i}{2}(\gamma^+_{i}+\gamma^+_{j})\sum_{k=1}^3\varepsilon_{ijk4}}
\Bigr)\eqncom\\
\bigl(\rho^{i}(\rho^{\dagger}_j)^{\T}\bigr)^A{}_{B}=
(\rho^{i})^{AC}(\rho^{\dagger}_j)_{BC}
&=-2\Bigl(\delta^i_j\tau^A_B\e^{\complexi
	\gamma^+_{i}\sum_{k=1}^3\varepsilon_{Aik4}
	}
	-\delta^A_j\delta^i_B\e^{\frac{i}{2}	(\gamma^+_{i}-\gamma^+_{j})\sum_{k=1}^3\varepsilon_{jik4}}
\Bigr)\eqncom\\
\bigl(\tilde\rho^i\tilde\rho^{\dagger}_j\bigr)_A{}^{B}=
(\tilde\rho^i)_{AC}(\tilde\rho^{\dagger}_j)^{CB}
&=2\Bigl(\delta^4_A\delta^B_4\delta^i_j+\delta_A^i\delta_j^B\e^{-\frac{i}{2}(\gamma_i^--\gamma_j^-)}\Bigr)\eqncom\\
\bigl(\tilde\rho^i(\tilde\rho^{\dagger}_j)^{\T}\bigr)_A{}^{B}=
(\tilde\rho^i)_{AC}(\tilde\rho^{\dagger}_j)^{BC}
&=-2\Bigl(
\delta^4_A\delta^B_4\delta^i_j\e^{\complexi\gamma_i^m}
+\delta_i^A\delta_B^j\e^{-\frac{\complexi}{2}(\gamma_i^-+\gamma_j^-)}
\Bigr)
\eqncom
\end{aligned}
\end{equation}
where the index $C$ is summed over. The analogous relations with switched \su{4} indices are obtained using $\bigl(\rho_j^\dagger\rho^i\bigr)^{\T}=\bigl(\rho^i\rho_j^\dagger\bigr)^\ast$ and $\bigl(\tilde\rho_j^\dagger\tilde\rho^i\bigr)^{\T}=\bigl(\tilde\rho^i\tilde\rho_j^\dagger\bigr)^\ast$, where the operators $\T$ acts in the \su{4} spinor index space as indicated in \eqref{eq:rho_relation1}.
With these results, the traces of two Yukawa coupling tensors that appear in the 
one-loop self-energies evaluate to
\begin{equation}
\begin{aligned}\label{rhorhotr}
\tr\big[\rho^i\rho^{\dagger}_j\bigr]
&=(\rho^{i})^{AC}(\rho^{\dagger}_j)_{CA}
=4\delta_i^j\eqncom\\
\tr\big[\rho^i(\rho^{\dagger}_j)^{\T}\bigr]
&=(\rho^{i})^{AC}(\rho^{\dagger}_j)_{AC}
=-4\delta_i^j\cos\gamma_i^+
\eqncom\\
\tr\big[\tilde\rho^i\tilde\rho^{\dagger}_j\bigr]
&=(\tilde\rho^i)_{AC}(\tilde\rho^{\dagger}_j)^{CA}
=4\delta_i^j\eqncom\\
\tr\big[\tilde\rho^i(\tilde\rho^{\dagger}_j)^{\T}\bigr]
&=(\tilde\rho^i)_{AC}(\tilde\rho^{\dagger}_j)^{AC}
=-4\delta_i^j\cos\gamma_i^-
\eqndot
\end{aligned}
\end{equation}
In addition, we need traces of four Yukawa-type vertices for the evaluation of the fermion-box contributions to the double-trace couplings \eqref{eq:running_dt_coupling}:
\begin{equation}
\begin{aligned}\label{l4rhotraces}
\tr\big[\rho^i(\rho^{\dagger}_i)^{\T}(\tilde\rho^{\dagger}_i)^{\T}\tilde\rho^i\big]
&=0
\eqncom\\
\tr\big[\tilde\rho^i(\tilde\rho^{\dagger}_i)^{\T}(\rho^{\dagger}_i)^{\T}\rho^i\big]
&=0
\eqncom\\
\tr\big[\rho^i(\rho^{\dagger}_i)^{\T}\rho^i(\rho^{\dagger}_i)^{\T}\big]
&=8\cos2\gamma_i^+
\eqncom\\
\tr\big[\tilde\rho^i(\tilde\rho^{\dagger}_i)^{\T}\tilde\rho^i(\tilde\rho^{\dagger}_i)^{\T}\big]
&=8\cos2\gamma_i^-
\eqncom
\end{aligned}
\end{equation}
where the index $i\in\{1,2,3\}$ is fixed. For the one-loop interaction with four scalars with identical field flavours by the means of two F-term-type interactions, we need the following contractions of coupling tensors
\begin{equation}
\begin{aligned}\label{eq:F_tensor_id}
\sum_{r=1}^3F^{ir}_{ri}F^{ir}_{ri}
=4\bigl(\e^{2\complexi \gamma_{i+1}}+\e^{-2\complexi \gamma_{i+2}}\bigr)
\eqncom\qquad
\sum_{r=1}^3F^{ri}_{ir}F^{ri}_{ir}
=4\bigl(\e^{-2\complexi \gamma_{i+1}}+\e^{2\complexi \gamma_{i+2}}\bigr)
\eqncom
\end{aligned}
\end{equation}
where $i\in\{1,2,3\}$ is fixed and the cyclic identification $i+3\sim i$ is understood. Note that the second identity is the complex conjugate of the first one, consistent with the conjugation rules \eqref{eq:coupling_tensor_conjugation}. Combining both results in \eqref{eq:F_tensor_id} yields
\begin{equation}
\begin{aligned}\label{FFsum}
\sum_{r=1}^3(F^{ir}_{ri}F^{ir}_{ri}+F^{ri}_{ir}F^{ri}_{ir})
&=8(\cos 2\gamma_{i+1}+\cos 2\gamma_{i+2})
=16\cos 2\gamma_{i}^+\cos 2\gamma_{i}^-
\eqndot
\end{aligned}
\end{equation}

In \secref{sec:non-conformal_double_trace_coupling}, we work in slightly different conventions for the scalar interactions. For the Feynman integrals in that section we combine the quartic scalar single trace terms in \eqref{eq:Lagrangian_Feynman_rules_3} and sort the result with respect to occurring trace structures as 
\begin{equation}\label{eq:F_to_Qhat} F^{ij}_{lk}\tr\bigl(\phi_i\phi_j\ol{\phi}^k\ol{\phi}^l\bigr)-\frac{1}{2}\tr\bigl([\phi_i,\ol{\phi}^i][\phi_j,\ol{\phi}^j]\bigr)=\hat{Q}^{ij}_{lk}\tr\bigl(\phi_i\phi_j\ol{\phi}^k\ol{\phi}^l\bigr)
+\tilde{Q}^{ij}_{lk}\tr\bigl(\ol{\phi}^k\phi_i\ol{\phi}^l\phi_j\bigr)\eqncom
\end{equation} 
with the two coupling tensors
\begin{equation}\label{eq:Qhat}
\hat{Q}^{ij}_{lk}=(F^{ij}_{lk}+\delta^i_l\delta^j_k)=2\delta^i_k\delta^j_l\e^{\complexi\mathbf{q}_{\phi_i}\wedge\mathbf{q}_{\phi_j}}
-\delta^j_k\delta^i_l\eqncom\qquad
\tilde{Q}^{ij}_{lk}=-\frac 12(\delta^i_k\delta^j_l+\delta^i_l\delta^j_k)\eqndot
\end{equation}
For the evaluation of the purely scalar integrals in \eqref{wrapLdiags} we need the following identities involving products of $L$ coupling tensors
\begin{equation}
\begin{aligned}\label{eq:Qidentity}
\sum_{j=1}^3 (\hat Q_{ij}^{ji})^L
&=2^{L+1}\e^{-iL\gamma_i^-}\cos{L\gamma_i^+}+1
\eqncom\qquad
\sum_{j=1}^3 (\hat Q_{ji}^{ij})^L
=2^{L+1}\e^{iL\gamma_i^-}\cos{L\gamma_i^+}+1
\eqncom
\end{aligned}
\end{equation}
where the index $i\in\{1,2,3\}$ is fixed. For the remaining diagrams, we first note that the following products can be written as
\begin{equation}
\begin{aligned}
\Bigl(\rho^\dagger_i(\rho^i)^{\T}\Bigr)^A_{\phan{A}B}&=(\rho^\dagger_i)^{AC}(\rho^i)_{BC}=
-2\delta^A_B\sum_{j=1}^3 (\varepsilon^{Aij4})^2 \,{\e}^{i\varepsilon^{Aij4}\gamma_i^+} \eqncom \\
\Bigl(\tilde\rho^\dagger_i(\tilde\rho^i)^{\T}\Bigr)_A^{\phan{A}B}&=(\tilde\rho^\dagger_i)_{AC}(\tilde\rho^i)^{BC}
=-2\Bigl(\delta^A_4\delta^4_B\,{\e}^{i\gamma_i^-}+\delta^A_i\delta^i_B\,{\e}^{-i\gamma_i^-}\Bigr) \eqndot
\end{aligned}
\end{equation}
We can now multiply $L$ packages of $\rho$ or $\tilde{\rho}$ tensors by simply contracting the open spinor indices. When we trace over all spinor indices in such a product we find the two identities needed to evaluate the diagrams in \eqref{wrapLdiags} that involve fermions:
\begin{equation}\label{eq:rhoidentity}
\begin{aligned}
\tr\bigl[\bigl(\rho^\dagger_i(\rho^i)^{\T}\bigr)^L\bigr] 
&= \sum_{A,j} \left(-2 (\varepsilon^{Aij4})^2 \e^{i\varepsilon^{Aij4}\gamma_i^+} \right)^L 
= -(-2)^{L+1} \cos L\gamma_i^+ \eqncom \\
\tr\bigl[\bigl(\tilde\rho^{\dagger}_i(\tilde\rho^i)^{\T}\bigr)^L\bigr] 
&= \sum_{A=1}^4\Bigl(-2\delta^A_4\,{\e}^{i\gamma_i^-}-2\delta^A_i\,{\e}^{-i\gamma_i^-}\Bigr)^L
=-(-2)^{L+1} \cos L\gamma_i^- \eqncom
\end{aligned}
\end{equation}
where $i\in\{1,2,3\}$ is again fixed.

\section{Calculation of \texorpdfstring{$\ev{P\mathfrak{D}^{L\geq3}_2(w,y)}$}{<PD2(L>=3)(w,y)>}}\label{app: PD2 calculation}
In \secref{sec:ingredients}, we briefly discussed how $\ev{P\mathfrak{D}^{L\geq3}_2(w,y)}$ is computed. In this appendix, we present the meat and bones of this computations starting from \eqref{eq: first eqation of PD2} in the three steps we also discussed in the main part. This appendix closely follows \cite{Fokken:2014moa}.

First, we use employ that the harmonic action $c_{\text{h}}$ does not depend on the kind of oscillator which hops from one site to the other. We rewrite the bosonic summations in terms of the variables
\begin{equation}
\begin{aligned}
\akindsite{i}=\sum_{\alpha=1}^2\akindsite[\alpha]{i}\eqncom\quad
a=\sum_{\alpha=1}^2\akind{}^{\alpha}\eqncom\quad
\bkindsite{i}&=\sum_{\dot\alpha=\dot1}^{\dot2}\bkindsite[\dot\alpha]{i}\eqncom\quad
b=\sum_{\dot\alpha=\dot1}^{\dot2}\bkind{}^{\dot\alpha}\eqncom
\end{aligned}
\end{equation}
with $i=1,2$. For a generic function $f$, we use the summation identity
\begin{equation}\label{eq:sum_order}
\sum_{a_{\bullet}^1,a_{\bullet}^2 = 0}^\infty f(a_{\bullet}^1,a_{\bullet}^2) = \sum_{a_{\bullet}= 0}^\infty \sum_{\tilde a_{\bullet}=0}^{a_{\bullet}} f(\tilde a_{\bullet},a_{\bullet}-\tilde a_{\bullet})\eqncom\qquad
(a_{\bullet}^1,a_{\bullet}^2)\in\bigl\{(\akindsite[1]{1},\akindsite[2]{1}),
(\akindsite[1]{2},\akindsite[2]{2}),(\akind[1],\akind[2])\bigr\}\eqncom
\end{equation}
to express the occurrences of all $\akind{}_\bullet^\alpha$ in terms of $a_\bullet$ and $\tilde a_\bullet$. In the resulting expressions, we perform the sums over $\tilde{\akind}_{(i)}$ via
\begin{equation}\label{eq:sum_binomial}
\sum_{\tilde{\akind{}}_{(i)}=0}^{\akindsite{i}} \binom{\tilde{\akind{}}_{(i)}}{\tilde{\akind{}}} \binom{\akindsite{i}-\tilde{\akind{}}_{(i)}}{\akind{}-\tilde{\akind{}}}= \binom{\akindsite{i}+1}{\akind{}+1}\eqndot
\end{equation}
The remaining sum over $\tilde{\akind{}}$ is then independent of the summation variable, yielding a factor of $(\akind{}+1)$. In this manner we can directly perform three of the six sums over $\aosc$-type oscillators. For $\bosc$-type oscillators we follow the analogous procedure and eliminate another three sums, leading to a total of six remaining infinite sums. For the fermionic oscillators, note that the coefficient $c_{\text{h}}$ is independent of the kind of $\cosc$-type oscillator that changes its site, but the phase factor introducing the deformation parameters is not. We simply absorb the dependence on these $\cosc$-type oscillator hoppings into
\begin{equation} 
\begin{aligned}\label{eq: def G(c1,c2,c)}
G^{\gamma^\pm_i}(\ckindsite1,\ckindsite2,\ckind)&= \prod_{e=1}^4\bigg(\sum_{\ckindsite[e]{1},\ckindsite[e]{2},\ckind[e]=0}^1 \binom{\ckindsite[e]{1}}{{\ckind[e]}}\binom{\ckindsite[e]{2}}{\ckind[e]}(-1)^{\ckind[e]}\bigg) 
\e^{-\complexi \sum_{l,m=1}^4 \ckindsite[l]{1}\ckindsite[m]{2}\bq_{\lambda_{l}}\times\bq_{\lambda_{m}}} \\
&\qquad \times \delta_{\left(\ckindsite{1}-\sum_{e=1}^4\ckindsite[e]{1}\right)}
\delta_{\left(\ckindsite{2}-\sum_{e=1}^4\ckindsite[e]{2}\right)}
\delta_{\left(\ckind-\sum_{e=1}^4\ckind[e]\right)}\eqndot
\end{aligned}
\end{equation}
With the above reformulations, \eqref{eq: first eqation of PD2} is cast into
\begin{equation}\label{eq: PD2 pre dual way}
\begin{aligned}
\ev{P\mathfrak{D}^{L\geq3}_2(w,y)}&= \sum_{\akindsite{1},\akindsite{2},\akind=0}^\infty \sum_{\bkindsite{1},\bkindsite{2},\bkind=0}^\infty \sum_{\ckindsite{1},\ckindsite{2},\ckind=0}^4
\delta_{(\akindsite1-\bkindsite1+\ckindsite1-2)}\delta_{(\akindsite2-\bkindsite2+\ckindsite2-2)}\\
&\phaneqtimes w^{\frac12\left(2+\akindsite{1}+\bkindsite{1}\right)} 
y^{\frac12\left(2+\akindsite{2}+\bkindsite{2}\right)}\, G^{\gamma^\pm_i}(\ckindsite1, \ckindsite2, \ckind) \\
&\phaneqtimes (\akind+1)\binom{\akindsite{1}+1}{\akind+1}\binom{\akindsite{2}+1}{\akind+1} (\bkind+1)\binom{\bkindsite{1}+1}{\bkind+1}\binom{\bkindsite{2}+1}{\bkind+1} \\
&\phaneqtimes
c_{\mathrm{h}}\Bigl[\textstyle \sum_{i=1}^2(a_{(i)}+b_{(i)}+c_{(i)}),\\ 
&\phaneq\qquad\quad\, (a_{(1)}-a)+(b_{(1)}-b)+(c_{(1)}-c),\\
&\phaneq\qquad\quad\, (a_{(2)}-a)+(b_{(2)}-b)+(c_{(2)}-c)\Bigr]\eqndot
\end{aligned}
\end{equation}
At this point, we use the Kronecker-$\delta$'s from the central charge constraint to eliminate two of the remaining sums six, say those over $b_{(1)}$ and $b_{(2)}$.

Second, we use the integral representation \eqref{def:Harmonic_action_integral} of the harmonic action to replace \eqref{eq: PD2 pre dual way} by the respective integrand defined in \eqref{eq:integral_representation_PD2}:
\begin{equation}\label{eq: PD2 after delta summation}
\begin{aligned}
\ev{ P \mathfrak{D}_2(w,y) }_{\text{int}}&= -2 \sum_{\ckindsite{1},\ckindsite{2},\ckind=0}^4 \sum_{\akind,\bkind=0}^\infty  \sum_{\akindsite{1}=\maxset{0,2-\ckindsite{1}}}^\infty \sum_{\akindsite{2}=\maxset{0,2-\ckindsite{2}}}^\infty
G^{\gamma^\pm_i}(\ckindsite{1}, \ckindsite{2}, \ckind)
 \\
&\phan{=}  (\akind+1)\binom{\akindsite{1}+1}{\akind+1} \binom{\akindsite{2}+1}{\akind+1}
(\bkind+1)\binom{\akindsite{1}+\ckindsite{1}-1}{\bkind+1} \binom{\akindsite{2}+\ckindsite{2}-1}{\bkind+1} \\
&\phan{=} \times 
w^{\akindsite{1} + \frac12\ckindsite{1}} y^{\akindsite{2} + \frac12\ckindsite{2}} 
 t^{\akindsite{1}+\ckindsite{1}+\akindsite{2}+\ckindsite{2}-3} \left(\frac{t-1}{t}\right)^{\akind+\bkind+\ckind}\eqndot
\end{aligned}
\end{equation}

In the third step, we rewrite the combinatorial coefficients in terms of the differential and integral operators\footnote{In a slight abuse of notation, we have labelled the integration variable with the same symbol as the upper integration boundary.}
\begin{equation}
\begin{aligned}\label{eq:div_int_operators}
\hat{I}_{(c)}(x)&=
 x^{-\frac12c} \IntOp{x}{0}{x} x^{\frac12c-1}  \eqncom\qquad
\hat{D}_{(a,c)}(x)&=
 x^{-c} x^{a}\Diff{x}{a} x^{c} \eqncom
\end{aligned}
\end{equation}
where $x$ is a thermal weight as introduced below \eqref{eq: partition function intro D}. Using these operators, we can rewrite \eqref{eq: PD2 after delta summation} into
\begin{equation}
\begin{aligned}
\ev{ P \mathfrak{D}_2(w,y) }_{\text{int}}&= -2 \sum_{\ckindsite{1},\ckindsite{2},\ckind=0}^4  \sum_{\akind,\bkind=0}^\infty  \sum_{\akindsite{1}=\maxset{0,2-\ckindsite{1}}}^\infty \sum_{\akindsite{2}=\maxset{0,2-\ckindsite{2}}}^\infty \\
&\qquad \times  G^{\gamma^\pm_i}(\ckindsite{1}, \ckindsite{2}, \ckind) \frac{1}{\akind!(\akind+1)!}\frac{1}{\bkind!(\bkind+1)!} \\
&\qquad \times 
\Bigl(
 \hat{I}_{(\ckindsite{1})}(w)\hat{D}_{(\bkind+2,\frac12\ckindsite{1})}(w)
\hat{D}_{(\akind+1,1-\frac12\ckindsite{1})}(w)\,
 w^{\akindsite{1} + \frac12\ckindsite{1}} 
 \Bigr)
 \\
&\qquad \times 
\Bigl(
\hat{I}_{(\ckindsite{2})}(y)\hat{D}_{(\bkind+2,\frac12\ckindsite{2})}(y)
\hat{D}_{(\akind+1,1-\frac12\ckindsite{2})}(y)\,
  y^{\akindsite{2} +\frac12\ckindsite{2}}
  \Bigr) \\
&\qquad \times  t^{\akindsite{1}+\ckindsite{1}+\akindsite{2}+\ckindsite{2}-3} \left(\frac{t-1}{t}\right)^{\akind+\bkind+\ckind}\eqncom
\end{aligned}
\end{equation}
where all differential and integral operators act on everything to their right. Since these operators do not explicitly depend on $\akindsite{1}$ and $\akindsite{2}$, we can now perform the two corresponding sums and find
\begin{equation}\label{eq:PD2_diff_Operators}
\begin{aligned} 
\ev{ P \mathfrak{D}_2(w,y) }_{\text{int}}&= -2 \sum_{\ckindsite{1},\ckindsite{2},\ckind=0}^4  \sum_{\akind,\bkind=0}^\infty    G^{\gamma^\pm_i}(\ckindsite{1}, \ckindsite{2}, \ckind) \frac{ \left(\frac{t-1}{t}\right)^{\akind+\bkind+\ckind} }{\akind!(\akind+1)!}\frac{t^{\ckindsite{1}+\ckindsite{2}-3}}{\bkind!(\bkind+1)!} 
\\
&\qquad \times
\left(
\hat{I}_{(\ckindsite{1})}(w)\hat{D}_{(\bkind+2,\frac12\ckindsite{1})}(w)
\hat{D}_{(\akind+1,1-\frac12\ckindsite{1})}(w)\,
\frac{w^{\frac12\ckindsite{1}}(w t)^{\maxset{0,2-\ckindsite{1}}}}{1-w t}
\right)
\\&\qquad \times 
\left(
\hat{I}_{(\ckindsite{2})}(y)\hat{D}_{(\bkind+2,\frac12\ckindsite{2})}(y)
\hat{D}_{(\akind+1,1-\frac12\ckindsite{2})}(y)\,
\frac{y^{\frac12\ckindsite{2}}(y t)^{\maxset{0,2-\ckindsite{2}}}}{1-y t}
\right) 
  \eqndot
\end{aligned}
\end{equation}
Note that we can evaluate the action of the rightmost derivative operator on the remaining terms in the second and third line explicitly. We find
\begin{equation}\label{eq:Diff_Operator_action}
\begin{aligned}
O(y,\akind,\ckindsite{2})&=
\hat{D}_{(\akind+1,1-\frac12\ckindsite{2})}(y)\,
\frac{y^{\frac12\ckindsite{2}}(y t)^{\maxset{0,2-\ckindsite{2}}}}{1-y t}\\
&=
y^{-\frac 12\ckindsite{2}}\Bigl(\frac{(\akind+1)!(t y)^{\akind}}{(1- t y)^{\akind+2}}y^{\ckindsite{2}}
-\delta_{\ckindsite{2}}
\left(\delta_{\akind} +2 t y\delta_{\akind}+2 t y\delta_{(\akind-1)} \right)
-\delta_{(\ckindsite{2}-1)}\delta_{\akind}y 
\Bigr)\eqncom
\end{aligned}
\end{equation}
and the analogous result for the other combination $O(w,\akind,\ckindsite{1})$. This allows us to perform the sum over $\akind$, using the following identity\footnote{We obtained this identity by the means of \tt{Mathematica}.}
\begin{equation} 
\begin{aligned}\label{eq:Operator_Sum_identity}
\sum_{\akind=0}^\infty \frac{\left(\frac{t-1}{t}\right)^{\akind}}{\akind!(\akind+1)!} & 
O(w,\akind,\ckindsite{1})O(y,\akind,\ckindsite{2})
=w^{-\frac 12\ckindsite{1}} y^{-\frac 12\ckindsite{2}}\Biggl(\frac{w^{\ckindsite{1}} y^{\ckindsite{2}}}{(1-t(w+y-wy))^2}\\
&\quad-\frac{y^{\ckindsite{2}}}{(1-t y)^3}
\left[
\delta_{\ckindsite{1}}\left(1+t(2w-y-2wy)\right)
-\delta_{(\ckindsite{1}-1)}w(1-ty)
\right]\\
&\quad-\frac{w^{\ckindsite{1}}}{(1-tw)^3}
\left[
\delta_{\ckindsite{2}}\left(1+t(2y-w-2wy)\right)
-\delta_{(\ckindsite{2}-1)}y(1-tw)
\right]\\
&\quad
+wy\delta_{(c_{(1)}-1)}\delta_{(c_{(2)}-1)}
+(1+2t(w+y-wy+twy))
\delta_{c_{(1)}}\delta_{c_{(2)}}\\
&\quad
+y(1+2tw)\delta_{c_{(1)}}\delta_{(c_{(2)}-1)}
+w(1+2ty)\delta_{c_{(2)}}\delta_{(c_{(1)}-1)}
\Biggr)
\eqndot
\end{aligned}
\end{equation}
When we act with the remaining two $\bkind$-dependent differential operators of \eqref{eq:PD2_diff_Operators} on this sum, the last four lines of \eqref{eq:Operator_Sum_identity} drop out\footnote{The common prefactor in $w$ and $y$ cancels a corresponding factor in the differential operators.}, as they are at most linear in either $w$ or $y$. With this, \eqref{eq:PD2_diff_Operators} turns into
\begin{equation}\label{eq: PD2 int pro op 2}
\begin{aligned}
\ev{ P \mathfrak{D}_2(w,y) }_{\text{int}}&= -2 \sum_{\ckindsite{1},\ckindsite{2},\ckind=0}^4  \sum_{\bkind=0}^\infty   G^{\gamma^\pm_i}(\ckindsite{1}, \ckindsite{2}, \ckind) \frac{t^{\ckindsite{1}+\ckindsite{2}-3}}{\bkind!(\bkind+1)!}
\hat{I}_{(\ckindsite{1})}(w)\hat{I}_{(\ckindsite{2})}(y) \\
&\qquad \times 
\left(\frac{t-1}{t}\right)^{\bkind+\ckind}
\hat{D}_{(\bkind+2,\frac12\ckindsite{1})}(w)
\hat{D}_{(\bkind+2,\frac12\ckindsite{2})}(y)
\frac{w^{\frac 12\ckindsite{1}} y^{\frac 12\ckindsite{2}}}{(1-t (w+y-wy))^2} \eqncom
\end{aligned}
\end{equation}
Writing the last factor in \eqref{eq: PD2 int pro op 2} as a power series in the variables $\hat w=w-1$ and $\hat y=y-1$ as
\begin{equation}
\begin{aligned}\label{eq: B40}
\frac{w^{\ckindsite{1}} y^{\ckindsite{2}}}{(1-t (w+y-wy))^2}=
\sum_{\alpha,\beta=0}^4 \binom{\ckindsite{1}}{\alpha}\binom{\ckindsite{2}}{\beta} 
\sum_{n=0}^\infty\frac{n+1}{(1-t)^2}
\left(\frac{t}{t-1}\right)^n 
{\hat w}^{n+\alpha} \hat{y}^{n+\beta}\eqncom
\end{aligned}
\end{equation}
we can apply the remaining derivative operators. We can now combine all $b$-dependent terms and substitute $l=b+2$, to finally obtain
\begin{equation}\label{eq: PD2 final}
\begin{aligned}
\ev{ P D_2(w,y) }_{\text{int}} 
&= -2 \sum_{\ckindsite{1},\ckindsite{2},\ckind=0}^4  \sum_{\alpha,\beta=0}^4  \binom{\ckindsite{1}}{\alpha}\binom{\ckindsite{2}}{\beta} G^{\gamma^\pm_i}(\ckindsite{1}, \ckindsite{2}, \ckind)  \\
&\qquad \times  \left( w^{-\frac12\ckindsite{1}} \IntOp{w}{0}{w} w^{-1} \right)\left( y^{-\frac12\ckindsite{2}} \IntOp{y}{0}{y} y^{-1} \right) \\
& \qquad \times  t^{\ckindsite{1}+\ckindsite{2}-\ckind-1} (t-1)^{\ckind-4}  (w-1)^{\alpha}(y-1)^{\beta} \\
& \qquad \times \xi_{\alpha\beta}\left(\frac{t}{t-1}(w-1) (y-1),\frac{t-1}{t}\frac{w y}{(w-1) (y-1)}\right)\eqncom
\end{aligned}
\end{equation}
where we used the fundamental definitions \eqref{eq:div_int_operators} for the remaining two integral operators. The function $\xi_{\alpha\beta}(X,Y)$ is defined as
\begin{equation}
\xi_{\alpha \beta}(X,Y)=\sum_{l=0}^\infty  \sum_{n=0}^\infty (l-1)l^2(n+1) \binom{n+\alpha}{l} \binom{n+\beta}{l} Y^n X^l
\end{equation}
and its evaluation is discussed in \appref{app: summation identities}. 

By the means of \appref{app: summation identities}, all infinite sums have vanished from the final expression \eqref{eq: PD2 final}. The remaining finite sums and integrals can be evaluated with the help of {\tt Mathematica}. Upon combining everything in \eqref{eq:integral_representation_PD2_2}, we find the result \eqref{eq:Result_PD2}--\eqref{eq:result_functions_last}.

\section{Calculation of \texorpdfstring{$Z_{\text{f.s.c.}}^{(1)}(x)$}{Z(f.s.c.)(x)}}\label{app: corrections}
In this appendix, which is based on \cite{Fokken:2014moa}, we compute the $L=2$ finite-size corrections to the single-trace partition function of the $\beta$-deformation with gauge group \SUN. They occur in \eqref{eq:fsc}, when the correct dilatation-operator density \eqref{eq:density_intro} for $L=2$ states is used instead of the asymptotic one from \eqref{eq: deformation of D_2}. The calculation of $\ev{P\mathfrak{D}^{L=2}_2(w,y)}$ is only different from the calculation in \appref{app: PD2 calculation} for the asymptotic density in the sums over fermionic oscillator occupation numbers. Therefore, we restrict this appendix to give the appropriate definitions of the finite-size corrected fermionic occupation number function $G^\beta_{L=2}(\ckindsite1,\ckindsite2,\ckind)$, which replaces the asymptotic version $G^{\gamma^\pm_i}(\ckindsite1,\ckindsite2,\ckind)$ in the derivation of \appref{app: PD2 calculation}.

In \secref{sec:beta_paper} we found that the finite-size-corrected dilatation-operator density $\mathfrak{D}^{L=2}_2$ from \eqref{eq:density_intro} can be obtained from the asymptotic version $\mathfrak{D}^{L\geq3}$ from \eqref{eq: deformation of D_2}. This is done by setting the deformation parameter $\beta$ in $\mathfrak{D}^{L\geq3}$ to zero whenever the fields $A_{i}$ at sites $i=1$ and $i=2$ are either taken from the matter subalphabet $\mathcal{A}_{\text{matter}}$ or from the anti-matter subalphabet $\ol{\mathcal{A}}_{\text{matter}}$, which were given in \eqref{eq:subalphabet_intro}. In the oscillator picture these restrictions can be turned into the following constraints 
\begin{equation}
\begin{aligned}
A_i\in \mathcal{A}_{\mathrm{matter}}\Leftrightarrow \sum_{e=1}^3c_{(i)}^e=1\eqncom
\qquad\qquad
A_i\in \bar{\mathcal{A}}_{\mathrm{matter}}	\Leftrightarrow \sum_{e=1}^3c_{(i)}^e=2\eqndot
\end{aligned}
\end{equation}
Including these constraints into the fermionic occupation number function $G^{\gamma^\pm_i}(\ckindsite1,\ckindsite2,\ckind)$ of \eqref{eq: def G(c1,c2,c)}, we find
\begin{equation} 
\begin{aligned}
G_{L=2}^{\beta}(\ckindsite1,\ckindsite2,\ckind)&= \prod_{e=1}^4\bigg(\sum_{\ckindsite[e]{1},\ckindsite[e]{2},\ckind[e]=0}^1 \binom{\ckindsite[e]{1}}{{\ckind[e]}}\binom{\ckindsite[e]{2}}{\ckind[e]}(-1)^{\ckind[e]}\bigg) \\
&\qquad \times \e^{-\complexi \sum_{l,m=1}^4 \ckindsite[l]{1}\ckindsite[m]{2}\bq_{\lambda_{l}}\times\bq_{\lambda_{m}}}\Big|_{
	\substack{
		\beta=0 \text{ if } \sum_{e=1}^3c_{(1)}^e=\sum_{e=1}^3c_{(2)}^e=1 \\
		\text{or if } \sum_{e=1}^3c_{(1)}^e=\sum_{e=1}^3c_{(2)}^e=2}
} \\
&\qquad \times \delta_{\left(\ckindsite{1}-\sum_{e=1}^4\ckindsite[e]{1}\right)}
\delta_{\left(\ckindsite{2}-\sum_{e=1}^4\ckindsite[e]{2}\right)}
\delta_{\left(\ckind-\sum_{e=1}^4\ckind[e]\right)}\eqndot
\end{aligned}
\end{equation}
To obtain the finite-size corrected generalised expectation value $\ev{P\mathfrak{D}^{L=2}_2(w,y)}$, we insert $\G_{L=2}^{\beta}(\ckindsite1,\ckindsite2,\ckind)$ into \eqref{eq: PD2 pre dual way} and follow the remaining derivation of \appref{app: PD2 calculation}. Finally, we obtain finite-size correction \eqref{eq: Z_corr in full theory} for the $\beta$-deformation with gauge group \SUN by combining this expression with the asymptotic version $\ev{P\mathfrak{D}^{L\geq3}_2(w,y)}$.

\section{Summation identities}\label{app: summation identities}
This appendix is based on \cite{Fokken:2014moa}. We derive the summation identities needed in \appref{app: PD2 calculation} for
\begin{equation}\label{eq:xi_alpha_beta}
\xi_{\alpha \beta}(X,Y)=\sum_{j=0}^\infty  \sum_{i=0}^\infty (i-1)i^2(j+1) \binom{j+\alpha}{i} \binom{j+\beta}{i} Y^i X^j \eqncom
\end{equation}
with $\alpha, \beta \in \{0,1,2,3,4\}$. This can be achieved by applying a finite number of derivative and integral operators and using the following identities:
\begin{equation}\label{eq:P_identity}
\sum_{i=0}^j \binom{j}{i}^2 t^i=(1-t)^j P_j\left(\frac{1+t}{1-t}\right)\eqncom\qquad
\sum_{n=0}^\infty P_n(x) z^n =  (1 - 2xz + z^2)^{-1/2} \eqncom
\end{equation}
where $P_n(x)$ denotes the $n^{\text{th}}$ Legendre polynomial. Since $\xi_{\alpha \beta}$ is symmetric in $\alpha$ and $\beta$, we assume that $\alpha\geq\beta$. In this case, the summand in \eqref{eq:xi_alpha_beta} can be written as
\begin{equation} 
\begin{aligned}\label{eq: xi in operators}
\text{summand}&= (i-1)i^2(j+1)\binom{j+\alpha}{i} \binom{j+\beta}{i} Y^i X^j \\
&= i^2(i-1)(j+1)  \prod_{\gamma=\beta+1}^{\alpha}\left(1-\frac{i}{j+\gamma}\right) \binom{j+\alpha}{i}^2 Y^i X^j \\
&= \hat{D}_{(2,0)}(Y)\hat{D}_{(1,0)}(Y)\hat{D}_{(1,0)}(X) \prod_{\gamma=\beta+1}^{\alpha}\left[1-\hat{D}_{(1,0)}(Y)\hat{I}_{(2\gamma)}(X)\right] \binom{j+\alpha}{i}^2 Y^i X^j \eqncom
\end{aligned}
\end{equation}
where we employed the differential and integral operators of \eqref{eq:div_int_operators}. In contrast to the original summand, these operators are independent of the summation variables $i$ and $j$ and hence we can evaluate the two infinite sums in \eqref{eq:xi_alpha_beta} using the identities \eqref{eq:P_identity}. We find
\begin{equation} 
\begin{aligned}\label{eq: xi Legendre}
&\sum_{i,j=0}^{\infty} \binom{j+\alpha}{i}^2 Y^i X^j 
= \frac{1}{X^\alpha}\sum_{j=0}^{\infty}  (X(1-Y))^{j+\alpha}P_{j+\alpha}\left(\frac{1+Y}{1-Y}\right) \\
&= \frac{1}{X^\alpha} \left[ \frac{1}{\left(1-2X(1+Y)+X^2(1-Y)^2\right)^{1/2}}-\sum_{k=0}^{\alpha-1}  (X(1-Y))^k P_k\left(\frac{1+Y}{1-Y}\right) \right] \eqndot
\end{aligned}
\end{equation}
Thus, combining \eqref{eq: xi in operators} with \eqref{eq: xi Legendre} allows to express $\xi_{\alpha \beta}(X,Y)$ in a form explicitly solvable with {\tt{Mathematica}}:
\begin{equation} 
\begin{aligned}
\xi_{\alpha \beta}(X,Y)&=
\hat{D}_{(2,0)}(Y)\hat{D}_{(1,0)}(Y)\hat{D}_{(1,0)}(X) 
\prod_{\gamma=\beta+1}^{\alpha}\left[1-\hat{D}_{(1,0)}(Y)\hat{I}_{(2\gamma)}(X)\right]
\frac{1}{X^\alpha} \Biggl[ 
\\
&\phaneq  \frac{1}{\left(1-2X(1+Y)+X^2(1-Y)^2\right)^{1/2}}-\sum_{k=0}^{\alpha-1}  (X(1-Y))^k P_k\left(\frac{1+Y}{1-Y}\right) \Biggr] \eqndot \\
\end{aligned}
\end{equation}
For example for $(\alpha,\beta)=(2,1)$, we obtain
\begin{equation} 
\begin{aligned}
\xi_{2 1}\left(X,Y\right)=& \frac{24 X Y^2}{\left(1-2 X (1+Y)+X^2 (1-Y)^2\right)^{9/2}}\\
&{}\Bigl[{}1-X^2(5-12Y+5Y^2)
+X^3(5-6Y-6Y^2+5Y^3)\\
&-9 X^4(1-Y)^2Y
-X^5(1-Y)^4(1+Y)\Bigr]\eqndot
\end{aligned}
\end{equation}
The remaining expressions follow analogously and we do not show them explicitly.\footnote{`So long and thanks for all the fish' \cite{Hitchhiker}.}

\phantomsection
\backmatter



{
\small
\renewcommand\bibname{References}
\bibliographystyle{JHEPJAN} 
\bibliography{ThesisINSPIRE,Thesis_gamma_i}{\protect\thispagestyle{back}} %
\addcontentsline{toc}{chapter}{References}
}


\end{fmffile}
\end{document}